\documentclass[12pt]{ociamthesis}

\usepackage[utf8]{inputenc}
\usepackage[T1]{fontenc} %

\usepackage{libertine}

\usepackage{microtype}
\usepackage{amsfonts}
\usepackage{amssymb}
\usepackage{amsthm}
\usepackage{amsmath}
\allowdisplaybreaks[1]

\usepackage{mathtools} %

\usepackage{mathabx}
\usepackage{stmaryrd}
\usepackage{extarrows}

\usepackage{physics}

\usepackage[inline]{enumitem}

\usepackage[dvipsnames]{xcolor}

\usepackage{suffix}

\usepackage{verbatim} %

\usepackage{graphicx}

\usepackage{bussproofs}

\usepackage{tikz}
\usepackage{tikzit}
\usetikzlibrary{cd, babel}
\usepackage{quiver}

\newcommand*{\secref}[1]{\S\ref{#1}}

\usepackage[backend=biber,style=numeric-comp,sorting=nyt,date=short]{biblatex}

\usepackage{hyperref}

\bibliography{bibliography}

\newcommand*{\citep}[1]{\parencite{#1}}
\newcommand*{\citet}[1]{\textcite{#1}}

\usepackage[capitalize]{cleveref} %

\def\mdash{{\hbox{-}}}

\makeatletter
\newcommand{\adjunction}{\@ifstar\named@adjunction\normal@adjunction}
\newcommand{\normal@adjunction}[4]{%
  #1\colon #2%
  \mathrel{\vcenter{%
    \offinterlineskip\m@th
    \ialign{%
      \hfil$##$\hfil\cr
      \longrightharpoonup\cr
      \noalign{\kern-.3ex}
      \smallbot\cr
      \longleftharpoondown\cr
    }%
  }}%
  #3 \noloc #4%
}
\newcommand{\named@adjunction}[4]{%
  #2%
  \mathrel{\vcenter{%
    \offinterlineskip\m@th
    \ialign{%
      \hfil$##$\hfil\cr
      \scriptstyle#1\cr
      \noalign{\kern.1ex}
      \longrightharpoonup\cr
      \noalign{\kern-.3ex}
      \smallbot\cr
      \longleftharpoondown\cr
      \scriptstyle#4\cr
    }%
  }}%
  #3%
}
\newcommand{\longrightharpoonup}{\relbar\joinrel\rightharpoonup}
\newcommand{\longleftharpoondown}{\leftharpoondown\joinrel\relbar}
\newcommand\noloc{%
  \nobreak
  \mspace{6mu plus 1mu}
  {:}
  \nonscript\mkern-\thinmuskip
  \mathpunct{}
  \mspace{2mu}
}
\newcommand{\smallbot}{%
  \begingroup\setlength\unitlength{.15em}%
  \begin{picture}(1,1)
  \roundcap
  \polyline(0,0)(1,0)
  \polyline(0.5,0)(0.5,1)
  \end{picture}%
  \endgroup
}
\makeatother

\newcommand\bdag[1]{\ensuremath{#1_{(\cdot)}^\dag}}

\let\op=\relax

\let\innerprod=\relax
\newcommand{\innerprod}[2]{\left\langle #1 {,} \: #2 \right\rangle}

\def\op{\ensuremath{^{\,\mathrm{op}}}}
\def\coop{\ensuremath{^{\,\mathrm{co\,op}}}}

\newcommand{\nn}{{\mathbb{N}}}
\newcommand{\rr}{{\mathbb{R}}}
\newcommand{\Tt}{{\mathbb{T}}}

\newcommand{\from}{\leftarrow}
\newcommand{\xfrom}[1]{\xleftarrow{#1}}

\newcommand{\Cat}[1]{\mathbf{#1}}
\newcommand{\cat}[1]{\mathcal{#1}}
\newcommand{\Fun}[1]{\mathsf{#1}}

\newcommand{\Kl}{\mathcal{K}\mspace{-2mu}\ell}
\def\coKl{\mathrm{co}\Kl}

\let\Pr\relax
\DeclareMathOperator{\Pr}{P}
\DeclareMathOperator*{\E}{\mathbb{E}}

\renewcommand{\d}{\mathrm{d}}

\DeclareMathOperator{\dom}{dom}
\DeclareMathOperator{\cod}{cod}

\newcommand{\para}{\Cat{Para}}
\newcommand{\copara}[1][]{\Cat{Copara}^{#1}}
\newcommand{\ccopara}[1][]{{\Cat{Copara}^{#1}_2}}

\newcommand{\Da}{{\mathcal{D}}}
\newcommand{\Ea}{{\mathcal{E}}}
\newcommand{\Fa}{{\mathcal{F}}}
\newcommand{\Ga}{{\mathcal{G}}}

\newcommand{\Pa}{{\mathcal{P}}}

\newcommand{\Pow}{{\mathcal{P}}}

\newcommand{\Dst}{{\mathcal{D}}}

\newcommand{\Giry}{{\mathcal{G}}}

\newcommand{\Zb}{{\mathbb{Z}}}

\DeclareMathOperator{\im}{\mathsf{im}}

\DeclareMathOperator{\id}{\mathsf{id}}

\DeclareMathOperator{\colim}{\mathrm{colim}}

\DeclareMathOperator{\Set}{\Cat{Set}}

\DeclareMathOperator{\Alg}{\Cat{Alg}}

\font\maljapanese=dmjhira at 2ex %
\DeclareRobustCommand{\yo}{\textrm{\maljapanese\char"48}\!}
\DeclareRobustCommand{\coyo}{\text{\reflectbox{$\yo$}}}

\newcommand{\xto}[2][]{\xrightarrow[#1]{#2}}

\makeatletter
\newcommand{\mathoverlap}[2]{\mathpalette\mathoverlap@{{#1}{#2}}}
\newcommand{\mathoverlap@}[2]{\mathoverlap@@{#1}#2}
\newcommand{\mathoverlap@@}[3]{\ooalign{$\m@th#1#2$\crcr\hidewidth$\m@th#1#3$\hidewidth}}
\makeatother

\newcommand{\klcirc}{\bullet} %
\newcommand*{\smallklcirc}{\raisebox{0.18ex}{\scalebox{0.66}{$\klcirc$}}}
\newcommand{\klto}{\mathoverlap{\rightarrow}{\smallklcirc\,}}

\newcommand{\xklto}[2][]{\mathoverlap{\xrightarrow[#1]{#2}}{\smallklcirc\,}}

\def\lenscirc{\baro}
\newcommand{\lensto}{\mathrel{\ooalign{\hfil$\mapstochar\mkern5mu$\hfil\cr$\to$\cr}}}
\newcommand{\xlensto}[2][]{\mathoverlap{\xrightarrow[#1]{#2}}{\raisebox{0.375ex}{\scalebox{1.0}[0.33]{$|$}}\,}}

\makeatletter
\providecommand*{\xmapstofill@}{%
  \arrowfill@{\mapstochar\relbar}\relbar\rightarrow
}
\providecommand*{\xmapsto}[2][]{%
  \ext@arrow 0395\xmapstofill@{#1}{#2}%
}

\makeatletter
\def\slashedarrowfill@#1#2#3#4#5{%
  $\m@th\thickmuskip0mu\medmuskip\thickmuskip\thinmuskip\thickmuskip
   \relax#5#1\mkern-7mu%
   \cleaders\hbox{$#5\mkern-2mu#2\mkern-2mu$}\hfill
   \mathclap{#3}\mathclap{#2}%
   \cleaders\hbox{$#5\mkern-2mu#2\mkern-2mu$}\hfill
   \mkern-7mu#4$%
}
\def\rightslashedarrowfill@{%
  \slashedarrowfill@\relbar\relbar\mapstochar\rightarrow}
\newcommand\xslashedrightarrow[2][]{%
  \ext@arrow 0055{\rightslashedarrowfill@}{#1}{#2}}
\makeatother

\theoremstyle{definition}
\newtheorem{defn}{Definition}[section]
\newtheorem{notation}[defn]{Notation}

\newtheorem{ex}[defn]{Example}

\newtheorem{rmk}[defn]{Remark}

\newtheorem*{rmk*}{Remark}

\newtheorem{prop}[defn]{Proposition}
\newtheorem{prop*}{Proposition}

\newtheorem{lemma}[defn]{Lemma}
\newtheorem{thm}[defn]{Theorem}
\newtheorem{cor}[defn]{Corollary}

\newtheorem*{thm*}{Theorem}
\newtheorem*{cor*}{Corollary}

\definecolor{darkblue}{rgb}{0,0,0.7} 
\newcommand{\red}[1]{{\color{red} #1}}
\newcommand{\green}[1]{{\color{ForestGreen} #1}}

\newcommand{\fuchsia}[1]{{\color{Fuchsia} #1}}

\tikzstyle{xshiftu}=[shift={(#1, 0)}]
\tikzstyle{yshiftu}=[shift={(0, #1)}]
\tikzstyle{dot}=[inner sep=0.0mm, outer sep=0.0mm, minimum size=1mm, draw, shape=circle]
\tikzstyle{copier}=[dot, fill=white, scale=2.0]
\tikzstyle{black copier}=[dot, fill=black, scale=2.0]
\tikzstyle{white dot}=[dot, fill=white, draw=black]
\tikzstyle{action}=[dot, fill=white, scale=0.667, inner sep=0.5mm]
\tikzstyle{box}=[fill=white, draw=black, shape=rectangle]
\tikzstyle{medium box}=[fill=white, draw=black, shape=rectangle, minimum width=1.5cm, minimum height=0.66cm]
\tikzstyle{arrow box}=[fill=white, draw, shape=rectangle, minimum height=5mm, yshift=-0.5mm, minimum width=5mm]
\tikzstyle{effect}=[regular polygon, regular polygon sides=3, draw]
\tikzstyle{state0}=[regular polygon, regular polygon sides=3, draw, shape border rotate=0]
\tikzstyle{state90}=[regular polygon, regular polygon sides=3, draw, shape border rotate=90]
\tikzstyle{state180}=[regular polygon, regular polygon sides=3, draw, shape border rotate=180]
\tikzstyle{state270}=[regular polygon, regular polygon sides=3, draw, shape border rotate=270]
\tikzstyle{scalar}=[diamond, draw, inner sep=1pt]
\tikzstyle{discarder}=[my ground, draw, inner sep=0pt, minimum width=4.2pt, minimum height=11.2pt, anchor=input, rotate=90]
\tikzstyle{discarder0}=[my ground, draw, inner sep=0pt, minimum width=4.2pt, minimum height=11.2pt, anchor=input, rotate=0]

\tikzstyle{pointy1}=[->]
\tikzstyle{midpoint1}=[-, {postaction={decorate,decoration={markings, mark=at position .5 with {\arrow{>}}}}}]
\tikzstyle{midpointy1pointy}=[->, {postaction={decorate,decoration={markings, mark=at position .5 with {\arrow{>}}}}}]
\tikzstyle{dashed1}=[-, dashed]
\tikzstyle{dash-fill1}=[-, dashed, fill={rgb,255: red,157; green,180; blue,180}]
\tikzstyle{dotted1}=[-, dotted]
\tikzstyle{dash-pointy}=[->, dashed]

\input{strings.tikzdefs}

\ifexternalizetikz\tikzexternaldisable\fi
\newsavebox\sbground
\savebox\sbground{%
  \begin{tikzpicture}[baseline=0pt]
    \draw (0,-.1ex) to (0,.85ex)
    node[ground IEC,draw,anchor=input,inner sep=0pt,
    minimum width=3.15pt,minimum height=8.4pt,rotate=90] {};
  \end{tikzpicture}%
}
\newcommand{\ground}{\mathord{\usebox\sbground}}
\newcommand{\smallground}{\scalebox{0.5}{$\mathord{\usebox\sbground}$}}

\newsavebox\sbcopier
\savebox\sbcopier{%
  \begin{tikzpicture}[baseline=0pt]
    \node[copier,scale=0.7] (a) at (0,3.8pt) {};
    \draw (a) -- +(-90:.21);
    \draw (a) -- +(45:.21);
    \draw (a) -- +(135:.21);
  \end{tikzpicture}}
\newcommand{\copier}{\mathord{\usebox\sbcopier}}
\ifexternalizetikz\tikzexternalenable\fi

\newsavebox\bsbcopier
\savebox\bsbcopier{%
  \begin{tikzpicture}[baseline=0pt]
    \node[black copier,scale=0.7] (a) at (0,3.8pt) {};
    \draw (a) -- +(-90:.21);
    \draw (a) -- +(45:.21);
    \draw (a) -- +(135:.21);
  \end{tikzpicture}}
\newcommand{\bcopier}{\mathord{\usebox\bsbcopier}}
\newcommand{\smallbcopier}{\scalebox{0.5}{$\mathord{\usebox\bsbcopier}$}}
\ifexternalizetikz\tikzexternalenable\fi

\title{Mathematical Foundations for a Compositional Account of the Bayesian Brain}

\author{Toby St Clere Smithe}
\college{St Edmund Hall}

\degreedate{Trinity 2023}

\hypersetup{
  pdftitle={Mathematical Foundations for a Compositional Account of the Bayesian Brain},
  pdfsubject={DPhil Thesis},
  pdfauthor={Toby St Clere Smithe}
}

\def\Alpha{\mathrm{A}}
\def\Rho{\mathrm{P}}

\def\deloop{\mathbf{B}}

\newcommand{\Sum}{\sum\limits}

\def\List{\mathsf{List}}

\def\Stat{\mathsf{Stat}}
\def\SStat{{\Stat_2}}
\def\disc{\Fun{disc}\,}

\def\Lens{\Cat{Lens}}

\newcommand{\blhom}[1]{\ldbrack{#1}\rdbrack}

\newcommand{\BLens}[1]{{\Cat{BayesLens}_{\cat{#1}}}} %

\newcommand{\SGame}[1]{{\Cat{SGame}_{#1}}}         %

\def\Loss{\mathsf{Loss}}
\def\MonLoss{\mathsf{MonLoss}}
\def\pLoss{{\pi_{\Loss}}}
\def\pLens{{\pi_{\mathsf{Lens}}}}
\def\pLensDag{{\pi_{\mathsf{Lens}}^\dag}}

\def\KL{\mathsf{KL}}
\def\MLE{\mathsf{MLE}}
\def\FE{\mathsf{FE}}
\def\LFE{\mathsf{LFE}}

\def\Poly{\Cat{Poly}}

\newcommand{\pMCoalgT}[3]{\Cat{Coalg}_{#2}^{#3}({#1})}

\newcommand{\PMCoalgT}[1]{\pMCoalgT{p}{\cat{C}}{#1}}

\newcommand{\Coalg}{\Cat{Coalg}}

\makeatletter
\newcommand{\Cilia}{\Cat{Cilia}}
\newcommand{\Hier}{\Cat{Cilia}}
\newcommand{\HierE@nostar}{{\Cat{Cilia}|_{\cat{E}}}}
\newcommand{\HierE@star}[1]{{\Cat{Cilia}_{#1}|_{\cat{E}}}}
\newcommand{\HierE}{\@ifstar{\HierE@star}{\HierE@nostar}}
\makeatother

\makeatletter
\DeclareRobustCommand\Equiv{\mathrel{%
  \mathchoice
    {\Equiv@\textfont\displaystyle{.43}}
    {\Equiv@\textfont\textstyle{.43}}
    {\Equiv@\scriptfont\scriptstyle{.45}}
    {\Equiv@\scriptscriptfont\scriptscriptstyle{.5}}
}}
\newcommand{\Equiv@}[3]{%
  \rlap{\raisebox{#3\fontdimen5#12}{$\m@th#2 = $}}%
  \raisebox{-#3\fontdimen5#12}{$\m@th#2 = $}%
}
\makeatother

\def\MonCat{\Cat{MonCat}}

\def\Mon{\Cat{Mon}}
\def\CMon{\Cat{CMon}}
\def\Comon{\Cat{Comon}}
\def\CComon{\Cat{CComon}}
\def\Alg{\Cat{Alg}}
\def\Span{\Cat{Span}}
\def\DOpfib{\Cat{DOpfib}}

\def\LinCirc{\Cat{LinCirc}}
\def\OLinCirc{\cat{O}\LinCirc}

\def\inj{\mathsf{inj}}
\def\proj{\mathsf{proj}}

\def\Tan{\Fun{T}}

\def\gauss{\Cat{Gauss}}
\def\FdGauss{\Cat{FdGauss}}

\def\DiffSys{\Cat{DiffSys}}
\def\DiffHier{\Cat{DiffCilia}}
\def\DiffCilia{\Cat{DiffCilia}}

\def\H{\Cat{H}}

\def\eP{\Cat{P}}

\def\Laplace{\Fun{L}}
\def\Hebb{\Fun{H}}

\newcommand{\DynT}[1]{\Cat{Coalg}_{\id}^{\mathbb{#1}}}
\newcommand{\RDyn}{\Cat{RDyn}}
\newcommand{\RDynT}[1]{\RDyn^{\mathbb{#1}}}

\newcommand{\vlens}[2]{\ensuremath{\begin{pmatrix} #1 \\ #2 \end{pmatrix}}}
\WithSuffix\newcommand\vlens*[2]{\ensuremath{\Bigl(\negthinspace\begin{smallmatrix}#1\\#2\end{smallmatrix}\Bigr)}}

\def\bkwd{\overleftarrow}

\newcommand{\dhess}[1]{\overline{\partial^2_{#1}}}

\begin{document}

\baselineskip=20pt plus1pt

\setcounter{secnumdepth}{3}
\setcounter{tocdepth}{3}

\maketitle
\begin{acknowledgements}
  This thesis would not exist in anything like this form without the marvellous \textit{Applied Category Theory} community, a more welcoming and thoughtful group of researchers one could not wish to find.
  This community makes a serious and thoroughgoing effort to be inclusive and outward-looking, and it was in this spirit that they set up the \textit{Applied Category Theory Adjoint School}, which I attended in 2019, and to which I recommend any category-theory-curious thinker to apply.
  Without that experience, and the group of friends I made there, none of this would have been possible.

  Before I attended the Adjoint School, I was trying to understand too much about the brain, and seeking a mathematically coherent unifying framework with which I could organize my thoughts.
  In Oxford, I was a member of the Department of Experimental Psychology, but had become aware of the work being done on cognition and linguistics in the Quantum Group, in the Department of Computer Science, and so I began attending lectures and hanging around there.
  It was there that I attended the \textit{Open Games} workshop in 2018, at which I realized that predictive coding and open games had the same abstract structure; a fact that took me longer than it should have to formalize, but about which I started telling anyone who listened.
  The first individuals who took me seriously were Jules Hedges and Brendan Fong, and I thank them heartily for their encouragement and assistance: it was after discussion with Jules (and Bruno Gavranović) at the Sixth Symposium on Compositional Structures (SYCO 6, in Leicester) that I proved abstractly that ``Bayesian updates compose optically''; and it was Brendan Fong who let me know about the Adjoint School, at which we (Brendan, Bruno, David Spivak, David Jaz Myers, and Sophie Libkind, as well as others occasionally, including Jules, Eliana Lorch, and davidad) discussed autopoiesis from a categorical perspective.

  After these meetings, and through my Quantum Group interactions, I acquired some funding from the Foundational Questions Institute to concentrate on the category theory of predictive coding and approximate inference, which was distributed through the Topos Institute.
  I thank everyone who made these interactions possible and delightful, including (in no particular order) the following individuals that I have not yet named: Samson Abramsky; Bob Coecke; Johannes Kleiner; Tim Hosgood; Owen Lynch; Valeria de Paiva; Evan Patterson; Sam Staton; Juliet Szatko; Tish Tanski; Sean Tull; and Vincent Wang-Maścianica.

  Outside of Oxford, I have been fortunate to be part of some wonderful interactions through the Active Inference and Strathclyde MSP (Mathematically Structured Programming) communities.
  I first spoke about categorical active inference to Karl Friston's group in March 2020, shortly after my first visit to Glasgow at the end of 2019; and I found Glasgow so appealing that I now find myself living there.
  For these interactions, besides those named above, I must recognize: Dylan Braithwaite; Matteo Capucci; Lance da Costa; Neil Ghani; Maxwell Ramstead; Riu Rodríguez Sakamoto; and Dalton Sakthivadivel.

  I would not have had the opportunity to pursue this research at all had I not been granted a position in the Oxford Experimental Psychology department, where I have been a member of the Oxford Centre for Theoretical Neuroscience and Artificial Intelligence (OCTNAI), under the direction of Simon Stringer.
  I thank Simon for his patience and latitude, particularly when my plans were not quite as he would have expected, and I thank my Oxford co-supervisor (and present director of graduate studies), Mark Buckley, and my previous director of graduate studies, Brian Parkinson, for their always excellent advice.
  Thanks also to the other student members of OCTNAI (particularly Dan, Hannah, Harry, James, Nas, and Niels) for being so welcoming to an oddball such as myself.
  And at this point, it would be remiss not to thank also the administrative staff of the Department, and my college, St Edmund Hall, who are always helpful and wise; in particular, Rebecca Cardus and Vinca Boorman, who have guided me through much of Oxford's strange bureaucracy.

  Finally, and most of all, I thank my family and my beloved wife, Linda, who in particular has suffered through this long journey with me with beyond-infinite patience, love, and understanding (so much patience, in fact, that she humoured the category-theoretic content of my wedding speech!).
  Thank you, to you all.
  It takes a village!
\end{acknowledgements}

\begin{abstractlong}
  This dissertation reports some first steps towards a compositional account of active inference and the Bayesian brain.
  Specifically, we use the tools of contemporary applied category theory to supply \textit{functorial semantics} for approximate inference.
  To do so, we define on the `syntactic' side the new notion of \textit{Bayesian lens} and show that Bayesian updating composes according to the compositional lens pattern.
  Using Bayesian lenses, and inspired by compositional game theory, we define fibrations of \textit{statistical games} and classify various problems of statistical inference as corresponding sections: the chain rule of the relative entropy is formalized as a strict section, while maximum likelihood estimation and the free energy give lax sections.
  In the process, we introduce a new notion of `copy-composition'.

  On the `semantic' side, we present a new formalization of general open dynamical systems (particularly: deterministic, stochastic, and random; and discrete- and continuous-time) as certain coalgebras of polynomial functors, which we show collect into monoidal opindexed categories (or, alternatively, into algebras for multicategories of generalized polynomial functors).
  We use these opindexed categories to define monoidal bicategories of \textit{cilia}: dynamical systems which control lenses, and which supply the target for our functorial semantics.
  Accordingly, we construct functors which explain the bidirectional compositional structure of predictive coding neural circuits under the free energy principle, thereby giving a formal mathematical underpinning to the bidirectionality observed in the cortex.
  Along the way, we explain how to compose rate-coded neural circuits using an algebra for a multicategory of \textit{linear circuit diagrams}, showing subsequently that this is subsumed by lenses and polynomial functors.

  Because category theory is unfamiliar to many computational neuroscientists and cognitive scientists, we have made a particular effort to give clear, detailed, and approachable expositions of all the category-theoretic structures and results of which we make use.
  We hope that this dissertation will prove helpful in establishing a new ``well-typed'' science of life and mind, and in facilitating interdisciplinary communication.
\end{abstractlong}

\begin{romanpages}
\tableofcontents
\end{romanpages}

\chapter{Introduction} \label{chp:intro}

The work of which this dissertation is a report began as a project to understand the brain's ``cognitive map'', its internal representation of the structure of the world.
Little of that work is reported here, for it rapidly became clear at the outset that there was no coherent framework in which such a project should most profitably be undertaken.
This is not to say that no progress on understanding the cognitive map can be made, a claim which would be easily contradicted by the evidence.
Rather, each research group has its own language and its own research questions, and it is not always evident how to translate concepts from one group, or even one moment in time, faithfully to another; what translation is done is performed at best highly informally.

If the aim of science\footnote{%
Or indeed, ``if the aim of scientists'', as science itself may not have volition of its own.}
is to tell just-so stories, or if the aim is only to answer one's own research questions in isolation, then this state of affairs may be perfectly satisfactory.
But the brain and the behaviours that brains produce are so marvellous and so complex, and the implications of a finer understanding so monumental, that one cannot but hope that science could do better.
Of course, of late, science has not been doing better, with disciplines as socially important as psychology \parencite{OpenScienceCollaboration2015Estimating} and medicine \parencite{Ioannidis2005Contradicted,Begley2012Raise,Mobley2013Survey} and machine learning \parencite{Hutson2018Artificial,Kapoor2022Leakage} struck by crises of reproducibility.
At the same time, as broadband internet has spread across the globe, the sheer amount of output produced by scientists and other researchers has ballooned, contributing to the impossibility of verification and the aforementioned translational difficulties, at least if one desires to do other than simply following the herd.
In some sense, although scientists all now speak English, science still lacks a \textit{lingua franca}, or at least a sufficiently precise one.

As luck would have it, while mainstream science has been suffering from this loss of faith, the first phrases of a potentially adequate precise new language have begun to spread, with the coalescence of a new community of researchers in \textit{applied category theory}\footnote{%
The first major interdisciplinary meeting of applied category theorists (or at least the first meeting sufficiently confident to take \textit{Applied Category Theory} as its name) was held in 2018 in Leiden, although categorical methods have for some time been used in computer science \parencite{Pierce1991Basic} and physics \parencite{Baez1995Higher}, and especially at their nexus \parencite{Abramsky2004categorical,Coecke2015Categorical,Coecke2016Categorical}.
More sporadically, category theory had shown up elsewhere, such as in biology \parencite{Rosen1958Representation,Ehresmann2006Memory}, network theory \parencite{Fong2013Causal,Fong2015Decorated,Fong2016Algebra}, game theory \parencite{Ghani2016Compositional,Abramsky2012Coalgebraic,Escardo2010Selection}, cognitive science \parencite{Phillips2010Categorial,Ehresmann2007Memory,AlMehairi2016Compositional,Bolt2019Interacting} and linguistics \parencite{Coecke2010Mathematical,Heunen2013Quantum,Coecke2019Mathematics}, and in 2014 a workshop was held at Dagstuhl bringing together some of these researchers \parencite{Abramsky2014Categorical}, in what was to be a precursor to the Applied Category Theory meetings; many of those researchers still work in this new interdisciplinary field.}.
One part of the present difficulty of scientific translation is that each research group has not only its own language, but also its own perspective; and another part of the difficulty is that these languages and perspectives are not well connected, with the English language a very lossy medium through which to make these connections.
Fortunately, the language of category theory---being a mathematical rather than a natural language---resolves both of these difficulties.

Category theory is the mathematics of pattern, composition, connection, and interaction; its concepts are as crisp and clear as the water of a mountain pool; its simplicity lends it great power.
Categories describe how objects can be constructed from parts, and such compositional descriptions extend to categories themselves: as a result, the language of category theory is `homoiconic', and can be used to translate constructions between contexts.
One is able to abstract away from irrelevant details, and show precisely how structures give rise to phenomena; and by choosing the abstractions carefully, it becomes possible to see that, sometimes, important constructions are `universal', able to be performed in any relevant context.
As a result, category theory resolves both problems of scientific translation indicated above: concepts expressed categorically are inevitably expressed in context, and not in isolation; and these contexts are naturally interconnected as if by a categorical web (with the connections also expressed categorically).
Moreover, not being English, categorical definitions tend to be extremely concise and information-dense; and since the basic concepts of category theory are themselves simple, concepts so expressed are not biased by geography or geopolitics.

From the middle of the 20\textsuperscript{th} century, the concepts of category theory began to revolutionize much of mathematics\footnote{%
The basic concepts of category theory were originally written down by Eilenberg and Mac Lane in order to formalize processes of translation, and so clarify structures in the ways indicated in the main text above, in the field of algebraic topology.
This occurred at the end of the first half of the 20\textsuperscript{th} century, in 1945 \parencite{Eilenberg1945General}.
The ideas soon spread beyond algebraic topology, gathering momentum rapidly from the 1950s, in which Cartan defined the concept of sheaf \parencite{Cartan1953Varietes,Cartan1956Homological} and Grothendieck reconceived the foundations of algebraic geometry \parencite{Grothendieck1957Quelques}.
By the mid-1960s, and especially through the work of Lawvere on logic \parencite{Lawvere1963Functorial} and set theory \parencite{Lawvere1964Elementary}, it was clear that category theory would be able to supply supple but sturdy new foundations for all of mathematics.},
and applied category theorists such as the present author believe that the time is nigh for this revolution to spread throughout the sciences and alleviate some of their struggles.
Just as the internet constitutes physical infrastructure that fundamentally accelerates human communications, we expect category theory to constitute conceptual infrastructure of similar catalytic consequence.
This thesis is a contribution to building this infrastructure, in the specific domain of computational neuroscience and the general domain of (what was once, and will be again, called) cybernetics\footnote{%
Owing to its affinity for pattern and abstraction, it is hard to do interesting domain-specific work in category theory without there being at least some more general results to be found, and indeed this is the case here: what began as a project in theoretical neuroscience swiftly became a study of adaptive and cybernetic systems more broadly, of which the brain is of course the prime exemplar.}.
In particular, we show that a prominent theory of brain function---\textit{predictive coding}---has a clear compositional structure, that explains the bidirectional circuitry observed in the brain \parencite{Bastos2012Canonical}, and that renders precise connections to the structure of statistical and machine learning systems \parencite{Whittington2017Approximation,Millidge2022Predictive,Rosenbaum2022Relationship}, as well as to the structure of much larger scale adaptive systems traditionally modelled by economic game theory \parencite{Ghani2016Compositional}.

Predictive coding models were originally developed in the neuroscience of vision to explain observations that neural activity might decrease as signals became less surprising \parencite{Rao1999Predictive} (rather than increase as signals became more `preferred'), as well as to explain the robustness of sensory processing to noise \parencite{Srinivasan1982Predictive} and as a source of metabolic efficiency \parencite{Boerlin2013Predictive}\footnote{%
If the prediction is good, then communicating the difference between prediction and actuality can be done much more efficiently than transmitting the whole incoming signal, which would contain much redundant information.
This is the principle underlying most data compression algorithms.}.
The typical form of these models involves a neuron or neural ensemble representing the system's current prediction of (or expectation about) its input, alongside another neuron or ensemble representing the difference between this prediction and the actual input (\textit{i.e.}, representing the prediction error).
We can think of the former ensemble as directed from within the brain towards the sensory interface (such as the retina), and the latter ensemble as carrying information from the world into the brain: this is the aforementioned bidirectionality.

Another important observation about visual processing in the brain is that its circuitry seems to be roughly hierarchical \parencite{Marr1982Vision}, with regions of cortex further from the retina being involved in increasingly abstract representation \parencite{Poggio2013Models}.
Given a model of predictive coding at the level of a single circuit, accompanied by models of how sensory circuits are coupled (and their representations transformed), a natural next step is to construct hierarchical predictive coding models, in an attempt to extend the benefits of the single circuit to a whole system; and indeed such hierarchical circuits were prominently proposed in the literature \parencite{Rao1999Predictive,Friston2009Predictive}.

This hierarchical structure is a hint of compositionality, and thus a sign that a categorical approach may be helpful and enlightening.
This impression is strengthened when one considers a particularly influential class of predictive coding models, obtained in the context of the ``free energy principle'' \parencite{Friston2005theory,Friston2007Free,Friston2009Predictive}, where the underlying equations themselves exhibit a form of compositionality which is (more or less explicitly) used to obtain the hierarchical models\footnote{%
That is to say, the dynamics of each level of hierarchy $i$ are governed by a quantity $\Fa_i$, and the dynamics of two adjacent levels $i$ and $i+1$ are governed by $\Fa_i+\Fa_{i+1}$; see \textcite[{Eq. 72}]{Buckley2017free}.}.
Despite this hint of compositionality, the equations of motion for these hierarchical systems are typically derived from scratch each time \parencite{Buckley2017free,DaCosta2020Active,Tschantz2019Learning,Kaplan2018Planning,Ueltzhoeffer2018Deep,Friston2010Action,Bastos2012Canonical}, a redundant effort that would not be required had a compositional formalism such as category theory been used from the start.
This thesis supplies such a categorical formalism and exemplifies it with hierarchical predictive coding under the free energy principle.

The ``free energy'' framework not only underpins a modern understanding of predictive coding, but has more broadly been proposed as a unified theory of brain function \parencite{Friston2005theory}, and latterly of all adaptive or living systems \parencite{Friston2019free,Parr2019Markov,Kirchhoff2018Autopoiesis,Boonstra2019Dialectics}.
In the neuroscientific context, it constitutes a theory of the \textit{Bayesian brain}, by which most or all brain function can be understood as implementing approximate Bayesian inference \parencite{Knill2004Bayesian}; in the more broadly biological (or even metaphysical) contexts, this claim is generalized to state that all life can be understood in this way.
However, despite these claims to universality, these proposals have to date been quite informally specified, leading to confusion \parencite{Biehl2020technical,Friston2020Some} and charges of unfalsifiability \parencite{Colombo2021First,Boonstra2019Dialectics,Williams2022Brain}.
As we will see, category theory has a rich formal vocabulary for precisely describing universal constructions, and so not only does a categorical formulation of the free energy framework promise to clarify the current confusions, but it may be expected also to shed light on its potential universality.
In particular, as we discuss in Chapter \ref{chp:future}, we will be able to make precise the questions of whether any dynamical system of the appropriate type can universally be seen as performing approximate inference (in our language, ``playing a statistical game''), and of whether any cybernetic system (such as an economic game player) can be expressed as an active inference system.

The notion of \textit{active inference} is closely related to the free energy framework: an active inference model of a system describes both the processes by which it updates its internal states on the basis of incoming signals, and the processes by which it chooses how to act, using approximate Bayesian inference.
In this thesis, we do not get as far as a completely general formulation of active inference, but we hope that our development of \textit{statistical games} and their ``dynamical semantics'' in \textit{approximate inference doctrines} will provide a useful starting point for such a formulation, and in our final chapter (\ref{chp:future}) we sketch how we might expect this formulation to go.
Because active inference models, and the free energy framework more broadly, are descriptions of systems that are `open' to an environment, interacting with it, and therefore situated ``in context'', they are particularly suited to a category-theoretic reformulation.
Likewise, Bayesianism and the free energy framework lend themselves to a subjectivist metaphysics \parencite{Fuchs2009QuantumBayesian,Galavotti2001Subjectivism,Friston2019free}, which is itself in alignment with the unavoidable perspective-taking of categorical models, and which is not dissimilar from the emerging `biosemiotic' reconceptualization of biological information-processing \parencite{Barbieri2007Introduction}.
As we have indicated, categorical tools help us to draw connections between concepts, and we see our efforts as a contribution to this endeavour.

It is through these connections that we hope eventually to make contact again with the cognitive map.
As noted above, the state of the art is fragmented, but there exist current models that are expressed in the language of approximate (variational) inference \parencite{Whittington2019Tolman}, models expressed in the language of reinforcement learning \parencite{Stachenfeld2016hippocampus}, and models that attempt to combine the two \parencite{Millidge2022Successor}.
We will see throughout the thesis that reinforcement learning (and its cousin, game theory) is closely related to approximate inference, and so we expect that the foundations developed here, along with the extensions proposed in \secref{sec:cog-cart}, will help us unify these accounts.
The key observation that we expect to drive such a development is that learning a cognitive map (alternatively, learning a ``world model'') means internalizing a representation of the structure of the environment; and comparing and translating structures is category theory's forte.

Of course, even if the theory that we develop is sufficient to unify these computational-phenomenological models, this is not to say it will satisfy all neuroscientists, many of which may be expected to desire more biologically detailed models.
In the contemporary undergraduate neuroscience curriculum, one is taught informally to relate models at a high `computational' level to lower level models concerned with biological `implementation', following Marr's ``three levels of explanation'' \parencite{Marr1982Vision}.
As we discuss in \secref{sec:syn-sem}, this story is a shadow of the categorical notion of \textit{functorial semantics}, by which structures are translated precisely between contexts formalized as categories.
Although we concentrate on the more abstract computational level in this thesis, our discussion of functorial semantics foreshadows the introduction of formal algebraic tools for building biologically plausible neural circuit models (\secref{sec:sys-circ}).

Our treatment of cognitive and neural systems is not the first to adopt categorical methods, but we do believe that it is the first to do so in a comprehensively integrated and wide-ranging way, taking functorial semantics seriously.
Categorical concepts have been variously proposed in biology as early as 1958 \parencite{Rosen1958Representation}, and in cognitive science (with one eye toward the brain) since at least 1987 \parencite{Ehresmann1987Hierarchical,Ehresmann2007Memory}; more recently, category theory has been used to study classic cognitive-science concepts such as systematicity \parencite{Phillips2010Categorial}.
While inspirational, these studies do not make the most of the translational power of categories, using only some concepts or methods in isolation.
Moreover, by working almost purely categorically, these works were invariably rather abstract, and did not make direct contact with the tools and concepts of mainstream mathematical science.
As a result, they did not have the unifying impact or adoption that we hope the new wave of applied category theoretical developments to have.

Our primary motivation in writing this thesis is to lay the groundwork for \textit{well-typed} cognitive science and computational neuroscience.
`Types' are what render categorical concepts so precise, and what allow categorical models to be so cleanly compositional: two systems can only ``plug together'' if their interface types match.
Because every concept in category theory has a type (\textit{i.e.}, every object is an object of some category), categorical thinking is forced to be very clear.
As we will sketch in \secref{sec:closed-cats}, the ``type theories'' (or ``internal languages'') of categories can be very richly structured, but still the requirement to express concepts with types is necessarily burdensome.
But this burden is only the burden of thinking clearly: if one is not able to supply a detailed type, one can resort to abstraction.
And, to avoid the violence of declaring some object to be identified as of some type\footnote{%
A perspective for which we must thank Brendan Fong.},
it is necessary to understand the relationships between types; fortunately, as we will soon make clear, and as we have attempted to emphasize, category theory is fundamentally the mathematics of relationship.

Contemporary science is unavoidably computational, and the notion of `type' that we invoke here is closely related to (though not identical with) the informal notion of type that is used in computer programming.
Just as one of the strategies adopted to overcome the crises of modern science that we invoked at the opening of this introduction is the making available of the code and data that underlie scientific studies, we can envisage a near future in which accompanying these is a formal specification of the types of the concepts that each study is about\footnote{%
One might think of this specification as akin to a scientifically elaborated version of the notion of \textit{header file} in programming languages such as C or C++: these files specify the types of functions and data structures, typically without instantiating these types with detailed implementations.
We can thus think of category theory as a very rich metaprogramming language for the mathematical sciences (and this analogy goes quite far, as categorical proofs are typically `constructive' and hence correspond to computable functions, as we also sketch in \secref{sec:closed-cats}).}.
Some work along these lines has already begun, particularly with the development of the Algebraic Julia ecosystem \parencite{Halter2020Compositional}.

The free energy framework, like the structurally adjacent framework of compositional game theory, has a strong flavour of teleology (that follows directly from its mathematics): systems act in order to make their predictions come true.
We therefore hope that, although we do not quite get as far as a full compositional theory of active inference, the contributions reported in this dissertation may in some small way help to make this particular prediction (of a well-typed science) come true, and thereby help to overcome some of the aforenoted crises of scientific faith---as well as to shed light not only on the form and function of `Bayesian' brains, but also other complex adaptive systems, such as the whole scientific community itself.

\section{Overview of the dissertation}

Category theory being quite alien to most researchers in computational neuroscience (and the cognitive sciences more broadly), we begin the work of this dissertation in Chapter \ref{chp:basic-ct} with a comprehensive review of the concepts and results needed to understand our mathematical contributions.
Using three hopefully familiar examples, we introduce categories as contrapuntal to graphs, which are more familiar to scientists, but which lack important features of categories such as composition and, somehow, dynamism.
We then explain how enriched categories allow us to ``connect the connections'' of categories, and attach extra data to them, and we exemplify these concepts with the 2-category of categories, functors, and natural transformations---as well as a more formal discussion of functorial `translation' and semantics.
The remainder of Chapter \ref{chp:basic-ct} is dedicated to introducing the remaining key concepts of basic category theory: universal constructions, and the Yoneda Lemma (categories' fundamental theorem).
All of these ideas are very well known to category theorists.

In Chapter \ref{chp:algebra}, we begin to reapproach neural modelling, and more generally the `algebraic' modelling of the structure of interacting systems.
We explain how `monoidal' categories allow us to consider processes ``in parallel'' (as well as just sequentially), and how this gives us a formal account of the concept of `parameterized' system.
We then change the perspective a little, and introduce our first piece of original work: an account of how to connect neural circuits into larger-scale systems, using `multicategorical' algebra.
The remainder of the chapter is dedicated to developing the theory of such algebra to the point needed later in the thesis, ending with the introduction of \textit{polynomial functors} which will supply a rich syntax for the interaction of systems, as well as a language in which to express their dynamical semantics.

Chapter \ref{chp:buco} presents our first main result, that Bayesian updating composes according to the categorical `lens' pattern.
This result is abstractly stated, and so applies to whichever compositional model of probability one might be interested in---but because we are later interested in concrete models, we spend much of the chapter recapitulating compositional probability theory using the tools introduced in Chapters \ref{chp:basic-ct} and \ref{chp:algebra} and instantiating it in discrete and continuous settings.
We also introduce and contextualize the lens pattern, in order to define our new notion of \textit{Bayesian lens}, which provides a mathematical formalization of the bidirectionality of predictive coding circuits.

Our main aim in this thesis is to formalize predictive coding through functorial semantics, and Bayesian lenses will provide an important part of the `syntax' of statistical models that we need.
But the Bayesian lenses that satisfy the main result of Chapter \ref{chp:buco} are `exact', while natural systems are inherently approximate.
In order to measure the performance of such approximate systems, Chapter \ref{chp:sgame} introduces our next new notion, the concept of \textit{statistical game}, which attaches loss functions to lenses.
These statistical games collect into a categorical structure known as a fibration (a kind of categorified fibre bundle), and we can use the sections of this fibration to classify well-behaved systems of approximate inference into \textit{loss models}.
These loss models include well-known quantities such as the relative entropy, (maximum) likelihood, the free energy, and the Laplace approximation of the latter.
However, in order to make this classification work, we first introduce a new kind of categorical composition, which we call \textit{copy-composition}, and which seems to cleave the basic process of composition in categories of stochastic channels, which typically proceeds first by copying and then by marginalization (`discarding').

Having developed the syntactic side of predictive coding, we turn in Chapter \ref{chp:coalg} to the semantics, which is found in a new abstract formalization of the concept of \textit{open dynamical system}.
We make much use here of the language of polynomial functors: these will represent the interfaces of interacting systems, and the dynamical systems themselves will be defined as particular classes of morphisms of polynomials.
We extend the traditional notion of polynomial functor to a setting which allows for non-determinism, and thereby obtain new categories of open Markov process and random dynamical system, both in discrete and continuous time.
We then synthesize these developments with the algebraic structures of Chapter \ref{chp:algebra}, to define monoidal bicategories of `hierarchical' cybernetic systems that we call \textit{cilia}, as they control lenses.

Connecting these pieces together, Chapter \ref{chp:brain} presents our functorial formalization of predictive coding, using a new notion of \textit{approximate inference doctrine}, by which statistical models are translated into dynamical systems.
This formalizes the process by which research in active inference turns the abstract specification of a ``generative model'' into a dynamical system that can be simulated and whose behaviours can then be compared with experimentally observed data.
We explain how this functorial process is decomposed into stages, and then exhibit them in two ways: first, with the basic `Laplacian' form of predictive coding; and then by introducing `Hebbian' plasticity.

Finally, Chapter \ref{chp:future} reviews the prospects for future work, from the mathematics of the cognitive map (a programme that we call \textit{compositional cognitive cartography}), to the composition of multi-agent systems and ecosystems and the connections with compositional game theory, categorical cybernetics, and categorical systems theory.
We close with some speculation on a new mathematics of life, along with associated developments of fundamental theory.

\section{Contributions}

The main individual contribution of this thesis is the formalization of models of predictive coding circuits as functorial semantics, and the associated development and exemplification of fibrations of statistical games, as well as the introduction of Bayesian lenses and the proof that Bayesian updates compose optically.
We believe our presentation of general open dynamical systems as certain polynomial coalgebras also to be novel, along with the concept of \textit{cilia} and their associated monoidal bicategories.
The categories of statistical games (and of Bayesian lenses) supply the syntax, and the monoidal bicategories of cilia the semantics, for our functorial treatment of predictive coding, and hence the basis for our compositional active inference framework.
Each of these structures is to our knowledge new, although of course inspired by much work that has gone before, and by interactions with the beneficent community of researchers of which this author finds himself a member.

Each of these strands of work has in some way been exhibited through publication, principally as refereed presentations at the conference on Applied Category Theory (ACT) in 2020 \parencite{Smithe2020Cyber}, 2021 \parencite{Smithe2021Polynomial}, and 2022 \parencite{Smithe2022Open} (each published in the conference proceedings); but also in preliminary form at the NeurIPS 2019 \textit{Context and Compositionality} workshop \parencite{Smithe2019Radically}, through a number of more informal invited talks (\textit{e.g.} \parencite{Smithe2021Compositional}), as one main theme of a full-day workshop at the 2022 Cognitive Science Society conference \parencite{Anderson2022Category}, and our ongoing series of preprints on compositional active inference \parencite{Smithe2021Compositional1,Smithe2022Compositional}.
Our work on Bayesian lenses, in collaboration with Dylan Braithwaite and Jules Hedges \parencite{Braithwaite2023Compositional}\footnote{
See Remark \ref{rmk:blens-history} for the scholarly history.},
has been accepted for publication at MFCS 2023; and we are presently preparing for journal publication an account of our compositional framework for predictive coding aimed explicitly at computational neuroscientists.

Besides these specific novel contributions, we hope that this dissertation contributes to a renaissance of cognitive and computational (neuro)science through the adoption of categorical methods; it is for this reason that we have been so diligent in our exposition of the basic theory.
We hope that this exposition proves itself a useful contribution for interested researchers, and that its cognitive-neuroscientific framing is sufficiently novel to be interesting.

Some work performed during the author's DPhil studies is not included in this dissertation.
In particular, there has unfortunately not been the scope to include our simulation results on a fragment of the circuitry underlying the cognitive map---a study on the development of place and head-direction cells, which was published as \parencite{Smithe2022Role}---although this did motivate our algebra of rate-coded neural circuits (\secref{sec:sys-circ}), which is to the best of our knowledge novel (though much inspired by earlier work on wiring-diagram algebras \parencite{Spivak2013Operad,Yau2018Operads}).
We have also not exhibited our work on Bayesian optics (as an alternative to Bayesian lenses) \parencite{Smithe2020Bayesian}, as this would require a digression through some unnecessarily complicated theory; and we have not presented in detail the examples of ``polynomial life'' presented at ACT 2021 \parencite{Smithe2021Polynomial}.

A first draft of this thesis was produced in December 2022, at which point the author intended to submit it.
However, shortly before submission, the author realized that the then-current treatment of statistical games could be much improved.
This led to the present fibrational account, and the new notion of loss model (which formalizes the chain rule of the relative entropy), but which also demanded a corresponding revision of the treatment of predictive coding.
At the cost of some higher-categorical machinery, we believe these changes amount to a substantial improvement, worth the delay in submission.
The new account of statistical games has been accepted as a proceedings paper at ACT 2023.

\chapter{Basic category theory for computational and cognitive (neuro)scientists} \label{chp:basic-ct}

This chapter constitutes a comprehensive review of the concepts and results from basic category theory that scaffold the rest of the thesis, written for the computational neuroscientist or cognitive scientist who has noticed the `network' structure of complex systems like the brain and who wonders how this structure relates to the systems' function.
Category theory gives us a mathematical framework in which precise answers to such questions can be formulated, and reveals the interconnectedness of scientific ideas.
After introducing the notions of category and diagram (\secref{sec:cats-nets}), we swiftly introduce the notions of enriched category, functor, and adjunction (\secref{sec:enrich}), with which we can translate and compare mathematical concepts.
We then explain how category theory formalizes pattern as well as translation, using the concept of universal construction (\secref{sec:univ-const}), which we exemplify with many common and important patterns.
Finally, we introduce the fundamental theorem of category theory, the Yoneda Lemma, which tells us that to understand a thing is to see it from all perspectives (\secref{sec:yoneda}).

Category theory is well established in the foundations of mathematics, but not yet explicitly in the foundations of science.
As a result, although the only slightly original part of this chapter is its presentation, we have given proofs of most results and plentiful examples, in order to familiarize the reader with thinking categorically.

\section{Categories, graphs, and networks} \label{sec:cats-nets}

We begin by motivating the use of category theory by considering what is missing from a purely graph-theoretic understanding of complex computational systems.
Later in the thesis, we will see how each of the diagrams depicted below can be formalized categorically, incorporating all the scientifically salient information into coherent mathematical objects.

\subsection{Three examples}

\subsubsection{Neural circuits: dynamical networks of neurons} \label{sec:mtv-neur}

In computational and theoretical neuroscience, it is not unusual to encounter diagrams depicting proposed architectures for neural circuits, such as on the left or right below:
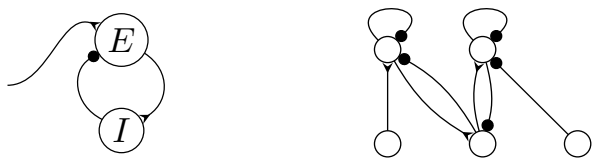
\[ \scalebox{1.2}{\input{img/EI-network-1.tikz}} \qquad\quad\qquad \scalebox{1.25}{\input{img/bogacz-circuit-1.tikz}} \]
On the left, we have depicted a standard ``excitatory-inhibitory circuit'' motif, in which one neuron or ensemble of neurons $E$ receives input from an external source as well as from a counterposed inhibitory circuit $I$ which itself is driven solely by $E$.
On the right, we have reproduced a figure depicting a ``predictive coding'' circuit from \textcite{Bogacz2017tutorial}, and we see that the $E$-$I$ circuit is indeed motivic, being recapitulated twice:
we could say that the predictive coding circuit is \textit{composed} from interconnected $E$-$I$ motifs, in a sense similarly to the composition of the $E$-$I$ circuit from the subnetworks $E$ and $I$ of neurons.

Both circuits have evident graphical structure --- the nodes are the white circles, and the edges the black wires between them --- but of course there is more to neural circuits than these graphs: not only do graphs so defined omit the decorations on the wires (indicating whether a connection is excitatory or inhibitory), but they miss perhaps the more important detail, that these are circuits \textit{of dynamical systems}, which have their own rich structure and behaviours.
Moreover, mere graphs miss the aforementioned \textit{compositionality} of neural circuits: we can fill in the white circles with neurons or ensembles or other circuits and we can wire circuits together, and at the end of doing so we have another `composite' neural circuit.

Working only with graphs means we have to treat the decorations, the dynamics, and the compositionality informally, or at least in some other data structure, thereby increasing the overhead of this accounting.

\subsubsection{Bayesian networks: belief and dependence} \label{sec:mtv-bayes}

In computational statistics, one often begins by constructing a model of the causal dependence between events, which can then be interrogated for the purposes of inference or belief-updating.
Such models are typically graphical, with representations as shown below; the nodes are again the circles, and the dashed edge implies the repetition of the depicted motif:
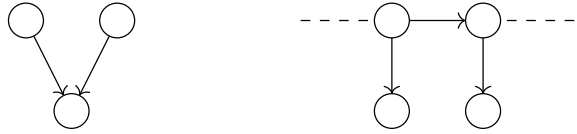
\[ \scalebox{1.2}{\begin{tikzpicture}
	\begin{pgfonlayer}{nodelayer}
		\node [style=dot] (0) at (-1, 1) {$\quad$};
		\node [style=dot] (1) at (0, -1) {$\quad$};
		\node [style=dot] (2) at (1, 1) {$\quad$};
	\end{pgfonlayer}
	\begin{pgfonlayer}{edgelayer}
		\draw [->] (0) to (1);
		\draw [->] (2) to (1);
	\end{pgfonlayer}
\end{tikzpicture}
} \qquad\quad\qquad \scalebox{1.2}{\input{img/bayesnet-ex-2.tikz}} \]
On the left, the graph represents a model of an event with two possible antecedents; on the right, a set of events (or an event, repeated) with a recurrent cause.
Although these graphical models --- otherwise known as \textit{Bayesian networks} --- may encode useful information about causal structure, in themselves they do not encode the information about \textit{how} events are caused; this is data that must be accounted for separately.
And once again, mere graphs are unlike causality in that they are non-compositional: the structure does not explain how, given the causal dependence of $B$ on $A$ and $A'$ and of $C$ on $B$, one might model the dependence of $C$ on $A$.

\subsubsection{Computations: sets and functions} \label{sec:mtv-func}

In a similar way, pure computations --- in the sense of transformations between sets of data --- are often depicted graphically:
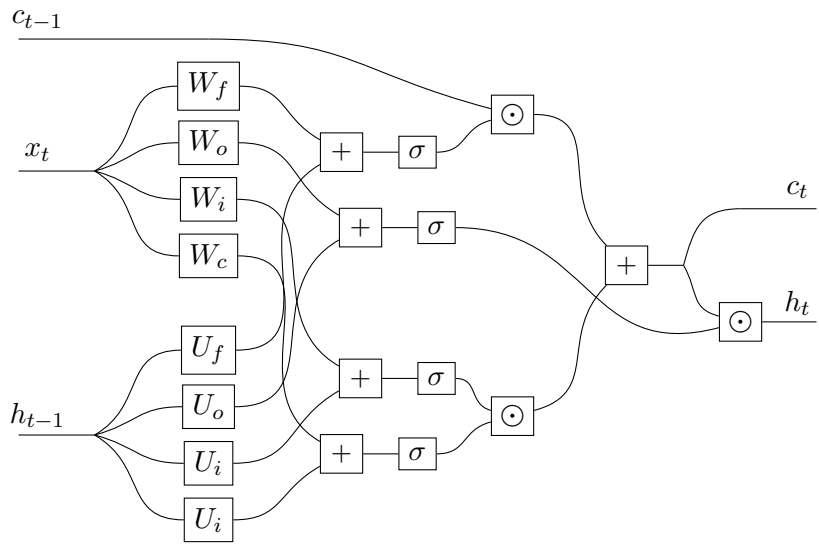
\[ \input{img/lstm-cell.tikz} \]
Here, we have depicted a single `cell' from a long short-term memory network \parencite{Hochreiter1997Long}: a function that ingests three variables ($c_{t-1}$, an internal state; $x_t$, an external input; and $h_{t-1}$, an internal `memory'), and emits two ($c_t$, a new internal state; and $h_t$, an updated memory).
This function is itself composed from other functions, depicted above as boxes.
(One typically takes the variables $c_t,x_t,h_t$ as vectors of given dimension for all $t$, so that the domain and codomain of the function are products of vector spaces; the boxes $W_i$ and $U_i$ represent matrices which act on these vectors; the boxes $+$ and $\odot$ denote elementwise sum and product; the box $\sigma$ represents the elementwise application of a logisitic function; and the splitting of wires represents the copying of data.)
The nodes of the graph in this instance are the functions (boxes), and the edges encode the flow of information.
Once more, however, a purely graphical model does not account for the compositional structure of the computation: we could fill in the boxes with other graphs (representing elaborations of the computations implied), and we could adjoin another such diagram beside and connect the wires where the types match.
To account for this compositionality --- here and in the examples above --- we will need to add something to the structure: we need to move from graphs to categories.

\subsection{From graphs to categories} \label{sec:grph-to-cat}

%
%
%
%
%
%

A category is a directed graph in which edges can be composed: whenever the target of an edge $f$ is the source of another edge $g$, then there must be a composite edge denoted $g\circ f$ whose source is the source of $f$ and whose target is the target of $g$, as in the following diagram.
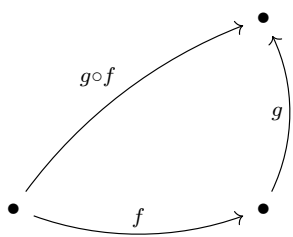
\[\begin{tikzcd}
  &&& \bullet \\
  \\
	\\
	\bullet &&& \bullet
	\arrow["f", curve={height=12pt}, from=4-1, to=4-4]
	\arrow["g", curve={height=12pt}, from=4-4, to=1-4]
	\arrow["{g\circ f}", curve={height=-12pt}, from=4-1, to=1-4]
\end{tikzcd}\]
This composition rule incorporates into the structure a way to allow systems with compatible interfaces to connect to each other, and for the resulting composite system also to be a system of the same `type'; but as we will see, it has some other important consequences.
Firstly, every (`small') category has an underlying directed graph: but because of the composition rule, this underlying graph typically has more edges than the graphs of the examples above, in order to account for the existence of composites.
Secondly, it is the edges, which in a categorical context we will call \textit{morphisms}, that compose: the nodes, which we will call \textit{objects}, represent something like the `interfaces' at which composition is possible.
This means that we cannot just interpret a circuit diagram ``as a category'', whose objects are ensembles of neurons and whose morphisms are their axons: as we will see in \secref{sec:sys-circ}, we need to do something a bit more sophisticated.

Before we get to that, however, we must first define categories precisely.
We will take a graphical approach, with a view to interpreting the above examples categorically, starting with the diagram demonstrating the composition of $g\circ f$: how should we interpret this in a category?
To answer this question, we first need to specify exactly what we mean by `graph'.

\begin{defn} \label{def:graph}
  A \textit{directed graph} $\cat{G}$ is a set $\cat{G}_0$ of \textit{nodes} along with a set $\cat{G}(a, b)$ of \textit{edges} from $a$ to $b$ for each pair $a,b : \cat{G}_0$ of nodes.
  We will sometimes write $\cat{G}_1$ to denote the disjoint union of the sets of edges, $\cat{G}_1 := \sum_{a,b} \cat{G}(a,b)$.
  If $e : \cat{G}(a,b)$ is an edge from $a$ to $b$, we will write this as $e : a\to b$ and call $a$ its \textit{source} or \textit{domain} and $b$ its \textit{target} or \textit{codomain}.
  This assignment of domain and codomain induces a pair of functions, $\dom,\cod : \cat{G}_1 \to \cat{G}_0$ respectively, such that for $e : a\to b$ we have $\dom(e) = a$ and $\cod(e) = b$.
\end{defn}

A category is a graph whose edges can be `associatively' composed together, and where every node has a special edge from itself to itself called its `identity'.

\begin{defn} \label{def:small-cat}
  A \textit{(small) category} $\cat{C}$ is a directed graph whose nodes are each assigned a corresponding \textit{identity} edge and whose edges are equipped with a \textit{composition} operation $\circ$ that is \textit{associative} and \textit{unital} with respect to the identities.
  In the context of categories, we call the nodes $\cat{C}_0$ the \textit{objects} or \textit{0-cells}, and the edges $\cat{C}_1$ the \textit{morphisms} or \textit{1-cells}.

  Identities are assigned by a function $\id : \cat{C}_0 \to \cat{C}_1$ satisfying $\dom(\id_a) = a = \cod(\id_a)$ for every object $a$.
  The composition operation is a family of functions $\circ_{a,b,c} : \cat{C}(b,c)\times\cat{C}(a,b) \to \cat{C}(a,c)$ for each triple of objects $a,b,c$.
  The notation $\cat{C}(a,b)$ indicates the set of all morphisms $a\to b$, for each pair of objects $a$ and $b$; we call this set the \textit{hom set} from $a$ to $b$.

  Given morphisms $f:a\to b$ and $g:b\to c$, their composite $a\xto{f}b\xto{g}c$ is written $g\circ f$, which we can read as ``$g$ after $f$''.

  Associativity means that $h\circ(g\circ f) = (h\circ g)\circ f$, and so we can omit the parentheses to write $h\circ g\circ f$ without ambiguity.
  Unitality means that, for every morphism $f : a\to b$, we have $\id_b\circ f = f = f\circ\id_a$.
\end{defn}

\begin{rmk}
  We say \textit{small} category to mean that both the collection of objects $\cat{C}_0$ and the collection of morphisms $\cat{C}_1$ is a true set, rather than a proper class.
  We will say a category is \textit{locally small} if, for every pair $a,b$ of objects in $\cat{C}$, the hom set $\cat{C}(a,b)$ is a set (rather than a proper class); this allows for the collection of objects still to be a proper class, while letting us avoid ``size issues'' such as Russell's paradox in the course of normal reasoning.

  More precisely, we can fix a `universe' of sets, of size assumed to be smaller than a hypothesized (and typically inaccessible) cardinal $\aleph_i$.
  Then we say that a category is locally small with respect to $\aleph_i$ if every hom set is within this universe, or small if both $\cat{C}_0$ and $\cat{C}_1$ are.
  We say that a category is \textit{large} if it is not small, but note that the `set' of objects or morphisms of a large category may still be a `set', just in a larger universe: a universe whose sets are of cardinality at most $\aleph_{i+1} > \aleph_i$.

  In the remainder of this thesis, we will typically assume categories to be locally small with respect to a given (but unspecified) universe.
\end{rmk}

Our first example of a category is in some sense the foundation of basic category theory, and supplies a classic illustration of size issues.

\begin{ex}
  The category $\Set$ has sets as objects and functions as morphisms.
  The identity on a set $A$ is the identity function $\id_A:A\to A:a\mapsto a$.
  Composition of morphisms in $\Set$ is function composition: given $f:A\to B$ and $g:B\to C$, their composite is the function $g\circ f:A\to C$ defined for each $a:A$ by $(g\circ f)(a) = g(f(a))$; it is easy to check that function composition is associative.

  Note that $\Set$ is a large category: the set $\Set_0$ of all sets of at most size $\aleph_i$ must live in a larger universe.
\end{ex}

Not all categories are large, of course.
Some are quite small, as the following examples demonstrate.

\begin{ex} \label{ex:graph-schema}
  There is a category with only two objects $0$ and $1$ and four morphisms: the identities $\id_0:0\to 0$ and $\id_1:1\to 1$, and two non-identity morphisms $s,t:0\to 1$, as in the following diagram:
  
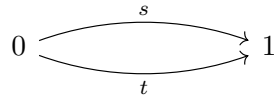
\[\begin{tikzcd}
  0 &&& 1
  \arrow["s", curve={height=-12pt}, from=1-1, to=1-4]
  \arrow["t"', curve={height=12pt}, from=1-1, to=1-4]
  \end{tikzcd}\]
  When depicting categories graphically, we often omit identity morphisms as they are implied by the objects.
\end{ex}

\begin{ex} \label{ex:terminal-cat}
  There is a category, denoted $\Cat{1}$, with a single object $\ast$ and a single morphism, its identity.
\end{ex}

\begin{ex} \label{ex:nat-num-cat}
  The natural numbers $\nn$ form the morphisms of another category with a single object $\ast$: here, composition is addition and the identity morphism $\id_\ast:\ast\to\ast$ is the number $0$.
  Since addition is associative and unital with respect to $0$, this is a well-defined category.
\end{ex}

Since a category is a directed graph equipped with a composition operation, we can `forget' the latter to recover the underlying graph on its own.

\begin{prop} \label{prop:forget-cat}
  Given a category $\cat{C}$, we can obtain a directed graph $(\cat{C}_0,\cat{C}_1)$ by keeping the objects $\cat{C}_0$ and morphisms $\cat{C}_1$ and forgetting the composition and identity functions.
  \begin{proof}
    Take the objects to be the nodes and the morphisms to be the edges.
  \end{proof}
\end{prop}

However, in the absence of other data, obtaining a category from a given graph is a little more laborious, as we must ensure the existence of well-defined composite edges.
The following proposition tells us how we can do this.

\begin{prop} \label{prop:free-cat}
  Given a directed graph $\cat{G}$, we can construct the \textit{free category} generated by $\cat{G}$, denoted $F\cat{G}$, as follows.
  The objects of $F\cat{G}$ are the nodes $\cat{G}_0$ of $\cat{G}$.
  The morphisms $F\cat{G}(a,b)$ from $a$ to $b$ are the \textit{paths} in $\cat{G}$ from $a$ to $b$: finite lists $(e,f,g)$ of edges in which the domain of the first edge is $a$, the codomain of any edge equals the domain of its successor (if any), and the codomain of the last edge is $b$.
  Composition is by concatenation of lists, and the identity morphism for any node is the empty list $()$.
  \begin{proof}
    Let $f := (f_1, \dots, f_l) : a\to b$, $g := (g_1, \dots, g_m) : b\to c$, and $h := (h_1, \dots, h_n) : c\to d$ be paths.
    Then
    \begin{align*}
      h\circ(g\circ f)
      &= (h_1, \dots, h_n)\circ (f_1, \dots, f_l, g_1, \dots, g_m) \\
      &= (f_1, \dots, f_l, g_1, \dots, g_m, h_1, \dots, h_n) \\
      &= (g_1, \dots, g_m, h_1, \dots, h_n) \circ (f_1, \dots, f_l) = (h\circ g)\circ f
    \end{align*}
    so concatenation of lists is associative.
    Concatenation is trivially unital on both right and left: $()\circ (f_1, \dots, f_l) = (f_1, \dots, f_l) = (f_1, \dots, f_l)\circ ()$.
    So the free category as defined is a well-defined category.
  \end{proof}
\end{prop}

\begin{rmk}
  Observe that the underlying graph of $F\cat{G}$ is not in general the same as the original graph $\cat{G}$:
  because the edges of $\cat{G}$ have no composition information (even if, given a pair of edges $a\to b$ and $b\to c$, there is an edge $a\to c$), we needed a canonical method to generate such information, without any extra data.
  Since there is a notion of path in any graph, and since paths are naturally composable, this gives us the canonical method we seek.
\end{rmk}

We begin to see some important differences between categories and graphs, as foreshadowed above.
Categories are somehow more `dynamical' objects, more concerned with movement and change than graphs; later in Chapter \ref{chp:coalg}, we will even see how a general definition of \textit{dynamical system} emerges simply from some of the examples we have already seen.

At this point, to emphasize that categories allow us to study not just individual structures themselves but also the relationships and transformations between structures, we note that directed graphs themselves form a category.

\begin{ex} \label{ex:graph-cat}
  Directed graphs $(\cat{G}_0,\cat{G}_1,\dom_{\cat{G}},\cod_{\cat{G}})$ are the objects of a category, denoted $\Cat{Graph}$.
  Given directed graphs $\cat{G}:=(\cat{G}_0,\cat{G}_1,\dom_{\cat{G}},\cod_{\cat{G}})$ and $\cat{H}:=(\cat{H}_0,\cat{H}_1,\dom_{\cat{H}},\cod_{\cat{H}})$, a morphism $f:\cat{G}\to\cat{H}$ is a \textit{graph homomorphism} from $\cat{G}$ to $\cat{H}$: a pair of functions $f_0:\cat{G}_0\to\cat{G}_0$ and $f_1:\cat{G}_1\to\cat{H}_1$ that preserve the graphical structure in the sense that for every edge $e$ in $\cat{G}$, $f_0(\dom_{\cat{G}}(e)) = \dom_{\cat{H}}(f_1(e))$ and $f_0(\cod_{\cat{G}}(e)) = \cod_{\cat{H}}(f_1(e))$.
  Since graph homomorphisms are pairs of functions, they compose as functions, and the identity morphism on a graph $\cat{G}$ is the pair $(\id_{\cat{G}_0},\id_{\cat{G}_1})$ of identity functions on its sets of nodes and edges.
\end{ex}

In large part, the power of category theory derives from its elevation of relationship and transformation to mathematical prominence: objects are represented and studied in context, and one we gain the ability to compare patterns of relationships across contexts.
By expressing these patterns categorically, we are able to abstract away irrelevant detail, and focus on the fundamental structures that drive phenomena of interest; and since these patterns and abstract structures are again expressed in the same language, we can continue to apply these techniques, to study phenomena from diverse perspectives.
Indeed, as we will soon see, category theory is `homoiconic', able to speak in its language about itself.

Accordingly, it is often helpful to apply graphical or diagrammatic methods to reason about categories: for example, to say that two (or more) morphisms are actually equal.
We can illustrate this using the category $\Cat{Graph}$: the definition of graph homomorphism requires two equalities to be satisfied.
These equalities say that two (composite) pairs of functions are equal; since functions are morphisms in $\Set$, this is the same as saying that they are equal as morphisms there.
Using the fact that $\Set$ has an underlying graph, we can represent these morphisms graphically, as in the following two diagrams:
\begin{equation} \label{eq:graph-hom-nat} \begin{tikzcd}
  {\cat{G}_1} && {\cat{H}_1} \\
  \\
  {\cat{G}_0} && {\cat{H}_0}
  \arrow["{f_1}", from=1-1, to=1-3]
  \arrow["{f_0}"', from=3-1, to=3-3]
  \arrow["{\dom_{\cat{G}}}"', from=1-1, to=3-1]
  \arrow["{\dom_{\cat{H}}}", from=1-3, to=3-3]
\end{tikzcd}
\qquad\qquad
\begin{tikzcd}
  {\cat{G}_1} && {\cat{H}_1} \\
  \\
  {\cat{G}_0} && {\cat{H}_0}
  \arrow["{f_1}", from=1-1, to=1-3]
  \arrow["{f_0}"', from=3-1, to=3-3]
  \arrow["{\cod_{\cat{G}}}"', from=1-1, to=3-1]
  \arrow["{\cod_{\cat{H}}}", from=1-3, to=3-3]
\end{tikzcd} \end{equation}
Then to say that $f_0\circ\dom_{\cat{G}} = \dom_{\cat{H}}\circ f_1$ and $f_0\circ\cod_{\cat{G}} = \cod_{\cat{H}}\circ f_1$ is to say that these diagrams \textit{commute}.

\begin{defn}
  We say that two paths in a graph are \textit{parallel} if they have the same start and end nodes.
  We say that a diagram in a category $\cat{C}$ \textit{commutes} when every pair of parallel paths in the diagram corresponds to a pair of morphisms in $\cat{C}$ that are equal.
\end{defn}

To clarify this definition, we can use category theory to formalize the concept of diagram, which will have the useful side-effect of simultaneously rendering it more general and more precise.

\subsubsection{Diagrams in a category, functorially}

The richness of categorical structure is reflected in the variety of diagrammatic practice, and in this thesis we will encounter a number of formal diagram types.
Nonetheless, there is one type that is perhaps more basic than the rest, which we have already begun to call \textit{diagrams in a category}: these are the categorical analogue of equations in algebra.
Often in category theory, we will be interested in the relationships between more than two morphisms at once, and expressing such relationships by equations quickly becomes cumbersome;
instead, one typically starts with a directed graph and interprets its nodes as objects and its edges as morphisms in one's category of interest.

Formally, this interpretation is performed by taking the category generated by the graph and mapping it `functorially' into the category of interest.
However, in order to account for relationships such as equality between the morphisms represented in the graph, the domain of this mapping cannot be as `free' as in Proposition \ref{prop:free-cat}, as it needs to encode these relationships.
To do this, we can quotient the free category by the given relationships, as we now show.

\begin{prop}[{\textcite[{Prop. II.8.1}]{MacLane1998Categories}}] \label{prop:cat-gen-rel}
  Let $\cat{G}$ be a directed graph, and suppose we are given a relation ${\sim}_{a,b}$ on each set $F\cat{G}(a,b)$ of paths $a\to b$; write $\sim$ for the whole family of relations, and call it a \textit{relation on the category} $\cat{C}$.
  Then there is a category $F\cat{G}/{\sim}$, the \textit{quotient} of the free category $F\cat{G}$ by $\sim$, which we call \textit{the category generated by} $\cat{G}$ \textit{with relations} $\sim$ or simply \textit{generated by} $(\cat{G},{\sim})$.

  The objects of $F\cat{G}/{\sim}$ are again the nodes $\cat{G}_0$.
  The morphisms are equivalence classes of paths according to $\sim$, extended to a congruence:
  suppose $p \sim_{a,b} p'$; then they both belong to the same equivalence class $[p]$, and correspond to the same morphism $[p] : a\to b$ in $F\cat{G}/{\sim}$.
\end{prop}

Before we can make sense of and prove this proposition, and thus establish that composition in $F\cat{G}/{\sim}$ does what we hope, we need to define \textit{congruence}.

\begin{defn}
  Suppose $\sim$ is a relation on the category $\cat{C}$.
  We call $\sim$ a \textit{congruence} when its constituent relations $\sim_{a,b}$ are equivalence relations compatible with the compositional structure of $\cat{C}$.
  This means that
  \begin{enumerate}
  \item if $f \sim_{a,b} f':a\to b$ and $g \sim_{b,c} g':b\to c$, then $g\circ f \sim_{a,c} g'\circ f'$; and
  \item for each pair of objects $a,b:\cat{C}$, $\sim_{a,b}$ is a symmetric, reflexive, transitive relation.
  \end{enumerate}
\end{defn}

The notion of congruence is what allows us to extend the family of relations $\sim$ to composites of morphisms and thus ensure that it is compatible with the categorical structure; constructing the most parsimonious congruence from $\sim$ is the key to the following proof.

\begin{proof}[Proof sketch for Proposition \ref{prop:cat-gen-rel}]
  First of all, we extend $\sim$ to a congruence; we choose the smallest congruence containing $\sim$, and denote it by $\cong$.
  Explicitly, we can construct $\cong$ in two steps.
  First, define an intermediate relation $\simeq$ as the symmetric, reflexive, transitive closure of $\sim$.
  This means that if $f\simeq f'$, then either $f\sim f'$, or $f'\sim f$ (symmetry), or $f = f'$ (reflexivity), or there exists some $\phi:a\to c$ such that $f \sim\phi$ and $\phi\sim f'$ (transitivity).
  Next, define $\cong$ as the closure of $\simeq$ under composition.
  This means that if $\varphi\cong\varphi':a\to c$, then either $\varphi\simeq\varphi'$, or there exist composable pairs $f,f':a\to b$ and $g,g':b\to c$ such that $f\simeq f'$ and $g\simeq g'$, and such that $\varphi = g\circ f$ and $\varphi' = g'\circ f'$.
  To see that $\cong$ is the least congruence on $F\cat{G}$, observe that every congruence must contain it by definition.

  Having constructed the congruence $\cong$, we can form the quotient of $F\cat{G}$ by it, which we denote by $F\cat{G}/{\sim}$ in reference to the generating relation ${\sim}$.
  As in the statement of the proposition, the objects of $F\cat{G}/{\sim}$ are the nodes of $\cat{G}$ and the morphisms are equivalence classes of paths, according to $\cong$; since $\cong$ is by definition an equivalence relation, these equivalence classes are well-defined.
  Moreover, the composite of two equivalence classes of morphisms $[f]:a\to b$ and $[g]:b\to c$ coincides with the equivalence class $[g\circ f]$.
\end{proof}

\begin{ex} \label{ex:FJ}
  To exemplify the notion of category generated with relations, let $\cat{J}$ denote the following directed graph
  \[\begin{tikzcd}
    {G_1} && {H_1} \\
    \\
    {G_0} && {H_0}
    \arrow["{\varphi_1}", from=1-1, to=1-3]
    \arrow["{\varphi_0}"', from=3-1, to=3-3]
    \arrow["{\delta_G}"', from=1-1, to=3-1]
    \arrow["{\delta_H}", from=1-3, to=3-3]
  \end{tikzcd}\]
  and let $\sim$ be the relation $\varphi_0\circ\delta_G\sim\delta_H\circ\varphi_1$.
  Then the category $F\cat{J}/{\sim}$ generated by $(\cat{J},{\sim})$ has four objects ($G_1,G_0,H_1,H_0$) and nine morphisms: an identity for each of the four objects; the morphisms $\varphi_0:G_0\to H_0$, $\varphi_1:G_1\to H_1$, $\delta_G:G_1\to G_0$, and $\delta_H:H_1\to H_0$; and a single morphism $G_1\to H_0$, the equivalence class consisting of $\varphi_0\circ\delta_G$ and $\delta_H\circ\varphi_1$.
\end{ex}

The category $F\cat{J}/{\sim}$ generated in this example expresses the commutativity of one of the diagrams defining graph homomorphisms, but as things stand, it is simply a category standing alone: to say that any particular pair of functions $(f_0,f_1)$ satisfies the property requires us to interpret the morphisms $\varphi_0$ and $\varphi_1$ accordingly \textit{as those functions}.
That is, to interpret the diagram, we need to translate it, by mapping $F\cat{J}/{\sim}$ into $\Set$.
Such a mapping of categories is known as a \textit{functor}.

\begin{defn} \label{def:functor}
  A \textit{functor} $F : \cat{C}\to\cat{D}$ from the category $\cat{C}$ to the category $\cat{D}$ is a pair of functions $F_0 : \cat{C}_0 \to \cat{D}_0$ and $F_1 : \cat{C}_1 \to \cat{D}_1$ between the sets of objects and morphisms that preserve domains, codomains, identities and composition, meaning that $F_0(\dom_{\cat{C}}(f)) = \dom_{\cat{D}}(F_1(f))$ and $F_0(\cod_{\cat{C}}(f)) = \cod_{\cat{D}}(F_1(f))$ for all morphisms $f$, $F_1(\id_a) = \id_{F(a)}$ for all objects $a$, and $F_1(g\circ f) = F_1(g)\circ F_1(f)$ for all composites $g\circ f$ in $\cat{C}$.
\end{defn}

\begin{rmk}
  Note that we could equivalently say that a functor $\cat{C}\to\cat{D}$ is a homomorphism from the underlying graph of $\cat{C}$ to that of $\cat{D}$ that is additionally \textit{functorial}, meaning that it preserves identities and composites.
\end{rmk}

\begin{notation}
  Although a functor $F$ consists of a pair of functions $(F_0,F_1)$, we will typically write just $F$ whether it is applied to an object or a morphism, since the distinction will usually be clear from the context.
  Since function composition (and hence application) is associative, we will also often omit brackets, writing $Fa$ for $F(a)$, except where it is helpful to leave them in.
\end{notation}

For each object $c$ in a category $\cat{C}$, there are two very important functors, the \textit{hom} functors, which exhibit $\cat{C}$ in $\Set$ ``from the perspective'' of $c$ by returning the hom sets out of and into $c$.

\begin{defn} \label{def:covar-hom}
  Given an object $c:\cat{C}$, its \textit{covariant hom functor} $\cat{C}(c,-):\cat{C}\to\Set$ is defined on objects $x$ by returning the hom sets $\cat{C}(c,x)$ and on morphisms $g:x\to y$ by returning the \textit{postcomposition} function $\cat{C}(c,g):\cat{C}(c,x)\to\cat{C}(c,y)$ defined by mapping morphisms $f:c\to x$ in the set $\cat{C}(c,x)$ to the composites $g\circ f:c\to y$ in $\cat{C}(c,y)$.
  To emphasize the action of $\cat{C}(c,g)$ by postcomposition, we will sometimes write it simply as $g\circ(-)$.
  (That $\cat{C}(c,-)$ is a well-defined functor follows immediately from the unitality and associativity of composition in $\cat{C}$.)
\end{defn}

The covariant hom functor $\cat{C}(c,-)$ ``looks forward'' along morphisms emanating \textit{out of} $c$, in the direction that these morphisms point, and therefore in the direction of composition in $\cat{C}$: it is for this reason that we say it is \textit{covariant}.
Dually, it is of course possible to ``look backward'' at morphisms pointing \textit{into} $c$.
Since this means looking contrary to the direction of composition in $\cat{C}$, we say that the resulting backwards-looking hom functor is \textit{contravariant}.
To define it as a functor in the sense of Definition \ref{def:functor}, we perform the trick of swapping the direction of composition in $\cat{C}$ around and then defining a covariant functor accordingly.

\begin{defn} \label{def:op-cat}
  For any category $\cat{C}$ there is a corresponding \textit{opposite category} $\cat{C}\op$ with the same objects as $\cat{C}$ and where the hom set $\cat{C}\op(a,b)$ is defined to be the `opposite' hom set in $\cat{C}$, namely $\cat{C}(b,a)$.
  Identity morphisms are the same in $\cat{C}\op$ as in $\cat{C}$, but composition is also reversed.
  If we write $\circ$ for composition in $\cat{C}$ and $\mathbin{{\circ}^{\mathrm{op}}}$ for composition in $\cat{C}\op$, then, given morphisms $g:c\to b$ and $f:b\to a$ in $\cat{C}\op$ corresponding to morphisms $g:b\to c$ and $f:a\to b$ in $\cat{C}$, their composite $f\mathbin{{\circ}^{\mathrm{op}}}g:c\to a$ in $\cat{C}\op$ is the morphism $g\circ f:a\to c$ in $\cat{C}$.
  (Observe that this makes $\cat{C}\op$ a well-defined category whenever $\cat{C}$ is.)
\end{defn}

\begin{rmk} \label{rmk:dual-co}
  Because we can always form opposite categories in this way, categorical constructions often come in two forms: one in $\cat{C}$, and a `dual' one in $\cat{C}\op$.
  Typically, we use the prefix \textit{co-} to indicate such a dual construction: so if we have a construction in $\cat{C}$, then its dual in $\cat{C}\op$ would be called a \textit{co}construction.
\end{rmk}

The dual of the covariant hom functor $\cat{C}(c,-):\cat{C}\to\Set$ is the contravariant hom functor.

\begin{defn} \label{def:contra-hom}
  Given an object $c:\cat{C}$, its \textit{contravariant hom functor} $\cat{C}(-,c):\cat{C}\op\to\Set$ is defined on objects $x$ by returning the hom sets $\cat{C}(x,c)$.
  Given a morphism $f:x\to y$ in $\cat{C}$, we define the \textit{precomposition} function $\cat{C}(f,c):\cat{C}(y,c)\to\cat{C}(x,c)$ by mapping morphisms $g:y\to c$ in the set $\cat{C}(y,c)$ to the composites $g\circ f:x\to c$ in $\cat{C}(x,c)$.
  To emphasize the action of $\cat{C}(f,c)$ by precomposition, we will sometimes write it simply as $(-)\circ f$.
  (That $\cat{C}(-,c)$ is a well-defined functor again follows from the unitality and associativity of composition in $\cat{C}$ and hence in $\cat{C}\op$.)
\end{defn}

\begin{rmk}
  A \textit{contravariant} functor on $\cat{C}$ is a (covariant) functor on $\cat{C}\op$.
\end{rmk}

\begin{notation} \label{not:pull-push}
  In line with other mathematical literature, we will also occasionally write the precomposition function $(-)\circ f$ as $f^\ast$; dually, we can write the postcomposition function $g\circ(-)$ as $g_\ast$.
  In these forms, the former action $f^\ast$ is also known as \textit{pullback} along $f$, as it ``pulls back'' morphisms along $f$, and the latter action $g_\ast$ is also known as \textit{pushforward} along $g$, as it ``pushes forward'' morphisms along $g$.
  There is a close relationship between the pulling-back described here and the universal construction also known as pullback (Example \ref{ex:pullback}): $f^\ast(-)$ defines a functor which acts by the universal construction on objects and by precomposition on morphisms, which we spell out in Definition \ref{def:self-idx}.
\end{notation}

Functors are the homomorphisms of categories, and just as graphs and their homomorphisms form a category, so do categories and functors.

\begin{ex} \label{ex:cat}
  The category $\Cat{Cat}$ has categories for objects and functors for morphisms.
  The identity functor $\id_{\cat{C}}$ on a category $\cat{C}$ is the pair $(\id_{\cat{C}_0},\id_{\cat{C}_1})$ of identity functions on the sets of objects and morphisms.
  Since functors are pairs of functions, functor composition is by function composition, which is immediately associative and unital with respect to the identity functors so defined.
  Note that, without a restriction on size, $\Cat{Cat}$ is a large category, like $\Set$.
\end{ex}

As an example, we observe that the construction of the category $F\cat{G}/{\sim}$ generated by $(\cat{G},{\sim})$ from the free category $F\cat{G}$ is functorial.

\begin{ex} \label{ex:quot-cat-proj}
  There is a `projection' functor $[\cdot]:F\cat{G}\to F\cat{G}/{\sim}$.
  It maps every object to itself, and every morphism to the corresponding equivalence class.
  The proof of Proposition \ref{prop:cat-gen-rel} demonstrated the functoriality:
  identities are preserved by definition, and we have $[g\circ f] = [g]\circ[f]$ by construction.
\end{ex}

With the notion of functor to hand, we can formalize the concept of diagram simply as follows.

\begin{defn} \label{def:J-diagram}
  A $J$\textit{-shaped diagram} in a category $\cat{C}$ is a functor $D : J\to\cat{C}$.
  Typically, $J$ is a small category generated from a graph with some given relations, and the functor $D$ interprets $J$ in $\cat{C}$.
\end{defn}

\begin{ex}
  The diagrams expressing the commutativity conditions for a graph homomorphism \eqref{eq:graph-hom-nat} are therefore witnessed by a pair of functors $F\cat{J}/{\sim}\to\Set$ from the category $F\cat{J}/{\sim}$ generated in Example \ref{ex:FJ} into $\Set$: each functor interprets $\varphi_0$ and $\varphi_1$ as $f_0$ and $f_1$ respectively, while one functor interprets $\delta_G$ as $\dom_{\cat{G}}$ and $\delta_H$ as $\dom_{\cat{H}}$ and the other interprets $\delta_G$ as $\cod_{\cat{G}}$ and $\delta_H$ as $\cod_{\cat{H}}$.
  The fact that there is only a single morphism $G_1\to H_0$ in $F\cat{J}/{\sim}$ (even though there are two in $F\cat{J}$) encodes the requirements that $f_0\circ\dom_{\cat{G}} = \dom_{\cat{H}}\circ f_1$ and $f_0\circ\cod_{\cat{G}} = \cod_{\cat{H}}\circ f_1$.
\end{ex}

Throughout this thesis, we will see the utility of diagrams as in Definition \ref{def:J-diagram}:
not only will they be useful in reasoning explicitly about categorical constructions, but in \secref{sec:limits} they will also be used to formalize `universal constructions', another concept which exhibits the power of category theory.

Despite this, `mere' categories and their diagrams are in some ways not expressive enough:
often we will want to encode looser relationships than strict equality, or to compose diagrams together by `pasting' them along common edges; we may even want to consider morphisms between morphisms!
For this we will need to `enrich' our notion of category accordingly.

\section{Connecting the connections} \label{sec:enrich}

As we have indicated, basic category theory is not sufficient if we want to encode information about the relationships between morphisms into the formal structure.
In this section, we will see how to enrich the notion of category by letting the morphisms collect into more than just sets, and how this leads naturally to higher category theory, where we have morphisms between the morphisms, and from there to the notion of adjunction, with which we can translate concepts faithfully back and forth between contexts.
Amidst the development, we discuss the concept of ``functorial semantics'' from a scientific perspective, considering how categorical tools let us supply rich semantics for structured models of complex systems such as the brain.

\subsection{Enriched categories}

We can think of the condition that a diagram commutes --- or equivalently the specification of an equivalence relation on its paths --- as a `filling-in' of the diagram with some extra data.
For example, we can `fill' the diagram depicting the graph homomorphism condition $f_0\circ\dom_{\cat{G}} = \dom_{\cat{H}}\circ f_1$ with some annotation or data witnessing this relation, as follows:
\[\begin{tikzcd}
  {\cat{G}_1} && {\cat{H}_1} \\
  \\
  {\cat{G}_0} && {\cat{H}_0}
  \arrow["{f_1}", from=1-1, to=1-3]
  \arrow["{f_0}"', from=3-1, to=3-3]
  \arrow["{\dom_{\cat{G}}}"', from=1-1, to=3-1]
  \arrow["{\dom_{\cat{H}}}", from=1-3, to=3-3]
  \arrow[shorten <=24pt, shorten >=24pt, Rightarrow, no head, from=3-1, to=1-3]
\end{tikzcd}\]
If we have a composite graph homomorphism $g\circ f:\cat{G}\to\cat{I}$, we should be able to paste the commuting diagrams of the factors together and fill them in accordingly:
\[\begin{tikzcd}
  {\cat{G}_1} && {\cat{H}_1} && {\cat{I}_1} \\
  \\
  {\cat{G}_0} && {\cat{H}_0} && {\cat{I}_0}
  \arrow["{\dom_{\cat{G}}}"', from=1-1, to=3-1]
  \arrow["{\dom_{\cat{H}}}"{description}, from=1-3, to=3-3]
  \arrow["{\dom_{\cat{I}}}", from=1-5, to=3-5]
  \arrow["{f_1}", from=1-1, to=1-3]
  \arrow["{g_1}", from=1-3, to=1-5]
  \arrow["{f_0}"', from=3-1, to=3-3]
  \arrow["{g_0}"', from=3-3, to=3-5]
  \arrow[shorten <=24pt, shorten >=24pt, Rightarrow, no head, from=3-1, to=1-3]
  \arrow[shorten <=24pt, shorten >=24pt, Rightarrow, no head, from=3-3, to=1-5]
\end{tikzcd}\]
and we should be able to `compose' the filler equalities to obtain the diagram for the composite:
\[\begin{tikzcd}
  {\cat{G}_1} && {\cat{H}_1} && {\cat{I}_1} \\
  \\
  {\cat{G}_0} && {\cat{H}_0} && {\cat{I}_0}
  \arrow["{\dom_{\cat{G}}}"', from=1-1, to=3-1]
  \arrow["{\dom_{\cat{I}}}", from=1-5, to=3-5]
  \arrow["{f_1}", from=1-1, to=1-3]
  \arrow["{g_1}", from=1-3, to=1-5]
  \arrow["{f_0}"', from=3-1, to=3-3]
  \arrow["{g_0}"', from=3-3, to=3-5]
  \arrow[shorten <=48pt, shorten >=48pt, Rightarrow, no head, from=3-1, to=1-5]
\end{tikzcd} \; .\]

The extra data with which we have filled these diagrams sits `between' the morphisms, and so if we wish to incorporate it into the categorical structure, we must move beyond mere sets, for sets are just collections of elements, with nothing ``in between''.
What we will do is allow the hom sets of a category to be no longer sets, but objects of another `enriching' category.
Now, observe that, in pasting the two diagrams above together, we had to place them side by side: this means that any suitable enriching category must come equipped with an operation that allows us to place its objects side by side; in the basic case, where our categories just have hom sets, the enriching category is $\Set$, and this side-by-side operation is the product of sets.

\begin{defn} \label{def:prod-set}
  Given sets $A$ and $B$, their \textit{product} is the set $A\times B$ whose elements are pairs $(a,b)$ of an element $a:A$ with an element $b:B$.
\end{defn}

We have already made use of the product of sets above, when we defined the composition operation for (small) categories in Definition \ref{def:small-cat}.
In general, however, we don't need precisely a product; only something weaker, which we call \textit{tensor}.
In order to define it, we need the notion of \textit{isomorphism}.

\begin{defn} \label{def:isomorphism}
  A morphism $l : c\to d$ in a 1-category is an \textit{isomorphism} if there is a morphism $r : d\to c$ such that $l\circ r = \id_d$ and $\id_c = r\circ l$.
  We say that $l$ and $r$ are mutually \textit{inverse}.
\end{defn}

\begin{defn} \label{def:tensor}
  We will say that a category $\cat{C}$ has a \textit{tensor product} if it is equipped with a functor $\otimes:\cat{C}\times\cat{C}\to\cat{C}$ along with an object $I:\cat{C}$ called the \textit{tensor unit} and three families of isomorphisms:
  \begin{enumerate}
  \item \textit{associator} isomorphisms $\alpha_{a,b,c} : (a\otimes b)\otimes c\xto{\sim} a\otimes(b\otimes c)$ for each triple of objects $a,b,c\,$;
  \item \textit{left unitor} isomorphisms $\lambda_a : I\otimes a \xto{\sim} a$ for each object $a$; and
  \item \textit{right unitor} isomorphisms $\rho_a : a\otimes I \xto{\sim} a$ for each object $a$.
  \end{enumerate}
\end{defn}

\begin{rmk}
  The notion of tensor product forms part of the definition of \textit{monoidal category}, which we will come to in \secref{sec:mon-cat}.
  Beyond having a tensor product, a monoidal category must have structure isomorphisms that are \textit{coherent} with respect to the ambient categorical structure, which itself satisfies properties of associativity and unitality; this is an echo of the \textit{microcosm principle} which we discuss in Remark \ref{rmk:pseudomonoid}.
  However, to give the full definition the notion of monoidal category requires us to introduce the notion of \textit{natural transformation}, which we otherwise do not need until Definition \ref{def:nat-trans}; moreover, questions of coherence of tensor products will not yet arise.
\end{rmk}

Unsurprisingly, the product of sets gives us our first example of a tensor product structure.

\begin{ex} \label{ex:prod-set}
  The product of sets gives us a tensor product $\times:\Set\times\Set\to\Set$.
  To see that it is functorial, observe that, given a product of sets $A\times B$ and a function $f:A\to A'$, we naturally obtain a function $f\times B:A\times B\to A\times A'$ by applying $f$ only to the $A$-components of the elements of the product $A\times B$; likewise given a function $g:B\to B'$.
  The unit of the tensor product structure is the set $1$ with a single element $\ast$.
  The associator and unitors are almost trivial: for associativity, map $((a,b),c)$ to $(a,(b,c))$.
\end{ex}

Using the tensor product to put morphisms side by side, we can define the notion of \textit{enriched category}.

\begin{defn} \label{def:enriched-cat}
  Suppose $(\cat{E},\otimes,I,\alpha,\lambda,\rho)$ is a category equipped with a tensor product.
  An $\cat{E}$\textit{-category} $\cat{C}$, or \textit{category} $\cat{C}$ \textit{enriched in} $\cat{E}$, constitutes
  \begin{enumerate}
  \item a set $\cat{C}_0$ of \textit{objects};
  \item for each pair $(a,b)$ of $\cat{C}$-objects, an $\cat{E}$-object $\cat{C}(a,b)$ of \textit{morphisms} from $a$ to $b$;
  \item for each object $a$ in $\cat{C}$, an $\cat{E}$-morphism $\id_a : I\to\cat{C}(a,a)$ witnessing \textit{identity}; and
  \item for each triple $(a,b,c)$ of $\cat{C}$-objects, an $\cat{E}$-morphism $\circ_{a,b,c}:\cat{C}(b,c)\otimes\cat{C}(a,b)\to\cat{C}(a,c)$ witnessing \textit{composition};
  \end{enumerate}
  such that composition is unital, \textit{i.e.} for all $a,b:\cat{C}$
  \begin{equation*}
    \begin{tikzcd}
	    {\cat{C}(a,b)\otimes I} && {\cat{C}(a,b)\otimes\cat{C}(a,a)} \\
	    \\
	    && {\cat{C}(a,b)}
	    \arrow["{\rho_{\cat{C}(a,b)}}"', from=1-1, to=3-3]
	    \arrow["{{\cat{C}(a,b)}\otimes{\id_a}}", from=1-1, to=1-3]
	    \arrow["{\circ_{a,a,b}}", from=1-3, to=3-3]
    \end{tikzcd}
    \text{and}
    \begin{tikzcd}
	    {\cat{C}(a,b)\otimes\cat{C}(a,a)} && {I\otimes\cat{C}(a,b)} \\
	    \\
	      {\cat{C}(a,b)}
	      \arrow["{\lambda_{\cat{C}(a,b)}}", from=1-3, to=3-1]
	      \arrow["{{\id_b}\otimes{\cat{C}(a,b)}}"', from=1-3, to=1-1]
	      \arrow["{\circ_{a,b,b}}"', from=1-1, to=3-1]
    \end{tikzcd}
    \, ,
  \end{equation*}
  and associative, \textit{i.e.} for all $a,b,c,d:\cat{C}$
  \[
  \begin{tikzcd}
	  {\bigl(\cat{C}(c,d)\otimes\cat{C}(b,c)\bigr)\otimes\cat{C}(a,b)} && {\cat{C}(c,d)\otimes\bigl(\cat{C}(b,c)\otimes\cat{C}(a,b)\big)} \\
	  \\
    {\cat{C}(b,d)\otimes\cat{C}(a,b)} && {\cat{C}(c,d)\otimes\cat{C}(a,c)} \\
	  \\
	  & {\cat{C}(a,d)}
	  \arrow["\alpha_{a,b,c,d}", from=1-1, to=1-3]
	  \arrow["{\circ_{b,c,d}\otimes\cat{C}(a,b)}"', from=1-1, to=3-1]
	  \arrow["{\cat{C}(c,d)\otimes\circ_{a,b,c}}", from=1-3, to=3-3]
	  \arrow["{\circ_{a,b,d}}"', from=3-1, to=5-2]
	  \arrow["{\circ_{a,c,d}}", from=3-3, to=5-2]
  \end{tikzcd}
  \, . \]
\end{defn}

Our first example of enriched categories validates the definition.

\begin{ex}
  A locally small category is a category enriched in $(\Set,\times,1)$.
\end{ex}

\begin{rmk} \label{rmk:element}
  In $\Set$, morphisms $1\to A$ out of the unit set $1$ correspond to elements of $A$: each such morphism is a function mapping the unique element $\ast:1$ to its corresponding element of $A$.
  This is why identities in enriched category theory are given by morphisms $I\to\cat{C}(a,a)$, and it is also why we will call morphisms out of a tensor unit \textit{generalized elements}.
  (Even more generally, we might say that morphisms $X\to A$ are generalized elements \textit{of shape} $X$, reflecting our use of the word `shape' to describe the domain of a diagram.)
\end{rmk}

To incorporate nontrivial fillers into our diagrams, we move instead to enrichment in \textit{pro}sets.

\begin{ex}
  A \textit{preordered set} or \textit{proset} is a category where there is at most one morphism between any two objects.
  The objects of such a `thin' category are the points of the proset, and the morphisms encode the (partial) ordering of the points; as a result, they are often written $a\leq a'$.
  Functors between prosets are functions that preserve the ordering, and the restriction of $\Cat{Cat}$ to prosets produces a category that we denote by $\Cat{Pro}$.
  The product of sets extends to prosets as follows: if $A$ and $B$ are prosets, then their product is the proset $A\times B$ whose points are the points of the product set $A\times B$ and a morphism $(a,b)\leq (a',b')$ whenever there are morphisms $a\leq a'$ and $b\leq b'$ in $A$ and $B$ respectively.

  A category enriched in $\Cat{Pro}$ is therefore a category whose hom sets are (pre)ordered and whose composition operation preserves this ordering, which we can illustrate as follows:
  \[\begin{tikzcd}
    A && B && C
    \arrow[""{name=0, anchor=center, inner sep=0}, "f", curve={height=-12pt}, from=1-1, to=1-3]
    \arrow[""{name=1, anchor=center, inner sep=0}, "g", curve={height=-12pt}, from=1-3, to=1-5]
    \arrow[""{name=2, anchor=center, inner sep=0}, "{f'}"', curve={height=12pt}, from=1-1, to=1-3]
    \arrow[""{name=3, anchor=center, inner sep=0}, "{g'}"', curve={height=12pt}, from=1-3, to=1-5]
    \arrow["{\rotatebox[origin=c]{270}{$\leq$}}"{description}, draw=none, from=0, to=2]
    \arrow["{\rotatebox[origin=c]{270}{$\leq$}}"{description}, draw=none, from=1, to=3]
  \end{tikzcd}
  \quad\; \xmapsto{\circ} \;\quad
  \begin{tikzcd}
	  A &&&& C
	  \arrow[""{name=0, anchor=center, inner sep=0}, "{g\circ f}", curve={height=-18pt}, from=1-1, to=1-5]
	  \arrow[""{name=1, anchor=center, inner sep=0}, "{g'\circ f'}"', curve={height=18pt}, from=1-1, to=1-5]
	  \arrow["{\rotatebox[origin=c]{270}{$\leq$}}"{description}, draw=none, from=0, to=1]
  \end{tikzcd}\]
  We can see how enrichment in $\Cat{Pro}$ generalizes the situation with which we introduced this section, where we considered filling diagrams with data witnessing the equality of morphisms: here we have \textit{inequality} data, and it is not hard to see how enriched composition encompasses the pasting-and-composing discussed there (just replace the cells here by the squares above).
\end{ex}

In order to make these filled diagrams precise, we need to extend the notion of functor to the enriched setting; and so we make the following definition.

\begin{defn} \label{def:enr-func}
  Suppose $\cat{C}$ and $\cat{D}$ are $\cat{E}$-categories.
  Then an $\cat{E}$\textit{-functor} $F$ constitutes
  \begin{enumerate}
  \item a function \(F_0 : \cat{C}_0 \to \cat{D}_0\) between the sets of objects; and
  \item for each pair \((a,b) : \cat{C}_0 \times \cat{C}_0\) of objects in \(\cat{C}\), an \(\cat{E}\)-morphism \(F_{a,b} : \cat{C}(a,b) \to \cat{D}(F_0 a, F_0 b)\)
  \end{enumerate}
  which preserve identities
  \[ \begin{tikzcd}
	  & I \\
	  \\
	    {\cat{C}(a,a)} && {\cat{D}(F_0a,F_0a)}
	    \arrow["{\id_a}"', from=1-2, to=3-1]
	    \arrow["{\id_{F_0a}}", from=1-2, to=3-3]
	    \arrow["{F_{a,a}}"', from=3-1, to=3-3]
  \end{tikzcd} \]
  and composition
  \[ \begin{tikzcd}
	  {\cat{C}(b,c)\otimes\cat{C}(a,b)} && {\cat{C}(a,c)} \\
	  \\
	    {\cat{D}(F_0b,F_0c)\otimes\cat{D}(F_0a,F_0b)} && {\cat{D}(F_0a,F_0c)}
	    \arrow["{F_{b,c}\otimes F_{a,b}}"', from=1-1, to=3-1]
	    \arrow["{\circ_{a,b,c}}", from=1-1, to=1-3]
	    \arrow["{F_{a,c}}", from=1-3, to=3-3]
	    \arrow["{\circ_{F_0a,F_0b,F_0c}}"', from=3-1, to=3-3]
  \end{tikzcd}
  \, . \]
\end{defn}

A diagram in an $\cat{E}$-enriched category $\cat{C}$ is therefore a choice of $\cat{E}$-enriched category $J$ (the diagram's shape) and an $\cat{E}$-functor $J\to\cat{C}$.
$J$ encodes the objects, morphisms and relationships of the diagram, and the functor interprets it in $\cat{C}$.
In this enriched setting, we need not quotient parallel paths in the shape of a diagram (which destroys their individuality); instead, we have extra data (the fillers) encoding their relationships.

\subsection{2-categories}

We have seen that filling the cells of a diagram with inequalities pushes us to consider enrichment in $\Cat{Pro}$.
Since $\Cat{Pro}$ is the category of categories with at most one morphism (\textit{i.e.}, the inequality) between each pair of objects, a natural generalization is to allow a broader choice of filler: that is, to allow there to be morphisms between morphisms.
This means moving from enrichment in $\Cat{Pro}$ to enrichment in $\Cat{Cat}$, and hence to the notion of \textit{2-category}.
We therefore make the following definition.

\begin{defn}
  A \textit{strict 2-category} is a category enriched in the 1-category $\Cat{Cat}$.
  This means that, instead of hom sets, a 2-category has hom \textit{categories}: the objects of these hom categories are the \textit{1-cells} of the 2-category, and the morphisms of the hom categories are the \textit{2-cells}; the \textit{0-cells} of the 2-category are its objects.
  To distinguish between the composition defined by the enriched category structure from the composition within the hom categories, we will sometimes call the former \textit{horizontal} and the latter \textit{vertical} composition.
\end{defn}

\begin{rmk}
  We say \textit{1-category} above to refer to the `1-dimensional' notion of category defined in Definition \ref{def:small-cat}.
\end{rmk}

\begin{rmk} \label{rmk:weak-enr}
  We say \textit{strict} to mean that the associativity and unitality of composition hold up to equality; later, it will be helpful to weaken this so that associativity and unitality only hold up to ``coherent isomorphism'', meaning that instead of asking the diagrams in Definition \ref{def:enriched-cat} simply to commute (and thus be filled by equalities), we ask for them to be filled with `coherently' defined isomorphism.
  Weakening 2-categorical composition in this way leads to the notion of \textit{bicategory} (\secref{sec:bicat}).
\end{rmk}

In order to give a well-defined notion of enrichment in $\Cat{Cat}$, we need to equip it with a suitable tensor product structure; for this, we can extend the product of sets to categories, as follows.

\begin{prop} \label{prop:prod-cat}
  Given categories $\cat{C}$ and $\cat{D}$, we can form the \textit{product} category $\cat{C}\times\cat{D}$.
  Its set of objects $(\cat{C}\times\cat{D})_0$ is the product set $\cat{C}_0\times\cat{D}_0$.
  Similarly, a morphism $(c,d)\to(c',d')$ is a pair $(f,g)$ of a morphism $f : c\to c'$ in $\cat{C}$ with a morphism $g:d\to d'$ in $\cat{D}$; hence $(\cat{C}\times\cat{D})_1 = \cat{C}_1\times\cat{D}_1$.
  Composition is given by composing pairwise in $\cat{C}$ and $\cat{D}$: $(f',g')\circ(f,g) := (f'\circ f,g'\circ g)$.
  \begin{proof}
    That composition is associative and unital in $\cat{C}\times\cat{D}$ follows immediately from those properties in the underlying categories $\cat{C}$ and $\cat{D}$.
  \end{proof}
\end{prop}

\begin{rmk} \label{rmk:hom-bifunc}
  Using the product of categories, we can gather the co- and contravariant families of hom functors $\cat{C}(c,-)$ and $\cat{C}(-,c)$ into a single \textit{hom functor} $\cat{C}(-,=):\cat{C}\op\times\cat{C}\to\Set$, mapping $(x,y):\cat{C}\op\times\cat{C}$ to $\cat{C}(x,y)$.
\end{rmk}

\begin{prop} \label{prop:prod-func}
  The product of categories extends to a functor $\times:\Cat{Cat}\times\Cat{Cat}\to\Cat{Cat}$.
  Given functors $F:\cat{C}\to\cat{C}'$ and $G:\cat{D}\to\cat{D}'$, we obtain a functor $F\times G$ by applying $F$ to the left factor of the product $\cat{C}\times\cat{D}$ and $G$ to the right.
  \begin{proof} Sufficiently obvious that we omit it. \end{proof}
\end{prop}

The archetypal 2-category is $\Cat{Cat}$ itself, as we will now see: morphisms between functors are called \textit{natural transformation}, and they will play an important rôle throughout this thesis.

\begin{defn} \label{def:nat-trans}
  Suppose $F$ and $G$ are functors $\cat{C}\to\cat{D}$.
  A \textit{natural transformation} $\alpha : F\Rightarrow G$ is a family of morphisms $\alpha_c : F(c)\to G(c)$ in $\cat{D}$ and indexed by objects $c$ of $\cat{C}$, such that for any morphism $f : c\to c'$ in $\cat{C}$, the following diagram --- called a \textit{naturality square} for $\alpha$ --- commutes:
  \[ \begin{tikzcd}
	Fc && Gc \\
	\\
	{Fc'} && {Gc'}
	\arrow["{\alpha_c}", from=1-1, to=1-3]
	\arrow["{\alpha_{c'}}"', from=3-1, to=3-3]
	\arrow["Ff"', from=1-1, to=3-1]
	\arrow["Gf", from=1-3, to=3-3]
  \end{tikzcd} \, . \]
  When the component 1-cells of a natural transformation $\alpha$ are all isomorphisms, then we call $\alpha$ a \textit{natural isomorphism}.
\end{defn}

\begin{ex} \label{ex:hom-nat-trans}
  Every morphism $f:a\to b$ in a category $\cat{C}$ induces a (contravariant) natural transformation $\cat{C}(f,-):\cat{C}(b,-)\Rightarrow\cat{C}(a,-)$ between covariant hom functors, acting by precomposition.
  Dually, every morphism $h:c\to d$ induces a (covariant) natural transformation $\cat{C}(-,h):\cat{C}(-,c)\Rightarrow\cat{C}(-,d)$ between contravariant hom functors, acting by postcomposition.
  To see that these two families are natural, observe that the square below left must commute for all objects $a,b,c:\cat{C}$ and morphisms $f:a\to b$ and $h:c\to d$, by the associativity of composition in $\cat{C}$ (as illustrated on the right)
  \[\begin{tikzcd}
    {\cat{C}(b,c)} && {\cat{C}(a,c)} \\
    \\
    {\cat{C}(b,d)} && {\cat{C}(a,d)}
    \arrow["{\cat{C}(f,c)}", from=1-1, to=1-3]
    \arrow["{\cat{C}(b,h)}"', from=1-1, to=3-1]
    \arrow["{\cat{C}(f,d)}"', from=3-1, to=3-3]
    \arrow["{\cat{C}(a,h)}", from=1-3, to=3-3]
  \end{tikzcd}
  \qquad\qquad
  \begin{tikzcd}
    g && {g\circ f} \\
    \\
    {h\circ g} && {h\circ g\circ f}
    \arrow[maps to, from=1-1, to=1-3]
    \arrow[maps to, from=1-3, to=3-3]
    \arrow[maps to, from=1-1, to=3-1]
    \arrow[maps to, from=3-1, to=3-3]
  \end{tikzcd}\]
  and that it therefore constitutes a naturality square for both $\cat{C}(f,-)$ and $\cat{C}(-,h)$.
  Note also that we can take either path through this square as a definition of the function $\cat{C}(f,h):\cat{C}(b,c)\to\cat{C}(a,d)$ which thus acts by mapping $g:b\to c$ to $h\circ g\circ f:a\to d$.
\end{ex}

\begin{rmk}
  We will see in \secref{sec:mon-cat} that the families of structure morphisms for a tensor product (and hence used in the definition of enriched category) are more properly required to be natural transformations.
\end{rmk}

The existence of morphisms between functors implies that the collection of functors between any pair of categories itself forms a category, which we now define.

\begin{prop} \label{prop:functor-cat}
  The functors between two categories $\cat{C}$ and $\cat{D}$ constitute the objects of a category, called the \textit{functor category} and denoted by $\Cat{Cat}(\cat{C},\cat{D})$ or $\cat{D}^{\cat{C}}$, whose morphisms are the natural transformations between those functors.
  The identity natural transformation on a functor is the natural transformation whose components are all identity morphisms.
  \begin{proof}
    First, observe that the identity natural transformation is well-defined, as the following diagram commutes for any morphism $f:c\to c'$:
    \[ \begin{tikzcd}
	  Fc && Fc \\
	  \\
	    {Fc'} && {Fc'}
	    \arrow["{\id_{Fc}}", Rightarrow, no head, from=1-1, to=1-3]
	    \arrow["{\id_{Fc'}}"', Rightarrow, no head, from=3-1, to=3-3]
	    \arrow["Ff"', from=1-1, to=3-1]
	    \arrow["Ff", from=1-3, to=3-3]
    \end{tikzcd} \]
    (Note that in general, we will depict an identity morphism in a diagram as an elongated equality symbol, as above.)
    Given two natural transformations $\alpha:F\Rightarrow G$ and $\beta:G\Rightarrow H$, their composite is the natural transformation defined by composing the component functions: $(\beta\circ\alpha)_c := \beta_c\circ\alpha_c$.
    We can see that this gives a well-defined natural transformation by pasting the component naturality squares:
    \[
    \begin{tikzcd}
	    Fc && Gc && Hc \\
	    \\
	      {Fc'} && {Gc'} && {Hc'}
	      \arrow["{\alpha_c}", from=1-1, to=1-3]
	      \arrow["{\alpha_{c'}}"', from=3-1, to=3-3]
	      \arrow["Ff"', from=1-1, to=3-1]
	      \arrow["Gf", from=1-3, to=3-3]
	      \arrow["{\beta_c}", from=1-3, to=1-5]
	      \arrow["{\beta_{c'}}"', from=3-3, to=3-5]
	      \arrow["Hf", from=1-5, to=3-5]
    \end{tikzcd}
    \]
    Since the two inner squares commute, so must the outer square.
    And since the composition of natural transformations reduces to the composition of functions, and the identity natural transformation has identity function components, the composition of natural transformations inherits strict associativity and unitality from composition in $\Set$.
  \end{proof}
\end{prop}

This gives us our a first nontrivial example of a 2-category.

\begin{ex}
  Functor categories constitute the hom categories of the strict 2-category $\Cat{Cat}$, and henceforth we will write $\Cat{Cat}_1$ to denote the 1-category of categories and functors; we can therefore say that $\Cat{Cat}$ is enriched in $\Cat{Cat}_1$.
  The 0-cells of $\Cat{Cat}$ are categories, the 1-cells are functors, and the 2-cells are natural transformations.
  If $\alpha$ is a natural transformation $F\Rightarrow G$, with $F$ and $G$ functors $\cat{C}\to\cat{D}$, then we can depict it as filling the cell between the functors:
  \[\begin{tikzcd}
    {\cat{C}} && {\cat{D}}
	  \arrow[""{name=0, anchor=center, inner sep=0}, "F", curve={height=-18pt}, from=1-1, to=1-3]
	  \arrow[""{name=1, anchor=center, inner sep=0}, "G"', curve={height=18pt}, from=1-1, to=1-3]
	  \arrow["\alpha", shorten <=5pt, shorten >=5pt, Rightarrow, from=0, to=1]
  \end{tikzcd}\]
  (More generally, we will depict 2-cells in this way, interpreting such depictions as diagrams of enriched categories in the sense discussed above.)

  Since $\Cat{Cat}$ is a 2-category, it has both vertical composition (composition within hom-categories) and horizontal (composition between them).
  In Proposition \ref{prop:functor-cat}, we introduced the vertical composition, so let us now consider the horizontal, which we will denote by $\diamond$ to avoid ambiguity.
  The horizontal composition of 1-cells is the composition of functors (as morphisms in $\Cat{Cat}_1$), but by the definition of enriched category, it must also extend to the 2-cells (here, the natural transformations).
  Suppose then that we have natural transformations $\varphi$ and $\gamma$ as in the following diagram:
  \[\begin{tikzcd}
    {\cat{B}} && {\cat{C}} && {\cat{D}}
	  \arrow[""{name=0, anchor=center, inner sep=0}, "F", curve={height=-18pt}, from=1-1, to=1-3]
	  \arrow[""{name=1, anchor=center, inner sep=0}, "G", curve={height=-18pt}, from=1-3, to=1-5]
	  \arrow[""{name=2, anchor=center, inner sep=0}, "{F'}"', curve={height=18pt}, from=1-1, to=1-3]
	  \arrow[""{name=3, anchor=center, inner sep=0}, "{G'}"', curve={height=18pt}, from=1-3, to=1-5]
	  \arrow["\varphi", shorten <=5pt, shorten >=5pt, Rightarrow, from=0, to=2]
	  \arrow["\gamma", shorten <=5pt, shorten >=5pt, Rightarrow, from=1, to=3]
  \end{tikzcd}\]
  The horizontal composite $\gamma\diamond\varphi$ is the natural transformation $GF\Rightarrow G'F'$ with components
  \[ GFb\xto{G\varphi_b}GF'b\xto{\gamma_{F'b}}G'F'b \; . \]
\end{ex}

\begin{notation}[Whiskering]
  It is often useful to consider the horizontal composite of a natural transformation $\alpha:F\Rightarrow G$ with (the identity natural transformation on) a functor, as in the following diagrams, with precomposition on the left and postcomposition on the right:
  \[\begin{tikzcd}
    {\cat{D}} && {\cat{C}} && {\cat{C'}}
    \arrow[""{name=0, anchor=center, inner sep=0}, "L", curve={height=-18pt}, from=1-1, to=1-3]
    \arrow[""{name=1, anchor=center, inner sep=0}, "F", curve={height=-18pt}, from=1-3, to=1-5]
    \arrow[""{name=2, anchor=center, inner sep=0}, "G"', curve={height=18pt}, from=1-3, to=1-5]
    \arrow[""{name=3, anchor=center, inner sep=0}, "L"', curve={height=18pt}, from=1-1, to=1-3]
    \arrow["\alpha", shorten <=5pt, shorten >=5pt, Rightarrow, from=1, to=2]
    \arrow["{\id_L}", shorten <=5pt, shorten >=5pt, Rightarrow, from=0, to=3]
  \end{tikzcd}
  \qquad\qquad
  \begin{tikzcd}
    {\cat{C}} && {\cat{C'}} && {\cat{D'}}
	  \arrow[""{name=0, anchor=center, inner sep=0}, "F", curve={height=-18pt}, from=1-1, to=1-3]
    \arrow[""{name=1, anchor=center, inner sep=0}, "G"', curve={height=18pt}, from=1-1, to=1-3]
    \arrow[""{name=2, anchor=center, inner sep=0}, "R", curve={height=-18pt}, from=1-3, to=1-5]
    \arrow[""{name=3, anchor=center, inner sep=0}, "R"', curve={height=18pt}, from=1-3, to=1-5]
    \arrow["\alpha", shorten <=5pt, shorten >=5pt, Rightarrow, from=0, to=1]
    \arrow["{\id_R}", shorten <=5pt, shorten >=5pt, Rightarrow, from=2, to=3]
  \end{tikzcd}\]
  We will often write the left composite $\alpha\diamond L:FL\Rightarrow GL$ as $\alpha_L$, since its components are $\alpha_{Ld} : FLd\to GLd$ for all $d:\cat{D}$; and we will often write the right composite $R\diamond\alpha:RF\Rightarrow RG$ as $R\alpha$, since its components are $R\alpha_c:RFc\to RGc$ for all $c:\cat{C}$.
  This use of notation is called \textit{whiskering}.
\end{notation}

\subsection{On functorial semantics} \label{sec:syn-sem}

At this point, we pause to consider category theory from the general perspective of our motivating examples, to reflect on how category theory might surprise us: as we indicated in \secref{sec:grph-to-cat}, categories are more `dynamical' than graphs, more preoccupied with change, and so behave differently; in fact, they have a much richer variety of behaviours, and just as categories can often be very well-behaved, they can also be quite unruly.
Through its homoiconicity---its ability to describe itself---the use of category theory impels us to consider not only how individual systems are constructed, nor only how systems of a given type can be compared, but also how to compare different classes of system.
In this way, category theory rapidly concerns itself with notions not only of connection and composition, but also of pattern and translation.

Scientifically, this is very useful: in the computational, cognitive, or otherwise cybernetic sciences, we are often concerned with questions about when and how natural systems `compute'.
Such questions amount to questions of translation, between the abstract realm of computation to the more concrete realms inhabited by the systems of interest and the data that they generate; one often asks how natural structures might correspond to `algorithmic' details, or whether the behaviours of systems correspond to computational processes.
It is for this reason that we chose our motivating examples, which exhibited (abstract) natural structure as well as two kinds of informational or computational structure: a central question in contemporary neuroscience is the extent to which neural circuits can be understood as performing computation (particularly of the form now established in machine learning).
This question is in some way at the heart of this thesis, which aims to establish a compositional framework in which the theories of predictive coding and active inference may be studied.

The dynamism of categories is a hint that it is possible to describe both the structure of systems and their function categorically, with a `syntax' for systems on the one hand and `semantics' on the other.
This is the notion of \textit{functorial semantics} \parencite{Lawvere1963Functorial}, by which we translate syntactic structures in one category to another category which supplies semantics: the use of functors means that this translation preserves basic compositional structure, and we often ask for these functors to preserve other structures, too; a typical choice, that we will adopt in Chapter \ref{chp:algebra} is to use \textit{lax monoidal} functors, which preserve composition in two dimensions, allowing us to place systems ``side by side'' as well as ``end to end''.

Of course, the particular choices of syntactic and semantic category will depend upon the subject at hand---in this thesis we will be particularly interested in supplying dynamical semantics for approximate inference problems---but typically the syntactic category will have some `nice' algebraic structure that is then preserved and interpreted by the functorial semantics.
This is, for instance, how functorial semantics lets us understand processes that ``happen on graphs'', and as a simple example, we can consider diagrams in $\Set$: the shape of the diagram tells us how to compose the parts of a system together, while the diagram functor gives us, for each abstract part, a set of possible components that have a compatible interface, as well as functions realizing their interconnection.

In categorical `process' theory, and the more general categorical theory of systems, one therefore often considers the objects of the `syntactic' category as representing the shapes or interfaces of systems and the morphisms as representing how the different shapes can plug together.
This is an algebraic approach to systems design: mathematically, the syntactic structure is encoded as a \textit{monad}, and the functorial semantics corresponds to a \textit{monad algebra}, as we explain in Chapter \ref{chp:algebra}; and the desire for composition richer than merely end-to-end is another motivation for venturing into higher category theory.
In Chapter \ref{chp:coalg}, we will `unfold' a combination of these ideas, to construct bicategories whose objects represent interfaces, whose 1-cells are processes `between' the interfaces that can be composed both sequentially and in parallel, and whose 2-cells are homomorphisms of processes.
This bicategory will then in Chapter \ref{chp:brain} supply the semantics for models of predictive coding.

In science, there is rarely only one way to study a phenomenon, and our collective understanding of phenomena is therefore a patchwork of perspectives.
At the end of this chapter, we will discuss the Yoneda Lemma, which formalizes this observation that to understand a thing is to see it from all perspectives, and it is for this reason that we expect category theory to supply a \textit{lingua franca} for the mathematical sciences.
In computational neuroscience specifically, an influential methodological theory is David Marr's ``three levels of explanation'' \parencite{Marr1982Vision}, in which complex cognitive systems are profitably studied at the levels of `computation', `algorithm', and `implementation'.
These levels are only very informally defined, and the relationships between them not at all clear.
We hope that functorial semantics and other categorical approaches can replace such methodologies so that instead of a simplistic hierarchical understanding of systems, we can progressively and clearly expose the web of relationships between models.

\subsection{Adjunction and equivalence}%

We discussed above the use of functors to translate between mathematical contexts.
Often, we are interested not only in translation in one direction, but also in translating back again.
When we have a pair of functors---or 1-cells more generally---in opposite directions and when the two translations are somehow reversible, we often find ourselves with an \textit{adjunction}; for example, the functorial mappings of graphs to categories and back are adjoint (Example \ref{ex:grph-cat-adjunc} below), and we conjecture in Chapter \ref{chp:future} that the mapping of ``statistical games'' to dynamical systems forms part of an adjunction, too.
Adjunctions are particularly well-behaved `dual' translations, and they will therefore be of much use throughout this thesis.
For its conceptual elegance, we begin with an abstract definition, which exhibits the fundamental essence.

\begin{defn} \label{def:adjunction}
  Suppose $L : \cat{C}\to\cat{D}$ and $R : \cat{D}\to\cat{C}$ are 1-cells of a 2-category.
  We say that they are \textit{adjoint} or form an \textit{adjunction}, denoted $L\dashv R$, if there are 2-cells $\eta : \id_{\cat{C}} \Rightarrow RL$ and $\epsilon : LR\Rightarrow\id_{\cat{D}}$, called respectively the \textit{unit} and \textit{counit} of the adjunction, which satisfy the \textit{triangle} equalities $\epsilon_L\circ L\eta = \id_L$ and $R\epsilon\circ\eta_R = \id_R$, so called owing to their diagrammatic depictions:
  \[\begin{tikzcd}
    L && LRL \\
    \\
    && L
    \arrow["L\eta", from=1-1, to=1-3]
    \arrow["{\epsilon_L}", from=1-3, to=3-3]
    \arrow[Rightarrow, no head, from=1-1, to=3-3]
  \end{tikzcd}
  \qquad\text{and}\qquad
  \begin{tikzcd}
    R && RLR \\
    \\
    && R
    \arrow["{\eta_R}", from=1-1, to=1-3]
    \arrow["R\epsilon", from=1-3, to=3-3]
	  \arrow[Rightarrow, no head, from=1-1, to=3-3]
  \end{tikzcd}\]
\end{defn}

The unit and counit of the adjunction measure `how far' the round-trip composite functors $LR:\cat{C}\to\cat{C}$ and $RL:\cat{D}\to\cat{D}$ leave us from our starting place, as indicated in the following diagrams:
\[\begin{tikzcd}
  & {\cat{D}} \\
  {\cat{C}} && {\cat{C}}
  \arrow["L", from=2-1, to=1-2]
  \arrow["R", from=1-2, to=2-3]
  \arrow[""{name=0, anchor=center, inner sep=0}, "{\id_{\cat{C}}}"', Rightarrow, no head, from=2-1, to=2-3]
  \arrow["\eta", shorten <=3pt, Rightarrow, from=0, to=1-2]
\end{tikzcd}
\qquad\text{and}\qquad
\begin{tikzcd}
  & {\cat{C}} \\
  {\cat{D}} && {\cat{D}}
  \arrow["R", from=2-1, to=1-2]
  \arrow["L", from=1-2, to=2-3]
  \arrow[""{name=0, anchor=center, inner sep=0}, "{\id_{\cat{D}}}"', from=2-1, to=2-3]
  \arrow["\epsilon"', shorten >=3pt, Rightarrow, from=1-2, to=0]
\end{tikzcd}\]
The triangle identities then ensure that the round-trips have an isomorphic `core', so that it is possible to translate morphisms on one side to the other losslessly (which we will exemplify in Proposition \ref{prop:adjoint-adjunct}), and that the adjunction has a natural `algebraic' interpretation (which we will encounter in Proposition \ref{prop:monad-from-adj}).

In the specific case of the 2-category $\Cat{Cat}$, we can make the following alternative characterization of adjunctions.
Here we see that the ``isomorphic core'' of the adjunction can be characterized by saying that morphisms into objects in $\cat{C}$ that come from $\cat{D}$ via $R$ are in bijection with morphisms out of objects in $\cat{D}$ that come from $\cat{C}$ via $L$.

\begin{defn} \label{def:adjoint-func}
  Suppose $L : \cat{C}\to\cat{D}$ and $R : \cat{D}\to\cat{C}$ are functors between categories $\cat{C}$ and $\cat{D}$.
  We say that they are \textit{adjoint functors} when there is an isomorphism between the hom-sets $\cat{D}(Lc, d) \cong \cat{C}(c, Rd)$ that is natural in $c : \cat{C}$ and $d : \cat{D}$.

  Given a morphism $f : Lc\to d$ in $\cat{D}$, we denote its \textit{(right) adjunct} in $\cat{C}$ by $f^\sharp : c\to Rd$.
  Inversely, given a morphism $g  : c\to Rd$ in $\cat{C}$, we denote its \textit{(left) adjunct} in $\cat{D}$ by $g^\flat : Lc\to d$.
  The existence of the isomorphism means that ${f^\sharp}^\flat = f$ and $g = {g^\flat}^\sharp$.
\end{defn}

\begin{ex} \label{ex:grph-cat-adjunc}
  The functor $F:\Cat{Graph}\to\Cat{Cat}$ mapping a graph to the corresponding free category (Proposition \ref{prop:free-cat}) is left adjoint to the forgetful functor $U:\Cat{Cat}\to\Cat{Graph}$ mapping a category to its underlying graph (Proposition \ref{prop:forget-cat}).
  To see this, we need to find a natural isomorphism $\Cat{Cat}(F\cat{G},\cat{C})\cong\Cat{Graph}(\cat{G},U\cat{C})$.
  A graph homomorphism $\cat{G}\to U\cat{C}$ is a mapping of the nodes of $\cat{G}$ to the objects of $\cat{C}$ and of the edges of $\cat{G}$ to the morphisms of $\cat{C}$ that preserves sources (domains) and targets (codomains).
  A functor $F\cat{G}\to\cat{C}$ is a mapping of the nodes of $\cat{G}$ to the objects of $\cat{C}$ along with a mapping of paths in $\cat{G}$ to morphisms in $\cat{C}$ that preserves domains, codomains, identities and composites.
  A path in $\cat{G}$ is a list of `composable' edges, with the identity path being the empty list, so such a mapping of paths is entirely determined by a mapping of edges to morphisms that preserves domains and codomains.
  That is to say, a functor $F\cat{G}\to\cat{C}$ is determined by, and determines, a graph homomorphism $\cat{G}\to U\cat{C}$, and so the two sets are isomorphic: in some sense, functors between free categories \textit{are} graph homomorphisms.
  To see that the isomorphism is natural, observe that it doesn't matter if we precompose a graph homomorphism $\cat{G}'\to\cat{G}$ (treated as a functor between free categories) or postcompose a functor $\cat{C}\to\cat{C}'$ (treated as a graph homomorphism): because graph homomorphisms compose preserving the graph structure, we would still have an isomorphism $\Cat{Cat}(F\cat{G'},\cat{C'})\cong\Cat{Graph}(\cat{G'},U\cat{C'})$.
\end{ex}

Before we can properly say that adjoint functors form an adjunction, we need to prove it.
As the following proof shows, the mappings $(-)^\sharp$ and $(-)^\flat$ define and are defined by the unit and counit of the adjunction.

\begin{prop} \label{prop:adjoint-adjunct}
  Functors that form an adjunction in $\Cat{Cat}$ are exactly adjoint functors.
  \begin{proof}
    We need to show that functors that form an adjunction are adjoint, and that adjoint functors form an adjunction; that is, we need to show that any pair of functors $L:\cat{C}\to\cat{D}$ and $R:\cat{D}\to\cat{C}$ satisfying the definition of adjunction in Definition \ref{def:adjunction} necessarily constitute adjoint functors according to Definition \ref{def:adjoint-func}, and that if $L$ and $R$ are adjoint according to Definition \ref{def:adjoint-func} then they form an adjunction according to Definition \ref{def:adjunction}: \textit{i.e.}, the two definitions are equivalent.

    We begin by showing that if $L\dashv R$, then $L$ and $R$ are adjoint functors.
    This means we need to exhibit a natural isomorphism $\cat{D}(Lc,d)\cong\cat{C}(c,Rd)$.
    We define a function $(-)^\sharp : \cat{D}(Lc,d)\to\cat{C}(c,Rd)$ by setting
    \[ f^\sharp := c\xto{\eta_c}RLc\xto{Rf}Rd \]
    and a function $(-)^\flat : \cat{C}(c,Rd)\to\cat{D}(Lc,d)$ by setting
    \[ g^\flat := Lc\xto{Lg}LRd\xto{\epsilon_d}d \, . \]
    We then use naturality and the triangle equalities to show that ${f^\sharp}^\flat = f$ and ${g^\flat}^\sharp = g$:
    \begin{gather*}
      \begin{aligned}
        {f^\sharp}^\flat
        &= Lc \xto{Lf^\sharp} LRd \xto{\epsilon_d} d \\
        &= Lc \xto{L\eta_c} LRLc \xto{LRf} LRd \xto{\epsilon_d} d \\
        &= Lc \xto{L\eta_c} LRLc \xto{\epsilon_{Lc}} Lc \xto{f} d \\
        &= Lc \xto{f} d
      \end{aligned}
      \qquad\qquad
      \begin{aligned}
        {g^\flat}^\sharp
        &= c \xto{\eta_c} RLc \xto{Rg^\flat} Rd \\
        &= c \xto{\eta_c} RLc \xto{RLc} RLRd \xto{R\epsilon_d} Rd \\
        &= c \xto{g} Rd \xto{\eta_{Rd}} RLRd \xto{R\epsilon_d} Rd \\
        &= c \xto{g} Rd
      \end{aligned}
    \end{gather*}
    In each case the first two lines follow by definition, the third by naturality, and the fourth by the triangle equality; hence we have an isomorphism $\cat{D}(Lc,d)\cong\cat{C}(c,Rd)$.
    The naturality of this isomorphism follows from the naturality of $\eta$ and $\epsilon$.
    We first check that the isomorphisms $(-)^\sharp$ are natural in $c$, which means that the following squares commute for all $\phi:c'\to c$ in $\cat{C}$:
    \[\begin{tikzcd}
	    {\cat{D}(Lc,d)} && {\cat{C}(c,Rd)} \\
	    \\
	      {\cat{D}(Lc',d)} && {\cat{C}(c',Rd)}
	      \arrow["{\cat{D}(L\phi,d)}"', from=1-1, to=3-1]
	      \arrow["{(-)^\sharp_{c',d}}"', from=3-1, to=3-3]
	      \arrow["{\cat{C}(\phi,Rd)}", from=1-3, to=3-3]
	      \arrow["{(-)^\sharp_{c,d}}", from=1-1, to=1-3]
    \end{tikzcd}\]
    This requires in turn that $(f\circ L\phi)^\sharp = f^\sharp\circ\phi$, which we can check as follows:
    \begin{align*}
      (f\circ L\phi)^\sharp
      &= c' \xto{\eta_{c'}} RLc' \xto{RL\phi} RLc \xto{Rf} Rd \\
      &= c' \xto{\phi} c \xto{\eta_c} RLc \xto{Rf} Rd \\
      &= c' \xto{\phi} c \xto{f^\sharp} Rd
    \end{align*}
    where the second equality holds by the naturality of $\eta$.
    The naturality of $(-)^\sharp$ in $d$ requires that $(\phi'\circ f)^\sharp = R\phi'\circ f^\sharp$ for all $\phi' : d\to d'$, which can be checked almost immediately:
    \begin{align*}
      (\phi'\circ f)^\sharp
      &= c\xto{\eta_c}RLc\xto{Rf}Rd\xto{R\phi'}Rd' \\
      &= c\xto{f^\sharp}Rd\xto{R\phi'}Rd'
    \end{align*}
    Dually, the naturality of $(-)^\flat : \cat{C}(c,Rd)\to\cat{D}(Lc,d)$ in $d$ requires that $(R\phi'\circ g)^\flat = \phi'\circ g^\flat$ for all $\phi':d\to d'$, which obtains by the naturality of $\epsilon$:
    \begin{align*}
      (R\phi'\circ g)^\flat
      &= Lc\xto{Lg}LRd\xto{LR\phi'}LRd'\xto{\epsilon_{d'}}d' \\
      &= Lc\xto{Lg}LRd\xto{\epsilon_d}d\xto{\phi'}d'\\
      &= Lc\xto{g^\flat}d\xto{\phi'}d'
    \end{align*}
    The naturality of $(-)^\flat$ in $c$, which requires that $(g\circ\phi)^\flat = g^\flat\circ L\phi$, obtains similarly immediately:
    \begin{align*}
      (g\circ \phi)^\flat
      &= Lc'\xto{L\phi}Lc\xto{Lg}LRd\xto{\epsilon_d}d \\
      &= Lc'\xto{L\phi}Lc\xto{g^\flat}d
    \end{align*}
    Thus $(-)^\sharp$ and $(-)^\flat$ are both natural in $c$ and $d$, and hence $L$ and $R$ are adjoint functors.

    To show the converse, that if $L:\cat{C}\to\cat{D}$ and $R:\cat{D}\to\cat{C}$ are adjoint functors then $L\dashv R$, we need to establish natural transformations $\eta:\id_{\cat{C}}\Rightarrow RL$ and $\epsilon : LR\Rightarrow\id_{\cat{D}}$ from the natural isomorphisms $(-)^\sharp$ and $(-)^\flat$, such that the triangle equalities $\epsilon_L\circ L\eta = \id_L$ and $R\epsilon\circ\eta_R = \id_R$ are satisfied.
    We first define $\eta$ componentwise, by observing that $\eta_c$ must have the type $c\to RLc$, and that the image of $\id_{Lc} : Lc\to Lc$ under $(-)^\sharp$ is of this type, and therefore defining $\eta_c := (\id_{Lc})^\sharp$.
    Dually, we define $\epsilon$ by observing that $\epsilon_d$ must have the type $LRd\to d$, and that the image of $\id_{Rd}$ under $(-)^\flat$ has this type.
    We therefore define $\epsilon_d := (\id_{Rd})^\flat$.
    To see that these definitions constitute natural transformations, observe that they are themselves composed from natural transformations.
    Explicitly, the naturality of $\eta$ means that for any $f:c\to c'$, we must have $RLf\circ\eta_c = \eta_{c'}\circ f$, and the naturality of $\epsilon$ means that for any $g:d\to d'$, we must have $g\circ\epsilon_d = \epsilon_{d'}\circ LRg$.
    These obtain as follows:
    \begin{gather*}
      \begin{aligned}
        RLf\circ\eta_c
        &= c\xto{(\id_{Lc})^\sharp}RLc\xto{RLf}RLc' \\
        &= c\xto{(Lf\circ\id_{Lc})^\sharp}RLc' \\
        &= c\xto{(\id_{Lc'}\circ Lf)^\sharp}RLc' \\
        &= c\xto{f}c'\xto{(\id_{Lc'})^\sharp}RLc' \\
        &= \eta_{c'}\circ f
      \end{aligned}
      \qquad\qquad
      \begin{aligned}
        g\circ\epsilon_d
        &= LRd\xto{(\id_{Rd})^\flat}d\xto{g}d' \\
        &= LRd\xto{(Rg\circ\id_{Rd})^\flat}d' \\
        &= LRd\xto{(\id_{Rd'}\circ Rg)^\flat}d' \\
        &= LRd\xto{LRg}LRd'\xto{(\id_{Rd'})^\flat}d' \\
        &= \epsilon_{d'}\circ LRg
      \end{aligned}
    \end{gather*}
    In each case, the first equality holds by definition, the second by naturality of $(-)^\sharp$ and $(-)^\flat$ (left and right, respectively) in $d$, the third by naturality of $\id$, the fourth by naturality in $c$, and the last by definition.
    It remains to check that $\eta$ and $\epsilon$ so defined satisfy the triangle equalities.
    Expressed componentwise, we demonstrate that $\epsilon_{Lc}\circ L\eta_c = \id_{Lc}$ and that $R\epsilon_d\circ\eta_{Rd} = \id_{Rd}$ as follows:
    \begin{gather*}
      \begin{aligned}
        \epsilon_{Lc}\circ L\eta_c
        &= Lc\xto{L(\id_{Lc})^\sharp}LRLc\xto{(\id_{RLc})^\flat}Lc \\
        &= Lc\xto{(\id_{RLc}\circ(\id_{Lc})^\sharp)^\flat}Lc \\
        &= Lc\xto{{(\id_{Lc})^\sharp}^\flat}Lc \\
        &= Lc\xto{\id_{Lc}}Lc
      \end{aligned}
      \quad\quad
      \begin{aligned}
        R\epsilon_d\circ\eta_{Rd}
        &= Rd\xto{(\id_{LRd})^\sharp}RLRd\xto{R(\id_{Rd})^\flat}Rd \\
        &= Rd\xto{((\id_{Rd})^\flat\circ\id_{LRd})^\sharp}Rd \\
        &= Rd\xto{{(\id_{Rd})^\flat}^\sharp}Rd \\
        &= Rd\xto{\id_{Rd}}Rd
      \end{aligned}
    \end{gather*}
    The first equality (on each side) holds by definition, the second (on the left) by naturality of $(-)^\flat$ in $c$ and (on the right) by naturality of $(-)^\sharp$ in $d$, the third by unitality of composition, and the fourth by the $\sharp$/$\flat$ isomorphism.
    This establishes that $L\dashv R$, and hence the result.
  \end{proof}
\end{prop}

Sometimes, the `distances' measured by the unit and counit are so small that the categories $\cat{C}$ and $\cat{D}$ are actually `equivalent': this happens when the unit and counit are natural isomorphisms, meaning that the isomorphic core of the adjunction extends to the whole of $\cat{C}$ and $\cat{D}$.
This gives us the following definition.

\begin{defn}
  Suppose $L\dashv R$ in a 2-category.
  When the unit and counit of the adjunction are additionally isomorphisms, we say that $L$ and $R$ form an \textit{adjoint equivalence}.
\end{defn}

\begin{rmk}
  More generally, an \textit{equivalence} of categories is a pair of functors connected by natural isomorphisms of the form of the unit and counit of an adjunction, but which may not necessarily satisfy the triangle identities; however, given any such equivalence, it is possible to modify the unit or counit so as to upgrade it to an adjoint equivalence.
  Henceforth, we will have no need to distinguish equivalences from adjoint equivalences, so we will say simply `equivalence' for both.
  If there is an equivalence between a pair of categories, then we will say that the two categories are \textit{equivalent}.

  Note that the notion of equivalence of categories can be generalized to \textit{equivalence in a 2-category}, by replacing the categories by 0-cells, the functors by 1-cells, and the natural isomorphisms by invertible 2-cells.
\end{rmk}

The structure of an equivalence of categories can alternatively be specified as properties of the functors concerned, which in some situations can be easier to verify.

\begin{defn}
  We say that a functor $F:\cat{C}\to\cat{D}$ is
  \begin{enumerate}
  \item \textit{full} when it is surjective on hom sets, in the sense that the functions $F_{a,b}:\cat{C}(a,b)\to\cat{D}(Fa,Fb)$ are surjections;
  \item \textit{faithful} when it is injective on hom sets, in the sense that the functions $F_{a,b}$ are injections;
  \item \textit{fully faithful} when it is both full and faithful (\textit{i.e.}, isomorphic on hom sets); and
  \item \textit{essentially surjective} when it is surjective on objects up to isomorphism, in the sense that for every object $d:\cat{D}$ there is an object $c:\cat{C}$ such that $Fc\cong d$.
  \end{enumerate}
\end{defn}

\begin{prop}
  Categories $\cat{C}$ and $\cat{D}$ are equivalent if and only if there is a functor $F:\cat{C}\to\cat{D}$ that is fully faithful and essentially surjective.
  \begin{proof}[{Proof {\parencite[{Lemma 9.4.5}]{UniFoundationsProgram2013Homotopy}}}]
    First, we show that if $F\dashv G:\cat{D}\to\cat{C}$ is an equivalence of categories, then $F:\cat{C}\to\cat{D}$ is fully faithful and essentially surjective.
    For the latter, observe that $G$ gives us, for any $d:\cat{D}$, an object $Gd:\cat{C}$ and $\epsilon_d$ is by definition an isomorphism $FGd\xto{{\sim}}d$; hence $F$ is essentially surjective.
    To show that $F$ is fully faithful means showing that each function $F_{a,b}:\cat{C}(a,b)\to\cat{D}(Fa,Fb)$ is an isomorphism; we can define the inverse $F^{-1}_{a,b}$ as the following composite:
    \[\begin{matrix}
      \cat{D}(Fa,Fb) & \xto{G_{Fa,Fb}} & \cat{C}(GFa,GFb) & \xto{\cat{C}(\eta_a,\eta^{-1}_b)} & \cat{C}(a,b) \\
      g & \mapsto & Gg & \mapsto & \bigl( a\xto{\eta_a}GFa\xto{Gg}GFb\xto{\eta^{-1}_b}b \bigr)
    \end{matrix}\]
    Here, the function $\cat{C}(\eta_a,\eta^{-1}_b)$ is the function $f\mapsto \eta^{-1}_b\circ f\circ\eta_a$ obtained from the hom functor (Remark \ref{rmk:hom-bifunc}).
    Hence $F^{-1}_{a,b}(g) := \eta^{-1}_b\circ Gg\circ\eta_a$.
    To see that this is indeed the desired inverse, consider applying the functor $F$ to the morphism $F^{-1}_{a,b}(g)$; we have the following equalities:
    \begin{gather*}
      Fa\xto{F\eta_a}FGFa\xto{FGg}FGFb\xto{F\eta^{-1}_b}Fb \\
      = Fa\xto{g}Fb\xto{F\eta_b}FGFb\xto{F\eta^{-1}_b}Fb \\
      = Fa\xto{g}Fb
    \end{gather*}
    where the first equality holds by the naturality of $\eta$ and the second equality holds since $\eta_b$ is an isomorphism.
    Since $F$ is therefore isomorphic on hom sets, it is fully faithful.

    Next, we show that if $F:\cat{C}\to\cat{D}$ is fully faithful and essentially surjective, then there is a functor $G:\cat{D}\to\cat{C}$ and natural isomorphisms $\eta:\id_{\cat{C}}\Rightarrow GF$ and $\epsilon:FG\Rightarrow\id_{\cat{D}}$.
    On objects $d:\cat{D}$, we can define $Gd:\cat{C}$ as any choice of object such that $FGd\xto{\sim}d$: such an object must exist since $F$ is essentially surjective.
    We then define $\epsilon_d$ to be the associated isomorphism $FGd\to d$; it is easy to verify that $\epsilon$ so defined is natural.
    On morphisms, let the functions $G_{d,e}$ be defined as the composite functions
    \[\begin{matrix}
    \cat{D}(d,e) & \xto{\cat{D}(\epsilon_d,\epsilon^{-1}_e)} & \cat{D}(FGd,FGe) & \xto{F^{-1}_{Gd,Ge}} & \cat{C}(Gd,Ge) \\
    g & \mapsto & \bigl( FGd\xto{\epsilon_d}d\xto{g}e\xto{\epsilon^{-1}_e}FGe \bigr) & \mapsto & F^{-1}_{Gd,Ge}\bigl( \epsilon^{-1}_e\circ g\circ\epsilon_d \bigr)
    \end{matrix} \; .\]
    Since $F$ is a fully faithful functor and $\epsilon$ is a natural isomorphism, this makes $G$ a well-defined functor.
    Finally, we define $\eta$ as having the components $\eta_c := F_{c,GFc}^{-1}\left(\epsilon_{Fc}^{-1}\right)$; since $\epsilon$ is a natural isomorphism, so is $\epsilon^{-1}$, which is thus preserved as such by the inverse action of $F$ in defining $\eta$.
    This establishes all the data of the equivalence.

    (Note that we can actually prove a little more: it is not hard to verify additionally that the two constructions are inverse, so that equivalences are themselves equivalent to fully faithful essentially surjective functors.)
  \end{proof}
\end{prop}

\begin{rmk}
  In the above proof, we assumed the axiom of choice, defining $Gd$ as a choice of object such that $FGd\xto{\sim}d$.
  It is possible to avoid making this assumption, by asking for the surjection on objects $F_0:\cat{C}_0\to\cat{D}_0$ to be `split' in the sense that it comes with a function $s:\cat{D}_0\to\cat{C}_0$ such that $F_0(s(d))\cong d$ in $\cat{D}$ for every object $d:\cat{D}$; then we just set $Gd := s(d)$.
\end{rmk}

\section{Universal constructions} \label{sec:univ-const}

In the preceding sections, we have used diagrams to represent some patterns in a categorical context, and we have discussed how functors allow us to translate patterns and structures between contexts; indeed, we used functors to formalize diagrams themselves.
But an important facet of the notion of pattern is replication across contexts, and in many important situations, we will encounter patterns that apply to all objects in a category.
We call such patterns \textit{universal}, and much of science is a search for such universal patterns: for example, much of physics, and by extension much of the theory of the Bayesian brain, is a study of the universal principle of stationary action.
In this section, we introduce the formal characterization of universality and exemplify it with some examples that will be particularly important later on --- as well as some examples that we have encountered already.

\subsection{The universality of common patterns}

We begin with some basic examples of universal patterns.

\subsubsection{Disjunctions, or coproducts}

Our first example of a universal pattern is the \textit{coproduct}, which captures the essence of the following examples --- situations like disjunction, where there is an element of choice between alternatives.

\begin{ex}
  Given two propositions, such as $\mathsf{P}_1 := \text{ ``}-\text{ is flat''}$ and $\mathsf{P}_2 := \text{ ``}-\text{ is sharp''}$, we can form their disjunction $\mathsf{P}_1\vee \mathsf{P}_2$ (meaning $\text{ ``}-\text{ is flat or sharp''}$).
  Similarly, given two subsets $P_1,P_2 \subseteq X$, we can form their \textit{join} or \textit{union}, $P_1\cup P_2$: an element $x$ is an element of $P_1\cup P_2$ if (and only if) it is an element of $P_1$ or an element of $P_2$.
\end{ex}

\begin{ex}
  Given two numbers, we can form their \textit{sum}; for instance, $1+2=3$.
  More generally, given two sets $A$ and $B$, we can form their \textit{disjoint union}, denoted $A+B$.
  The elements of $A+B$ are pairs $(i,x)$ where $x$ is an element of $A$ or of $B$ and $i$ indicates which set $x$ is drawn from (this ensures that if an element $x$ of $A$ is the same as an element of $B$, it is added twice to the disjoint union).
  Therefore, if $A$ has $1$ element and $B$ has $2$, then $A+B$ has $3$ elements.
\end{ex}

\begin{rmk}
  The preceding example illustrates how we can think of numbers equivalently as sets of the indicated cardinality.
  Many operations on sets are generalizations of familiar operations on numbers in this way.
\end{rmk}

\begin{ex}
  Given two graphs, $G$ and $G'$, we can form the sum graph $G+G'$, whose set of nodes is $G_0+G'_0$ and whose set of edges is $G_1+G'_1$.
\end{ex}

\begin{ex}
  Given two vector spaces $V$ and $V'$, we can form their \textit{direct sum} $V\oplus V'$, whose vectors are linear combinations of vectors either in $V$ or in $V'$.
\end{ex}

Each of these is an example of a coproduct, which we now define.

\begin{defn}
  Given objects $A$ and $B$ in a category $\cat{C}$, their \textit{coproduct} (if it exists) is an object, canonically denoted $A+B$, equipped with two morphisms $\inj_A:A\to A+B$ and $\inj_B:B\to A+B$ such that, for any object $Q$ equipped with morphisms $f:A\to Q$ and $g:B\to Q$, there is a unique morphism $u:A+B\to Q$ such that $f = u\circ\inj_A$ and $g = u\circ\inj_B$.
  The morphisms $\inj_A$ and $\inj_B$ are called \textit{injections}, and the morphism $u$ is called the \textit{copairing} and often denoted by $[f,g]$.
\end{defn}

\begin{ex}
  Morphisms of subsets are inclusions, so given subsets $P_1,P_2 \subseteq X$, there are evident inclusions $P_1 \subseteq P_1\cup P_2$ and $P_2 \subseteq P_1\cup P_2$.
  Moreover, given a subset $Q$ such that $P_1\subseteq Q$ and $P_2\subseteq Q$, it is clearly the case that $P_1\subseteq P_1+P_2\subseteq Q$ and $P_2\subseteq P_1+P_2\subseteq Q$.

  Similarly, morphisms of propositions are implications, so given $\mathsf{P}_1$ and $\mathsf{P}_2$ such that $\mathsf{P}_1\to Q$ and $\mathsf{P}_2\to Q$, then it is necessarily the case that $\mathsf{P}_1\to \mathsf{P}_1\vee\mathsf{P}_2\to Q$ and $\mathsf{P}_2\to \mathsf{P}_1\vee\mathsf{P}_2\to Q$: clearly, both $\mathsf{P}_1$ and $\mathsf{P}_2$ imply $\mathsf{P}_1\vee\mathsf{P}_2$ by definition, and if both $\mathsf{P}_1$ and $\mathsf{P}_2$ imply $Q$, then so does $\mathsf{P}_1\vee\mathsf{P}_2$.
\end{ex}

\begin{ex}
  Given sets $A$ and $B$, the injections $\inj_A:A\to A+B$ and $\inj_B:B\to A+B$ are the corresponding inclusions: $\inj_A$ maps $a$ to $(1,a)$ and $\inj_B$ maps $b$ to $(2,b)$, where $1$ tags an element as coming from $A$, and $2$ tags an element as coming from $B$.
  Given $f:A\to Q$ and $g:B\to Q$ the copairing $[f,g]:A+B\to Q$ is the function that takes an element $(i,x)$ and returns $f(x)$ if $i=1$ or $g(x)$ if $i=2$; it is from this that the `choice' interpretation arises for the coproduct.
\end{ex}

\begin{ex}
  Morphisms of vectors spaces are linear maps, and if the spaces are finite-dimensional, then we can represent these maps as matrices: if $V$ is $n$-dimensional and $W$ is $m$-dimensional, then a morphism $V\to W$ is a matrix of shape $(m,n)$; writing the elements of $V$ and $W$ as column vectors, such a matrix has $m$ rows and $n$ columns.
  Moreover, in this case, the direct sum $V\oplus W$ is $(n+m)$-dimensional.

  Therefore suppose that $V$, $V'$ and $W$ have respective dimensions $n$, $n'$ and $m$, and suppose we have linear maps $f:V\to W$ and $g:V'\to W$.
  The injection $V\to V\oplus V'$ is the block matrix $\begin{pmatrix} 1_n \\ 0_{n'} \end{pmatrix}$ where $1_n$ is the $n$-by-$n$ identity matrix and $0_{n'}$ is the $n'$-by-$n'$ zero matrix; similarly, the injection $V'\to V\oplus V'$ is the block matrix $\begin{pmatrix} 0_n \\ 1_{n'} \end{pmatrix}$.
  And the copairing $[f,g]:V\oplus V'\to W$ is the block matrix $\begin{pmatrix} f & g \end{pmatrix}$.
\end{ex}

\begin{rmk} \label{rmk:idx-sum}
  The coproducts we have considered so far have all been \textit{binary}, being coproducts of only two objects.
  More generally, we can often consider coproducts of more objects, by repeating the binary coproduct operation; typically, there is an isomorphism $(A+B)+C \cong A+(B+C)$.
  We can extend this further to finite (and, often, infinite) collections of objects.
  Suppose then that $\{A_i\}$ is a collection of objects indexed by $i:I$, where $I$ is a set, and form the iterated coproduct $\sum_{i:I} A_i$; we will call this object a \textit{dependent sum}, because the summands $A_i$ \textit{depend on} $i:I$.
  In the case where the objects $A_i$ are all sets, the dependent sum $\sum_i A_i$ is again a set, whose elements are pairs $(i,x)$ where $i$ is an element of $I$ and $x$ is an element of $A_i$.
  In other categories $\cat{C}$, typically the name \textit{dependent sum} is reserved for the case when all of the objects $A_i$ and the object $I$ are objects of $\cat{C}$.
  But when $I$ remains a set, we may still be able to form the $I$-indexed coproduct $\sum_i A_i$ in $\cat{C}$.
\end{rmk}

\subsubsection{Conjunctions, products, and sections}

Our next example of a universal pattern is the \textit{product}, which captures situations like conjunction, in which things come along in separable pairs of individuals.

\begin{ex}
  Given two propositions, such as $\mathsf{P}_1 := \text{ ``}-\text{ is small''}$ and $\mathsf{P}_2 := \text{ ``}-\text{ is connected''}$, we can form their conjunction $\mathsf{P}_1\wedge \mathsf{P}_2$ (meaning $\text{ ``}-\text{ is small and connected''}$).
  Similarly, given two subsets $P_1,P_2 \subseteq X$, we can form their \textit{meet} or \textit{intersection}, $P_1\cap P_2$: an element $x$ is an element of $P_1\cap P_2$ if (and only if) it is an element of $P_1$ and an element of $P_2$.
\end{ex}

\begin{ex}
  Given two numbers, we can form their \textit{product}; for instance, $2\times 3=6$.
  More generally, as we saw in Definition \ref{def:prod-set}, we can form the product of any two sets $A$ and $B$, denoted $A\times B$.
  The elements of $A\times B$ are pairs $(a,b)$ where $a$ is an element of $A$ and $b$ is an element of $B$.
  Therefore, if $A$ has $2$ elements and $B$ has $3$, then $A\times B$ has $6$ elements.
\end{ex}

\begin{rmk} \label{rmk:repeat-add}
  When all the summands of a dependent sum are the same set or object $A$ regardless of their associated index $i:I$, then the object $\sum_{i:I} A$ is isomorphic to the product $I\times A$: this is simply a categorification of the fact that ``multiplication is repeated addition''.
\end{rmk}

\begin{ex}
  Given vector spaces $V$ and $V'$ (of respective dimensions $n$ and $n'$), their product is again the direct sum $V\oplus V'$.
  Since the direct sum of vector spaces is both a product and a coproduct, it is also said to be a \textit{biproduct}.
\end{ex}

Categorically, the product is the dual of the coproduct.

\begin{defn} \label{def:product}
  Given objects $A$ and $B$ in a category $\cat{C}$, their \textit{product} (if it exists) is an object, canonically denoted $A\times B$, equipped with two morphisms $\proj_A:A\times B\to A$ and $\proj_B:A\times B\to B$ such that, for any object $Q$ equipped with morphisms $f:Q\to A$ and $g:Q\to B$, there is a unique morphism $u:Q\to A\times B$ such that $f = \proj_A\circ u$ and $g = \proj_B\circ u$. The morphisms $\proj_A$ and $\proj_B$ are called \textit{projections}, and the morphism $u$ is called the \textit{pairing} and often denoted by $(f,g)$.
\end{defn}

\begin{ex}
  Given subjects $P_1,P_2 \subseteq X$, there are evident inclusions $P_1\cap P_2\subseteq P_1$ and $P_1\cap P_2\subseteq P_2$.
  Moreover, given a subset $Q$ such that $Q\subseteq P_1$ and $Q\subseteq P_2$, it is clearly then the case that $Q\subseteq P_1\cap P_2\subseteq P_1$ and $Q\subseteq P_1\cap P_2\subseteq P_2$.

  Similarly, given propositions $\mathsf{P}_1$ and $\mathsf{P}_2$ such that $Q\to\mathsf{P}_1$ and $Q\to\mathsf{P}_2$, it is (by the definition of ``and'') the case that $Q\to\mathsf{P}_1\wedge\mathsf{P}_2\to\mathsf{P}_1$ and $Q\to\mathsf{P}_1\wedge\mathsf{P}_2\to\mathsf{P}_2$.
\end{ex}

\begin{ex}
  Given sets $A$ and $B$, the projections $\proj_A:A\times B\to A$ and $\proj_B:A\times B\to B$ are the functions $(a,b)\mapsto a$ and $(a,b)\mapsto b$ respectively.
  Given $f:Q\to A$ and $g:Q\to B$, the pairing $(f,g):Q\to A\times B$ is the function $x\mapsto \bigl(f(x),g(x)\bigr)$; note that this involves `copying' $x$, which will be relevant when we come to consider copy-discard categories in \secref{sec:cd-cats}.
\end{ex}

\begin{rmk}
  Above, we observed that a coproduct with constant summands $A$ is equivalently a product $I\times A$ of the indexing object $I$ with $A$; we therefore have a projection $\proj_I:I\times A\to I$.
  More generally, for any dependent sum $\sum_{i:I}A_i$, there is a projection $\sum_{i:I}A_i\to I$; in the case of dependent sums in $\Set$, this is unsurprisingly the function $(i,x)\mapsto i$.
\end{rmk}

\begin{ex}
  Suppose we have vector spaces $V$, $V'$ and $W$ of respective dimensions $n$, $n'$ and $m$.
  The projection $V\oplus V'\to V$ is the block matrix $\begin{pmatrix} 1_n & 0_{n'} \end{pmatrix}$, and the projection $V\oplus V'\to V'$ is the block matrix $\begin{pmatrix} 0_n & 1_{n'} \end{pmatrix}$.
  Given linear maps $f:W\to V$ and $g:W\to V'$, the pairing $(f,g):W\to V\oplus V'$ is the block matrix $\begin{pmatrix} f \\ g \end{pmatrix}$.
  Note that, in a sign of the duality between products and coproducts, the projections and the pairing are respectively the injections and the copairing transposed.
\end{ex}

\begin{rmk} \label{rmk:dep-prod}
  Just as in the case of coproducts, we can also consider products of more than two objects, by repeating the product operation; there is again typically an isomorphism $(A\times B)\times C\cong A\times(B\times C)$.
  If $\{A_i\}$ is a collection of objects indexed by $i:I$ (with $I$ again a set), we can form the \textit{dependent product}\footnote{%
  This set-indexed product is also known as an \textit{indexed product}, to emphasize that the factors are indexed by the set $I$; since $I$ has elements, we can properly think of these as indices, which may not be true for other kinds of object.
  We will see in Definition \ref{def:dep-prod} how to use categorical structure to abstract away the requirement that $I$ be a set.}
  $\prod_{i:I} A_i$.
  In the case where $I$ is finite and the objects $A_i$ are all sets, the dependent product $\prod_{i:I}A_i$ is again a set, whose elements can equivalently be seen as lists $(a_i,\dots)$ indexed by $i$ with each element $a_i$ drawn from the corresponding set $A_i$ or as functions $s$ with domain $I$ and codomain the dependent sum $\sum_{i:I}A_i$ such that each $s(i)$ is tagged by $i$.
  This situation is summarized by the commutativity of the diagram
  \[\begin{tikzcd}
    I && {\sum_{i:I}A_i} \\
    \\
    && I
    \arrow["s", from=1-1, to=1-3]
    \arrow["\pi", from=1-3, to=3-3]
    \arrow[Rightarrow, no head, from=1-1, to=3-3]
  \end{tikzcd}\]
  where $\pi$ is the projection and which therefore requires that $\pi\circ s = \id_I$.
  Such a function $s$ is known as a \textit{section} of $p$, and we can think of sections therefore as \textit{dependent functions}, since the types of their output values (\textit{i.e.}, $A_i$) may depend on the input values $i$.
\end{rmk}

The notion of section is important enough to warrant a general definition.

\begin{defn}
  Suppose $p:E\to B$ is a morphism.
  A \textit{section} of $p$ is a morphism $s:B\to E$ such that $p\circ s = \id_B$.
\end{defn}

\subsubsection{Subobjects and equalizers}

Our next examples of universal patterns do not involve pairing or grouping objects together to make new ones.
For instance, in many situations, it is of interest to restrict our attention to `subobjects' (of a single given object) that satisfy a certain property, which may not extend to the whole object at hand.

\begin{ex} \label{ex:subobj-class}
  In examples above, we saw that subsets and propositions behave similarly with respect to disjunctions and conjunctions.
  More broadly, there is a correspondence between subsets and propositions, if we think of propositions on a set $X$ as functions $X\to 2$, where $2$ is the 2-element set $\{\bot,\top\}$ of truth values (where we interpret $\bot$ as `false' and $\top$ as `true').
  Every subset $A\subseteq X$ has an associated injection, $A\hookrightarrow X$, and there is a correspondence between such injections and propositions $\mathsf{P}_A:X\to 2$, where $\mathsf{P}_A(x)$ is true whenever $x$ is an element of $A$.
  This situation can be summarized by the commutativity of the diagram
  \[\begin{tikzcd}
    A && 1 \\
    \\
    X && 2
    \arrow["{!}", from=1-1, to=1-3]
    \arrow["\top", from=1-3, to=3-3]
    \arrow[hook, from=1-1, to=3-1]
    \arrow["{\mathsf{P}_A}"', from=3-1, to=3-3]
  \end{tikzcd}\]
  where $1$ is the singleton set $\{\ast\}$, ${!}$ is the unique function sending every element of $A$ to $\ast$, and $\top$ is the function $\ast\mapsto\top$ picking the truth value $\top$.
  If, in a category $\cat{C}$, there is an object such that, for any subobject $A\hookrightarrow X$, there is a unique morphism $X\to 2$ such that the above diagram commutes (and moreover defines a \textit{pullback square} in the sense of Example \ref{ex:pullback} below), then we say that the object $2$ is a \textit{subobject classifier} in $\cat{C}$; in this case, we interpret $1$ as the `terminal' object in $\cat{C}$ (introduced below, in Example \ref{ex:term-init-obj}).
\end{ex}

A pattern that will be particularly common is that in which we care about a subset of elements of a set that make two functions equal.
This can be generalized to arbitrary categories using the following notion.

\begin{defn} \label{def:equalizer}
  Suppose $f$ and $g$ are both morphisms $X\to Y$.
  Their \textit{equalizer} is an object $E$ equipped with a function $e:E\to X$ such that $f\circ e = g\circ e$ (so that $e$ is said to \textit{equalize} $f$ and $g$) as in the commuting diagram
  \[\begin{tikzcd}
    E && X && Y
    \arrow["f", shift left=2, from=1-3, to=1-5]
    \arrow["g"', shift right=2, from=1-3, to=1-5]
    \arrow["e", from=1-1, to=1-3]
  \end{tikzcd}\]
  and such that, for any $d:D\to X$ equalizing $f$ and $g$, there is a unique morphism $u:D\to E$ such that $d = e\circ u$, as in the diagram
  \[\begin{tikzcd}
    D && E && X && Y
    \arrow["f", shift left=2, from=1-5, to=1-7]
    \arrow["g"', shift right=2, from=1-5, to=1-7]
    \arrow["e", from=1-3, to=1-5]
    \arrow["u", dashed, from=1-1, to=1-3]
    \arrow["d"', curve={height=18pt}, from=1-1, to=1-5]
  \end{tikzcd} \; .\]
\end{defn}

\begin{ex} \label{ex:intersect-eq}
  Via the correspondence between subsets and propositions, we can express the conjunction of propositions as an equalizer.
  Suppose have have two propositions $\mathsf{P}_A:X\to 2$ and $\mathsf{P}_B:X\to 2$, corresponding to subsets $A\hookrightarrow X$ and $B\hookrightarrow X$, whose inclusions we denote by $\iota_A$ and $\iota_B$ respectively.
  The equalizer of $\begin{tikzcd} {A\times B} && X \arrow["{\iota_A\circ\proj_A}", shift left=2, from=1-1, to=1-3] \arrow["{\iota_B\circ\proj_B}"', shift right=2, from=1-1, to=1-3] \end{tikzcd}$ is the subset of $A\times B$ whose elements are pairs $(a,b)$ in which $a=b$ in $X$.
  This subset is isomorphic to the meet $A\cap B$, which corresponds as a proposition to the conjunction $\mathsf{P}_A\wedge\mathsf{P}_B$.
\end{ex}

\subsubsection{Coequalizers and quotients}

We can also make objects `smaller' by dividing them into equivalence classes, as we did when quotienting free categories by given relations (\textit{cf.} Proposition \ref{prop:cat-gen-rel}).
In general, this pattern is captured by the notion of \textit{coequalizer}, which is dual to the notion of equalizer in the same way that coproducts are dual to products.

\begin{defn}
  Suppose $f$ and $g$ are both morphisms $X\to Y$.
  Their \textit{coequalizer} is an object $P$ equipped with a function $p:Y\to P$ such that $p\circ f = p\circ g$ (with $p$ said to \textit{coequalize} $f$ and $g$) as in the commuting diagram
  \[\begin{tikzcd}
    X && Y && P
    \arrow["f", shift left=2, from=1-1, to=1-3]
    \arrow["g"', shift right=2, from=1-1, to=1-3]
    \arrow["p", from=1-3, to=1-5]
  \end{tikzcd}\]
  and such that, for any $q:Y\to Q$ coequalizing $f$ and $g$, there is a unique morphism $u:P\to Q$ such that $q = u\circ p$, as in the diagram
  \[\begin{tikzcd}
    X && Y && P && Q
    \arrow["f", shift left=2, from=1-1, to=1-3]
    \arrow["g"', shift right=2, from=1-1, to=1-3]
    \arrow["p", from=1-3, to=1-5]
    \arrow["u", dashed, from=1-5, to=1-7]
    \arrow["q"', curve={height=18pt}, from=1-3, to=1-7]
  \end{tikzcd} \; .\]
\end{defn}

\begin{ex} \label{ex:quot-coeq}
  A relation $\sim$ on a set $X$ is a proposition on $X\times X$, and thus equivalently a subset $R \hookrightarrow X\times X$; let $\iota$ denote the inclusion.
  The coequalizer of $\begin{tikzcd} {R} && X \arrow["{\proj_1\circ\iota}", shift left=2, from=1-1, to=1-3] \arrow["{\proj_2\circ\iota}"', shift right=2, from=1-1, to=1-3] \end{tikzcd}$ is the set of equivalence classes of $X$ according to $\sim$, which is precisely the quotient $X/{\sim}$.
\end{ex}

\subsection{The pattern of universality}

There is a common pattern to the common patterns above:
in each case, we described an object $U$ equipped with some morphisms, such that, given any object $X$ with morphisms of a similar shape, there was a unique morphism $u$ relating $X$ and $U$.
The existence of such a unique morphism for any comparable $X$ makes the object $U$ a \textit{universal} representative of the situation at hand and has a number of powerful consequences: in particular, it entirely characterizes the object $U$ up to isomorphism.
Much of the power of category theory comes from the use of universal properties to classify, compare, and reason about situations of interest --- for the general notion of universality itself can be characterized categorically.

\begin{defn}
  Suppose $F:\cat{C}\to\cat{D}$ is a functor and $X : \cat{D}$ an object.
  We define two dual universal constructions.
  A \textit{universal morphism from} $X$ \textit{to} $F$ is a morphism $u : X\to FU$ for a corresponding \textit{universal object} $U:\cat{C}$ such that for any $f : X\to FV$ in $\cat{D}$ there exists a unique $e : U\to V$ such that $f = X \xto{u} FU \xto{Fe} FV$.

  Dually, a \textit{universal morphism from} $F$ \textit{to} $X$ is a morphism $u : FU\to X$ for a given $U:\cat{C}$ such that for any $f : FV\to X$ in $\cat{D}$ there exists a unique $e : V\to U$ such that $f = FV \xto{Fe} FU \xto{u} X$.
\end{defn}

We can now formalize the universal properties of the examples we met above, beginning with the coproduct.

\begin{ex}
  Let $\Delta:\cat{C}\to\cat{C}\times\cat{C}$ denote the functor $X\mapsto (X,X)$.
  A coproduct of $X$ and $Y$ in $\cat{C}$ is a universal morphism from the object $(X,Y)$ in $\cat{C}\times\cat{C}$ to $\Delta$:
  that is, an object $X+Y$ in $\cat{C}$ and a morphism $(\inj_X,\inj_Y):(X,Y)\to(X+Y,X+Y)$ in $\cat{C}\times\cat{C}$ such that, for any $(f,g):(X,Y)\to(Q,Q)$ in $\cat{C}\times\cat{C}$, the copairing $[f,g]:X+Y\to Q$ uniquely satisfies the equation $(f,g) = (X,Y)\xto{(\inj_X,\inj_Y)}(X+Y,X+Y)\xto{([f,g],[f,g])}(Q,Q)$.
\end{ex}

\begin{ex}
  Again let $\Delta:\cat{C}\to\cat{C}\times\cat{C}$ denote the functor $X\mapsto (X,X)$.
  A product of $X$ and $Y$ in $\cat{C}$ is a universal morphism from the object $(X,Y):\cat{C}\times\cat{C}$ to $\Delta$: that is, an object $X\times Y$ in $\cat{C}$ and a morphism $(\proj_X,\proj_Y):(X\times Y,X\times Y)\to(X,Y)$ in $\cat{C}\times\cat{C}$ such that, for any $(f,g):(Q,Q)\to(X,Y)$ in $\cat{C}\times\cat{C}$, the pairing $(f,g):Q\to X\times Y$ uniquely satisfies the equation $(f,g) = (Q,Q)\xto{((f,g),(f,g))}(X\times Y,X\times Y)\xto{(\proj_X,\proj_Y)}(X,Y)$.
\end{ex}

\begin{rmk} \label{rmk:cc-c-sq}
  If we let $\Cat{2}$ denote the two-object discrete category $\{\bullet\;\bullet\}$, then there is an equivalence $\cat{C}\times\cat{C}\cong\cat{C}^{\Cat{2}}$ and so a pair of morphisms in $\cat{C}$ is equivalently a natural transformation in $\cat{C}^{\Cat{2}}$.
  (This is a categorification of the familiar fact that ``exponentiation is repeated multiplication'', which we will explore in \secref{sec:closed-cats}.)
\end{rmk}

Consequently, the functor $\Delta$ from the preceding examples can equivalently be defined as a functor $\cat{C}\to\cat{C}^{\Cat{2}}$.
Letting the exponent take a more general shape, we obtain a family of constant functors.

\begin{prop}
  Suppose $\cat{C}$ and $\cat{D}$ are categories, and $d:\cat{D}$ is an object.
  Then there is a \textit{constant functor} on $d$, denoted $\Delta_d : \cat{C}\to\cat{D}$, which takes each object $c:\cat{C}$ to $d:\cat{D}$ and each morphism $f:c\to c'$ to $\id_d$; note that $Fc = d = Fc'$.
  The assignment $d \mapsto \Delta_d$ is itself trivially functorial, giving a functor $\Delta : \cat{D}\to\cat{D}^{\cat{C}}$ which we call the \textit{constant functor functor}.
\end{prop}

\begin{ex} \label{ex:eq-univ-mor}
  Let $\cat{J}$ be the category with two objects, $1$ and $2$, and two non-identity morphisms $\alpha,\beta:1\to 2$, as in the diagram $\begin{tikzcd} 1 & 2 \arrow["{\alpha}", shift left=1, from=1-1, to=1-2] \arrow["{\beta}"', shift right=1, from=1-1, to=1-2] \end{tikzcd}$, and let $\Delta$ be the constant functor functor $\cat{C}\to\cat{C}^{\cat{J}}$.
  Now suppose $f$ and $g$ are two morphisms $X\to Y$ in $\cat{C}$.
  To construct their equalizer as a universal morphism, let $D$ be the diagram $\cat{J}\to\cat{C}$ mapping $\alpha\mapsto f$ and $\beta\mapsto g$.
  Then an equalizer of $f$ and $g$ is a universal morphism from $\Delta$ to $D$ (with $D$ being an object of the functor category $\cat{C}^{\cat{J}}$): that is, an object $E:\cat{C}$ equipped with a natural transformation $\epsilon:\Delta_E\Rightarrow D$ satisfying the universal property that, for any $\varphi:\Delta_F\Rightarrow D$ there exists a unique morphism $u:F\to E$ such that $\varphi = \epsilon\circ\Delta_u$.

  Unraveling this definition, we find that such a natural transformation $\epsilon$ consists of a pair of morphisms $\epsilon_1:E\to X$ and $\epsilon_2:E\to Y$ making the following naturality squares commute:
  \[\begin{tikzcd}
    E & X \\
    E & Y
    \arrow["{\epsilon_1}", from=1-1, to=1-2]
    \arrow[Rightarrow, no head, from=1-1, to=2-1]
    \arrow["{\epsilon_2}"', from=2-1, to=2-2]
    \arrow["f", from=1-2, to=2-2]
  \end{tikzcd}
  \qquad\qquad
  \begin{tikzcd}
    E & X \\
    E & Y
    \arrow["{\epsilon_1}", from=1-1, to=1-2]
    \arrow[Rightarrow, no head, from=1-1, to=2-1]
    \arrow["{\epsilon_2}"', from=2-1, to=2-2]
    \arrow["g", from=1-2, to=2-2]
  \end{tikzcd}\]
  We can therefore set $\epsilon_1 = e$, where $e$ is the equalizing morphism $E\to X$.
  The commutativity of the naturality squares enforces that $f\circ e = \epsilon_2 = g\circ e$ and hence that $f\circ e = g\circ e$, which is the first condition defining the equalizer.
  Unwinding the universal property as expressed here shows that the morphisms $\varphi_1$ and $u$ correspond exactly to the morphisms $d$ and $u$ of Definition \ref{def:equalizer}.
\end{ex}

\begin{ex}
  The case of a coequalizer is precisely dual to that of an equalizer.
  Therefore, let $\cat{J}$, $\Delta$, and $D$ be defined as above.
  A coequalizer of $f,g:X\to Y$ is then a universal morphism from $D$ to $\Delta$.
\end{ex}

\begin{ex}
  In Proposition \ref{prop:cat-gen-rel}, we constructed a category generated with relations  $F\cat{G}/{\sim}$ as a quotient of a free category on a graph $F\cat{G}$.
  Since this category $F\cat{G}/{\sim}$ is a quotient and quotients are coequalizers (by Example \ref{ex:quot-coeq}), the projection functor $F\cat{G}\to F\cat{G}/{\sim}$ (Example \ref{ex:quot-cat-proj}) constitutes the associated universal morphism, in the sense dual to the morphism $\epsilon_1$ of Example \ref{ex:eq-univ-mor}.
\end{ex}

\begin{ex} \label{ex:free-cat-univ-prop}
  The free category construction itself (Proposition \ref{prop:free-cat}) exhibits a universal property, as a consequence of the free-forgetful adjunction (Example \ref{ex:grph-cat-adjunc}): given a graph $\cat{G}$ and a category $\cat{C}$, any functor $F\cat{G}\to\cat{C}$ is uniquely determined by a graph homomorphism $\cat{G}\to U\cat{C}$ from $\cat{G}$ to the underlying graph of $\cat{C}$.
  More precisely, there is a universal morphism from the singleton set $1$ to the functor $\Cat{Cat}(F-, \cat{C})$ for every category $\cat{C}$.
  This means that, for any graph $\cat{G}$, every functor $f:F\cat{G}\to\cat{C}$ factors as $F\cat{G}\xto{Fh} FU\cat{C}\xto{u}\cat{C}$ where $u$ is the universal morphism and $h$ is the unique graph homomorphism.
  This universal property follows abstractly from facts that we will soon encounter: that adjoint functors are `representable' (Proposition \ref{prop:adjoint-rep}); and that representable functors are universal (Proposition \ref{prop:univers-repr}).
  We hinted at this property in Example \ref{ex:grph-cat-adjunc}, where we observed that functors between free categories `are' graph homomorphisms: the homomorphism $h$ here is the graph homomorphism corresponding to the functor $f$, and $u$ renders it as a functor into $\cat{C}$.
\end{ex}

When an object satisfies a universal property, then this property characterizes the object uniquely: as a result, universal properties are powerful constructions, telling us that for certain questions, there can be only one possible answer.
Characterizing an object by a universal property abstracts away from contextually irrelevant details (for example, the particular elements making up a set), and crystallizes its essence.

The uniqueness of universal morphisms is formalized by the following proposition.

\begin{prop}[Universal constructions are unique up to unique isomorphism] \label{prop:univs-unique}
  Suppose $u:X\to FU$ and $u':X\to FU'$ are both universal morphisms from $X:\cat{C}$ to $F:\cat{C}\to\cat{D}$.
  Then there is a unique isomorphism $i:U\to U'$.
\end{prop}

To prove this, we need to know that functors preserve isomorphisms.

\begin{prop}
  If $F:\cat{C}\to\cat{D}$ is a functor and $f:x\to y$ is an isomorphism in $\cat{C}$, then $Ff:Fx\to Fy$ is an isomorphism in $\cat{D}$.
  \begin{proof}
    For $f$ to be an isomorphism, there must be a morphism $f^{-1}:y\to x$ such that $f^{-1}\circ f = \id_x$ and $f\circ f^{-1} = \id_y$.
    We have $\id_{Fx} = F(\id_x) = F(f^{-1}\circ f) = Ff^{-1}\circ Ff$, where the first and third equations hold by the functoriality of $F$ and the second equation holds \textit{ex hypothesi}.
    Similarly, $\id_{Fy} = F(\id_y) = F(f\circ f^{-1}) = Ff\circ Ff^{-1}$.
    Therefore $Ff^{-1}$ is both a right and left inverse for $Ff$, and so $Ff$ is an isomorphism.
  \end{proof}
\end{prop}

\begin{proof}[Proof of Proposition \ref{prop:univs-unique}]
  Since $u'$ is a morphism from $X$ to $F$, the universal property of $u$ says that there exists a unique morphism $i:U\to U'$ such that $u' = Fi\circ u$.
  Similarly, the universal property of $u'$ stipulates that there exists a unique morphism $i':U'\to U$ such that $u = Fi'\circ u'$.
  We can substitute the latter into the former and the former into the latter:
  \begin{gather*}
    \begin{aligned}
      u'
      &= X\xto{u'}FU'\xto{i'}FU\xto{Fi}FU' \\
      &= X\xto{u'}FU'\xto{F(i\circ i')}FU' \\
      &= X\xto{u'}FU'\xto{F\id_{U'}}FU'
    \end{aligned}
    \qquad\qquad
    \begin{aligned}
      u
      &= X\xto{u}FU\xto{Fi}FU'\xto{Fi'}FU \\
      &= X\xto{u}FU\xto{F(i'\circ i)}FU \\
      &= X\xto{u}FU\xto{F\id_U}FU
    \end{aligned}
  \end{gather*}
  and since functors preserve isomorphism, we have $i\circ i' = \id_{U'}$ and $i'\circ i = \id_U$.
  Therefore, $i$ is an isomorphism which is unique by definition.
\end{proof}

\subsection{Limits and colimits: mapping in to and out of diagrams} \label{sec:limits}

Many of the universal constructions above\footnote{%
Products, coproducts, equalizers, and coequalizers.}
fall into their own general pattern, in which a diagram of objects and morphisms is specified, and a universal morphism is produced which encodes the data of mapping into or out of that diagram, in a sufficiently parsimonious way that any other way of mapping into or out of the diagram factors through it.
In the case of the (co)product, the diagram is simple: simply a pair of objects, with no morphisms between them.
In the case of the (co)equalizer, the diagram is a little more complex, being a `fork' of the form $\begin{tikzcd} 1 & 2 \arrow["{\alpha}", shift left=1, from=1-1, to=1-2] \arrow["{\beta}"', shift right=1, from=1-1, to=1-2] \end{tikzcd}$.
We can generalize these examples further, to consider the most parsimonious ways of mapping into or out of diagrams of arbitrary shape:
these universal constructions are called \textit{colimits} and \textit{limits} respectively, and to formalize them, we need to define what it means to map into or out of a diagram.
For this purpose, we use the following notion of \textit{cone} over a diagram.

\begin{defn}
  A \textit{cone} over the $J$-shaped diagram $D$ in $\cat{C}$ is a natural transformation $\Delta_c \Rightarrow D$ for a given object $c:\cat{C}$ which we call its \textit{apex}.
  Dually, a \textit{cocone} under $D$ with apex $c$ is a natural transformation $D \Rightarrow \Delta_c$.
  We say that $J$ is the \textit{shape} of the cone.
\end{defn}

With this definition to hand, the notions of limit and colimit are easy to define.

\begin{defn}
  A \textit{limit} is a universal cone, and a \textit{colimit} is a universal cocone.
  More explicitly, if $D$ is a $J$-shaped diagram in $\cat{C}$, then the limit of $D$ is a universal morphism from the constant diagram functor functor $\Delta : \cat{C}\to\cat{C}^J$ to $D$ (considered as an object of the functor category), and the colimit of $D$ is a universal morphism from $D$ to $\Delta$; alternatively, a colimit in $\cat{C}$ is a limit in $\cat{C}\op$.
  In both cases, the apex of the cone is the universal object of the construction, which in the case of the limit of $D$ we denote by $\lim D$, and in the case of the colimit, $\colim D$.

  Note that we will often say `(co)limit' to refer to the apex of the universal (co)cone, even though the (co)limit is properly the whole universal construction.
  We are entitled to say ``\textit{the} (co)limit'' thanks to the uniqueness of universal constructions.

  We will often denote a universal cone by $\proj$ and call its component morphisms \textit{projections}; dually, we will often denote a universal cocone by $\inj$ and call its morphisms \textit{injections}.
\end{defn}

\begin{ex}
  We can now exemplify the pattern of the limiting examples above.
  We will draw diagrams to depict the shape categories, with each symbol $\bullet$ indicating a distinct object and each arrow $\to$ indicating a distinct non-identity morphism.
  \begin{enumerate}
  \item A coproduct is a colimit of shape $\left\{ \begin{tikzcd} {\bullet} & {\bullet} \end{tikzcd} \right\}$;
  \item a product is a limit of shape $\left\{ \begin{tikzcd} {\bullet} & {\bullet} \end{tikzcd} \right\}$;
  \item an equalizer is a limit of shape $\left\{ \begin{tikzcd} {\bullet} & {\bullet} \arrow[shift left=1, from=1-1, to=1-2] \arrow[shift right=1, from=1-1, to=1-2] \end{tikzcd} \right\}$; and
  \item a coequalizer is a colimit of shape $\left\{ \begin{tikzcd} {\bullet} & {\bullet} \arrow[shift left=1, from=1-1, to=1-2] \arrow[shift right=1, from=1-1, to=1-2] \end{tikzcd} \right\}$.
  \end{enumerate}
\end{ex}

Of course, these are not the only possible shapes of limits and colimits.
Some others will be particularly important, too.

\begin{ex} \label{ex:term-init-obj}
  Let $\Cat{0}$ denote the category with no objects or morphisms.
  A limit of shape $\Cat{0}$ is known as a \textit{terminal object}.
  This is an object $1$ such that, for every object $X$, there is a unique morphism $!:X\to 1$.
  The terminal object in $\Set$ is a singleton set $\{\ast\}$.

  Dually, a colimit of shape $\Cat{0}$ is known as an \textit{initial object}: an object $0$ such that, for every object $X$, there is a unique morphism \rotatebox[origin=c]{180}{$!$}$\,:0\to X$.
  The initial object in $\Set$ is the empty set.
\end{ex}

\begin{rmk}
  In Remark \ref{rmk:element}, we noted that morphisms $1\to A$ in $\Set$ correspond to elements of $A$.
  In general categories $\cat{C}$ with a terminal object, one sometimes calls morphisms out of the terminal object \textit{global elements}.
  The word `global' emphasizes the special position of the terminal object in a category, which has a unique view of every object.
\end{rmk}

\begin{ex} \label{ex:pullback}
  A \textit{pullback} is a limit of shape $\left\{ \begin{tikzcd} {\bullet} & {\bullet} & {\bullet} \arrow[from=1-1, to=1-2] \arrow[from=1-3, to=1-2] \end{tikzcd} \right\}$.
  That is, given morphisms $f:A\to X$ and $g:B\to X$, their pullback is an object $P$ and morphisms $\proj_A:P\to A$ and $\proj_B:P\to B$ making the following diagram commute
  \[\begin{tikzcd}
    P & B \\
    A & X
    \arrow["g", from=1-2, to=2-2]
    \arrow["f"', from=2-1, to=2-2]
    \arrow["{\proj_B}", from=1-1, to=1-2]
    \arrow["{\proj_A}"', from=1-1, to=2-1]
    \arrow["\lrcorner"{anchor=center, pos=0.125}, draw=none, from=1-1, to=2-2]
  \end{tikzcd}\]
  in the universal sense that, for any object $Q$ and morphisms $\pi_A:Q\to A$ and $\pi_B:Q\to B$ such that $f\circ\pi_A = g\circ\pi_B$, then there is a unique morphism $u:Q\to P$ such that $\pi_A = \proj_A\circ u$ and $\pi_B = \proj_B\circ u$.
  We indicate a \textit{pullback square} using the symbol $\lrcorner$ as in the diagram above, and will variously denote the limiting object $P$ by $A\times_X B$, $f^*B$, or $g^*A$, depending on the context.

  The interpretation of the pullback is something like a generalized equation:
  in the category $\Set$, the pullback $A\times_X B$ is the subset of the product $A\times B$ consisting of elements $(a,b)$ for which $f(a) = g(b)$.
  Alternatively, it can be understood as a kind of generalized intersection: given two objects $A$ and $B$ and ``ways of assigning them properties in $X$'' $f$ and $g$, the pullback $A\times_X B$ is the generalized intersection of $A$ and $B$ according to these $X$-properties.
  In fact, we already saw this latter interpretation in Example \ref{ex:intersect-eq}, where we exhibited an intersection as an equalizer; now we can see that that equalizer was `secretly' a pullback.
\end{ex}

\begin{rmk}
  Dually, a colimit of shape $\left\{ \begin{tikzcd} {\bullet} & {\bullet} & {\bullet} \arrow[from=1-1, to=1-2] \arrow[from=1-3, to=1-2] \end{tikzcd} \right\}$ is known as a \textit{pushout}.
  Whereas a pullback has an interpretation as a subobject of a product, a pushout has an interpration as a quotient of a coproduct.
  In this work, we will make far more use of pullbacks than pushouts.
\end{rmk}

The observation that pullbacks can be interpreted as subobjects of products (and dually that pushouts can be interpreted as quotients of coproducts) is a consequence of the more general result that all limits can be expressed using products and equalizers (and hence dually that colimits can be expressed using coproducts and coequalizers).

\begin{prop} \label{prop:lim-prod-eq}
  Let $D:J\to\cat{C}$ be a diagram in $\cat{C}$, and suppose the products $\prod_{j:J_0} D(j)$ and $\prod_{f:J_1} D(\cod f)$ exist.
  Then, if it exists, the equalizer of
  \[\begin{tikzcd}
    {\prod_{j:J_0}D(j)} &&& {\prod_{f:J_1}D(\cod f)}
    \arrow["{\prod_{f:J_1}\left(Df\,\circ\,\proj_{\dom f}\right)}", shift left=1, curve={height=-3pt}, from=1-1, to=1-4]
    \arrow["{\prod_{f:J_1} \proj_{\cod f}}"', shift right=1, curve={height=3pt}, from=1-1, to=1-4]
  \end{tikzcd}\]
  is the limit of $D$.
  \begin{proof}[Proof sketch]
    Observe that the equalizer of the diagram above is an object $L$ such that, for every morphism $f:j\to j'$ in $J$, the diagram
    \[\begin{tikzcd}
      & L \\
      \\
      Dj && {Dj'}
      \arrow["{\proj_j}"', from=1-2, to=3-1]
	    \arrow["{\proj_{j'}}", from=1-2, to=3-3]
	    \arrow["Df"', from=3-1, to=3-3]
    \end{tikzcd}\]
    commutes, and such that any cone over $D$ factors through it.
    This is precisely the universal property of the limit of $D$, and so by Proposition \ref{prop:univs-unique}, $(L,\proj)$ is the limit.
  \end{proof}
\end{prop}

\begin{rmk}
  As we indicated above, a dual result holds expressing colimits using coequalizers and coproducts.
  Because results obtained for limits in $\cat{C}$ will hold for colimits in $\cat{C}\op$, we will henceforth not always give explicit dualizations.
\end{rmk}

\subsubsection{Functoriality of taking limits}

In the statement of Proposition \ref{prop:lim-prod-eq}, we used the fact that taking products extends to morphisms, too: a fact that was exemplified concretely in Example \ref{ex:prod-set}, and which follows from the fact that a pair of morphisms in $\cat{C}$ is equivalently a morphism in $\cat{C}\times\cat{C}$.
We then saw in Remark \ref{rmk:cc-c-sq} that $\cat{C}\times\cat{C} \cong \cat{C}^{\Cat{2}}$.
By letting the exponent again vary, the functoriality of taking products generalizes to the functoriality of taking limits, as long as $\cat{C}$ has all limits of the relevant shape.

\begin{prop}[Taking limits is functorial] \label{prop:lim-func}
  Suppose $\cat{C}$ has all limits of shape $J$ (\textit{i.e.}, for any diagram $D:J\to\cat{C}$, the limit $\lim D$ exists in $\cat{C}$).
  Then $\lim$ defines a functor $\Cat{Cat}(J,\cat{C})\to\cat{C}$.
  \begin{proof}
    We only need to check the assignment is well-defined on morphisms and functorial.
    Suppose $D$ and $D'$ are two diagrams $J\to\cat{C}$ with corresponding limiting cones $u:\Delta_{\lim D}\Rightarrow D$ and $u':\Delta_{\lim D'}\Rightarrow D'$, and suppose $\delta : D\Rightarrow D'$ is a natural transformation.
    Observe that the composite natural transformation $\Delta_{\lim D}\xRightarrow{u}D\xRightarrow{\delta}D'$ is a cone on $D'$, and that cones on $D'$ are in bijection with morphisms in $\cat{C}$ into the apex object $\lim D'$.
    Therefore, by the universal property of the limit, there is a unique morphism $d:\lim D\to\lim D'$ such that $\delta\circ u = u'\circ\Delta_d$.
    This situation is summarized by the commutativity of the following diagram, where the dashed arrow indicates the uniqueness of $\Delta_d$:
    \[\begin{tikzcd}
	    {\Delta_{\lim D}} && {\Delta_{\lim D'}} \\
	    \\
	    D && {D'}
	    \arrow["{\Delta_d}", Rightarrow, dashed, from=1-1, to=1-3]
	    \arrow["\delta", Rightarrow, from=3-1, to=3-3]
	    \arrow["u", Rightarrow, from=1-1, to=3-1]
	    \arrow["{u'}", Rightarrow, from=1-3, to=3-3]
    \end{tikzcd}\]
    We define the action of the functor $\lim:\Cat{Cat}(J,\cat{C})\to\cat{C}$ on the natural transformation $\delta$ by this unique morphism $d$, setting $\lim\delta := d : \lim D\to\lim D'$.

    It therefore only remains to check that this assignment is functorial (\textit{i.e.}, that it preserves identities and composites).
    To see that $\lim$ preserves identities, just take $\delta = \id_D$ in the situation above; clearly, by the uniqueness of $d$, we must have $\lim\id_D = \id_{\lim D}$.
    Now suppose $\delta':D'\to D''$ is another natural transformation.
    To see that $\lim(\delta'\circ\delta) = \lim\delta'\circ\lim\delta$, consider the pasting of the associated diagrams:
    \[\begin{tikzcd}
	    {\Delta_{\lim D}} && {\Delta_{\lim D'}} && {\Delta_{\lim D''}} \\
	    \\
	    D && {D'} && {D''}
	    \arrow["{\Delta_d}", Rightarrow, dashed, from=1-1, to=1-3]
	    \arrow["\delta", Rightarrow, from=3-1, to=3-3]
	    \arrow["u", Rightarrow, from=1-1, to=3-1]
	    \arrow["{u'}", Rightarrow, from=1-3, to=3-3]
	    \arrow["{\Delta_{d'}}", Rightarrow, dashed, from=1-3, to=1-5]
	    \arrow["{u''}", Rightarrow, from=1-5, to=3-5]
	    \arrow["{\delta'}", Rightarrow, from=3-3, to=3-5]
	    \arrow["{\Delta_{d'd}}", curve={height=-30pt}, Rightarrow, dashed, from=1-1, to=1-5]
    \end{tikzcd}\]
    We have $\lim(\delta'\circ\delta) = d'd$, which is unique by definition.
    Therefore we must have $d'd = d'\circ d = \lim(\delta')\circ\lim(\delta)$, and hence $\lim(\delta'\circ\delta) = \lim(\delta')\circ\lim(\delta)$ as required.
  \end{proof}
\end{prop}

\subsubsection{(Co)limits as adjoints}

Since taking limits is functorial, it makes sense to ask if the functor $\lim$ has an adjoint, and indeed it does, in a familiar form.

\begin{prop} \label{prop:limit-adjoint}
  The functor $\lim:\cat{C}^J\to\cat{C}$ is right adjoint to the constant diagram functor functor $\Delta:\cat{C}\to\cat{C}^J$, \textit{i.e.} $\Delta \dashv \lim$.
  \begin{proof}
    We need to show that $\cat{C}^J(\Delta_c,D) \cong \cat{C}(c,\lim D)$ naturally in $c:\cat{C}$ and $D:J\to\cat{C}$.
    It is sufficient to demonstrate naturality in each argument separately, by the universal property of the product in $\Cat{Cat}$.
    We have already established naturality in $c:\cat{C}$ in Lemma \ref{lemma:nat-into-lim} and shown that taking limits is functorial (Proposition \ref{prop:lim-func}).
    So it only remains to show that this extends to naturality in $D:J\to\cat{C}$, which requires the commutativity of the following diagram for any $\delta:D\to D'$, where we write $\alpha_D$ for the isomorphism $\cat{C}(c,\lim D)\xto{\sim}\cat{C}^J(\Delta_c,D)$:
    \[\begin{tikzcd}
	    {\cat{C}(c,\lim D)} && {\cat{C}^J(\Delta_c,D)} \\
	    \\
      {\cat{C}(c,\lim D')} && {\cat{C}^J(\Delta_c,D')}
	    \arrow["{\alpha_D}", from=1-1, to=1-3]
	    \arrow["{\alpha_{D'}}", from=3-1, to=3-3]
	    \arrow["{\cat{C}^J(\Delta_c,\delta)}", from=1-3, to=3-3]
	    \arrow["{\cat{C}(c,\lim\delta)}", from=1-1, to=3-1]
    \end{tikzcd}\]
    Chasing a morphism $\beta:c\to\lim D$ around this diagram, we find that its commutativity amounts to the commutativity of the following diagram of cones for all $\varphi:i\to j$ in $J$, where by definition $\alpha_D(\beta)_i = \pi_i\circ\beta$ and $\alpha_{D'}(\lim\delta\circ\beta)_i = \pi'_i\circ\lim\delta\circ\beta$:
    \[\begin{tikzcd}
	    &&&& Di &&& {D'i} \\
	    c && {\lim D} &&& {\lim D'} \\
	    &&&& Dj &&& {D'j}
	    \arrow["D\varphi"{pos=0.2}, from=1-5, to=3-5]
	    \arrow["{\delta_i}", from=1-5, to=1-8]
	    \arrow["{D'\varphi}"{pos=0.2}, from=1-8, to=3-8]
	    \arrow["{\delta_j}"', from=3-5, to=3-8]
	    \arrow["{\pi_i}"{description}, from=2-3, to=1-5]
	    \arrow["{\pi_j}"{description}, from=2-3, to=3-5]
	    \arrow["{\pi'_i}"{description}, from=2-6, to=1-8]
	    \arrow["{\pi'_j}"{description}, from=2-6, to=3-8]
	    \arrow["\lim\delta"{description, pos=0.4}, dashed, from=2-3, to=2-6]
	    \arrow["\beta"{description}, dashed, from=2-1, to=2-3]
	    \arrow["{\pi_i\circ\beta}"{description}, curve={height=-12pt}, from=2-1, to=1-5]
	    \arrow["{\pi_j\circ\beta}"{description}, curve={height=12pt}, from=2-1, to=3-5]
    \end{tikzcd}\]
    This diagram commutes by definition, so the isomorphism is natural in $D$, which therefore establishes the desired adjunction.
  \end{proof}
\end{prop}

\begin{rmk}
  Dually, if all colimits of shape $J$ exist in $\cat{C}$, then $\colim$ is left adjoint to $\Delta$.
\end{rmk}

Later, we will see that every adjoint functor exhibits a universal property (Propositions \ref{prop:adjoint-rep} and \ref{prop:univers-repr}, results that we've already seen exemplified in Example \ref{ex:free-cat-univ-prop}), and this therefore gives us another perspective on the universality of limits.

\subsubsection{Hom preserves limits}

We end this section with a useful result on the interaction between the covariant hom functors $\cat{C}(c,-):\cat{C}\to\Set$ and taking limits.

\begin{prop}[Hom functor preserves limits] \label{prop:hom-preserves-limits}
  Suppose $D:J\to\cat{C}$ is a diagram in the category $\cat{C}$.
  There is an isomorphism $\cat{C}(c,\lim D) \cong \lim\cat{C}(c, D(-))$ which is natural in $c:\cat{C}$.
\end{prop}

To prove this proposition, it helps to have the following lemma, which establishes a natural isomorphism between the set of morphisms into a limit and the set of cones on the corresponding diagram.

\begin{lemma} \label{lemma:nat-into-lim}
  $\cat{C}(c,\lim D) \cong \cat{C}^J(\Delta_c,D)$, naturally in $c:\cat{C}$.
  \begin{proof}
    For a given $c:\cat{C}$, the isomorphism $\cat{C}(c,\lim D) \cong \cat{C}^J(\Delta_c,D)$ follows directly from the universal property of the limit: morphisms from $c$ into the limiting object $\lim D$ are in bijection with cones $\Delta_c\Rightarrow D$.
    So it only remains to show that this isomorphism is natural in $c:\cat{C}$.
    Writing $\alpha : \cat{C}(-,\lim D) \Rightarrow \cat{C}^J(\Delta_{(-)},D)$ for the natural transformation that takes each morphism into the limit to the corresponding cone on $D$, naturality amounts to the commutativity of the following square for each $f:c'\to c$ in $\cat{C}$:
    \[\begin{tikzcd}
	    {\cat{C}(c,\lim D)} && {\cat{C}^J(\Delta_c,D)} \\
	    \\
	      {\cat{C}(c',\lim D)} && {\cat{C}^J(\Delta_{c'},D)}
	      \arrow["{\cat{C}(f,\lim D)}"', from=1-1, to=3-1]
	      \arrow["{\cat{C}^J(\Delta_f, D)}", from=1-3, to=3-3]
	      \arrow["{\alpha_c}", from=1-1, to=1-3]
	      \arrow["{\alpha_{c'}}", from=3-1, to=3-3]
    \end{tikzcd}\]
    Commutativity of this naturality square witnesses the fact that, given a morphism $g:c\to\lim D$, you can either take the corresponding cone $\alpha_c(g)$ and pull it back along $\Delta_f$ (at its apex) to obtain the cone $\alpha_c(g)\circ\Delta_f$, or you can form the composite morphism $g\circ f$ and take its cone $\alpha_{c'}(g\circ f)$, and you'll have the same cone: $\alpha_c(g)\circ\Delta_f = \alpha_{c'}(g\circ f)$.
    This is illustrated by the commutativity of the following diagram, which shows fragments of the limiting cone denoted $\pi$, the cone $\alpha_c(g)$, and the cone $\alpha_{c'}(g\circ f)$, for a morphism $\varphi : i\to j$ in $J$:
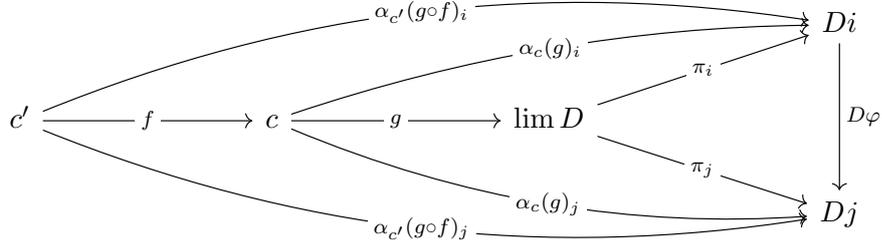
    \[\begin{tikzcd}
	    &&&&&&&&& {Di} \\
	    {c'} &&& c &&& {\lim D} \\
	    &&&&&&&&& {Dj}
	    \arrow["{D\varphi}", from=1-10, to=3-10]
	    \arrow["{\pi_i}"{description}, from=2-7, to=1-10]
	    \arrow["{\pi_j}"{description}, from=2-7, to=3-10]
	    \arrow["g"{description}, from=2-4, to=2-7]
	    \arrow["f"{description}, from=2-1, to=2-4]
	    \arrow["{\alpha_c(g)_i}"{description}, curve={height=-12pt}, from=2-4, to=1-10]
	    \arrow["{\alpha_{c'}(g\circ f)_i}"{description}, curve={height=-30pt}, from=2-1, to=1-10]
	    \arrow["{\alpha_c(g)_j}"{description}, curve={height=18pt}, from=2-4, to=3-10]
	    \arrow["{\alpha_{c'}(g\circ f)_j}"{description}, curve={height=30pt}, from=2-1, to=3-10]
    \end{tikzcd}\]
    By the universal property of the limit, we must have $\alpha_{c'}(g\circ f)_i = \alpha_c(g)_i \circ f = \pi_i\circ g\circ f$ naturally in $i$, and hence $\alpha_{c'}(g\circ f) = \alpha_c(g)\circ\Delta_f$.
  \end{proof}
\end{lemma}

\begin{proof}[Proof of Proposition \ref{prop:hom-preserves-limits}]
  By Lemma \ref{lemma:nat-into-lim}, we have a natural isomorphism $\cat{C}^J(\Delta_c,D) \cong \lim\cat{C}(c, D(-))$, so it suffices to establish a natural isomorphism $\cat{C}^J(\Delta_c,D) \cong \lim\cat{C}(c, D(-))$.
  This says that cones on $D$ with apex $c$ are isomorphic to the limit of $\cat{C}(c,D(-)):J\to\Set$, naturally in $c$.
  First, note that this limiting cone in $\Set$ is constituted by a family of functions $\{p_i : \lim\cat{C}(c,D(-))\to\cat{C}(c,Di)\}_{i:J}$, as in the following commuting diagram:
  
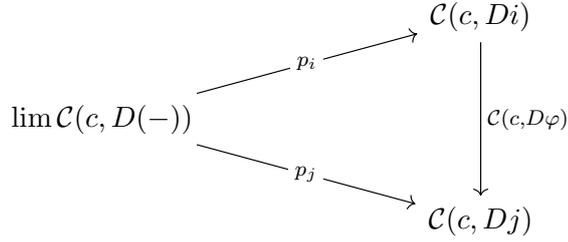
\[\begin{tikzcd}
	  &&& {\cat{C}(c,Di)} \\
	  {\lim \cat{C}(c,D(-))} \\
	  &&& {\cat{C}(c,Dj)}
	  \arrow["{\cat{C}(c,D\varphi)}", from=1-4, to=3-4]
	  \arrow["{p_i}"{description}, from=2-1, to=1-4]
	  \arrow["{p_j}"{description}, from=2-1, to=3-4]
  \end{tikzcd}\]
  Next, note there is a bijection between cones $\Delta_c\Rightarrow D$ on $D$ in $\cat{C}$ with apex $c$, as in the commuting diagram below-left, and cones $\Delta_1\Rightarrow \cat{C}(c,D(-))$ in $\Set$, as in the commuting diagram below-right.
  
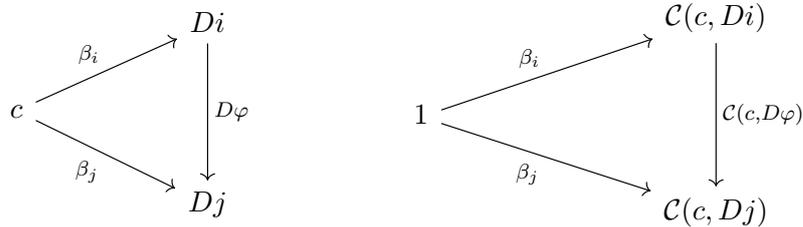
\[
  \begin{tikzcd}
	  && Di \\
	  c \\
	  && Dj
	  \arrow["{\beta_i}", from=2-1, to=1-3]
	  \arrow["D\varphi", from=1-3, to=3-3]
	  \arrow["{\beta_j}"', from=2-1, to=3-3]
  \end{tikzcd}
  \qquad\quad\qquad
  \begin{tikzcd}
    &&& {\cat{C}(c,Di)} \\
	  1 \\
	  &&& {\cat{C}(c,Dj)}
	  \arrow["{\cat{C}(c,D\varphi)}", from=1-4, to=3-4]
	  \arrow["{\beta_i}", from=2-1, to=1-4]
	  \arrow["{\beta_j}"', from=2-1, to=3-4]
  \end{tikzcd}
  \]
  By the univeral property of the limit, any cone $\{\beta_i\}$ as on the right factors uniquely through $\lim\cat{C}(c,D(-))$, as in the following commuting diagram.
  Similarly, any element $\beta$ of $\lim\cat{C}(c,D(-))$ induces a corresponding cone $\{p_i(\beta)\}$, by composition with the limiting cone $p$.
  To see that this correspondence is an isomorphism, observe that the element of the set $\lim\cat{C}(c,D(-))$ assigned to the cone $\{p_i(\beta)\}$ must be exactly $\beta$, since the universal property of $\lim\cat{C}(c,D(-))$ ensures that $\beta$ is uniquely determined.
  
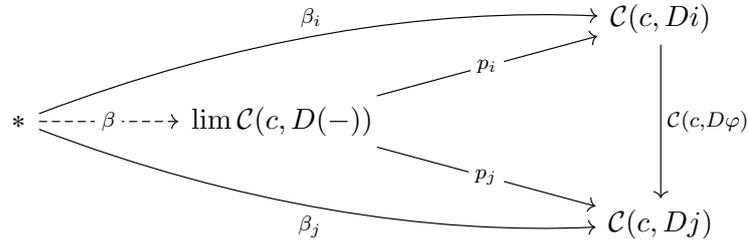
\[\begin{tikzcd}
	  &&&&& {\cat{C}(c,Di)} \\
	  \ast && {\lim \cat{C}(c,D(-))} \\
	  &&&&& {\cat{C}(c,Dj)}
	  \arrow["{\cat{C}(c,D\varphi)}", from=1-6, to=3-6]
	  \arrow["{p_i}"{description}, from=2-3, to=1-6]
	  \arrow["{p_j}"{description}, from=2-3, to=3-6]
	  \arrow["\beta"{description}, dashed, from=2-1, to=2-3]
	  \arrow["{\beta_i}", curve={height=-18pt}, from=2-1, to=1-6]
	  \arrow["{\beta_j}"', curve={height=18pt}, from=2-1, to=3-6]
  \end{tikzcd}\]
  It only remains to check that this correspondence is natural in $c$, so suppose $f$ is any morphism $c'\to c$ in $\cat{C}$.
  If we write $p_{-}:\lim\cat{C}(c,D(-))\to\cat{C}^J(\Delta_c,D)$ to denote the function $\beta\mapsto \{p_i(\beta)\}$, and $p'_{-}$ to denote the corresponding function for $c'$, naturality requires the following square to commute:
  \[\begin{tikzcd}
	  {\lim\cat{C}(c,D(-))} && {\cat{C}^J(\Delta_c,D)} \\
	  \\
	    {\lim\cat{C}(c',D(-))} && {\cat{C}^J(\Delta_{c'},D)}
	    \arrow["{p_{-}}", from=1-1, to=1-3]
	    \arrow["{p'_{-}}", from=3-1, to=3-3]
	    \arrow["{\cat{C}^J(\Delta_f,D)}", from=1-3, to=3-3]
	    \arrow["{\lim\cat{C}(f,D(-))}"', from=1-1, to=3-1]
  \end{tikzcd}\]
  The commutativity of this square in turn corresponds to the commutativity of the following diagram in $\Set$, for any cone $\beta$:
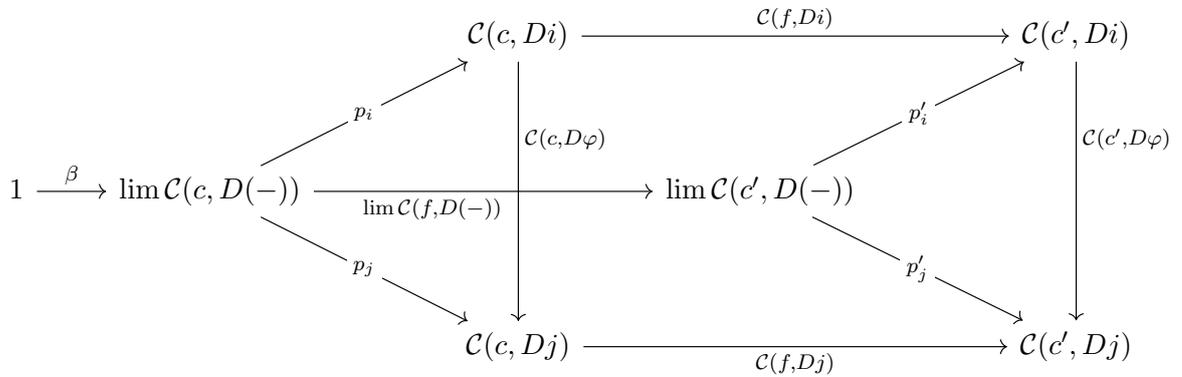
  \[\begin{tikzcd}
	  &&& {\cat{C}(c,Di)} &&& {\cat{C}(c',Di)} \\
	  \\
	  1 & {\lim\cat{C}(c,D(-))} &&& {\lim\cat{C}(c',D(-))} \\
	  \\
	  &&& {\cat{C}(c,Dj)} &&& {\cat{C}(c',Dj)}
	  \arrow["{\cat{C}(f,Di)}", from=1-4, to=1-7]
	  \arrow["{\cat{C}(f,Dj)}"', from=5-4, to=5-7]
	  \arrow["{\cat{C}(c',D\varphi)}"{pos=0.3}, from=1-7, to=5-7]
	  \arrow["{p'_i}"{description}, from=3-5, to=1-7]
	  \arrow["{p'_j}"{description}, from=3-5, to=5-7]
	  \arrow["{\cat{C}(c,D\varphi)}"{pos=0.3}, from=1-4, to=5-4]
	  \arrow["{\lim\cat{C}(f,D(-))}"'{pos=0.35}, from=3-2, to=3-5]
	  \arrow["{p_i}"{description}, from=3-2, to=1-4]
	  \arrow["{p_j}"{description}, from=3-2, to=5-4]
	  \arrow["\beta", from=3-1, to=3-2]
  \end{tikzcd}\]
  By the correspondence between cones $\Delta_c\Rightarrow D$ in $\cat{C}$ and cones $\Delta_1\Rightarrow\cat{C}(c,D(-))$ in $\Set$, this diagram commutes if and only if the following diagram commutes:
  
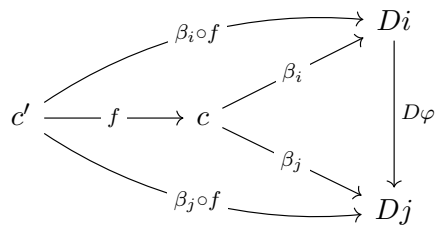
\[\begin{tikzcd}
    &&&& Di \\
	  {c'} && c \\
	  &&&& Dj
	  \arrow["{\beta_i}"{description}, from=2-3, to=1-5]
	  \arrow["D\varphi", from=1-5, to=3-5]
	  \arrow["{\beta_j}"{description}, from=2-3, to=3-5]
	  \arrow["f"{description}, from=2-1, to=2-3]
	  \arrow["{\beta_i\circ f}"{description}, curve={height=-18pt}, from=2-1, to=1-5]
	  \arrow["{\beta_j\circ f}"{description}, curve={height=18pt}, from=2-1, to=3-5]
  \end{tikzcd}\]
  This diagram commutes by the definition of $\beta$ and of the composites $\{\beta_i\circ f\}$, thereby establishing the naturality of the isomorphism $\lim\cat{C}(c,D(-))\cong\cat{C}^J(\Delta_c,D)$.
  Since we also have a natural isomorphism $\cat{C}^J(\Delta_c,D)\cong\cat{C}(c,\lim D)$, we have established the result.
\end{proof}

The preceding proof established more than just the hom functor's preservation of limits:
it gave us another useful natural isomorphism, this time betwen the set of cones $\Delta_c\Rightarrow D$ in $\cat{C}$ and the set of cones $\Delta_1\Rightarrow\cat{C}(c,D)$ on the diagram $\cat{C}(c,D):J\to\Set$ with apex the terminal set $1$.

\begin{cor} \label{cor:cone-presheaf}
  There is an isomorphism $\cat{C}^J(\Delta_c,D)\cong\Set^J(\Delta_1,\cat{C}(c,D))$, natural in $c:\cat{C}$.
\end{cor}

\begin{rmk} \label{rmk:contra-hom-lim-colim}
  Since limits in $\cat{C}\op$ are colimits in $\cat{C}$, Proposition \ref{prop:hom-preserves-limits} implies that the contravariant hom functors $\cat{C}(-,c)$ turn limits into colimits; \textit{i.e.} $\cat{C}(\lim D,c)\cong\colim\cat{C}(D(-),c)$.
\end{rmk}

\subsection{Closed categories and exponential objects} \label{sec:closed-cats}

A distinguishing feature of adaptive systems such as the brain is that they contain processes which themselves control other processes, and so it is useful to be able to formalize this situation compositionally.
When a category contains objects which themselves represent the morphisms of the category, we say that the category is \textit{closed}: in such categories, we may have processes whose outputs are again processes, and we may think of the latter as controlled by the former.

A basic instance of this mathematical situation is found amidst the natural numbers, where repeated multiplication coincides with exponentiation, as in $2\times 2\times 2 = 2^3$.
If we think of numbers as sets of the corresponding size, and let $2^3$ denote the set of functions $3\to 2$, then it is not hard to see that there are $8$ such distinct functions.
If we generalize this situation from numbers to arbitrary objects, and from functions to morphisms, we obtain the following general definition of exponentiation.

\begin{defn} \label{def:exponential}
  Let $\times$ denote the product in a category $\cat{C}$.
  When there is an object $e:\cat{C}$ such that $\cat{C}(x, e) \cong \cat{C}(x\times y, z)$ naturally in $x$, we say that $e$ is an \textit{exponential object} and denote it by $z^y$.
  The image of $\id_{z^y}$ under the isomorphism is called the \textit{evaluation map} and is written $\mathsf{ev}_{y,z} : z^y\times y \to z$.
\end{defn}

\begin{ex}
  In $\Set$, given sets $A$ and $B$, the exponential object $B^A$ is the set of functions $A\to B$.
  Given a function $f:A\to B$, the evaluation map $\mathsf{ev}_{B,A}$ acts by applying $f$ to elements of $A$: \textit{i.e.}, $\mathsf{ev}_{B,A}(f,a) = f(a)$.
\end{ex}

Typically, we are most interested in situations where every pair of objects is naturally exponentiable, which induces the following adjunction, formalizing the idea that exponentiation is repeated multiplication.

\begin{prop} \label{prop:prod-exp-adj}
  When the isomorphism $\cat{C}(x\times y, z) \cong \cat{C}(x, z^y)$ is additionally natural in $z$, we obtain an adjunction $(-)\times y \dashv (-)^y$ called the \textit{product-exponential adjunction}, and this uniquely determines a functor $\cat{C}\op\times\cat{C} \to \cat{C} : (y,z) \mapsto z^y$ that we call the \textit{internal hom} for $\cat{C}$.
  \begin{proof}
    That the natural isomorphism induces an adjunction is immediate from Proposition \ref{prop:adjoint-adjunct}; the counit of this adjunction is the family of evaluation maps $\mathsf{ev} : (-)^y\times y\Rightarrow\id_{\cat{C}}$.
    The uniqueness of the internal hom follows from the uniqueness of adjoint functors (which we will establish in Corollary \ref{cor:adjoints-unique}).
  \end{proof}
\end{prop}

\begin{defn} \label{def:cartes-clos}
  A category $\cat{C}$ in which every pair of objects has a product is called \textit{Cartesian}.
  A Cartesian category $\cat{C}$ with a corresponding internal hom is called \textit{Cartesian closed}.
\end{defn}

\begin{ex}
  We've already seen that $\Set$ is Cartesian closed.
  So is $\Cat{Cat}$: the internal hom $\cat{C}^{\cat{B}}$ is the category of functors $\cat{B}\to\cat{C}$.
\end{ex}

\begin{ex}[{A non-example}]
  The category $\Cat{Meas}$ of measurable spaces with measurable functions between them is Cartesian but not Cartesian closed: the evaluation function is not always measurable \parencite{Aumann1961Borel}.
  In this context, we will introduce \textit{quasi-Borel spaces} (originally due to \textcite{Heunen2017Convenient}) in \secref{sec:prob-monads}, in order to construct stochastic processes which emit functions.
\end{ex}

It is not hard to prove the following result, which says that Cartesian closed categories can ``reason about themselves''.

\begin{prop} \label{prop:ccc-self-enr}
  A Cartesian closed category is enriched in itself.
\end{prop}

This `internalization' is witnessed by the hom functors, which in the case of a Cartesian closed enriching category $\cat{E}$ become $\cat{E}$-functors.

\begin{prop} \label{prop:ccc-enr-hom}
  Suppose $\cat{C}$ is an $\cat{E}$-category where $\cat{E}$ is Cartesian closed.
  Then the hom functor $\cat{C}(-,=)$ is an $\cat{E}$-functor $\cat{C}\op\times\cat{C}\to\cat{E}$.
  On objects $(c,d)$, the hom functor returns the object $\cat{C}(c,d)$ in $\cat{E}$ of morphisms $c\to d$.
  Then, for each quadruple $(b,c,a,d)$ of objects in $\cat{C}$, we define an $\cat{E}$-morphism $\cat{C}\op(b,a)\times\cat{C}(c,d)\to\cat{E}\bigl(\cat{C}(b,c),\cat{C}(a,d)\bigr)$ as the image of the composite
  \begin{align*}
    & \bigl(\cat{C}(a,b)\times\cat{C}(c,d)\bigr)\times\cat{C}(b,c)
    \xto{\alpha} \cat{C}(a,b)\times\bigl(\cat{C}(c,d)\times\cat{C}(b,c)\bigr) \; \cdots \\
    & \cdots \; \xto{\cat{C}(a,b)\times{\circ_{b,c,d}}} \cat{C}(a,b)\times\cat{C}(b,d)
    \xto{\sigma}\cat{C}(b,d)\times\cat{C}(a,b)\xto{\circ_{a,b,d}}\cat{C}(a,d)
  \end{align*}
  under the product-exponential isomorphism
  \[ \cat{E}\bigl(\cat{C}(a,b)\times\cat{C}(c,d), \, \cat{C}(a,d)^{\cat{C}(b,c)}\bigr)
  \cong
  \cat{E}\Bigl(\bigl(\cat{C}(a,b)\times\cat{C}(c,d)\bigr)\times\cat{C}(b,c), \, \cat{C}(a,d)\Bigr) \]
  where $\alpha$ is the associativty of the product and $\sigma$ is its symmetry $X\times Y\cong Y\times X$, and where we have used that $\cat{C}\op(b,a) = \cat{C}(a,b)$.
  \begin{rmk}
    The rôle of the symmetry here is testament to the fact that we can read a composite morphism $g\circ f$ as either ``$g$ after $f$'' or ``$f$ before $g$''.
  \end{rmk}
  \begin{proof}[Proof sketch]
    To give an $\cat{E}$-functor (Definition \ref{def:enr-func}) is to give a function on objects and a family of $\cat{E}$-morphisms (corresponding to the hom objects of $\cat{C}$) such that identities and composites are preserved.
    We have given such a function and such a family in the statement of the proposition, and so it remains to check the axioms: these follow by the unitality and associativity of composition in an $\cat{E}$-category (Definition \ref{def:enriched-cat}).
  \end{proof}
  When $\cat{E}$ is Cartesian closed, then as a corollary its hom functor $\cat{E}(-,=)$ is an $\cat{E}$-functor.
\end{prop}

When a diagram commutes, every parallel path is equal when interpreted as a morphism.
If a diagram commutes up to some 2-cell or 2-cells, then parallel paths can be transformed into each other using the 2-cell(s).
Much categorical reasoning therefore consists in using morphisms in the base of enrichment to translate between different hom objects; the simplest such of course being pre- and post-composition.
In the next section, we will see many explicit examples of this kind of reasoning when we prove the Yoneda Lemma---which says that the hom objects contain all the data of the category---but we have already seen examples of it above, when we considered adjunctions: after all, adjunctions are families of isomorphisms between hom objects.

When a category is Cartesian closed, it is its own base of enrichment, and so one does not have to move to an external perspective to reason categorically about it: one can do so using its `internal language'.
We have already seen a correspondence between the language of logic and that of sets, in which we can think of elements of sets as witnesses to the proof of propositions represented by those sets, and where logical operations such as conjunction and disjunction correspond to operations on sets.
This correspondence extends to Cartesian closed categories generally: universal constructions such as those we have introduced above can be interpreted as encoding the logic of the internal language.

More precisely, Cartesian closed categories are said to provide the semantics for \textit{dependent type theory}: a higher-order logic in which propositions are generalized by `types'\footnote{%
A type is something like a proposition in which we're `allowed' to distinguish between its witnesses, which we call \textit{terms} of the given type.}.
One can construct a `syntactic' category representing the logic of the type theory, and then interpret it functorially in a Cartesian closed category.
This correspondence is known as the \textit{Curry-Howard-Lambek correspondence}, which says that logical proofs correspond to morphisms in a Cartesian closed category, and that such morphisms can equally be seen as representing the functions computed by deterministic computer programs.
(In general, the correspondence is an adjoint one: dually, one can construct from a given category a `syntactic' category encoding the logic of its internal language.)

When a category moreover has (internal) dependent sums and products, then it can be interpreted as a model of \textit{dependent type theory}, in which types themselves may depend on values;
for instance, one might expect that the type of a weather forecast should depend on whether one is on land or at sea.
We will not say much more about dependent type theory, although we will make implicit use of some of its ideas later in the thesis.
Therefore, before moving on to the Yoneda Lemma, we will say just enough to define the notion of dependent product `universally', without reference to sets.

\subsubsection{Dependent products} \label{sec:dep-prod}

In Remark \ref{rmk:dep-prod}, we discussed products where the factors were indexed by an arbitrary set and explained how they correspond to sets of generalized `dependent' functions, where the codomain type may vary with the input.
In that case, we were restricted to considering products indexed by sets, but with the machinery of limits at hand, we can `internalize' the definition to other Cartesian closed categories.

\begin{defn} \label{def:dep-prod}
  Suppose $\cat{C}$ is Cartesian closed and has all limits, and suppose $p:E\to B$ is a morphism in $\cat{C}$.
  The \textit{dependent product} of $p$ along $B$ is the pullback object $\prod_B p$ as in the diagram
  \[\begin{tikzcd}
    {\prod_B p} & {E^B} \\
    1 & {B^B}
    \arrow[from=1-1, to=1-2]
    \arrow[from=1-1, to=2-1]
    \arrow["{\id_B}"', from=2-1, to=2-2]
    \arrow["{p^B}", from=1-2, to=2-2]
    \arrow["\lrcorner"{anchor=center, pos=0.125}, draw=none, from=1-1, to=2-2]
  \end{tikzcd}\]
  where $1$ is the terminal object, $\id_B$ is the element picking the identity morphism $B\to B$, and $p^B$ is the postcomposition morphism induced by the functoriality of exponentiation.
\end{defn}

\begin{rmk}
  When $p$ is the projection $\sum_{b:B}P_b\to B$ out of a dependent sum, we will write its dependent product as $\prod_{b:B}P_b$.
  Since a product $B\times C$ is isomorphic to the dependent sum $\sum_{b:B} C$, note that this means we can alternatively write the exponential object $C^B$ as $\prod_{b:B} C$.
\end{rmk}

To understand how Definition \ref{def:dep-prod} generalizes Remark \ref{rmk:dep-prod}, we can interpret the former in $\Set$ and see that the two constructions coincide.
The set $E^B$ is the set of functions $s:B\to E$, and $p^B$ acts by $s\mapsto p\circ s$.
The indicated pullback therefore selects the subset of $E^B$ such that $p\circ s = \id_B$.
This is precisely the set of sections of $p$, which is in turn the dependent product of $p$ in $\Set$.

\begin{rmk}
  Definition \ref{def:dep-prod} is entirely \textit{internal} to $\cat{C}$: it depends only on structure that is available within $\cat{C}$ itself, and not on `external' structures (such as indexing sets) or knowledge (such as knowledge of the make-up of the objects of $\cat{C}$).
  It is epistemically parismonious: a purely categorical definition, stated entirely in terms of universal constructions.
\end{rmk}

\begin{rmk}
  Under the Curry-Howard-Lambek correspondence, exponential objects represent the propositions that one proposition implies another; in type theory, they represent the type of functions from one type to another.
  As dependent exponential objects, dependent products could therefore be seen as representing `dependent' implications; as we have already seen, they do represent the type of dependent functions.
  However, dependent products and sums have another kind of logical interpretation: as \textit{quantifiers}.
  That is, the logical proposition represented by $\prod_{b:B} P_b$ is $\forall{b:B}.P(b)$: an element of $\prod_{b:B} P_b$ is a proof that, for all $b:B$, the proposition $P(b)$ is satisfied.
  Dually, the proposition represented by $\sum_{b:B} P_b$ is $\exists{b:B}.P(b)$: an element of $\sum_{b:B} P_b$ is a pair $(b,x)$ of a witness to $B$ and a witness $x$ of the satisfaction of $P(b)$.
\end{rmk}

\section{The Yoneda Lemma: a human perspective} \label{sec:yoneda}

We end this chapter by introducing the fundamental theorem of category theory, the Yoneda Lemma,
which expresses mathematically the idea that to know how a thing is related to other things is to know the identity of the thing itself.
The notion of relational identity is recognized throughout human endeavour.
In linguistics, it underlies the observation of Firth \parencite{Firth1957Synopsis} that ``you shall know a word by the company it keeps!'', which in turn is the foundation of distributional semantics and thus much of contemporary natural language processing in machine learning.
In culture, it is illustrated by the ancient parable of the blind men and the elephant, in which the identity of the creature is only known by stitching together evidence from many perspectives.
In society, it is reflected in the South African philosophy of \textit{ubuntu} (meaning ``I am because we are'') and the Māori notion of \textit{whanaungatanga} (in which personal identity is developed through kinship), and the observation that ``actions speak louder than words''.
Finally, the Yoneda Lemma is manifest in science, where our understanding of phenomena derives from the accumulation across contexts of results and their interpretation and translation: no single individual understands the totality of any subject, and no subject or phenomenon is understood in isolation.

\subsection{Formalizing categorical reasoning via the Yoneda embedding}

In \secref{sec:closed-cats}, we saw how Cartesian closed categories allow us to internalize categorical reasoning.
The category $\Set$ is the archetypal Cartesian closed category, and constitutes the base of enrichment for all locally small categories.
The Yoneda embedding allows us to move from reasoning about the objects in any given category $\cat{C}$ to reasoning about the morphisms between its hom sets: the natural transformations between hom functors.
In this context, the hom functors constitute special examples of functors into the base of enrichment, which we call `presheaves' (contravariantly) and `copresheaves' (covariantly), and which can be thought of as $\cat{C}$-shaped diagrams in $\Set$.

\begin{defn}
  Let $\cat{C}$ be a category.
  A \textit{presheaf} on $\cat{C}$ is a functor $\cat{C}\op\to\Set$.
  Dually, a \textit{copresheaf} is a functor $\cat{C}\to\Set$.
  The corresponding functor categories are the categories of \textit{(co)presheaves} on $\cat{C}$.
\end{defn}

\begin{rmk}
  In the enriched setting, when $\cat{C}$ is enriched in $\cat{E}$, an $\cat{E}$\textit{-presheaf} is an $\cat{E}$-functor $\cat{C}\op\to\cat{E}$ and an $\cat{E}$\textit{-copresheaf} is an $\cat{E}$-functor $\cat{C}\to\cat{E}$.
\end{rmk}

As a first example of a presheaf, we have an alternative definition of the notion of directed graph.

\begin{ex} \label{ex:graph-psh}
  Let $\mathbb{G}$ denote the category of Example \ref{ex:graph-schema} containing two objects $0$ and $1$ and two morphisms $s,t:0\to 1$.
  Then a \textit{directed graph} is a presheaf on $\mathbb{G}$.
\end{ex}

This definition is justified by the following proposition.

\begin{prop} \label{prop:grph-psh-cat}
  There is an equivalence of categories $\Cat{Graph} \cong \Set^{\mathbb{G}\op}$, where $\Cat{Graph}$ is the category of directed graphs introduced in Example \ref{ex:graph-cat}.
  \begin{proof}
    To each graph $\cat{G}$ we can associate a presheaf $G:\mathbb{G}\op\to\Set$ by defining $G(0) := \cat{G}_0$, $G(1) := \cat{G}_1$, $G(s) := \dom_{\cat{G}}$ and $G(t) := \cod_{\cat{G}}$; and to each presheaf we can likewise associate a graph, so that we have defined a bijection on objects.
    It therefore only remains to show that there is a bijection between graph homomorphisms and natural transformations accordingly: but this is easy to see once we have observed that the graph homomorphism axioms are precisely the law of naturality, as illustrated diagrammatically in \eqref{eq:graph-hom-nat}.
  \end{proof}
\end{prop}

Taking again a general perspective, the Yoneda embedding is the embedding of a category $\cat{C}$ into its presheaf category, obtained by mapping $c:\cat{C}$ to the presheaf $\cat{C}(-,c)$; and there is of course a dual `coYoneda' embedding.

\begin{rmk} \label{rmk:embedding}
  We say `embedding' to mean a functor that is injective on objects and faithful (injective on hom sets).
  The Yoneda embedding will turn out to be fully faithful (bijective on hom sets), as a consequence of the Yoneda lemma.
\end{rmk}

Owing to its importance, we make a formal definition of the Yoneda embedding.

\begin{defn}
  Let $\cat{C}$ be a category.
  By applying the product-exponential adjunction in $\Cat{Cat}$ to the hom functor $\cat{C}(-,=):\cat{C}\op\times\cat{C}\to\Set$, we obtain a functor $\yo : \cat{C}\to\Set^{\cat{C}\op} : c \mapsto \cat{C}(-,c)$ of $\cat{C}$ into its presheaf category, and dually a functor $\coyo : \cat{C}\op\to\Set^{\cat{C}} : c \mapsto \cat{C}(c,=)$ into the copresheaf category.
  We call the former functor the \textit{Yoneda embedding} and the latter the \textit{coYoneda embedding}.
  When $\cat{C}$ is an $\cat{E}$-category and $\cat{E}$ is Cartesian closed, then the Yoneda embedding is instead an $\cat{E}$-functor $\cat{C}\to\cat{E}^{\cat{C}\op}$ (and likewise for the coYoneda embedding).
\end{defn}

\begin{rmk}
  This abstract definition does not make explicit how $\yo\;$ acts on morphisms.
  However, we have already seen this action, when we first exemplified natural transformations in Example \ref{ex:hom-nat-trans}.
\end{rmk}

As we discussed in \secref{sec:closed-cats}, much categorical reasoning corresponds to following morphisms between hom objects, and often the reasoning is agnostic either to where one starts, or to where one ends up.
The Yoneda embedding witnesses such proofs as morphisms in the (co)presheaf categories.
As an example, consider the proof of Proposition \ref{prop:rapl} below: each step corresponds to the application of a natural transformation.

\begin{rmk}
  It also so happens that every (co)presheaf category is very richly structured, inheriting its structure from the base of enrichment.
  For example, this means that the presheaf category $\Set^{\cat{C}\op}$ has all limits, is Cartesian closed, has a subobject classifier, and dependent sums and products, even when $\cat{C}$ has none of these.
  (Interestingly, this means that the category of directed graphs is accordingly richly structured, being a presheaf category by Proposition \ref{prop:grph-psh-cat}.)
  As a result, (co)presheaf categories are very powerful places to do categorical reasoning.
\end{rmk}

\subsection{Knowing a thing by its relationships}

The Yoneda lemma says that every (co)presheaf on $\cat{C}$ is determined by ``how it looks from $\cat{C}$''.
Since under the (co)Yoneda embedding every object gives rise to a (co)presheaf, a corollary of the Yoneda lemma is that every object can be identified by its relationships.

\begin{rmk}
  If the base of enrichment of a category is Cartesian closed, then one can prove an analogous enriched version of the Yoneda lemma.
  We will only prove the standard $\Set$-enriched case here.

  We will also only prove the Yoneda lemma for presheaves; there is of course a dual coYoneda lemma for copresheaves, which follows simply by swapping $\cat{C}$ for $\cat{C}\op$.
\end{rmk}

\begin{thm}[Yoneda lemma]
  Let $F:\cat{C}\op\to\Set$ be a presheaf on $\cat{C}$.
  Then for each $c:\cat{C}$, there is an isomorphism $Fc \cong \Set^{\cat{C}\op}(\cat{C}(-,c),F)$.
  Moreover, this isomorphism is natural in both $F:\cat{C}\op\to\Set$ and $c:\cat{C}$.
  \begin{proof}
    We first define a mapping $\gamma : Fc\to\Set^{\cat{C}\op}(\cat{C}(-,c),F)$ as follows.
    Given $h:Fc$, we define the natural transformation $\gamma(h) : \cat{C}(-,c)\Rightarrow F$ to have components $\gamma(h)_b : \cat{C}(b,c)\to Fb : f \mapsto Ff(h)$; note that since $h:Fc$ and $f:b\to c$, we have $Ff : Fc\to Fb$ and hence $Ff(h) : Fb$.
    To check that this definition makes $\gamma(h)$ into a natural transformation, suppose $g:a\to b$.
    We need to check $Fg\circ\gamma(h)_b = \gamma(h)_a\circ\cat{C}(g,c)$.
    Since $\cat{C}(g,c)(f) = f\circ g$, this means verifying $Fg\circ Ff(h) = F(f\circ g)(h)$.
    But $F$ is a contravariant functor, so $F(f\circ g) = Fg\circ Ff$, thereby establishing naturality.

    Conversely, we define a mapping $\gamma' : \Set^{\cat{C}\op}(\cat{C}(-,c),F)\to Fc$ as follows.
    Suppose $\alpha$ is a natural transformation $\cat{C}(-,c)\Rightarrow F$, so that its component at $c$ is the function $\alpha_c : \cat{C}(c,c)\to Fc$.
    We define $\gamma'(\alpha) := \alpha_c(\id_c)$.

    Next, we need to establish that $\gamma$ and $\gamma'$ are mutually inverse.
    First, we check that $\gamma'\circ\gamma = \id_{Fc}$.
    Given $h:Fc$, we have
    \[ \gamma'(\gamma(h)) = \gamma(h)_c(\id_c) = F(\id_c)(h) = \id_{Fc}(h) = h \]
    as required.
    We now check that $\gamma\circ\gamma' = \id_{\Set^{\cat{C}\op}(\cat{C}(-,c),F)}$.
    Given $\alpha : \cat{C}(-,c)\Rightarrow F$, we have $\gamma'(\alpha) = \alpha_c(\id_c)$ by definition.
    Hence $\gamma(\gamma'(\alpha)) : \cat{C}(-,c)\Rightarrow F$ has components $\gamma(\gamma'(\alpha))_b : \cat{C}(b,c)\to Fb$ which act by $f \mapsto Ff(\alpha_c(\id_c))$.
    So we need to show that $Ff(\alpha_c(\id_c)) = \alpha_b(f)$.
    This follows directly from the naturality of $\alpha$.
    The commutativity of the naturality square on the left in particular holds at $\id_c:\cat{C}(c,c)$ as on the right:
    \[\begin{tikzcd}
	    {\cat{C}(c,c)} && Fc \\
	    \\
	      {\cat{C}(b,c)} && Fb
	      \arrow["{\alpha_c}", from=1-1, to=1-3]
	      \arrow["Ff", from=1-3, to=3-3]
	      \arrow["{\cat{C}(f,c)}"', from=1-1, to=3-1]
	      \arrow["{\alpha_b}"', from=3-1, to=3-3]
    \end{tikzcd}
    \qquad\qquad
    \begin{tikzcd}
	    {\id_c} && {\alpha_c(\id_c)} \\
	    \\
	    f && {\alpha_b(f) = Ff(\alpha_c(\id_c))}
	    \arrow[maps to, from=1-1, to=3-1]
	    \arrow[maps to, from=1-1, to=1-3]
	    \arrow[maps to, from=1-3, to=3-3]
	    \arrow[maps to, from=3-1, to=3-3]
    \end{tikzcd}\]
    Note that $\cat{C}(f,c)(\id_c) = \id_c\circ f = f$.
    This establishes that $\gamma\circ\gamma' = \id_{\Set^{\cat{C}\op}(\cat{C}(-,c),F)}$, and since $\gamma'\circ\gamma = \id_{Fc}$, we have $Fc \cong \Set^{\cat{C}\op}(\cat{C}(-,c),F)$.

    It remains to verify that this isomorphism is natural in $F$ and $c$.
    Suppose $\varphi:F\Rightarrow F'$ is a natural transformation, and write $\gamma'_{Fc}$ for the function $\gamma'$ defined above, and $\gamma'_{F'c}$ for the corresponding function for $F'$.
    Naturality in $F$ means that the diagram on the left below commutes, which we can see by chasing the natural transformation $\alpha$ as on the right:
    \[\begin{tikzcd}
	    {\Set^{\cat{C}\op}(\cat{C}(-,c),F)} && Fc \\
      \\
      {\Set^{\cat{C}\op}(\cat{C}(-,c),F')} && {F'c}
	    \arrow["{\Set^{\cat{C}\op}(\cat{C}(-,c),\varphi)}", from=1-1, to=3-1]
	    \arrow["{\gamma'_{Fc}}", from=1-1, to=1-3]
	    \arrow["{\gamma'_{F'c}}", from=3-1, to=3-3]
	    \arrow["{\varphi_c}", from=1-3, to=3-3]
    \end{tikzcd}
    \qquad\quad
    \begin{tikzcd}
	    \alpha && {\gamma'_{Fc}(\alpha)} \\
	    \\
	    \varphi\circ\alpha && {\gamma'_{F'c}(\varphi\circ\alpha)=\varphi_c\circ\gamma'_{Fc}(\alpha)}
	    \arrow[maps to, from=1-1, to=1-3]
	    \arrow[maps to, from=1-3, to=3-3]
	    \arrow[maps to, from=1-1, to=3-1]
	    \arrow[maps to, from=3-1, to=3-3]
    \end{tikzcd}\]
    Since $\gamma'_{Fc}(\alpha)_c:=\alpha_c(\id_c)$ and $\gamma'_{F'c}(\varphi\circ\alpha)_c:=\varphi_c\circ\alpha_c(\id_c)$, the equation $\gamma'_{F'c}(\varphi\circ\alpha)=\varphi_c\circ\gamma'_{Fc}(\alpha)$ holds by definition, thereby establishing naturality in $F$.
    Finally, suppose $f:b\to c$ in $\cat{C}$, and write $\gamma_{Fc}$ for the function $\gamma$ defined above and $\gamma_{Fb}$ for the corresponding function for $b:\cat{C}$.
    Naturality in $c$ means the commutativity of the following diagram:
    \[\begin{tikzcd}
	    Fc && {\Set^{\cat{C}\op}(\cat{C}(-,c),F)} \\
	    \\
	    Fb && {\Set^{\cat{C}\op}(\cat{C}(-,b),F)}
	    \arrow["{\Set^{\cat{C}\op}(\cat{C}(-,f),F)}", from=1-3, to=3-3]
	    \arrow["{\gamma_{Fc}}", from=1-1, to=1-3]
	    \arrow["{\gamma_{Fb}}", from=3-1, to=3-3]
	    \arrow["Ff", from=1-1, to=3-1]
    \end{tikzcd}\]
    Suppose $h:Fc$.
    The component of $\gamma_{Fc}(h)$ at $a:\cat{C}$ is the function $\gamma_{Fc}(h)_a : \cat{C}(a,c)\to Fa$ defined by $g \mapsto Fg(h)$.
    The component of $\Set^{\cat{C}\op}(\cat{C}(-,f),F)\circ\gamma_{Fc}(h)$ at $a:\cat{C}$ is thus the function $\gamma_c(h)_a\circ\cat{C}(a,f) : \cat{C}(a,b)\to Fa$ taking $g:a\to b$ to $F(f\circ g)(h)$.
    On the other hand, the component of $\gamma_{Fb}(Ff(h))$ at $a:\cat{C}$ is the function $\gamma_{Fb}(Ff(h))_a : \cat{C}(a,b)\to Fa$ taking $g$ to $Fg(Ff(h))$.
    Since $F$ is a contravariant functor, we have $F(f\circ g)(h) = Fg(Ff(h))$.
    This establishes the commutativity of the naturality square, and thus naturality in $c$ as well as $F$.
  \end{proof}
\end{thm}

The identification of an object with its collection of hom sets is formalized by the following corollary.

\begin{cor}[Representables are unique up to isomorphism] \label{cor:uniq-rep}
  Suppose there is an isomorphism of presheaves $\cat{C}(-,a)\cong\cat{C}(-,b)$.
  Then $a\cong b$ in $\cat{C}$.
\end{cor}

This corollary follows from the next one, which expresses that the image of the Yoneda embedding is isomorphic with $\cat{C}$ itself.

\begin{cor} \label{cor:yo-ff}
  The Yoneda embedding is fully faithful.
  \begin{proof}
    The Yoneda embedding defines a family of functions on the hom sets of $\cat{C}$:
    \begin{align*}
      \yo_{b,c} : \cat{C}(b,c)&\to\Set^{\cat{C}\op}(\cat{C}(-,b),\cat{C}(-,c)) \\
      f &\mapsto \cat{C}(-,f)
    \end{align*}
    By the Yoneda lemma, we immediately have $\Set^{\cat{C}\op}(\cat{C}(-,b),\cat{C}(-,c)) \cong \cat{C}(b,c)$, which is the required isomorphism of hom sets.
  \end{proof}
\end{cor}

Next, we have the following fact, that fully faithful functors transport isomorphisms in their codomain to their domain (they `reflect' them).

\begin{prop}[Fully faithful functors reflect isomorphisms] \label{prop:ff-refl-iso}
  Suppose $F : \cat{C}\to\cat{D}$ is a fully faithful functor.
  If $f : a\to b$ is a morphism in $\cat{C}$ such that $Ff$ is an isomorphism in $\cat{D}$, then $f$ is an isomorphism in $\cat{C}$.
  \begin{proof}
    $Ff:Fa\to Fb$ being an isomorphism means that there is a morphism $g':Fb\to Fa$ in $\cat{D}$ such that $g'\circ Ff = \id_{Fa}$ and $Ff\circ g' = \id_{Fb}$.
    By the functoriality of $F$, we have $\id_{Fa} = F\id_a$ and $\id_{Fb} = F\id_b$.
    Hence $g'\circ Ff = F\id_a$ and $Ff\circ g' = F\id_b$.
    Since $F$ is isomorphic on hom sets, there is a unique $g:b\to a$ such that $g' = Fg$.
    Hence $Fg\circ Ff = F\id_a$ and $Ff\circ Fg = F\id_b$.
    By the functoriality of $F$, we have $Fg\circ Ff = F(g\circ f)$ and $Ff\circ Fg = F(f\circ g)$.
    Hence $F(g\circ f) = F\id_a$ and $F(f\circ g) = F\id_b$.
    Finally, since $F$ is isomorphic on hom sets, we must have $g\circ f = \id_a$ and $f\circ g = \id_b$, and hence $f$ is an isomorphism in $\cat{C}$.
  \end{proof}
\end{prop}

And this gives us the proof we seek:

\begin{proof}[Proof of Corollary \ref{cor:uniq-rep}]
  Since the Yoneda embedding is fully faithful (Corollary \ref{cor:yo-ff}), it reflects isomorphisms by Proposition \ref{prop:ff-refl-iso}.
\end{proof}

Presheaves in the image of the Yoneda embedding consequently play a special rôle in category theory: to show that an arbitrary presheaf $F$ is isomorphic to $\cat{C}(-,c)$ is to identify it with the object $c$ itself, and in this case, we can say that $F$ is \textit{represented} by $c$.
We therefore make the following definition.

\begin{defn} \label{def:repr}
  Suppose $F$ is a presheaf on $\cat{C}$.
  We say that it is \textit{representable} if there is a natural isomorphism $F \cong \cat{C}(-,c)$ for some object $c:\cat{C}$ which we call its \textit{representing object}; we call the natural isomorphism $\cat{C}(-,c)\Rightarrow F$ its \textit{representation}.
  Dually, if $F$ is instead a copresheaf, we call it \textit{corepresentable} if there is a natural isomorphism $F \cong \cat{C}(c,=)$, with $c$ being the \textit{corepresenting object}; we call the natural isomorphism $\cat{C}(c,=)\Rightarrow F$ its \textit{corepresentation}.
\end{defn}

\begin{rmk}
  Corepresentable copresheaves will play an important rôle later in this thesis: their coproducts are called \textit{polynomial functors} (\secref{sec:poly}), and these will be used to formalize the interfaces of interacting adaptive systems.
\end{rmk}

Via the uniqueness of representables, the Yoneda lemma underlies universal constructions, since knowing the morphisms into or out of an object is enough to identify that object.
The definition of a limit, notably, is the statement that morphisms into it correspond to morphisms into a diagram; and this in turn is equivalently the statement that $\lim$ is right adjoint to $\Delta$.
Indeed, adjointness is itself a certain kind of representability: the definition of adjoint functor (\ref{def:adjoint-func}) is precisely a natural characterization of morphisms into and out of objects, as related by the adjunction!

\begin{prop}[Adjoints are representable] \label{prop:adjoint-rep}
  Suppose $R:\cat{D}\to\cat{C}$ is right adjoint to $L$.
  Then for every $d:\cat{D}$, the presheaf $\cat{D}(L-,d):\cat{C}\op\to\Set$ is represented by the object $Rd:\cat{C}$.
  Dually, the copresheaf $\cat{C}(c,R-):\cat{D}\to\Set$ is corepresented by the object $Lc:\cat{D}$.
  \begin{proof}
    Since $L\vdash R$, we have an isomorphism $\cat{D}(Lc,d)\cong\cat{C}(c,Rd)$ natural in $c$ and $d$.
    Therefore in particular we have a natural isomorphism of presheaves $\cat{C}(-,Rd)\Rightarrow\cat{D}(L-,d)$ and a natural isomorphism of copresheaves $\cat{D}(Lc,-)\Rightarrow\cat{C}(c,R-)$; the former is a representation and the latter a corepresentation.
  \end{proof}
\end{prop}

From this, we can formalize the representability of limits and colimits.

\begin{cor}[Limits are representations] \label{cor:limit-rep}
  Suppose $D:J\to\cat{C}$ is a diagram in $\cat{C}$.
  A limit of $D$ is a representation of $\cat{C}^J(\Delta_{(-)},D):\cat{C}\op\to\Set$, or equivalently of $\Set^J(\Delta_1,\cat{C}(-,D))$.
  Dually, a colimit of $D$ is a corepresentation of $\cat{C}^J(D,\Delta_{(-)}):\cat{C}\to\Set$, or equivalently of $\Set^J(\Delta_1,\cat{C}(D,-))$.
  \begin{proof}
    If $\cat{C}$ has all limits of shape $J$, then this follows directly from the facts that $\lim$ is right adjoint to $\Delta$ (Proposition \ref{prop:limit-adjoint}) and that adjoints are representable (Proposition \ref{prop:adjoint-rep}); the dual result follows similarly from the fact that $\colim$ is left adjoint to $\Delta$.

    Otherwise, the limit case follows immediately from Lemma \ref{lemma:nat-into-lim} (or equivalently Corollary \ref{cor:cone-presheaf}) and the definition of representation (\ref{def:repr}); the colimit case is formally dual.
  \end{proof}
\end{cor}

Accordingly, we recover the uniqueness of universal constructions.

\begin{cor} \label{cor:adjoints-unique}
  Adjoint functors are unique up to unique isomorphism.
\end{cor}

\begin{cor}
  Limits and colimits are unique up to unique isomorphism.
\end{cor}

Using these ideas, we obtain the following useful result relating limits and adjoint functors.

\begin{prop}[Right adjoints preserve limits] \label{prop:rapl}
  Suppose $D:J\to\cat{D}$ is a diagram in $\cat{D}$ and $L\vdash R:\cat{D}\to\cat{C}$ is an adjunction.
  Then $R\lim D \cong \lim RD$ in $\cat{C}$.
  \begin{proof}
    We have the following chain of natural isomorphisms:
    \begin{align*}
      \cat{C}(c,R\lim D)
      &\cong \cat{D}(Lc, \lim D) &\text{since }R\text{ is right adjoint to }L \\
      &\cong \lim\cat{D}(Lc, D)  &\text{since hom preserves limits}\\
      &\cong \lim\cat{C}(c, RD)  &\text{since }R\text{ is right adjoint to }L \\
      &\cong \cat{C}(c,\lim RD)  &\text{since hom preserves limits}
    \end{align*}
    Since representables are unique up to isomorphism and we have established an isomorphism of presheaves $\cat{C}(-,R\lim D)\cong\cat{C}(-,\lim RD)$, we must have $R\lim D \cong \lim RD$ in $\cat{C}$.
  \end{proof}
\end{prop}

\begin{rmk} \label{rmk:lapc}
  There is of course a dual result that left adjoints preserve colimits.
\end{rmk}

\begin{rmk}
  One might speculate about the converse: is it the case that the preservation of limits by a functor is enough to guarantee the existence of its left adjoint?
  The answer to this question is, ``under certain conditions'' on the size and structure of the categories and functors involved, and a positive answer is called an \textit{adjoint functor theorem}.
  The ``certain conditions'' hold quite generally, and so it is often sufficient just to check whether a functor preserves limits (or colimits) to see that it is a right (or left) adjoint.
\end{rmk}

We end this chapter by closing the loop between universality and representability.

\begin{prop}[Universality of representability] \label{prop:univers-repr}
  Representable presheaves $F:\cat{C}\op\to\Set$ correspond bijectively to universal morphisms from $1:\Set$ to $F$.
  \begin{proof}
    A representation of $F$ is a choice of object $c:\cat{C}$ and a natural isomorphism $\upsilon:\cat{C}(-,c)\Rightarrow F$.
    We construct a bijection between the set of representations of $F$ and the set of universal morphisms from $1$ to $F$.
    Therefore suppose given a representation $\upsilon:\cat{C}(-,c)\Rightarrow F$ of $F$; its component at $c:\cat{C}$ is the isomorphism $\upsilon_c:\cat{C}(c,c)\to Fc$.
    The Yoneda lemma assigns to $\upsilon$ an element $\gamma'(\upsilon) : 1\to Fc$ satisfying $\gamma'(\upsilon) = \upsilon_c(\id_c)$.
    We now show that this element $\upsilon_c(\id_c)$ satisfies the universal property that for all $f:1\to Fb$ there exists a unique morphism $h:b\to c$ in $\cat{C}$ such that $f = Fh\circ \upsilon_c(\id_c)$.
    Therefore let $f$ be any such element $1\to Fb$.
    Since $\upsilon$ is a natural isomorphism, it has an inverse component at $b:\cat{C}$, denoted $\upsilon'_b : Fb\to\cat{C}(b,c)$, and so we obtain by composition an element $h := 1\xto{f}Fb\xto{\upsilon'_b}\cat{C}(b,c)$ of $\cat{C}(b,c)$.
    Such an element is precisely a morphism $h:b\to c$ in $\cat{C}$.
    Consider now the following diagram:
    \[\begin{tikzcd}
      1 && {\cat{C}(c,c)} && {\cat{C}(b,c)} \\
	    \\
	    && Fc && Fb
	    \arrow["{\id_c}", from=1-1, to=1-3]
	    \arrow["{\cat{C}(h,c)}", from=1-3, to=1-5]
	    \arrow["{\upsilon_b}", from=1-5, to=3-5]
	    \arrow["{\upsilon_c}", from=1-3, to=3-3]
	    \arrow["Fh", from=3-3, to=3-5]
	    \arrow["{\upsilon_c(\id_c)}"', from=1-1, to=3-3]
    \end{tikzcd}\]
    The triangle on the left commutes by definition and the square on the right commutes by the naturality of $\upsilon$, so that the whole diagram commutes.
    The composite morphism $\cat{C}(h,c)\circ\id_c$ along the top of the diagram picks out the element $\id_c\circ h$ of $\cat{C}(b,c)$.
    By the unitality of composition, this element is equal to $h$ itself, so we can rewrite the diagram as follows:
    \[\begin{tikzcd}
	    1 &&&& {\cat{C}(b,c)} \\
	    \\
	    Fc &&&& Fb
	    \arrow["{\upsilon_b}", from=1-5, to=3-5]
	    \arrow["Fh", from=3-1, to=3-5]
	    \arrow["{\upsilon_c(\id_c)}"', from=1-1, to=3-1]
	    \arrow["h", from=1-1, to=1-5]
    \end{tikzcd}\]
    Next, we can substitute the definition $h := \upsilon'_b\circ f$, and observe that $\upsilon_b\circ \upsilon'_b = \id_{Fb}$ (since $\upsilon_b$ is an isomorphism with $\upsilon'_b$ its inverse):
    \[\begin{tikzcd}
	    1 && Fb && {\cat{C}(b,c)} \\
	    \\
	    Fc &&&& Fb
	    \arrow["{\upsilon_b}", from=1-5, to=3-5]
	    \arrow["Fh", from=3-1, to=3-5]
	    \arrow["{\upsilon_c(\id_c)}"', from=1-1, to=3-1]
	    \arrow["f", from=1-1, to=1-3]
	    \arrow["{\upsilon'_b}", from=1-3, to=1-5]
	    \arrow[Rightarrow, no head, from=1-3, to=3-5]
    \end{tikzcd}\]
    The commutativity of this diagram means that $f = Fh\circ \upsilon_c(\id_c)$.
    Moreover, since $h = \upsilon'_b\circ f$ and $\upsilon'_b$ is an isomorphism, $h$ is unique for a given $f$.
    Therefore $\upsilon_c(\id_c) : 1\to Fc$ is a universal morphism from $1$ to $F$.

    Next, suppose given a universal morphism $u:1\to Fc$.
    The Yoneda lemmea associates to this element a natural transformation $\gamma(u)$ whose component at $b$ is the function $\gamma(u)_b : \cat{C}(b,c)\to Fb$ which acts by $f\mapsto Ff(u)$.
    We need to show that this function is an isomorphism for every $b:\cat{C}$, so that $\gamma(u) : \cat{C}(-,c)\Rightarrow F$ is a natural isomorphism and hence $F$ is represented by $c$.
    We therefore need to define an inverse function $\varphi_b : Fb\to\cat{C}(b,c)$, which we do using the universal property of $u$: for each element $f:1\to Fb$, we have a unique morphism $h:b\to c$ such that $f = Fh(u)$.
    This unique $h$ is an element of $\cat{C}(b,c)$, and so we can simply define $\varphi_b(f) := h$.
    The uniqueness of $h$ ensures that $\varphi_b$ is an inverse of $\gamma(u)_b$: observe that $\gamma(u)_b\circ\varphi_b$ acts by $f \mapsto h \mapsto Fh(u)$ and $f = Fh(u)$ by definition; in the opposite direction, we necessarily have $f\mapsto Ff(u)\mapsto f$.

    We have constructed mappings between the set of representations of $F$ and universal morphisms from $1$ to $F$, so it remains to show that these mappings are mutually inverse.
    This again follows directly from the Yoneda lemma: the mapping of representations to universal morphisms takes a representation $\upsilon$ to the element $\gamma'(\upsilon)$ induced by the Yoneda lemma; and the mapping of universal morphisms to representations takes a universal morphism $u$ to the natural transformation $\gamma(u)$ induced by the Yoneda lemma.
    Since the functions $\gamma$ and $\gamma'$ are mutually inverse, so must these mappings be: $\gamma\circ\gamma'(\upsilon) = \upsilon$ and $\gamma'\circ\gamma(u) = u$.
  \end{proof}
\end{prop}

Using the universality of representability, and the uniqueness of universal morphisms, and the representability of limits and adjoints, we therefore obtain alternative proofs of the uniqueness of those universal constructions.

%
%

\chapter{Algebraic connectomics} \label{chp:algebra}

In Chapter \ref{chp:basic-ct}, we motivated applied category theory in the context of complex systems like brains by its abilities to relate structure and function, to translate between models and frameworks, and to distil phenomena to their essences.
However, the focus in that chapter was on `one-dimensional' morphisms, which can be understood as connecting one interface to another, with the composition of 1-cells representing something like the `end-to-end' composition of processes; although we considered some higher-dimensional category theory, this was largely restricted to weakening equalities and thus comparing morphisms.

Because systems can be placed `side-by-side' as well as end-to-end, and because two systems placed side by side may be nontrivially wired together, in this chapter we extend the higher-dimensional categorical language accordingly, with a particular focus once more on the graphical and diagrammatic representation of systems and processes.
In line with the distinction made in \secref{sec:syn-sem} between syntax and semantics, our treatment here of the syntax of wiring---of \textit{connectomics}---is largely `algebraic'.
Later, in Chapter \ref{chp:coalg}, we will see how our semantic focus will be `coalgebraic'.

We will begin therefore by introducing the graphical calculus of monoidal categories, which allow us to depict and reason about sequential and parallel composition simultaneously.
We follow this with the formal underpinnings of the structure---to use the term from Chapter \ref{chp:basic-ct}, a monoidal structure is a `well-behaved' tensor product---before explaining how monoidal categories relate to the higher category theory of Chapter \ref{chp:basic-ct} using the notion of bicategory.
We then make use of the extra freedom afforded by bicategories to consider parameterized systems, with which we can model systems that not only act but also learn.

By this point, we will find ourselves ready to apply our new toolkit, and so in \secref{sec:sys-circ}, we use functorial semantics to define a graphical algebra for neural circuits, revisiting our first example from Chapter \ref{chp:basic-ct}.
This involves a change of perspective from the graphical calculus with which we begin the chapter: instead of using the composition of morphisms to encode the plugging-together of systems at the same `scale' or ``level of hierarchy'', we use composition to encode the wiring of circuits at one level into systems at a higher level.
Although formally closely related to monoidal categories, this `hierarchical' perspective is strictly speaking \textit{multicategorical} and allows morphisms' domains to take very general shapes.

After this extended example, we return to algebra, explaining what makes monoidal categories \textit{monoidal}, and using the related concept of \textit{monad} to explain how we think of them as algebraic; monads will later prove to be of importance in categorical approaches to probability theory.
Finally, we end the chapter by introducing the richly structured category of polynomial functors $\Set\to\Set$, which we will use in Chapter \ref{chp:coalg} both to formalize a wide variety of open dynamical systems as well as to specify the shapes of those systems' interfaces.

Excepting the extended example of \secref{sec:sys-circ}, the content of this chapter is well known to category-theoreticians.
However, since it is not well known to mathematical scientists, we have again endeavoured to supply detailed motivations for the concepts and results that we introduce.

\section{Categories and calculi for process theories} \label{sec:cd-cats}

In this section, we introduce an alternative way of depicting morphisms and their composites in categories equipped with notions of both sequential and parallel composition.
Such categories are useful for representing processes in which information flows: we formalize the processes as morphisms, and consider the flow as from domain to codomain, even when the categories themselves are quite abstract and lack a notion of time with which to make sense of `flow'.
In such contexts, the categories are often not only \textit{monoidal}, but also \textit{copy-discard} categories, since a distinctive feature of classical information is that it can be copied and deleted.
Monoidal categories will therefore be important not only in depicting composite computations (as indicated in \secref{sec:mtv-func}), but also in depicting and manipulating the factorization of probabilistic models (as indicated in \secref{sec:mtv-bayes}).

\subsection{String diagrams} \label{sec:str-diag}

Rather than beginning with the formal definition of ``monoidal category'', we start with the associated graphical calculus of \textit{string diagrams} and its intuition.

\paragraph{Sequential and parallel composition}

Diagrams in the graphical calculus depict morphisms as boxes on strings:
the strings are labelled with objects, and a string without a box on it can be interpreted as an identity morphism.
Sequential composition is represented by connecting strings together, and parallel composition by placing diagrams adjacent to one another; sequential composition distributes over parallel, and so we can of course compose parallel boxes in sequence.

Because monoidal structures are ``well-behaved tensor products'', we will typically denote them using the same symbols that we adopted in Chapter \ref{chp:basic-ct}, with sequential composition denoted by $\circ$ and parallel composition (tensor) denoted by $\otimes$.
Diagrams will be read in the direction of information flow, which will be either bottom-to-top or left-to-right; we will adopt the former convention in this section.

In this way, \(c : X \to Y\), \(\id_X : X \to X\), \(d \circ c : X \xto{c} Y \xto{d} Z\), and \(f \otimes g : X \otimes Y \to A \otimes B\) are depicted respectively as:
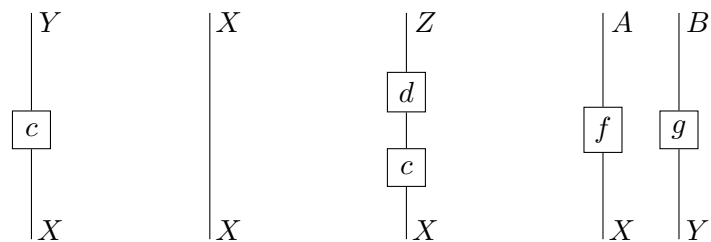
\[
\hspace{0.125\linewidth} \input{img/channel-c.tikz}
\hspace{0.125\linewidth} \begin{tikzpicture}
	\begin{pgfonlayer}{nodelayer}
		\node [style=none] (0) at (0, -3) {};
		\node [style=none] (2) at (0, 3) {};
		\node [style=none] (3) at (0.5, -2.75) {$X$};
		\node [style=none] (4) at (0.5, 2.75) {$X$};
	\end{pgfonlayer}
	\begin{pgfonlayer}{edgelayer}
		\draw (0.center) to (2.center);
	\end{pgfonlayer}
\end{tikzpicture}

\hspace{0.125\linewidth} \input{img/channel-dc.tikz}
\hspace{0.125\linewidth} \input{img/channel-f_g.tikz}
\hspace{0.125\linewidth}
\]
A monoidal structure comes with a \textit{monoidal unit}, which we will also continue to call a \textit{tensor unit}, and which will be not be depicted in diagrams, but rather left implicit.
(Alternatively, it is depicted as the ``empty diagram''.)
This is justified, as we will see, by the requirement that $I\otimes X\cong X\cong X\otimes I$ naturally in $X$.

\paragraph{States and costates}

In Remark \ref{rmk:element}, we called a morphism $I\to X$ out of the tensor unit a \textit{generalized element}, but owing to the many rôles they play, such morphisms go by many names.
When we think of $X$ as representing a system, we will also call such morphisms \textit{states} of $X$.
Dually, morphisms $X\to I$ can be called \textit{costates}, or sometimes \textit{effects}.
When the unit object is the terminal object (such as when the monoidal structure is given by the categorical product), then these costates are trivial.
In other categories, costates may be more effectful, and so carry more information: for example, in a category of vector spaces, states are vectors, costates are linear functionals, and so the composite of a state with a costate is an inner product.

Graphically, states \(\eta : I \to X\) and costates \(\epsilon : X \to I\) will be represented respectively as follows:
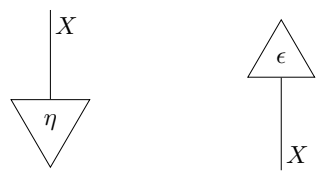
\begin{gather*}
\scalebox{0.85}{\begin{tikzpicture}
	\begin{pgfonlayer}{nodelayer}
		\node [style=state180] (0) at (0, -2) {$\eta$};
		\node [style=none] (1) at (0, 1.5) {};
		\node [style=none] (2) at (0.5, 1) {$X$};
	\end{pgfonlayer}
	\begin{pgfonlayer}{edgelayer}
		\draw (0) to (1.center);
	\end{pgfonlayer}
\end{tikzpicture}
}
\hspace{0.125\linewidth}
\scalebox{0.85}{\begin{tikzpicture}
	\begin{pgfonlayer}{nodelayer}
		\node [style=state0] (0) at (0, 0) {$\epsilon$};
		\node [style=none] (1) at (0, -3.5) {};
		\node [style=none] (2) at (0.5, -3) {$X$};
	\end{pgfonlayer}
	\begin{pgfonlayer}{edgelayer}
		\draw (0) to (1.center);
	\end{pgfonlayer}
\end{tikzpicture}
}
\end{gather*}

\paragraph{Discarding, marginalization, and causality}

In a category with only trivial effects, we can think of these as witnessing the `discarding' of information: in electronics terms, they ``send the signal to ground''.
For this reason, we will denote such trivial effects by the symbol $\ground$, writing $\ground_X:X\to I$ for each object $X$.

We can use discarding to depict \textit{marginalization}.
Given a `joint' state (a state of a tensor product) $\omega:I\to X\otimes Y$, we can discard either $Y$ or $X$ to obtain `marginal' states $\omega_1$ of $X$ and $\omega_2$ of $Y$ respectively, as in the following depiction:
\[
\input{img/marginalization-X.tikz}
\hspace{0.06\linewidth}
\text{and}
\hspace{0.06\linewidth}
\input{img/marginalization-Y.tikz}
\, . \]
We will see in Chapter \ref{chp:buco} how this corresponds to the marginalization familiar from probability theory.

To make the notion of discarding more mathematically precise, we can use it to encode a causality condition: physically realistic processes should not be able to affect the past.

\begin{defn} \label{def:causal}
  Whenever a morphism $c$ satisfies the equation
  \[
  \input{img/causality-condition.tikz}
  \]
  we will say that $c$ is \textit{causal}: the equation says that, if you do $c$ and throw away the result, the effect is of not having done $c$ at all---and so $c$ could not have had an anti-causal effect on its input.
\end{defn}

\begin{rmk}
  If in a category every morphism is causal, then this is equivalently a statement of the naturality of family of discarding morphisms $\ground_X:X\to I$, which implies that there is only one such morphism $X\to I$ for every object $X$, and which therefore means that $I$ must be a terminal object.
\end{rmk}

Some categories of interest will have nontrivial costates, yet we will still need notions of discarding and marginalization.
In these categories, it suffices to ask for each object $X$ to be equipped with a `comonoid' structure (to be elaborated in \secref{sec:comon}), of which one part is a `counit' morphism $X\to I$ which can play a discarding rôle, and which we will therefore also denote by $\ground_X$.

\paragraph{Copying}

The other part of a comonoid structure on $X$ is a `copying' map $\bcopier_X:X\to X\otimes X$, which has an intuitive graphical representation.
As we will see in \secref{sec:comon}, the \textit{comonoid laws} say that copying must interact nicely with the discarding maps:
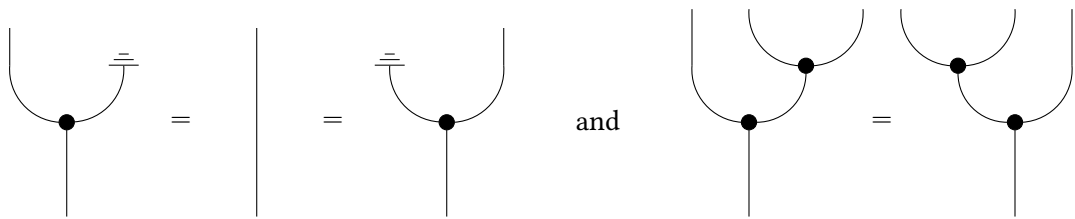
\begin{equation*}%
\input{img/copy-delete-identity.tikz}
\hspace{0.06\linewidth}
\text{and}
\hspace{0.06\linewidth}
\input{img/copy-copy-identity.tikz}
\end{equation*}
These equations say that making a copy and throwing it away is the same as not making a copy (left, \textit{counitality}; and that in copying a copy, it doesn't matter which copy you copy (right, \textit{coassociativity}).

\begin{defn} \label{def:cd-cat}
  A category with a comonoid structure \((\bcopier_X, \ground_X)\) for every object \(X\) is called a \textit{copy-discard category} \parencite{Cho2017Disintegration}.%
\end{defn}

\paragraph{Symmetry}

In all our applications, the tensor product structure will be \textit{symmetric}, meaning that $X\otimes Y$ can reversibly be turned into $Y\otimes X$ simply by swapping terms around.
In the graphical calculus, we depict this by the swapping of wires, which we ask to satisfy the following equations:
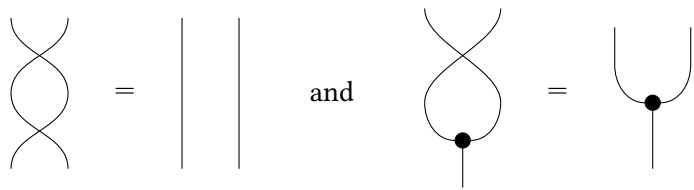
\begin{equation*} \label{eq:grph-comon-comm}
\input{img/swap-swap-identity.tikz}
\hspace{0.06\linewidth}
\text{and}
\hspace{0.06\linewidth}
\input{img/copy-swap-identity.tikz}
\end{equation*}
The equations say that swapping is self-inverse (on the left), and that copying is invariant under the symmetry (on the right).
(Strictly speaking, the right equation is an axiom called \textit{cocommutativity} that we additionally ask the comonoid structure to satisfy in the presence of a symmetric tensor.)

\subsection{Monoidal categories} \label{sec:mon-cat}

It being important to use tools appropriate for the jobs at hand, we will not always work just with the graphical calculus: we will need to translate between string diagrams and the symbolic algebra of Chapter \ref{chp:basic-ct}.
In the first instance, this means making mathematical sense of the graphical calculus itself, for which the key definition is that of the monoidal category.

\begin{defn} \label{def:monoidal-cat}
  We will call a category $\cat{C}$ \textit{monoidal} if it is equipped with a functor $\otimes:\cat{C}\times\cat{C}\to\cat{C}$ called the \textit{tensor} or \textit{monoidal product} along with an object $I:\cat{C}$ called the \textit{monoidal unit} and three natural isomorphisms
  \begin{enumerate}
  \item an \textit{associator} $\alpha : ((-)\otimes(-))\otimes(-)\Rightarrow(-)\otimes((-)\otimes(-))$;
  \item a \textit{left unitor} $\lambda : I\otimes(-)\Rightarrow(-)$; and
  \item a \textit{right unitor} $\rho : (-)\otimes I\Rightarrow(-)$
  \end{enumerate}
  such that the unitors are compatible with the associator, \textit{i.e.} for all $a,b:\cat{C}$ the diagram
  \[\begin{tikzcd}
	  {(a\otimes I)\otimes b} && {a\otimes(I\otimes b)} \\
	  \\
	  & {a\otimes b}
	  \arrow["{\rho_a\otimes \id_b}"', from=1-1, to=3-2]
	  \arrow["{\alpha_{a,I,b}}", from=1-1, to=1-3]
	  \arrow["{\id_a\otimes\lambda_b}", from=1-3, to=3-2]
  \end{tikzcd}\]
  commutes, and such that the associativity is `order-independent', \textit{i.e.} for all $a,b,c,d:\cat{C}$ the diagram
  \[\begin{tikzcd}
	  {(a\otimes(b\otimes c))\otimes d} && {a\otimes((b\otimes c)\otimes d)} \\
	  \\
    {((a\otimes b)\otimes c)\otimes d} && {a\otimes(b\otimes(c\otimes d))} \\
	  \\
	  & {(a\otimes b)\otimes(c\otimes d)}
	  \arrow["{\alpha_{a\otimes b,c,d}}"', from=3-1, to=5-2]
	  \arrow["{\alpha_{a,b,c\otimes d}}"', from=5-2, to=3-3]
	  \arrow["{\alpha_{a,b,c}\otimes \id_d}", from=3-1, to=1-1]
	  \arrow["{\alpha_{a,b\otimes c,d}}", from=1-1, to=1-3]
	  \arrow["{\id_a\otimes\alpha_{b,c,d}}", from=1-3, to=3-3]
  \end{tikzcd}\]
  commutes.

  We call $\cat{C}$ \textit{strict monoidal} if the associator and unitors are equalities rather than isomorphisms; in this case, the diagrams above commute by definition.
\end{defn}

\begin{ex}
  Any category equipped with a tensor product in the sense of Definition \ref{def:tensor} where the structure isomorphisms are additionally natural and satisfy the axioms of compatibility and order-independence is a monoidal category.
\end{ex}

\begin{ex}
  If $(\cat{C},\otimes,I)$ is a monoidal category, then so is $(\cat{C}\op,{\otimes}^{\mathrm{op}},I)$, where ${\otimes}^{\mathrm{op}}$ is the induced opposite functor $\cat{C}\op\times\cat{C}\op\to\cat{C}\op$.
\end{ex}

The associativity of the tensor is what allows us to depict string diagrams ``without brackets'' indicating the order of tensoring, and the unitality is what allows us to omit the monoidal unit from the diagrams.
Note that the functoriality of the tensor means that $\otimes$ distributes over $\circ$ as in $(f'\circ f)\otimes(g'\circ g) = (f'\otimes g')\circ(f\otimes g)$, both of which expressions are therefore depicted as
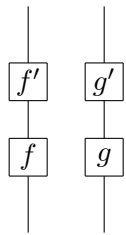
\[\input{img/channel-ff_gg.tikz} \quad . \]

The symmetry of a monoidal structure is formalized as follows.

\begin{defn}
  A \textit{symmetric monoidal} category is a monoidal category $(\cat{C},\otimes,I,\alpha,\lambda,\rho)$ that is additionally equipped with a natural isomorphism $\sigma : (-)\otimes(=)\Rightarrow(=)\otimes(-)$, called the \textit{symmetry}, such that $\sigma_{b,a}\circ\sigma_{a,b} = \id_{a\otimes b}$ for all $a,b:\cat{C}$, and whose compatibility with the associator is witnessed by the commutativity of the following diagram:
  \[\begin{tikzcd}
    {(a\otimes b)\otimes c} && {a\otimes(b\otimes c)} && {(b\otimes c)\otimes a} \\
	  \\
    {(b\otimes a)\otimes c} && {b\otimes(a\otimes c)} && {b\otimes(c\otimes a)}
	  \arrow["{\alpha_{a,b,c}}", from=1-1, to=1-3]
	  \arrow["{\sigma_{a,b\otimes c}}", from=1-3, to=1-5]
	  \arrow["{\alpha_{b,c,a}}", from=1-5, to=3-5]
	  \arrow["{\sigma_{a,b}\otimes\id_c}"', from=1-1, to=3-1]
	  \arrow["{\alpha_{b,a,c}}"', from=3-1, to=3-3]
	  \arrow["{\id_b\otimes\sigma_{a,c}}"', from=3-3, to=3-5]
  \end{tikzcd}\]
\end{defn}

Here is a familiar family of examples of symmetric, but not strict, monoidal categories.

\begin{ex}
  Any category within which every pair of objects has a product is said to \textit{have finite products}, and any category with finite products and a terminal object is a monoidal category.
  This includes the Cartesian products of sets (Definition \ref{def:prod-set} and Example \ref{ex:prod-set}) and of categories (Propositions \ref{prop:prod-cat} and \ref{prop:prod-func}).

  To see that the Cartesian product of sets is not strictly associative, observe that the elements of $A\times(B\times C)$ are tuples $(a,(b,c))$ whereas the elements of $(A\times B)\times C$ are tuples $((a,b),c)$; evidently, these two sets are isomorphic, but not equal, and the same holds for the product of categories.

  At the same time, it is easy to see that a Cartesian product is symmetric: we have $A\times B\cong B\times A$ by the mapping $(a,b)\leftrightarrow(b,a)$.
\end{ex}

And here is a family of examples of strict, but not symmetric, monoidal categories.

\begin{ex} \label{ex:endofunc-monoid}
  If $\cat{C}$ is any category, then the category $\cat{C}^{\cat{C}}$ of \textit{endofunctors} $\cat{C}\to\cat{C}$ is a strict monoidal category, where the monoidal product is given by composition $\circ$ of endofunctors and the monoidal unit is the identity functor $\id_{\cat{C}}$ on $\cat{C}$.
  That the monoidal structure here is strict follows from the fact that composition in a category is strictly associative and unital.
\end{ex}

In practice, we will tend to encounter strict monoidal categories only when the monoidal structure derives from the composition operator of a category, as in the preceding example.
However, when we work with the graphical calculus, we are often implicitly working with strict monoidal structure, as a result of the following important theorem.

\begin{thm}[{\textcite[{Theorem XI.3.1}]{MacLane1998Categories}}] \label{thm:mon-coh}
  Every monoidal category is strong monoidally equivalent to a strict monoidal one.
\end{thm}

As a consequence of this \textit{coherence theorem}, any two string diagrams where one can be transformed into the other by a purely topological transformation are equal, as in the following example (read from left to right):
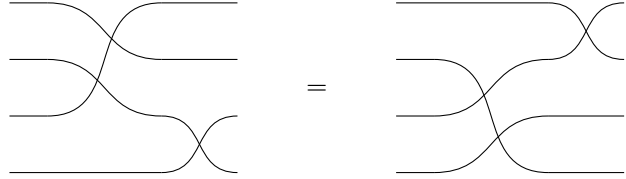
\[ \input{img/string-deform-1.tikz} \qquad = \qquad \input{img/string-deform-2.tikz} \; \]
This follows because the coherence theorem renders parallel morphisms entirely constructed from identities, associators and unitors (and the symmetry, as long as it is strictly self-inverse) equal ``on the nose''\footnote{%
This process of turning natural isomorphisms into equalities is called \textit{strictification}.}.

To make sense of the notion of \textit{strong monoidal equivalence}, we need a notion of functor that preserves monoidal structure; we define the `weak' case first.

\begin{defn} \label{def:lax-mon-func}
  Suppose $(\cat{C},\otimes_{\cat{C}},I_{\cat{C}})$ and $(\cat{D},\otimes_{\cat{D}},I_{\cat{D}})$ are monoidal categories.
  A \textit{lax monoidal functor} $(\cat{C},\otimes_{\cat{C}},I_{\cat{C}})\to(\cat{D},\otimes_{\cat{D}},I_{\cat{D}})$ is a triple of
  \begin{enumerate}
  \item a functor $F:\cat{C}\to\cat{D}$;
  \item a state $\epsilon:I_{\cat{D}}\to F(I_{\cat{C}})$ called the \textit{unit}; and
  \item a natural transformation, the \textit{laxator}, $\mu : F(-)\otimes_{\cat{D}}F(=)\Rightarrow F((-)\otimes_{\cat{C}}(=))$
  \end{enumerate}
  satisfying the axioms of
  \begin{enumerate}[label=(\alph*)]
  \item associativity, in that the following diagram commutes
    \[\begin{tikzcd}
      {(F(a)\otimes_{\cat{D}}F(b))\otimes_{\cat{D}}F(c)} && {F(a)\otimes_{\cat{D}}(F(b)\otimes_{\cat{D}}F(c))} \\
	    \\
      {F(a\otimes_{\cat{C}}b)\otimes_{\cat{D}}F(c)} && {F(a)\otimes_{\cat{D}}F(b\otimes_{\cat{C}}c)} \\
	    \\
      {F((a\otimes_{\cat{C}}b)\otimes_{\cat{C}}c)} && {F(a\otimes_{\cat{C}}(b\otimes_{\cat{C}}c))}
	    \arrow["{\alpha^{\cat{D}}_{F(a),F(b),F(c)}}", from=1-1, to=1-3]
	    \arrow["{F(a)\otimes_{\cat{D}}\mu_{b,c}}", from=1-3, to=3-3]
	    \arrow["{\mu_{a,b}\otimes_{\cat{D}} F(c)}"', from=1-1, to=3-1]
	    \arrow["{\mu_{a\otimes_{\cat{C}}b,c}}"', from=3-1, to=5-1]
	    \arrow["{F(\alpha^{\cat{C}}_{a,b,c})}", from=5-1, to=5-3]
	    \arrow["{\mu_{F(a),b\otimes_{\cat{C}}c}}", from=3-3, to=5-3]
    \end{tikzcd}\]
    where $\alpha^{\cat{C}}$ and $\alpha^{\cat{D}}$ are the associators of the respective monoidal structures on $\cat{C}$ and $\cat{D}$; and
  \item (left and right) unitality, in that the following diagrams commute
    \[\begin{tikzcd}
      {I_{\cat{D}}\otimes_{\cat{D}}F(a)} && {F(I_{\cat{C}})\otimes_{\cat{D}}F(a)} \\
	    \\
      {F(a)} && {F(I_{\cat{C}}\otimes_{\cat{C}}a)}
	    \arrow["{\lambda^{\cat{D}}_{F(a)}}", from=1-1, to=3-1]
	    \arrow["{\mu_{I_{\cat{C}},a}}", from=1-3, to=3-3]
	    \arrow["{\epsilon\otimes_{\cat{D}}F(a)}", from=1-1, to=1-3]
	    \arrow["{F(\lambda^{\cat{C}}_a)}", from=3-3, to=3-1]
    \end{tikzcd}
    \;\text{and}\;
    \begin{tikzcd}
      {F(a)\otimes_{\cat{D}}I_{\cat{D}}} && {F(a)\otimes_{\cat{D}}F(I_{\cat{C}})} \\
	    \\
      {F(a)} && {F(a\otimes_{\cat{C}}I_{\cat{C}})}
	    \arrow["{\rho^{\cat{D}}_{F(a)}}", from=1-1, to=3-1]
	    \arrow["{\mu_{a,I_{\cat{C}}}}", from=1-3, to=3-3]
	    \arrow["{F(a)\otimes_{\cat{D}}\epsilon}", from=1-1, to=1-3]
	    \arrow["{F(\rho^{\cat{C}}_a)}", from=3-3, to=3-1]
    \end{tikzcd}\]
    where $\lambda^{\cat{C}}$ and $\lambda^{\cat{D}}$ are the left, and $\rho^{\cat{C}}$ and $\rho^{\cat{D}}$ the right, unitors of the respective monoidal structures on $\cat{C}$ and $\cat{D}$.
  \end{enumerate}
  A \textit{strong monoidal functor} is a lax monoidal functor for which the unit and laxator are isomorphisms.
  A strong monoidal equivalence is therefore an equivalence of categories in which the two functors are strong monoidal.
\end{defn}

\begin{rmk} \label{rmk:lax-emerg}
  Laxness can be read as a sign of an ``emergent property'': if $F$ is lax monoidal, then this means there are systems of type $F(X\otimes Y)$ that do not arise simply by placing a system of type $F(X)$ beside a system of type $F(Y)$ using $\otimes$; whereas if $F$ is strong monoidal, then there are no such `emergent' systems.
  More generally, we can think of emergence as an indication of higher-dimensional structure that is hidden when one restricts oneself to lower dimensions (and hence can appear mysterious).
  In this example, the higher-dimensional structure is the 2-cell of the laxator.
\end{rmk}

There is of course a notion of monoidal natural transformation, making monoidal categories, lax monoidal functors, and monoidal natural transformations into the constituents of a 2-category.

\begin{defn} \label{def:mon-nat-trans}
  If $(F,\mu,\epsilon)$ and $(F',\mu',\epsilon')$ are lax monoidal functors $(\cat{C},\otimes_{\cat{C}},I_{\cat{C}})\to(\cat{D},\otimes_{\cat{D}},I_{\cat{D}})$, then a \textit{monoidal natural transformation} $\alpha:(F,\mu,\epsilon)\Rightarrow (F',\mu',\epsilon')$ is a natural transformation $\alpha:F\Rightarrow F'$ that is compatible with the unitors
  \[\begin{tikzcd}
    & {I_{\cat{D}}} \\
    \\
    {F(I_{\cat{C}})} && {F'(I_{\cat{C}})}
    \arrow["\epsilon"', from=1-2, to=3-1]
    \arrow["{\epsilon'}", from=1-2, to=3-3]
    \arrow["{\alpha_{I_{\cat{C}}}}"', from=3-1, to=3-3]
  \end{tikzcd}\]
  and the laxators
  \[\begin{tikzcd}
    {Fa\otimes_{\cat{D}}Fb} && {F'a\otimes_{\cat{C}}F'b} \\
    \\
    {F(a\otimes_{\cat{C}}b)} && {F'(a\otimes_{\cat{C}}b)}
	  \arrow["{\alpha_a\otimes_{\cat{D}}\alpha_b}", from=1-1, to=1-3]
	  \arrow["{\alpha_{a\otimes_{\cat{C}}b}}", from=3-1, to=3-3]
	  \arrow["{\mu_{a,b}}"', from=1-1, to=3-1]
	  \arrow["{\mu'_{a,b}}", from=1-3, to=3-3]
  \end{tikzcd}\]
  for all $a,b:\cat{C}$.
\end{defn}

\begin{prop}
  Monoidal categories, lax monoidal functors, and monoidal natural transformations form the 0-cells, 1-cells, and 2-cells of a 2-category, denoted $\MonCat$.
  \begin{proof}
    Given composable lax monoidal functors $(F,\epsilon,\mu):(\cat{C},\otimes_{\cat{C}},I_{\cat{C}})\to(\cat{D},\otimes_{\cat{D}},I_{\cat{D}})$ and $(F',\epsilon',\mu'):(\cat{D},\otimes_{\cat{D}},I_{\cat{D}})\to(\cat{E},\otimes_{\cat{E}},I_{\cat{E}})$, form their horizontal composite as follows.
    The functors compose as functors, $G\circ F$.
    The composite state is given by $I_{\cat{E}}\xto{\epsilon'}F'(I_{\cat{D}})\xto{F'\epsilon}F'F(I_{\cat{C}})$.
    The laxator is given by
    \[
    F'F(-)\otimes_{\cat{E}}F'F(=)
    \xRightarrow{\mu'_{F(-),F(=)}}
    F'(F(-)\otimes_{\cat{D}}F(=))
    \xRightarrow{F'\mu_{a,b}}
    F'F((-)\otimes_{\cat{C}}(=)) \, .
    \]
    The identity lax monoidal functor on $\cat{C}$ is given by $(\id_{\cat{C}},\id_{I_{\cat{C}}},\id_{(-)\otimes_{\cat{C}}(=)})$.
    Unitality and associativity of composition of lax monoidal functors follow straightforwardly from unitality and associativity of composition of morphisms, functors, and natural transformations.
    Monoidal natural transformations compose vertically as natural transformations, and it is easy to see that the composites satisfy the compatibility conditions by pasting the relevant diagrams.
  \end{proof}
\end{prop}

\subsection{Closed monoidal categories}

Since one source of monoidal structures is the generalization of the categorical product, it is no surprise that there is a corresponding generalization of exponentials: a `tensor-hom' adjunction that induces a concept of closed monoidal category.
Such categories will be important later in the thesis when we consider learning and adaptive systems: our compositional model of predictive coding, for example, will be built on a certain generalized exponential (see Remark \ref{rmk:ext-hom}).

\begin{defn} \label{def:internal-hom}
  Let $(\cat{C},\otimes,I)$ be a monoidal category.
  When there is an object $e:\cat{C}$ such that $\cat{C}(x, e) \cong \cat{C}(x\otimes y, z)$ naturally in $x$, we say that $e$ is an \textit{internal hom object} and denote it by $[y,z]$.
  The image of $\id_{[y,z]}$ under the isomorphism is called the \textit{evaluation map} and is written $\mathsf{ev}_{y,z} : [y,z]\otimes y \to z$.
\end{defn}

\begin{prop} \label{prop:tens-hom-adj}
  When the isomorphism $\cat{C}(x\otimes y, z) \cong \cat{C}(x, [y,z])$ is additionally natural in $z$, we obtain an adjunction $(-)\otimes y \dashv [y,-]$ called the \textit{tensor-hom adjunction}, which uniquely determines a functor $\cat{C}\op\times\cat{C} \to \cat{C} : (y,z) \mapsto [y,z]$ that we call the \textit{internal hom} for $\cat{C}$.
  \begin{proof}
    A direct generalization of the Cartesian case (Proposition \ref{prop:prod-exp-adj}).
  \end{proof}
\end{prop}

\begin{defn} \label{def:mon-clos}
  A monoidal category $\cat{C}$ with a corresponding internal hom is called \textit{monoidal closed}.
\end{defn}

\begin{ex}
  The category of finite-dimensional real vector spaces and linear maps between them is monoidal closed with respect to the tensor product of vector spaces, as each space of linear maps is again a vector space and the tensor is necessarily bilinear.
\end{ex}

As in the Cartesian case, monoidal closed categories can reason about themselves.

\begin{prop} \label{prop:mon-clos-self-enr}
  A monoidal closed category is enriched in itself.
\end{prop}

And when a category is enriched in a symmetric monoidal category, then its hom functor is likewise enriched.

\begin{prop}
  Suppose $\cat{C}$ is an $\cat{E}$-category where $\cat{E}$ is symmetric monoidal closed.
  Then the hom functor $\cat{C}(-,=)$ is an $\cat{E}$-functor.
  \begin{proof}
    A direct generalization of Proposition \ref{prop:ccc-enr-hom}.
  \end{proof}
\end{prop}

\begin{rmk}
  Since Cartesian closed categories have a rich internal logic, via the Curry-Howard-Lambek correspondence, one might wonder if there is an analogous situation for monoidal closed categories.
  To a certain intricate extent there is: the internal logic of monoidal closed categories is generally known as \textit{linear logic}, and its corresponding language \textit{linear type theory}.
  These are `refinements' of intuitionistic logic and type theory which of course coincide in the Cartesian case, but which more generally clarify certain logical interactions; we shall say no more in this thesis, except that such logics find application in quantum mechanics, owing to the monoidal closed structure of vector spaces, where the linear structure constrains the use of resources (in relation, for example, to the famous quantum `no-cloning' and `no-deleting' theorems).

  With respect to \textit{dependent} types, the situation is a little more vexed, as the existence of well-behaved dependent sums and products classically depends on the existence of pullbacks and their coherence with products (and, for example, the tensor product of vector spaces is not a categorical product); this means that classical dependent data is somehow not resource-sensitive.
  Nonetheless, various proposals have been made to unify linear logic with dependent type theory\parencite{Vakar2015Syntax,McBride2016I,Lundfall2018diagram,Atkey2018Syntax,Fu2020Linear}: the simplest of these proceed by requiring dependence to be somehow Cartesian, which is the approach we will take in Chapter \ref{chp:coalg} when we face a similar quandary in the context of defining a category of polynomial functors with non-deterministic feedback.
  (We will see in Chapter \ref{chp:buco} that the property of Cartesianness is equally closely related to determinism.)
\end{rmk}

\subsection{Bicategories} \label{sec:bicat}

Monoidal categories are not the first two-dimensional categorical structures we have so far encountered, the other primary example being 2-categories.
These two classes of examples are closely related: a strict monoidal category is a 2-category with one object; and so just as a monoidal category is a correspondingly weakened version, a bicategory is a `weak 2-category'.

\begin{defn} \label{def:bicat}
  A \textit{bicategory} $\cat{B}$ is constituted by
  \begin{enumerate}
  \item a set $\cat{B}_0$ of \textit{objects} or \textit{0-cells};
  \item for each pair $(A,B)$ of $\cat{B}$-objects, a category $\cat{B}(A,B)$ called the \textit{hom category}, the objects of which are the \textit{morphisms} or \textit{1-cells} from $A$ to $B$, and the morphisms of which are the \textit{2-cells} between those 1-cells;
  \item for each 0-cell $A$, a 1-cell $\id_a : \cat{B}(A,A)$ witnessing \textit{identity}; and
  \item for each triple $(A,B,C)$ of 0-cells, a functor $\diamond_{A,B,C}:\cat{B}(B,C)\times\cat{B}(A,B)\to\cat{B}(A,C)$ witnessing \textit{horizontal composition} (with \textit{vertical composition} referring to composition within each hom category);
  \item for each pair $(A,B)$ of 0-cells, natural isomorphisms $\rho_{A,B}$ (the \textit{right unitor}) and $\lambda_{A,B}$ (the \textit{left unitor}) witnessing the unitality of horizontal composition, as in the diagrams
  \begin{equation*}
    \begin{tikzcd}
	    {\cat{B}(A,B)\times \Cat{1}} && {\cat{B}(A,B)\times\cat{B}(A,A)} \\
	    \\
	    && {\cat{B}(A,B)}
	    \arrow[""{name=0, anchor=center, inner sep=0},"{\Rho_{\cat{B}(A,B)}}"', from=1-1, to=3-3]
	    \arrow["{{\cat{B}(A,B)}\times{\id_A}}", from=1-1, to=1-3]
	    \arrow["{\diamond_{A,A,B}}", from=1-3, to=3-3]
      \arrow["\rho_{A,B}"', shorten >=6pt, Rightarrow, from=1-3, to=0]
    \end{tikzcd}
    \text{and}
    \begin{tikzcd}
	    {\cat{B}(B,B)\times\cat{B}(A,B)} && {\Cat{1}\times\cat{B}(A,B)} \\
	    \\
	      {\cat{B}(A,B)}
	      \arrow[""{name=0, anchor=center, inner sep=0},"{\Lambda_{\cat{B}(A,B)}}", from=1-3, to=3-1]
	      \arrow["{{\id_B}\times{\cat{B}(A,B)}}"', from=1-3, to=1-1]
	      \arrow["{\diamond_{A,B,B}}"', from=1-1, to=3-1]
        \arrow["\lambda_{A,B}", shorten >=6pt, Rightarrow, from=1-1, to=0]
    \end{tikzcd}
  \end{equation*}
  where $\Lambda:\Cat{1}\times(-)\Rightarrow(-)$ and $\Rho:(-)\times\Cat{1}\Rightarrow(-)$ are the (almost trivial) left and right unitors of the product $\times$ on $\Cat{Cat}$; and
  \item for each quadruple $(A,B,C,D)$ of 0-cells, a natural isomorphism $\alpha_{A,B,C,D}$ witnessing the associativity of horizontal composition, as in the diagram
  \[\begin{tikzcd}
	  {\bigl(\cat{B}(C,D)\times\cat{B}(B,C)\bigr)\times\cat{B}(A,B)} && {\cat{B}(C,D)\times\bigl(\cat{B}(B,C)\times\cat{B}(A,B)\big)} \\
	  \\
	  {\cat{B}(B,D)\times\cat{B}(A,B)} && {\cat{B}(C,D)\times\cat{B}(A,C)} \\
	  \\
	  & {\cat{B}(A,D)}
	  \arrow["\Alpha_{\cat{B}(C,D),\cat{B}(B,C),\cat{B}(A,B)}", from=1-1, to=1-3]
	  \arrow[""{name=0, anchor=center, inner sep=0},"{\diamond_{B,C,D}\times\cat{B}(A,B)}"', from=1-1, to=3-1]
	  \arrow[""{name=1, anchor=center, inner sep=0},"{\cat{B}(C,D)\times\diamond_{A,B,C}}", from=1-3, to=3-3]
	  \arrow["{\diamond_{A,B,D}}"', from=3-1, to=5-2]
	  \arrow["{\diamond_{A,C,D}}", from=3-3, to=5-2]
    \arrow["\alpha_{A,B,C,D}", shorten <=26pt, shorten >=26pt, Rightarrow, from=0, to=1]
  \end{tikzcd}\]
  where $\Alpha:((-)\times(-))\times(-)\Rightarrow(-)\times((-)\times(-))$ is the (almost trivial) associator of the product $\times$ on $\Cat{Cat}$;
  \end{enumerate}
  such that the unitors are compatible with the associator, \textit{i.e.} for all 1-cells $a:\cat{B}(A,B)$ and $b:\cat{B}(B,C)$ the diagram
  \[\begin{tikzcd}
	  {(b\diamond \id_B)\diamond a} && {b\diamond(\id_B\diamond a)} \\
	  \\
	  & {b\diamond a}
	  \arrow["{\rho_{b}\diamond \id_a}"', from=1-1, to=3-2]
	  \arrow["{\alpha_{b,\id_B,a}}", from=1-1, to=1-3]
	  \arrow["{\id_b\diamond\lambda_{a}}", from=1-3, to=3-2]
  \end{tikzcd}\]
  commutes (where we have omitted the subscripts indexing the 0-cells on $\alpha$, $\rho$, and $\lambda$);
  and such that the associativity is `order-independent', \textit{i.e.} for all 1-cells $a:\cat{B}(A,B)$, $b:\cat{B}(B,C)$, $c:\cat{B}(C,D)$, and $d:\cat{B}(D,E)$ the diagram
  \[\begin{tikzcd}
	  {(a\diamond(b\diamond c))\diamond d} && {a\diamond((b\diamond c)\diamond d)} \\
	  \\
    {((a\diamond b)\diamond c)\diamond d} && {a\diamond(b\diamond(c\diamond d))} \\
	  \\
	  & {(a\diamond b)\diamond(c\diamond d)}
	  \arrow["{\alpha_{a\diamond b,c,d}}"', from=3-1, to=5-2]
	  \arrow["{\alpha_{a,b,c\diamond d}}"', from=5-2, to=3-3]
	  \arrow["{\alpha_{a,b,c}\diamond \id_d}", from=3-1, to=1-1]
	  \arrow["{\alpha_{a,b\diamond c,d}}", from=1-1, to=1-3]
	  \arrow["{\id_a\diamond\alpha_{b,c,d}}", from=1-3, to=3-3]
  \end{tikzcd}\]
  commutes (where we have again omitted the subscripts indexing the 0-cells on $\alpha$).
\end{defn}

\begin{rmk}
  Just as a 2-category is a category enriched in $\Cat{Cat}$, a bicategory is a category \textit{weakly enriched} in $\Cat{Cat}$.
  This is easy to see by comparing Definition \ref{def:bicat} with Definition \ref{def:enriched-cat}: the former is obtained from the latter by taking $\cat{E}$ to be $\Cat{Cat}$ and filling the unitality and associativity diagrams with nontrivial fillers which are required to satisfy coherence laws generalizing those of the monoidal category structure (Definition \ref{def:monoidal-cat}).
  Conceptually, we can see this weakening in the context of our brief discussion of emergence above (Remark \ref{rmk:lax-emerg}): we recognize the property of axiom-satisfaction as a shadow of a higher-dimensional structure (the fillers), which we categorify accordingly.
\end{rmk}

Bicategories will appear later in this thesis when we construct categories of dynamical hierarchical inference systems: the construction proceeds by using polynomial functors to ``wire together'' categories of dynamical systems, and the composition of polynomials distributes weakly but naturally over the categories of systems, thereby producing a category weakly enriched in $\Cat{Cat}$.
Before then, we will encounter bicategories in the abstract context of general parameterized morphisms, where the 2-cells witness changes of parameter.

For now, our first examples of bicategories are induced by monoidal categories, which are equivalently single-object bicategories.

\begin{prop} \label{prop:deloop}
  Suppose $(\cat{C},\otimes,I)$ is a monoidal category.
  Then there is a bicategory $\deloop\cat{C}$ with a single 0-cell, $\ast$, and whose category of 1-cells $\deloop\cat{C}(\ast,\ast)$ is $\cat{C}$.
  The identity 1-cell is $I$, and horizontal composition is given by the monoidal product $\cat{C}$; vertical composition is just the composition of morphisms in $\cat{C}$.
  The unitors and associator of the bicategory structure are the unitors and associator of the monoidal structure.
  We call $\deloop\cat{C}$ the \textit{delooping} of $\cat{C}$.
  \begin{proof}
    The bicategory axioms are satisfied immediately, because the structure morphisms satisfy the (in this case identical) monoidal category axioms.
  \end{proof}
\end{prop}

In the opposite direction, the equivalence is witnessed by the following proposition.

\begin{prop}
  Suppose $\cat{B}$ is a bicategory with a single 0-cell, $\ast$, and whose horizontal composition is denoted $\diamond$.
  Then $\bigl(\cat{B}(\ast,\ast),\diamond,\id_\ast\bigr)$ is a monoidal category.
\end{prop}

\begin{rmk}
  It is possible to define a notion of monoidal bicategory, as something like a monoidal category weakly enriched in $\Cat{Cat}$, or as a one-object `tricategory', and in many cases the bicategories considered below are likely to have such structure.
  We will say a little more about this in Remark \ref{rmk:pseudomonoid} below, but will not define or make formal use of this higher structure in this thesis.

  More generally, there are analogues of the other structures and results of basic category theory introduced both in this chapter and in Chapter \ref{chp:basic-ct} that are applicable to higher-dimensional categories such as bicategories, but they too will not play an important rôle in this thesis.
\end{rmk}

\section{Parameterized systems} \label{sec:para-sys}

A category does not have to be monoidal closed for us to be able to talk about ``controlled processes'' in it: its being monoidal is sufficient, for we can consider morphisms of the form $P\otimes X\to Y$ and treat the object $P$ as an object of adjustable parameters.
Parameterized morphisms of this form can easily be made to compose: given another morphism $Q\otimes Y\to Z$, we can straightforwardly obtain a composite parameterized morphism $(Q\otimes P)\otimes X\to Z$, as we elaborate in \secref{sec:int-para} below.

Categories of such parameterized morphisms play a central rôle in the compositional modelling of cybernetic systems\parencite{Capucci2021Towards,Smithe2020Cyber}, where we typically see the parameter as controlling the choice of process, and understand learning as a `higher-order' process by which the choice of parameter is adjusted.
More concretely, consider the synaptic strengths or weights of a neural network, which change as the system learns about the world, affecting the predictions it makes and actions it takes; or consider the process of Bayesian inference, where the posterior is dependent on a parameter that is typically called the `prior'.

In this section, we introduce two related formal notions of parameterization: `internal', where the parameter object constitutes a part of the domain of morphisms in a category; and `external', where the parameters remain outside of the category being parameterized and the choice of morphism is implemented as a morphism in the base of enrichment.
We will make use of both kinds of parameterization in this thesis.

\begin{rmk}
  Parameterization can be understood as introducing a new dimension into a category of processes.
  Consequently, the parameterization (either internal or external) of a category will produce a bicategory.
  When representing processes graphically, such as when using the string diagram calculus, this extra dimension becomes particularly explicit, and although we won't make use of graphical representations of parameterized processes in this thesis, they are typical in the applied-categorical literature, particularly in the literature on categorical cybernetics; for example, see \textcite[{Fig. 1}]{Capucci2021Towards}, \textcite[{pp.1--2}]{Cruttwell2022Categorical}, and \textcite[{Fig. 1}]{Capucci2022Diegetic}.
\end{rmk}

\subsection{Internal parameterization} \label{sec:int-para}

Internal parameterization generalizes the case with which we opened this section, of morphisms $P\otimes X\to Y$, to a situation in which the parameterization may have different structure to the processes at hand, so that the parameterizing objects live in a different category.
For this reason, we describe the `actegorical' situation in which a category of parameters \(\cat{M}\) \textit{acts on} on a category of processes \(\cat{C}\) to generate a category of parameterized processes.
Nonetheless, even in this case, the parameter ends up constituting part of the domain of the morphism representing the parameterized process.

The first concept we need is that of an `actegory', which categorifies the better known mathematical notion of monoid action\footnote{%
For a comprehensive reference on actegory theory, see \textcite{Capucci2022Actegories}.}.

\begin{defn}[\(\cat{M}\)-actegory] \label{def:m-act}
  Suppose \(\cat{M}\) is a monoidal category with tensor \(\boxtimes\) and unit object \(I\).
  We say that \(\cat{C}\) is a \textit{left} \(\cat{M}\)\textit{-actegory} when there is a functor \(\odot : \cat{M} \times \cat{C} \to \cat{C}\) called the \textit{action} along with natural unitor and associator isomorphisms \(\lambda^\odot_X : I \odot X \xto{\sim} X\) and \(a^\odot_{M,N,X} : (M \boxtimes N) \odot X \xto{\sim} M \odot (N \odot X)\) compatible with the monoidal structure of \((\cat{M}, \boxtimes, I)\), in a sense analogous to the coherence data of a monoidal category (Definition \ref{def:monoidal-cat}).
  This means that the following triangle and pentagon diagrams must commute, where $\rho$ and $\alpha$ are the right unitor and the associator of the monoidal structure on $\cat{M}$.
    \[\begin{tikzcd}
	  {(M\boxtimes I) \odot C} && {M\odot(I\odot C)} \\
	  \\
	  & {M\odot C}
	  \arrow["{a^\odot_{M,I,C}}", from=1-1, to=1-3]
	  \arrow["{\rho_M\otimes \id_C}"', from=1-1, to=3-2]
	  \arrow["{\id_M\odot\lambda^\odot_C}", from=1-3, to=3-2]
  \end{tikzcd}\]
  \[\begin{tikzcd}
	  {(K\boxtimes(M\boxtimes N))\odot C} && {K\odot((M\boxtimes N)\odot C)} \\
	  \\
    {((K\boxtimes M)\boxtimes N)\odot C} && {K\odot(M\odot(N\odot C))} \\
	  \\
	  & {(K\boxtimes M)\odot(N\odot C)}
	  \arrow["{a^\odot_{K,M\boxtimes N,C}}", from=1-1, to=1-3]
	  \arrow["{\alpha_{K,M,N}\otimes \id_C}", from=3-1, to=1-1]
	  \arrow["{\id_K\otimes a^\odot_{M,N,C}}", from=1-3, to=3-3]
	  \arrow["{a^\odot_{K\boxtimes M,N,C}}"', from=3-1, to=5-2]
	  \arrow["{a^\odot_{K,M,N\odot C}}"', from=5-2, to=3-3]
  \end{tikzcd}\]
\end{defn}

Given an actegory, we can define a category of correspondingly parameterized morphisms.

\begin{prop}[\textcite{Capucci2021Towards}] \label{prop:para-bicat}
  Let \((\cat{C}, \odot, \lambda^\odot, a^\odot)\) be an \((\cat{M}, \boxtimes, I)\)-actegory.
  Then there is a bicategory of \(\cat{M}\)-parameterized morphisms in \(\cat{C}\), denoted \(\para(\odot)\).
  Its objects are those of \(\cat{C}\).
  For each pair of objects \(X,Y\), the set of 1-cells is defined as \(\para(\odot)(X, Y) := \sum_{M:\cat{M}} \cat{C}(M \odot X, Y)\); we denote an element \((M, f)\) of this set by \(f : X \xto{M} Y\).
  Given 1-cells \(f : X \xto{M} Y\) and \(g : Y \xto{N} Z\), their composite
  \(g \circ f : X \xto{N \boxtimes M} Z\)
  is the following morphism in \(\cat{C}\):
  \[ (N \boxtimes M) \odot X \xto{a^\odot_{N,M,X}} N \odot (M \odot X) \xto{\id_N \odot f} N \odot Y \xto{g} Z \]
  Given 1-cells \(f : X \xto{M} Y\) and \(f' : X \xto{M'} Y\), a 2-cell \(\alpha : f \Rightarrow f'\) is a morphism \(\alpha : M \to M'\) in \(\cat{M}\) such that \(f = f' \circ (\alpha \odot \id_X)\) in \(\cat{C}\); identities and composition of 2-cells are as in \(\cat{C}\).
\end{prop}

And when the action is `strong' and the monoidal structure on $\cat{C}$ is symmetric, these parameterized categories inherit a monoidal structure.

\begin{defn} \label{def:strength}
  Suppose $\cat{C}$ is a monoidal category and $F:\cat{C}\to\cat{C}$ is an endofunctor.
  A \textit{right strength} for $F$ is a natural transformation $\mathsf{str}^r_{X,Y}:FX\otimes Y\to F(X\otimes Y)$ making the following diagrams commute:
  \[\begin{tikzcd}[sep=scriptsize]
	{FX\otimes(Y\otimes Z)} &&&& {F(X\otimes(Y\otimes Z))} \\
	\\
	{(FX\otimes Y)\otimes Z} && {F(X\otimes Y)\otimes Z} && {F((X\otimes Y)\otimes Z)}
	\arrow["{\alpha_{FX,Y,Z}}", from=3-1, to=1-1]
	\arrow["{\mathsf{str}^r_{X,Y}\otimes\id_Z}"', from=3-1, to=3-3]
	\arrow["{\mathsf{str}^r_{X\otimes Y,Z}}"', from=3-3, to=3-5]
	\arrow["{F(\alpha_{X,Y,Z})}"', from=3-5, to=1-5]
	\arrow["{\mathsf{str}^r_{X,Y\otimes Z}}", from=1-1, to=1-5]
  \end{tikzcd}\]
  \[\begin{tikzcd}[sep=scriptsize]
	{FX\otimes I} && {F(X\otimes I)} \\
	\\
	& FX
	\arrow["{\mathsf{str}^r_{X,I}}", from=1-1, to=1-3]
	\arrow["{F\rho_X}", from=1-3, to=3-2]
	\arrow["{\rho_{FX}}"', from=1-1, to=3-2]
  \end{tikzcd}\]
  An action $\odot:\cat{M}\times\cat{C}\to\cat{C}$ induces a family of functors $M\odot(-):\cat{C}\to\cat{C}$, natural in $M:\cat{M}$.
  If each of these is equipped with a right strength, also natural in $M:\cat{M}$, then we call the resulting transformation $\mathsf{str}^r_{M,X,Y}$ a right strength for $\odot$.
  Dually, there are notions of \textit{left strength}, $\mathsf{str}^l_{X,Y}:X\otimes FY\to F(X\otimes Y)$ and \textit{costrength}, with the latter obtained in the usual way as a strength in $\cat{C}\op$ (reverse all the defining arrows).

  Note that, if $\cat{C}$ is symmetric monoidal, then a left strength induces a right strength (by swapping) and likewise a right strength induces a left strength.
\end{defn}

\begin{prop}[{\textcite[\S2.1]{Capucci2021Towards}}] \label{prop:para-tensor}
  When \(\cat{C}\) is equipped with both a symmetric monoidal structure \((\otimes, I)\) and an \((\cat{M}, \boxtimes, I)\)-actegory structure $\odot$, and these are compatible in that the action $\odot$ has a strength isomorphism, the symmetric monoidal structure \((\otimes, I)\) lifts to \(\para(\odot)\).

  The tensor of objects in \(\para(\odot)\) is then defined as the tensor of objects in \(\cat{C}\), and the tensor of morphisms (1-cells) \(f : X \xto{M} Y\) and \(g : A \xto{N} B\) is given by the composite
  \[
  f \otimes g : X \otimes A \xto{M \boxtimes N} Y \otimes B
  \; := \;
  (M \boxtimes N) \odot (X \otimes A) \xto{\iota_{M,N,X,A}}
  (M \odot A) \otimes (N \odot A) \xto{f \otimes g} Y \otimes B
  \]
  where the \textit{interchanger} $\iota_{M,N,X,A} : (M\boxtimes N)\odot(X\otimes A) \xto{\sim} (M\odot X)\otimes(N\odot A)$ is obtained using the associator of the actegory structure and the costrengths:
  \begin{align*}
    \iota_{M,N,X,A} & :=  (M\boxtimes N)\odot(X\otimes A) \xto{a^\odot_{M,N,(X\otimes A)}} M\odot(N\odot(X\otimes A)) \; \cdots \\ & \quad \cdots \, \xto{M\odot\mathsf{costr}^l_{N,X,A}} M\odot(X\otimes(N\odot A)) \xto{\mathsf{costr}^r_{M,X,N\odot A}} (M\odot X)\otimes(N\odot A) \; .
  \end{align*}
  (Note that the costrengths are obtained as the inverses of the strengths.)
\end{prop}

We can see a monoidal product $\otimes:\cat{C}\times\cat{C}\to\cat{C}$ as an action of $\cat{C}$ on itself, and this induces the self-parameterization of $\cat{C}$.

\begin{prop}[Self-parameterization] \label{prop:para-self}
  If \((\cat{C}, \otimes, I)\) is a monoidal category, then it induces a parameterization \(\para(\otimes)\) on itself.
  For each \(M,X,Y : \cat{C}\), the morphisms \(X \xto{M} Y\) of \(\para(\otimes)\) are the morphisms \(M \otimes X \to Y\) in \(\cat{C}\).
\end{prop}

\begin{notation}
  When considering the self-paramterization induced by a monoidal category \((\cat{C}, \otimes, I)\), we will often write \(\para(\cat{C})\) instead of \(\para(\otimes)\).
\end{notation}

It will frequently be the case that we do not in fact need the whole bicategory structure.
The following proposition tells us that we can also just work 1-categorically, as long as we work with equivalence classes of isomorphically-parameterized maps, in order that composition is suffiently strictly associative.

\begin{prop} \label{prop:para-1cat}
  Each bicategory \(\para(\odot)\) induces a 1-category \(\para(\odot)_1\) by forgetting the bicategorical structure.
  The hom sets \(\para(\odot)_1(X, Y)\) are given by \(U \para(\odot)(X, Y) / \sim\) where \(U\) is the forgetful functor \(U : \Cat{Cat} \to \Set\) and \(f \sim g\) if and only if there is some 2-cell \(\alpha : f \Rightarrow g\) that is an isomorphism.
  We call \(\para(\odot)_1\) the \textit{1-categorical truncation} of \(\para(\odot)\).
  When \(\para(\odot)\) is monoidal, so is \(\para(\odot)_1\).
\end{prop}

\begin{rmk}
  We can understand the 1-categorical truncation of $\para(\odot)$ as grouping the objects of each hom-category into their isomorphism-connected components.
\end{rmk}

\subsection{External parameterization} \label{sec:ext-para}

In a monoidal closed category, morphisms $P\otimes X\to Y$ correspond bijectively to morphisms $P\to[X,Y]$.
The fact that monoidal closed categories are enriched in themselves presents an opportunity for generalization in a different direction to the actegorical approach taken above: that is, given a category of processes $\cat{C}$ enriched in $\cat{E}$, we can think of an externally parameterized process from $X$ to $Y$ as a morphism $P\to\cat{C}(X,Y)$ in $\cat{E}$.

This notion of external parameterization can be operationally more faithful to the structure of systems of interest, even though in the case of monoidal closed categories it is equivalent.
For example, the improvement of the performance of a system of inference due to learning is often treated `externally' to the inference process itself: the learning process might proceed by observing (but not interfering with) the inference process, and updating the parameters accordingly; and, if treated dynamically, the two processes might be assumed to exhibit a separation of timescales such that the parameters are stationary on the timescale of inference.
We will make such assumptions when we formalize learning in Chapter \ref{chp:brain}, and so we will make use of external parameterization.

The definition of external parameterization is simplified by using the `slice' construction.

\begin{defn} \label{def:slice-cat}
  Suppose $X$ is an object of a category $\cat{E}$.
  We define the \textit{slice} of $\cat{E}$ over $X$, denoted $\cat{E}/X$, as the category of `bundles' over $X$ in $\cat{E}$:
  its objects are morphisms $p:A\to X$ into $X$ for any $A:\cat{E}$, which we call \textit{bundles} over $X$ and write as $(A,p)$.
  The morphisms $f:(A,p)\to(B,q)$ in $\cat{E}/X$ are morphisms $f:A\to B$ in $\cat{E}$ such that $q\circ f = p$, as in the diagram
  \[\begin{tikzcd}
    A && B \\
    & X
	  \arrow["f", from=1-1, to=1-3]
	  \arrow["p"', from=1-1, to=2-2]
	  \arrow["q", from=1-3, to=2-2]
  \end{tikzcd} \; .\]
\end{defn}

We therefore define external parameterization using slices over hom objects.

\begin{defn} \label{def:ext-para}
  Given a category $\cat{C}$ enriched in $(\cat{E},\times,1)$, we define the \textit{external parameterization} $\eP\cat{C}$ of $\cat{C}$ in $\cat{E}$ as the following bicategory.
  0-cells are the objects of $\cat{C}$, and each hom-category $\eP\cat{C}(A,B)$ is given by the slice category $\cat{E}/\cat{C}(A,B)$.
  The composition of 1-cells is by composing in $\cat{C}$ after taking the product of parameters:
  given $f:\Theta\to\cat{C}(A,B)$ and $g:\Omega\to\cat{C}(B,C)$, their composite $g\circ f$ is
  \[ g\circ f := \Omega\times\Theta \xto{g\times f} \cat{C}(B,C)\times\cat{C}(A,B) \xto{\klcirc} \cat{C}(A,C) \]
  where $\klcirc$ is the composition map for $\cat{C}$ in $\cat{E}$.
  The identity 1-cells are the points on the identity morphisms in $\cat{C}$.
  For instance, the identity 1-cell on $A$ is the corresponding point $\id_A : 1\to\cat{C}(A,A)$.
  We will denote 1-cells using our earlier notation for parameterized morphisms: for instance, $f : A\xto{\Theta}B$ and $\id_A : A\xto{1}A$.
  The horizontal composition of 2-cells is given by taking their product.
\end{defn}

\begin{rmk} \label{rmk:ext-para-enrch}
  External parameterization is alternatively obtained as the change-of-enrichment induced by the \textit{covariant self-indexing}, the functor $\cat{E}/(-):\cat{E}\to\cat{E}\mdash\Cat{Cat}$, given on objects by $X\mapsto\cat{E}/X$ and on morphisms by the functor induced by post-composition\footnote{
  Later, in Definition \ref{def:self-idx}, we will encounter the \textit{contravariant} self-indexing, which has the same action on objects but is given on morphisms by pullback.
  Whereas the covariant self-indexing is always well-defined, the contravariant self-indexing is therefore only well-defined in the more restricted situation where $\cat{E}$ has all pullbacks.}.
  A base of enrichment must \textit{a fortiori} be a monoidal category, and in this case $\cat{E}/(-)$ is a lax monoidal functor.
  A lax monoidal functor out of the base of enrichment induces a corresponding change-of-enrichment pseudofunctor\footnote{
  A \textit{pseudofunctor} is a kind of `weakened' functor, for which functoriality only needs to hold up to isomorphism; see Definition \ref{def:pseudofunctor}.},
  and $\eP$ is obtained precisely as the change-of-enrichment induced by $\cat{E}/(-)$.

  One important consequence of this is that $\eP$ defines a pseudofunctor $\eP:\cat{E}\mdash\Cat{Cat}\to(\cat{E}\mdash\Cat{Cat})\mdash\Cat{Cat}$.
  Note that we take enrichment here to mean \textit{weak} enrichment, in the sense indicated by Remark \ref{rmk:weak-enr}.
  In the case of locally small categories, where $\cat{E} = \Set$, this means that $\eP$ has the type $\Cat{Cat}\to\Cat{Bicat}$, as suggested above.
  (We will discuss the definition of $\Cat{Bicat}$ in \secref{sec:lax-func}, where we also define pseudofunctors between bicategories.)
\end{rmk}

\begin{rmk}
  In prior work, this external parameterization has been called `proxying' \parencite{Capucci2021Parameterized}.
  We prefer the more explicit name `external parameterization', reserving `proxying' for a slightly different double-categorical construction to appear in future work by the present author.
\end{rmk}

\begin{rmk}
  Both internal and external parameterization are jointly generalized by the notion of \textit{locally graded category} \parencite{Levy2019Locally}, which can be understood to mean ``presheaf-enriched category''.
  If $\cat{M}$ acts on $\cat{C}$ by $\odot$, then the hom category $\para(\odot)(A,B)$ is the category of elements of the presheaf $\cat{C}(-\odot A,B):\cat{M}\op\to\Set$.
  Similarly, the hom category $\eP\cat{C}(A,B)$ is the category of elements of the presheaf $\cat{E}\bigl(-,\cat{C}(A,B)\bigr):\cat{E}\op\to\Set$.
  We will see in \secref{sec:idx-cat} that the category of elements construction yields an equivalence between presheaves and categories-of-elements, and so we may as well consider $\para(\odot)$ to be enriched in the presheaf category $[\cat{M}\op,\Set]$ and $\eP\cat{C}$ to be enriched in $[\cat{E}\op,\Set]$.
  The phrase ``locally graded'' indicates that the `hom sets' of $\para(\odot)$ and $\eP\cat{C}$ are `graded' by the objects of $\cat{M}$ and $\cat{E}$ respectively.
  We learnt about locally graded categories from Dylan Braithwaite.
\end{rmk}

\section{Systems from circuits} \label{sec:sys-circ}

The dominant motivation for the use of monoidal categories so far has been in modelling the compositional structure of processes, on the basis of the observation that processes may generally be composed both sequentially and in parallel, and so 1-dimensional category theory alone is insufficient.
The processes for which this kind of structure is most suited are those that exhibit a flow of information.
For example, if we take the morphisms of the category $\Set$ as computable functions, then we see that the corresponding ``process theory'' is adequate for interpreting diagrams of the form of \secref{sec:mtv-func}; and we will encounter in Chapter \ref{chp:buco} a process-theoretic framework formalizing probabilistic graphical models of the kind discussed in \secref{sec:mtv-bayes}.

In these monoidal categories, processes are represented by morphisms, with composition used to connect processes together: the composite of two processes is again a process.
However, some morphisms are purely `structural', implementing the plumbing of information flow---such as copying, discarding, and swapping---and so these categories somewhat blur the boundary between syntax and semantics.
At the same time, it is strange to think of something like a neural circuit as a `process': although it might reify some process in its behaviour, it is rather a \textit{system}.

To sharpen the syntax-semantics boundary, one can show that every monoidal category arises as an algebra for a certain monad.
We will make these notions precise in \secref{sec:monoid-monad} below, and here it will suffice to provide some intuition: the monad defines the syntax, and the algebra supplies a compatible semantics.
Algebra in this sense is a vast generalization of the abstract algebra of familiar mathematics, and typically involves defining symbolic operations and rules by which they can be combined, substituted, compared, and reduced.

In this section, although we do not explicitly make use of the technology of monads, we exemplify this approach with an example of compositional connectomics: on the syntactic side, we will introduce a `multicategory' of \textit{linear circuit diagrams} which govern patterns of neural connectivity; while on the semantic side, we will equip this multicategory with a functorial algebra of rate-coded neural circuits\footnote{%
In the Appendix (\secref{sec:a-monad-operad}), we sketch the connection between this multicategorical story and the monadic one.}.
We will find that this more explicitly algebraic approach resolves the dilemma observed above between the compositional structure of processes and that of systems: algebraic syntax is in some sense about substitution, and so circuit diagrams will have `holes' into which can be substituted other circuit diagrams.
That is to say, a circuit diagram is a morphism which takes a given pattern of holes and connects them together into a single circuit, as in the following diagram, which brings us back to our first motivating example from \secref{sec:mtv-neur} and which we formalize below.
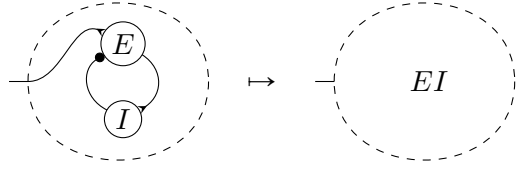
\[ \input{img/EI-network-2.tikz} \quad \mapsto \quad \input{img/EI-network-2b.tikz} \]

We will use a similar approach when we supply dynamical semantics for approximate inference, although there, for our general syntax of systems, we will use categories of polynomial functors, which we introduce in \secref{sec:poly} at the end of this chapter.
In any case, it will turn out that linear circuit diagrams embed naturally into polynomials, and so the circuits below can be understood as providing a sample of what is to come.

\subsection{Multicategorical algebra for hierarchical systems}

A multicategory is like a category, but where morphisms may have a `complex' domain, such as a list of objects \parencite{Leinster2004Higher}.
A morphism whose domain is an $n$-length list is called `$n$-ary', and we can abstractly think of such morphisms as `$n$-ary operations':  for example, we will use them to model connecting $n$ circuits together into a single system.
Because these morphisms effect a kind of `zooming-out', we can use them to construct hierarchical or `nested' systems-of-systems.

\begin{defn} \label{def:multicat}
  A \textit{multicategory} $\cat{O}$ consists of
  \begin{enumerate}
  \item a set $\cat{O}_0$ of objects;
  \item a set $\cat{O}_1$ of morphisms, equipped with
    \begin{enumerate}
    \item a codomain function $\cod : \cat{O}_1\to\cat{O}_0$, and
    \item a domain function $\dom : \cat{O}_1\to\List(\cat{O}_0)$, where $\List(\cat{O}_0)$ is the set of finite lists of objects $(o_1, \dots, o_n)$,
    \end{enumerate}
    so that each $n$\textit{-ary} morphism $f$ has a list of $n$ objects as its domain and a single object as its codomain, written $f:(o_1, \dots, o_n)\to p$;
  \item an \textit{identity} function $\id:\cat{O}_0\to\cat{O}_1$ such that $\cod(\id_o) = o$ and $\dom(\id_o) = (o)$, so that the identity on $o$ is written $\id_o : o\to o$;
  \item a family of \textit{composition} functions
    \begin{gather*}
      \circ_{p,(o_i),(o_i^j)} : \cat{O}(o_1,\dots,o_n;p) \times \cat{O}(o_1^1,\dots,o_1^{k_1};o_1) \times \cdots \times \cat{O}(o_n^1,\dots,o_n^{k_n}; o_n) \\
      \to \cat{O}(o_1^1,\dots,o_1^{k_1},\dots,o_n^1,\dots,o_n^{k_n}; p)
    \end{gather*}
    written as
    \[ (f, f_1, \dots, f_n) \mapsto f\circ(f_1,\dots,f_n) \]
    for each object $p$, $n$-ary list objects $(o_1,\dots,o_n)$, and $n$ $k_i$-ary lists of objects $(o_i^1,\dots,o_i^{k_i})$;
  \end{enumerate}
  satisfying the equations of \textit{associativity}
  \begin{gather*}
    f \circ \bigl(f_1\circ(f_1^1,\dots,f_1^{k_1}),\dots,f_n\circ(f_n^1,\dots,f_n^{k_n})\bigr) \\
    = \bigl(f\circ(f_1,\dots,f_n)\bigr)\circ(f_1^1,\dots,f_1^{k_1},\dots,f_n^1,\dots,f_n^{k_n})
  \end{gather*}
  whenever such composites make sense, and \textit{unitality}
  \[ f \circ (\id_{o_1},\dots,\id_{o_n}) = f = \id_p\circ f \]
  for every $f:(o_1, \dots, o_n)\to p$.
\end{defn}

For our purposes, the order of objects in the lists will not matter, which we formalize with the notion of \textit{symmetric} multicategory---analogous to the symmetric monoidal categories of \secref{sec:mon-cat}.

\begin{defn}
  Let $S_n$ be the symmetric group on $n$ elements.
  A \textit{symmetric multicategory} $\cat{O}$ is a multicategory $\cat{O}$ which is additionally equipped, for each $n:\nn$, with an action $\sigma_n$ of $S_n$ on the set $\cat{O}_1^n$ of $n$-ary morphisms
  \[ \sigma_n : S_n\times\cat{O}_1^n\to\cat{O}_1^n \]
  such that composition $\circ$ preserves this action.
\end{defn}

\begin{rmk}
  In other applied-category-theoretical contexts, multicategories of this kind are sometimes called \textit{operads} (\textit{cf}. \textit{e.g.} \parencite{Fong2018Seven,Yau2018Operads,Baez2017Network,Vagner2015Algebras,Spivak2015String,Lerman2016algebra,Spivak2015Nesting,Spivak2013Operad,Rupel2013Operad,Fong2019Hypergraph,Patterson2021Wiring,Shapiro2022Dynamic}).
  Traditionally, an operad is the same as a multicategory with one object\parencite{Leinster2004Higher}; sometimes therefore, multicategories are called \textit{coloured} or \textit{typed} operads\parencite{Leinster2004Higher,Fong2019Hypergraph,Baez1997Higher,Cheng2003Weak}.
  In order to avoid confusion, we will stick with `multicategory'.
\end{rmk}

Although the multicategorical intuition---of hierarchically constructing complex systems---is valuable, the following fact means that there is a close connection between multicategories and monoidal categories, for in a monoidal category, we can interpret an $n$-ary tensor $x_1\otimes\cdots\otimes x_n$ as an $n$-ary list of objects.

\begin{prop} \label{prop:multicat-smc}
  Any monoidal category $(\cat{C},\otimes,I)$ induces a corresponding multicategory $\cat{OC}$.
  The objects $\cat{OC}_0$ are the objects $\cat{C}_0$ of $\cat{C}$.
  The $n$-ary morphisms $(c_1,\dots,c_n)\to d$ are the morphisms $c_1\otimes\dots\otimes c_n\to d$; \textit{i.e.}, $\cat{OC}(c_1,\dots,c_n;d) := \cat{C}(c_1\otimes\dots\otimes c_n, d)$.
  Identities are as in $\cat{C}$, and composition is defined by $(f, f_1, \dots, f_n) \mapsto f\circ (f_1\otimes\dots\otimes f_n)$.
  When $\cat{C}$ is symmetric, so is $\cat{OC}$.
\end{prop}

\begin{ex}
  An example that will soon become important is the operad $\Cat{Sets}$ of sets and $n$-ary functions, which is obtained from the symmetric monoidal category $\Set$ by $\Cat{Sets} := \cat{O}\Set$.
\end{ex}

As we discussed above, we will consider multicategories as supplying a syntax for the composition of systems, and so actually to compose systems requires the extra data of what those systems are and how they can be composed according to the syntax.
This extra semantic data is called an `algebra' for the multicategory.

\begin{defn}
  An \textit{algebra for a multicategory} $\cat{M}$ is a \textit{multifunctor} $\cat{M}\to\Cat{Sets}$.
\end{defn}

Multifunctors are the multicategorical analogues of functors; but fortunately (even though the definition is not a hard one), we will not need to define them, owing to the following result, which relates multifunctors and lax monoidal functors.

\begin{prop}[{\textcite[{Example 4.3.3, Definition 2.1.12}]{Leinster2004Higher}}]
  If the multicategory $\cat{M}$ arises from a monoidal category $(\cat{C},\otimes,I)$ as $\cat{M} = \cat{OC}$, then an algebra for $\cat{M}$ is determined by a lax monoidal functor $(\cat{C},\otimes,I)\to(\Set,\times,1)$.
\end{prop}

\begin{rmk}
  In \secref{sec:monoid-monad}, we will encounter the concept of ``algebra for a monad'', which is perhaps the more familiar concept in mathematics and computer science.
  One might therefore wonder what the relationship between the two notions of `algebra' is: why do they both have this name?
  The answer is provided by \textcite{Leinster2004Higher}: every `shape' of multicategory corresponds to a certain monad; and every multicategory algebra corresponds to an algebra for a monad derived (in the context of the particular multicategory at hand) from this multicategory-shape monad.
  For the interested reader, we review these results in the Appendix (\secref{sec:a-monad-operad}).
  In \secref{sec:lin-circ}, we will exemplify the notion of monad algebra with the more basic result that every small category corresponds to an algebra for a certain monad.
  Monad algebras will also prove useful later in the thesis in the context of compositional probability theory.
\end{rmk}

\subsection{Linear circuit diagrams} \label{sec:lin-circ}

Let us now exhibit the multicategory formalizing circuit diagrams of the type with which we opened this section.
Although our motivation is multicategorical, for simplicity we will proceed by defining a symmetric monoidal category.
Its objects will represent the `output-input' dimensions of a circuit, written as pairs of numbers $(n_o,n_i)$, and its morphisms $(n_o,n_i)\to(m_o,m_i)$ encode how to wire a circuit with $n_o$ outputs and $n_i$ inputs together to produce a circuit of $m_o$ outputs and $m_i$ inputs: this may involve connecting some of the $n_o$ outputs to the $m_o$ outputs; or connecting some of the $m_i$ inputs, or (to allow recurrence) the $n_o$ outputs, to the $n_i$ inputs.
The definition may seem somewhat mysterious at first, but its form is owed to a more abstract structure (lenses) that we will define later, in \secref{sec:bidi}.

\begin{ex} \label{ex:lin-circ}
  We define a symmetric monoidal category $\bigl(\LinCirc,+,(0,0)\bigr)$ of \textit{linear circuit diagrams} and consider the induced multicategory $\OLinCirc$.
  The objects of $\LinCirc$ are pairs $(n_o,n_i)$ of natural numbers.
  A morphism $(n_o,n_i)\to(m_o,m_i)$ is a pair of real-valued matrices $(A,B)$ with $A$ of shape $(m_o,n_o)$ and semi-orthogonal (\textit{i.e.}, such that $AA^T = 1_{m_o}$) and $B$ of shape $(n_i,n_o+m_i)$; equivalently, $A$ is a semi-orthogonal linear map $\rr^{n_o}\to\rr^{m_o}$ and $B$ is a linear map $\rr^{n_o+m_i}\to\rr^{n_i}$.
  The identity morphism $\id_{(n_o,n_i)}$ on $(n_o,n_i)$ is the pair of matrices $(1_{n_o},01_{n_o})$
  where $01_{n_o}$ is the block matrix $\begin{pmatrix} 0_{n_o} & 1_{n_o} \end{pmatrix}$.
  Given morphisms $(A,B):(n_o,n_i)\to(m_o,m_i)$ and $(A',B'):(m_o,m_i)\to(k_o,k_i)$, their composite is the pair $(A'A,BB'_A)$ where $A'A$ is the usual matrix product and $BB'_A$ is defined as the following block matrix multiplication:
  \[ BB'_A := B
  \begin{pmatrix} 1_{n_o} & 0 \\ 0 & B' \end{pmatrix}
  \begin{pmatrix} 1_{n_o} & 0 \\ A & 0 \\ 0 & 1_{k_i} \end{pmatrix} \]
  Unitality and associativity of composition follow from those properties of matrix multiplication, and $AA'$ is easily seen to be semi-orthogonal (by $AA'(AA')^T = AA'{A'}^TA^T = AA^T = 1_{m_o}$), so $\LinCirc$ is a well-defined category.

  We now turn to the monoidal structure.
  The monoidal unit is the pair $(0,0)$; note that $\rr^o\cong 1$.
  The monoidal product $+$ is defined on objects as the pointwise sum: $(n_o,n_i)+(m_o,m_i) := (n_o+m_o,n_i+m_i)$; note that $\rr^{n_o+m_o} \cong \rr^{n_o}\times\rr^{m_o}$.
  Given morphisms $(A,B):(n_o,n_i)\to(m_o,m_i)$ and $(A',B'):(n'_o,n'_i)\to(m'_o,m'_i)$, their monoidal product $(A,B)+(A',B')$ is defined as the pair $(A\oplus A',B\oplus B'):(n_o+n'_o,n_i+n'_i)\to(m_o+m'_o,m_i+m'_i)$ with
  \[
  A\oplus A' :=
  \begin{pmatrix} A & 0 \\ 0 & A' \end{pmatrix}
  \;\quad\text{and}\quad\;
  B\oplus B' :=
  \begin{pmatrix} B & 0 \\ 0 & B' \end{pmatrix}
  \begin{pmatrix} 1_{n_o} & 0 & 0 & 0 \\ 0 & 0 & 1_{m_i} & 0 \\ 0 & 1_{n'_o} & 0 & 0 \\ 0 & 0 & 0 & 1_{m'_i} \end{pmatrix} \; .
  \]
  For each pair of objects $(n_o,n_i)$ and $(m_o,m_i)$, the symmetry $\sigma_{(n_o,n_i),(m_o,m_i)}:(n_o,n_i)+(m_o,m_i)\to(m_o,m_i)+(n_o,n_i)$ is defined as the pair of matrices $(\sigma^o_{n,m},\sigma^i_{n,m})$,
  \[
  \sigma^o_{n,m} := \begin{pmatrix} 0 & 1_{m_o} \\ 1_{n_o} & 0 \end{pmatrix}
  \;\quad\text{and}\quad\;
  \sigma^i_{n,m} := \begin{pmatrix} 0 & 0 & 0 & 1_{n_i} \\ 0 & 0 & 1_{m_i} & 0 \end{pmatrix} \; .
  \]
  That this definition produces a well-defined symmetric monoidal structure follows from more abstract considerations that we explain in Remark \ref{rmk:lincirc-poly} and Corollary \ref{cor:lens-smc}: $\LinCirc$ is a subcategory of Cartesian lenses, with the monoidal structure inherited accordingly.
\end{ex}

The category of linear circuit diagrams is a syntactic category: on its own, it does not do anything.
We need to equip it with semantics.

\subsection{An algebra of rate-coded neural circuits}

We begin by defining a notion of `rate-coded' neural circuit.

\begin{defn} \label{def:rate-circ}
  An $n_o$-dimensional \textit{rate-coded neural circuit} with $n_i$-dimensional input is an ordinary differential equation
  \[ \dot{x} = -\lambda \odot x + h\bigl(W (x\oplus i); \alpha,\beta,\gamma\bigr) \]
  where $x,\lambda,\alpha,\beta,\gamma$ are real vectors of dimension $n_o$, $i$ a real vector of dimension $n_i$, $W$ a real matrix of shape $(n_o,n_o+n_i)$, $\odot$ elementwise multiplication, $\oplus$ the direct sum (so that $x\oplus i$ is the concatenation $\begin{pmatrix} x \\ i \end{pmatrix}$), and $h$ the logistic function
  \[ h(x; \alpha,\beta,\gamma) = \frac{\gamma}{1 + \exp\bigl(-\beta(x - \alpha)\bigr)} \]
  applied elementwise.
  We summarize the data of such a circuit as the tuple $(\lambda,\alpha,\beta,\gamma,W)$.
\end{defn}

\begin{rmk}
  Rate-coded neural circuits are a coarse phenomenological model of neural dynamics.
  The state variable $x$ represents the \textit{firing rates} of an ensemble of neurons, either averaged over time or over subpopulations.
  Neural activity is of course not so simple: neurons communicate by the transmission of discrete `action potentials' along their axons.
  The emission of an action potential is governed by the electrical potential of its cellular membrane: if this potential crosses a threshold, then the neuron `fires' an action potential down its axon.
  The axon crosses the dendrites of other neurons at junctions called synapses, which modulate and transmit the activity accordingly: it is these afferent signals which in large part determine the neurons' membrane potentials.

  There are of course detailed physiological models of this process (\textit{cf. e.g.} \parencite{Hodgkin1952quantitative,Mascagni1989Numerical,Dayan2001Theoretical,Saudargiene2004How}), as well as many models which aim to capture its statistics and phenomenology in a more explicitly computational setting (\textit{cf. e.g.} \parencite{Gerstner2008Spike,Jolivet2003Spike,Izhikevich2000Neural,Ostojic2011Spiking,Deneve2012Making,Deneve2016Efficient,Schaffer2013Complex,Vreeswijk1996Chaos}), but in some situations, one can simply model neural firing as an inhomogeneous Poisson process: in this case the variable $x$ encodes the rate parameters of the processes.
  We expect there to be functorial connections between the different classes of models: in particular, we expect adjoint functors between certain spike-emission models and firing rate models of the class defined above; and in the specific case of `efficient balanced' networks\parencite{Boerlin2013Predictive,Deneve2016Efficient}, the relationships are expected to be quite simple.
  Nonetheless, we leave the exploration of such functors to future work.

  The parameters of a rate-coded neural circuit---the terms $\lambda,\alpha,\beta,\gamma,W$--- have a neurological interpretation, even though this dynamical model is not physiologically faithful.
  The term $\lambda$ represents the `leak' of voltage from the neuron's membrane, which has the effect of determining the timescale of its memory or signal-sensitivity (effectively, the voltage leak entails a process of filtering).
  The term $\alpha$ represents an abstraction of the neuron's firing threshold, and the term $\beta$ its sensitivity (\textit{i.e.}, how much its firing rate increases with incoming signals); the term $\gamma$ determines the maximum firing rate of the neuron (and is typically normalized to $1$).
  Finally, the matrix $W$ records the strengths of the synaptic connections within the circuit: positive coefficients represent excitatory connections, while negative coefficients represent inhibitory connections.
\end{rmk}

Rate-coded neural circuits can be organized into complex `hierarchical' systems using linear circuit diagrams: the linear connectivity of the diagrams is used to define the synaptic connection matrix of the complex, algebraically.
The proof that the following construction does actually constitute an algebra ensures that composing systems from circuits using diagrams is predictably well-behaved, as we will subsequently exemplify.

\begin{prop}[Algebra of rate-coded neural circuits] \label{prop:rate-circ-alg}
  There is a $\LinCirc$-algebra $(R,\mu,\epsilon):\bigl(\LinCirc,+,(0,0)\bigr)\to(\Set,\times,1)$ of rate-coded neural circuits.
  On objects $(n_o,n_i)$, define $R(n_o,n_i)$ to be the set of $n_o$-dimensional rate-coded neural circuits with $n_i$-dimensional input.
  Then, given a linear circuit diagram $(A,B):(n_o,n_i)\to(m_o,m_i)$, define a function
  \begin{align*}
    R(A,B):R(n_o,n_i)&\to R(m_o,m_i) \\
    (\lambda,\alpha,\beta,\gamma,W) &\mapsto (A\lambda,A\alpha,A\beta,A\gamma,W_{AB})
  \end{align*}
  where $W_{AB}$ is the following block matrix product:
  \[ W_{AB} := AW
  \begin{pmatrix} 1_{n_o} & 0 \\ 0 & B \end{pmatrix}
  \begin{pmatrix} 1_{n_o} & 0 \\ 1_{n_o} & 0 \\ 0 & 1_{m_i} \end{pmatrix}
  \begin{pmatrix} A^T & 0 \\ 0 & 1_{m_i} \end{pmatrix} \; . \]
  The laxator $\mu$ is defined componentwise as the family of functions
  \[ \mu_{(n_o,n_i),(m_o,m_i)} : R(n_o,n_i)\times R(m_o,m_i)\to R\bigl((n_o,n_i)+(m_o,m_i)\bigr) \]
  taking a pair of circuits $(\lambda,\alpha,\beta,\gamma,W):R(n_o,n_i)$ and $(\lambda',\alpha',\beta',\gamma',W'):R(m_o,m_i)$
  to the circuit $(\lambda\oplus\lambda',\alpha\oplus\alpha',\beta\oplus\beta',\gamma\oplus\gamma',WW')$ where $x\oplus y$ is again the direct sum $\begin{pmatrix} x \\ y \end{pmatrix}$ and where the matrix $WW'$ is defined as
  \[ WW' :=
  \begin{pmatrix} W & 0 \\ 0 & W' \end{pmatrix}
  \begin{pmatrix} 1_{n_o} & 0 & 0 & 0 \\ 0 & 0 & 1_{n_i} & 0 \\ 0 & 1_{m_o} & 0 & 0 \\ 0 & 0 & 0 & 1_{m_i} \end{pmatrix} \; .
  \]
  The unitor $\epsilon$ is the isomorphism $\epsilon : 1\xto{\sim}R(0,0)$.
  \begin{proof}
    We need to check that $R$ is a lax monoidal functor, and begin by verifying functoriality.
    So suppose $(A',B')$ is a linear circuit diagram $(m_o,m_i)\to(k_o,k_i)$.
    On the terms $\lambda,\alpha,\beta,\gamma$, the functoriality of $R$ is immediate from matrix multiplication, so we concentrate on the action of $R$ on $W$.
    We need to show that $R\bigl((A',B')\circ(A,B)\bigr)(W) = R(A',B')\circ R(A,B)(W)$, where $R(A,B)(W) = W_{AB}$ as defined above.
    Note that we can alternatively write $W_{AB}$ as the following composite linear map
    \[ m_o + m_i \xto{A^T + m_i} n_o + m_i \xto{\bcopier + m_i} n_o + n_o + m_i \xto{n_o + B} n_o + n_i \xto{W} n_o \xto{A} m_o \; . \]
    We can therefore write $R(A',B')(W_{AB})$ as
    \begin{align*}
      & k_o + k_i \xto{{A'}^T + k_i} m_o + k_i \xto{A^T + k_i} n_o + k_i \xto{\bcopier + k_i} n_o + n_o + k_i \; \cdots \\
      & \cdots \; \xto{n_o + \bcopier + k_i} n_o + n_o + n_o + k_i \xto{n_o + n_o + A + k_i} n_o + n_o + m_o + k_i \; \cdots \\
      & \cdots \; \xto{n_o + n_o + B'} n_o + n_o + m_i \xto{n_o + B} n_o + n_i \xto{W} n_o \xto{A} m_o \xto{A'} k_o
    \end{align*}
    and $R\bigl((A',B')\circ(A,B)\bigr)(W)$ as
    \begin{align*}
      & k_o + k_i \xto{{A'}^T + k_i} m_o + k_i \xto{\bcopier + k_i} m_o + m_o + k_i \xto{m_o + B'} m_o + m_i \xto{A^T + m_i} \; \cdots \\
      & \cdots \; n_o + m_i \xto{\bcopier + m_i} n_o + n_o + m_i \xto{n_o + B} n_o + n_i \xto{W} n_o \xto{A} m_o \xto{A'} k_o
    \end{align*}
    so it suffices to check that
    \begin{gather*}
      \begin{aligned}
        & m_o + k_i \xto{A^T + k_i} n_o + k_i \xto{\bcopier + k_i} n_o + n_o + k_i \xto{n_o + \bcopier + k_i} n_o + n_o + n_o + k_i \; \cdots \\
        & \cdots \; \xto{n_o + n_o + A + k_i} n_o + n_o + m_o + k_i \xto{n_o + n_o + B'} n_o + n_o + m_i
      \end{aligned}
      \\ = \\
      m_o + k_i \xto{\bcopier + k_i} m_o + m_o + k_i \xto{m_o + B'} m_o + m_i \xto{A^T + m_i} n_o + m_i \xto{\bcopier + m_i} n_o + n_o + m_i
    \end{gather*}
    which we can do using the graphical calculus:
    \begin{gather*}
      \input{img/rate-func-1.tikz} \quad \overset{\mathclap{\text{(1)}}}= \quad
      \input{img/rate-func-2.tikz} \quad \cdots \\ \cdots \quad \overset{\mathclap{\text{(2)}}}= \quad
      \input{img/rate-func-3.tikz} \quad \overset{\mathclap{\text{(3)}}}= \quad
      \input{img/rate-func-4.tikz} \quad \cdots \\ \cdots \quad \overset{\mathclap{\text{(4)}}}= \quad
      \input{img/rate-func-5.tikz} \quad \overset{\mathclap{\text{(5)}}}= \quad
      \input{img/rate-func-6.tikz}
    \end{gather*}
    where the equality (1) holds because $A^T$ is a comonoid morphism (Definition \ref{def:comon-morph})\footnote{This in turn because $\oplus$ is the Cartesian product, and so every morphism is a $\oplus$-comonoid morphism.}, (2) likewise, (3) because $A$ is semi-orthogonal, (4) by the coassociativity of copying, and (5) again because $A^T$ is a comonoid morphism.
    Finally, we observe that the last string diagram depicts the linear map
    \[ m_o + k_i \xto{\bcopier + k_i} m_o + m_o + k_i \xto{A^T + B'} n_o + m_i \xto{\bcopier + m_i} n_o + n_o + m_i \]
    which equals the required map
    \[ m_o + k_i \xto{\bcopier + k_i} m_o + m_o + k_i \xto{m_o + B'} m_o + m_i \xto{A^T + m_i} n_o + m_i \xto{\bcopier + m_i} n_o + n_o + m_i \]
    by the unitality of composition.
    This establishes that $R$ preserves composites; it remains to check that it preserves identities.
    Once again, this follows immediately on the terms $\lambda,\alpha,\beta,\gamma$, so we concentrate on the action on $W$.
    We have
    \[ R(1_{n_o},01_{n_o})(W) =
    1_{n_o}W
    \begin{pmatrix} 1_{n_o} & 0 & 0 \\ 0 & 0 & 1_{n_o} \end{pmatrix}
    \begin{pmatrix} 1_{n_o} & 0 \\ 1_{n_o} & 0 \\ 0 & 1_{m_i} \end{pmatrix}
    \begin{pmatrix} 1_{n_o} & 0 \\ 0 & 1_{m_i} \end{pmatrix} \]
    which is easily seen to be equal to $W$ itself.
    Therefore $R$ defines a functor.

    We now need to verify that the unitor and laxator satisfy the unitality and associativity axioms of a lax monoidal functor.
    We begin by checking associativity, so suppose that we are given three circuits: $(\lambda,\alpha,\beta,\gamma,W):R(n_o,n_i)$, and $(\lambda',\alpha',\beta',\gamma',W'):R(m_o,m_i)$, and $(\lambda'',\alpha'',\beta'',\gamma'',W''):R(k_o,k_i)$.
    Associativity on all the terms but $W,W',W''$ follows from the associativity of the direct sum $\oplus$, and so we just need to check that $\mu(W,\mu(W',W'')) = \mu(\mu(W,W'),W'')$ where $\mu(W,W') = WW'$ and $\mu(W',W'') = W'W''$, according to the definition above.
    Once more, we use the graphical calculus.
    Observe that we can depict $WW'$ and $W'W''$ as
    \[ \input{img/neur-WW-nm.tikz} \qquad\text{and}\qquad \input{img/neur-WW-mk.tikz} \]
    respectively.
    Hence $\mu(W,\mu(W',W''))$ satisfies the equality
    \[ \mu(W,\mu(W',W'')) \quad = \quad \input{img/neur-WWW-nmk-1.tikz} \qquad = \qquad \input{img/neur-WWW-nmk-2.tikz} \]
    and likewise $\mu(\mu(W,W'),W'')$ satisfies
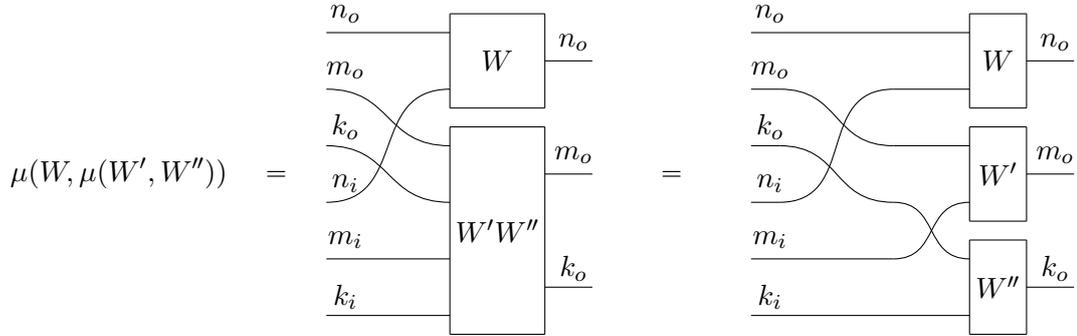
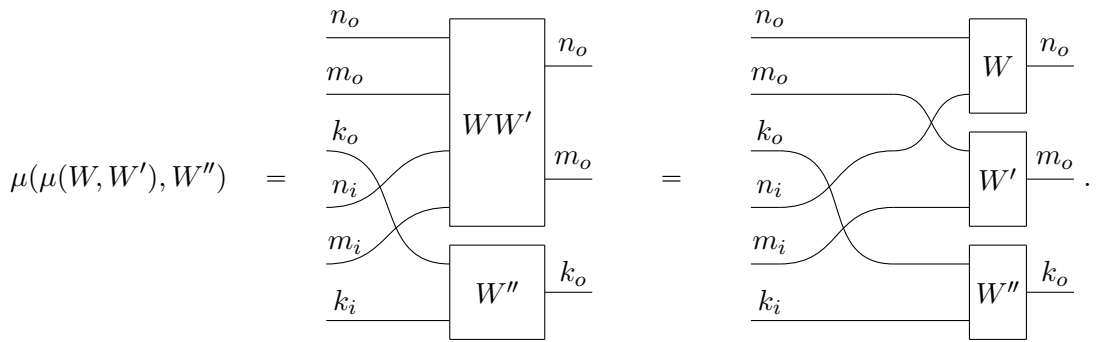
    \[ \mu(\mu(W,W'),W'') \quad = \quad \input{img/neur-WWW-nmk-3.tikz} \qquad = \qquad \input{img/neur-WWW-nmk-4.tikz} \; . \]
    The two diagrams on the right hand side are equal up to a topological deformation, and so the depicted morphisms are equal by the coherence theorem for monoidal categories.
    This establishes the associativity of the laxator.
    It remains to establish unitality: but this follows immediately, because $R(0,0)\cong\rr^0$ and the 0-dimensional space is the unit for the direct sum $\oplus$.
    Hence $(R,\mu,\epsilon)$ is a lax monoidal functor, and hence an algebra for $\bigl(\LinCirc,+,(0,0)\bigr)$.
  \end{proof}
\end{prop}

\begin{rmk}
  At points in the preceding proof, we used the fact that a linear map is a \textit{comonoid morphism}, which implies that it commutes with copying.
  We will define the notion of comonoid morphism in \secref{sec:comon} below; meanwhile, the fact that $A^T$ is one follows from the fact that $\oplus$ is the categorical product of vector spaces, and so every linear map is a $\oplus$-comonoid morphism.
\end{rmk}

\begin{rmk}
  Let us return briefly to the distinction made at the beginning of this section between processes and systems, and their corresponding categorical incarnations.
  One might be tempted to try constructing a symmetric monoidal category of neural circuits using this algebra whose objects would be natural numbers and whose morphisms $i\to o$ would be circuits in $R(o,i)$, treated thus as `processes'.
  But this won't work, because there is no neural circuit that will function as an identity morphism!
  Later in the thesis (\secref{sec:mon-bicat-coalg}), we will see one way around this problem, building monoidal categories of hierarchical dynamical systems that are in some sense analogous to these circuits (while being more general): there, we will use the rich structure of polynomial functors to define both the syntax of composition as well as the hom categories (for our construction will be bicategorical) of dynamical systems, and the extra generality will mean we will have identity systems (that compose appropriately unitally).
  Until then, we note that the moral of this observation might be that it affirms that the composition of neural circuits is multicategory-algebraic (formalizing a notion of hierarchy), rather than merely categorical.
\end{rmk}

The weight matrices resulting from the linear circuit algebra encode the pattern of connectivity specified by the diagram, as we now exemplify.

\begin{ex} \label{ex:EI-circ-1}
  Let us consider the circuit example from the beginning of this section, the wiring of an inhibitory circuit to an excitatory circuit, as in the diagram
  \[ \input{img/EI-network-2.tikz} \quad \mapsto \quad \input{img/EI-network-2b.tikz} \]
  which depicts a linear circuit diagram $E+I\to EI$.
  In such a diagram, the input dimension of an object (such as $E$) must have dimension equalling the sum of the dimensions of the incoming wires.
  Dually, the dimension along a wire emanating from an object must have dimension equal to the output dimension of that object.
  To distinguish the source and target of a wire, we decorate the target ends: a filled circle denotes an inhibitory connection, interpreted in the linear circuit as the negative identity matrix $-1$ of the appropriate dimension; and an inverted arrowhead denotes an excitatory connection, interpreted as the positive identity $1$ of the appropriate dimension.
  We will write the dimensions of the object $E$ as $(o_E,i_E)$, of $I$ as $(o_I,i_I)$, and of $EI$ as $(o_{EI},i_{EI})$.
  Therefore, in this example, the following equalities must hold: $i_E = o_I + i_{EI}$; $i_I = o_E$; and $o_{EI} = o_E + o_I$.
  The last equation holds because the circuit $EI$ is formed from the sum of the circuits $E$ and $I$.

  To give a circuit diagram $(A,B):(o_E,i_E)+(o_I,i_I)\to(o_{EI},i_{EI})$ is to give a semi-orthogonal real matrix $A$ of shape $(o_{EI},o_E+o_I)$ and a real matrix $B$ of shape $(i_E+i_I,o_E+o_I+i_{EI})$.
  Using the preceding equalities, these are equivalently shaped as $(o_E+o_I,o_E+o_I)$ and $(o_I+i_{EI}+o_E,o_E+o_I+i_{EI})$, and we just choose the identity matrix $1_{o_E+o_I}$ for $A$.
  To define $B$, we read it off from the diagram as
  \[ B := \begin{pmatrix} 0 & -1_{o_I} & 0 \\ 0 & 0 & 1_{i_{EI}} \\ 1_{o_E} & 0 & 0 \end{pmatrix} \; . \]
  Now suppose $(\lambda_E,\alpha,_E,\beta_E,\gamma_E,W_E)$ and $(\lambda_I,\alpha,_I,\beta_I,\gamma_I,W_I)$ are two rate-coded neural circuits, the former of type $R(o_E,i_E)$ and the latter of type $R(o_I,i_I)$.
  How does $R(A,B)$ act upon them to give our composite circuit?

  On all the parameters but the weight matrices, $R(A,B)$ acts trivially (since $A$ is just the identity matrix), and so we will concentrate on the action on $W_E,W_I$.
  Firstly, we need to form the monoidal product of the weight matrices, $\mu(W_E,W_I)$, which is defined by
  \begin{align*}
    \mu(W_E,W_I) &= \begin{pmatrix} W_E & 0 \\ 0 & W_I \end{pmatrix}
    \begin{pmatrix} 1_{o_E} & 0 & 0 & 0 \\ 0 & 0 & 1_{i_E} & 0 \\ 0 & 1_{o_I} & 0 & 0 \\ 0 & 0 & 0 & 1_{i_I} \end{pmatrix} \\
    &= \begin{pmatrix} W_E & 0 \\ 0 & W_I \end{pmatrix}
    \begin{pmatrix} 1_{o_E} & 0 & 0 & 0 & 0 \\ 0 & 0 & 1_{o_I} & 0 & 0 \\ 0 & 0 & 0 & 1_{i_{EI}} & 0 \\ 0 & 1_{o_I} & 0 & 0 & 0 \\ 0 & 0 & 0 & 0 & 1_{o_E} \end{pmatrix}
  \end{align*}
  where the second equality holds by applying the equalities between the dimensions defined above.
  The weight matrix $R(A,B)(\mu(W_E,W_I))$ is then defined as
  \[ A\mu(W_E,W_I)
  \begin{pmatrix} 1_{o_{E}+o_{I}} & 0 \\ 0 & B \end{pmatrix}
  \begin{pmatrix} 1_{o_{E}+o_{I}} & 0 \\ 1_{o_{E}+o_{I}} & 0 \\ 0 & 1_{i_{EI}} \end{pmatrix}
  \begin{pmatrix} A^T & 0 \\ 0 & 1_{i_{EI}} \end{pmatrix} \; . \]
  Since $A = 1_{o_E+o_I}$, and by substituting the definition of $\mu(W_E,W_I)$, this reduces to
  \[ \begin{pmatrix} W_E & 0 \\ 0 & W_I \end{pmatrix}
  \begin{pmatrix} 1_{o_E} & 0 & 0 & 0 & 0 \\ 0 & 0 & 1_{o_I} & 0 & 0 \\ 0 & 0 & 0 & 1_{i_{EI}} & 0 \\ 0 & 1_{o_I} & 0 & 0 & 0 \\ 0 & 0 & 0 & 0 & 1_{o_E} \end{pmatrix}
  \begin{pmatrix} 1_{o_{E}+o_{I}} & 0 \\ 0 & B \end{pmatrix}
  \begin{pmatrix} 1_{o_{E}+o_{I}} & 0 \\ 1_{o_{E}+o_{I}} & 0 \\ 0 & 1_{i_{EI}} \end{pmatrix} \; . \]
  Then, by substitution and matrix multiplication, we have the following equalities:
  \begin{align*}
    & \begin{pmatrix} 1_{o_E} & 0 & 0 & 0 & 0 \\ 0 & 0 & 1_{o_I} & 0 & 0 \\ 0 & 0 & 0 & 1_{i_{EI}} & 0 \\ 0 & 1_{o_I} & 0 & 0 & 0 \\ 0 & 0 & 0 & 0 & 1_{o_E} \end{pmatrix}
    \begin{pmatrix} 1_{o_{E}+o_{I}} & 0 \\ 0 & B \end{pmatrix}
    \begin{pmatrix} 1_{o_{E}+o_{I}} & 0 \\ 1_{o_{E}+o_{I}} & 0 \\ 0 & 1_{i_{EI}} \end{pmatrix} \\
    &= \begin{pmatrix} 1_{o_E} & 0 & 0 & 0 & 0 \\ 0 & 0 & 1_{o_I} & 0 & 0 \\ 0 & 0 & 0 & 1_{i_{EI}} & 0 \\ 0 & 1_{o_I} & 0 & 0 & 0 \\ 0 & 0 & 0 & 0 & 1_{o_E} \end{pmatrix}
    \begin{pmatrix} 1_{o_E} & 0 & 0 & 0 & 0 \\ 0 & 1_{o_I} & 0 & 0 & 0 \\ 0 & 0 & 0 & -1_{o_I} & 0 \\ 0 & 0 & 0 & 0 & 1_{i_{EI}} \\ 0 & 0 & 1_{o_E} & 0 & 0 \end{pmatrix}
    \begin{pmatrix} 1_{o_E} & 0 & 0 \\ 0 & 1_{o_I} & 0 \\ 1_{o_E} & 0 & 0 \\ 0 & 1_{o_I} & 0 \\ 0 & 0 & 1_{i_{EI}} \end{pmatrix} \\
    &= \begin{pmatrix} 1_{o_E} & 0 & 0 & 0 & 0 \\ 0 & 0 & 1_{o_I} & 0 & 0 \\ 0 & 0 & 0 & 1_{i_{EI}} & 0 \\ 0 & 1_{o_I} & 0 & 0 & 0 \\ 0 & 0 & 0 & 0 & 1_{o_E} \end{pmatrix}
    \begin{pmatrix} 1_{o_I} & 0 & 0 \\ 0 & 1_{o_I} & 0 \\ 0 & -1_{o_I} & 0 \\ 0 & 0 & 1_{i_{EI}} \\ 1_{o_E} & 0 & 0 \end{pmatrix} \\
    &= \begin{pmatrix} 1_{o_E} & 0 & 0 \\ 0 & -1_{o_I} & 0 \\ 0 & 0 & 1_{i_{EI}} \\ 0 & 1_{o_I} & 0 \\ 1_{o_E} & 0 & 0 \end{pmatrix}
  \end{align*}
  so that the resulting weight matrix $R(A,B)(\mu(W_E,W_I))$ is
  \[ \begin{pmatrix} W_E & 0 \\ 0 & W_I \end{pmatrix}
  \begin{pmatrix} 1_{o_E} & 0 & 0 \\ 0 & -1_{o_I} & 0 \\ 0 & 0 & 1_{i_{EI}} \\ 0 & 1_{o_I} & 0 \\ 1_{o_E} & 0 & 0 \end{pmatrix} \; . \]
  Reading off this weight matrix, we see that the $E$ neurons receive external input plus recurrent excitatory input from themselves as well as inhibitory input from $I$, and that the $I$ neurons receive only recurrent excitatory input plus excitatory input from $E$.
  This is exactly as it should be, given the diagram: by formalizing these computations, we render them mechanical (and hence computer-implementable).
  In particular, we can treat the resulting $EI$ circuit as a ``black box'' and substitute it into other diagrams to construct still larger-scale systems.
\end{ex}

Since linear circuit diagrams allow for any linear pattern of connectivity, we can of course generalize the picture above to allow for more subtle interconnections.

\begin{ex}
  Suppose that instead of incorporating only excitatory or inhibitory connections, we sought something a little more complex, as in the following circuit diagram:
  \[ \input{img/EI-network-3.tikz} \quad \mapsto \quad \input{img/EI-network-3b.tikz} \]
  Now, we have decorated the wires with fleches, to indicate the flow of activity; and besides the circular boxes (representing circuits), we have incorporated square boxes (representing linear patterns of interconnection).
  Using the same notation for the dimensions of the circuits $E$,$I$ and $EI$ as in Example \ref{ex:EI-circ-1}, this means that the boxes square boxes represent matrices $C$ of shape $(i_{E}+i_{I},n+i_{EI})$ and $D$ of shape $(n,o_E+o_I)$, where $n$ is the dimension of the $D$-$C$ wire.
  If we again write $(A,B)$ for the implied circuit diagram, and we can again set $A$ to be the identity matrix, and read $B$ from the diagram as the composite matrix
  \[ B := o_E + o_I + i_{EI} \xto{D \oplus 1_{i_{EI}}} n + i_{EI} \xto{C} i_E + i_I \; . \]
  The rest of the calculation follows mechanically, just as before.
\end{ex}

One feature missing from the construction in this section is synaptic plasticity: although we have shown how to compose circuits into systems, it is only the neural firing rates that are dynamical; the connection matrices remain fixed.
In the preceding section, we motivated the introduction of parameterized categories by their application to learning problems, and indeed one could factorize the linear circuit algebra above by extracting the connection matrices into parameters; if one wanted to retain a choice of initial weight matrix, this could also be incorporated into a `pointed' version of the structure.

This parameterized construction would be bicategorical, and so a faithful semantics for it would no longer land in $\Set$, but rather in $\Cat{Cat}$: we would have categories of circuits related by reparameterizations of the weight matrices, and with the dynamics also incorporating plasticity\footnote{%
An even more faithful dynamical semantics would land in ``bundle dynamical systems'', of the form that we introduce in Chapter \ref{chp:coalg}: two two levels of the bundle would witness the dynamics of the firing activity and the plasticity, and the bundles themselves would witness the timescale separation.}.

With a sufficiently sophisticated algebra, it would even be possible to allow the circuit diagrams themselves to be dynamical and subject to learning.
We will not pursue this line of enquiry further here, but we will return to it when we introduce plasticity into approximate inference doctrines: there, our structures will be sufficiently supple to incorporate all of the concepts sketched here.

\section{From monoids to monads} \label{sec:monoid-monad}

In order to reach the level of suppleness required by plastic dynamical approximate inference, it will help to understand the structures underlying the definitions and constructions introduced so far in this chapter---in particular, we will need a firm grasp of the concepts of monad and comonoid---and so at this point we return to an expository mode.

The fundamental concept underlying many of the structures we have seen so far is the \textit{monoid}: an object equipped with two operations, one binary and one `nullary', with the latter acting as a `unit' for the former, and although the major operation is only binary, it can be chained in order to form $n$-ary operations.
For this reason, monoids are fundamental to abstract algebra: categories themselves are ``monoids with many objects'' (in the same way that a multicategory is an operad with many objects).
Both monads and comonoids can be defined using monoids.

Even though monoids are fundamental and intimately familiar to mathematicians and computer scientists, they remain underappreciated in computational and cognitive neuroscience.
For this reason, we once again take a fairly pedagogical approach in this section.

\begin{defn} \label{def:monoid}
  Suppose $(\cat{C},\otimes,I)$ is a monoidal category.
  A \textit{monoid object} in $\cat{C}$ is an object $m$ equipped with a \textit{multiplication} morphism $\mu : m\otimes m\to m$ and a \textit{unit} morphism $\eta : I\to m$, satisfying the axioms of (left and right) unitality
  \[ \input{img/monoid-unitality.tikz} \]
  and associativity
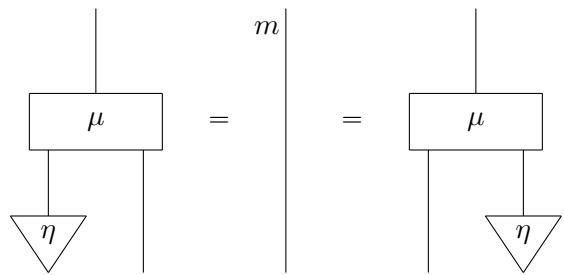
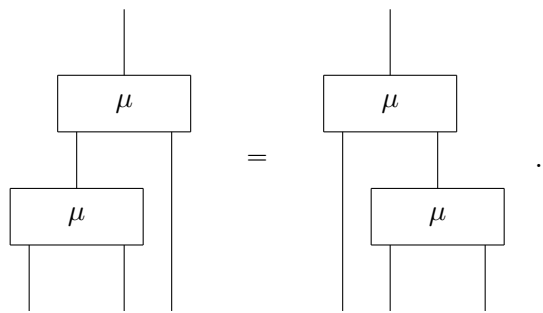
  \[ \input{img/monoid-associativity.tikz} \quad . \]
  If $\cat{C}$ is symmetric monoidal then we say that the monoid $(m,\mu,\eta)$ is \textit{commutative} if $\mu$ commutes with the symmetry as in
  \[ \input{img/monoid-commutativity.tikz} \quad . \]
\end{defn}

Since we are doing category theory, it is important to understand morphisms of monoids.

\begin{defn} \label{def:monoid-morph}
  Suppose $(m,\mu,\eta)$ and $(m',\mu',\eta')$ are monoids in $(\cat{C},\otimes,I)$.
  A \textit{monoid morphism} $(m,\mu,\eta)\to(m',\mu',\eta')$ is a morphism $f:m\to m'$ in $\cat{C}$ that is compatible with the monoidal structures, \textit{i.e.} by satisfying the axioms
  \[
  \input{img/monoid-morph-unit.tikz}
  \quad\quad\text{and}\quad\quad
  \input{img/monoid-morph-assoc.tikz}
  \quad .
  \]
  Monoids and their morphisms in $\cat{C}$ constitute a category $\Mon(\cat{C})$; composition and identities are as in $\cat{C}$, and it is easy to check that the composite of two monoid morphisms is again a monoid morphism.

  If $\cat{C}$ is symmetric monoidal, then there is a subcategory $\CMon(\cat{C})\hookrightarrow\Mon(\cat{C})$ of commutative monoids and their morphisms.

  In the names $\Mon(\cat{C})$ and $\CMon(\cat{C})$, we leave the monoidal structure implicit; should it be necessary to be explicit, we write $\Mon_{\otimes}(\cat{C})$ and $\CMon_{\otimes}(\cat{C})$.
\end{defn}

Let us consider some first examples of monoids in monoidal categories.

\begin{ex}
  The natural numbers $\nn$ equipped with addition $+:\nn\times\nn\to\nn$ and zero $0$ constitute a monoid in $\Set$. (In fact, $(\nn,+,0)$ is the \textit{free monoid} generated by a single element.)
\end{ex}

\begin{ex} \label{ex:list-monoid}
  If $A$ is a set, then there is a monoid $\bigl(\List(A),\circ,()\bigr)$ of \textit{lists of elements} of $A$: the elements of the set $\List(A)$ are finite lists $(a,b,\dots)$ of elements of $A$; the multiplication $\circ : \List(A)\times\List(A)\to\List(A)$ is given by concatenation of lists $(b_1,\dots)\circ(a_1,\dots) = (a_1,\dots,b_1,\dots)$; and the unit $1\to\List(A)$ is given by then empty list $()$. We saw in the proof of Proposition \ref{prop:free-cat} that list concatenation is associative and unital.
\end{ex}

\begin{ex}
  A monoid $(m,\circ,\ast)$ in $\Set$ is a category with a single object, denoted $\ast$.
  We already saw an example of this, in Example \ref{ex:nat-num-cat}: the monoid $(\nn,+,0)$, treated as a category.
  More generally, a monoid in a monoidal category $(\cat{C},\otimes,I)$ is a $\cat{C}$-enriched category with a single object.
\end{ex}

\begin{ex} \label{ex:strict-mon-cat}
  A monoid $(\cat{C},\otimes,I)$ in the monoidal category $(\Cat{Cat},\times,\Cat{1})$ of categories and functors is a strict monoidal category: the tensor is the monoid multiplication, and its unit is the monoid unit.
  In fact, this explains the name ``monoidal category'': a (strict) monoidal category is a monoid object in $\Cat{Cat}$.
\end{ex}

\begin{rmk} \label{rmk:pseudomonoid}
  Non-strict monoidal categories are `weak' in the same sense that bicategories are weak 2-categories; after all, a monoidal category is a one-object bicategory.
  In this way, we can also weaken the notion of monoid object in a bicategory, so that the axioms of unitality and associativity only hold up to `coherent isomorphism': that is, up to isomorphisms that cohere with the weak unitality and associativity of the ambient bicategory.
  Such weak monoid objects are called \textit{pseudomonoids}\footnote{%
  One often uses the prefix `pseudo-' in category theory to denote a weak structure.},
  and when interpreted in the monoidal 2-category $(\Cat{Cat},\times,\Cat{1})$ their formal definition\parencite[\S3]{Day1997Monoidal} yields exactly the non-strict monoidal categories.

  But note that to make sense in general of the notion of pseudomonoid, we first need to have a notion of monoidal bicategory.
  Abstractly, such a thing should be a one-object tricategory, but this often doesn't help: in those cases, we need something more concrete.
  Informally, then, a monoidal bicategory is a bicategory equipped with a monoidal structure that is coherent with the 2-cells, but as we have begun to see here, to specify all this coherence data quickly becomes quite verbose, and to prove their satisfaction by any given structure quite arduous, so we will only make use informally in this thesis of the notions of monoidal bicategory and pseudomonoid --- and when we do, it will be by reference to the familiar structures on and in $\Cat{Cat}$: its Cartesian monoidal structure; and (non-strict) monoidal categories.

  Finally, we note that the general phenomenon, of which we observe an instance here, wherein algebraic structures (such as monoids) may be defined internally to categories equipped with higher-dimensional analogues of that same structure is known as the \textit{microcosm principle}\parencite{Baez1997Higher}.
\end{rmk}

In Example \ref{ex:endofunc-monoid}, we saw that categories of endofunctors are strict monoidal categories.
Following Example \ref{ex:strict-mon-cat}, this means that endofunctor categories are equivalently monoid objects.
In fact, since categories are monoids with many objects\footnote{%
This pattern---of extending structures to ``many objects''---is sometimes called \textit{horizontal categorification}, to distinguish it from the `vertical' categorification of adding an extra dimension of morphism.},
this means we can consider any object of endomorphisms as an appropriately typed monoid object.

\begin{ex} \label{cor:endomorph-monoid}
  If $c:\cat{C}$ is any object in any category $\cat{C}$, then the hom-set $\cat{C}(c,c)$ is a monoid $\bigl(\cat{C}(c,c),\circ,\id_c\bigr)$ in $\Set$.
  More generally, if $\cat{C}$ is enriched in $\cat{E}$, then $\bigl(\cat{C}(c,c),\circ,\id_c\bigr)$ is a monoid in $\cat{E}$.
  In each case, we call the monoid the \textit{endomorphism monoid} on $c$.
\end{ex}

In the case when the endomorphism objects are categories, as in the case of Example \ref{ex:endofunc-monoid}, the monoidal structure makes them into monoidal categories, and so we can consider monoids objects defined internally to them.
More generally, we can do this inside any bicategory, and the resulting monoids will play an important rôle subsequently.

\begin{rmk} \label{rmk:monoid-bicat}
  Just as a monoidal category is a bicategory with a single object, the hom-category $\cat{B}(b,b)$ for any 0-cell $b$ in a bicategory $\cat{B}$ is a monoidal category: the objects are the 1-cells $b\to b$, the morphisms are the 2-cells between them, composed vertically; the tensor is horizontal composition of 1-cells, and its unit is the identity 1-cell $\id_b$.
  We can therefore define a \textit{monoid in a bicategory} $\cat{B}$ to be a monoid in $\cat{B}(b,b)$ for some 0-cell $b:\cat{B}$, using this induced monoidal structure.
\end{rmk}

Since $\Cat{Cat}$ is morally a 2-category (and \textit{a fortiori} a bicategory), and thus to avoid confusion with monoid objects in $(\Cat{Cat},\times,1)$ (\textit{i.e.} strict monoidal categories), we will introduce a new term for monoids in the bicategory $\Cat{Cat}$.

\begin{defn}
  A \textit{monad} on the category $\cat{C}$ is a monoid object in the strict monoidal category $(\cat{C}^{\cat{C}},\circ,\id_{\cat{C}})$.
\end{defn}

Monads are often defined in a more explicit way, by expressing the monoid structures and axioms directly and diagrammatically.

\begin{prop}
  A monad on $\cat{C}$ is equivalently a triple $(T,\mu,\eta)$ of
  \begin{enumerate}
  \item a functor $T:\cat{C}\to\cat{C}$;
  \item a natural transformation $\mu:TT \Rightarrow T$ called the \textit{multiplication}; and
  \item a natural transformation $\eta:\id_{\cat{C}}\Rightarrow T$ called the \textit{unit};
  \end{enumerate}
  such that, for all $c:\cat{C}$, the following diagrams commute:
  \[
  \begin{tikzcd}
	  TTTc && TTc \\
	  \\
	  TTc && Tc
	  \arrow["{\mu_{Tc}}"', from=1-1, to=3-1]
	  \arrow["{T\mu_c}", from=1-1, to=1-3]
	  \arrow["{\mu_c}", from=1-3, to=3-3]
	  \arrow["{\mu_c}", from=3-1, to=3-3]
  \end{tikzcd}
  \qquad\text{and}\qquad
  \begin{tikzcd}
	  Tc && TTc && Tc \\
	  \\
	  && Tc
	  \arrow["{\mu_c}", from=1-3, to=3-3]
	  \arrow["{\eta_{Tc}}", from=1-1, to=1-3]
	  \arrow["{T\eta_c}"', from=1-5, to=1-3]
	  \arrow[Rightarrow, no head, from=1-1, to=3-3]
	  \arrow[Rightarrow, no head, from=1-5, to=3-3]
  \end{tikzcd}
  \]
\end{prop}

A monad is like a monoidal structure for composition: instead of taking two objects and constructing a single object representing their conjunction (like the tensor of a monoidal category), a monad takes two levels of nesting and composes them into a single level; this is the source of the connection between multicategory algebras and monad algebras.

\begin{ex} \label{ex:list-monad}
  Recall the list monoid from Example \ref{ex:list-monoid}.
  The mapping $A \mapsto \List(A)$ defines the functor part of a monad $\List:\Set\to\Set$;
  given a function $f:A\to B$, $\List(f):\List(A)\to\List(B)$ is defined by applying $f$ to each element of the lists: $(a_1,a_2,\dots)\mapsto(f(a_1),f(a_2),\dots)$.
  The monad multiplication $\mu:\List^2\Rightarrow\List$ is given by ``removing inner brackets'' from lists of lists: $\mu_A\bigl((a_1^1,a_1^2,\dots),(a_2^1,\dots),\dots\bigr) = (a_1^1,a_1^2,\dots,a_2^1,\dots,\dots)$; equivalently, form the perspective of Example \ref{ex:list-monoid}, this is the concatenation of the `inner' lists into a single list.
  The monad unit $\eta:\id_{\Set}\Rightarrow\List$ is defined by returning `singleton' lists: $\eta_A:A\to\List(A):a\mapsto(a)$.
\end{ex}

There is a close connection between monads and adjunctions: every adjunction induces a monad.

\begin{prop} \label{prop:monad-from-adj}
  Suppose $L\dashv R:\cat{D}\to\cat{C}$ is an adjunction, with unit $\eta:\id_{\cat{C}}\Rightarrow RL$ and counit $\epsilon:LR\Rightarrow\id_{\cat{D}}$.
  Then $(RL,R\epsilon_L,\eta)$ is a monad.
  \begin{proof}
    To see that the associativity axiom is satisfied, observe that $R\epsilon_{LRL} = RL\epsilon_{RL} = RLR\epsilon_L$ by naturality.
    Right unitality follows by the triangle identity $\epsilon_L\circ L\eta = \id_L$, which entails the required equation $R\epsilon_L\circ RL\eta = \id_{RL}$; and left unitality follows from right unitality by naturality, as $\eta_{RL} = RL\eta$.
  \end{proof}
\end{prop}

It is also true that every monad arises from an adjunction: in fact, there are typically multiple adjunctions inducing the same monad, and we will exhibit one extremal case in \secref{sec:comp-prob}.

\begin{rmk}
  This dual correspondence is itself an example of an adjunction---in the quite general bicategorical sense, following the definition of monad as a monoid in a bicategory---though we leave the demonstration of this to the reader.
\end{rmk}

Before we show in generality how every monad arises from an adjunction, we can exhibit the list monad as a classic special case.

\begin{ex}[Lists are free monoids] \label{ex:list-free-monoid}
  There is a forgetful functor $U:\Mon(\Set)\to\Set$, taking each monoid $(M,\circ,\ast)$ (or monoid morphism $f$) and forgetting the monoid structure to return just the set $M$ (or the morphism $f$).
  This functor has a left adjoint $F:\Set\to\Mon(\Set)$, which takes each set $A$ to the \textit{free monoid} on $A$; this free monoid $F(A)$ is precisely the monoid $\bigl(\List(A),\circ,()\bigr)$ of lists in $A$, equipped with concatenation as multiplication and the empty list as unit, as described in Example \ref{ex:list-monoid}.
  The induced monad $(\List,\mu,\eta)$, described in Example \ref{ex:list-monad}, is then precisely the monad induced by this adjunction, with $\List = UF$.
\end{ex}

At this point, with an example of a monad to hand, we can start to explore their connection to algebra.

\begin{defn} \label{def:monad-alg}
  Suppose $(T,\mu,\eta)$ is a monad on $\cat{C}$.
  A $T$\textit{-algebra} is a choice of object $A:\cat{C}$ and a morphism $a:TA\to A$ such that the following diagrams commute:
  \[\begin{tikzcd}
	  A && TA \\
	  \\
	  && A
	  \arrow[Rightarrow, no head, from=1-1, to=3-3]
	  \arrow["{\eta_A}", from=1-1, to=1-3]
	  \arrow["a", from=1-3, to=3-3]
  \end{tikzcd}
  \qquad\text{and}\qquad
  \begin{tikzcd}
	  TTA && TA \\
	  \\
	  TA && A
	  \arrow["a", from=3-1, to=3-3]
	  \arrow["a", from=1-3, to=3-3]
	  \arrow["{\mu_A}"', from=1-1, to=3-1]
	  \arrow["Ta", from=1-1, to=1-3]
  \end{tikzcd}\]
\end{defn}

Once again, this being category theory, we are interested less in individual $T$-algebras than in their category.

\begin{defn}
  A \textit{morphism of} $T$\textit{-algebras} $(A,a)\to(B,b)$ is a morphism $f:A\to B$ that preserves the $T$-algebra structures, in the sense that the following diagram commutes:
  \[\begin{tikzcd}
	  TA && TB \\
	  \\
	  A && B
	  \arrow["f", from=3-1, to=3-3]
	  \arrow["b", from=1-3, to=3-3]
	  \arrow["a"', from=1-1, to=3-1]
	  \arrow["Tf", from=1-1, to=1-3]
  \end{tikzcd}\]
  $T$-algebras and their morphisms constitute a category, denoted $\Alg(T)$ and called the \textit{category of} $T$\textit{-algebras} or the \textit{Eilenberg-Moore category} for $T$.
  (Algebra morphisms compose by the composition of morphisms; a composite morphism of $T$-algebras is again a morphism of $T$-algebras by pasting. Identities are the usual identity morphisms in $\cat{C}$.)
\end{defn}

We now demonstrate the `algebra' of monad algebras using two familiar examples.

\begin{ex}
  The category of monoids in $(\Set,\times,1)$ is equivalent to the category of $\List$-algebras.
  A $\List$-algebra is a pair of a set $A$ and a function $a:\List(A)\to A$ satisfying the algebra axioms, which mean that $a$ must map singleton lists to their corresponding elements, and that $a$ must respect the ordering of elements in the list (so that it doesn't matter whether you apply $a$ to the lists in a lists of lists, or to the collapsed list resulting from the monad multiplication).
  To obtain a monoid, we can simply take the set $A$.
  The monoid multiplication is given by the action of $a$ on 2-element lists; and the monoid unit is given by the action of $a$ on the empty list.
  Since $a$ satisfies the monad algebra laws, the resulting multiplication and unit satisfy the monoid axioms: the monad laws are a categorification of the monoid axioms, and the algebra laws ensure compatibility with them.

  Dually, given a monoid $(A,m,e)$, we can construct a $\List$ algebra $a$ by induction: on empty lists, return $e$; on singleton lists, return their elements; on 2-element lists, apply $m$; on lists of length $n$, apply $m$ to the first two elements to obtain a list of length $n-1$ repeatedly until reaching the 2-element case.
  The monoid laws then ensure that the monad axioms are satisfied.
\end{ex}

\begin{ex} \label{ex:path-algebras}
  Recall from Proposition \ref{prop:forget-cat} that one can obtain from any category a directed graph by forgetting the compositional structure and retaining only the objects and morphisms as nodes and edges.
  Recall also from Proposition \ref{prop:free-cat} that one can obtain from any directed graph $\cat{G}$ a category $F\cat{G}$, the free category on $\cat{G}$ whose objects are the nodes of $\cat{G}$ and whose morphisms are paths in $\cat{G}$.
  These two constructions form a free-forgetful adjunction, $F\dashv U:\Cat{Cat}\to\Cat{Graph}$, and the induced monad $UF:\Cat{Graph}\to\Cat{Graph}$ is called the \textit{path monad}: on objects, it takes a graph $\cat{G}$ and returns a graph with the same nodes but whose edges are the paths in $\cat{G}$.
  The category $\Alg(UF)$ of algebras of $UF$ is equivalent to the category $\Cat{Cat}$ of (small) categories.

  To see this, note that a $UF$-algebra is a graph homomorphism $UF\cat{G}\to\cat{G}$, for some graph $\cat{G}$: a mapping of nodes in $UF\cat{G}$ to nodes in $\cat{G}$, and a mapping of edges in $UF\cat{G}$ to edges in $\cat{G}$ that preserves domains and codomains.
  Since $UF\cat{G}$ and $\cat{G}$ have the same nodes, the simplest choice is to map each node to itself: we will consider the nodes as the objects of the resulting category.
  The mapping of paths to edges induces a composition operation on the edges of $\cat{G}$, which we henceforth think of as morphisms.
  The reasoning proceeds inductively, much like the $\List$-algebra case: we take paths of length $0$ to be identity morphisms; paths of length $1$ are taken to their constituent morphisms; paths of length $2$ are taken to their composites; and one obtains the composites of longer paths by induction.
  Associativity and unitality then follow easily from the monad algebra laws.
\end{ex}

\begin{rmk} \label{rmk:alg-ind-coalg}
  Both the preceding examples suggest a connection between monad algebras and inductive reasoning, and indeed one can formalize inductive reasoning (as \textit{inductive types}) in terms of algebras.
  Dually, there is a close connection between `coalgebras' and `coinduction', which can be used to formalize the behaviours of systems that can be iterated, such as dynamical systems.
  As an informal example, the \textit{coinductive type} corresponding to $\List$ is the type of ``streams'': possibly infinite lists of the states or outputs of transition systems.
  In Chapter \ref{chp:coalg}, we use coalgebra to formalize the compositional structure of `open' (\textit{i.e.}, interacting) dynamical systems quite generally.
\end{rmk}

In the Appendix (\secref{sec:a-monad-operad}), we pursue the monad algebra story a little further, to demonstrate the connection with multicategory algebra.
However, since that connection is not strictly germane to the rest of the thesis, and with the suggested notion of coalgebra to whet our appetite, we now turn to monoids' duals, comonoids.

\subsection{Comonoids} \label{sec:comon}

We introduced comonoids graphically at the beginning of \secref{sec:str-diag}, as a structural manifestation of copying and discarding, but in the fullest of generality, comonoids are simply monoids in opposite categories.

\begin{defn} \label{def:comonoid}
  A \textit{comonoid} in $(\cat{C},\otimes,I)$ is a monoid in $\cat{C}\op$, when $\cat{C}\op$ is equipped with the opposite monoidal structure induced by $(\otimes,I)$.
  Explicitly, this means an object $c:\cat{C}$ equipped with a \textit{comultiplication} $\delta:c\to c\otimes c$ and \textit{counit} $\epsilon:c\to I$, satisfying \textit{counitality} and \textit{coassociativity} laws formally dual to the corresponding unitality and associativity laws of monoids: read the diagrams of Definition \ref{def:monoid} top-to-bottom, rather than bottom-to-top.
  Likewise, if $\cat{C}$ is symmetric monoidal, we say that a comonoid in $\cat{C}$ is \textit{cocommutative} if its comultiplication commutes with the symmetry.
\end{defn}

\begin{ex} \label{ex:prod-comon}
  Every object in a category with finite products $\times$ and a terminal object $1$ is a comonoid with respect to the monoidal structure $(\times,1)$.
  The comultiplications $\delta_X:X\to X\times X$ are defined by the pairing $(\id_X,\id_X)$ (recall Definition \ref{def:product}) and the counits $\epsilon_X:X\to 1$ are (necessarily) the unique morphisms into the terminal object.

  Coassociativity follows because $(\id_X,(\id_X,\id_X)) = \alpha_{X,X,X}\circ((\id_X,\id_X),\id_X)$, where $\alpha$ is the associator of the product.
  Counitality follows by the naturality of pairing, $(\id_X\times\,!)\circ(\id_X,\id_X) = (\id_X,\,!)$, and because $\proj_X\circ(\id_X,\,!) = \id_X$ by the universal property of the product; note that $\proj_X$ is the $X$ component of the right unitor of the monoidal structure, and $(\id_X,\,!)$ is its inverse.

  Instantiating this example in $\Set$, we see that the comultiplication is given by copying, \textit{i.e.}, $x\mapsto(x,x)$; and the counit is the unique map $x\mapsto\ast$ into the singleton set.
  This justifies our writing of the comonoid structure in copy-discard style as $(\bcopier_X,\ground_X)$.
\end{ex}

In general, when a comonoid structure is to be interpreted as a copy-discard structure, we will therefore write the struture morphisms as $(\bcopier, \ground)$ and depict them accordingly in the graphical calculus, rather than using the boxed forms of Definition \ref{def:monoid}.
However, copy-discard structures are not the only important comonoids that we will encounter.
In the next section, we introduce the category of polynomial functors $\Set\to\Set$, and since these are endofunctors, their category inherits a monoidal structure given by functor composition.
Comonoids for this monoidal structure in $\Poly$ give us another definition for a now multifariously familiar concept: they are again small categories, although their morphisms are not functors but rather \textit{cofunctors}.

Of course, a morphism of comonoids is much like a morphism of monoids.

\begin{defn} \label{def:comon-morph}
  A \textit{comonoid morphism} $f:(c,\delta,\epsilon)\to(c',\delta',\epsilon')$ in $(\cat{C},\otimes,I)$ is a morphism $f:c\to c'$ that is compatible with the comonoid structures, in the sense of satisfying axioms dual to those of Definition \ref{def:monoid-morph}.
  There is thus a category $\Comon(\cat{C})$ of comonoids in $\cat{C}$ and their morphisms, as well as a subcategory $\CComon(\cat{C})\hookrightarrow\Comon(\cat{C})$ of commutatitve comonoids.
\end{defn}

In the more familiar copy-discard setting, comonoid morphisms also play an important rôle.
In the next chapter, we will see concretely that, in the context of stochastic maps, comonoid morphisms (with respect to the tensor product) correspond to the deterministic functions.
This result is closely related to the following fact.

\begin{prop} \label{prop:comon-morph-prod}
  If every morphism in the monoidal category $(\cat{C},\otimes,I)$ is a comonoid morphism, then $a\otimes b$ satisfies the universal property of the product for every $a,b:\cat{C}$, and hence $\otimes$ is the categorical product and $I$ the terminal object in $\cat{C}$ (up to isomorphism).
  \begin{proof}
    If every morphism is a comonoid morphism, then every object $a:\cat{C}$ carries a comonoid structure; assume a choice of comonoid structure $(\bcopier_a:a\to a\otimes a,\ground_a:a\to I)$ for every $a:\cat{C}$.
    The universal property of the product says that every morphism $f:x\to a\otimes b$ factors as
    \[ \input{img/f-x-ab.tikz} \qquad = \qquad \input{img/f_copy-x-ab.tikz} \]
    where $f_a:x\to a$ and $f_b:x\to b$ are uniquely defined as
    \[ f_a \; := \; \input{img/f_a.tikz} \qquad\text{and}\qquad f_b \; := \; \input{img/f_b.tikz} \quad . \]
    Since $f$ is \textit{ex hypothesi} a comonoid morphism, we have
    \[ \input{img/f-x-ab.tikz} \; = \; \input{img/f-x-ab-cd.tikz} \; = \; \input{img/f_a-f_b.tikz} \; = \; \input{img/f_copy-x-ab.tikz} \]
    where the first equality holds by counitality, the second since $f$ commutes with $\bcopier_x$ \textit{ex hypothesi}, and the third by definition.
    This establishes that $a\otimes b$ satisfies the universal property, and hence that $\otimes$ is the categorical product.

    To see that $I$ is the terminal object up to isomorphism, suppose that $1$ is the terminal object.
    Since $\otimes$ is the categorical product, there is an isomorphism $a\xto{\sim}a\otimes 1$ for any $a:\cat{C}$, by the universal property.
    In particular, there is an isomorphism $I\xto{\sim}I\otimes 1$.
    But since $I$ is the monoidal unit for $\otimes$, the component of the left unitor at $1$ is an isomorphism $I\otimes 1\xto{\sim}1$.
    Hence we have a composite isomorphism $I\xto{\sim}I\otimes 1\xto{\sim}1$, and so $I\cong 1$.
  \end{proof}
\end{prop}

\begin{rmk}
  The preceding proposition gives us another way to look at comonoids: we can think of them as ``products without the universal property''.
  The reason for this is that, since products are characterized by their (universal) projections, we can use the counits to define projections for the monoidal product of comonoids: that is, if $a$ and $b$ are comonoids in $\cat{C}$, then we can define (non-universal) projections $a\xfrom{\proj_a}a\otimes b\xto{\proj_b}b$ by
  \[ a \xfrom{\rho_a} a\otimes I \xfrom{\id_a\otimes\ground_b} a\otimes b \xto{\ground_a\otimes\id_b} I\otimes b \xto{\lambda_b} b \]
  where $\rho$ and $\lambda$ denote the right and left unitors of the monoidal structure respectively.
  The failure of universality means that the family of projections $\{\proj_a\}_{a:\cat{C}}$ in $\cat{C}$ does not constitute a natural transformation.
\end{rmk}

\begin{rmk} \label{rmk:nat-determ}
  Abstractly, we can use naturality as a way to characterize deterministic morphisms: the naturality law for $\bcopier$ requires that
  \[ a \xto{f} b \xto{\bcopier_b} b\otimes b \, = \, a \xto{\bcopier_a} a\otimes a \xto{f\otimes f} b\otimes b \]
  and this says that first doing $f$ and the copying its output is the same as copying the input and feeding each copy into $f$.
  If $f$ were non-deterministic, then there would be a correlation between the copies in the former case but not in the latter, and so this equation would not hold.
  Therefore, we can think of those morphisms $f$ for which copying \textit{is} natural as the deterministic morphisms in $\cat{C}$.
  We will return to this perspective in Remark \ref{rmk:comon-determ}.
\end{rmk}

Finally, there is also a notion of comonad, dual to monad: a comonad is quite generally a comonoid in a bicategory, in the sense of Remark \ref{rmk:monoid-bicat}, or, less generally, a comonoid with respect to the composition product in a category of endofunctors.
This means that the polynomial comonoids we discussed above are by definition comonads.

In Remark \ref{rmk:alg-ind-coalg}, we introduced the notion of `coalgebra', and indeed there is a notion of comonad coalgebra that is dual to the notion of monad algebra; and indeed we will use coalgebras later to formalize dynamical systems.
But although these coalgebras will be morphisms of the form $FX\to X$, for $F$ an endofunctor and $X$ an object, the endofunctor $F$ will not necessarily have a comonad structure, and so the coalgebras will be more general than the algebras we considered above: there will be no comonad compatibility axioms to satisfy.

In many cases, the endofunctor $F$ will be a polynomial functor, so let us now introduce these.

\section{Polynomial functors} \label{sec:poly}

In order to be considered \textit{adaptive}, a system must have something to adapt to.
This `something' is often what we call the system's \textit{environment}, and we say that the system is \textit{open} to its environment.
The interface or boundary separating the system from its environment can be thought of as `inhabited' by the system: the system is embodied by its interface of interaction; the interface is animated by the system.
In this way, the system can affect the environment, by changing the shape or configuration of its interface\footnote{
Such changes can be very general: consider for instance the changes involved in producing sound (\textit{e.g.}, rapid vibration of tissue) or light (\textit{e.g.}, connecting a luminescent circuit, or the molecular interactions involved therein).
}; through the coupling, these changes are propagated to the environment.
In turn, the environment may impinge on the interface: its own changes, mediated by the coupling, arrive at the interface as immanent signals; and the type of signals to which the system is alive may depend on the system's configuration (as when an eye can only perceive if its lid is open).
Thus, information flows across the interface.

The mathematical language capturing this kind of inhabited interaction is that of \textit{polynomial functors}, which we adopt following \textcite{Spivak2021Polynomial}.
We will see that this language---or rather, its category---is sufficiently richly structured to provide both a satisfactory syntax for the patterns of interaction of adaptive systems, generalizing the circuit diagrams of \secref{sec:lin-circ}, as well as a home for the dynamical semantics that we will seek.

Polynomial functors are so named because they are a categorification of polynomial functions: functions built from sums, products, and exponentials, of the form $y\mapsto \sum_{i:I} b_i\, y^{a_i}$.
To categorify a function of this kind, we can simply interpret the coefficients and exponents and the variable $y$ as standing for sets rather than mere numbers.
In this way, we reinterpret the term $y^{a_i}$ as the representable copresheaf $\Set(a_i,-)$, so that we can substitute in any set $X$ and obtain the exponential $X^{a_i}$ (just as in the classical case).
To categorify the sums and products, we can simply use the universal constructions available in the copresheaf category $\Set^{\Set}$: these are still available in the subcategory $\Poly$, since $\Poly$ is by definition the subcategory of the copresheaf category on sums of representables (and as we have seen, products are equivalently iterated coproducts).

\begin{rmk}
  Limits and colimits in (co)presheaf categories are computed `pointwise'.
  Therefore, if $F$ and $G$ are two copresheaves $\cat{C}\to\Set$, then their sum $F+G$ is the copresheaf defined by $x\mapsto F(x)+G(x)$ and their product is the copresheaf defined by $x\mapsto F(x)\times G(x)$.
\end{rmk}

We will adopt the standard notation for polynomial functors of \textcite{Spivak2021Polynomial}, so that if $p$ is a polynomial, we will expand it as $\sum_{i:p(1)} y^{p[i]}$.
When treating $p$ as encoding the type of a system's interface, we will interpret $p(1)$ as encoding the set of possible configurations (or `shapes') that the system may adopt, and for each configuration $i:p(1)$, the set $p[i]$ is the set of possible immanent signals (`inputs') that may arrive on the interface in configuration $i$.

\begin{defn} \label{def:poly-set}
  First, if $A$ be any set, we will denote by $y^A$ its representable copresheaf $y^A := \Set(A, -) : \Set\to\Set$.
  A \emph{polynomial functor} $p:\Set\to\Set$ is then an indexed coproduct of such representable copresheaves, written $p := \sum_{i : p(1)} y^{p_i}$, where $p(1)$ denotes the indexing set and $p[i]$ the representing set for each $i$.
  The category of polynomial functors is the full subcategory $\Poly \hookrightarrow \Set^{\Set}$ of the copresheaf category spanned by coproducts of representables.
  A morphism of polynomials is thus a natural transformation.
\end{defn}

\begin{rmk}
  Note that, given a polynomial functor $p:\Set\to\Set$, the indexing set $p(1)$ is indeed obtained by applying $p$ to the terminal set $1$.
\end{rmk}

We will make much use of the following `bundle' representation of polynomial functors and their morphisms.

\begin{prop} \label{prop:poly-bundles} %
  Every polynomial functor $\sum_{i:p(1)} y^{p_i}$ corresponds to a bundle (a function) $p:\sum_{i:p(1)} p_i \to p(1)$, where the set $\sum_{i:p(1)} p_i$ is the $p(1)$-indexed coproduct of the representing objects $p_i$ and $p$ is the projection out of the coproduct onto the indexing set $p(1)$.

  Every morphism of polynomials $f:p\to q$ corresponds to a pair $(f_1,f^\sharp)$ of a function $f_1:p(1)\to q(1)$ and a $p(1)$-indexed family of functions $f^\sharp_i:q[f_1(i)]\to p[i]$ making the diagram below commute. We adopt the notation $p[i] := p_i$, and write $f^\sharp$ to denote the coproduct $\sum_{i} f^\sharp_i$.
  \[\begin{tikzcd}%
    {\sum_{i:p(1)}p[i]} & {\sum_{i:p(1)}q[f_1(i)]} & {\sum_{j:q(1)}q[j]} \\
    {p(1)} & {p(1)} & {q(1)}
    \arrow["{f^\sharp}"', from=1-2, to=1-1]
    \arrow[from=1-2, to=1-3]
    \arrow["q", from=1-3, to=2-3]
    \arrow["p"', from=1-1, to=2-1]
    \arrow[from=2-1, to=2-2, Rightarrow, no head]
    \arrow["{f_1}", from=2-2, to=2-3]
    \arrow[from=1-2, to=2-2]
    \arrow["\lrcorner"{anchor=center, pos=0.125}, draw=none, from=1-2, to=2-3]
  \end{tikzcd}\]
  Given $f : p \to q$ and $g : q \to r$, their composite $g \circ f : p \to r$ is as marked in the diagram
  \[\begin{tikzcd}%
    {\sum_{i:p(1)}p[i]} & {\sum_{i:p(1)}r[g_1\circ f_1(i)]} & {\sum_{k:r(1)}r[k]} \\
    {p(1)} & {p(1)} & {r(1)}
    \arrow["{(gf)^\sharp}"', from=1-2, to=1-1]
    \arrow[from=1-2, to=1-3]
    \arrow["r", from=1-3, to=2-3]
    \arrow["p"', from=1-1, to=2-1]
    \arrow[from=2-2, to=2-1, Rightarrow, no head]
    \arrow["{g_1 \circ f_1}", from=2-2, to=2-3]
    \arrow[from=1-2, to=2-2]
    \arrow["\lrcorner"{anchor=center, pos=0.125}, draw=none, from=1-2, to=2-3]
  \end{tikzcd}\]
  where $(gf)^\sharp$ is the coproduct of the \(p(1)\)-indexed family of composite maps
  \[ r[g_1(f_1(i))] \xto{f^\ast g^\sharp} q[f_1(i)] \xto{f^\sharp} p[i] \; . \]
  The identity morphism on a polynomial $p$ is $(\id_{p(1)},\id)$.
  \begin{proof}
    We just need to show that every natural transformation between polynomial functors corresponds to a pair of maps $(f_1,f^\sharp)$ as defined above.
    The set of natural transformations $\sum_{i:p(1)}y^{p[i]}\Rightarrow\sum_{j:q(1)}y^{q[j]}$ is the hom-set $\Set^{\Set}\bigl(\sum_{i:p(1)}y^{p[i]},\sum_{j:q(1)}y^{q[j]}\bigr)$.
    Since the contravariant hom functor takes colimits to limits (Remark \ref{rmk:contra-hom-lim-colim}), this hom-set is isomorphic to $\prod_{i:p(1)}\Set^{\Set}(y^{p[i]},\sum_{j:q(1)}y^{q[j]})$.
    By the Yoneda lemma, this is in turn isomorphic to $\prod_{i:p(1)}\sum_{j:q(1)}p[i]^{q[j]}$.
    And since products distribute over sums, we can rewrite this as $\sum_{f_1:p(1)\to q(1)}\prod_{i:p(1)} p[i]^{q[f_1(i)]}$.
    The elements of this set are precisely pairs of a function $f_1:p(1)\to q(1)$ along with a family of functions $q[f_1(i)]\to p[i]$ indexed by $i:p(1)$, such that the diagram above commutes.
  \end{proof}
\end{prop}

We now recall a handful of useful facts about polynomials and their morphisms, each of which is explained in \textcite{Spivak2021Polynomial} and summarized in \textcite{Spivak2022reference}.

We will consider the unit polynomial $y$ to represent a `closed' system, since it has no nontrivial configurations and no possibility of external input.
For this reason, morphisms $p\to y$ will represent ways to make an open system closed, and in this context the following fact explains why: such morphisms correspond to a choice of possible input for each $p$-configuration; that is, they encode ``how the environment might respond to $p$''.

\begin{prop} \label{prop:poly-section}
  Polynomial morphisms $p \to y$ correspond to sections $p(1) \to \sum_i p[i]$ of the corresponding function $p:\sum_ip[i]\to p(1)$.
\end{prop}

The following embedding of $\Set$ into $\Poly$ will be useful in constructing `hierarchical' dynamical systems.

\begin{prop}
  There is an embedding of \(\Set\) into \(\Poly\) given by taking sets \(X\) to the linear polynomials \(Xy : \Poly\) and functions \(f : X\to Y\) to morphisms \((f,\id_X) : Xy \to Yy\).
\end{prop}

There are many monoidal structures on $\Poly$, but two will be particularly important for us.
The first represents the parallel composition of systems.

\begin{prop} \label{prop:poly-tensor}
  There is a symmetric monoidal structure \((\otimes, y)\) on \(\Poly\) that we call \textit{tensor}, and which is given on objects by \(p\otimes q := \sum_{i:p(1)}\sum_{j:q(1)} y^{p[i]\times q[j]}\) and on morphisms \(f := (f_1, f^\sharp) : p\to p'\) and \(g := (g_1, g^\sharp) : q\to q'\) by \(f\otimes g := \bigl(f_1 \times g_1, \Sigma(f^\sharp, g^\sharp)\bigr)\), where $\Sigma(f^\sharp, g^\sharp)$ is the family of functions
  \[ \Sigma(f^\sharp, g^\sharp)_{i,j} := p'[f_1(i)]\times q'[g_1(j)] \xto{f^\sharp_i\times g^\sharp_j} p[i]\times q[j] \]
  indexed by $(i,j):p(1)\times q(1)$.
  This is to say that the `forwards' component of $f\otimes g$ is the product of the forwards components of $f$ and $g$, while the `backwards' component is the \textit{pointwise} product of the respective backwards components.
\end{prop}

\begin{prop}
  $(\Poly, \otimes, y)$ is symmetric monoidal closed, with internal hom denoted $[{-},{=}]$.
  Explicitly, we have $[p,q] = \sum_{f:p\to q} y^{\sum_{i:p(1)} q[f_1(i)]}$.
  Given an set $A$, we have $[Ay, y] \cong y^A$.
\end{prop}

The second important monoidal structure is that inherited from the composition of endofunctors.
To avoid confusion with other composition operators, we will in this context denote the operation by $\lhd$.

\begin{prop}
  The composition of polynomial functors $q \circ p : \cat{E}\to\cat{E}\to\cat{E}$ induces a monoidal structure on $\Poly_\Ea$, which we denote $\lhd$, and call `composition' or `substitution'.
  Its unit is again $y$.
\end{prop}

Comonoids with respect to $\lhd$ play a particularly important rôle in the theory of polynomial functors, and we will make accordingly much use of them.

\begin{prop}[{\textcite[{\S3.2}]{Ahman2016Directed}}]
  Comonoids in $(\Poly,\lhd,y)$ correspond to small categories.
  If $(c,\delta,\epsilon)$ is a comonoid, then the shapes $c(1)$ are the objects of the corresponding category $\cat{C}$.
  For each object $x:c(1)$, $c[i]$ is the set $\sum_{y:c(1)}\cat{C}(x,y)$ of morphisms out of $x$.
  The counit morphism $\epsilon:c\to y$ is, following Proposition \ref{prop:poly-section}, a section of $c$, and assigns to each $x:c(1)$ its identity morphism $\id_x:x\to x$.
  The comultiplication $\delta:c\to c\lhd c$ encodes morphisms' codomains (its forward action) and their composition (its backward action).
  Finally, the comonoid laws ensure that the category is well defined.
\end{prop}

\begin{rmk}
  $\lhd$-comonoid homomorphisms are not, as one might expect, functors; rather, they are `cofunctors': they act backwards on morphisms.
  We will not explore the theory of cofunctors any further in this thesis, although we will make frequent use of them later in the context of dynamical systems.
\end{rmk}

The following $\lhd$-comonoids will play a prominent rôle in our dynamical developments.

\begin{prop}
  If $\Tt$ is a monoid in $(\Set,\times,1)$, then the comonoid structure on $y^\Tt$ witnesses it as the category $\deloop{\Tt}$.
\end{prop}

\begin{prop}
  Monomials of the form $Sy^S$ can be equipped with a canonical comonoid structure witnessing the codiscrete groupoid on $S$: the category with an object for every element $s$ of $S$ and a morphism $s\to t$ for every pair of elements $(s,t)$.
\end{prop}

\chapter{The compositional structure of Bayesian inference} \label{chp:buco}

This chapter introduces the fundamental concepts and structures needed for the development of statistical games in Chapter \ref{chp:sgame}, and proves the crucial result that Bayesian updating composes according to the `lens' pattern.
To make sense of this statement, we first introduce compositional probability (\secref{sec:comp-prob}), motivating it as a resolution of some imprecision that arises when one works informally with probability and statistics, particularly in the context of `hierarchical' models.
We exhibit categorical probability theory both abstractly (\secref{sec:abstract-bayes} and \secref{sec:density-functions}) and concretely (using discrete probability in \secref{sec:fin-prob} and `continuous' probability in \secref{sec:sfKrn}).
We then move on to construct categories of bidirectional processes in \secref{sec:bidi}, by first categorifying our earlier discussion of dependent data using the Grothendieck construction (\secref{sec:idx-cat}) and then using this to introduce the lens pattern (\secref{sec:groth-lens}).

In \secref{sec:bidi-bayes}, we present our novel results.
First, we introduce the indexed category of ``state-dependent channels'' in \secref{sec:stat-chan}.
These formalize the type of Bayesian inversions, and so in \secref{sec:bayes-lens} we define the associated notion of \textit{Bayesian lens}, and show in \secref{sec:buco} that Bayesian updating composes according to the lens pattern.
We end with a brief discussion of the `lawfulness' of Bayesian lenses.

\begin{rmk} \label{rmk:cinema}
  To gain some intuition about the hierarchical compositional structure of Bayesian inference, consider sitting close to the screen at a cinema.
  Your neural activity encodes a belief about where the characters are on the screen and what they are doing, but your visual field can only capture a part of the image at any one time.
  These incoming visual signals contain ``low-level'' information, about the light intensity over the patch of screen you can see, and the first job of the visual system is to infer what this means for what's going on in this patch.
  Of course, having been following the film so far, your brain encodes a high-level belief about what is going on across the whole screen, and it uses this to predict what to expect in the patch.
  This intermediate-level belief is then updated using the received visual signals, through a process of (approximate) Bayesian inference.
  The resulting intermediate-level posterior then supplies the input for a second inference process, updating the prior high-level belief accordingly.
  Notice that this means that the process of prediction in such a hierarchical inference system points from inside an agent ``towards the world''; and the belief-updating process points the other way, from the world into the agent.
\end{rmk}

\section{Compositional probability} \label{sec:comp-prob}

In informal literature, Bayes' rule is often written in the following form:
\begin{equation*}
\Pr(A|B) = \frac{\Pr(B|A) \cdot \Pr(A)}{\Pr(B)}
\end{equation*}
where \(\Pr(A)\) is the probability of the event \(A\), and \(\Pr(A|B)\) is the probability of the event \(A\) given that the event \(B\) occurred; and \emph{vice versa} swapping \(A\) and \(B\).
Unfortunately, this notation obscures that there is in general no unique assignment of probabilities to events: different observers can hold different beliefs.
Moreover, we are usually less interested in the probability of particular events than in the process of assigning probabilities to arbitrarily chosen beliefs; and what should be done if \(\Pr(B) = 0\) for some \(B\)?
The aim in this section is to exhibit a general, precise, and compositional, form of Bayes' rule; we begin, as before, by introducing the intuition.

In the categorical setting, the assignment of probabilities or beliefs to events will formally be the task of a \textit{state} (in the sense of \secref{sec:str-diag}) on the space from which the events are drawn; we should think of states as generalizing distributions or measures.
With this notion to hand, we can write \(\Pr_\pi(A)\) to denote the probability of \(A\) according to the state \(\pi\).

The formalization of \textit{conditional} probability will be achieved by morphisms that we will call \textit{channels}, meaning that we can write \(\Pr_c(B|A)\) to denote the probability of \(B\) given \(A\) according to the channel \(c\).
We can think of the channel \(c\) as taking events \(A\) as inputs and emitting states \(c(A)\) as outputs.
This means that we can alternatively write \(\Pr_c(B|A) = \Pr_{c(A)}(B)\).

If the input events are drawn from the space \(X\) and the output states encode beliefs about \(Y\), then the channel \(c\) will be a morphism \(X \klto Y\).
Given a channel \(c : X \klto Y\) and a channel \(d : Y \klto Z\), we will understand their composite \(d \klcirc c : X \klto Z\) as marginalizing (averaging) over the possible outcomes in \(Y\).
We will see precisely how this works in various settings below.

\begin{notation}
  In a stochastic context, we will denote channels by the arrow $\klto$, and write their composition operator as $\klcirc$.
  We do this to distinguish stochastic channels from deterministic functions, which we will continue to write as $\to$ with composition $\circ$; in a number of situations, it will be desirable to work with both kinds of morphism and composition.
\end{notation}

Given two spaces \(X\) and \(Y\) of events, we can form beliefs about them jointly, represented by states on the product space denoted \(X \otimes Y\).
The numerator in Bayes' rule represents such a joint state, by the law of conditional probability or `product rule':
\begin{equation} \label{eq:pr-cond-prob}
\Pr_\omega(A, B) = \Pr_c(B|A) \cdot \Pr_\pi(A)
\end{equation}
where \(\cdot\) is multiplication of probabilities, \(\pi\) is a state on \(X\), and \(\omega\) denotes the joint state on \(X \otimes Y\).
By composing \(c\) and \(\pi\) to form a state \(c \klcirc \pi\) on \(Y\), we can write
\begin{equation*}
\Pr_{\omega'}(B, A) = \Pr_{c^\dag_\pi}(A|B) \cdot \Pr_{c \klcirc \pi}(B)
\end{equation*}
where \(c^\dag_\pi\) will denote the Bayesian inversion of \(c\) with respect to \(\pi\).

Joint states in classical probability theory are symmetric---and so the tensor $\otimes$ is symmetric---meaning that there is a family of isomorphisms \(\mathsf{swap} : X \otimes Y \xklto{\sim} Y \otimes X\), as in \secref{sec:str-diag}, and which will satisfy the symmetric monoidal category axioms (Definition \ref{def:monoidal-cat}).
Consequently, we have \(\Pr_\omega(A, B) = \Pr_{\omega'}(B, A)\) where \(\omega' = \mathsf{swap} \klcirc \omega\), and thus
\begin{equation} \label{eq:pr-disintegrations}
\Pr_c(B|A) \cdot \Pr_\pi(A) = \Pr_{c^\dag_\pi}(A|B) \cdot \Pr_{c \klcirc \pi}(B)
\end{equation}
where both left- and right-hand sides are called \emph{disintegrations} of the joint state \(\omega\) \citep{Cho2017Disintegration}. From this equality, we can write down the usual form of Bayes' theorem, now with the sources of belief indicated:
\begin{equation} \label{eq:pr-bayes}
\Pr_{c^\dag_\pi}(A|B) = \frac{\Pr_c(B|A) \cdot \Pr_\pi(A)}{\Pr_{c \klcirc \pi}(B)} \, .
\end{equation}
As long as \(\Pr_{c \klcirc \pi}(B) \neq 0\), this equality defines the inverse channel \(c^\dag_\pi\). If the division is undefined, or if we cannot guarantee \(\Pr_{c \klcirc \pi}(B) \neq 0\), then \(c^\dag_\pi\) can be any channel satisfying \eqref{eq:pr-disintegrations}.

There is therefore generally no unique Bayesian inversion \(c^\dag : Y \klto X\) for a given channel \(c : X \klto Y\): rather, we have an inverse \(c^\dag_\pi : Y \klto X\) for each prior state \(\pi\) on \(X\).
Moreover, \(c^\dag_\pi\) is not a ``posterior distribution'' (as written in some literature), but a process which emits a posterior distribution, given an observation in \(Y\).
If we denote our category of stochastic channels by $\cat{C}$, then, by allowing \(\pi\) to vary, we obtain a map of the form \(\bdag{c} : \Pow X \to \cat{C}(Y, X)\), where \(\Pow X\) denotes a space of states on \(X\).
Note that here we are not assuming the object $\Pow X$ to be an object of $\cat{C}$ itself (though it often will be), but rather an object in its base of enrichment, so that here we can think of $\bdag{c}$ as a kind of externally parameterized channel (in the sense of \secref{sec:ext-para}).
Making the type of this `state-dependent' channel $\bdag{c}$ precise is the task of \secref{sec:idx-cat}.

\begin{rmk}
  There are two easily confused pieces of terminology here.
  We will call the channel \(c^\dag_\pi\) the \textit{Bayesian inversion} of the channel \(c\) with respect to \(\pi\).
  Then, given some \(y \in Y\), the state \(c^\dag_\pi (y)\) is a new `posterior' distribution on X. We will call \(c^\dag_\pi(y)\) the \textit{Bayesian update} of \(\pi\) along \(c\) given \(y\).
\end{rmk}

In the remainder of this section, we instantiate the ideas above in categories of stochastic channels of various levels of generality, beginning with the familiar case of discrete (\textit{i.e.}, finitely supported, or `categorical') probability.

\subsection{Discrete probability, algebraically}
\label{sec:fin-prob}

Interpreting the informal Bayes' rule \eqref{eq:pr-bayes} is simplest in the case of discrete or \emph{finitely-supported} probability.
Here, every event is a set, generated as the disjoint union of so many atomic (singleton) events, which one can therefore take as the elements of the set.
A finitely-suported probability distribution is then simply an assignment of nonzero probabilities to finitely many elements, such that the sum of all the assignments is $1$.
This condition is a \textit{convexity} condition, and so in this subsection we will introduce discrete compositional probability theory from a geometric perspective, using the algebraic tools of the previous chapter.

\begin{defn}
  Suppose $X$ is a set.
  A function $c:X\to[0,1]$ such that $c(x) > 0$ for only finitely many $x:X$ and $\sum_{x:X}c(x) = 1$ will be called a \textit{discrete} or \textit{finitely-supported} \textit{distribution} on $X$.
  We write $\Dst X$ to denote the set of discrete distributions on $X$.
  A (real-valued) \textit{convex set} is a set $X$ equipped with a function $\varsigma:\Dst X\to X$ called its \textit{evaluation}.
\end{defn}

Convex sets $X$ are sets in which we can form convex combinations of elements.
Algebraically, we can model these convex combinations as distributions on $X$, and the evaluations realize the convex combinations (distributions) as elements again of $X$: geometrically, the evaluation returns the barycentre of the distribution.

In light of Chapter \ref{chp:algebra}, this situation may seem familiar.
Indeed, the assignment $X\mapsto\Dst X$ is the functor part of a monad on $\Set$, whose algebras are convex sets.
This monad arises from a free-forgetful adjunction between the category of convex sets (the category of algebras of the monad) and the category $\Set$.
Later, we will find that the category of finitely-supported conditional probability distributions---the category of discrete stochastic channels---is equivalent to the category of free convex sets and their morphisms: a free convex set on $X$ is equivalently a distribution on $X$.

Let us first formalize the functor $\Dst$.

\begin{prop}
  The mapping of sets $X\mapsto\Dst X$ is functorial.
  Given a function $f:X\to Y$, we obtain a function $\Dst f:\Dst X\to\Dst Y$ mapping $c:\Dst X$ to the distribution $Df(c):\Dst Y$,
  \begin{align*}
    \Dst f(c) : Y &\to [0,1] \\
    y &\mapsto \sum_{x:f(x)=y} c(x) \; .
  \end{align*}
  \begin{proof}
    Given $f:X\to Y$ and $g:Y\to Z$, we have
    \begin{align*}
      \Dst g(\Dst f(c)) : Z &\to [0,1] \\
      z &\mapsto \sum_{y:g(y)=z} \sum_{x:f(x)=y} c(x) \\
      &= \sum_{x:g\circ f(x) = z} c(x)
    \end{align*}
    hence $\Dst g\circ \Dst f = \Dst (g\circ f)$. We also have
    \begin{align*}
      \Dst(\id_X)(c) : X&\to [0,1] \\
      x&\mapsto \sum_{x':\id_X(x')=x} c(x') \\
       &= c(x)
    \end{align*}
    and hence $\Dst(\id) = \id_\Dst$.
  \end{proof}
\end{prop}

To obtain the monad structure of $\Dst$, we will exhibit the free-forgetful adjunction.
We start by defining the category of convex sets, and the specific case of a \textit{free} convex set.

\begin{defn}
  The \textit{category of (real-valued) convex sets} $\Cat{Conv}$ has convex sets $(X,\varsigma_X)$ as objects.
  Its morphisms $(X,\varsigma_X)\to(Y,\varsigma_Y)$ are functions $f:X\to Y$ that preserve the convex structure, in the sense that the following square commutes:
  \[\begin{tikzcd}
    {\Dst X} && {\Dst Y} \\
	  \\
	  X && Y
	  \arrow["{\Dst f}", from=1-1, to=1-3]
	  \arrow["f", from=3-1, to=3-3]
	  \arrow["{\varsigma_X}"', from=1-1, to=3-1]
	  \arrow["{\varsigma_Y}", from=1-3, to=3-3]
  \end{tikzcd}\]
\end{defn}

\begin{defn} \label{def:cvx-eval}
  If $X$ is any set, then the \textit{free convex set} on $X$ is the set $\Dst X$ equipped with the evaluation $\mu_X:\Dst\Dst X\to\Dst X$ which maps $\alpha:\Dst\Dst X$ to the distribution $\mu_X(\alpha):\Dst X$,
  \begin{align*}
    \mu_X(\alpha) : X &\to [0,1] \\
    x &\mapsto \sum_{c:\Dst X} \alpha(c)\cdot c(x) \; .
  \end{align*}
\end{defn}

\begin{notation}
  To emphasize the algebraic nature of finitely-supported distributions $\pi:\Dst X$, instead of writing them as functions $x\mapsto\pi(x)$, we can write them as \textit{formal sums} or \textit{formal convex combinations} $\sum_{x:X}\pi(x)\ket{x}$, with each element $x:X$ corresponding to a formal ``basis vector'' $\ket{x}$ with the coefficient $\pi(x)$.
  If $X$ is a convex set, then the evaluation realizes this formal sum as an actual element (`vector') in $X$.
\end{notation}

We are now in a position to exhibit the adjunction: the reasoning behind the following proposition follows the lines of Example \ref{ex:path-algebras} and Proposition \ref{ex:grph-cat-adjunc} (on the free-forgetful adjunction between graphs and categories).

\begin{prop}
  The mapping of $X:\Set$ to the free convex set $(\Dst X,\mu_X)$ defines a functor $F:\Set\to\Cat{Conv}$ which takes functions $f:X\to Y$ to morphisms $Ff:(\Dst X,\mu_X)\to(\Dst Y,\mu_Y)$ defined by $\Dst f:\Dst X\to\Dst Y$.
  This functor $F$ is left adjoint to the forgetful functor $U:\Cat{Conv}\to\Set$ which acts by $(X,\varsigma_X)\mapsto X$.
\end{prop}

Using Proposition \ref{prop:monad-from-adj}, the adjunction gives us a monad.

\begin{cor}
  The functor $\Dst:\Set\to\Set$ is equivalently the functor part of the monad induced by the free-forgetful adjunction on convex sets.
  It therefore acquires a monad structure $(\mu,\eta)$ where the components of the multiplication $\mu$ are the free evaluations $\mu_X:\Dst\Dst X\to\Dst X$, and the unit $\eta$ has components $\eta_X:X\to\Dst X$ which return the `Dirac' distributions, as in
  \begin{align*}
    \eta_X(x) : X &\to [0,1] \\
    x' &\mapsto \begin{cases}
      1 & \text{iff } x = x' \\
      0 & \text{otherwise.}
    \end{cases}
  \end{align*}
\end{cor}

And $\Cat{Conv}$ is the category of algebras for this monad.

\begin{cor} \label{cor:conv-alg-dst}
  $\Cat{Conv} \cong \Alg(\Dst)$.
\end{cor}

Using Corollary \ref{cor:conv-alg-dst}, we can actually exhibit the relationship between the monad $\Dst$ and its defining adjunction tautologously: every monad $T$ on a category $\cat{C}$ induces an free-forgetful adjunction between its category of algebras $\Alg(T)$ and $\cat{C}$ itself, such that the monad generated by this adjunction is again $T$.
This is precisely the situation here.

\begin{prop} \label{prop:monad-alg-adjunction}
  Suppose $(T,\mu,\eta)$ is a monad on the category $\cat{C}$.
  There is a forgetful functor $U:\Alg(T)\to\cat{C}$ which has a left adjoint $F:\cat{C}\to\Alg(T)$ taking each object $X:\cat{C}$ to the \textit{free} $T$\textit{-algebra} $(TX,\mu_X)$ on $X$, where $\mu_X:TTX\to TX$ is the component of the monad multiplication $\mu$ at $X$.
  The unit of the adjunction is the monadic unit $\eta$, the counit $\epsilon$ is defined by $\epsilon_{(X,\varsigma_X)} := \varsigma_X$, and the monad induced by the adjunction is $(T,\mu,\eta)$.
  \begin{proof}[Proof sketch]
    The proof that $F$ is left adjoint to $U$ is standard (see \textcite[{Prop. 4.1.4}]{Borceux1994Handbook2}), and that the adjunction generates the monad follows almost immediately from Proposition \ref{prop:monad-from-adj}.
  \end{proof}
\end{prop}

\begin{rmk}
  It must be emphasized that, although every monad arises from such a free-forgetful adjunction, not every adjunction does!
  (Consider for example the adjunction $\Delta\dashv\lim$ of Proposition \ref{prop:limit-adjoint}: $\Delta$ does not assign to each $c:\cat{C}$ the ``free $J$-shaped diagram on $c$'', and $\lim$ does not simply forget diagrammatic structure.)
  Those adjunctions which do arise from monads in this way are called \textit{monadic}.
\end{rmk}

There is a special name for subcategories of \textit{free} algebras.

\begin{defn}
  Suppose $(T,\mu,\eta)$ is a monad on $\cat{C}$.
  The subcategory of $\Alg(T)$ on the free $T$-algebras $(TX,\mu_X)$ is called the \textit{Kleisli category} for $T$, and denoted $\Kl(T)$.
\end{defn}

The following proposition gives us an alternative presentation of $\Kl(T)$ which, when applied to the monad $\Dst$, will yield a computationally meaningful category of finitely-supported stochastic channels.

\begin{prop} \label{prop:kl-t}
  The objects of $\Kl(T)$ are the objects of $\cat{C}$.
  The morphisms $X\klto Y$ of $\Kl(T)$ are the morphisms $X\to TY$ of $\cat{C}$.
  Identity morphisms $\id_X:X\klto X$ are given by the monadic unit $\eta_X:X\to TX$.
  Composition is defined by \textit{Kleisli extension}: given $g:Y\klto Z$, we form its Kleisli extension $g^\rhd:TY\klto Z$ as the composite $TY\xto{Tg}TTZ\xto{\mu_Z}TZ$ in $\cat{C}$.
  Then, given $f:X\klto Y$, we form the composite $g\klcirc f:X\klto Z$ as $g^\rhd\circ f$: $X\xto{f}TY\xto{Tg}TTZ\xto{\mu_Z}TZ$.
  \begin{proof}
    Observe that there is a bijection between the objects $X$ of $\cat{C}$ and the free $T$-algebras $(TX,\mu_X)$.
    We therefore only need to establish a bijection between the hom-sets $\Alg(T)\bigl((TX,\mu_X),(TY,\mu_Y)\bigr)$ and $\Kl(T)(X,Y)$, with the latter defined as in the statement of the proposition.

    First, we demonstrate that Kleisli extension defines a surjection
    \[ \Kl(T)(X,Y)\to\Alg(T)\bigl((TX,\mu_X),(TY,\mu_Y)\bigr) \; . \]
    Suppose $\phi$ is any algebra morphism $(TX,\mu_X)\to(TY,\mu_Y)$; we show that it is equal to the Kleisli extension of the Kleisli morphism $X\xto{\eta_X}TX\xto{\phi}TY$:
    \begin{align*}
      TX \xto{(\phi\circ\eta_X)^\rhd} TY
      &= TX \xto{T\eta_X} TTX \xto{T\phi} TTY \xto{\mu_{TY}} TY \\
      &= TX \xto{T\eta_X} TTX \xto{\mu_{TX}} TX \xto{\phi} TY \\
      &= TX \xlongequal{\id_{TX}} TX \xto{\phi} TY \\
      &= TX\xto{\phi}TY
    \end{align*}
    where the first equality holds by definition, the second line by naturality of $\mu$, and the third by the unitality of the monad $(T,\mu,\eta)$.
    Hence every free algebra morphism is in the image of Kleisli extension, and so Kleisli extension defines a surjection.

    Next, we show that this surjection is additionally injective.
    Suppose $f,g$ are two Kleisli morphisms $X\to TY$ such that their Kleisli extensions are equal
    \[ TX\xto{Tf}TTY\xto{\mu_{TY}}TY = TX\xto{Tg}TTY\xto{\mu_{TY}}TY \]
    and recall that the identity in $\Kl(T)$ is $\eta$.
    We therefore have the following equalities:
    \begin{align*}
        X\xto{\eta_X}TX\xto{Tf}TTY\xto{\mu_{TY}}TY &= X\xto{\eta_X}TX\xto{Tg}TTY\xto{\mu_{TY}}TY \\
      = X\xto{f}TY\xto{T\eta_y}TTY\xto{\mu_{TY}}TY &= X\xto{g}TY\xto{T\eta_y}TTY\xto{\mu_{TY}}TY \\
      = X\xto{f}Y \qquad\qquad\qquad\qquad\quad\;\;\, &= X\xto{g}Y \; .
    \end{align*}
    where the equality in the first line holds \textit{ex hypothesi}, the second by naturality, and the third by monadic unitality.
    Since $f = g$ when their Kleisli extensions are equal, Kleisli extension is injective.
    Since it is also surjective, we have an isomorphism between $\Kl(T)(X,Y)$ and $\Alg(T)\bigl((TX,\mu_X),(TY,\mu_Y)\bigr)$.
    Hence $\Kl(T)$ is the subcategory of $\Alg(T)$ on the free algebras.
  \end{proof}
\end{prop}

If $T$ is a monad on $\cat{C}$, there is a canonical embedding of $\cat{C}$ into $\Kl(T)$.
In the case of $\Kl(\Dst)$, this will yield the subcategory of \textit{deterministic} channels: those which do not add any uncertainty.

\begin{prop}
  Suppose $T$ is a monad on $\cat{C}$.
  Then there is an identity-on-objects embedding $\cat{C}\hookrightarrow\Kl(T)$ given on morphisms by mapping $f:X\to Y$ in $\cat{C}$ to the Kleisli morphism $X\xto{\eta_X}TX\xto{Tf}TY$.
  \begin{proof}[Proof sketch]
    Functoriality follows from the unitality of $\eta$ in the monad structure, since Kleisli composition involves post-composing the monad multiplication, and $\mu_T\circ T\eta = \id$.
  \end{proof}
\end{prop}

\subsubsection{Stochastic matrices} \label{sec:stoch-mat}

At this point, let us exhibit $\Kl(\Dst)$ a little more concretely, by instantiating Proposition \ref{prop:kl-t}.
Since a distribution $\pi$ on the set $X$ is a function $X\to[0,1]$, and following the ``formal sum'' intuition, we can alternatively think of $\pi$ as a vector, whose coefficients are indexed by elements of $X$ (the basis vectors $\ket{x}$).
Morphisms $X\klto Y$ in $\Kl(\Dst)$ are functions $X\to\Dst Y$, and so we can similarly think of these as stochastic matrices, by the Cartesian closure of $\Set$: a function $X\to\Dst Y$ is equivalently a function $X\to[0,1]^Y$, which in turn corresponds to a function $X\times Y\to[0,1]$, which we can read as a (left) stochastic matrix, with only finitely many nonzero coefficients and each of whose columns must sum to $1$.
We will adopt `conditional probability' notation for the coefficients of these matrices: given \(p : X \klto Y\), \(x \in X\) and \(y \in Y\), we write \(p(y|x) := p(x)(y) \in [0, 1]\) for ``the probabilty of \(y\) \emph{given} \(x\), according to \(p\)''.

Composition in $\Kl(\Dst)$ is then matrix multiplication:
given \(p : X \to \Dst Y\) and \(q : Y \to \Dst Z\), we compute  their composite \(q \klcirc p : X \to \Dst Z\) by `averaging over' or `marginalizing out' \(Y\) via the Chapman-Kolmogorov equation:
\begin{equation*}
q \klcirc p : X \to \Dst Z := x \mapsto \sum_{z : Z} \, \boxed{\sum_{y : Y} q(z|y) \cdot p(y|x)} \, \ket{z} .
\end{equation*}
Here we have again used the formal sum notation, drawing a box to indicate the coefficients (\textit{i.e.}, the probabilities returned by the conditional distribution $q\klcirc p(x)$ for each atomic event $z$ in $Z$).

Via the monadic unit, identity morphisms \(\id_X : X \klto X\) in \(\Kl(\Dst)\) take points to `Dirac delta' distributions: \(\id_X := x \mapsto 1 \ket{x}\).
The embedding $\Set\hookrightarrow\Kl(\Dst)$ makes any function \(f : Y \to X\) into a (deterministic) channel \(\overline{f} = \eta_X \circ f : Y \to \Dst X\) by post-composing with \(\eta_X\).

\subsubsection{Monoidal structure}

We will want to equip $\Kl(\Dst)$ with a copy-discard category structure, in order to represent joint states (joint distributions) and their marginalization, as well as the copying of information.
The first ingredient making a copy-discard category, after the category itself, is a monoidal structure.
Once again, in the case of $\Kl(\Dst)$, this can be obtained abstractly from a more fundamental structure---the categorical product $(\times, 1)$ on $\Set$---as a consequence of $\Dst$ being a `monoidal' monad.
We will write the induced tensor product on $\Kl(\Dst)$ as $\otimes$; its monoidal unit remains the object $1$.

\begin{defn} \label{def:monoidal-monad}
  A \textit{monoidal monad} is a monad in $\MonCat$.
  This means that it is a monad $(T,\mu,\eta)$ in $\Cat{Cat}$ whose functor $T:\cat{C}\to\cat{C}$ is additionally equipped with a lax monoidal structure $(\alpha,\epsilon)$ such that the monad multiplication $\mu$ and unit $\eta$ are monoidal natural transformations accordingly.
\end{defn}

With this extra structure, it is not hard to verify that the following proposition makes $\Kl(T)$ into a well-defined monoidal category.

\begin{prop}
  The Kleisli category $\Kl(T)$ of a monoidal monad $(T,\alpha,\epsilon,\mu,\eta)$ is a monoidal category.
  The monoidal product is given on objects by the monoidal product $\otimes$ of the base category $\cat{C}$.
  On Kleisli morphisms $f:X\klto Y$ and $f':X'\klto Y'$, their tensor $f\otimes g$ is given by the following composite in $\cat{C}$:
  \[ X\otimes X' \xto{f\otimes f'} TX\otimes TX' \xto{\alpha_{X,X'}} T(X\otimes X') \]
  The monoidal unit is the monoidal unit $I$ in $\cat{C}$.
  The associator and unitor of the monoidal category structure are inherited from $\cat{C}$ under the embedding $\cat{C}\hookrightarrow\Kl(T)$.
  When $(\cat{C},\otimes,I)$ is symmetric monoidal, then so is $(\Kl(T),\otimes,I)$.
\end{prop}

In the specific case of $\Kl(\Dst)$, the tensor product $\otimes$ is given on objects by the product of sets and on stochastic channels \(f : X \to \Dst A\) and \(g : Y \to \Dst B\) as
\begin{equation*}
X \times Y \xto{f \times g} \Dst A \times \Dst B \xrightarrow{\alpha_{A,B}} \Dst(A \times B) \, .
\end{equation*}
Note that because not all joint states have independent marginals, the monoidal product \(\otimes\) is not Cartesian:
that is, given an arbitrary \(\omega : \Dst (X \otimes Y)\), we do not necessarily have \(\omega = (\rho, \sigma)\) for some \(\rho : \Dst X\) and \(\sigma : \Dst Y\).
The laxator takes a pair of distributions \((\rho, \sigma)\) in \(\Dst X \times \Dst Y\) to the joint distribution on \(X \times Y\) given by \((x, y) \mapsto \rho(x) \cdot \sigma(y)\); \(\rho\) and \(\sigma\) are then the (independent) marginals of this joint distribution.
(Of course, the joint distribution $(\rho,\sigma)$ is not the only joint distribution with those marginals: other joint states may have these marginals but also correlations between them, and this is what it means for not all joint states to have independent marginals.)

Since $(\Set,\times,1)$ is symmetric monoidal, \(\Kl(\Dst)\) is too, with swap isomorphisms  \(\mathsf{swap}_{X,Y} : X \otimes Y \xklto{\sim} Y \otimes X\) similarly inherited form those of $\times$.

\subsubsection{Copy-discard structure}

The copy-discard structure in $\Kl(\Dst)$ is inherited from $\Set$ through its embedding:
since every object in $\Kl(\Dst)$ is an object in $\Set$, and every object in $\Set$ is a comonoid (Example \ref{ex:prod-comon}), and since functors preserve equalities, these comonoid structures are preserved under the embedding.
More explicitly, the discarding channels $\ground_X$ are given by $x\mapsto 1\ket{\ast}$, and the copiers $\bcopier_X$ by $x\mapsto 1\ket{x,x}$.
Note that the copiers are not natural in \(\Kl(\Dst)\): in general, $\bcopier\klcirc f\neq f\otimes f\klcirc\bcopier$, as a result of the possibility of correlations.

Since the projections $\proj_X:X\times Y\to X$ in $\Set$ satisfy $\proj_X = \rho_X\circ(\id_X\times\ground_Y)$ where $\rho_X:X\times1\to X$ is component of the right unitor, we can see how discarding and projection give us marginalization, thereby explaining the string diagrams of \secref{sec:str-diag}.
Given some joint state $\omega:1\klto X\otimes Y$, its $X$-marginal $\omega_X:1\klto X$ is given by $\proj_X\klcirc\omega$, which in $\Kl(\Dst)$ is given by the formal sum formula $\sum_{x:X}\boxed{\textstyle{\sum_{y:Y}\omega(x,y)}}\ket{x}$, where we have again drawn a box to distinguish the probability assigned to $\ket{x}$, which we note coincides with the classical rule for marginal discrete probability. (The $Y$-marginal is of course symmetric.)

\begin{rmk} \label{rmk:semicartesian}
  A \textit{semicartesian} category is a monoidal category in which the monoidal unit is terminal.
  In a semicartesian monoidal category, every tensor product $X\otimes Y$ is equipped with a natural family of projections $\proj_X:X\otimes Y\to X$ and $\proj_Y:X\otimes Y\to Y$ given by `discarding' one of the factors and using the unitor; the existence of such projections is not otherwise implied by a monoidal structure (though of course it does follow when the tensor is the product).

  A related notion is that of an \textit{affine} functor, which is one that preserves the terminal object, and of which $\Dst$ is an example.
  As a result, and following the discussion above, we can see that $\Kl(\Dst)$ is an example of a semicartesian category.

  Semicartesian copy-discard categories are also known as \textit{Markov categories}, following \textcite{Fritz2019synthetic}.
\end{rmk}

\begin{rmk} \label{rmk:comon-determ}
  Since $1$ is therefore terminal in $\Kl(\Dst)$, Proposition \ref{prop:comon-morph-prod} tells us that those channels $f$ that do commute with copying (\textit{i.e.}, for which $\bcopier$ is natural; Remark \ref{rmk:nat-determ}), and which are therefore comonoid morphisms, are precisely the \textit{deterministic} channels: those in the image of the embedding of $\Set$ (and which therefore emit Dirac delta distributions).
  As a result, we can think of $\Comon(\Kl(\Dst))$ as the subcategory of deterministic channels, and write $\Comon(\Kl(\Dst))\cong\Set$.
  (Intuitively, this follows almost by definition: a deterministic process is one that has no informational side-effects; that is to say, whether we copy a state before performing the process on each copy, or perform the process and then copy the resulting state, or whether we perform the process and then marginalize, or just marginalize, makes no difference to the resulting state.)
\end{rmk}

\subsubsection{Bayesian inversion}

We can now instantiate Bayesian inversion in \(\Kl(\Dst)\), formalizing Equation \eqref{eq:pr-bayes}.
Given a channel \(p : X \to \Dst Y\) satisfying the condition in Remark \ref{rmk:bayes-supp} below, its Bayesian inversion is given by the function
\begin{equation} \label{eq:bayes-discr}
p^\dag : \Dst X \times Y \to \Dst X := (\pi, y) \mapsto \sum_{x : X} \, \boxed{\frac{p(y|x) \cdot \pi(x)}{\sum_{x':X} p(y|x') \cdot \pi(x')}} \, \ket{x} = \sum_{x : X} \, \boxed{\frac{p(y|x) \cdot \pi(x)}{(p \klcirc \pi)(y)}} \, \ket{x}
\end{equation}
so that the Bayesian update of $p$ along $\pi$ is the conditional distribution defined by
\[ p^\dag_\pi(x|y) = \frac{p(y|x) \cdot \pi(x)}{(p \klcirc \pi)(y)} \; . \]
Note that here we have used the Cartesian closure of $\Set$, writing the type of $p^\dag$ as $\Dst X\times Y\to\Dst X$ rather than $\Dst X\to\Kl(\Dst)(Y,X)$, where $\Kl(\Dst)(Y,X) = (\Dst X)^Y$.

\begin{rmk} \label{rmk:bayes-supp}
  In the form given above, $p^\dag$ is only well-defined when the support of $p\klcirc\pi$ is the whole of $Y$, so that, for all $y$, $(p\klcirc\pi)(y) > 0$; otherwise, the division is ill-defined.
  Henceforth, in the context of Bayesian inversion, we will therefore assume that $p\klcirc\pi$ has full support (see Definition \ref{def:admit-bayes}).

  To avoid this (rather ugly) condition, one can replace it by the assumption that the notion of `support' is well-defined, and modify the type of $p^\dag$ accordingly: this is the refinement made by \textcite{Braithwaite2022Dependent}, and were it not for the presently-uncertain nature of support objects in general, it would now be this author's preferred approach.
  This leads to writing the type of the inversion $p^\dag$ as $\sum_{\pi:\Dst X}\mathsf{supp}(p\klcirc \pi)\to\Dst X$, where $\mathsf{supp}(p\klcirc \pi)$ is the subobject of $Y$ on which $p\klcirc\pi$ is supported: with this type, $p^\dag_\pi$ is always a well-defined channel.
  One can then proceed with the definition of `dependent' Bayesian lenses accordingly; for the details, we refer the reader to \textcite{Braithwaite2022Dependent}.
  In this thesis, for simplicity of exposition and faithfulness to this author's earlier work, we will proceed under the full-support assumption.
\end{rmk}

\subsection{Abstract Bayesian inversion}
\label{sec:abstract-bayes}

Beyond the concerns of Remark \ref{rmk:bayes-supp}, in a more general setting it is not always possible to define Bayesian inversion using an equation like Equation \eqref{eq:bayes-discr} or Equation \eqref{eq:pr-bayes}: the expression $p(y|x)$ might not be well-defined, or there might not be a well-defined notion of division.
Instead being guided by Equation \eqref{eq:pr-bayes} in defining Bayesian inversion, we can use Equation \eqref{eq:pr-disintegrations}.
Therefore, supposing a channel $c:X\klto Y$ and a state $\pi:I\klto X$ in an ambient copy-discard category $\cat{C}$, we can ask for the Bayesian inversion $c^\dag_\pi$ to be any channel satisfying the graphical equality \parencite[eq. 5]{Cho2017Disintegration}:
\begin{equation} \label{eq:bayes-abstr}
\input{img/joint-c-pi.tikz} \quad = \quad \input{img/joint-cdag-c-pi.tikz}
\end{equation}
This diagram can be interpreted as follows.
Given a prior \(\pi : I \klto X\) and a channel \(c : X \klto Y\), we form the joint distribution \(\omega := (\id_X \otimes \, c) \klcirc \bcopier_X \klcirc \pi : I \klto X \otimes Y\) shown on the left hand side:
this formalizes the product rule, \(\Pr_\omega(A, B) = \Pr_c(B | A) \cdot \Pr_\pi(A)\), and \(\pi\) is the corresponding \(X\text{-marginal}\).
As in the concrete case of \(\Kl(\Dst)\), we seek an inverse channel \(Y \klto X\) witnessing the `dual' form of the rule, \(\Pr_\omega(A, B) = \Pr(A | B) \cdot \Pr(B)\); this is depicted on the right hand side.
By discarding \(X\), we see that \(c \klcirc \pi : I \klto Y\) is the \(Y\text{-marginal}\) witnessing \(\Pr(B)\).
So any channel \(c^\dag_\pi : Y \klto X\) witnessing \(\Pr(A | B)\) and satisfying the equality above is a Bayesian inverse of \(c\) with respect to \(\pi\).

In light of Remark \ref{rmk:bayes-supp}, we therefore make the following definition.

\begin{defn} \label{def:admit-bayes}
  We say that a channel \(c : X \klto Y\) \textit{admits Bayesian inversion} with respect to \(\pi : I \klto X\) if there exists a channel \(c^\dag_\pi : Y \klto X\) satisfying equation \eqref{eq:bayes-abstr}. We say that \(c\) admits Bayesian inversion \emph{tout court} if \(c\) admits Bayesian inversion with respect to all states \(\pi : I \klto X\).
\end{defn}

\begin{rmk} \label{rmk:zero-inversions}
  We need to be careful about the existence of inversions as a consequence of the fact that $c\klcirc\pi$ may not always be fully supported on $Y$ (recall Remark \ref{rmk:bayes-supp}).
  In this thesis we will henceforth assume that $c\klcirc\pi$ \textit{is} always fully supported, in order to keep the exposition clear.
  This is justified in two ways: first, because we can always restrict to a wide subcategory all of whose channels do admit inversion; and second, because we may equally work with dependent Bayesian lenses (as described by \textcite{Braithwaite2022Dependent} and noted in Remark \ref{rmk:bayes-supp}).
\end{rmk}

\subsection{Density functions}
\label{sec:density-functions}

Abstract Bayesian inversion \eqref{eq:bayes-abstr} generalizes the product rule form of Bayes' theorem \eqref{eq:pr-disintegrations} but in most applications, we are interested in a specific channel witnessing \(\Pr(A | B) = \Pr(B | A) \cdot \Pr(A) / \Pr(B)\).
In the typical measure-theoretic setting, this is often written informally as
\begin{equation} \label{eq:bayes-density-informal}
p(x|y) 
= \frac{p(y|x) \cdot p(x)}{p(y)}
= \frac{p(y|x) \cdot p(x)}{\int_{x':X} p(y|x') \cdot p(x') \, \d x'}
\end{equation}
but the formal semantics of such an expression are not trivial: for instance, what is the object \(p(y|x)\), and how does it relate to a channel \(c : X \klto Y\)?

Following \textcite{Cho2017Disintegration}, we can interpret \(p(y|x)\) as a \textit{density function} for a channel, abstractly witnessed by an effect \(X \otimes Y \klto I\) in our ambient category \(\cat{C}\).
Consequently, \(\cat{C}\) cannot be semicartesian---as this would trivialize all density functions---though it must still supply comonoids.
We can think of this as expanding the collection of channels in the category to include acausal or `partial' maps and unnormalized distributions or states.

\begin{ex}
  An example of such a category is \(\Kl(\Dst_{\leq 1})\), whose objects are sets and whose morphisms \(X \klto Y\) are functions \(X \to \Dst(Y + 1)\).
  Then a stochastic map is partial if it sends any probability to the added element \(\ast\), and the subcategory of total (equivalently, causal) maps is \(\Kl(\Dst)\) (see \parencite{Cho2015Introduction} for more details).

  A morphism $X\klto 1$ in $\Kl(\Dst_{\leq 1})$ is therefore a function $X\to\Dst(1+1)$.
  Now, a distribution $\pi$ on $1+1$ is the same as a number $\bar{\pi}$ in $[0,1]$: note that $1+1$ has two points, and so $\pi$ assigns $\bar{\pi}$ to one of them and $1-\bar{\pi}$ to the other.
  Therefore an effect $X\klto 1$ is equivalently a function $X\to[0,1]$, which is precisely the type we expect for a density function.
\end{ex}

We therefore adopt the following abstract definition.

\begin{defn}[{Density functions \parencite[{Def. 8.1}]{Cho2017Disintegration}}] \label{def:density-functions}
  A channel \(c : X \klto Y\) is said to be \textit{represented by an effect} \(p : X \otimes Y \klto I\) with respect to \(\mu : I \klto Y\) if
  \[ \input{img/def-density-function-c.tikz}. \]
  In this case, we call \(p\) a \textit{density function} for \(c\).
\end{defn}

We will also need the concepts of almost-equality and almost-invertibility.

\begin{defn}[{Almost-equality, almost-invertibility \parencite[{Def. 8.2}]{Cho2017Disintegration}}] \label{def:almost-eq}
  Given a state \(\pi : I \klto X\), we say that two channels \(c : X \klto Y\) and \(d : X \klto Y\) are \(\mathbf{\pi}\textit{-almost-equal}\), denoted \(c \overset{\pi}{\sim} d\), if
  \[ \input{img/joint-c-pi.tikz} \ =\ \input{img/joint-d-pi.tikz} \]
  and we say that an effect \(p : X \klto I\) is \(\mathbf{\pi}\textit{-almost-invertible}\) with \(\mathbf{\pi}\textit{-almost-inverse } q : X \klto I\) if
  \[ \input{img/almost-invertibility.tikz}. \]
\end{defn}

The following basic results about almost-equality will prove helpful.

\begin{prop}[Composition preserves almost-equality] \label{prop:comp-preserve-almost-eq}
  If \(c \overset{\pi}{\sim} d\), then \(f \klcirc c \overset{\pi}{\sim} f \klcirc d\).
  \begin{proof}
    Immediate from the definition of almost-equality.
  \end{proof}
\end{prop}

\begin{prop}[Almost-inverses are almost-equal] \label{prop:almost-inverse-almost-equal}
  Suppose \(q : X \klto I\) and \(r : X \klto I\) are both \(\pi\text{-almost-inverses}\) for the effect \(p : X \klto I\).
  Then \(q \overset{\pi}{\sim} r\).
  \begin{proof}
    By assumption, we have
    \[ \input{img/almost-inv-almost-eq-1.tikz} \; . \]
    Then, by the definition of almost-equality (Definition \ref{def:almost-eq}):
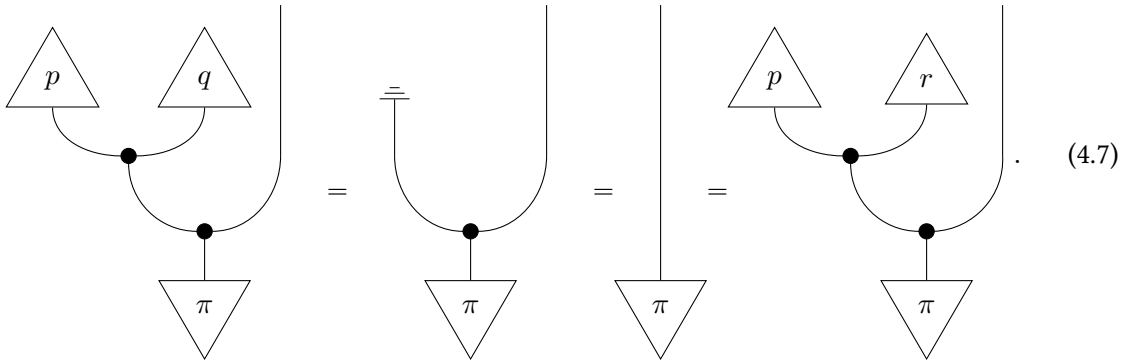
    \begin{equation} \label{eq:almost-inverse-almost-equal-1}
      \input{img/almost-inv-almost-eq-2.tikz} \; .
    \end{equation}
    We seek to show that
    \begin{equation} \label{eq:almost-inverse-almost-equal-2}
      \input{img/almost-inv-almost-eq-3.tikz} \; .
    \end{equation}
    Substituting the right-hand-side of \eqref{eq:almost-inverse-almost-equal-1} for $\pi$ in the left-hand-side of \eqref{eq:almost-inverse-almost-equal-2}, we have that
    \[ \input{img/almost-inv-almost-eq-4.tikz} \]
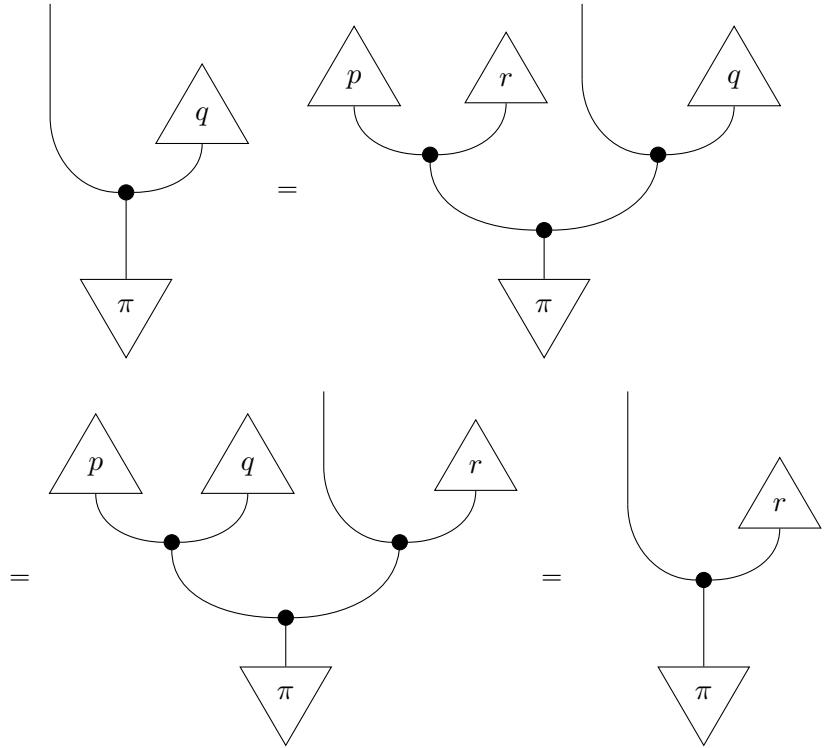
    \[ \input{img/almost-inv-almost-eq-5.tikz} \]
    which establishes the result.
    The second equality follows by the coassociativity of $\bcopier$ and the third by its counitality.
\end{proof}
\end{prop}

With these notions, we can characterise Bayesian inversion via density functions.
The result is due to \textcite{Cho2017Disintegration}, but we include the graphical proof for expository completeness, as an example of string-diagrammatic reasoning.

\begin{prop}[{Bayesian inversion via density functions \parencite[{Thm. 8.3}]{Cho2017Disintegration}}] \label{prop:bayes-density-graph}
  Suppose \(c : X \klto Y\) is represented by the effect \(p\) with respect to \(\mu\).
  The Bayesian inverse \(c^\dag_\pi : Y \klto X\) of \(c\) with respect to \(\pi : I \klto X\) is given by
  \[ \input{img/channel-density-cdag-pi.tikz} \]
  where \(p^{-1} : Y \klto I\) is a \(\mu\text{-almost-inverse}\) for the effect
  \[ \input{img/likelihood-p-pi.tikz} \]
  \begin{proof}
    We seek to establish the relation \eqref{eq:bayes-abstr} characterizing Bayesian inversion.
    By substituting the density function representations for $c$ and $c^\dag_\pi$ into the right-hand-side of \eqref{eq:bayes-abstr}, we have
    \[ \input{img/bayes-density-graph-1.tikz} \]
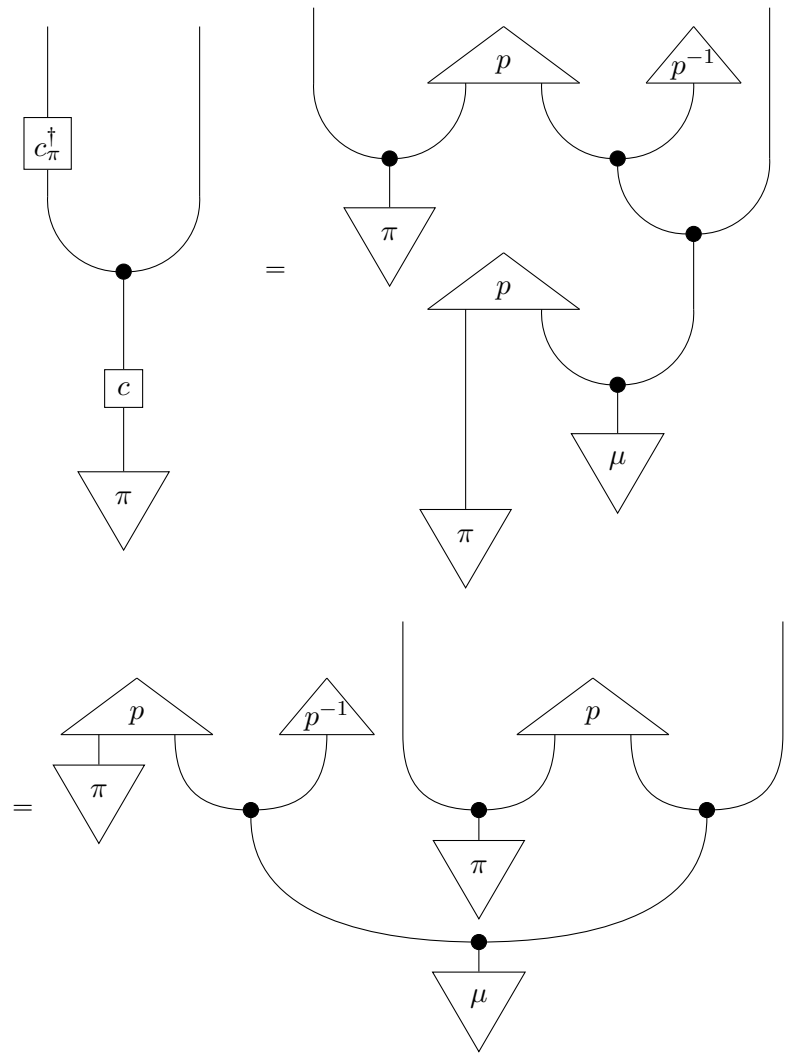
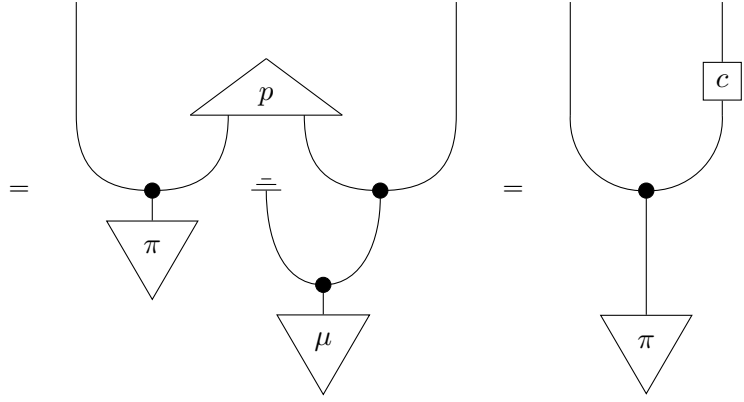
    \[ \input{img/bayes-density-graph-2.tikz} \]
    \[ \input{img/bayes-density-graph-3.tikz} \]
    as required.
    The second equality holds by the coassociativity of $\bcopier$, the third since $p^{-1}$ is an almost-inverse \emph{ex hypothesi}, and the fourth by the counitality of $(\bcopier,\ground)$ and the density function representation of $c$.
\end{proof}
\end{prop}

The following proposition is an immediate consequence of the definition of almost-equality and of the abstract characterisation of Bayesian inversion \eqref{eq:bayes-abstr}.
We omit the proof.
\begin{prop}[Bayesian inverses are almost-equal] \label{prop:bayes-almost-equal}
  Suppose \(\alpha : Y \klto X\) and \(\beta : Y \klto X\) are both Bayesian inversions of the channel \(c : X \klto Y\) with respect to \(\pi : I \klto X\).
  Then \(\alpha \overset{c \klcirc \pi}{\sim} \beta\).
\end{prop}

\subsection{S-finite kernels}
\label{sec:sfKrn}

To represent channels by concrete density functions, we can work in the category  \(\Cat{sfKrn}\) of measurable spaces and s-finite kernels.
We will only sketch the structure of this category, and refer the reader to \citet{Cho2017Disintegration,Staton2017Commutative} for elaboration.

Objects in \(\Cat{sfKrn}\) are measurable spaces \((X, \Sigma_X)\); often we will just write \(X\), and leave the \(\sigma\text{-algebra } \Sigma_X\) implicit.
Morphisms \((X, \Sigma_X) \klto (Y, \Sigma_Y)\) are s-finite kernels.
A \emph{kernel} \(k\) from \(X\) to \(Y\) is a function \(k : X \times \Sigma_Y \to [0, \infty]\) satisfying the following conditions:
\begin{itemize}
\item for all \(x \in X\), \(k(x, -) : \Sigma_Y \to [0, \infty]\) is a measure; and
\item for all \(B \in \Sigma_Y\), \(k(-, B) : X \to [0, \infty]\) is measurable.
\end{itemize}
A kernel \(k : X \times \Sigma_Y \to [0, \infty]\) is \emph{finite} if there exists some \(r \in [0, \infty)\) such that, for all \(x \in X\), \(k(x, Y) \leq r\).
And \(k\) is \emph{s-finite} if it is the sum of at most countably many finite kernels \(k_n\), \(k = \sum_{n : \nn} k_n\).

Identity morphisms \(\id_X : X \klto X\) are Dirac kernels \(\delta_X : X \times \Sigma_X \to [0, \infty] := x \times A \mapsto 1\) iff \(x \in A\) and 0 otherwise.
Composition is given by a Chapman-Kolmogorov equation, analogously to composition in \(\Kl(\Dst)\).
Suppose \(c : X \klto Y\) and \(d : Y \klto Z\).
Then
\[
d \klcirc c : X \times \Sigma_Z \to [0, \infty]
:= x \times C \mapsto \int_{y:Y} d(C|y) \, c(\d y| x)
\]
where we have again used the `conditional probability' notation \(d(C|y) := d \circ (y \times C)\).
Reading \(d(C|y)\) from left to right, we can think of this notation as akin to reading the string diagrams from top to bottom, \emph{i.e.} from output(s) to input(s).

\paragraph{Monoidal structure on \(\Cat{sfKrn}\)}

There is a monoidal structure on \(\Cat{sfKrn}\) analogous to that on \(\Kl(\Dst)\).
On objects, \(X \otimes Y\) is the Cartesian product \(X \times Y\) of measurable spaces.
On morphisms, \(f \otimes g : X \otimes Y \klto A \otimes B\) is given by
\[
f \otimes g : (X \times Y) \times \Sigma_{A \times B}
:= (x \times y) \times E \mapsto \int_{a:A} \int_{b:B} \delta_{A \otimes B}(E|x, y) \, f(\d a|x) \, g(\d b|y)
\]
where, as above, \(\delta_{A \otimes B}(E|a, b) = 1\) iff \((a, b) \in E\) and 0 otherwise.
Note that \((f \otimes g)(E|x, y) = (g \otimes f)(E|y, x)\) for all s-finite kernels (and all \(E\), \(x\) and \(y\)), by the Fubini-Tonelli theorem for s-finite measures \citep{Cho2017Disintegration,Staton2017Commutative}, and so \(\otimes\) is symmetric on \(\Cat{sfKrn}\).

The monoidal unit in \(\Cat{sfKrn}\) is again \(I = 1\), the singleton set.
Unlike in \(\Kl(\Dst)\), however, we do have nontrivial effects \(p : X \klto I\), given by kernels \(p : (X \times \Sigma_1) \cong X \to [0, \infty]\), with which we will represent density functions.

\paragraph{Comonoids in \(\Cat{sfKrn}\)}

Every object in \(\Cat{sfKrn}\) is a comonoid, analogously to \(\Kl(\Dst)\).
Discarding is given by the family of effects \(\ground_X : X \to [0, \infty] := x \mapsto 1\), and copying is again Dirac-like: \(\bcopier_X : X \times \Sigma_{X \times X} := x \times E \mapsto 1\) iff \((x, x) \in E\) and 0 otherwise.
Because we have nontrivial effects, discarding is only natural for causal or `total' channels:
if \(c\) satisfies \(\ground \klcirc c = \ground\), then \(c(-|x)\) is a probability measure for all \(x\) in the domain\footnote{%
This means that the subcategory of total maps in \(\Cat{sfKrn}\) is equivalent to the Kleisli category $\Kl(\Giry)$ of the \emph{Giry monad} $\Giry$ taking each measurable space \(X\) to the space \(\Giry X\) of measures over \(X\); see Example \ref{ex:giry-monad} for more details.}.
And, once again, copying is natural (that is, \(\bcopier \klcirc c = (c \otimes c) \klcirc \bcopier\)) if and only if the channel is deterministic.

\paragraph{Channels represented by effects}

We can interpret the string diagrams of \secref{sec:density-functions} in \(\Cat{sfKrn}\), and we will do so by following the intuition of the conditional probability notation and reading the string diagrams from outputs to inputs.
Hence, if \(c : X \klto Y\) is represented by the effect \(p : X \otimes Y \klto I\) with respect to the measure \(\mu : I \klto Y\), then
\[
c : X \times \Sigma_Y \to [0, \infty]
:= x \times B \mapsto \int_{y:B} \mu(\d y) \, p(y | x) .
\]
Note that we also use conditional probability notation for density functions, and so \(p(y|x) := p \circ (x \times y)\).

Suppose that \(c : X \klto Y\) is indeed represented by \(p\) with respect to \(\mu\), and that \(d : Y \klto Z\) is represented by \(q : Y \otimes Z \klto I\) with respect to \(\nu : I \klto Z\).
Then in \(\Cat{sfKrn}\), \(d \klcirc c : X \klto Z\) is given by
\[
d \klcirc c : X \times \Sigma_Z 
:= x \times C \mapsto \int_{z:C} \nu(\d z) \, \int_{y:Y} q(z|y) \, \mu(\d y) \, p(y|x)
\]
Alternatively, by defining the effect \((p \mu q) : X \otimes Z \klto I\) as
\[
(p \mu q) : X \times Z \to [0, \infty]
:= x \times z \mapsto \int_{y:Y} q(z|y) \, \mu(\d y) \, p(y|x),
\]
we can write \(d \klcirc c\) as
\[
d \klcirc c : X \times \Sigma_Z
:= x \times C \mapsto \int_{z:C} \nu(\d z) \, (p \mu q)(z|x) .
\]

\paragraph{Bayesian inversion via density functions}

Once again writing \(\pi : I \klto X\) for a prior on X, and interpreting the string diagram of Proposition \ref{prop:bayes-density-graph} for \(c^\dag_\pi : Y \klto X\) in \(\Cat{sfKrn}\), we have
\begin{equation} \label{eq:bayes-density-krn}
\begin{aligned}
c^\dag_\pi : Y \times \Sigma_X \to [0, \infty]
:= y \times A & \mapsto \left( \int_{x:A} \pi(\d x) \, p(y|x) \right) p^{-1}(y) \\
&= p^{-1}(y) \int_{x:A} p(y|x) \, \pi(\d x) ,
\end{aligned}
\end{equation}
where \(p^{-1} : Y \klto I\) is a \(\mu\text{-almost-inverse}\) for effect \(p \klcirc (\pi \otimes \id_Y)\), and is given up to \(\mu\text{-almost-equality}\) by
\[
p^{-1} : Y \to [0, \infty]
:= y \mapsto \left( \int_{x:X} p(y|x) \, \pi(\d x) \right)^{-1} \, .
\]
Note that from this we recover the informal form of Bayes' rule for measurable spaces \eqref{eq:bayes-density-informal}. Suppose \(\pi\) is itself represented by a density function \(p_\pi\) with respect to the Lebesgue measure \(\d x\). Then
\[
c^\dag_\pi (A|y) = \int_{x:A} \, \frac{p(y|x) \, p_\pi(x)}{\int_{x':X} \, p(y|x') \, p_\pi(x') \, \d x'} \, \d x.
\]

\subsection{On probability monads} \label{sec:prob-monads}

Later, it will at times be helpful to work in a category of stochastic channels that is the Kleisli category for a monad, without fixing that monad in advance; in this case we will speak of a \textit{probability monad}.
Unfortunately, an abstract characterization of probability monads is not presently known to the author, and so we use this term informally.
However, when we do so, we have in mind a monoidal monad that maps spaces to spaces of measures or valuations on them, and that maps morphisms to the corresponding pushforwards.
In the setting of finitary probability, we have already seen one example, the monad $\Da$ explored in \secref{sec:fin-prob}.
Here we note the existence of others.

\begin{ex}[{Giry monad \parencite{Giry1982Categorical}}] \label{ex:giry-monad}
  Let $\Cat{Meas}$ denote the category of measurable spaces, whose objects are sets equipped with $\sigma$-algebras and whose morphisms are measurable functions.
  The \textit{Giry monad} $\Ga:\Cat{Meas}\to\Cat{Meas}$ maps each measurable space $(X,\Sigma_X)$ to the space $\Ga X$ of probability measures $\alpha:\Sigma_X\to[0,1]$ over it, equipped with the smallest $\sigma$-algebra making the evaluation functions
  \begin{align*}
    \mathsf{ev}_U : \Ga X &\to [0,1] \\
    \alpha &\mapsto \alpha(U)
  \end{align*}
  measurable for all $U \in \Sigma_X$.
  Given a measurable function $f:X\to Y$, the function $\Ga f : \Ga X\to \Ga Y$ is defined by pushforwards: that is, for each $\alpha : \Ga X$, we define
  \begin{align*}
    \Ga f(\alpha) : \Sigma_Y &\to [0,1] \\
    V &\mapsto \alpha\bigl(f^{-1}(V)\bigr) \; .
  \end{align*}
  (We may also write $f_\ast\alpha$ to denote $\Ga f(\alpha)$.)
  The unit of the monad $\eta$ has components $\eta_X : X\to \Ga X$ mapping each point $x$ to the corresponding Dirac measure $\delta_x$, which is defined by $\delta_x(U) = 1$ iff $x\in U$ and $\delta_x(U) = 0$ otherwise.
  Finally, the multiplication $\mu$ has components $\mu_X : \Ga\Ga X\to \Ga X$ defined by integration, analogous to the `evaluation' of $\Da$ (Def. \ref{def:cvx-eval}): for each $\nu : \Ga\Ga X$, define
  \begin{align*}
    \mu_X(\nu) : \Sigma_X &\to [0,1] \\
    U &\mapsto \int_{\alpha:\Ga X} \alpha(U) \, \d\nu \; .
  \end{align*}
  Note that the subcategory of total morphisms in $\Cat{sfKrn}$ is equivalent to $\Kl(\Ga)$.
\end{ex}

The category $\Cat{Meas}$ has all finite limits (it has products and equalizers), and this will mean that we will be able in Chapter \ref{chp:coalg} to define ``effectful polynomials'' in $\Kl(\Ga)$, and hence obtain categories of continuous-time continuous-space open Markov processes.
However, $\Cat{Meas}$ does not have exponentials and is therefore not Cartesian closed, because the evaluation function $\mathsf{ev}_{\rr,\rr}:\Cat{Meas}(\rr,\rr)\times\rr\to\rr:(f,x)\mapsto f(x)$ is not measurable, for any choice of $\sigma$-algeba on the function space $\Cat{Meas}(\rr,\rr)$ \parencite{Aumann1961Borel}.
This means that $\Kl(\Ga)$ cannot be enriched in $\Cat{Meas}$, and so we cannot define Bayesian lenses internally to $\Cat{Meas}$.
Circumnavigating this obstruction would complicate our construction of cilia --- dynamical systems that control lenses --- which are central to our formalization of predictive coding.
This is because the output maps of stochastic dynamical systems are deterministic functions: in the case of systems in $\Kl(\Ga)$, this means they are morphisms in $\Cat{Meas}$; for a general probability monad $\Pa:\cat{E}\to\cat{E}$, they are morphisms in $\cat{E}$.
For a system to be able to emit a lens, therefore, the hom objects of $\BLens{}$ must be objects in $\cat{E}$, and this in turn requires $\Kl(\Pa)$ to be enriched in $\cat{E}$.
Fortunately, as the following example notes, a suitable probability monad does exist.

\begin{ex}[{Quasi-Borel spaces \parencite{Heunen2017Convenient}}] \label{ex:qbs}
  A \textit{quasi-Borel space} is a set $X$ equipped with a set $M_X$ of `random variables' on $X$ taking samples from the real line, $M_X \subset X^{\rr}$.
  The set $M_X$ is taken to satisfy three closure properties: (i) $M_X$ contains all constant functions; (ii) $M_X$ is closed under composition with measurable functions, such that if $\rho \in M_X$ and $f:\rr\to\rr$ is measurable with respect to the standard Borel structure on $\rr$, then $\rho\circ f \in M_X$; and (iii) $M_X$ is closed under gluing `disjoint Borel domains', meaning that if $\rr$ is countably partitioned by $\rr \cong \sum_{i:\nn} S_i$, and if $\{\alpha_i\}_{i:\nn} \subset M_X$, then the function $(x\in S_i)\mapsto\alpha_i(x)$ is in $M_X$.
  A function $f:X\to Y$ is a morphism of quasi-Borel spaces if for all $\rho\in M_X$, $f\circ\rho \in M_Y$.
  Quasi-Borel spaces and their morphisms form a category, $\Cat{QBS}$, and this category is Cartesian closed: if $X$ and $Y$ are quasi-Borel spaces, then $\Cat{QBS}(X,Y)$ can be given a quasi-Borel structure $M_{X^Y}$ by defining
  \[ M_{X^Y} := \{ \rho : \rr\to\Cat{QBS}(X,Y) \mid \bigl(\rho^\flat : \rr\times X\to Y\bigr) \in \Cat{QBS}(\rr\times X, Y) \} \; . \]
  A probability measure on a quasi-Borel space $X$ is defined to be a pair of a (standard) probability measure $\nu$ on $\rr$ and a random variable $\rho\in M_X$.
  Since two different pairs $(\nu,\rho)$ and $(\mu,\tau)$ may produce equal pushforward measures, $\rho_\ast\nu = \tau_\ast\mu$, it makes sense to consider two such QBS measures to be equivalent if their pushforwards are equal.
  The set $\Pa X$ of such equivalence classes of QBS measures on $X$ can then be equipped with the structure of a quasi-Borel space, and the assignment $\Pa$ is made functorial by the pushforwards action.
  Finally, the functor $\Pa:\Cat{QBS}\to\Cat{QBS}$ can be equipped with the structure of a (monoidal) monad in a manner analogous to the Giry monad: the unit yields Dirac measures, and the multiplication acts by integration.
\end{ex}

We end this section by noting that the notions of s-finite measure and s-finite kernel can be reconstructed within $\Cat{QBS}$, so that we may interpret $\Cat{sfKrn}$ to be enriched accordingly \parencite[\S11]{Vakar2018S}.
Moreover, \textcite{Vakar2018S} show that the set $TX$ of s-finite measures on $X$ can be given a quasi-Borel structure, and this assignment actually yields a monad $T:\Cat{QBS}\to\Cat{QBS}$ (by analogy with the `continuation' monad).
This licenses us to take $\Cat{sfKrn}$ to be instead defined as $\Kl(T)$.

For further examples of probability monads, we refer the reader to \textcite{Jacobs2018probability}.

\section{Dependent data and bidirectional processes} \label{sec:bidi}

Two properties of Bayesian inversion are particularly notable.
Firstly, given a channel $X\klto Y$, its inversion yields a channel in the opposite direction, $Y\klto X$.
Secondly, this inverse channel does not exist in isolation, but rather depends on a supplied `prior' distribution.
In Chapter \ref{chp:brain} we will want to assign functorially to stochastic channels dynamical systems that invert them, and to do this requires understanding how inversions compose.
The general pattern for the composition of dependent bidirectional processes is called the \textit{lens} pattern, and this section is dedicated to introducing it.
The more fundamental aspect is that of dependence, which we began to explore in the context of dependent sums and products in Chapter \ref{chp:basic-ct}: we therefore begin this section by introducing the \textit{Grothendieck construction}, a `fibrational' framework for composing dependent processes.

\subsection{Indexed categories and the Grothendieck construction} \label{sec:idx-cat}

At various point above, we have encountered `dependent' objects and morphisms: indexed and dependent sums (Remark \ref{rmk:idx-sum}); indexed products (Remark \ref{rmk:dep-prod}); dependent products (\secref{sec:dep-prod}); hints at dependent type theory (end of \secref{sec:closed-cats}); parameterized morphisms (\secref{sec:ext-para}); circuit algebras (\secref{sec:sys-circ}); and, of course, Bayesian inversions.
The Grothendieck construction classifies each of these as examples of a common pattern, allowing us to translate between `indexed' and `fibrational' perspectives:
from the indexed perspective, we consider functors from an indexing object into a category (think of diagrams); from the fibrational perspective, we consider bundles as projection maps.
The correspondence is then, roughly speaking, between ``the object indexed by $i$'' and ``the subobject that projects to $i$'', which is called the `fibre' of the bundle over $i$.

For this reason, categories of bundles are an important part of the story, from which much else is generalized.
Recall from Definition \ref{def:slice-cat} that these categories of bundles are slice categories: the category of bundles over $B$ in $\cat{C}$ is the slice $\cat{C}/B$, whose objects are pairs $(E,p)$ of an object $E$ and a morphism $p:E\to B$; and whose morphisms $(E,p)\to(E',p')$ are morphisms $\alpha:E\to E'$ of $\cat{C}$ such that $p = p'\circ\alpha$.
We call this the category of bundles over $B$ as a generalization of the notion of ``fibre bundle'', from which we inherit the notion of `fibre'.

\begin{defn}
  Suppose $\cat{C}$ is a category with finite limits.
  Given a bundle $p:E\to B$ in $\cat{C}$, its \textit{fibre over} $b:B$ is the subobject $E_b$ of $E$ such that $p(e) = b$ for all $e:E_b$.
  The fibre $E_b$ can be characterized as the following pullback object, where $1$ is the terminal object in $\cat{C}$:
  \[\begin{tikzcd}
    {E_b} & E \\
	  1 & B
	  \arrow[from=1-1, to=2-1]
	  \arrow["b"', from=2-1, to=2-2]
	  \arrow["p", from=1-2, to=2-2]
	  \arrow[from=1-1, to=1-2]
	  \arrow["\lrcorner"{anchor=center, pos=0.125}, draw=none, from=1-1, to=2-2]
  \end{tikzcd}\]
\end{defn}

In the case where $\cat{C} = \Set$, there is an equivalence between the slice $\Set/B$ and a certain presheaf category: the category of $B$-diagrams in $\Set$, which we can equivalently think of as the category of $B$-indexed sets.

\begin{defn} \label{def:disc-cat}
  Suppose $B$ is a set.
  The \textit{discrete category} on $X$ is the category whose objects are the elements of $B$ and whose only morphisms are identity morphisms $\id_b:b\to b$ for each element $b:B$.
  We will denote the discrete category on $B$ simply by $B$.
\end{defn}

\begin{prop} \label{prop:groth-equiv--1}
  For each set $B$, there is an equivalence $\Set/B \cong \Set^B$.
  \begin{proof}
    In the direction $\Set/B\to\Set^B$, let $p:E\to B$ be a bundle over B.
    We construct a functor $P:B\to\Set$ by defining $P(b) := E_b$, where $E_b$ is the fibre of $p$ over $b$; there are no nontrivial morphisms in $B$, so we are done.
    Now suppose $f:(E,p)\to(F,q)$ is a morphism of bundles.
    A natural transformation $\varphi:P\Rightarrow Q$ in $\Set^B$ is just a family of functions $\varphi_b:Pb\to Qb$ indexed by $b$.
    Hence, given $f$, we define $\varphi_b$ as the restriction of $f$ to $E_b$ for each $b:B$.

    In the direction $\Set^B\to\Set/B$, let $P:B\to\Set$ be a functor.
    We define $E$ as the coproduct $\sum_{b:B}P(b)$, and the bundle $p:E\to B$ as the projection $(b,x)\mapsto b$ for every $(b,x)$ in $\sum_{b:B}P(b)$.
    Now suppose $\varphi:P\Rightarrow Q$ is a natural transformation in $\Set^B$.
    We define the function $f:(E,p)\to(F,q)$ by the coproduct of the functions $\varphi_b$, as $f := \sum_{b:B}\varphi_b$.

    These two constructions are easily verified as mutually inverse.
  \end{proof}
\end{prop}

If the $B$ in $\Set^B$ is not just a set, but rather a category, then there is a correspondingly categorified notion of the category of bundles.

\begin{defn} \label{def:cat-el}
  Suppose $F:\cat{C}\to\Set$ is a copresheaf on $\cat{C}$.
  Its \textit{category of elements} $\cat{C}/F$ has for objects pairs $(X,x)$ of an object $X:\cat{C}$ and an element $x:FX$.
  A morphism $(X,x)\to(Y,y)$ is a morphism $f:X\to Y$ in $\cat{C}$ such that $Ff(x)=y$, as in the following diagram, where the top triangle commutes in $\Set$:
  \[\begin{tikzcd}
	  & 1 \\
	  FX && FY \\
	  X && Y
	  \arrow["x"', from=1-2, to=2-1]
	  \arrow["Ff", from=2-1, to=2-3]
	  \arrow[""{name=0, anchor=center, inner sep=0}, "y", from=1-2, to=2-3]
	  \arrow["f", from=3-1, to=3-3]
  \end{tikzcd} \; .\]
  Identities are given by identity morphisms in $\cat{C}$, and composition is composition of the underlying morphisms in $\cat{C}$.
  There is an evident forgetful functor $\pi_F:\cat{C}/F\to\cat{C}$, which acts on objects as $(X,x)\mapsto X$ and on morphisms as $f\mapsto f$.
\end{defn}

To validate that the category of elements construction is a good generalization of the slice category, we have the following example.

\begin{ex}
  The category of elements of a representable copresheaf $\cat{C}(C,-)$ is equivalent to the slice category $\cat{C}/C$, from which we derive the similar notation.
\end{ex}

\begin{rmk}
  Another way to look at the morphisms $(X,x)\to(Y,y)$ in $\cat{C}/F$ is as \textit{pairs} $(f,\iota)$, where $f$ is a morphism $X\to Y$ in $\cat{C}$ and $\iota$ is an identification $Ff(x)=y$.
  Then composition in $\cat{C}/F$ is not just composition of morphisms in $\cat{C}$, but also composition of identifications: given $(f,\iota):(X,x)\to(Y,y)$ and $(g,\kappa):(Y,y)\to(Z,z)$, the composite $(g,\kappa)\circ(f,\iota)$ is $(g\circ f,\kappa\circ Fg(\iota))$, where $\kappa\circ Fg(\iota)$ is the composite identification $F(g\circ f)(x)\xlongequal{Fg(\iota)}Fg(y)\xlongequal{\kappa}z$.
  We can think of these identifications as witnesses to the required equalities.
  This perspective on $\cat{C}/F$ is analogous to the process of categorification we considered in Chapter \ref{chp:basic-ct}, where we added witnesses (fillers) to equations and diagrams.
\end{rmk}

A better way to validate the category of elements construction is to generalize the Grothendieck correspondence, Proposition \ref{prop:groth-equiv--1}, which means we need something to correspond to $\Set^{\cat{B}}$: a category of categories of elements.
These generalized categories of elements are called ``discrete opfibrations'', and constitute our first examples of categorified bundles.

\begin{defn} \label{def:disc-opfib}
  A \textit{discrete opfibration} is a functor $F:\cat{E}\to\cat{B}$ such that, for every object $E$ in $\cat{E}$ and morphism $g:FE\to B$ in $\cat{B}$, there exists a unique morphism $h:E\to E'$ in $\cat{E}$ such that $Fh = g$ (called the \textit{lift} of $g$):
  \[\begin{tikzcd}
    E && {E'} \\
    \\
    FE && B
	  \arrow[""{name=0, anchor=center, inner sep=0}, "h", dashed, from=1-1, to=1-3]
	  \arrow[""{name=1, anchor=center, inner sep=0}, "g", from=3-1, to=3-3]
	  \arrow[shorten <=13pt, shorten >=13pt, dotted, no head, from=0, to=1]
  \end{tikzcd}\]
  Write $\DOpfib(\cat{B})$ to denote the full subcategory of $\Cat{Cat}/\cat{B}$ on those objects which are discrete opfibrations.
  The subcategory $\cat{E}_B$ of $\cat{E}$ all of whose objects are mapped by $F$ to $B$ and all of whose morphisms are mapped to $\id_B$ is called the \textit{fibre} of $F$ over $B$.
\end{defn}

\begin{ex}
  The forgetful functor $\pi_F:\cat{C}/F\to\cat{C}$ out of the category of elements of a copresheaf $F$ is a discrete opfibration: for any object $(X,x)$ in $\cat{C}/F$ and morphism $g:X\to Y$ in $\cat{C}$, there is a unique morphism $g:(X,x)\to(Y,y)$, namely where $y=Ff(x)$.
\end{ex}

And thus we obtain a Grothendieck correspondence at the next level of categorification.

\begin{prop} \label{prop:groth-equiv-0}
  For any category $\cat{B}$, there is an equivalence $\DOpfib(\cat{B})\cong\Set^{\cat{B}}$.
  \begin{proof}[Proof sketch]
    We only sketch the bijection on objects; the correspondence on morphisms subsequently follows quite mechanically.

    Given a discrete opfibration $p:\cat{E}\to\cat{B}$, it is easy to check that each fibre $E_b$ is a discrete category and hence a set.
    Given a morphism $f:b\to c$ in $\cat{B}$, we define a function $\varphi:E_b\to E_c$ by mapping each $e:E_b$ to the codomain of the unique lift $h$.
    This defines a functor $\cat{B}\to\Set$; functoriality follows from the uniqueness of lifts.

    In the inverse direction, given a copresheaf $F:\cat{B}\to\Set$, take the forgetful functor $\pi_F:\cat{B}/F\to\cat{B}$ out of its category of elements, which is a discrete opfibration by the example above.
    Given a natural transformation $\sigma:F\Rightarrow G$, define a functor $S:\cat{B}/F\to\cat{B}/G$ on objects as $S(X,x) = (X,\sigma_X(x))$ and on morphisms $f:(X,x)\to(Y,y)$ as $Sf = (X,\sigma_X(x))\xto{f}(Y,\sigma_Y(y))$; this is well-defined by the naturality of $\sigma$ and the definition of $f$, since $Gf\circ\sigma_X(x) = \sigma_Y\circ Ff(x)$ and $Ff(x) = y$.

    The verification that these two constructions are mutually inverse is straightforward.
  \end{proof}
\end{prop}

In many cases, the dependent data of interest will have more structure than that of mere sets.
For example, in \secref{sec:sys-circ} we introduced a rate-coded circuit diagrams as an indexing of sets of rate-coded circuits by a category of circuit diagrams; later, we will see that dynamical systems have a canonical notion of morphism, and so our dynamical semantics will take the form of an indexed collection of categories.
This requires us to categorify not only the domain of indexing (as we have done above), but also the codomain of values (as we do now).
As with monoidal categories---and as in the case of circuit algebras---in this higher-dimensional setting, it becomes necessary to work with weak composition, and the relevant notion of weak functor is the `pseudofunctor'.

\begin{defn} \label{def:pseudofunctor}
  Suppose $\cat{C}$ is a category and $\cat{B}$ is a bicategory.
  A \textit{pseudofunctor} $F:\cat{C}\to\cat{B}$ is constituted by
  \begin{enumerate}
  \item a function $F_0:\cat{C}_0\to\cat{B}_0$ on objects;
  \item for each pair of objects $a,b:\cat{C}$, a function $F_{a,b}:\cat{C}(a,b)\to\cat{B}(F_0a,F_0b)_0$ on morphisms;
  \item for each object $c:\cat{C}$, a 2-isomorphism $F_{\id_c}:\id_{F_0c}\Rightarrow F_{c,c}(\id_c)$ witnessing \textit{weak unity}, natural in $c$; and
  \item for each composable pair of morphisms $f:a\to b$ and $g:b\to c$ in $\cat{C}$, a 2-isomorphism $F_{g,f}:F_{b,c}(g)\diamond F_{a,b}(f)\Rightarrow F_{a,c}(g\circ f)$ witnessing \textit{weak functoriality}, natural in $f$ and $g$, where we have written composition in $\cat{C}$ as $\circ$ and horizontal composition in $\cat{B}$ as $\diamond$;
  \end{enumerate}
  satisfying the following conditions:
  \begin{enumerate}[label=(\alph*)]
  \item coherence with left and right unitality of horizontal composition, so that the respective diagrams of 2-cells commute:
  \[\begin{tikzcd}
	  {\id_{F_0b}\diamond F_{a,b}(f)} && {F_{a,b}(f)} \\
	  \\
	  {F_{b,b}(\id_b)\diamond F_{a,b}(f)} && {F_{a,b}(\id_b\circ f)}
	  \arrow["{\lambda_{F_{a,b}(f)}}", Rightarrow, from=1-1, to=1-3]
	  \arrow[Rightarrow, no head, from=1-3, to=3-3]
	  \arrow["{F_{\id_b}\diamond F_{a,b}(f)}"', Rightarrow, from=1-1, to=3-1]
	  \arrow["{F_{\id_b,f}}", Rightarrow, from=3-1, to=3-3]
  \end{tikzcd}
  \qquad
  \begin{tikzcd}
    {F_{a,b}(f)\diamond\id_{F_0a}} && {F_{a,b}(f)} \\
    \\
    {F_{a,b}(f)\diamond F_{a,a}(\id_a)} && {F_{a,b}(f\circ\id_a)}
	  \arrow["{\rho_{F_{a,b}(f)}}", Rightarrow, from=1-1, to=1-3]
	  \arrow[Rightarrow, no head, from=1-3, to=3-3]
	  \arrow["{F_{a,b}(f)\diamond F_{\id_a}}"', Rightarrow, from=1-1, to=3-1]
	  \arrow["{F_{f,\id_a}}", Rightarrow, from=3-1, to=3-3]
  \end{tikzcd}\]
  \item coherence with associativity of horizontal composition, so that the following diagram of 2-cells commutes:
  \[\begin{tikzcd}
    {(F_{c,d}(h)\diamond F_{b,c}(g))\diamond F_{a,b}(f)} &&& {F_{c,d}(h)\diamond(F_{b,c}(g)\diamond F_{a,b}(f))} \\
    \\
    {F_{b,d}(h\circ g)\diamond F_{a,b}(f)} &&& {F_{c,d}(h)\diamond F_{a,c}(g\circ f)} \\
	  \\
	  {F_{a,d}((h\circ g)\circ f)} &&& {F_{a,d}(h\circ(g\circ f))}
	  \arrow["{\alpha_{F_{c,d}(h),F_{b,c}(g),F_{a,b}(f)}}", Rightarrow, from=1-1, to=1-4]
	  \arrow["{F_{c,d}(h)\diamond F_{g,f}}", Rightarrow, from=1-4, to=3-4]
	  \arrow["{F_{h,g\circ f}}", Rightarrow, from=3-4, to=5-4]
	  \arrow["{F_{h,g}\diamond F_{a,b}(f)}"', Rightarrow, from=1-1, to=3-1]
	  \arrow["{F_{h\circ g,f}}"', Rightarrow, from=3-1, to=5-1]
	  \arrow[Rightarrow, no head, from=5-1, to=5-4]
  \end{tikzcd} \; .\]
  \end{enumerate}
\end{defn}

\begin{rmk} \label{rmk:pseudofunc}
  If $\cat{C}$ is in fact a nontrivial bicategory, then the definition of pseudofunctor is weakened accordingly: the functions $F_{a,b}$ are replaced by functors between the corresponding hom-categories, and the equalities in the functoriality conditions (a) and (b) are replaced by the relevant unitor or associator isomorphism.
  We will encounter this more general case in the next chapter, where we introduce the (yet weaker) concept of \textit{lax functor}: see Definition \ref{def:lax-functor}, and the associated footnote \ref{fn:pseudofunc} for the relationship with the present notion of pseudofunctor.
\end{rmk}

With pseudofunctors, we gain a notion of indexed category.

\begin{defn}
  An \textit{indexed category} is a pseudofunctor $F:\cat{C}\op\to\Cat{Cat}$, for some indexing category $\cat{C}$.
  An \textit{opindexed category} is a pseudofunctor $F:\cat{C}\to\Cat{Cat}$.
  Given an (op)indexed category $F$, we call the categories $Fc$ its \textit{fibres}, for each object $c:\cat{C}$.
\end{defn}

Working with indexed categories rather than indexed sets, the relevant notion of (op)fibration is no longer discrete, as there are now (non-trivial) morphisms to account for.
Following the Grothendieck logic, fibrations $p:\cat{E}\to\cat{B}$ should be in bijective correspondence with indexed categories $F:\cat{B}\op\to\Cat{Cat}$.
This means that we should be able to turn any indexed category into a fibration by appropriately gluing together its fibres; and conversely, given a fibration $p$, the assignment $b\mapsto\cat{E}_b$\footnote{
with $\cat{E}_b$ being the subcategory of $\cat{E}$ sometimes denoted $p^{-1}(b)$ all of whose objects are mapped by $p$ to $b$, as in the proof of Proposition \ref{prop:groth-equiv-0}.}
should define a pseudofunctor $\cat{B}\op\to\Cat{Cat}$.
These considerations yield the following definition.

\begin{defn}
  A \textit{fibration} is a functor $p:\cat{E}\to\cat{B}$ such that, for every pair of morphisms $f:E'\to E$ and $\phi:E''\to E$ in $\cat{E}$, and for every morphism $g:p(E'')\to p(E')$ such that $p(\phi) = p(f)\circ g$ in $\cat{B}$, there exists a unique morphism $h:E''\to E'$ in $\cat{E}$ such that $p(h) = g$ and $\phi = f\circ h$:
  \[\begin{tikzcd}[row sep=small]
    {E''} && {E'} \\
    \\
    &&& E \\
    {p(E'')} && {p(E')} \\
    \\
    &&& {p(E)}
    \arrow[""{name=0, anchor=center, inner sep=0}, "h", dashed, from=1-1, to=1-3]
    \arrow[""{name=1, anchor=center, inner sep=0}, "g", from=4-1, to=4-3]
    \arrow["f", from=1-3, to=3-4]
    \arrow["\phi"', from=1-1, to=3-4]
    \arrow["{p(\phi)}"', from=4-1, to=6-4]
    \arrow["{p(f)}", from=4-3, to=6-4]
    \arrow[shorten <=19pt, shorten >=19pt, dotted, no head, from=0, to=1]
  \end{tikzcd}\]
  The subcategory $\cat{E}_B$ of all those objects mapped by $p$ to $B:\cat{B}$ and all those morphisms mapped to $\id_B$ is called the \textit{fibre} of $p$ over $B$.
  An \textit{opfibration} is a functor $p:\cat{E}\to\cat{B}$ for which $p\op:\cat{E}\op\to\cat{B}\op$ is a fibration.
\end{defn}

\begin{rmk}
  Note that a discrete (op)fibration is a(n) (op)fibration each of whose fibres is a discrete category: this means that in each fibre, there are no non-identity morphisms, so that morphisms $f$ and $\phi$ in the definition above are trivialized, thereby recovering the form of Definition \ref{def:disc-opfib}.
\end{rmk}

The Grothendieck construction then tells us how to translate from (op)indexed categories to (op)fibrations: in some situations, it will be easier to work with the one, and in others the other.
In particular, categories of lenses (and polynomial functors) will be seen to arise as Grothendieck constructions.

\begin{defn} \label{def:grot-con}
  Suppose $F:\cat{C}\op\to\Cat{Cat}$ is a pseudofunctor.
  Its \textit{(contravariant) Grothendieck construction} is the category $\int F$ defined as follows.
  The objects of $\int F$ are pairs $(X,x)$ of an object $X:\cat{C}$ and an object $x:FX$.
  A morphism $(X,x)\to(Y,y)$ is a pair $(f,\varphi)$ of a morphism $f:X\to Y$ in $\cat{C}$ and a morphism $\varphi:x\to Ff(y)$ in $FX$, as in the following diagram, where the upper triangle is interpreted in $\Cat{Cat}$ (note the contravariance of $Ff$):
  \[\begin{tikzcd}
    & 1 \\
    \\
	  FX && FY \\
	  X && Y
	  \arrow[""{name=0, anchor=center, inner sep=0}, "x"', from=1-2, to=3-1]
	  \arrow["Ff"', from=3-3, to=3-1]
	  \arrow[""{name=1, anchor=center, inner sep=0}, "y", from=1-2, to=3-3]
	  \arrow["f"', from=4-1, to=4-3]
	  \arrow["\varphi", shift right=3, shorten <=6pt, shorten >=6pt, Rightarrow, from=0, to=1]
  \end{tikzcd}\]
  We can thus write the hom set $\int F\bigl((X,x),(Y,y)\bigr)$ as
  the dependent sum $\sum_{f\,:\,\cat{C}(X,Y)}FX\big(x,Ff(y)\big)$.
  The identity morphism on $(X,x)$ is $(\id_X,\id_x)$, and composition is defined as follows.
  Given $(f,\varphi):(X,x)\to(Y,y)$ and $(g,\gamma):(Y,y)\to(Z,z)$, their composite $(g,\gamma)\circ(f,\varphi)$ is the pair
  \[ (g\circ f, Ff(\gamma)\circ\varphi) \; . \]
\end{defn}

The following well-known result tells us that the Grothendieck construction yields fibrations.

\begin{prop}[{\textcite[{Prop. 10.1.10}]{Johnson20202Dimensional}}]
  Suppose $F:\cat{C}\op\to\Cat{Cat}$ is an indexed category.
  Then there is a `projection' functor $\pi_F:\int F\to\cat{C}$ mapping $(X,x)\mapsto X$ and $(f,\varphi)\mapsto f$, and this functor is a fibration.
\end{prop}

\begin{rmk}
  Dually, there is a \textit{covariant Grothendieck construction}, for opindexed categories $F:\cat{C}\to\Cat{Cat}$. The objects of $\int F$ are again pairs $(X:\cat{C},x:FX)$, but now the morphisms $(X,x)\to(Y,y)$ are pairs $(f,\varphi)$ with $f:X\to Y$ in $\cat{C}$ as before and now $\varphi:Ff(x)\to y$; all that we have done is swapped the direction of the arrow $Ff$ in the diagram in Definition \ref{def:grot-con} (compare the identifications in the category of elements of a copresheaf, in Definition \ref{def:cat-el}).
  As a result, we can write the hom set $\int F\bigl((X,x),(Y,y)\bigr)$ in this case as $\sum_{f\,:\,\cat{C}(X,Y)}FX\big(Ff(x),y\big)$.
\end{rmk}

\begin{rmk} \label{rmk:groth-equiv-1}
  The Grothendieck construction induces an analogue of Proposition \ref{prop:groth-equiv-0} between the bicategory of pseudofunctors $\cat{B}\op\to\Cat{Cat}$ and the bicategory of \textit{Grothendieck fibrations} on $\cat{B}$ \parencite[{Theorem 10.6.16}]{Johnson20202Dimensional}.
  Indeed there are analogues of Propositions \ref{prop:groth-equiv-0} and \ref{prop:groth-equiv--1} in any categorical dimension.
  Because fibrations are the higher-dimensional analogues of bundles, they have a base category (the codomain) and a `total' category (the domain), which is a kind of colimit of the fibres (constructed by the Grothendieck construction): strictly speaking, what we have called the Grothendieck construction above is total category of the full fibrational construction; the fibration itself is the corresponding forgetful (projection) functor.
  For a highly readable exposition of Grothendieck constructions, we refer the reader to \textcite{Loregian2018Categorical}.
\end{rmk}

\subsubsection{The monoidal Grothendieck construction}

When \(\cat{C}\) is a monoidal category with which \(F\) is appropriately compatible, then we can `upgrade' the notions of indexed category and Grothendieck construction accordingly.
In this chapter, we will restrict ourselves to locally trivial monoidal indexed categories, those for which the domain \(\cat{C}\) is only a category; \textcite{Moeller2018Monoidal} work out the structure for bicategorical $\cat{C}$.
(As noted in Remark \ref{rmk:pseudofunc}, in Chapter \ref{chp:sgame}, we will sketch a notion of monoidal indexed bicategory which amounts to a categorification of the present notion; but that will also in some sense be locally trivial.)

\begin{defn}[{After \textcite[{\S}3.2]{Moeller2018Monoidal}}] \label{def:mon-idx-cat}
  Suppose $(\cat{C},\otimes,I)$ is a monoidal category.
  We say that $F$ is a \textit{monoidal indexed category} when $F$ is a lax monoidal pseudofunctor
  $(F,\mu,\eta) : (\cat{C}\op,\otimes\op,I)\to(\Cat{Cat},\times,\Cat{1})$.
  This means that the laxator $\mu$ is given by a natural family of functors $\mu_{A,B} : FA\times FB \to F(A\otimes B)$ along with, for any morphisms $f:A\to A'$ and $g:B\to B'$ in $\cat{C}$, a natural isomorphism $\mu_{f,g} : \mu_{A',B'} \circ \left(Ff\times Fg\right) \Rightarrow F(f\otimes g) \circ \mu_{A,B}$.
  (This makes $\mu$ into a pseudonatural transformation in the sense of Definition \ref{def:lax-trans}.)
  The laxator and the unitor $\eta : \Cat{1}\to FI$ together satisfy axioms of associativity and unitality that constitute indexed versions of the associators and unitors of a monoidal category (Definition \ref{def:monoidal-cat}).

  Explicitly, this means that there must be three families of natural isomorphisms, indexed by objects $A,B,C:\cat{C}$,
  \begin{enumerate}
  \item an \textit{associator} family $\alpha_{A,B,C}:\mu_{A\otimes B,C}(\mu_{A,B}(-,-),-)\Rightarrow\mu_{A,B\otimes C}(-,\mu_{B,C}(-,-))$;
  \item a \textit{left unitor} $\lambda_A:\mu_{I,A}(\eta,-)\Rightarrow\id_{FA}$; and
  \item a \textit{right unitor} $\rho_A:\mu_{A,I}(-,\eta)\Rightarrow\id_{FA}$
  \end{enumerate}
  such that the unitors are compatible with the associator, \textit{i.e.} for all $A,B:\cat{C}$ the diagram
  \[\begin{tikzcd}
    {\mu_{A\otimes I,B}(\mu_{A,I}(-,\eta),-)} && {\mu_{A,I\otimes B}(-,\mu_{I,B}(\eta,-))} \\
    \\
    & {\mu_{A,B}(-,-)}
    \arrow["{\mu_{\Rho_A,B}(\rho_A,-)}"', Rightarrow, from=1-1, to=3-2]
    \arrow["{\alpha_{A,I,B}(-,\eta,-)}", Rightarrow, from=1-1, to=1-3]
    \arrow["{\mu_{A,\Lambda_B}(-,\lambda_B)}", Rightarrow, from=1-3, to=3-2]
  \end{tikzcd}\]
  commutes (where $\Rho$ and $\Lambda$ are the right and left associators of the monoidal structure $(\otimes,I)$ on $\cat{C}$), and such that the associativity is `order-independent', \textit{i.e.} for all $A,B,C,D:\cat{C}$, the diagram
  \hspace*{-2.5cm}\scalebox{0.95}{\vbox{
  \[\begin{tikzcd}[ampersand replacement=\&,column sep=tiny]
    {\mu_{A\otimes(B\otimes C),D}(\mu_{A,B\otimes C}(-,\mu_{B,C}(-,-)),-)} \&\& {\mu_{A,(B\otimes C)\otimes D}(-,\mu_{B\otimes C,D}(\mu_{B,C}(-,-),-))} \\
    \\
    {\mu_{(A\otimes B)\otimes C,D}(\mu_{A\otimes B,C}(\mu_{A,B}(-,-),-),-)} \&\& {\mu_{A,B\otimes(C\otimes D)}(-,\mu_{B,C\otimes D}(-,\mu_{C,D}(-,-)))} \\
    \\
    \& {\mu_{A\otimes B,C\otimes D}(\mu_{A,B}(-,-),\mu_{C,D}(-,-))}
    \arrow["{\alpha_{A,B\otimes C,D}}", Rightarrow, from=1-1, to=1-3]
    \arrow["{\mu_{\mathrm{A}_{A,B,C},D}(\alpha_{A,B,C},-)}", Rightarrow, from=3-1, to=1-1]
    \arrow["{\alpha_{A\otimes B,C,D}}"', Rightarrow, from=3-1, to=5-2]
    \arrow["{\alpha_{A,B,C\otimes D}}"', Rightarrow, from=5-2, to=3-3]
    \arrow["{\mu_{A,\mathrm{A}_{B,C,D}}(-,\alpha_{B,C,D})}", Rightarrow, from=1-3, to=3-3]
  \end{tikzcd}\]}} \\
  commutes (where $\mathrm{A}$ is the associator of the monoidal structure on $\cat{C}$).
\end{defn}

The following proposition exhibits the monoidal structure carried by the Grothendieck construction when the indexed category is monoidal.

\begin{prop}[{\textcite[{\S}6.1]{Moeller2018Monoidal}}] \label{prop:monoidal-gr}
  Suppose \((F, \mu, \eta) : (\cat{C}\op, \otimes\op, I) \to (\Cat{Cat}, \times, \Cat{1})\) is a monoidal indexed category.
  Then the total category of the Grothendieck construction \(\int F\) obtains a monoidal structure \((\otimes_\mu, I_\mu)\).
  On objects, define
  \[ (C, X) \otimes_\mu (D, Y) := \big(C \otimes D, \mu_{CD}(X, Y)\big) \]
  where \(\mu_{CD} : FC \times FD \to F(C \otimes D)\) is the component of \(\mu\) at \((C, D)\).
  On morphisms \((f, f^\dag) : (C, X) \lensto (C', X')\) and \((g, g^\dag) : (D, Y) \lensto (D', Y')\), define
  \[ (f,f^\dag) \otimes_\mu (g, g^\dag) := \big(f \otimes g, \mu_{CD}(f^\dag, g^\dag)\big) \, . \]
  The monoidal unit \(I_\mu\) is defined to be the object \(I_\mu := \big(I, \eta(\ast)\big)\).
  Writing \(\lambda : I \otimes (-) \Rightarrow (-)\) and \(\rho : (-) \otimes I \Rightarrow (-)\) for the left and right unitors of the monoidal structure on \(\cat{C}\), the left and right unitors in \(\int F\) are given by \((\lambda, \id)\) and \((\rho, \id)\) respectively.
  Writing \(\alpha\) for the associator of the monoidal structure on \(\cat{C}\), the associator in \(\int F\) is given by \((\alpha, \id)\).
\end{prop}

\begin{rmk} \label{rmk:fib-mon-idx-cat}
  Sometimes, rather than (or in addition to) an indexed category $F$ being lax monoidal as a pseudofunctor (which yields a `global' monoidal structure), it may in fact be \textit{fibrewise monoidal}, meaning that each fibre $FX$ is itself a monoidal category (yielding `local' monoidal structures); in this case, the pseudofunctor $F$ can be written with the type $\cat{C}\op\to\Cat{MonCat}$.
  In general, the fibrewise monoidal structures may be independent both of each other and of the lax monoidal structure on $F$ itself, but when $\cat{C}$ is in fact \textit{Cartesian} monoidal, the local and global monoidal structures coincide.
  For more reading on this, we refer the reader to \textcite[{\S}4]{Moeller2018Monoidal}.
\end{rmk}

\subsection{Grothendieck lenses} \label{sec:groth-lens}

Lenses formalize bidirectional processes in which the `backward' process depends on data in the domain of the `forward' process.
The name originates in database theory \parencite{Bohannon2006Relational,Foster2007Combinators}, where the forward process gives a zoomed-in `view' onto a database record, and the backward process is used to update it.
Following an observation of Myers and \textcite{Spivak2019Generalized}, lenses of this general shape can be given a concise definition using the Grothendieck construction.
In order to obtain the backward directionality of the dependent part, we use the ``pointwise opposite'' of a pseudofunctor.

\begin{defn}
  Suppose $F:\cat{C}\op\to\Cat{Cat}$ is a pseudofunctor.
  We define its \textit{pointwise opposite} $F^p:\cat{C}\op\to\Cat{Cat}$ to be the pseudofunctor $c\mapsto Fc\op$ returning the opposite of each category $Fc$; given $f:c\to c'$, $F^pf:Fc\op\to {Fc'}\op$ is defined as $(Ff)\op:Fc\op\to {Fc'}\op$.
\end{defn}

Categories of Grothendieck lenses are then obtained via the Grothendieck construction of pointwise opposites of pseudofunctors.

\begin{defn}[Grothendieck lenses \citep{Spivak2019Generalized}] \label{def:gr-lens}
  We define the category \(\Lens_F\) of Grothendieck lenses for a pseudofunctor \(F : \cat{C}\op \to \Cat{Cat}\) to be the total category of the Grothendieck construction for the pointwise opposite $F^p$ of \(F\).
  Explicitly, its objects \((\Lens_F)_0\) are pairs \((C, X)\) of objects \(C\) in \(\cat{C}\) and \(X\) in \(F(C)\), and its hom-sets \(\Lens_F \big( (C, X), (C', X') \big)\) are the dependent sums
  \begin{equation}
    \Lens_F \big( (C, X), (C', X') \big) = \sum_{f \, : \, \cat{C}(C, C')} F(C) \big( F(f)(X'), X \big)
  \end{equation}
  so that a morphism \((C, X) \lensto (C', X')\) is a pair \((f, f^\dag)\) of \(f : \cat{C}(C, C')\) and \(f^\dag : F(C) \big( F(f)(X'), X \big)\).
  We call such pairs \textit{Grothendieck lenses} for \(F\) or \(F\mathrm{\textit{-lenses}}\).
  We say that the morphism $f$ is the \textit{forward} component of the lens, and the morphism $f^\dag$ the \textit{backward} component.

  The identity Grothendieck lens on \((C, X)\) is \(\id_{(C, X)} = (\id_C, \id_X)\).
  Sequential composition is as follows.
  Given \((f, f^\dag) : (C, X) \lensto (C', X')\) and \((g, g^\dag) : (C', X') \lensto (D, Y)\), their composite \((g, g^\dag) \lenscirc (f, f^\dag)\) is defined to be the lens \(\big(g \klcirc f, f^\dag\circ F(f)(g^\dag) \big) : (C, X) \lensto (D, Y)\).
\end{defn}

\begin{notation}
  In the context of lenses, we will often write the backward map as $f^\dag$ or $f^\sharp$, with the former particularly used for Bayesian lenses.
  We will also use $\lensto$ to denote a lens, and $\lenscirc$ for lens composition.
  Above, we additionally used $\klcirc$ for composition in the base category and $\circ$ for composition in the fibres.

  Since lenses are bidirectional processes and English is read horizontally, when it comes to string diagrams for lenses, we will depict these horizontally, with the forwards direction read from left to right.
\end{notation}

Whenever $\cat{C}$ is a monoidal category, it gives rise to a canonical category of lenses, in which the forwards morphisms are comonoid morphisms in $\cat{C}$ and the backwards morphisms are (internally) parameterized by the domains of the forwards ones.
Comonoids and their morphisms are necessary to copy parameters during composition.
The resulting `monoidal' lenses are a natural generalization of the `Cartesian' lenses used in the database setting, and we will see that Bayesian lenses are similarly constructed using an indexed category of (externally) parameterized morphisms.

\begin{ex} \label{ex:mon-lens}
  Suppose $(\cat{C},\otimes,I)$ is a monoidal category and let $\Comon(\cat{C})$ be its subcategory of comonoids and comonoid morphisms.
  A \textit{monoidal lens} $(X,A)\lensto(Y,B)$ is a pair $(f,f^\sharp)$ of a comonoid morphism $f:X\to Y$ in $\Comon(\cat{C})$ and a morphism $f^\sharp:X\otimes B\to A$ in $\cat{C}$.
  Such lenses can be characterized as Grothendieck lenses, following \textcite[{\S}3.2]{Spivak2019Generalized}.

  First, define a pseudofunctor $\Fun{P}:\Comon(\cat{C})\op\to\Cat{Cat}$ as follows.
  On objects $X:\Comon(\cat{C})$, define $\Fun{P}X$ as the category with the same objects as $\cat{C}$ and with hom-sets given by $\Fun{P}X(A,B) := \cat{C}(X\otimes A, B)$; denote a morphism $f$ from $A$ to $B$ in $\Fun{P}X$ by $f:A\xto{X}B$.
  The identity $\id_A:A\xto{X}A$ is defined as the projection $\proj_A:X\otimes A\xto{\ground_X\otimes\id_A}I\otimes A\xto{\lambda_A}A$.
  Given $f:A\xto{X}B$ and $g:B\xto{X}C$, their composite $g\circ f:A\xto{X}C$ is given by the following string diagram in $\cat{C}$:
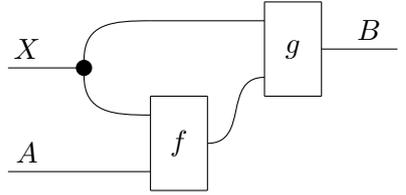
  \[ \input{img/copy-f-g.tikz} \; . \]
  Given $h:X\to Y$ in $\Comon(\cat{C})$, the functor $\Fun{P}h:\Fun{P}Y\to\Fun{P}X$ acts by precomposition on morphisms, taking $f:A\xto{Y}B$ to the morphism $\Fun{P}h(f):A\xto{X}B$ given by
  \[ X\otimes A\xto{h\otimes\id_A}Y\otimes A\xto{f}B \; . \]
  (An alternative way to obtain $\Fun{P}X$ is as the `coKleisli' category of the comonad $X\otimes(-)$.)

  The category of monoidal lenses is then defined to be the category of Grothendieck $\Fun{P}$-lenses.
  The objects of $\Lens_{\Fun{P}}$ are pairs $(X,A)$ of a comonoid $X$ and an object $A$ in $\cat{C}$, and the morphisms are monoidal lenses.
  Given lenses $(f,f^\sharp):(X,A)\to(Y,B)$ and $(g,g^\sharp):(Y,B)\to(Z,C)$, the composite lens has forward component given by $g\circ f:X\to Z$ and backward component given by $f^\sharp\circ \Fun{P}f(g^\sharp):C\xto{X}A$.

  We can depict monoidal lenses string-diagrammatically, with the forward and backward components oriented in opposite directions.
  To exemplify this, note that, because the forwards components are comonoid morphisms, the following equality holds for all composite monoidal lenses $(g,g^\sharp)\circ(f,f^\sharp)$:
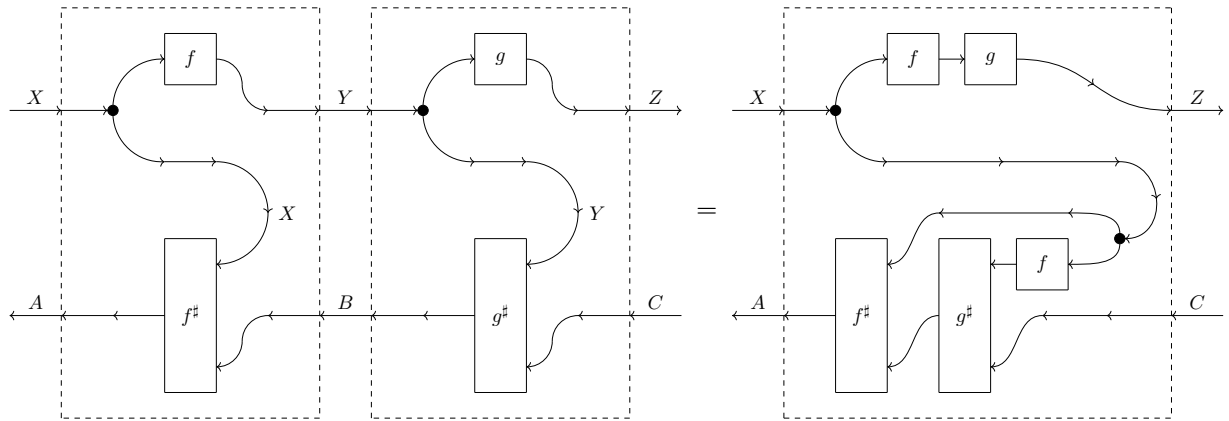
  \[\scalebox{0.68}{\input{img/lens-cartesian-gf-1.tikz}}
  =
  \scalebox{0.68}{\input{img/lens-cartesian-gf-2b.tikz}}\]
  Here, we have decorated the strings with fletches to indicate the direction of information-flow and disambiguate the bidirectionality, and drawn boxes around the pairs that constitute each lens.
  Note however that the parameterizing input to the backwards component of the first lens is not constrained to be a copy of the input to the forwards component; it is only for compositional convenience that we depict lenses this way.
\end{ex}

\begin{defn}
  When $\cat{C}$ is Cartesian monoidal, so that its monoidal structure $(\times,1)$ is the categorical product, we will call monoidal lenses in $\cat{C}$ \textit{Cartesian lenses}.
\end{defn}

\begin{rmk}
  The string-diagrammatic depictions of lenses above were not strictly formal, or at least we have not explain how they might be; we have not exhibited a coherence theorem such as \ref{thm:mon-coh}.
  In this case, the diagrams above are depictions in the graphical calculus of \textcite{Boisseau2020String}.
  An alternative graphical language for a generalization of lenses called \textit{optics}\parencite{Milewski2017Profunctor,Riley2018Categories} has been described by \textcite{Roman2020Open}.
\end{rmk}

Monoidal lenses find uses not only in database theory, but in many other situations, too: the general pattern is ``interacting systems where information flows bidirectionally''.
In economics (specifically, compositional game theory), lenses are used to model the pattern of interaction of economic games: the forward maps encode how players act in light of observations, and the backward maps encode how utility is passed ``backwards in time'' from outcomes, in order to assign credit\parencite{Ghani2016Compositional}.
In non-probabilistic machine learning, lenses can be used to formalize reverse differentiation and hence the backpropagation of error (another kind of credit assignment): the forwards maps represent differentiable processes (such as neural networks), and the backward maps are the reverse-derivatives used to pass error back (\textit{e.g.}, between neural network layers)\parencite{Fong2019Lenses,Cruttwell2022Categorical}.
Generalizations of lenses known as optics\parencite{Milewski2017Profunctor,Riley2018Categories} have also been used both to model economic games with uncertainty (`mixed' strategies)\parencite{Bolt2019Bayesian} and to model the process of dynamic programming (Bellman iteration) used in the related field of reinforcement learning\parencite{Hedges2022Value}, as well as to model client-server interactions in computing\parencite{Videla2022Lenses}.

In systems theory, lenses can be used to formalize various kinds of dynamical system: the forward maps encode their `outputs' or `actions', and the backward maps encode how states and inputs give rise to transitions\parencite{Myers2020Double}.
This latter application will be a particular inspiration to us, and is closely related to Example \ref{ex:lens-poly}, which expresses polynomial functors as lenses (thereby explaining Proposition \ref{prop:poly-bundles}), and for which we need the following canonical family of indexed categories.

\begin{defn} \label{def:self-idx}
  When a category $\cat{C}$ has pullbacks, its slice categories $\cat{C}/C$ collect into an indexed category $\cat{C}/(-):\cat{C}\op\to\Cat{Cat}$ called the \textit{(contravariant\footnote{
    `Contravariant' in contradistinction to the covariant self-indexing of Remark \ref{rmk:ext-para-enrch} (in the context of external parameterization as change-of-enrichment).})
    self-indexing} of $\cat{C}$, and defined as follows.
  On objects $C:\cat{C}$, the self-indexing unsurprisingly returns the corresponding slice categories $\cat{C}/C$.

  Given a morphism $f:A\to B$, the functor $\cat{C}/f:\cat{C}/B\to\cat{C}/A$ is defined by pullback.
  On objects $(E,p):\cat{C}/B$, we define $(\cat{C}/f)(E,p) := (f^\ast E,f^\ast p)$, where $f^\ast E$ is the pullback object $A\times_B E$ and $f^\ast p$ is the associated projection to $A$.
  On morphisms $\varphi:(E,p)\to(E',p')$ in $\cat{C}/B$, we define $(\cat{C}/f)(\varphi)$ as the morphism $f^\ast\varphi:(f^\ast E,f^\ast p)\to(f^\ast E',f^\ast p')$ induced by the universal property of the pullback $f^\ast E'$, as in the commuting diagram
  \[\begin{tikzcd}
    {f^\ast E} && E \\
	  \\
    {f^\ast E'} && {E'} \\
	  \\
	  A && B
	  \arrow[from=3-1, to=3-3]
	  \arrow["\varphi"', from=1-3, to=3-3]
	  \arrow["{f^\ast\varphi}", dashed, from=1-1, to=3-1]
	  \arrow[from=1-1, to=1-3]
	  \arrow["\lrcorner"{anchor=center, pos=0.125}, draw=none, from=1-1, to=3-3]
	  \arrow["{p'}"', from=3-3, to=5-3]
	  \arrow["p", curve={height=-24pt}, from=1-3, to=5-3]
	  \arrow["f", from=5-1, to=5-3]
	  \arrow["{f^\ast p'}", from=3-1, to=5-1]
	  \arrow["\lrcorner"{anchor=center, pos=0.125}, draw=none, from=3-1, to=5-3]
	  \arrow["{f^\ast p}"', curve={height=24pt}, from=1-1, to=5-1]
  \end{tikzcd} \; .\]
\end{defn}

\begin{rmk} \label{rmk:base-change}
  The functors $\cat{C}/f:\cat{C}/B\to\cat{C}/A$ are also known as \textit{base-change} functors, as they change the `base' of the slice category.
\end{rmk}

\begin{ex} \label{ex:lens-poly}
  The category $\Poly$ of polynomial functors (\secref{sec:poly}) is equivalent to the category of Grothendieck lenses for the self-indexing of $\Set$: that is, $\Poly \cong \Lens_{\Set/(-)}$.
  To see this, observe that the objects of $\Lens_{\Set/(-)}$ are bundles $p:E\to B$ of sets.
  If we define the set $p[i]$ to be the fibre $E_i$ of $p$ for each $i:B$, we have an isomorphism $E\cong\sum_{i:B}p[i]$.
  We can then define a polynomial functor $P:=\sum_{i:B}y^{p[i]}$, and then find that $P(1) = B$, which justifies writing the original bundle as $p:\sum_{i:p(1)}p[i]\to p(1)$.
  We saw in Proposition \ref{prop:poly-bundles} how to associate to any polynomial functor $P$ a bundle $p$, and it is easy to check that applying this construction to the $P$ defined here returns our original bundle $p$.
  This shows that the objects of $\Poly$ are in bijection with the objects of $\Lens_{\Set/(-)}$.
  What about the morphisms?

  A morphism $p\to q$ in $\Lens_{\Set/(-)}$, for $p:X\to A$ and $q:Y\to B$ is a pair of functions $f_1:A\to B$ and $f^\sharp:f_1^\ast Y\to X$ such that $f_1^\ast q = p\circ f^\sharp$, as in the following diagram:
  \[\begin{tikzcd}%
    {X} & {f_1^\ast Y} & {Y} \\
    {A} & {A} & {B}
    \arrow["{f^\sharp}"', from=1-2, to=1-1]
    \arrow[from=1-2, to=1-3]
    \arrow["q", from=1-3, to=2-3]
    \arrow["f_1^\ast q"', from=1-2, to=2-2]
    \arrow["p"', from=1-1, to=2-1]
    \arrow[from=2-1, to=2-2, Rightarrow, no head]
    \arrow["{f_1}", from=2-2, to=2-3]
    \arrow[from=1-2, to=2-2]
    \arrow["\lrcorner"{anchor=center, pos=0.125}, draw=none, from=1-2, to=2-3]
  \end{tikzcd}\]
  Replacing the bundles $p$ and $q$ by their polynomial representations $p:\sum_{i:p(1)}p[i]\to p(1)$ and $q:\sum_{j:q(1)}q[j]\to q(1)$, we see that the pair $(f_1,f^\sharp)$ is precisely a morphism of polynomials of the form established in Proposition \ref{prop:poly-bundles}, and that every morphism of polynomials corresponds to such a lens.
  This establishes an isomorphism of hom-sets, and hence $\Poly \cong \Lens_{\Set/(-)}$.
\end{ex}

Lenses are also closely related to wiring diagrams\parencite{Yau2018Operads,Spivak2020Poly} and our linear circuit diagrams (\secref{sec:lin-circ}).

\begin{ex}
  Let $\Cat{FVect}$ denote the category of finite-dimensional real vector spaces and linear maps between them; write $n$ for the object $\rr^n$.
  $\Cat{FVect}$ has a Cartesian monoidal product $(+,0)$ given by the direct sum of vectors ($n+m = \rr^n\oplus\rr^m = \rr^{n+m}$), and whose unit object is $0$.
  The category of monoidal lenses in $(\Cat{FVect},+,0)$ is the category of linear circuit diagrams (Example \ref{ex:lin-circ}).
\end{ex}

Cartesian lenses $(X,A)\lensto(Y,B)$ are in some sense `non-dependent' lenses, because the domain of the backwards map is a simple product $X\times B$, in which the object $B$ does not depend on $x:X$.
We can see polynomial functors as a dependent generalization of Cartesian lenses in $\Set$.

\begin{prop}
  The category of monoidal lenses in $(\Set,\times,1)$ is equivalently the full subcategory of $\Poly$ on the monomials $Xy^A$.
  \begin{proof}[Proof sketch]
    A morphism of monomials $(f_1,f^\sharp):Xy^A\to Yy^B$ is a pair of functions $f_1:X\to Y$ and $f^\sharp:X\times B\to A$; this is a Cartesian lens $(X,A)\to(Y,B)$.
    There is clearly a bijection of objects $Xy^A \leftrightarrow (X,A)$.
  \end{proof}
\end{prop}

In particular, this situation encompasses linear circuit diagrams, which embed into $\Poly$ accordingly.

\begin{rmk} \label{rmk:lincirc-poly}
  There is a forgetful functor from vector spaces to sets, $U:\Cat{FVect}\to\Set$.
  If we write $\Lens(\cat{C})$ to denote the category of monoidal lenses in $\cat{C}$ (with the relevant monoidal structure left implicit), this forgetful functor induces a `change of base' $\Lens(U):\Lens(\Cat{FVect})\to\Lens(\Set)$, since the Grothendieck construction is functorial by Remark \ref{rmk:groth-equiv-1}, and hence so is the $\Lens$ construction.
  There is therefore a canonical embedding of linear circuit diagrams into $\Poly$, $\Lens(\Cat{FVect})\xto{\Lens(U)}\Lens(\Set)\hookrightarrow\Poly$.
\end{rmk}

Our dynamical semantics for approximate inference (Chapter \ref{chp:brain}) can, if one squints a little, be therefore seen as a kind of probabilistic generalization of our algebra for rate-coded neural circuits: it will be an algebra for (a stochastic analogue of) the multicategory $\cat{O}\Poly$ with semantics in categories of (stochastic) dynamical systems.
One can see a morphism of polynomials therefore as a kind of `dependent' circuit diagram, with the forwards component transporting `outgoing' information from inside a (`nested') system to its boundary (its external interface), and the backward component transporting `incoming' information (``immanent signals'') from the boundary internally, depending on the configuration of the boundary.

Of course, to give an $\cat{O}\Poly$-algebra is to give a lax monoidal functor, which means knowing the relevant monoidal structure.
While we saw this in the case of polynomial functors of sets in Proposition \ref{prop:poly-tensor}, it will be helpful when it comes to generalizing $\Poly$ to see how this structure is obtained.
Moreover, we will want a monoidal structure on Bayesian lenses, in order to define joint approximate inference systems.
For these reasons, we now turn to monoidal categories of lenses.

\subsubsection{Monoidal categories of lenses}

The monoidal structures on categories of Grothendieck lenses---at least those of interest here---are a direct corollary of the monoidal Grothendieck construction, Proposition \ref{prop:monoidal-gr}.

\begin{cor} \label{cor:monoidal-grlens}
  When \(F : \cat{C}\op \to \Cat{Cat}\) is equipped with a monoidal indexed category structure \((\mu, \eta)\), its category of lenses \(\Lens_F\) becomes a monoidal category \((\Lens_F, \otimes'_\mu, I_\mu)\).
  On objects \(\otimes'_\mu\) is defined as \(\otimes_\mu\) in Proposition \ref{prop:monoidal-gr}, as is \(I_\mu\).
  On morphisms \((f, f^\dag) : (C, X) \lensto (C', X')\) and \((g, g^\dag) : (D, Y) \lensto (D', Y')\), define
  \[
  (f,f^\dag) \otimes'_\mu (g, g^\dag) := \big(f \otimes g, \mu_{CD}^{\mathrm{op}}(f^\dag, g^\dag)\big)
  \]
  where \(\mu_{CD}^{\mathrm{op}} : F(C)\op \times F(D)\op \to F(C \otimes D)\op\) is the pointwise opposite of \(\mu_{CD}\).
  The associator and unitors are defined as in Proposition \ref{prop:monoidal-gr}.
\end{cor}

As an example, this gives us the tensor product on $\Poly$, which is inherited by the category of Cartesian lenses in $\Set$.

\begin{ex}
  The tensor product structure $(\otimes,y)$ on $\Poly$ is induced by a monoidal indexed category structure $(\mu,\eta)$ on the self-indexing of $(\Set,\times,1)$.
  To define the unitor $\eta$, first note that $\Set/1 \cong \Set$, so that $\eta$ equivalently has the type $\Cat{1}\to\Set$; we thus make the natural choice for $\eta$, the terminal element $\ast\mapsto 1$.
  The laxator $\mu$ is defined for each $B,C:\Set$ by the functor
  \begin{align*}
    \mu_{B,C} \quad : \;\quad\quad \Set/B \;\quad \times \quad\quad \Set/C \quad\quad & \to \qquad \Set/(B\times C) \\
    \bigl(p:\sum_{i:B} p[i]\to B,\;\; q:\sum_{j:C} q[j]\to C \bigr) & \mapsto \sum_{(i,j):B\times C} p[i]\times q[j]
  \end{align*}
  the naturality and functoriality of which follow from the functoriality of $\times$.
  Applying Corollary \ref{cor:monoidal-grlens} to this structure, we obtain precisely the tensor product of polynomials introduced in Proposition \ref{prop:poly-tensor}.
\end{ex}

\begin{cor} \label{cor:lens-smc}
  Since the category of Cartesian lenses in $\Set$ is the monomial subcategory of $\Poly$, to which the tensor structure $(\otimes,y)$ restricts, the latter induces a symmetric monoidal structure on the former, the unit of which is the object $(1,1)$.
  Given objects $(X,A)$ and $(X',A')$, their tensor $(X,A)\otimes(X',A')$ is $(X\times X',A\times A')$.
  Given lenses $(f,f^\sharp):(X,A)\to(Y,B)$ and $(f',{f'}^\sharp):(X',A')\to(Y',B')$, their tensor has forward component $f\times f':X\times X'\to Y\times Y'$ and backward component
  \[ X\times X'\times B\times B' \xto{\id_X\times\mathsf{\sigma_{X',B}}\times\id_{B'}} X\times B\times X'\times B' \xto{f^\sharp\times{f'}^\sharp} A\times A' \]
  where $\sigma$ is the symmetry of the product $\times$.
\end{cor}

We will see that the monoidal structure on Bayesian lenses is defined similarly.
First of all, we need to define Bayesian lenses themselves.

\section{The bidirectional structure of Bayesian updating} \label{sec:bidi-bayes}

In this section, we define a collection of indexed categories, each denoted \(\Fun{Stat}\), whose morphisms can be seen as generalized Bayesian inversions.
Following Definition \ref{def:gr-lens}, these induce corresponding categories of lenses which we call \textit{Bayesian lenses}.
In \secref{sec:buco}, we show abstractly that, for the subcategories of \textit{exact} Bayesian lenses whose backward channels correspond to `exact' Bayesian inversions, the Bayesian inversion of a composite of forward channels is given (up to almost-equality) by the lens composite of the corresponding backward channels.
This justifies calling these lenses `Bayesian', and provides the foundation for the study of approximate (non-exact) Bayesian inversion in Chapter \ref{chp:sgame}.

\begin{rmk} \label{rmk:blens-history}
  Bayesian lenses, and the result that ``Bayesian updates compose optically'', were first introduced by the present author in \parencite{Smithe2020Bayesian}.
  \textcite{Braithwaite2022Dependent} then elaborated the structure to define \textit{dependent} Bayesian lenses, solving the `divide-by-zero' issue already indicated in Remark \ref{rmk:zero-inversions}.
  All three authors then joined forces to produce a paper \parencite{Braithwaite2023Compositional}, published at MFCS 2023, which we take to be a canonical summary of the definitions and basic results.
\end{rmk}

\subsection{State-dependent channels} \label{sec:stat-chan}

As we saw in \secref{sec:comp-prob}, a channel \(c : X \klto Y\) admitting a Bayesian inversion induces a family of inverse channels \(c^\dag_\pi : Y \klto X\), indexed by `prior' states \(\pi : 1 \klto X\).
Making the state-dependence explicit, in typical cases where \(c\) is a probability kernel we obtain a function \(c^\dag : \Giry X \times Y \to \Giry X\), under the assumption that $c\klcirc\pi$ is fully supported for all $\pi:\Giry X$ (see Remark \ref{rmk:bayes-supp} for our justification of this simplifying assumption).
In more general situations, and in light of the full-support assumption, we obtain a morphism \(c^\dag : \cat{C}(I, X) \to \cat{C}(Y, X)\) in the base of enrichment of the monoidal category \((\cat{C}, \otimes, I)\) of \(c\), which for simplicity we take to be $\Set$ (although the construction still succeeds for an arbitrary Cartesian base of enrichment).
We call morphisms of this general type \textit{state-dependent channels}, and structure the indexing as an indexed category.

\begin{defn} \label{def:stat-cat}
  Let \((\cat{C}, \otimes, I)\) be a monoidal category.
  Define the \(\cat{C}\)\textit{-state-indexed} category \(\Fun{Stat}: \cat{C}\op \to \Cat{Cat}\) as follows.
  \begin{align}
    \Fun{Stat} \;\; : \;\; \cat{C}\op \; & \to \; \Cat{Cat} \nonumber \\
    X & \mapsto \Fun{Stat}(X) := \quad \begin{pmatrix*}[l]
      & \Fun{Stat}(X)_0 & := \quad \;\;\; \cat{C}_0 \\
      & \Fun{Stat}(X)(A, B) & := \quad \;\;\; \Set\bigl(\cat{C}(I, X), \cat{C}(A, B)\bigr) \\
      \id_A \: : & \Fun{Stat}(X)(A, A) & := \quad
      \left\{ \begin{aligned}
        \id_A : & \; \cat{C}(I, X)     \to     \cat{C}(A, A) \\
        & \quad\;\;\: \rho \quad \mapsto \quad \id_A
      \end{aligned} \right. \label{eq:stat} \\
    \end{pmatrix*} \\ \nonumber \\
    f : \cat{C}(Y, X) & \mapsto \begin{pmatrix*}[c]
      \Fun{Stat}(f) \; : & \Fun{Stat}(X) & \to & \Fun{Stat}(Y) \vspace*{0.5em} \\
      & \Fun{Stat}(X)_0 & = & \Fun{Stat}(Y)_0 \vspace*{0.5em} \\
      & \Set(\cat{C}(I, X), \cat{C}(A, B)) & \to & \Set\bigl(\cat{C}(I, Y), \cat{C}(A, B)\bigr) \vspace*{0.125em} \\
      & \alpha & \mapsto & \big( \, \sigma : \cat{C}(I, Y) \, \big) \mapsto \big( \, \alpha_{f \klcirc \sigma} : \cat{C}(A, B) \, \big)
    \end{pmatrix*} \nonumber
  \end{align}
  Composition in each fibre \(\Fun{Stat}(X)\) is as in \(\cat{C}\).
  Explicitly, indicating morphisms \(\cat{C}(I, X) \to \cat{C}(A, B)\) in \(\Fun{Stat}(X)\) by \(A \xklto{X} B\), and given \(\alpha : A \xklto{X} B\) and \(\beta : B \xklto{X} C\), their composite $\beta \circ \alpha : A \xklto{X}$ is defined by \((\beta\circ\alpha)_\rho := \beta_\rho \klcirc \alpha_\rho\), where here we indicate composition in \(\cat{C}\) by \(\klcirc\) and composition in the fibres \(\Fun{Stat}(X)\) by \(\circ\).
  Given \(f : Y \klto X\) in \(\cat{C}\), the induced functor \(\Fun{Stat}(f) : \Fun{Stat}(X) \to \Fun{Stat}(Y)\) acts by pre-composition (compare Definition \ref{def:self-idx} of the functorial action of the self-indexing); for example:
  \[ \begin{matrix*}
    \Fun{Stat}(f)(\alpha) & : & \cat{C}(I,Y) & \xto{\cat{C}(I,f)} & \cat{C}(I,X) & \xto{\alpha} & \cat{C}(A,B) \\
    && \sigma & \mapsto & f\klcirc\sigma & \mapsto & \alpha_{f\klcirc\sigma}
  \end{matrix*} \; . \]
\end{defn}

\begin{rmk}
  If we do not wish to make the full-support assumption, and instead we know that the category $\cat{C}$ has a well-defined notion of \textit{support object}\parencite{Fritz2019synthetic,Stein2021Structural,Braithwaite2022Dependent}, then for a given general channel $c:X\klto Y$, we can write the type of its Bayesian inversion $c^\dag$ as $\prod_{\pi:\cat{C}(I,X)}\cat{C}\bigl(\Fun{supp}(c\klcirc\pi),Y\bigr)$.
  As \textcite{Braithwaite2022Dependent} show, this corresponds to a morphism in a certain fibration, and gives rise to a category of dependent Bayesian lenses; see Remark \ref{rmk:bayes-supp}.
\end{rmk}

\begin{notation}
  Just as we wrote \(X \xto{M} Y\) for an internally \(M\)-parameterized morphism in \(\cat{C}(M \odot X, Y)\) (see Proposition \ref{prop:para-bicat}) and $A\xto{\Theta}B$ for an externally $\Theta$-parameterized morphism in $\cat{E}\bigl(\Theta,\cat{C}(A,B)\bigr)$ (see Definition \ref{def:ext-para}), we write \(A \xklto{X} B\) for an \(X\)-state-dependent morphism in \(\Set\big(\cat{C}(I, X), \cat{C}(A, B)\big)\).
  Given a state \(\rho\) in \(\cat{C}(I, X)\) and an \(X\)-state-dependent morphism \(f : A \xklto{X} B\), we write \(f_\rho\) for the resulting morphism in \(\cat{C}(A,B)\).
\end{notation}

\begin{rmk} \label{rmk:stat-para-prox}
  The similarities between state-dependent channels and externally parameterized functions are no coincidence: the indexed category $\Fun{Stat}$ is closely related to an indexed category underlying external parameterization in $\Set$, which in previous work, reported by \textcite{Capucci2021Parameterized}, we called $\Cat{Prox}$ (for `proxies').

  When \(\cat{C}\) is a Kleisli category \(\Kl(T)\), it is of course possible to define a variant of \(\Fun{Stat}\) on the other side of the product-exponential adjunction, with state-dependent morphisms \(A\xklto{X}B\) having the types \(TX \times A \to TB\).
  This avoids the technical difficulties sketched in the preceding example at the cost of requiring a monad \(T\).
  However, the exponential form makes for better exegesis, and so we will stick to that.
\end{rmk}

We will want to place inference systems side-by-side, which means we want a monoidal category structure for Bayesian lenses.
Following Corollary \ref{cor:monoidal-grlens}, this means $\Fun{Stat}$ needs to be a monoidal indexed category.

\begin{prop} \label{prop:stat-lax}
  \(\Fun{Stat}\) is a monoidal indexed category, in the sense of Definition \ref{def:mon-idx-cat}.
  The components \(\mu_{XY} : \Fun{Stat}(X) \times \Fun{Stat}(Y) \to \Fun{Stat}(X \otimes Y)\) of the laxator are defined on objects by \(\mu_{XY}(A, A') := A \otimes A'\) and on morphisms \(f : A \xklto{X} B\) and \(f' : A' \xklto{Y} B'\) as the $X\otimes Y$-state-dependent morphism  denoted $f\otimes f'$ and given by the function
  \[\begin{aligned}
    \mu_{XY}(f,f') : \cat{C}(I, X\otimes Y) &\to \cat{C}(A\otimes A', B\otimes B') \\
    \omega & \mapsto f_{\omega_X} \otimes f'_{\omega_Y}
  \end{aligned} \quad . \]
  Here, \(\omega_X\) and \(\omega_Y\) are the \(X\) and \(Y\) marginals of \(\omega\), given by \(\omega_X := \mathsf{proj}_X \klcirc \omega\) and \(\omega_Y := \mathsf{proj}_Y \klcirc \omega\).
  (Note that this makes $\mu$ into a strict transformation in the sense of Definition \ref{def:lax-trans}.)
  The unit \(\eta : \Cat{1} \to \Fun{Stat}(I)\) of the lax monoidal structure is the functor mapping the unique object \(\ast : \Cat{1}\) to the unit object \(I : \Fun{Stat}(I)\).
\end{prop}

\begin{rmk}
  Note that $\Fun{Stat}$ is also \textit{fibrewise monoidal} in the sense of Remark \ref{rmk:fib-mon-idx-cat}, as an almost trivial consequence of $\cat{C}$ being monoidal.
  We will not make use of this structure in this chapter, but we will return to it in the construction of statistical games in \secref{sec:loss-funcs}.
\end{rmk}

At this point, we can turn to Bayesian lenses themselves.

\subsection{Bayesian lenses} \label{sec:bayes-lens}

We define the category of Bayesian lenses in \(\cat{C}\) to be the category of Grothendieck \(\Fun{Stat}\)-lenses.

\begin{defn} \label{def:stat-lens}
  The category \(\BLens{C}\) of Bayesian lenses in \(\cat{C}\) is the category \(\Lens_{\Fun{Stat}}\) of Grothendieck lenses for the functor \(\Fun{Stat}\).
  A \textit{Bayesian lens} is a morphism in \(\BLens{C}\).
  Where the category \(\cat{C}\) is evident from the context, we will just write \(\Cat{BayesLens}\).
\end{defn}

Unpacking this definition, we find that the objects of \(\BLens{C}\) are pairs \((X, A)\) of objects of \(\cat{C}\).
Morphisms (that is, Bayesian lenses) \((X,A) \lensto (Y,B)\) are pairs \((c, c^\dag)\) of a channel \(c : X \klto Y\) and a generalized Bayesian inversion \(c^\dag : B \xklto{X} A\); that is, elements of the hom objects
\begin{align*}
  \BLens{C}\big((X,A),(Y,B)\big)
  :&= \Lens_\Fun{Stat} \big((X,A),(Y,B)\big) \\
  &\cong \cat{C}(X, Y) \times \Set \big( \cat{C}(I, X), \cat{C}(B, A) \big) \, .
\end{align*}
The identity Bayesian lens on \((X, A)\) is \((\id_X, \id_A)\), where by abuse of notation \(\id_A : \cat{C}(I, Y) \to \cat{C}(A, A)\) is the constant map \(\id_A\) defined in Equation \eqref{eq:stat} that takes any state on \(Y\) to the identity on \(A\).

The sequential composite \((d, d^\dag) \lenscirc (c, c^\dag)\) of \((c, c^\dag) : (X, A) \lensto (Y, B)\) and \((d, d^\dag) : (Y, B) \lensto (Z, C)\) is the Bayesian lens \(\big( (d \klcirc c), (c^\dag \circ c^\ast d^\dag) \big) : (X, A) \lensto (Z, C)\) where \((c^\dag \circ c^\ast d^\dag) : C \xklto{X} A\) takes a state \(\pi : I \klto X\) to the channel \(c^\dag_{\pi} \klcirc \d^\dag_{c \klcirc \pi} : C \klto A\).

To emphasize the structural similarity between Bayesian and monoidal lenses, and visualize the channel $c^\dag_{\pi} \klcirc \d^\dag_{c \klcirc \pi}$, note that following Example \ref{ex:mon-lens}, we can depict Bayesian lens composition using the graphical calculus of \textcite{Boisseau2020String} as
\[\scalebox{0.68}{\input{img/lens-cartesian-dc-1.tikz}}
=
\scalebox{0.68}{\input{img/lens-cartesian-dc-2b.tikz}} \; .\]

\begin{rmk}
  Strictly speaking, these depictions are diagrams in \textcite{Boisseau2020String}'s calculus of string diagrams for optics, which means that they are not direct depictions of the Bayesian lenses themselves; rather they are depictions of the corresponding optics, which we define and elaborate in \parencite{Smithe2020Bayesian}.
  Briefly, these optics are obtained by embedding the categories of forrwards and backwards channels into their corresponding (co)presheaf categories and coupling them together along the `residual' category $\cat{C}$; in the depictions, the string diagrams in the forwards and backwards directions are thus interpreted in these different categories.
  This explains why we are allowed to `copy' the channel $c$ in the depiction above, producing the right-hand side by pushing $c$ through the copier as if it were a comonoid morphism:
  it is because the comonoids in question are $\cat{C}(I,X)$ and $\cat{C}(I,Y)$, and the function $\cat{C}(I,c)$ is indeed a comonoid morphism, even though $c$ is in general not!
\end{rmk}

\begin{rmk}
  Note that the definition of \(\Fun{Stat}\) and hence the definition of \(\BLens{C}\) do not require \(\cat{C}\) to be a copy-delete category, even though our motivating categories of stochastic channels are; all that is required for the definition is that \(\cat{C}\) is monoidal.
  On the other hand, as we can define Bayesian lenses in any copy-delete category, we can define them in $\Set$, where $\Set(1,X) \cong X$ for every set $X$: in this case, Bayesian lenses coincide with Cartesian lenses.
\end{rmk}

Of course, since $\Fun{Stat}$ is a monoidal indexed category, $\BLens{C}$ is a monoidal category.

\begin{prop} \label{prop:blens-monoidal}
  \(\BLens{C}\) is a monoidal category, with structure \(\big((\otimes, (I,I)\big)\) inherited from \(\cat{C}\).
  On objects, define \((A, A') \otimes (B, B') := (A \otimes A', B \otimes B')\).
  On morphisms \((f, f^\dag) : (X, A) \lensto (Y, B)\) and \((g, g^\dag) : (X', A') \lensto (Y', B')\), define \((f, f^\dag) \otimes (g, g^\dag) := (f \otimes g, f^\dag \otimes g^\dag)\), where \(f^\dag \otimes g^\dag : B \otimes B' \xklto{X \otimes X'} A \otimes A'\) acts on states \(\omega : I \klto X \otimes X'\) to return the channel \(f^\dag_{\omega_X} \otimes g^\dag_{\omega_{X'}}\), following the definition of the laxator \(\mu\) in Proposition \ref{prop:stat-lax}.
  The monoidal unit in \(\BLens{C}\) is the pair \((I,I)\) duplicating the unit in \(\cat{C}\).
  When \(\cat{C}\) is moreover symmetric monoidal, so is \(\BLens{C}\).
  \begin{proof}[Proof sketch]
    The main result is immediate from Proposition \ref{prop:stat-lax} and Corollary \ref{cor:monoidal-grlens}.
    When \(\otimes\) is symmetric in \(\cat{C}\), the symmetry lifts to the fibres of \(\Fun{Stat}\) and hence to \(\BLens{C}\).
  \end{proof}
\end{prop}

But $\BLens{C}$ is not in general a copy-discard category.

\begin{rmk}
  Although \(\BLens{C}\) is a monoidal category, it does not inherit a copy-discard structure from \(\cat{C}\), owing to the bidirectionality of its component morphisms.
  To see this, we can consider morphisms into the monoidal unit \((I,I)\), and find that there is generally no canonical discarding map.
  For instance, a morphism \((X,A)\lensto(I,I)\) consists in a pair of a channel \(X\klto I\) (which may indeed be a discarding map) and a state-dependent channel \(I \xklto{X} A\), for which there is generally no suitable choice satisfying the comonoid laws.
  Note, however, that a lens of the type \((X,I)\lensto(I,B)\) might indeed act by discarding, since we can choose the constant state-dependent channel \(B\xklto{X} I\) on the discarding map \(\ground : B\klto I\).
  By contrast, the Grothendieck category \(\int \Fun{Stat}\) \textit{is} a copy-delete category, as the morphisms \((X,A)\to(I,I)\) in \(\int\Fun{Stat}\) are pairs \(X\klto I\) and \(A\xklto{X} I\), and so for both components we can choose morphisms witnessing the comonoid structure.
\end{rmk}

\subsection{Bayesian updates compose optically} \label{sec:buco}

In this section we prove the fundamental result that justifies the development of \textit{statistical games} as hierarchical inference systems in Chapter \ref{chp:sgame}:
that the Bayesian inversion of a composite channel is given up to almost-equality by the lens composite of the backwards components of the associated `exact' Bayesian lenses.

\begin{defn} \label{def:exact-blens}
  Let \((c, c^\dag) : (X, X) \lensto (Y, Y)\) be a Bayesian lens.
  We say that \((c, c^\dag)\) is \textit{exact} if \(c\) admits Bayesian inversion and, for each \(\pi : I \klto X\) such that \(c \klcirc \pi\) has full support, \(c\) and \(c^\dag_\pi\) together satisfy equation \eqref{eq:bayes-abstr} (p. \pageref{eq:bayes-abstr}).
  Bayesian lenses that are not exact are said to be \textit{approximate}.
\end{defn}

\begin{thm} \label{thm:buco}
  Let \((c, c^\dag)\) and \((d, d^\dag)\) be sequentially composable exact Bayesian lenses.
  Then, for any state $\pi$ on the domain of $c$, the contravariant component $c^\dag \circ c^\ast d^\dag$ of the composite lens \((d, d^\dag) \lenscirc (c, c^\dag)\) is the Bayesian inversion of \(d \klcirc c\).
  That is to say, \emph{Bayesian updates compose optically}: \((d \klcirc c)^\dag_\pi \overset{d
    \klcirc c \klcirc \pi}{\sim} c^\dag_\pi \klcirc d^\dag_{c \klcirc \pi}\).
  \begin{proof}
    For any suitably-typed $\pi$, the state-dependent channel $c^\dag \circ c^\ast d^\dag$ returns the channel \(c^\dag_\pi \klcirc d^\dag_{c \klcirc \pi} : Z \klto X\), so to establish the result it suffices to show that
    \[
    \scalebox{0.875}{\input{img/joint-cdag-ddag-dc-pi.tikz}}
    \ = \ 
    \scalebox{0.875}{\input{img/joint-dc-pi.tikz}} .
    \]

    We have
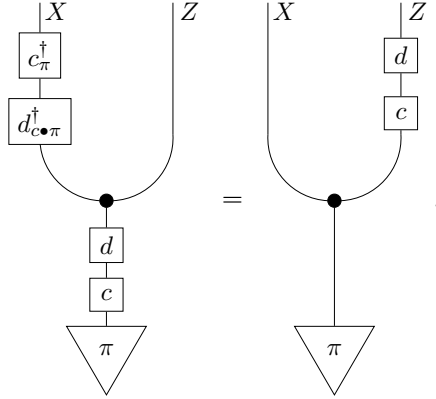
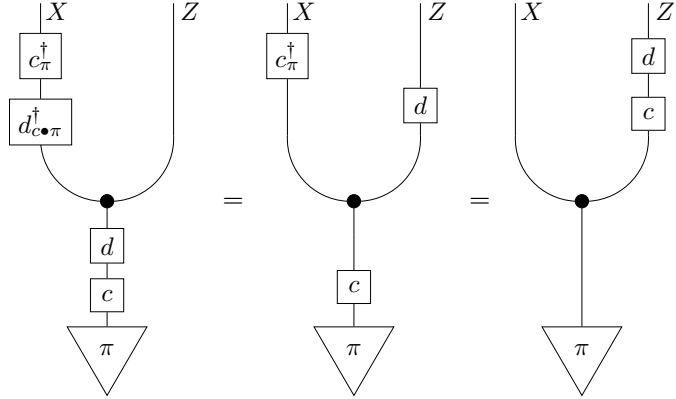
    \[
    \scalebox{0.875}{\input{img/joint-cdag-ddag-dc-pi.tikz}}
    \ = \ 
    \scalebox{0.875}{\input{\detokenize{img/joint-cdag_d-c-pi}.tikz}}
    \ = \ 
    \scalebox{0.875}{\input{img/joint-dc-pi.tikz}}
    \]
    where the first obtains because \(d^\dag_{c \klcirc \pi}\) is by assumption a Bayesian inverse of \(d\) with respect to \(c \klcirc \pi\), and the second because \(c^\dag_\pi\) is likewise a Bayesian inverse of \(c\) with respect to \(\pi\).
    Hence, \(c^\dag_\pi \klcirc d^\dag_{c \klcirc \pi}\) and \((d \klcirc c)^\dag_\pi\) are both Bayesian inversions of \(d \klcirc c\) with respect to \(\pi\).
    Since Bayesian inversions are almost-equal (Proposition \ref{prop:bayes-almost-equal}), we have \(c^\dag_\pi \klcirc d^\dag_{c\klcirc \pi} \overset{d \klcirc c \klcirc \pi}{\sim} (d \klcirc c)^\dag_\pi\), as required.
  \end{proof}
\end{thm}

This theorem has the following important immediate consequence.

\begin{cor} \label{cor:buco-as-func}
  Suppose $\cat{C}^\dag$ is a subcategory of $\cat{C}$ all of whose channels admit Bayesian inversion, and consider the restriction to $\cat{C}^\dag$ of the fibration $\pLens:\BLens{\cat{C}}\to\cat{C}$ of Bayesian lenses, denoted $\pLensDag$.
  Then there is an almost sure section $\dag:\cat{C}^\dag\to\BLens{\cat{C}}$ of $\pLensDag$ taking each object $X$ to $(X,X)$ and each channel $c:X\klto Y$ to a lens $(c,c^\dag):(X,X)\lensto(Y,Y)$ where $c^\dag$ is an almost-surely unique Bayesian inversion of $c$.
  Hence the composite $\cat{C}\xto{\dag}\BLens{\cat{C}}\xto{\pLensDag}\cat{C}$ is equal to the identity functor $\id_{\cat{C}}$.
\end{cor}

\begin{rmk}
  A morphism $\sigma:B\to E$ is a section of $\pi:E\to B$ when $\pi\circ\sigma = \id_B$.
  In standard category theory, a section of a fibration $\pi$ is therefore a functor: but, since Bayesian inversion is only defined up to almost-equality, the functoriality of the preceding corollary is accordingly weakened.
  This leads to the notion of \textit{almost sure section}, which we formalize by lifting the relation of almost-equality from $\cat{C}$ to $\BLens{\cat{C}}$, as follows.
  Suppose $(c,c^\sharp)$ and $(d,d^\sharp)$ are lenses $(X,X)\lensto(Y,Y)$.
  Then we may say that they are equivalent up to almost equality, denoted $(c,c^\sharp)\approx(d,d^\sharp)$, if for all states $\alpha:I\klto X$, we have $c\overset{\alpha}{\sim}d$ and $c^\sharp_\alpha\overset{c\klcirc\alpha}{\sim}d^\sharp_\alpha$.
  If additionally we have $c = d$, we write $(c,c^\sharp)\simeq(d,d^\sharp)$ and say that they are strongly equivalent.
  Note then that the mapping $\dag$ of the preceding corollary is functorial up to this strong equivalence: $\dag(d)\lenscirc\dag(c) \simeq \dag(d\klcirc c)$; this is what we mean by \textit{almost sure section}.
  We believe this notion (and the implicit more general one of \textit{almost sure functor}) to be new, but do not study it further here.
\end{rmk}

\begin{rmk}
  In the context of finitely-supported probability (\textit{i.e.}, in \(\Kl(\Dst)\)), almost-equality coincides with simple equality over the support, and so Bayesian inversions are then just equal (over the support).
  This suggests that, in this context, $\dag$ may be strengthened to a strict functor: but the qualification \textit{over the support} means we must use the machinery of \textit{dependent Bayesian lenses} (Remark \ref{rmk:bayes-supp}); then, $\dag$ does yield a strict functor.
\end{rmk}

\begin{rmk} \label{rmk:dag-not-monoidal}
  Note that the functor $\dag$ is not monoidal, because inverting the tensor of two channels with respect to a joint distribution is not the same as inverting the two channels independently with respect to the marginals and tensoring them together (unless the joint is already the product of two independent states); that is, $(c\otimes d)^\dag_\omega \neq c^\dag_{\omega_1}\otimes d^\dag_{\omega_2}$, where $\omega_1$ and $\omega_2$ are the two marginals of the joint state $\omega$.
  Technically, this situation obtains because there is no channel $X_1\otimes X_2 \klto X_1\otimes X_2$ that performs this marginalization-then-tensoring that could play the part of the laxator of $\dag$.
  (But note that typically a probability monad $\Pa$ will be `bimonoidal', with the `opmonoidal' structure $\Pa(X_1\times X_2)\to \Pa X_1\times \Pa X_2$ witnessing this joint-marginalization operation \parencite[{\S4}]{Fritz2018Bimonoidal}; the technical hurdle is that this structure typically interacts nicely with the monad structure, since the tensor of two Dirac deltas is again a Dirac delta.)

  In \secref{sec:mon-stat-games}, we will use the machinery of statistical games to measure the error produced by inverting two channels independently, versus inverting them jointly.
\end{rmk}

Historically, lenses have often been associated with `lens laws': additional axioms guaranteeing their well-behavedness.
These laws originate in the context of database systems, and we now investigate how well they are satisfied by Bayesian lenses, where one might see an inference system as a kind of uncertain database.
We will find that Bayesian lenses are not lawful in this traditional sense, because they `mix' information.

\subsection{Lawfulness of Bayesian lenses}
\label{sec:lawfulness}

The study of Cartesian lenses substantially originates in the context of bidirectional transformations of data in the computer science and database community \citep{Bohannon2006Relational,Foster2007Combinators}, where we can think of the \(\mathsf{view}\) (or \(\mathsf{get}\)) function as returning part of a database record, and the \(\mathsf{update}\) (or \(\mathsf{put}\)) function as `putting' a part into a record and returning the updated record.
In this setting, axioms known as \emph{lens laws} can be imposed on lenses to ensure that they are `well-behaved' with respect to database behaviour:
for example, that updating a record with some data is idempotent (the `put-put' law).

We might hope that well-behaved or ``very well-behaved'' lenses in the database context should roughly correspond to our notion of exact Bayesian lens: with the view that an inference system, formalized by a Bayesian lens, is something like a probabilistic database.
However, as we will see, even exact Bayesian lenses are only weakly lawful in the database sense: Bayesian updating mixes information in the prior state (the `record') with the observation (the `data'), rather than replacing the prior information outright.

We will concentrate on the three lens laws that have attracted recent study \citep{Riley2018Categories,Boisseau2020String}: \texttt{GetPut}, \texttt{PutGet}, and \texttt{PutPut}.
A Cartesian lens satisfying the former two is \emph{well-behaved} while a lens satisfying all three is \emph{very well-behaved}, in the terminology of \citet{Foster2007Combinators}. Informally, \texttt{GetPut} says that getting part of a record and putting it straight back returns an unchanged record; \texttt{PutGet} says that putting a part into a record and then getting it returns the same part that we started with; and \texttt{PutPut} says that putting one part and then putting a second part has the same effect on a record as just putting the second part (that is, \(\mathsf{update}\) completely overwrites the part in the record).
We will express these laws graphically, and consider them each briefly in turn.

Note first that we can lift any channel \(c\) in the base category \(\cat{C}\) into any state-dependent fibre \(\Fun{Stat}(A)\) using the constant (identity-on-objects) functor taking \(c\) to the constant-valued state-indexed channel \(\rho \mapsto c\) that maps any state \(\rho\) to \(c\).
We can lift string diagrams in \(\cat{C}\) into the fibres accordingly.

\paragraph{\texttt{GetPut}}

\begin{defn}
  A lens \((c,c^\dag)\) is said to satisfy the \texttt{GetPut} law if it satisfies the left equality in \eqref{eq:getput-law} below.
  Equivalently, because the copier induced by the Cartesian product is natural (\emph{i.e.}, \(\copier \circ f = (f \times f) \circ \copier\)), for any state \(\pi\), we say that \((c,c^\dag)\) satisfies \texttt{GetPut} with respect to \(\pi\) if it satisfies the right equality in \eqref{eq:getput-law} below.
  \begin{equation} \label{eq:getput-law}
    \input{img/bayes-getput1.tikz}
    \hspace{0.06\linewidth}
    \Longrightarrow
    \hspace{0.06\linewidth}
    \input{img/bayes-getput2.tikz}
  \end{equation}
  (Note that here we have written the copying map as $\copier$, since we are assuming an ambient Cartesian monoidal structure; hence for a Bayesian lens we interpret the left diagram above in the image of the Yoneda embedding.)
\end{defn}

\begin{prop}
  When \(c\) is causal, the exact Bayesian lens \((c,c^\dag)\) satisfies the \texttt{GetPut} law with respect to any state \(\pi\) for which \(c\) admits Bayesian inversion.
  \begin{proof}
    Starting from the right-hand-side of \eqref{eq:getput-law}, we have the following chain of equalities
    \[ \input{img/bayes-getput3.tikz} \]
    where the first holds by the counitality of \(\bcopier\), the second by the causality of \(c\), the third since \(c\) admits Bayesian inversion \eqref{eq:bayes-abstr} with respect to \(\pi\), and the fourth again by counitality.
  \end{proof}
  Note that by Bayes' law, exact Bayesian lenses only satisfy \texttt{GetPut} with respect to states.
  This result means that, if we think of \(c\) as generating a prediction \(c \klcirc \pi\) from a prior belief \(\pi\), then if our observation exactly matches the prediction, updating the prior \(\pi\) according to Bayes' rule results in no change.
\end{prop}

\paragraph{\texttt{PutGet}}

The \texttt{PutGet} law is characterized for a lens $(v,u)$ by the following equality:
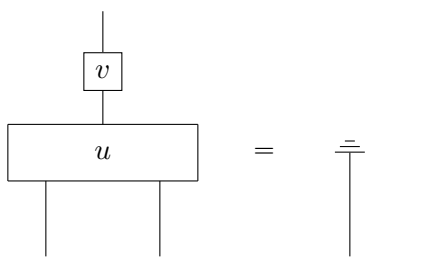
\[ \input{img/bayes-putget.tikz} \]
In general, \texttt{PutGet} does not hold for exact Bayesian lenses \((c,c^\dag)\).
However, because \texttt{GetPut} holds with respect to states \(\pi\), we do have \(c \klcirc c^\dag_\pi \klcirc c \klcirc \pi = c \klcirc \pi\); that is, \texttt{PutGet} holds for exact Bayesian lenses \((c,c^\dag)\) for the prior $\pi$ and `input' \(c \klcirc \pi\).

The reason \texttt{PutGet} fails to hold in general is that Bayesian updating mixes information from the prior and the observation, according to the strength of belief.
Consequently, updating a belief according to an observed state and then producing a new prediction need not result in the same state as observed; unless, of course, the prediction already matches the observation.

\paragraph{\texttt{PutPut}}

Finally, the \texttt{PutPut} law for a lens \((v,u)\) is characterized by the following equality:
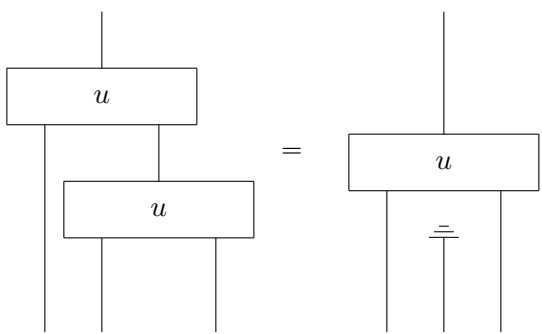
\[ \input{img/bayes-putput.tikz} \]
\texttt{PutPut} fails to hold for exact Bayesian lenses for the same reason that \texttt{PutGet} fails to hold in general: updates mix old and new beliefs, rather than entirely replace the old with the new.

\paragraph{Comment}

In the original context of computer databases, there is assumed to be no uncertainty, so a `belief' is either true or false.
Consequently, there can be no `mixing' of beliefs; and in database applications, such mixing may be highly undesirable.
Bayesian lenses, on the other hand, live in a fuzzier world: our present interest in Bayesian lenses originates in their application to describing cognitive and cybernetic processes such as perception and action, and here the ability to mix beliefs according to uncertainty is desirable.

Possibly it would be of interest to give analogous information-theoretic lens laws that characterize exact and approximate Bayesian lenses and their generalizations; and we might then expect the `Boolean' lens laws to emerge in the extremal case where there is no uncertainty and only Dirac states.
We leave such an endeavour for future work: Bayes' law \eqref{eq:bayes-abstr} is sufficiently concise and productive for our purposes here.

\chapter{Statistical games} \label{chp:sgame}

In this chapter, we characterize a number of well known systems of approximate inference as \textit{loss models} (defined in \secref{sec:loss-models}): lax sections of 2-fibrations of statistical games, themselves constructed (in \secref{sec:loss-funcs}) by attaching internally-defined loss functions to Bayesian lenses.
Our examples include the relative entropy (\secref{sec:rel-ent}), which constitutes a \textit{strict} section, and whose chain rule is formalized by the horizontal composition of the 2-fibration.
In order to capture this compositional structure, we first introduce the notion of `copy-composition' (in \secref{sec:copy-comp-copara}), alongside corresponding bicategories through which the composition of copy-discard categories factorizes.
These latter bicategories are obtained as a variant of the $\copara$ construction \parencite[\S2]{Capucci2021Towards} (dual to the internal parameterization of \secref{sec:int-para}), and so we additionally introduce coparameterized Bayesian lenses (\secref{sec:copara-blens}), proving that coparameterized Bayesian updates compose optically (\secref{sec:copara-buco}), as in the non-coparameterized case.

Besides the relative entropy, our other examples of loss models are given by maximum likelihood estimation (\secref{sec:mle}), the free energy (which gives us in \secref{sec:fe} a characterization of \textit{autoencoders}), and the `Laplace'  approximation to the free energy (\secref{sec:laplace}).
It is this latter loss model which will, in Chapter \ref{chp:brain}, finally yield the dynamical semantics for predictive coding.

We begin with a discussion of compositional approximate inference from the `lens' perspective, focusing on the relative entropy.

\section{Compositional approximate inference, via the chain rule for relative entropy}

In Chapter \ref{chp:buco}, we observed that the Bayesian inversion of a composite stochastic channel is (almost surely) equal to the `lens composite' of the inversions of the factors; that is, \textit{Bayesian updates compose optically} (`BUCO', Theorem \ref{thm:buco}).
Formalizing this statement for a given category $\cat{C}$ yields a fibration of \textit{Bayesian lenses} as a Grothendieck construction of the indexed category of state-dependent channels (Definition \ref{def:stat-lens}), and Bayesian inversion almost surely yields a section $\dag$ of the corresponding fibration (Corollary \ref{cor:buco-as-func}).
This section $\dag$ picks out a special class of Bayesian lenses, which we call \textit{exact} as they compute `exact' inversions (Definition \ref{def:exact-blens}), but although the category $\BLens{}(\cat{C})$ has many other morphisms, its construction is not extravagant:
by comparison, we think of the non-exact lenses as representing \textit{approximate} inference systems.
This is particularly necessary in computational applications, because computing exact inversions is usually intractable, but this creates a new problem: choosing an approximation, and measuring its performance.
In this chapter, we formalize this process, by attaching \textit{loss functions} to Bayesian lenses, thus creating another fibration, of \textit{statistical games}.
Sections of this latter fibration encode compositionally well-behaved systems of approximation that we call \textit{loss models}.

A classic example of a loss model will be supplied by the relative entropy, which in some sense measures the `divergence' between distributions: the game here is then to minimize the divergence between the approximate and exact inversions.
If $\pi$ and $\pi'$ are two distributions on a space $X$, with corresponding density functions $p_\pi$ and $p_{\pi'}$ (both with respect to a common measure),
then their relative entropy $D(\pi,\pi')$ is the real number given by $\E_{x\sim\pi}\left[\log p_\pi(x) - \log p_{\pi'}(x)\right]$\footnote{
  For details about this `expectation' notation $\E$, see \ref{nota:expectations}.}.
Given a pair of channels $\alpha,\alpha':A\klto B$ (again commensurately associated with densities), we can extend $D$ to a map $D_{\alpha,\alpha'}:A\to\rr_+$ in the natural way, writing $a\mapsto D\bigl(\alpha(a),\alpha'(a)\bigr)$.
We can assign such a map $D_{\alpha,\alpha'}$ to any such parallel pair of channels, and so, following the logic of composition in $\cat{C}$\footnote{
In which, following the Chapman-Kolmogorov rule, a composite channel $\beta\klcirc\alpha$ can be expressed as the expectation of $\beta$ under $\alpha$, \textit{i.e.} $a\mapsto\E_{b\sim\alpha(a)}\left[\beta(b)\right]$.},
we might hope for the following equation to hold for all $a:A$ and composable parallel pairs $\alpha,\alpha':A\klto B$ and $\beta,\beta':B\klto C$:
\[ D_{\beta\klcirc\alpha,\beta'\klcirc\alpha'}(a) = \E_{b\sim\alpha(a)} \left[ D_{\beta,\beta'}(b) \right] + D_{\alpha,\alpha'}(a) \]

The right-hand side is known as the \textit{chain rule} for relative entropy, but, unfortunately, the equation does \textit{not} hold in general, because the composites $\beta\klcirc\alpha$ and $\beta'\klcirc\alpha'$ each involve an extra expectation:
\begin{align*}
  D_{\beta\klcirc\alpha,\beta'\klcirc\alpha'}(a)
  &= \E_{c\sim\beta\klcirc\alpha(a)}\left[\log p_{\beta\klcirc\alpha(a)}(c) - \log p_{\beta'\klcirc\alpha'(a)}(c) \right] \\
  &= \E_{c\sim\beta\klcirc\alpha(a)}\left[\log\E_{b\sim\alpha(a)}\left[p_\beta(c|b)\right] - \log\E_{b\sim\alpha'(a)}\left[p_{\beta'}(c|b)\right] \right]
\end{align*}
However, we \textit{can} satisfy an equation of this form by using `copy-composition':
if we write $\bcopier_B$ to denote the canonical `copying' comultiplication on $B$, and define $\beta\klcirc^2\alpha := (\id_B\otimes\beta)\klcirc\bcopier_B\klcirc\alpha$, as depicted by the string diagram
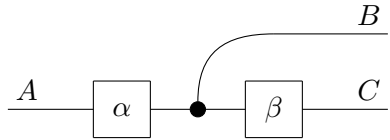
\[ \input{img/gen-model-alpha-beta-1.tikz} \]
then $D_{\beta\klcirc^2\alpha,\beta'\klcirc^2\alpha'}(a)$ \textit{does} equal the chain-rule form on the right-hand side:
\begin{align*}
  D_{\beta\klcirc^2\alpha,\beta'\klcirc^2\alpha'}(a)
  &= \E_{b\sim\alpha(a)} \E_{c\sim\beta(b)} \left[ \log p_\beta(c|b)p_\alpha(b|a) - \log p_{\beta'}(c|b)p_{\alpha'}(b|a) \right] \\
  &= \E_{b\sim\alpha(a)} \left[ \E_{c\sim\beta(b)} \left[ \log p_\beta(c|b) - \log p_{\beta'}(c|b) \right] + \log p_\alpha(b|a) - \log p_{\alpha'}(b|a) \right] \\
  &= \E_{b\sim\alpha(a)} \left[ D_{\beta,\beta'}(b) \right] + D_{\alpha,\alpha'}(a)
\end{align*}
where the second line follows by the linearity of expectation.
This result exhibits a general pattern about copy-discard categories (Definition \ref{def:cd-cat}) such as $\cat{C}$: composition $\klcirc$ can be decomposed into first copying $\bcopier$, and then discarding $\ground$.
If we don't discard, then we retain the `intermediate' variables, and this results in a functorial assignment of relative entropies to channels.

The rest of this chapter is dedicated to making use of this observation to construct loss models, including (but not restricted to) the relative entropy.
The first complication that we encounter is that copy-composition is not strictly unital, because composing with an identity retains an extra variable.
To deal with this, in \secref{sec:copy-comp}, we construct a \textit{bicategory} of copy-composite channels, extending the $\copara$ construction, and build coparameterized (copy-composite) Bayesian lenses accordingly; we also prove a corresponding BUCO result.
In \secref{sec:stat-games}, we then construct 2-fibrations of statistical games, defining loss functions internally to any copy-discard category $\cat{C}$ that admits ``bilinear effects''.
Because we are dealing with approximate systems, the 2-dimensional structure of the construction is useful: loss models are allowed to be \textit{lax} sections.
We then characterize the relative entropy, maximum likelihood estimation, the free energy, and the `Laplacian' free energy as such loss models.

Unsurprisingly, each of these loss functions are moreover (lax) monoidal, and, assuming $\cat{C}$ is symmetric monoidal, each of the constructions mentioned here result in monoidal (2-)fibrations.
We explore this monoidal structure in \secref{sec:mon-stat-games}.

\section{`Copy-composite' Bayesian lenses} \label{sec:copy-comp}

\subsection{Copy-composition by coparameterization} \label{sec:copy-comp-copara}

In a locally small copy-discard category $\cat{C}$ (Definition \ref{def:cd-cat}), every object $A$ is equipped with a canonical comonoid structure $(\bcopier_A,\ground_A)$, and so, by the comonoid laws (Definition \ref{def:comonoid}), it is almost a triviality that the composition functions $\klcirc:\cat{C}(B,C)\times\cat{C}(A,B)\to\cat{C}(A,C)$ factorize as
\begin{gather*}
  \cat{C}(B,C)\times\cat{C}(A,B)
  \xto{(\id_B\otimes-)\times\cat{C}\left(\id_A,\bcopier_B\right)}
  \cat{C}(B\otimes B,B\otimes C)\times\cat{C}(A,B\otimes B)
  \; \cdots \\ \cdots \; \xto{\klcirc}
  \cat{C}(A,B\otimes C)
  \xto{\cat{C}(\id_A,\proj_C)}
  \cat{C}(A,C)
\end{gather*}
where the first factor copies the $B$ output of the first morphism and tensors the second morphism with the identity on $B$, the second factor composes the latter tensor with the copies, and the third discards the extra copy of $B$\footnote{\label{fn:proj} We define $\proj_C := B\otimes C \xto{\ground_B\otimes\id_C} I\otimes C \xto{\lambda_C} C$, using the comonoid counit $\ground_B$ and the left unitor $\lambda_C$ of $\cat{C}$'s monoidal structure.}.
This is, however, only \textit{almost} trivial, since it witnesses the structure of Chapman-Kolmogorov style composition in categories of stochastic channels such as $\Kl(\Da)$, the Kleisli category of the (finitary) distributions monad $\Da:\Cat{Set}\to\Cat{Set}$ (\secref{sec:stoch-mat}).
There, given channels $c:A\klto B$ and $d:B\klto C$, the composite $d\klcirc c$ is formed first by constructing the `joint' channel, denoted $d\klcirc^2c$ and defined by $(d\klcirc^2c)(b,c|a) := d(c|b)c(b|a)$, and then discarding (marginalizing over) $b:B$, giving
\[ (d\klcirc c)(c|a) = \sum_{b:B} (d\klcirc^2c)(b,c|a) = \sum_{b:B} d(c|b)c(b|a) \, . \]
Of course, the channel $d\klcirc^2 c$ is not a morphism $A\klto C$, but rather $A\klto B\otimes C$; that is, it is \textit{coparameterized} by $B$, in a sense formally dual to the notion of parameterization of \secref{sec:int-para}.

Moreover, as noted above, $\klcirc^2$ is not strictly unital: given the composites $\id_B\overline{\klcirc}^2f$ and $f\overline{\klcirc}^2\id_A$, we need 2-cells that discard the coparameters introduced by the copying; and, inversely, we need 2-cells $f\mapsto\id_B\overline{\klcirc}^2f$ and $f\mapsto f\overline{\klcirc}^2\id_A$ that introduce them.
The former are of course given by the discarding structure
\begin{gather}
  \input{img/copara2-unit-left-1.tikz} \quad\mapsto\quad \input{img/copara2-unit-left-2.tikz} \quad=\quad \input{img/copara2-unit-left-3.tikz} \label{eq:copara2-unit-left-disc}
  \\
  \input{img/copara2-unit-right-1.tikz} \quad\mapsto\quad \input{img/copara2-unit-right-2.tikz} \quad=\quad \input{img/copara2-unit-left-3.tikz} \label{eq:copara2-unit-right-disc}
\end{gather}
while the latter are given by copying:
\begin{gather}
  \input{img/copara2-unit-left-3.tikz} \quad\mapsto\quad \input{img/copara2-unit-left-1.tikz} \label{eq:copara2-unit-left-copy}
  \\
  \input{img/copara2-unit-left-3.tikz} \quad\mapsto\quad \input{img/copara2-unit-right-1.tikz} \label{eq:copara2-unit-right-copy}
\end{gather}
These putative 2-cells clearly need access to copies of the domain and codomain of $f$, and hence are not available in the standard $\copara$ construction obtained by formally dualizing $\para$.
For this reason, we construct a bicategory $\ccopara(\cat{C})$ as a variant of the $\copara$ construction, in which a 1-cell $A\to B$ may be any morphism $A\klto M\otimes B$ in $\cat{C}$, and where horizontal composition is precisely copy-composition.
We will henceforth drop the cumbersome notation $\klcirc^2$, and write simply $\klcirc$ for horizontal composition in $\ccopara(\cat{C})$, matching the composition operator of $\cat{C}$ itself.
(Later, if we need to be explicit about horizontal composition, we will sometimes use the symbol $\diamond$.)

\begin{thm} \label{thm:copara2}
  Let $(\cat{C},\otimes,I)$ be a copy-discard category.
  Then there is a bicategory $\ccopara(\cat{C})$ as follows.
  Its 0-cells are the objects of $\cat{C}$.
  A 1-cell $f:A\xto[M]{}B$ is a morphism $f:A\to M\otimes B$ in $\cat{C}$.
  A 2-cell $\varphi:f\Rightarrow f'$, with $f:A\xto[M]{}B$ and $f':A\xto[M']{}B$, is a morphism $\varphi:A\otimes M\otimes B\to M'$ of $\cat{C}$, satisfying the \textit{change of coparameter} axiom:
  \[ \input{img/copara2-2cell-1.tikz} \quad\,=\quad\, \input{img/copara2-2cell-2b.tikz} \]

  Given 2-cells $\varphi:f\Rightarrow f'$ and $\varphi':f'\Rightarrow f''$, their vertical composite $\varphi'\odot\varphi:f\Rightarrow f''$ is given by the following string diagram:
  \[ \input{img/copara2-2cell-v-2.tikz} \]
  The identity 2-cell $\id_f:f\Rightarrow f$ on $f:A\xto[M]{}B$ is given by the projection morphism $\proj_M:A\otimes M\otimes B\to M$ obtained by discarding $A$ and $B$, as in footnote \ref{fn:proj}.

  The horizontal composite $g\circ f:A\xto[M\otimes B\otimes N]{}C$ of 1-cells $f:A\xto[M]{}B$ then $g:B\xto[N]{}C$ is given by the following string diagram in $\cat{C}$:
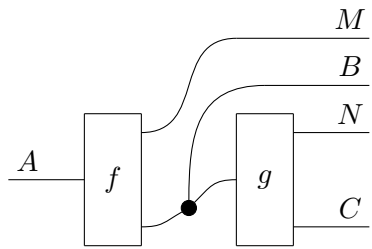
  \[ \input{img/gf-copara2.tikz} \]
  Strictly speaking, we define the coparameter of $g\circ f$ to be $(M\otimes B)\otimes N$.
  The identity 1-cell $\id_A$ on $A$ is given by the inverse of the left unitor of the monoidal structure on $\cat{C}$, \textit{i.e.} $\id_A := \lambda_A^{-1} : A\xto[I]{}A$, with coparameter thus given by the unit object $I$.

  The horizontal composite $\gamma\circ\varphi:(g\circ f)\Rightarrow (g'\circ f')$ of 2-cells $\varphi:f\Rightarrow f'$ and $\gamma:g\Rightarrow g'$ is given by the string diagram
  \[ \input{img/copara2-2cell-h-1.tikz} \quad . \]
  \begin{proof}
    To show that $\ccopara(\cat{C})$ is a bicategory, we need to establish the unitality and associativity of vertical composition; verify that horizontal composition is well-defined and functorial; establish the weak unitality and associativity of horizontal composition; and confirm that the corresponding unitors and associator satisfy the bicategorical coherence laws.
    Then, to prove that $\ccopara(\cat{C})$ is moreover monoidal, we need to demonstrate that the tensor as proposed satisfies the data of a monoidal bicategory.
    However, since the monoidal structure is inherited from that of $\cat{C}$, we will elide much of this latter proof, and demonstrate only that the tensor is functorial; the rest follows straightforwardly but tediously.

    We begin by confirming that vertical composition $\odot$ is unital and associative.
    To see that $\odot$ is unital, simply substitute the identity 2-cell (given by projection onto the coparameter) into the string diagram defining $\odot$ and then apply the comonoid counitality law twice (once on the left, once on the right).
    The associativity of $\odot$ requires that $\varphi''\odot(\varphi'\odot\varphi)=(\varphi''\odot\varphi')\odot\varphi$, which corresponds to the following graphical equation:
    \[ \scalebox{1.0}{\input{img/copara2-v-assoc-1.tikz}} \quad\, = \quad\, \scalebox{1.0}{\input{img/copara2-v-assoc-2.tikz}} \]
    To see that this equation is satisfied, simply apply the comonoid coassociativity law twice (once left, once right).

    Next, we check that horizontal composition $\circ$ is well-defined, which amounts to checking whether the horizontal composite of 2-cells satisfies the change of coparameter axiom.
    Again, we reason graphically.
    Given 2-cells $\varphi$ and $\gamma$ between composable pairs of 1-cells $f,f'$ and $g,g'$, our task is to verify that
    \[ \scalebox{1.0}{\input{img/copara2-h-1.tikz}} \quad = \quad \scalebox{1.0}{\input{img/copara2-h-4.tikz}} \quad . \]
    Since $\varphi$ and $\gamma$ satisfy change of coparameter \textit{ex hypothesi}, the left hand side is equal to the morphism
    \[ \scalebox{1.0}{\input{img/copara2-h-2.tikz}} \quad . \]
    By comonoid coassociativity, this is in turn equal to
    \[ \scalebox{1.0}{\input{img/copara2-h-3.tikz}} \]
    which, by the definition of $\circ$, is precisely equal to
    \[ \scalebox{1.0}{\input{img/copara2-h-4.tikz}} \]
    and so this establishes the result.

    We now verify that $\circ$ so defined is functorial on 2-cells, beginning with the preservation of composition.
    We need to validate the equation $(\gamma'\circ\varphi')\odot(\gamma\circ\varphi) = (\gamma'\odot\gamma)\circ(\varphi'\odot\varphi)$ (for appropriately composable 2-cells).
    This amounts to checking the following equation, which can be seen to hold by two applications of comonoid coassociativity:
    \[ \scalebox{1.0}{\input{img/copara2-h-func-1.tikz}} \quad\, = \quad\, \scalebox{1.0}{\input{img/copara2-h-func-2.tikz}} \]
    It is easy to verify that $\circ$ preserves identities, \textit{i.e.} that  $\id_g\circ\id_f = \id_{g\circ f}$; just substitute the identity 2-cells into the definition of $\circ$ on 2-cells, and apply comonoid counitality four times.

    Next, we establish that horizontal composition is weakly associative, which requires us to supply isomorphisms $\alpha_{f,g,h}:(h\circ g)\circ f\Rightarrow h\circ(g\circ f)$ natural in composable triples of 1-cells $h,g,f$.
    Supposing the three morphisms have the types $f:A\xto[M]{}B$, $g:B\xto[N]{}C$, and $h:C\xto[O]{}D$, we can choose $a_{f,g,h}$ to be the 2-cell represented by the morphism
    \begin{gather*}
      A\otimes\bigl((M\otimes B)\otimes((N\otimes C)\otimes O)\bigr)\otimes D \xto{\proj} (M\otimes B)\otimes((N\otimes C)\otimes O) \;\cdots \\ \cdots\; \xto{\alpha^{\cat{C}}_{(M\otimes B),(N\otimes C),O}} ((M\otimes B)\otimes(N\otimes C))\otimes O \;\cdots \\ \cdots\; \xto{\alpha^{\cat{C}}_{(M\otimes B),N,C}\otimes\id_O} (((M\otimes B)\otimes N)\otimes C)\otimes O
    \end{gather*}
    where the first factor is the projection onto the coparameter and $\alpha^{\cat{C}}$ denotes the associator of the monoidal structure $(\otimes,I)$ on $\cat{C}$.
    In the inverse direction, we can choose the component $\alpha^{-1}_{f,g,h}:h\circ(g\circ f)\Rightarrow(h\circ g)\circ f$ to be the 2-cell represented by the morphism
    \begin{gather*}
      A\otimes\bigl((((M\otimes B)\otimes N)\otimes C)\otimes O\bigr)\otimes D \xto{\proj} (((M\otimes B)\otimes N)\otimes C)\otimes O \;\cdots \\ \cdots\; \xto{\alpha^{\cat{C},-1}_{(M\otimes B),N,C}\otimes\id_O} ((M\otimes B)\otimes(N\otimes C))\otimes O \;\cdots \\ \cdots\; \xto{\alpha^{\cat{C},-1}_{(M\otimes B),(N\otimes C),O}} (M\otimes B)\otimes((N\otimes C)\otimes O)
    \end{gather*}
    where $\alpha^{\cat{C},-1}$ denotes the inverse of the associator on $(\cat{C},\otimes,I)$.
    That the pair of $\alpha_{f,g,h}$ and $\alpha_{f,g,h}^{-1}$ constitutes an isomorphism in the hom category follows from the counitality of the comonoid structures.
    That this family of isomorphisms is moreover natural follows from the naturality of the associator on $(\cat{C},\otimes,I)$.

    We come to the matter that motivated the construction of $\ccopara(\cat{C})$: the weak unitality of copy-composition, witnessed here by the weak unitality of horizontal composition.
    We need to exhibit two families of natural isomorphisms: the left unitors with components $\lambda_f:\id_B\circ f\Rightarrow f$, and the right unitors with components $\rho_f:f\circ\id_A\Rightarrow f$, for each morphism $f:A\xto[M]{}B$.
    Each such component will be defined by a projection morphism, and weak unitality will then follow from the counitality of the comonoid structures.
    More explicitly, $\lambda_f$ is witnessed by $\proj_M:A\otimes M\otimes B\otimes B\to M$; its inverse $\lambda^{-1}_f$ is witnessed by $\proj_{M\otimes B}:A\otimes M\otimes B\to M\otimes B$; $\rho_f$ is witnessed by $\proj_{M}:A\otimes A\otimes M\otimes B\to M$; and its inverse $\rho^{-1}_f$ is witnessed by $\proj_{A\otimes M}:A\otimes M\otimes B\to A\otimes M$.
    Checking that these definitions give natural isomorphisms is then an exercise in counitality that we leave to the reader.

    All that remains of the proof that $\ccopara(\cat{C})$ is indeed a bicategory is to check that the unitors are compatible with the associator (\textit{i.e.}, $(\id_g\circ\lambda_f)\odot\alpha_{g,\id_B,f} = \rho_g\circ\id_f$) and that associativity is order-dependent (\textit{i.e.}, the associator $\alpha$ satisfies the pentagon diagram).
    The latter follows immediately from the corresponding fact about the associator $\alpha^{\cat{C}}$ on $(\cat{C},\otimes,I)$.
    To demonstrate the former, it is easier to verify that $(\id_g\circ\lambda_f)\odot\alpha_{g,\id_B,f}\odot(\rho^{-1}_g\circ\id_f)=\id_{g\circ f}$.
    This amounts to checking that the following string diagram is equally a depiction of the morphism underlying $\id_{g\circ f}$:
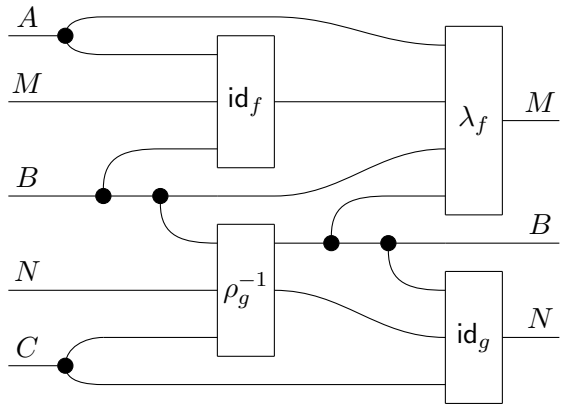
    \[ \scalebox{1.0}{\input{img/copara2-triangle-identity-1.tikz}} \]
    (Note that here we have elided the associator from the depiction. This is allowed by comonoid counitality, and because string diagrams are blind to bracketing.)
    Substituting the relevant morphisms into the boxes, we see that this diagram is equal to
    \[ \scalebox{1.0}{\input{img/copara2-triangle-identity-2.tikz}}\]
    and six applications of counitality give us $\id_{g\circ f}$.
    This establishes that $\ccopara(\cat{C})$ is a bicategory.
  \end{proof}
\end{thm}

\begin{rmk}
  When $\cat{C}$ is symmetric monoidal, $\ccopara(\cat{C})$ inherits a monoidal structure, elaborated in Proposition \ref{prop:mon-ccopara}.
\end{rmk}

\begin{rmk}
  In order to capture the bidirectionality of Bayesian inversion we will need to consider left- and right-handed versions of the $\ccopara$ construction.
  These are formally dual, and when $\cat{C}$ is symmetric monoidal (as in most examples) they are isomorphic.
  Nonetheless, it makes formalization easier if we explicitly distinguish $\ccopara[l](\cat{C})$, in which the coparameter is placed on the left of the codomain (as above), from $\ccopara[r](\cat{C})$, in which it is placed on the right.
  Aside from the swapping of this handedness, the rest of the construction is the same.
\end{rmk}

We end this section with three easy (and ambidextrous) propositions relating $\cat{C}$ and $\ccopara(\cat{C})$.

\begin{prop}
  There is an identity-on-objects lax embedding $(-)^{\smallbcopier}:\cat{C}\hookrightarrow\ccopara(\cat{C})$, mapping $f:X\to Y$ to $f^{\smallbcopier}:X\xto[I]{}Y$, which (in the left-handed case) has the underlying morphism $X\xto{f}Y\xto{\lambda_Y^{-1}}I\otimes Y$ (where $\lambda$ is the left unitor of the monoidal structure on $\cat{C}$).
  The laxator $\iota(g)\circ\iota(f)\to\iota(g\circ f)$ discards the coparameter obtained from copy-composition.
\end{prop}

\begin{rmk}
  We will define the notion of \textit{lax functor} in Definition \ref{def:lax-functor} below.
  A lax embedding is a lax functor that is an embedding in the sense of Remark \ref{rmk:embedding}: that is, a lax functor that is faithful on hom categories.
\end{rmk}

\begin{prop} \label{prop:discard-func}
  There is a `discarding' functor $(-)^{\smallground}:\ccopara(\cat{C})\to\cat{C}$, which takes any coparameterized morphism and discards the coparameter.
\end{prop}

\begin{prop}
  $(-)^{\smallbcopier}$ is a section of $(-)^{\smallground}$. That is, $\id_{\cat{C}} = \cat{C}\xhookrightarrow{(-)^{\smallbcopier}}\ccopara(\cat{C})\xto{(-)^{\smallground}}\cat{C}$; more suggestively, this can be written $(-) = {(-)^{\smallbcopier}}^{\smallground}$.
\end{prop}

\subsection{Lax functors, pseudofunctors, their transformations, and indexed bicategories} \label{sec:lax-func}

In order to define bicategories of statistical games, coherently with loss functions like the relative entropy that compose by copy-composition, we first need to define coparameterized (copy-composite) Bayesian lenses.
Analogously to non-coparameterized Bayesian lenses, these will be obtained by applying a Grothendieck construction to an indexed bicategory \parencite[{Def. 3.5}]{Bakovic2010Fibrations} of state-dependent channels, $\SStat$.
Constructing copy-composite Bayesian lenses in this way is the subject of \secref{sec:copara-blens}; in this section, we supply the necessary higher-categorical prerequisites.

An indexed category is a homomorphism of bicategories with codomain $\Cat{Cat}$ and locally trivial domain, and analogously an indexed bicategory will be a homomorphism of \textit{tricategories} with codomain $\Cat{Bicat}$ (appropriately defined) and locally `2-trivial' domain.
In order to stay as close as possible to the matter at hand, we do not give here an explicit definition of `tricategory' or indeed of `indexed bicategory', and instead refer the reader to \parencite[\S3]{Bakovic2010Fibrations}.
The definition of $\SStat$ will of course provide an example of an indexed bicategory, but in order to state it we will nonetheless need to extend the notion of pseudofunctor from Definition \ref{def:pseudofunctor} to the case where the domain is a true bicategory; and we will also need morphisms of pseudofunctors, called \textit{pseudonatural} (or \textit{strong}) \textit{transformations}.

We will begin by defining the notion of \textit{lax functor}, of which pseudofunctors constitute a special case.
Just as a lax monoidal functor $F$ is equipped with a natural family of morphisms $F(X)\otimes F(Y) \to F(X\otimes Y)$ (the laxator; \textit{cf.} Definition \ref{def:lax-mon-func}), a lax functor is a weak functor equipped with a natural family of 2-cells $F(g)\diamond F(f) \Rightarrow F(g\circ f)$; this \textit{lax functoriality} will be important when we come to study loss models in \secref{sec:loss-models}.

\begin{defn}[{\textcite[{Def. 4.1.2}]{Johnson20202Dimensional}}] \label{def:lax-functor}
  Suppose $\cat{B}$ and $\cat{C}$ are both bicategories.
  A \textit{lax functor} $F:\cat{B}\to\cat{C}$ is constituted by
  \begin{enumerate}
  \item a function $F_0 : \cat{B}_0 \to \cat{C}_0$ on 0-cells;
  \item for each pair of 0-cells $a,b:\cat{B}$, a functor $F_{a,b}:\cat{B}(a,b)\to\cat{C}(F_0a,F_0b)$;
  \item for each 0-cell $b:\cat{B}$, a natural transformation
    \[\begin{tikzcd}[sep=scriptsize]
      {\Cat{1}} && {\cat{B}(b,b)} \\
      \\
      && {\cat{C}(F_0b,F_0b)}
      \arrow[""{name=0, anchor=center, inner sep=0}, "{\id_{F_0b}}"', curve={height=18pt}, from=1-1, to=3-3]
      \arrow["{\id_b}", from=1-1, to=1-3]
      \arrow["{F_{b,b}}", from=1-3, to=3-3]
      \arrow["{F_1}", shorten <=11pt, shorten >=11pt, Rightarrow, from=0, to=1-3]
    \end{tikzcd}\]
    witnessing \textit{lax unity}, with component 2-cells $F_b:\id_{F_0b} \Rightarrow F_{b,b}(\id_b)$;
  \item for each triple of 0-cells $a,b,c:\cat{B}$, a natural transformation
    \[\begin{tikzcd}[sep=scriptsize]
      {\cat{B}(b,c)\times\cat{B}(a,b)} && {\cat{B}(a,c)} \\
      \\
      {\cat{C}(F_0b,F_0c)\times\cat{C}(F_0a,F_0b)} && {\cat{C}(F_0a,F_0c)}
      \arrow["\circ", from=1-1, to=1-3]
      \arrow["F_{a,c}", from=1-3, to=3-3]
      \arrow["{F_{b,c}\times F_{a,b}}"', from=1-1, to=3-1]
      \arrow["\diamond", from=3-1, to=3-3]
      \arrow["{F_2}", shorten <=23pt, shorten >=23pt, Rightarrow, from=3-1, to=1-3]
    \end{tikzcd}\]
    witnessing \textit{lax functoriality} and called the \textit{laxator}\footnote{
    By analogy with the laxator of a lax monoidal functor (Definition \ref{def:lax-mon-func}). Since \textit{monoidal category} is a special case of \textit{bicategory}, the notion of lax functor (between bicategories) generalizes the notion of lax monoidal functor (between monoidal categories).},
    with component 2-cells
    \[ F_{g,f} : F_{b,c}(g)\diamond F_{a,b}(f) \Rightarrow F_{a,c}(g\circ f) \]
    where $\circ$ and $\diamond$ denote horizontal composition in $\cat{B}$ and $\cat{C}$ respectively;
  \end{enumerate}
  satisfying the following conditions:
  \begin{enumerate}[label=(\alph*)]
  \item coherence with the left and right unitality of horizontal composition, so that for all 1-cells $f:a\to b$ the following diagrams commute:
  \[\begin{tikzcd}
	  {\id_{F_0b}\diamond F_{a,b}(f)} && {F_{a,b}(f)} \\
	  \\
	  {F_{b,b}(\id_b)\diamond F_{a,b}(f)} && {F_{a,b}(\id_b\circ f)}
	  \arrow["{\lambda^{\cat{C}}_{F_{a,b}(f)}}", Rightarrow, from=1-1, to=1-3]
	  \arrow["{F_{a,b}(\lambda^{\cat{B}}_f)}", Rightarrow, from=3-3, to=1-3]
	  \arrow["{F_{b}\diamond F_{a,b}(f)}"', Rightarrow, from=1-1, to=3-1]
	  \arrow["{F_{\id_b,f}}", Rightarrow, from=3-1, to=3-3]
  \end{tikzcd}
  \qquad
  \begin{tikzcd}
    {F_{a,b}(f)\diamond\id_{F_0a}} && {F_{a,b}(f)} \\
    \\
    {F_{a,b}(f)\diamond F_{a,a}(\id_a)} && {F_{a,b}(f\circ\id_a)}
	  \arrow["{\rho^{\cat{C}}_{F_{a,b}(f)}}", Rightarrow, from=1-1, to=1-3]
	  \arrow["{F_{a,b}(\rho^{\cat{B}}_f)}", Rightarrow, from=3-3, to=1-3]
	  \arrow["{F_{a,b}(f)\diamond F_{a}}"', Rightarrow, from=1-1, to=3-1]
	  \arrow["{F_{f,\id_a}}", Rightarrow, from=3-1, to=3-3]
  \end{tikzcd}\]
  where $\lambda^{\cat{B}},\lambda^{\cat{C}}$ and $\rho^{\cat{B}},\rho^{\cat{C}}$ are the left and right unitors for the horizontal composition in $\cat{B}$ and $\cat{C}$ respectively;
  \item coherence with the associativity of horizontal composition, so that for all 1-cells $f:a\to b$, $g:b\to c$, and $h:c\to d$, the following diagram commutes:
  \[\begin{tikzcd}
    {(F_{c,d}(h)\diamond F_{b,c}(g))\diamond F_{a,b}(f)} &&& {F_{c,d}(h)\diamond(F_{b,c}(g)\diamond F_{a,b}(f))} \\
    \\
    {F_{b,d}(h\circ g)\diamond F_{a,b}(f)} &&& {F_{c,d}(h)\diamond F_{a,c}(g\circ f)} \\
	  \\
	  {F_{a,d}((h\circ g)\circ f)} &&& {F_{a,d}(h\circ(g\circ f))}
	  \arrow["{\alpha^{\cat{C}}_{F_{c,d}(h),F_{b,c}(g),F_{a,b}(f)}}", Rightarrow, from=1-1, to=1-4]
	  \arrow["{F_{c,d}(h)\diamond F_{g,f}}", Rightarrow, from=1-4, to=3-4]
	  \arrow["{F_{h,g\circ f}}", Rightarrow, from=3-4, to=5-4]
	  \arrow["{F_{h,g}\diamond F_{a,b}(f)}"', Rightarrow, from=1-1, to=3-1]
	  \arrow["{F_{h\circ g,f}}"', Rightarrow, from=3-1, to=5-1]
	  \arrow["{F_{a,d}(\alpha^{\cat{B}}_{h,g,f})}", Rightarrow, from=5-1, to=5-4]
  \end{tikzcd}\]
  where $\alpha^{\cat{B}}$ and $\alpha^{\cat{C}}$ are the associators for the horizontal composition in $\cat{B}$ and $\cat{C}$ respectively.
  \end{enumerate}

  A \textit{pseudofunctor} is a lax functor $F$ for which $F_1$ and $F_2$ are natural isomorphisms\footnote{\label{fn:pseudofunc} Compare Definition \ref{def:pseudofunctor}, treating $\cat{C}$ there as a bicategory with discrete hom-categories.}.
\end{defn}

The variable laxness of lax functors is recapitulated in the laxness of their morphisms; again, we begin with the weakest case.

\begin{defn} \label{def:lax-trans}
  Suppose $F$ and $G$ are lax functors $\cat{B}\to\cat{C}$.
  A \textit{lax transformation} $\alpha:F\to G$ consists of
  \begin{enumerate}
  \item a 1-cell $\alpha_b : Fb\to Gb$ in $\cat{C}$ for each 0-cell $b:\cat{B}$;
  \item a natural transformation $\alpha_{b,c} : {\alpha_b}^\ast G \Rightarrow {\alpha_c}_* F$ (where ${\alpha_b}^\ast$ denotes pre-composition by $\alpha_b$, and ${\alpha_c}_*$ denotes post-composition by $\alpha_c$) for each pair $b,c$ of objects in $\cat{B}$, with component 2-cells
    \[\begin{tikzcd}[sep=scriptsize]
      Fb && Fc \\
      \\
      Gb && Gc
      \arrow["Ff", from=1-1, to=1-3]
      \arrow["{\alpha_b}"', from=1-1, to=3-1]
      \arrow["{\alpha_c}", from=1-3, to=3-3]
      \arrow["Gf", from=3-1, to=3-3]
      \arrow["{\alpha_f}", shorten <=11pt, shorten >=11pt, Rightarrow, from=3-1, to=1-3]
    \end{tikzcd}\]
    for each 1-cell $f:b\to c$ in $\cat{B}$;
  \end{enumerate}
  satisfying conditions of lax unity and lax naturality whose precise general form the reader may find in \textcite[{Def. 4.2.1}]{Johnson20202Dimensional}.

  A \textit{strong transformation} (or \textit{pseudonatural transformation}) is a lax transformation for which the component 2-cells constitute natural isomorphisms.
\end{defn}

It is notable that, unlike natural transformations, lax transformations do not compose, not even laxly; see \textcite[{Motivation 4.6.1}]{Johnson20202Dimensional}.
This means that there is no bicategory of bicategories, lax functors, and lax transformations, analogous to $\Cat{Cat}$.
However, \textit{strong} transformations between pseudofunctors \textit{do} compose, weakly, up to isomorphism.
These isomorphisms collect into 3-cells known as \textit{modifications}, producing a \textit{tricategory} $\Cat{Bicat}$ whose 0-cells are bicategories, 1-cells are pseudofunctors, 2-cells strong transformations, and 3-cells modifications.
This tricategory constitutes the codomain of an indexed bicategory.

\begin{rmk}
  There is another notion of composable morphism between \textit{lax} functors, called \textit{icon}, which yields a bicategory $\Cat{Bicat}_{\mathrm{ic}}$ of bicategories, lax functors, and icons.
  Icons are, however, more restrictive than lax transformations, as their 1-cell components must be identities.
  Nonetheless, this restriction is satisfied by loss models as defined in \secref{sec:loss-models}, and so morphisms of loss models will be icons.
\end{rmk}

\subsection{Coparameterized Bayesian lenses} \label{sec:copara-blens}

With that categorical background out of the way, we are able to define copy-composite Bayesian lenses, starting with the corresponding indexed bicategory.
Let $\disc$ denote the functor $\Set\to\Cat{Cat}$ taking sets to discrete categories (\textit{cf.} Definition \ref{def:disc-cat}).

\begin{defn}
  We define the indexed bicategory $\SStat:\ccopara[l](\cat{C})\coop\to\Cat{Bicat}$ fibrewise as follows.
  \begin{enumerate}[label=(\roman*)]
  \item The 0-cells $\SStat(X)_0 $ of each fibre $\SStat(X)$ are the objects $\cat{C}_0$ of $\cat{C}$.
  \item For each pair of 0-cells $A,B$, the hom-category $\SStat(X)(A,B)$ is defined to be the functor category $\Cat{Cat}\bigl(\disc\cat{C}(I,X),\ccopara[r](\cat{C})(A,B)\bigr)$.
  \item For each 0-cell $A$, the identity functor $\id_A : \Cat{1}\to\SStat(X)(A,A)$ is the constant functor on the identity on $A$ in $\ccopara[r](\cat{C})$; \textit{i.e.} $\disc\cat{C}(I,X) \xto{!} 1 \xto{\id_A} \ccopara[r](\cat{C})(A,A)$.
  \item For each triple $A,B,C$ of 0-cells, the horizontal composition functor $\circ_{A,B,C}$ is defined by
    \begin{gather*}
      \circ_{A,B,C} : \SStat(X)(B,C)\times\SStat(X)(A,B) \;\cdots \\
      \cdots\; \xto{=} \Cat{Cat}\bigl(\disc\cat{C}(I,X),\ccopara[r](\cat{C})(B,C)\bigr) \times \Cat{Cat}\bigl(\disc\cat{C}(I,X),\ccopara[r](\cat{C})(A,B)\bigr) \;\cdots \\
      \cdots\; \xto{\times} \Cat{Cat}\bigl(\disc\cat{C}(I,X)^2,\ccopara[r](\cat{C})(B,C)\times\ccopara[r](\cat{C})(A,B)\bigr) \;\cdots \\
      \cdots\; \xto{\Cat{Cat}\left(\bcopier,\circ\right)} \Cat{Cat}\bigl(\disc\cat{C}(I,X),\ccopara[r](\cat{C})(A,C)\bigr) \;\cdots \\
      \cdots\; \xto{=} \SStat(X)(A,C)
    \end{gather*}
    where $\Cat{Cat}\left(\bcopier,\circ\right)$ indicates pre-composition with the universal (Cartesian) copying functor in $(\Cat{Cat},\times,\Cat{1})$ and post-composition with the horizontal composition functor of $\ccopara[r](\cat{C})$.
  \end{enumerate}
  For each pair of 0-cells $X,Y$ in $\copara[l](\cat{C})$, we define the reindexing pseudofunctor $\SStat_{,X,Y}:\copara[l](\cat{C})(X,Y)\op\to\Cat{Bicat}\bigl(\SStat(Y),\SStat(X)\bigr)$ as follows.
  \begin{enumerate}[label=(\alph*)]
  \item For each 1-cell $f$ in $\copara[l](\cat{C})(X,Y)$, we obtain a pseudofunctor $\SStat(f):\SStat(Y)\to\SStat(X)$ which acts as the identity on 0-cells.
  \item For each pair of 0-cells $A,B$ in $\SStat(Y)$, the functor $\SStat(f)_{A,B}$ is defined as the precomposition functor $\Cat{Cat}\bigl(\disc\cat{C}(I,f^{\smallground}),\ccopara[r](\cat{C})(A,B)\bigr)$, where $(-)^{\smallground}$ is the discarding functor $\ccopara[l](\cat{C})\to\cat{C}$ of Proposition \ref{prop:discard-func}.
    \item For each 2-cell $\varphi:f\Rightarrow f'$ in $\ccopara[l](\cat{C})(X,Y)$, the pseudonatural transformation $\SStat(\varphi):\SStat(f')\Rightarrow\SStat(f)$ is defined on 0-cells $A:\SStat(Y)$ by the discrete natural transformation with components $\SStat(\varphi)_A := \id_A$, and on 1-cells $c:\SStat(Y)(A,B)$ by the substitution natural transformation with constitutent 2-cells $\SStat(\varphi)_c:\SStat(f)(c)\Rightarrow\SStat(f')(c)$ in $\SStat(X)$ which acts by replacing $\Cat{Cat}\bigl(\disc\cat{C}(I,f^{\smallground}),\ccopara[r](\cat{C})(A,B)\bigr)$ by $\Cat{Cat}\bigl(\disc\cat{C}(I,f'^{\smallground}),\ccopara[r](\cat{C})(A,B)\bigr)$; and which we might alternatively denote by $\Cat{Cat}\bigl(\disc\cat{C}(I,\varphi^{\smallground}),\ccopara[r](\cat{C})(A,B)\bigr)$.
  \end{enumerate}
\end{defn}

\begin{notation}
  We will write $f:A\xklto[M]{X}B$ to denote a state-dependent coparameterized channel $f$ with coparameter $M$ and state-dependence on $X$.
\end{notation}

\begin{rmk} \label{rmk:copara-stat-dep}
  We could give an alternative definition of $\SStat$, for which the definition above would give a sub-indexed bicategory, by defining the state-dependence on the whole hom-category $\ccopara(\cat{C})(I,X)$ rather than just $\cat{C}(I,X)$.
  However, doing this would cause the reindexing pseudofunctors to introduce coparameters (from the now-coparameterized priors), which would contradict the type signature of coparameterized Bayesian inversion: imagine the equation of Definition \ref{def:copara-bayes} below but without the discarding on the right-hand side.
\end{rmk}

\begin{rmk}
  Similarly, the same definitions would go through upon substituting $\copara(\cat{C})$ for $\ccopara(\cat{C})$; but, as noted above, we need copy-composition for the relative entropy.
\end{rmk}

As we saw in \secref{sec:groth-lens}, lenses in 1-category theory are morphisms in the fibrewise opposite of a fibration.
Analogously, our bicategorical Bayesian lenses are obtained as 1-cells in the bicategorical Grothendieck construction of (the pointwise opposite of) the indexed bicategory $\SStat$; this yields a 2-fibration.
So as not to over-complicate the presentation, we do not present all the details of this construction, and refer the reader instead to \textcite[\S6]{Bakovic2010Fibrations}.

\begin{defn}
  Fix a copy-discard category $(\cat{C},\otimes,I)$.
  We define the bicategory of coparameterized Bayesian lenses in $\cat{C}$, denoted $\BLens{}_2(\cat{C})$ or simply $\BLens{}_2$, to be the bicategorical Grothendieck construction of the pointwise opposite of the corresponding indexed bicategory $\SStat$, with the following data:
  \begin{enumerate}[label=(\roman*)]
  \item A 0-cell in $\BLens{}_2$ is a pair $(X,A)$ of an object $X$ in $\ccopara[l](\cat{C})$ and an object $A$ in $\SStat(X)$; equivalently, a 0-cell in $\BLens{}_2$ is a pair of objects in $\cat{C}$.
  \item The hom-category $\BLens{}_2\bigl((X,A),(Y,B)\bigr)$ is the product category $\ccopara[l](\cat{C})(X,Y)\times\SStat(X)(B,A)$.
  \item The identity on $(X,A)$ is given by the pair $(\id_X,\id_A)$.
  \item For each triple of 0-cells $(X,A),(Y,B),(Z,C)$, the horizontal composition functor is given by
    \begin{align*}
      & \BLens{}_2\bigl((Y,B),(Z,C)\bigr)\times\BLens{}_2\bigl((X,A),(Y,B)\bigr) \\
      =\quad& \ccopara[l](\cat{C})(Y,Z)\times\SStat(Y)(C,B) \times \ccopara[l](\cat{C})(X,Y)\times\SStat(X)(B,A) \\
      \xto{\sim}\quad& \sum_{g:\ccopara[l](\cat{C})(Y,Z)} \sum_{f:\ccopara[l](\cat{C})(X,Y)} \SStat(Y)(C,B) \times \SStat(X)(B,A) \\
      \xto{\sum_g \sum_f \SStat(f)_{C,B}\times\id}\quad& \sum_{g:\ccopara[l](\cat{C})(Y,Z)} \sum_{f:\ccopara[l](\cat{C})(X,Y)} \SStat(X)(C,B) \times \SStat(X)(B,A) \\
      \xto{\sum_{\circ^{\ccopara[l](\cat{C})}} \circ^{\SStat(X)}}\quad& \sum_{g\circ f:\ccopara[l](\cat{C})(X,Z)} \SStat(X)(C,A) \\
      \xto{\sim}\quad& \BLens{}_2\bigl((X,A),(Z,C)\bigr)
    \end{align*}
    where the functor in the penultimate line amounts to the product of the horizontal composition functors on $\ccopara[l](\cat{C})$ and $\SStat(X)$.
  \end{enumerate}
\end{defn}

\begin{prop}
  There is a projection pseudofunctor $\pLens:\BLens{}_2(\cat{C})\to\ccopara[l](\cat{C})$ mapping each 0-cell $(X,A)$ to $X$, each 1-cell $(f,f')$ to $f$, and each 2-cell $(\varphi,\varphi')$ to $\varphi$.
  This pseudofunctor is a 2-fibration in the sense of \textcite[{Def. 4.7}]{Bakovic2010Fibrations}.
  \begin{proof}
    The claim follows as a consequence of \textcite[{Theorem 6.2}]{Bakovic2010Fibrations}.
  \end{proof}
\end{prop}

\begin{rmk}
  When $\cat{C}$ is symmetric monoidal, $\SStat$ acquires the structure of a monoidal indexed bicategory (Definition \ref{def:mon-idx-bicat} and Theorem \ref{thm:monoidal-sstat}), and hence $\BLens{}_2$ becomes a monoidal bicategory (Corollary \ref{cor:mon-blens}).
\end{rmk}

\subsection{Coparameterized Bayesian updates compose optically} \label{sec:copara-buco}

So that our generalized Bayesian lenses are worthy of the name, we should also confirm that Bayesian inversions compose according to the lens pattern (`optically') in the coparameterized setting.
Such confirmation is the subject of the present section.

\begin{defn} \label{def:copara-bayes}
  We say that a coparameterized channel $\gamma:A\klto M\otimes B$ \textit{admits Bayesian inversion} if there exists a dually coparameterized channel $\rho_\pi:B\klto A\otimes M$ satisfying the graphical equation
  \[ \scalebox{1.0}{\input{img/bayes-copara-1.tikz}} \qquad = \qquad \scalebox{1.0}{\input{img/bayes-copara-2.tikz}} \quad . \]
  In this case, we say that $\rho_\pi$ is the \textit{Bayesian inversion of} $\gamma$ \textit{with respect to} $\pi$.
\end{defn}

With this definition, we can supply the desired result that ``coparameterized Bayesian updates compose optically''.

\begin{thm} \label{thm:copara2-buco}
  Suppose $(\gamma,\gamma^\dag):(A,A)\xlensto[M]{}(B,B)$ and $(\delta,\delta^\dag):(B,B)\xlensto[N]{}(C,C)$ are coparameterized Bayesian lenses in $\BLens{}_2$.
  Suppose also that $\pi:I\klto A$ is a state on $A$ in the underlying category of channels $\cat{C}$, such that $\gamma^\dag_\pi$ is a Bayesian inversion of $\gamma$ with respect to $\pi$, and such that $\delta^\dag_{\gamma\pi}$ is a Bayesian inversion of $\delta$ with respect to $(\gamma\pi)^{\smallground}$; where the notation $(-)^{\smallground}$ represents discarding coparameters.
  Then $\gamma^\dag_\pi\klcirc\delta^\dag_{\gamma\pi}$ is a Bayesian inversion of $\delta\klcirc\gamma$ with respect to $\pi$.
  (Here $\klcirc$ denotes copy-composition.)
  Moreover, if $(\delta\klcirc\gamma)^\dag_\pi$ is any Bayesian inversion of $\delta\klcirc\gamma$ with respect to $\pi$, then $\gamma^\dag_\pi\klcirc\delta^\dag_{\gamma\pi}$ is $(\delta\gamma\pi)^{\smallground}$-almost-surely equal to $(\delta\klcirc\gamma)^\dag_\pi$: that is, $(\delta\klcirc\gamma)^\dag_\pi \overset{(\delta\gamma\pi)^{\smallground}}{\sim} \gamma^\dag_\pi\klcirc\delta^\dag_{\gamma\pi}$.
  \begin{proof}
    We only need to show that $\gamma^\dag_\pi\klcirc\delta^\dag_{\gamma\pi}$ is a Bayesian inversion of $\delta\klcirc\gamma$ with respect to $\pi$; the `moreover' claim follows immediately because Bayesian inversions are almost surely unique (by Proposition \ref{prop:bayes-almost-equal}).
    Thus, $\delta\klcirc\gamma\klcirc\pi$ has the following depiction;
    \[ \scalebox{1.0}{\input{img/buco-copara-1.tikz}} \]
    Since $\gamma^\dag_\pi$ is a Bayesian inversion of $\gamma$ with respect to $\pi$, this is equal to
    \[ \scalebox{1.0}{\input{img/buco-copara-2.tikz}} \quad . \]
    By the coassociativity of copying, this in turn is equal to
    \[ \scalebox{1.0}{\input{img/buco-copara-3.tikz}} \quad . \]
    And since $\delta^\dag_{\gamma\pi}$ is a Bayesian inversion of $\delta$ with respect to $(\gamma\pi)^{\smallground}$, this is equal to
    \[ \scalebox{1.0}{\input{img/buco-copara-4.tikz}} \]
    which establishes the result.
  \end{proof}
\end{thm}

In order to satisfy this coparameterized Bayes' rule, a Bayesian lens must be of `simple' type.

\begin{defn}
  We say that a coparameterized Bayesian lens $(c,c')$ is \textit{simple} if its domain and codomain are `diagonal' (duplicate pairs of objects) and if the coparameter of $c$ is equal to the coparameter of $c'$.
  In this case, we can write the type of $(c,c')$ as $(X,X)\xlensto[M]{}(Y,Y)$ or simply $X\xlensto[M]{}Y$.
\end{defn}

\begin{rmk} \label{rmk:bl-dep-opt}
  In Remark \ref{rmk:copara-stat-dep}, we explained that we should restrict the type of state-dependent coparameterized morphisms so that they cohere with the coparameterized Bayes' rule of Definition \ref{def:copara-bayes}.
  The restriction here to simple lenses is by contrast not enforced by the type system, an oversight which (like the failure to restrict to supports noted in Remark \ref{rmk:bayes-supp}) is comparatively inelegant, but which is forced upon us by the Grothendieck construction, which does not have a place for such constraints.
  We expect that the use of (a bicategorical instance of) \textit{dependent optics} \parencite{Braithwaite2021Fibre,Vertechi2022Dependent,Capucci2022Seeing} would allow such a constraint to be enforced (alongside support objects), at the cost of requiring yet more high-powered categorical machinery, of which there is probably enough in this thesis.
  We therefore leave this avenue unexplored for now.
\end{rmk}  

By analogy with Corollary \ref{cor:buco-as-func}, we have the following important consequence of Theorem \ref{thm:copara2-buco}.

\begin{cor} \label{cor:copara2-buco-as-func}
  Suppose $\ccopara[l](\cat{C})^\dag$ is a subbicategory of $\ccopara[l](\cat{C})$ all of whose channels admit Bayesian inversion.
  Then there is almost surely a pseudofunctor $\dag:\ccopara[l](\cat{C})^\dag\to\BLens{}_2$ mapping each 1-cell to its almost-surely unique corresponding exact Bayesian lens.
  Moreover, $\dag$ is a section of the 2-fibration $\pLens:\BLens{}_2\to\ccopara[l](\cat{C})$ induced by the bicategorical Grothendieck construction.
\end{cor}

\section{Statistical games for local approximate inference} \label{sec:stat-games}

\subsection{Attaching losses to lenses} \label{sec:loss-funcs}

Statistical games are obtained by attaching to Bayesian lenses \textit{loss functions}, representing `local' quantifications of the performance of approximate inference systems.
Because this performance depends on the system's context (\textit{i.e.}, the prior $\pi:I\klto X$ and the observed data $b:B$), a loss function at its most concrete will be a function $\cat{C}(I,X)\times B\to\rr$.
To internalize this type in $\cat{C}$, we may recall that, when $\cat{C}$ is the category $\Cat{sfKrn}$ of s-finite kernels or the Kleisli category $\Kl(\Da_{\leq1})$ of the subdistribution monad\footnote{
Weaken the definition of the distribution monad $\Da:\Set\to\Set$ so that distributions may sum to any number in the unit interval.},
a density function $p_c:X\times Y\to[0,1]$ for a channel $c:X\klto Y$ corresponds to an \textit{effect} (or \textit{costate}) $X\otimes Y\klto I$.
In this way, we can see a loss function as a kind of \textit{state-dependent effect} $B\xklto{X}I$ (and not a coparameterized one).

Loss functions will compose by sum, and so we need to ask for the effects in $\cat{C}$ to form a monoid.
Moreover, we need this monoid to be `bilinear' with respect to channels, so that $\Stat$-reindexing (\textit{cf.} Definition \ref{def:stat-cat}) preserves sums.
These conditions are formalized in the following definition.

\begin{defn} \label{def:bilin-eff}
  Suppose $(\cat{C},\otimes,I)$ is a copy-discard category.
  We say that $\cat{C}$ \textit{has bilinear effects} if the following conditions are satisfied:
  \begin{enumerate}[label=(\roman*)]
  \item \textit{effect monoid}: there is a natural transformation $+:\cat{C}(-,I)\times\cat{C}(=,I)\Rightarrow\cat{C}({-}\otimes{=},I)$ making $\sum_{A:\cat{C}}\cat{C}(A,I)$ into a commutative monoid with unit $0:I\klto I$;
  \item \textit{bilinearity}: $(g+g')\klcirc\bcopier\klcirc f = g\klcirc f + g'\klcirc f$ for all effects $g,g'$ and morphisms $f$ such that $(g+g')\klcirc\bcopier\klcirc f$ exists.
  \end{enumerate}
\end{defn}

\begin{ex}
  A trivial example of a category with bilinear effects is supplied by any Cartesian category, such as $\Set$, in which there is a unique effect for each object, so the effect monoid structure is given only by the product of objects, and bilinearity follows from the terminality of the unit object $1$.
\end{ex}

\begin{ex}
  We might hope that $\Kl(\Da_{\leq1})$ has bilinear effects, but this is not the case, because the sum of two effects may exceed $1$: the effects only form a \textit{partial} monoid\footnote{
Indeed, an \textit{effect algebra} is a kind of partial commutative monoid \parencite[\S2]{Jacobs2015Effect}, but we do not need the extra complication here.}.
  But if $M$ is any monoid in $\Set$, then there is a monad $\Da_M$ taking each set $X$ to the set $\Da_M(X)$ of formal $M$-linear combinations of elements of $X$.
  This is the free $M$-module on $X$, just as traditionally $\Da X$ is the free convex space on $X$, and the monad structure is obtained from the adjunction in the same way \parencite[\S2]{Jacobs2010Convexity}.
  An effect $Y\klto I$ then corresponds to a function $Y\to M$, and the monoid structure follows from the monoid structure on $M$; bilinearity follows from the linearity of the (free) module structure:
  \begin{align*}
    (g+g')\klcirc\bcopier\klcirc f(x)
    &= \sum_{y} \bigl(g(y) + g'(y)\bigr) \cdot f(y|x) \\
    &= \sum_{y} g(y) \cdot f(y|x) + g'(y) \cdot f(y|x) \\
    &= \sum_{y} g(y) \cdot f(y|x) + \sum_{y} g'(y) \cdot f(y|x) \\
    &= g\klcirc f(x) + g'\klcirc f(x)
  \end{align*}
\end{ex}

\begin{ex} \label{ex:sfkrn-bilin-eff}
  The category $\Cat{sfKrn}$ of s-finite kernels \parencite{Vakar2018S} has bilinear effects.
  An effect $Y\klto I$ is a measurable function $Y\to[0,\infty]$, and bilinearity follows from the linearity of integration:
  \begin{align*}
    (g+g')\klcirc\bcopier\klcirc f(x)
    &= \int_{y} \bigl(g(y) + g'(y)\bigr) \, f(\d y|x) \\
    &= \int_{y} g(y) \, f(\d y|x) + g'(y) \, f(\d y|x) \\
    &= \int_{y} g(y) \, f(\d y|x) + \int_{y} g'(y) \, f(\d y|x) \\
    &= g\klcirc f(x) + g'\klcirc f(x)
  \end{align*}
  We will typically assume $\Cat{sfKrn}$ as our ambient $\cat{C}$ for the examples below.
\end{ex}

\begin{ex} \label{ex:stat-bilin-eff}
  Given a category $\cat{C}$ with bilinear effects, we can lift the natural transformation $+$, and hence the bilinear effect structure, to the fibres of $\Stat_{\cat{C}}$, using the universal property of the product of categories:
  \begin{align*}
    +_X &: \Stat(X)(-,I) \times \Stat(X)(=,I) \,\xlongequal{}\, \Set\bigl(\cat{C}(I,X),\cat{C}(-,I)\bigr) \times \Set\bigl(\cat{C}(I,X),\cat{C}(=,I)\bigr) \\
    &\xRightarrow{(\cdot,\cdot)} \Set\bigl(\cat{C}(I,X),\cat{C}(-,I)\times\cat{C}(=,I)\bigr) \\
    &\xRightarrow{\Set\bigl(\cat{C}(I,X),+\bigr)} \Set\bigl(\cat{C}(I,X),\cat{C}({-}\otimes{=},I)\bigr) \\
    &\xRightarrow{=} \Stat(X)({-}\otimes{=},I)
  \end{align*}
  Here, $(\cdot,\cdot)$ denotes the pairing operation obtained from the universal property.
  In this way, each $\Stat(X)$ has bilinear effects.
  Note that this lifting is (strictly) compatible with the reindexing of $\Stat$, so that $+_{(-)}$ defines an indexed natural transformation.
  This means in particular that \textit{reindexing distributes over sums}: given state-dependent effects $g,g':B\xklto{Y}I$ and a channel $c:X\klto Y$, we have $(g+_Y g')_c = g_c +_X g'_c$.
  We will thus generally omit the subscript from the lifted sum operation, and just write $+$.
\end{ex}

We are now ready to construct the bicategory of statistical games.

\begin{defn} \label{def:bicat-sgame}
  Suppose $(\cat{C},\otimes,I)$ has bilinear effects, and let $\BLens{}_2$ denote the corresponding bicategory of (copy-composite) Bayesian lenses.
  We will write $\SGame{\cat{C}}$ to denote the following \textit{bicategory of (copy-composite) statistical games} in $\cat{C}$:
  \begin{itemize}
  \item The 0-cells are the 0-cells $(X,A)$ of $\BLens{}_2$;
  \item the 1-cells, called \textit{statistical games}, $(X,A)\to(Y,B)$ are triples $(c,c',L^c)$ consisting of a 1-cell $(c,c'):(X,A)\lensto(Y,B)$ in $\BLens{}_2$ and a \textit{loss} $L^c:B\xklto{X}I$ in $\Stat(X)(B,I)$;
  \item given 1-cells $(c,c',L^c),(e,e',L^{e}):(X,A)\to(Y,B)$, the 2-cells $(c,L^c)\Rightarrow(e,L^{e})$ are pairs $(\alpha,K^\alpha)$ of a 2-cell $\alpha:(c,c')\Rightarrow (e,e')$ in $\BLens{}_2$ and a loss $K^\alpha:B\xklto{X}I$ such that $L^{c} = L^{e} + K^\alpha$;
  \item the identity 2-cell on $(c,c',L^c)$ is $(\id_{(c,c')},0)$;
  \item given 2-cells $(\alpha,K^\alpha):(c,c',L^c)\Rightarrow(d,d',L^{d})$ and $(\beta,K^{\beta}):(d,d',L^{d})\Rightarrow(e,e',L^{e})$, their vertical composite is $(\beta\circ\alpha, K^{\beta}+K^{\alpha})$, where $\circ$ here denotes vertical composition in $\BLens{}_2$;
  \item given 1-cells $(c,c',L^c):(X,A)\to(Y,B)$ and $(d,d',L^d):(Y,B)\to(Z,C)$, their horizontal composite is $\bigl((d,d')\lenscirc (c,c'), L^d_c + L^c\circ d'_c\bigr)$; and
    \begin{itemize}
    \item given 2-cells $(\alpha,K^\alpha):(c,c',L^c)\Rightarrow(e,e',L^{e})$ and $(\beta,K^\beta):(d,d',L^d)\Rightarrow(f,f',L^{f})$, their horizontal composite is $(\beta\lenscirc\alpha,K^\beta_c + K^\alpha\circ d'_c)$, where $\lenscirc$ here denotes horizontal composition in $\BLens{}_2$.
    \end{itemize}
  \end{itemize}
\end{defn}

\begin{rmk}
  In earlier work (such as versions 1 and 2 of our preprint \parencite{Smithe2021Compositional1}), we gave a more elaborate but less satisfying definition of ``statistical game'', as a Bayesian lens equipped with a function from its `context' to $\rr$ (which we also called a loss function).
  The construction given here shows that the complicated earlier notion of context, which follows the ideas of `Bayesian open games' \parencite{Bolt2019Bayesian}, is actually unnecessary for the purposes of statistical games.
  Considering a Bayesian lens in $\Kl(\Da)$ of type $(X,A)\to(Y,B)$, its `context' is an element of $\Da X\times\Set(\Da Y,\Da B)$.
  By comparison, a corresponding loss function of the type given above is equivalently a function with domain $\Da X\times B$, and so we have replaced the dependence on `continuations' in $\Set(\Da Y,\Da B)$ with a simple dependence on $B$.
\end{rmk}

\begin{thm} \label{thm:bicat-sgame-well-def}
  Definition \ref{def:bicat-sgame} generates a well-defined bicategory.
\end{thm}

The proof of this result is that $\SGame{\cat{C}}$ is obtained via a pair of bicategorical Grothendieck constructions: first to obtain Bayesian lenses; and then to attach the loss functions.
The proof depends on the following intermediate result that our effect monoids can be `upgraded' to monoidal categories; we then use the delooping of this structure to associate (state-dependent) losses to (state-dependent) channels, after discarding the coparameters of the latter.

\begin{lemma} \label{lemma:effect-mon-cats}
  Suppose $(\cat{C},\otimes,I)$ has bilinear effects.
  Then, for each object $B$, $\cat{C}(B,I)$ has the structure of a symmetric monoidal category.
  The objects of $\cat{C}(B,I)$ are its elements, the effects.
  If $g,g'$ are two effects, then a morphism $\kappa:g\to g'$ is an effect such that $g = g' + \kappa$; the identity morphism for each effect $\id_g$ is then the constant $0$ effect.
  Likewise, the tensor of two effects is their sum, and the corresponding unit is the constant $0$.
  Precomposition by any morphism $c:A\klto B$ preserves the monoidal category structure, making the presheaf $\cat{C}(-,I)$ into a fibrewise-monoidal indexed category $\cat{C}\op\to\Cat{MonCat}$ (\textit{cf.} Remark \ref{rmk:fib-mon-idx-cat}).
\end{lemma}

As already indicated, this structure lifts to the fibres of $\Stat$.

\begin{cor} \label{cor:stat-effect-mon-cats}
  For each object $X$ in a category with bilinear effects, and for each object $B$, $\Stat(X)(B,I)$ inherits the symmetric monoidal structure of $\cat{C}(B,I)$; note that morphisms of state-dependent effects are likewise state-dependent, and that tensoring (summing) state-dependent effects involves copying the parameterizing state.
  Moreover, $\Stat(-)(=,I)$ is a fibrewise-monoidal indexed category $\sum_{X:\cat{C}\op}\Stat(X)\op \to \Cat{MonCat}$.
\end{cor}

Using this corollary, we can give the abstract proof of Theorem \ref{thm:bicat-sgame-well-def}.
There are two further observations of note: first, that we can deloop a monoidal category into a bicategory with one object; second, that we can extend $\Stat(-)(=,I)$ to $\SStat$ via discarding.

\begin{proof}[Proof of \ref{thm:bicat-sgame-well-def}]
  Recall from Proposition \ref{prop:deloop} that every monoidal category $\cat{M}$ can be transformed into a one-object bicategory, its delooping $\deloop\cat{M}$, with the 1-cells and 2-cells being the objects and morphisms of $\cat{M}$, vertical composition being composition in $\cat{M}$, and horizontal composition being the tensor.
  This delooping is functorial, giving a 2-functor $\deloop:\Cat{MonCat}\to\Cat{Bicat}$ which, following Corollary \ref{cor:stat-effect-mon-cats}, we can compose after $\Stat(-)(=,I)$ (taking its domain as a locally discrete 2-category) to obtain indexed bicategories; we will assume this transformation henceforth.

  Next, observe that we can extend the domain of $\Stat(-)(=,I)$ to $\sum_{X:\ccopara[l](\cat{C})\coop}\SStat(X)\coop$ by discarding the coparameters of the (coparameterized) state-dependent channels as well as the coparameter on any reindexing, as in the following diagram of indexed bicategories:
  \[\begin{tikzcd}
    {\sum_{X:\ccopara[l](\cat{C})\coop}\SStat(X)\coop} \\
    \\
    {\sum_{X:\cat{C}\op}\Stat(X)\op} & {\Cat{Bicat}}
    \arrow["{\sum_{\smallground}\ground}"', from=1-1, to=3-1]
    \arrow[""{name=0, anchor=center, inner sep=0}, "{\SStat(-)(=,I)}", from=1-1, to=3-2]
    \arrow["{\Stat(-)(=,I)}"', from=3-1, to=3-2]
    \arrow[shorten <=9pt, shorten >=9pt, Rightarrow, from=0, to=3-1]
  \end{tikzcd}\]
  Here, the 2-cell indicates also discarding the coparameters of the `effects' in $\SStat(-)(=,I)$.

  If we let $L$ denote the composite functor in the diagram above, we can reason as follows:
  \begin{prooftree}
    \AxiomC{$L:\sum_{X:\ccopara[l](\cat{C})\coop}\SStat(X)\coop\to\Cat{Bicat}$}
    \RightLabel{\scriptsize sum/product}
    \UnaryInfC{$\prod_{X:\ccopara[l](\cat{C})\coop}\Cat{Bicat}^{\SStat(X)\coop}$}
    \RightLabel{\scriptsize $\prod\int$}
    \UnaryInfC{$\prod_{X:\ccopara[l](\cat{C})\coop}\Cat{2Fib}\left(\SStat(X)\right)$}
    \RightLabel{\scriptsize forget}
    \UnaryInfC{$\ccopara[l](\cat{C})\coop\to\Cat{Bicat}$}
    \RightLabel{\scriptsize op}
    \UnaryInfC{$G:\ccopara[l](\cat{C})\coop\to\Cat{Bicat}$}
  \end{prooftree}
  where the first step uses the adjointness of (dependent) sums and products; the second applies the bicategorical Grothendieck construction in the codomain; the third forgets the 2-fibrations, to leave only the total bicategory; and the fourth step takes the pointwise opposite.
  We can thus write the action of $G$ as $G(X) = \left(\int L(X,-)\right)\op$.

  Since each bicategory $L(X,B)$ has only a single 0-cell, the 0-cells of each $G(X)$ are equivalently just the objects of $\cat{C}$, and the hom-categories $G(X)(A,B)$ are equivalent to the product categories $\SStat(X)(B,A)\times\Stat(X)(B,I)$.
  That is to say, a 1-cell $A\to B$ in $G(X)$ is a pair of a state-dependent channel $B\xklto{X}A$ along with a correspondingly state-dependent effect on its domain $B$.
  We therefore seem to approach the notion of statistical game, but in fact we are already there: $\SGame{\cat{C}}$ is simply $\int G$, by the bicategorical Grothendieck construction.
  To see this is only a matter of further unfolding the definition.
\end{proof}

\begin{rmk}
  There are two notable details that the abstractness of the preceding proof obscures.
  Firstly, the horizontal composition of effects in $\SGame{\cat{C}}$ is strict.
  To see this, let $(c,\iota):A\to B$ and $(d,\kappa):B\to C$ and $(e,\lambda):C\to D$ be 1-cells in $G(X)$, and for concision write the horizontal composite of effects by concatenation, so that $\kappa\iota = \kappa + \iota\circ d^{\smallground}$ (by the Grothendieck construction).
  Then strict associativity demands that $\lambda(\kappa\iota) = (\lambda\kappa)\iota$.
  This obtains as follows:
  \[
  \begin{aligned}
    \lambda(\kappa\iota)
    &= \lambda + (\kappa\iota)\circ e^{\smallground} \\
    &= \lambda + (\kappa + \iota\circ d^{\smallground})\circ e^{\smallground} \\
    &= \lambda + (\kappa\circ e^{\smallground} + \iota \circ d^{\smallground} \circ e^{\smallground}) \\
    &= \lambda + (\kappa\circ e^{\smallground} + \iota \circ (d \circ e)^{\smallground}) \\
    &= \lambda + (\kappa\circ e^{\smallground} + \iota \circ (e\circ\op d)^{\smallground}) \\
    &= (\lambda + \kappa\circ e^{\smallground}) + \iota \circ (e\circ\op d)^{\smallground} \\
    &= (\lambda\kappa)\iota
  \end{aligned}
  \qquad
  \begin{aligned}
    &\text{by Grothendieck} \\
    &\text{by Grothendieck} \\
    &\text{by bilinearity} \\
    &\text{by functoriality} \\
    &\text{by ``pointwise opposite''} \\
    &\text{by monoid associativity} \\
    &\text{by Grothendieck}
  \end{aligned}
  \]
  Since the identity effect is the constant $0$, it is easy to see that horizontal composition is strictly unital on effects:
  \[ 0\kappa = 0 + \kappa\circ\id = \kappa = \kappa + 0\circ d^{\smallground} = \kappa 0 \]

  Secondly, note that the well-definedness of horizontal composition in $\SGame{\cat{C}}$ depends furthermore on the distributivity of reindexing over sums (\textit{cf.} Example \ref{ex:stat-bilin-eff}).
  Suppose we have 1-cells and 2-cells in $\SGame{\cat{C}}$ as in the following diagram:
  \[\begin{tikzcd}
    {(X,A)} && {(Y,B)} && {(Z,C)}
    \arrow[""{name=0, anchor=center, inner sep=0}, "{(c,L^c)}", curve={height=-24pt}, from=1-1, to=1-3]
    \arrow[""{name=1, anchor=center, inner sep=0}, "{(c',L^{c'})}"', curve={height=24pt}, from=1-1, to=1-3]
    \arrow[""{name=2, anchor=center, inner sep=0}, "{(d,L^d)}", curve={height=-24pt}, from=1-3, to=1-5]
    \arrow[""{name=3, anchor=center, inner sep=0}, "{(d',L^{d'})}"', curve={height=24pt}, from=1-3, to=1-5]
    \arrow["{(\alpha,K^\alpha)}"{description}, shorten <=6pt, shorten >=6pt, Rightarrow, from=0, to=1]
    \arrow["{(\beta,K^\beta)}"{description}, shorten <=6pt, shorten >=6pt, Rightarrow, from=2, to=3]
  \end{tikzcd}\]
  Then, writing $\lenscirc$ for horizontal composition in $\SGame{\cat{C}}$ and $\circ$ for composition in $\SStat$ (and leaving the discarding of coparameters implicit):
  \[
  \begin{aligned}
    L^{d'} \lenscirc L^{c'}
    &= (L^d + K^\beta) \lenscirc (L^c + K^\alpha) \\
    &= (L^d + K^\beta)_c + (L^c + K^\alpha)\circ\overline{d}_c \\
    &= L^d_c + K^\beta_c + (L^c\circ\overline{d}_c) + (K^\alpha\circ\overline{d}_c) \\
    &= L^d_c + (L^c\circ\overline{d}_c) + K^\beta_c + (K^\alpha\circ\overline{d}_c) \\
    &= L^d\lenscirc L^c + K^\beta \lenscirc K^\alpha
  \end{aligned}
  \qquad
  \begin{aligned}
    &\textit{ex hypothesi} \\
    &\text{by Grothendieck} \\
    &\text{by distributivity and bilinearity} \\
    &\text{by commutativity of the effect monoid} \\
    &\text{by Grothendieck}
  \end{aligned}
  \]
\end{rmk}

\begin{rmk}
  Of course, we don't strictly need to use $\BLens{}_2$ in the preceding; the structure equally makes sense if we work only with `marginalized' lenses in $\BLens{}$.
  In this case, although $\BLens{}$ is a 1-category, one still obtains 2-cells between statistical games, because it remains possible to consider their differences.
\end{rmk}

\subsection{Inference systems and loss models} \label{sec:loss-models}

In the context of approximate inference, one often does not have a single statistical model to evaluate, but a whole family of them.
In particularly nice situations, this family is actually a subcategory $\cat{D}$ of $\cat{C}$, with the family of statistical models being all those that can be composed in $\cat{D}$.
The problem of approximate inference can then be formalized as follows.
Since both $\BLens{}_2$ and $\SGame{\cat{C}}$ were obtained by bicategorical Grothendieck constructions, we have a pair of 2-fibrations $\SGame{\cat{C}}\xto{\pLoss}\BLens{}_2\xto{\pLens}\ccopara[l](\cat{C})$.
Each of $\pLoss$, $\pLens$, and the discarding functor $(-)^{\smallground}$ can be restricted to the subcategory $\cat{D}$.
The inclusion $(-)^{\smallbcopier}:\cat{D}\hookrightarrow\ccopara[l](\cat{D})$ restricts to a section of this restriction of $(-)^{\smallground}$; the assignment of inversions to channels in $\cat{D}$ then corresponds to a 2-section of the 2-fibration $\pLens$ (restricted to $\cat{D}$); and the subsequent assignment of losses is a further 2-section of $\pLoss$.
This situation is depicted in the following diagram of bicategories:
\[\begin{tikzcd}[row sep=scriptsize]
	{\SGame{\cat{D}}} & {\SGame{\cat{C}}} \\
	{\BLens{}_2|_{\cat{D}}} & {\BLens{}_2} \\
	{\ccopara[l](\cat{D})} & {\ccopara[l](\cat{C})} \\
	{\cat{D}} & {\cat{C}}
	\arrow["\pLoss", from=1-2, to=2-2]
	\arrow["\pLens", from=2-2, to=3-2]
	\arrow["\ground", from=3-2, to=4-2]
	\arrow[hook, from=4-1, to=4-2]
	\arrow[hook, from=3-1, to=3-2]
	\arrow[hook, from=2-1, to=2-2]
	\arrow[hook, from=1-1, to=1-2]
	\arrow["{\pLoss|_{\cat{D}}}", from=1-1, to=2-1]
	\arrow["{\pLens|_{\cat{D}}}", from=2-1, to=3-1]
	\arrow["{\ground|_{\cat{D}}}", from=3-1, to=4-1]
	\arrow[curve={height=-24pt}, from=2-1, to=1-1]
	\arrow[curve={height=-24pt}, from=3-1, to=2-1]
	\arrow[curve={height=-24pt}, hook', from=4-1, to=3-1]
\end{tikzcd}\]
This motivates the following definitions of \textit{inference system} and \textit{loss model}, although, for the sake of our examples, we will explicitly allow the loss-assignment to be \textit{lax}: if $L$ is a loss model and $c$ and $d$ are composable lenses, then rather than an equality or natural isomorphism $L(d)\diamond L(c) \cong L(d\lenscirc c)$, we will only require a natural transformation $L(d)\diamond L(c) \Rightarrow L(d\lenscirc c)$.

Before defining loss models and inference systems, it helps to recall the concept of \textit{essential image}: a generalization of the notion of \textit{image} from functions to functors.

\begin{defn}[\parencite{nLab2023EssentialImage}]
  Suppose $F:\cat{C}\to\cat{D}$ is an n-functor (a possibly weak homomorphism of weak n-categories).
  The \textit{image} of $F$ is the smallest sub-n-category of $\cat{D}$ that contains $F(\alpha)$ for all k-cells $\alpha$ in $\cat{C}$, along with any $(k+1)$-cells relating images of composites and composites of images, for all $0\leq k\leq n$.
  We say that a sub-n-category $\cat{D}$ is \textit{replete} if, for any k-cells $\alpha$ in $\cat{D}$ and $\beta$ in $\cat{C}$ (with $0\leq k<n$) such that $f:\alpha\Rightarrow\beta$ is a $(k+1)$-isomorphism in $\cat{C}$, then $f$ is also a $(k+1)$-isomorphism in $\cat{D}$.
  The \textit{essential image} of $F$, denoted $\im(F)$, is then the smallest replete sub-n-category of $\cat{D}$ containing the image of $F$.
\end{defn}

With these concepts in mind, we state our definitions.

\begin{defn}
  Suppose $(\cat{C},\otimes,I)$ is a copy-delete category.
  An \textit{inference system} in $\cat{C}$ is a pair $(\cat{D},\ell)$ of a subcategory $\cat{D}\hookrightarrow\cat{C}$ along with a section $\ell:\cat{D}^{\smallbcopier}\to\BLens{}_2|_{\cat{D}}$ of $\pLens|_{\cat{D}}$, where $\cat{D}^{\smallbcopier}$ is the essential image of the canonical lax inclusion $(-)^{\smallbcopier}:\cat{D}\hookrightarrow\ccopara[l](\cat{D})$.
\end{defn}

\begin{defn}
  Suppose $(\cat{C},\otimes,I)$ has bilinear effects and $\cat{B}$ is a subbicategory of $\BLens{}_2$.
  A \textit{loss model} for $\cat{B}$ is a lax section $L$ of the restriction $\pLoss|_{\cat{B}}$ of $\pLoss$ to $\cat{B}$.
  We say that $L$ is a \textit{strict} loss model if it is in fact a strict 2-functor, and a \textit{strong} loss model if it is in fact a pseudofunctor.
\end{defn}

\begin{rmk}
  We may often be interested in loss models for which $\cat{B}$ is in fact the essential image of an inference system, but we do not stipulate this requirement in the definition as it is not necessary for the following development.
\end{rmk}

In order for two loss models $F$ and $G$ to be comparable, they must both be sections of the same fibration of statistical games.
One consequence of this is that both $F$ and $G$ must map each 0-cell $(X,A)$ in the bicategory of lenses to the same 0-cell in the bicategory of games, which (by the definition of the bicategory of games) must again be $(X,A)$.
In such circumstances, the relevant type of morphism of lax functors is the \textit{icon}, whose definition we now review.

\begin{defn}[{\textcite[{Def. 4.6.2}]{Johnson20202Dimensional}}] \label{def:icon}
  Suppose $F$ and $G$ are lax functors $\cat{B}\to\cat{C}$ such that, for all $b:\cat{B}$, $Fb = Gb$.
  An \textit{icon} (or \textit{identity component oplax natural transformation}) $\alpha:F\to G$ consists of a family of natural transformations
  \[\begin{tikzcd}
    {\cat{B}(a,b)} &&& {\hspace*{0.4cm} \cat{C}(Fa,Fb) = \cat{C}(Ga,Gb)}
    \arrow[""{name=0, anchor=center, inner sep=0}, "{F_{a,b}}", curve={height=-24pt}, from=1-1, to=1-4]
    \arrow[""{name=1, anchor=center, inner sep=0}, "{G_{a,b}}"', curve={height=24pt}, from=1-1, to=1-4]
    \arrow["{\alpha_{a,b}}"', shorten <=6pt, shorten >=6pt, Rightarrow, from=0, to=1]
  \end{tikzcd}\]
  for each pair $a,b$ of 0-cells in $\cat{B}$, satisfying coherence conditions corresponding to unity and oplax naturality, and whose component 2-cells we write as $\alpha_f : Ff \Rightarrow Gf$ for each 1-cell $f$ in $\cat{B}$.
\end{defn}

Lax functors $\cat{B}\to\cat{C}$ and icons between them constitute the objects and morphisms of a category, $\Cat{Bicat}_{\mathrm{ic}}(\cat{B},\cat{C})$, which we can use to construct categories of loss models.
Moreover, owing to the monoidality of $+$, this category will be moreover monoidal: a property that we will use to define the free energy loss model below.
(Note that this monoidal structure, on the \textit{category} of loss models, is distinct from the monoidal structure that we will attach to loss models themselves in \secref{sec:mon-stat-games}.)

\begin{prop} \label{prop:loss-smc}
  Loss models for $\cat{B}$ constitute the objects of a symmetric monoidal category $\bigl(\Loss(\cat{B}), +, 0\bigr)$.
  The morphisms of $\Loss(\cat{B})$ are icons between the corresponding lax functors, and they compose accordingly.
  The monoidal structure is given by sums of losses.
  \begin{proof}[Proof sketch]
    From \textcite[{Theorem 4.6.13}]{Johnson20202Dimensional}, we know that icons compose, forming the morphisms of a category.
    Next, note that for any two loss models $F$ and $G$ and any $k$-cell $\alpha$ (for any $k\in\{0,1,2\}$), $F(\alpha)$ and $G(\alpha)$ must only differ on the loss component, and so we can sum the losses; this gives the monoidal product.
    The monoidal unit is necessarily the constant $0$ loss.
    Finally, observe that the structure is symmetric becauase effect monoids are commutative (by Definition \ref{def:bilin-eff}).
  \end{proof}
\end{prop}

\subsection{Examples} \label{sec:sgame-examples}

Each of our examples involves taking expectations of log-densities, and so to make sense of them it first helps to understand what we mean by ``taking expectations''.

\begin{notation}[Expectations] \label{nota:expectations}
  Written as a function, a density $p$ on $X$ has the type $X\to \rr_+$; written as an effect, the type is $X\klto I$.
  Given a measure or distribution $\pi$ on $X$ (equivalently, a state $\pi:I\klto X$), we can compute the expectation of $p$ under $\pi$ as the composite $p\klcirc\pi$.
  We write the resulting quantity as $\E_\pi[p]$, or more explicitly as $\E_{x\sim\pi}\bigl[p(x)\bigr]$.
  We can think of this expectation as representing the `validity' (or truth value) of the `predicate' $p$ given the state $\pi$ \parencite{Jacobs2018Logical}.
\end{notation}

\subsubsection{Relative entropy and Bayesian inference} \label{sec:rel-ent}

For our first example, we return to the subject with which we opened this paper: the compositional structure of the relative entropy.
We begin by giving a precise definition.

\begin{defn}
  Suppose $\alpha,\beta$ are both measures on $X$, with $\alpha$ absolutely continuous with respect to $\beta$.
  Then the \textit{relative entropy} or \textit{Kullback-Leibler divergence} from $\alpha$ to $\beta$ is the quantity $D_{KL}(\alpha,\beta) := \E_\alpha \left[ \log \frac{\alpha}{\beta} \right]$, where $\frac{\alpha}{\beta}$ is the Radon-Nikodym derivative of $\alpha$ with respect to $\beta$.
\end{defn}

\begin{rmk}
  When $\alpha$ and $\beta$ admit density functions $p_\alpha$ and $p_\beta$ with respect to the same base measure $\d x$, then $D_{KL}(\alpha,\beta)$ can equally be computed as $\E_{x\sim\alpha} \bigl[ \log p_\alpha(x) - \log p_\beta(x) \bigr]$.
  It it this form that we will adopt henceforth.
\end{rmk}

\begin{prop} \label{prop:kl-loss-model}
  Let $\cat{B}$ be a subbicategory of simple lenses in $\BLens{}_2$, all of whose channels admit density functions with respect to a common measure and whose forward channels admit Bayesian inversion (and whose forward and backward coparameters coincide), and with only structural 2-cells.
  Then the relative entropy defines a strict loss model $\KL:\cat{B}\to\SGame{}$.
  Given a lens $(c,c'):(X,X)\lensto(Y,Y)$, $\KL$ assigns the loss function $\KL(c,c'):Y\xklto{X}I$ defined, for $\pi:I\klto X$ and $y:Y$, by the relative entropy $\KL(c,c')_\pi(y) := D_{KL}\bigl(c'_\pi(y),c^\dag_\pi(y)\bigr)$,
  where $c^\dag$ is the exact inversion of $c$.
  \begin{proof}
    Being a section of $\pLoss|_{\cat{B}}$, $\KL$ leaves lenses unchanged, only acting to attach loss functions.
    It therefore suffices to check that this assignment of losses is strictly functorial.
    Writing $\klcirc$ for composition in $\cat{C}$, $\circ$ for horizontal composition in $\SStat$, $\lenscirc$ in $\BLens{}_2$, and $\diamond$ for horizontal composition of losses in $\SGame{}$, we have the following chain of equalities:
    \begin{align*}
      \KL\bigl((d,d')\lenscirc(c,c')\bigr)_\pi(z)
      &= \E_{(x,m,y,n)\sim(c'\circ d'_c)_\pi(z)} \Big[ \log p_{(c'\circ d'_c)_\pi}(x,m,y,n|z) \\ & \hspace*{4.8cm} - \log p_{(c^\dag\circ d^\dag_c)_\pi}(x,m,y,n|z) \Big] \\
      &= \E_{(y,n)\sim d'_{c\klcirc\pi}(z)} \E_{(x,m)\sim c'_\pi(y)} \Big[ \log p_{c'_\pi}(x,m|y) \, p_{d'_{c\klcirc\pi}}(y,n|z) \\ & \hspace*{4.8cm} - \log p_{c^\dag_\pi}(x,m|y) \, p_{d^\dag_{c\klcirc\pi}}(y,n|z) \Big] \\
      &= \E_{(y,n)\sim d'_{c\klcirc\pi}(z)} \Big[ \log p_{d'_{c\klcirc\pi}}(y,n|z) - \log p_{d^\dag_{c\klcirc\pi}}(y,n|z) \\ & \hspace*{3cm} + \E_{(x,m)\sim c'_\pi(y)} \left[ \log p_{c'_\pi}(x,m|y) - \log p_{c^\dag_\pi}(x,m|y) \right] \Big] \\
      &= D_{KL}\bigl( d'_{c\klcirc\pi}(z), d^\dag_{\klcirc\pi}(z) \bigr) + \E_{(y,n)\sim d'_{c\klcirc\pi}(z)} \left[ D_{KL}\bigl( c'_\pi(y), c^\dag_\pi(y) \bigr) \right] \\
      &= \KL(d,d')_{c\klcirc\pi}(z) + \bigl(\KL(c,c')\circ d'_c\bigr)_\pi(z) \\
      &= \bigl(\KL(d,d')\diamond\KL(c,c')\bigr)_\pi(z)
    \end{align*}
    The first line obtains by definition of $\KL$ and $\lenscirc$; the second by definition of $\circ$; the third by the $\log$ adjunction ($\log ab = \log a +\log b$) and by linearity of $\E$; the fourth by definition of $D_{KL}$; the fifth by definition of $\KL$ and of $\circ$; and the sixth by definition of $\diamond$.

    This establishes that $\KL\bigl((d,d')\lenscirc(c,c')\bigr) = \KL(d,d')\diamond\KL(c,c')$ and hence that $\KL$ is strictly functorial on 1-cells.
    Since we have assumed that the only 2-cells are the structural 2-cells (\textit{e.g.}, the horizontal unitors), which do not result in any difference between the losses assigned to the corresponding 1-cells, the only loss 2-cell available to be assigned is the 0 loss; which assignment is easily seen to be vertically functorial.
    Hence $\KL$ is a strict 2-functor, and moreover a section of $\pLoss|_{\cat{B}}$ as required.
  \end{proof}
\end{prop}

Successfully playing a relative entropy game entails minimizing the divergence from the approximate to the exact posterior.
This divergence is minimized when the two coincide, and so $\KL$ represents a form of approximate Bayesian inference.

\begin{rmk} \label{rmk:kl-chain-rule}
  We opened the chapter by observing that the relative entropy satisfies a chain rule defined not on Bayesian lenses, but simply on pairs of channels:
  to formalize this simpler case, we do not need the full machinery of statistical games (which is useful when we have bidirectional inference systems); but we do need some of it.

  If $c$ and $c'$ are parallel channels $X\klto Y$, then $D_{KL}\left(c(-),c'(-)\right)$ defines an effect $X\klto I$.
  This means we can use the statistical games idea to equip parallel (copy-composite) channels in $\cat{C}$ with such non-state-dependent loss functions; and the relative entropy will again form a strict section of the resulting Grothendieck fibration.

  Therefore, let $\cat{B}$ be the bicategory whose 0-cells are the objects of $\cat{C}$, but whose 1-cells and 2-cells are parallel pairs of 1-cells and 2-cells in $\ccopara(\cat{C})$; equivalently, the subbicategory of $\ccopara(\cat{C})^{\Cat{2}}$ which is diagonal on 0-cells.

  Next, let $K$ denote the indexed bicategory $\cat{B}\coop\to\Cat{Bicat}$ obtained as the composite
  \[ \cat{B}\coop \xto{\proj_1} \ccopara(\cat{C})\coop \xto{\ground\coop} \cat{C}\op \xto{\cat{C}(-,I)} \Cat{MonCat} \xto{\deloop} \Cat{Bicat} \]
  where $\proj_1$ indicates the projection of the 1st factor of the parallel pairs of 1-cells and 2-cells.
  Applying the Grothendieck construction to $K$ yields a 2-fibration $\int K \xto{\pi_K} \cat{B}$.
  The 0-cells of $\int K$ are the objects of $\cat{C}$.
  The 1-cells $X\to Y$ are triples $(c,c',L)$ where $c$ and $c'$ are parallel coparameterized channels $X\klto Y$ and $L$ is an effect (loss function) $X\klto I$.
  Given composable 1-cells $(c,c',L):X\to Y$ and $(d,d',M):Y\to Z$, their horizontal composite is defined on the parallel channels as copy-composition, and on the loss functions as $M\klcirc c^{\smallground} + L$ (where $\klcirc$ here is composition in $\cat{C}$).
  2-cells are pairs of 2-cells in $\ccopara(\cat{C})$ and differences of losses.

  Finally, the relative entropy $D_{KL}$ defines a strict section of $\pi_K$, mapping the parallel pair $(c,c')$ to $\bigl(c,c',D_{KL}(c,c')\bigr)$.
  Its chain rule is thus formalized by the horizontal composition in $\int K$.
  \qed
\end{rmk}

\subsubsection{Maximum likelihood estimation} \label{sec:mle}

A statistical system may be more interested in predicting observations than updating beliefs.
This is captured by the process of \textit{maximum} or \textit{marginal likelihood estimation}.

\begin{defn}
  Let $(c,c'):(X,X)\lensto(Y,Y)$ be a simple lens whose forward channel $c$ admits a density function $p_c$.
  Then its \textit{log marginal likelihood} is the loss function given by the marginal log evidence $\MLE(c,c')_\pi(y) := -\log p_{c^{\smallground}\klcirc\pi}(y)$.
\end{defn}

\begin{prop} \label{prop:mle-loss-model}
  Let $\cat{B}$ be a subbicategory of lenses in $\BLens{}_2$ all of which admit density functions with respect to a common measure, and with only structural 2-cells.
  Then the assignment $(c,c') \mapsto \MLE(c,c')$ defines a lax loss model $\MLE:\cat{B}\to\SGame{}$.
  \begin{proof}
    We adopt the notational conventions of the proof of Proposition \ref{prop:kl-loss-model}.
    Observe that
    \[ \MLE\bigl((d,d')\lenscirc(c,c')\bigr)_\pi(z) = -\log p_{d^{\smallground}\klcirc c^{\smallground}\klcirc \pi}(z) = \MLE(d,d')_{c\klcirc \pi}(z) \; . \]
    By definition, we have
    \[ \bigl(\MLE(d,d')\diamond\MLE(c,c')\bigr)_\pi(z) = \MLE(d,d')_{c\klcirc \pi}(z) + \bigl(\MLE(c,c')\circ d'_c\bigr)_\pi(z) \]
    and hence by substitution
    \[ \bigl(\MLE(d,d')\diamond\MLE(c,c')\bigr)_\pi(z) = \MLE\bigl((d,d')\lenscirc(c,c')\bigr)_\pi(z) + \bigl(\MLE(c,c')\circ d'_c\bigr)_\pi(z) \; . \]
    Therefore, $\MLE(c,c')\circ d'_c$ constitutes a 2-cell from $\MLE(d,d')\diamond\MLE(c,c')$ to $\MLE\bigl((d,d')\lenscirc(c,c')\bigr)$, and hence $\MLE$ is a lax functor.
    It is evidently moreover a section of $\pLoss|_{\cat{B}}$, and, like $\KL$, acts trivially on the (purely structural) 2-cells.
  \end{proof}
\end{prop}

Successfully playing a maximum likelihood game involves maximizing the log-likelihood that the system assigns to its observations $y:Y$.
This process amounts to choosing a channel $c$ that assigns high likelihood to likely observations, and thus encodes a valid model of the data distribution.

\subsubsection{Autoencoders via the free energy} \label{sec:fe}

Many adaptive systems neither just infer nor just predict: they do both, building a model of their observations that they also invert to update their beliefs.
In machine learning, such systems are known as \textit{autoencoders}, as they `encode' (infer) and `decode' (predict), `autoassociatively' \parencite{Kramer1991Nonlinear}.
In a Bayesian context, they are known as \textit{variational autoencoders} \parencite{Kingma2013Auto}, and their loss function is the \textit{free energy} \parencite{Dayan1995Helmholtz}.

\begin{defn} \label{def:fe-loss-model}
  The \textit{free energy} loss model is the sum of the relative entropy and the likelihood loss models: $\FE := \KL + \MLE$.
  Given a simple lens $(c,c'):(X,X)\lensto(Y,Y)$ admitting Bayesian inversion and with densities, $\FE$ assigns the loss function
  \begin{align*}
    \FE(c,c')_\pi(y) &= (\KL + \MLE)(c,c')_\pi(y) \\
    &= D_{KL}\bigl(c'_\pi(y),c^\dag_\pi(y)\bigr) - \log p_{c^{\smallground}\klcirc\pi}(y)
  \end{align*}
\end{defn}

Note that this means that optimizing the free energy is not guaranteed to optimize either $\KL$ or $\MLE$ individually, although by definition $\FE$ is an upper bound on them both (and hence often known in machine learning by the alternative name, the \textit{evidence upper bound}, thinking of $\MLE$ as encoding a measure of `evidence').

\begin{rmk} \label{rmk:fe-tractability}
  Beyond its autoencoding impetus, another important property of the free energy is its improved computational tractability compared to either the relative entropy or the likelihood loss.
  This property is a consequence of the following fact:
  although obtained as the sum of terms which both depend on an expensive marginalization\footnote{
  Evaluating the pushforward $c^{\smallground}\klcirc\pi$ involves marginalizing over the intermediate variable; and evaluating $c^\dag_\pi(y)$ also involves evaluating $c^{\smallground}\klcirc\pi$.},
  the free energy itself does not.
  This can be seen by expanding the definitions of the relative entropy and of $c^\dag_\pi$ and rearranging terms:
  \begin{align}
    \FE(c,c')_\pi(y)
    &= D_{KL}\bigl(c'_\pi(y),c^\dag_\pi(y)\bigr) - \log p_{c^{\smallground}\klcirc\pi}(y) \nonumber \\
    &= \E_{(x,m)\sim c'_\pi(y)} \bigl[ \log p_{c'_\pi}(x,m|y) - \log p_{c^\dag_\pi}(x,m|y) \bigr] - \log p_{c^{\smallground}\klcirc\pi}(y) \nonumber \\
    &= \E_{(x,m)\sim c'_\pi(y)} \bigl[ \log p_{c'_\pi}(x,m|y) - \log p_{c^\dag_\pi}(x,m|y) - \log p_{c^{\smallground}\klcirc\pi}(y) \bigr] \nonumber \\
    &= \E_{(x,m)\sim c'_\pi(y)} \bigl[ \log p_{c'_\pi}(x,m|y) - \log \frac{p_c(m,y|x) p_\pi(x)}{p_{c^{\smallground}\klcirc\pi}(y)} - \log p_{c^{\smallground}\klcirc\pi}(y) \bigr] \nonumber \\
    &= \E_{(x,m)\sim c'_\pi(y)} \bigl[ \log p_{c'_\pi}(x,m|y) - \log p_c(m,y|x) - \log p_\pi(x) \bigr] \label{eq:fe-expanded} \\
    &= D_{KL}\bigl(c'_\pi(y), \pi\otimes\mathbf{1}\bigr) - \E_{(x,m)\sim c'_\pi(y)} \bigl[ \log p_c(m,y|x) \bigr] \nonumber
  \end{align}
  Here, $\mathbf{1}$ denotes the measure with density $1$ everywhere.
  Note that when the coparameter is trivial, $\FE(c,c')_\pi(y)$ reduces to
  \[ D_{KL}\bigl(c'_\pi(y), \pi\bigr) - \E_{x\sim c'_\pi(y)} \bigl[ \log p_c(y|x) \bigr] \; . \]
\end{rmk}

\begin{rmk}
  The name \textit{free energy} is due to an analogy with the Helmholtz free energy in thermodynamics, as we can write it as the difference between an (expected) energy and an entropy term:
  \begin{align*}
    \FE(c,c')_\pi(y)
    &= \E_{(x,m) \sim c'_\pi(y)} \bigl[ - \log p_c(m,y|x) - \log p_\pi (x) \bigr]
       - S_{X\otimes M} \bigl[ c'_\pi(y) \bigr] \\
    &= \E_{(x,m) \sim c'_\pi(y)} \left[ E_{(c,\pi)}(x,m,y) \right] - S_{X\otimes M} \bigl[ c'_\pi(y) \bigr]
  \; = U - TS
  \end{align*}
  where we call $E_{(c,\pi)} : X\otimes M \otimes Y \xklto{X} I$ the \textit{energy}, and where $S_{X\otimes M} : I\xklto{X\otimes M}I$ is the Shannon entropy.
  The last equality makes the thermodynamic analogy: \(U\) here is the \textit{internal energy} of the system; \(T = 1\) is the \textit{temperature}; and \(S\) is again the entropy.
\end{rmk}

\subsubsection{The Laplace approximation} \label{sec:laplace}

Although optimizing the free energy does not necessitate access to exact inversions, it does still entail computing an expectation under the approximate inversion (\textit{cf.} equation \eqref{eq:fe-expanded} of Remark \ref{rmk:fe-tractability} above), which may remain non-trivial.
When one is interested in optimizing a model by gradient descent, this becomes particularly pressing, as one needs to form an estimate of the gradient of this expectation with respect to the parameters (which is not in general equal to the expectation of the gradient of the energy).
In machine learning, optimizing variational autoencoders typically involves a ``reparameterization trick'' \parencite[\S2.5]{Kingma2017Variational} to circumvent this difficulty, but in the context of neuroscientific modelling (where one is concerned with biological plausibility), this option is not generally available.

An alternative strategy is to make simplifying assumptions, enabling the desired computations without totally sacrificing biological realism.
In the context of predictive coding, a typical such assumption is that all measures are Gaussian \parencite{Rao1999Predictive,Friston2009Predictive,Bastos2012Canonical,Bogacz2017tutorial,Buckley2017free}.
This is motivated not only by hypotheses about the nature of biological noise (related to the Central Limit Theorem), but also by expediency, as a Gaussian distribution is determined by just two sufficient statistics: its mean and variance.
If one first restricts to lenses with Gaussian channels, and then to lenses whose inversion morphisms are constrained to emit `tightly-peaked' Gaussians (\textit{i.e.}, with small variances), then one can eliminate the expectation from the expected energy, and simply evaluate the energy at the posterior mean.

The conceptual justification for this approximation is due to Laplace \parencite[p.367]{Laplace1986Memoir}, who observed that, given a function of the form $f(x) = e^{n\, h(x)}$ with $h$ having a maximum at $x_0$, the only non-negligible contributions to its integral as $n\to\infty$ are those near to $x_0$\footnote{A demonstration of this can be found on Wikipedia at \url{https://en.wikipedia.org/w/index.php?title=Laplace\%27s_method&oldid=1154930495}.}.
Consequently, the function $h$ can be approximated by the quadratic form obtained from its its 2nd-order Taylor expansion about $x_0$, so that, in the one-dimensional (univariate) case,
\[ \int f(x) \, \d x \approx e^{n\, h(x_0)} \int e^{-\frac{1}{2\sigma^2}(x-x_0)^2} \, \d x \]
for $\sigma = \bigl(n\, \partial^2_x h(x_0)\bigr)^{-1/2}$.
Notably, the integrand on the right-hand side is a Gaussian function: it has the form of the density of a normal distribution.

In the present context, we are generally interested in expectations of the form $\E_{x\sim\pi}\bigl[g(x)\bigr]$, which correspond to integrals $\int g(x)\, e^{\log p_\pi(x)} \, \d x$.
It is possible to extend the foregoing reasoning to this case: supposing that $\log p_\pi(x) \, \propto \, n\, h(x)$ for some function $h$ with a maximum at $x_0$, then as $n\to\infty$, we can approximate both $g$ and $h$ by their 2nd-order expansions, thereby approximating $\pi$ by a Gaussian and $g$ by a quadratic form.

This method of approximating integrals is known as \textit{Laplace's method}, and it has been widely applied in statistics\footnote{%
In statistics, making Gaussian assumptions about Bayesian posteriors, or equivalently using second-order approximations to log posteriors, is also known as \textit{variational Laplace} \parencite{Friston2007Variational}.}
\parencite{Kass1995Bayes,Williams1998Bayesian,Beal2003Variational,Friston2007Variational} \parencite[{Chp. 27}]{MacKay2003Information}, in some circumstances, even yielding exact posteriors \parencite[\S10.2]{Vaart2007Asymptotic}.
For further exposition (and more rigour) in the finite-dimensional case, we refer the reader to \textcite[{Ch. 4}]{Bruijn1981Asymptotic} and \textcite[\S3.7]{Olver1997Asymptotics}; for the general case in Banach spaces, the reader may consult \textcite{Piterbarg1995Laplace}.
And for an analysis of the specific case of approximating Bayesian posteriors (beyond the exact case), with consideration of the approximation errors, one may refer to \textcite{Kass1990Validity} or the technical report accompanying \textcite{Tierney1986Accurate}.

This latter situation is of course closely related to the matter at hand.
Here, rather than approximating the posterior by a Gaussian, we \textit{assume} it to have Gaussian form.

\begin{rmk} \label{rmk:gauss-chan}
  We say that a channel $c:X\klto Y$ is \textit{Gaussian} if $c(x)$ is a Gaussian measure for every $x$ in its domain.
  We denote the mean and variance of $c(x)$ by $\mu_c(x)$ and $\Sigma_c(x)$ respectively, and write its (log) density function as
  \[ \log p_c (y | x) = \frac{1}{2} \innerprod{\epsilon_c(y,x)}{{\Sigma_c(x)}^{-1} {\epsilon_c(y,x)}} - \log \sqrt{(2 \pi)^n \det \Sigma_c(x) } \]
  having also defined the `error' function $\epsilon_c : Y \times X \to Y$ by $\epsilon_c(y, x) = y - \mu_c(x)$.
  In \secref{sec:doctrines-gauss}, we give a full definition of a category of (nonlinear) Gaussian channels.
\end{rmk}

We will still be concerned with approximating expectations $\E_{x\sim d(y)}\bigl[g(x)\bigr]$ by the quadratic expansion of $g$, and so to license Laplace's method we need an analogue of the condition $n\to\infty$.
This will be supplied by the further assumption that $\Sigma_d(y)$ has small eigenvalues: that is, we work in the limit $\tr\left(\Sigma_d(y)\right)\to 0$.
With these two assumptions, we can write
\[ \E_{x\sim d(y)}\bigl[g(x)\bigr] \; \propto \; \int_{x:X} g(x) \, \exp \innerprod{\epsilon_d(x,y)}{{\Sigma_d(y)}^{-1} {\epsilon_d(x,y)}} \d x \]
and observe that as $\tr\left(\Sigma_d(y)\right)\to 0$, we must have $\tr\left({\Sigma_d(y)}^{-1}\right)\to\infty$.
Thus, by Laplace's reasoning, the contributions to the integral are only appreciably non-zero near the mean $\mu_d(y)$.
This licenses the approximation of $g$ by its quadratic expansion around $\mu_d(y)$, and leads to the following approximation of the free energy, known in the predictive coding literature as the \textit{Laplace approximation} \parencite{Friston2007Variational}.
(Consistent with the other examples in this chapter, we consider the coparameterized case.)

\begin{defn} \label{def:cart-spc}
  A \textit{Cartesian space} is an object $X$ that is isomorphic to $\rr^n$ for some $n:\nn$.
\end{defn}

\begin{prop}[{Laplacian free energy}] \label{lemma:laplace-approx} \label{prop:laplace-approx}
  Suppose $(\gamma,\rho):(X,X)\lensto(Y,Y)$ is a Bayesian lens with Gaussian channels between finite-dimensional Cartesian spaces, for which, for all $y:Y$ and Gaussian priors $\pi:I\klto X$, the eigenvalues of $\Sigma_{\rho_\pi}(y)$ are small.
  Then the free energy $\FE(\gamma,\rho)_\pi(y)$ can be approximated by the \textit{Laplacian free energy}
  \begin{align} \label{eq:laplace-energy}
    \FE(\gamma,\rho)_\pi(y) & \approx \LFE(\gamma,\rho)_\pi(y) \\
    & := E_{(\gamma,\pi)}\bigl(\mu_{\rho_\pi}(y), y\bigr) - S_{X\otimes M} \bigl[ \rho_\pi(y) \bigr] \\
    & = -\log p_\gamma(\mu_{\rho_\pi}(y),y) - \log p_\pi(\mu_{\rho_\pi}(y)|_X) - S_{X\otimes M} \bigl[ \rho_\pi(y) \bigr] \nonumber
  \end{align}
  where we have written the argument of the density $p_\gamma$ in `function' style; where $(-)_X$ denotes the projection onto $X$; and where \(S_{X\otimes M}[\rho_\pi(y)] = \E_{(x,m) \sim \rho_\pi(y)} [ -\log p_{\rho_\pi}(x,m|y) ]\) is the Shannon entropy of \(\rho_\pi(y)\).
  The approximation is valid when \(\Sigma_{\rho_\pi}\) satisfies
  \begin{equation} \label{eq:laplace-sigma-rho-pi}
    \Sigma_{\rho_\pi} (y) = \left(\partial_{(x,m)}^2 E_{(\gamma,\pi)}\right)\left( \mu_{\rho_\pi}(y), y\right)^{-1} \, .
  \end{equation}
  We call $E_{(\gamma,\pi)}$ the \textit{Laplacian energy}.
\begin{proof}
  Recall that we can write the free energy \(\FE(\gamma,\rho)_\pi(y)\) as the difference between expected energy and entropy:
  \begin{align*}
    \FE(\gamma,\rho)_\pi(y)
    &= \E_{(x,m) \sim \rho_\pi(y)} \bigl[ - \log p_\gamma(m,y|x) - \log p_\pi (x) \bigr]
    - S_{X\otimes M} \bigl[ \rho_\pi(y) \bigr] \\
    &= \E_{(x,m) \sim \rho_\pi(y)} \bigl[ E_{(\gamma,\pi)}(x,m,y) \bigr] - S_X \bigl[ \rho_\pi(y) \bigr]
  \end{align*}
  Next, since the eigenvalues of \(\Sigma_{\rho_\pi}(y)\) are small for all \(y : Y\), we can approximate the expected energy by its second-order Taylor expansion around the mean \(\mu_{\rho_\pi}(y)\), following Laplace:
  \begin{align*}
    \FE(\gamma,\rho)_\pi(y)
    \approx & \; \E_{(x,m) \sim \rho_\pi(y)} \Biggl[ E_{(\gamma,\pi)}(\mu_{\rho_\pi}(y), y)
      + \innerprod{\epsilon_{\rho_\pi}(x,m,y)}{\left(\partial_{(x,m)} E_{(\gamma,\pi)}\right)\left( \mu_{\rho_\pi}(y), y\right)}
      \\ & \hspace*{2cm}
      + \frac{1}{2} \innerprod{\epsilon_{\rho_\pi}(x,m,y)}{\left(\partial_{(x,m)}^2 E_{(\gamma,\pi)}\right)\left( \mu_{\rho_\pi}(y), y\right) \cdot \epsilon_{\rho_\pi}(x,m,y)} \Biggr] \\
    & \qquad - S_{X\otimes M} \big[ \rho_\pi(y) \big] \\
    \overset{(a)}{=} & \; E_{(\gamma,\pi)}(\mu_{\rho_\pi}(y), y) + \innerprod{\E_{(x,m) \sim \rho_\pi(y)}\bigl[\epsilon_{\rho_\pi}(x,m,y)\bigr]}{\left(\partial_{(x,m)} E_{(\gamma,\pi)}\right)\left( \mu_{\rho_\pi}(y), y\right)}
    \\ & \qquad + \frac{1}{2} \tr \left[ \left(\partial_{(x,m)}^2 E_{(\gamma,\pi)}\right)\left( \mu_{\rho_\pi}(y), y\right) \, \Sigma_{\rho_\pi}(y) \right] - S_{X\otimes M} \big[ \rho_\pi(y) \big] \\
    \overset{(b)}{=} & \; E_{(\gamma,\pi)}(\mu_{\rho_\pi}(y), y) + \frac{1}{2} \tr \left[ \left(\partial_{(x,m)}^2 E_{(\gamma,\pi)}\right)\left( \mu_{\rho_\pi}(y), y\right) \, \Sigma_{\rho_\pi}(y) \right] - S_{X\otimes M} \big[ \rho_\pi(y) \big]
  \end{align*}
  where \(\left(\partial_{(x,m)}^2 E_{(\gamma,\pi)}\right)\left( \mu_{\rho_\pi}(y), y\right)\) is the Hessian of \(E_{(\gamma,\pi)}\) with respect to $(x,m)$ evaluated at \((\mu_{\rho_\pi}(y), y)\).
  The equality marked $(a)$ holds first by the linearity of expectations and second because
  \begin{align}
      & \E_{(x,m) \sim \rho_\pi(y)} \Biggl[ \innerprod{\epsilon_{\rho_\pi}(x,m,y)}{\left(\partial_{(x,m)}^2 E_{(\gamma,\pi)}\right)\left( \mu_{\rho_\pi}(y), y\right) \cdot \epsilon_{\rho_\pi}(x,m,y)} \Biggr] \nonumber \\
      & = \E_{(x,m) \sim \rho_\pi(y)} \Biggl[ \tr \left[ \left(\partial_{(x,m)}^2 E_{(\gamma,\pi)}\right)\left( \mu_{\rho_\pi}(y), y\right) \, \epsilon_{\rho_\pi}(x,m,y) \, \epsilon_{\rho_\pi}(x,m,y)^T \right] \Biggr] \nonumber \\
      & = \tr \left[ \left(\partial_{(x,m)}^2 E_{(\gamma,\pi)}\right)\left( \mu_{\rho_\pi}(y), y\right) \, \E_{(x,m) \sim \rho_\pi(y)} \Bigl[ \epsilon_{\rho_\pi}(x,m,y) \, \epsilon_{\rho_\pi}(x,m,y)^T \Bigr] \right] \nonumber \\
      & = \tr \left[ \left(\partial_{(x,m)}^2 E_{(\gamma,\pi)}\right)\left( \mu_{\rho_\pi}(y), y\right) \, \Sigma_{\rho_\pi}(y) \right] \label{eq:laplace-trace-sigma}
  \end{align}
  where the first equality obtains because the trace of an outer product equals an inner product; the second by linearity of the trace; and the third by the definition of the covariance $\Sigma_{\rho_\pi}(y)$.
  The equality marked $(b)$ above then holds because $\E_{(x,m) \sim \rho_\pi(y)}\bigl[\epsilon_{\rho_\pi}(x,m,y)\bigr] = 0$.

  Next, note that the entropy of a Gaussian measure depends only on its covariance,
  \[
  S_{X\otimes M} \big[ \rho_\pi(y) \big]
  = \frac{1}{2} \log \det \left( 2 \pi \, e \, \Sigma_{\rho_\pi}(y) \right) \, ,
  \]
  and that the energy \(E_{(\gamma,\pi)}(\mu_{\rho_\pi}(y), y)\) does not depend on \(\Sigma_{\rho_\pi}(y)\).
  We can therefore write down directly the covariance \(\Sigma_{\rho_\pi}^\ast(y)\) minimizing \(\FE(\gamma,\rho)_\pi(y)\) as a function of \(y\). We have
  \[
  \partial_{\Sigma_{\rho_\pi}} \FE(\gamma,\rho)_\pi(y) \overset{(b)}{\approx}
  \frac{1}{2} \left(\partial_{(x,m)}^2 E_{(\gamma,\pi)}\right)\left( \mu_{\rho_\pi}(y), y\right)
  + \frac{1}{2} {\Sigma_{\rho_\pi}}^{-1}
  \]
  by equation $(b)$ above.
  Setting \(\partial_{\Sigma_{\rho_\pi}} \FE(\gamma,\rho)_\pi(y) = 0\), we find the optimum as expressed by equation \eqref{eq:laplace-sigma-rho-pi}:
  \begin{equation*} \label{eq:laplace-sigma-opt}
    \Sigma_{\rho_\pi}^\ast (y) = \left(\partial_x^2 E_{(\gamma,\pi)}\right)\left( \mu_{\rho_\pi}(y), y\right)^{-1} \, .
  \end{equation*}
  Finally, by substituting \(\Sigma_{\rho_\pi}^\ast (y)\) in equation \eqref{eq:laplace-trace-sigma}, we obtain the desired expression, equation \eqref{eq:laplace-energy}:
  \[
  \FE(\gamma,\rho)_\pi(y) \approx E_{(\gamma,\pi)}\left(\mu_{\rho_\pi}(y), y\right) - S_{X\otimes M} \left[ \rho_\pi(y) \right] =: \LFE(\gamma,\rho)_\pi(y) \, .
  \]
\end{proof}
\end{prop}

\begin{rmk}
  The usual form of the Laplace model in the literature omits the coparameters.
  It is of course easy to recover the non-coparameterized form by taking $M=1$.
\end{rmk}

As well as being an approximation to a particular statistical game, the Laplacian free energy defines a lax loss model.

\begin{prop} \label{prop:lfe-loss-model}
  Let $\cat{B}$ be a subbicategory of $\BLens{}_2$ of Gaussian lenses between Cartesian spaces whose backward channels have small variance, and with only structural 2-cells\footnote{
  An example of $\cat{B}$ here is obtained by restricting $\BLens{}_2$ to the category $\FdGauss$ of Definition \ref{def:fd-gauss}, and by excluding all but the structural 2-cells}.
  Then $\LFE$ defines a lax loss model $\cat{B}\to\SGame{}$.
  \begin{proof}
    Again we follow the notational conventions of the proof of Proposition \ref{prop:kl-loss-model}.
    Additionally, if $\omega$ is a state on a tensor product such as $X\otimes Y$, we will write $\omega_X$ and $\omega_Y$ to denote its $X$ and $Y$ marginals.
    We will continue to write $c^{\smallground}$ to denote the result of discarding the coparameters of a coparameterized channel $c$.

    Observe that, by repeated application of the linearity of $\E$, the $\log$ adjunction, and the definitions of $\klcirc$ and $\circ$,
    \begin{align*}
      & \bigl(\LFE(d,d')\diamond \LFE(c,c')\bigr)_\pi(z) \\
      &= \LFE(d,d')_{c\klcirc\pi}(z) + \bigl(\LFE(c,c')\circ d'_{c}\bigr)_\pi(y) \\
      &= \LFE(d,d')_{c\klcirc\pi}(z) + \E_{(y,n)\sim d'_{c\klcirc\pi}(z)} \bigl[ \LFE(c,c')_\pi(y) \bigr] \\
      &= - \log p_d\bigl(\mu_{d'_{c\klcirc\pi}}(z),z\bigr) - \log p_{c^{\smallground}\klcirc\pi}\bigl(\mu_{d'_{c\klcirc\pi}}(z)_Y\bigr) \\ &\hspace*{1cm} + \E_{(y,n)\sim d'_{c\klcirc\pi}(z)} \Big[ \log p_{d'_{c\klcirc\pi}}(y,n|z) - \log p_c\bigl(\mu_{c'_\pi}(y),y\bigr) - \log p_\pi\bigl(\mu_{c'_\pi}(y)_X\bigr) \\ &\hspace*{4cm} + \E_{(x,m)\sim c'_\pi(y)} \bigl[ \log p_{c'_\pi}(x,m|y) \bigr] \Big] \\
      &= - \log p_d\bigl(\mu_{d'_{c\klcirc\pi}}(z),z\bigr) - \log p_{c^{\smallground}\klcirc\pi}\bigl(\mu_{d'_{c\klcirc\pi}}(z)_Y\bigr) \\ &\hspace*{1cm} + \E_{(y,n)\sim d'_{c\klcirc\pi}(z)} \bigl[ - \log p_c\bigl(\mu_{c'_\pi}(y),y\bigr) - \log p_\pi\bigl(\mu_{c'_\pi}(y)_X\bigr) \bigr] \\ &\hspace*{1cm} + \E_{(y,n)\sim d'_{c\klcirc\pi}(z)} \Bigl[ \log p_{d'_{c\klcirc\pi}}(y,n|z) + \E_{(x,m)\sim c'_\pi(y)} \bigl[ \log p_{c'_\pi}(x,m|y) \bigr] \Bigr] \\
      &= - \log p_d\bigl(\mu_{d'_{c\klcirc\pi}}(z),z\bigr) - \log p_{c^{\smallground}\klcirc\pi}\bigl(\mu_{d'_{c\klcirc\pi}}(z)_Y\bigr) \\ &\hspace*{1cm} + \E_{(y,n)\sim d'_{c\klcirc\pi}(z)} \bigl[ - \log p_c\bigl(\mu_{c'_\pi}(y),y\bigr) - \log p_\pi\bigl(\mu_{c'_\pi}(y)_X\bigr) \bigr] \\ &\hspace*{1cm} + \E_{(y,n)\sim d'_{c\klcirc\pi}(z)} \E_{(x,m)\sim c'_\pi(y)} \bigl[ \log p_{d'_{c\klcirc\pi}}(y,n|z) + \log p_{c'_\pi}(x,m|y) \bigr] \\
      &= - \log p_d\bigl(\mu_{d'_{c\klcirc\pi}}(z),z\bigr) - \log p_{c^{\smallground}\klcirc\pi}\bigl(\mu_{d'_{c\klcirc\pi}}(z)_Y\bigr) \\ &\hspace*{1cm} + \E_{(y,n)\sim d'_{c\klcirc\pi}(z)} \bigl[ - \log p_c\bigl(\mu_{c'_\pi}(y),y\bigr) - \log p_\pi\bigl(\mu_{c'_\pi}(y)_X\bigr) \bigr] \\ &\hspace*{1cm} + \E_{(x,m,y,n)\sim (c'\circ d'_{c})_\pi(z)} \bigl[ -\log p_{(c'\circ d'_c)_\pi}(x,m,y,n|z) \bigr] \\
      &= - \log p_d\bigl(\mu_{d'_{c\klcirc\pi}}(z),z\bigr) - \log p_{c^{\smallground}\klcirc\pi}\bigl(\mu_{d'_{c\klcirc\pi}}(z)_Y\bigr) \\ &\hspace*{1cm} + \E_{(y,n)\sim d'_{c\klcirc\pi}(z)} \bigl[ - \log p_c\bigl(\mu_{c'_\pi}(y),y\bigr) - \log p_\pi\bigl(\mu_{c'_\pi}(y)_X\bigr) \bigr] - S_{XMYN} \bigl[ (c'\circ d'_c)_\pi(z) \bigr] \\
      &= - \log p_d\bigl(\mu_{d'_{c\klcirc\pi}}(z),z\bigr) - \log p_{c^{\smallground}\klcirc\pi}\bigl(\mu_{d'_{c\klcirc\pi}}(z)_Y\bigr) \\ &\hspace*{1cm} + \E_{(y,n)\sim d'_{c\klcirc\pi}(z)} \bigl[ E_{(c,\pi)}\bigl(\mu_{c'_\pi}(y),y\bigr) \bigr] - S_{XMYN} \bigl[ (c'\circ d'_c)_\pi(z) \bigr] \\
      &= E_{(d,c\klcirc\pi)}(\mu_{d'_{c\klcirc\pi}}(z),z) + \E_{(y,n)\sim d'_{c\klcirc\pi}(z)} \bigl[ E_{(c,\pi)}\bigl(\mu_{c'_\pi}(y),y\bigr) \bigr] - S_{XMYN} \bigl[ (c'\circ d'_c)_\pi(z) \bigr]
    \end{align*}
    where $XMYN$ is shorthand for $X\otimes M\otimes Y\otimes N$.

    Now, writing $E^\mu_{(c,\pi)}(y) := E_{(c,\pi)}\bigl(\mu_{c'_\pi}(y),y\bigr)$, by the Laplace assumption, we have
    \[
      \E_{(y,n)\sim d'_{c\klcirc\pi}(z)} \bigl[ E^\mu_{(c,\pi)}(y) \bigr]
      \approx E^\mu_{(c,\pi)}(\mu_{d'_{c\klcirc\pi}}(z)_Y) + \frac{1}{2} \tr\left[ \left(\partial_{y}^2 E^\mu_{(c,\pi)}\right)\bigl(\mu_{d'_{c\klcirc\pi}}(z)_Y\bigr) \, \Sigma_{d'_{c\klcirc\pi}}(z)_{YY} \right]
      \]
    and so we can write
    \begin{align*}
      & \bigl(\LFE(d,d')\diamond \LFE(c,c')\bigr)_\pi(z) \\
      &\approx E_{(d,c\klcirc\pi)}(\mu_{d'_{c\klcirc\pi}}(z),z) + E^\mu_{(c,\pi)}(\mu_{d'_{c\klcirc\pi}}(z)_Y) - S_{XMYN} \bigl[ (c'\circ d'_{c})_\pi(z) \bigr] \\ &\hspace*{1cm} + \frac{1}{2} \tr\left[ \left(\partial_{y}^2 E^\mu_{(c,\pi)}\right)\bigl(\mu_{d'_{c\klcirc\pi}}(z)_Y\bigr) \, \Sigma_{d'_{c\klcirc\pi}}(z)_{YY} \right] \\
      &= - \log p_d\bigl(\mu_{d'_{c\klcirc\pi}}(z),z\bigr) - \log p_c\bigl(\mu_{c'_\pi}(\mu_{d'_{c\klcirc\pi}}(z)_Y),\mu_{d'_{c\klcirc\pi}}(z)_Y\bigr) - \log p_\pi\bigl(\mu_{c'_\pi}(\mu_{d'_{c\klcirc\pi}}(z)_Y)_X\bigr) \\&\hspace*{1cm} - S_{XMYN} \bigl[ (c'\circ d'_c)_\pi(z) \bigr] - \log p_{c^{\smallground}\klcirc\pi}\bigl(\mu_{d'_{c\klcirc\pi}}(z)_Y\bigr) \\&\hspace*{1cm} + \frac{1}{2} \tr\left[ \left(\partial_{y}^2 E^\mu_{(c,\pi)}\right)\bigl(\mu_{d'_{c\klcirc\pi}}(z)_Y\bigr) \, \Sigma_{d'_{c\klcirc\pi}}(z)_{YY} \right] \\
      &= E_{(d\klcirc c,\pi)}\bigl(\mu_{(c'\circ d'_c)_\pi}(z),z\bigr) - S_{XMYN} \bigl[ (c'\circ d'_c)_\pi(z) \bigr] \\&\hspace*{1cm} - \log p_{c^{\smallground}\klcirc\pi}\bigl(\mu_{d'_{c\klcirc\pi}}(z)_Y\bigr) + \frac{1}{2} \tr\left[ \left(\partial_{y}^2 E^\mu_{(c,\pi)}\right)\bigl(\mu_{d'_{c\klcirc\pi}}(z)_Y\bigr) \, \Sigma_{d'_{c\klcirc\pi}}(z)_{YY} \right] \\
      &= \LFE\bigl((d,d')\lenscirc(c,c')\bigr)_\pi(z) - \log p_{c^{\smallground}\klcirc\pi}\bigl(\mu_{d'_{c\klcirc\pi}}(z)_Y\bigr) \\&\hspace*{1cm} + \frac{1}{2} \tr\left[ \left(\partial_{y}^2 E^\mu_{(c,\pi)}\right)\bigl(\mu_{d'_{c\klcirc\pi}}(z)_Y\bigr) \, \Sigma_{d'_{c\klcirc\pi}}(z)_{YY} \right] \; .
    \end{align*}
    Therefore, if we define a loss function $\kappa$ by
    \begin{align*}
      \kappa_\pi(z) :=&\, \frac{1}{2} \tr\left[ \left(\partial_{y}^2 E^\mu_{(c,\pi)}\right)\bigl(\mu_{d'_{c\klcirc\pi}}(z)_Y\bigr) \, \Sigma_{d'_{c\klcirc\pi}}(z)_{YY} \right] - \log p_{c^{\smallground}\klcirc\pi}\bigl(\mu_{d'_{c\klcirc\pi}}(z)_Y\bigr)
    \end{align*}
    then $\kappa$ constitutes a 2-cell $\LFE(d,d')\diamond\LFE(c,c') \Rightarrow \LFE\bigl((d,d')\lenscirc(c,c')\bigr)$, as required.
  \end{proof}
\end{prop}

Effectively, this proposition says that, under the stated conditions, the free energy and the Laplacian free energy coincide.
Consequently, successfully playing a Laplacian free energy game has the same autoencoding effect as playing a free energy game in the sense of \secref{sec:fe}.

\begin{rmk}
  We formalized the idea of a Gaussian having \textit{small} or \textit{tightly-peaked} variance as meaning its covariance matrix $\Sigma$ has small eigenvalues.
  We do not specify precisely what `small' means here: only, it must be enough to license the use of Laplace's method.
  Of course, as the eigenvalues approach $0$, the Gaussian approaches a Dirac delta distribution.
  In this case, one may truncate the approximating expansion at first order and just work with the means --- in fact, the inversions become deterministic --- and indeed, this is the choice made in some of the predictive coding literature \parencite{Bogacz2017tutorial}.
\end{rmk}

\section{Monoidal statistical games} \label{sec:mon-stat-games}

In Remark \ref{rmk:dag-not-monoidal}, we noted that the canonical section $\dag$ taking a channel $c$ to the lens equipped with its exact inversion $c^\dag$ is not monoidal, because inverting the tensor of two channels with respect to a joint state is in general not the same as inverting the two channels independently with respect to the marginals, owing to the possibility of correlations.
At the same time, we know from Proposition \ref{prop:blens-monoidal} that the category $\BLens{\cat{C}}$ of non-coparameterized Bayesian lenses in $\cat{C}$ is nonetheless a monoidal category (and it is moreover symmetric monoidal when $\cat{C}$ is); and we saw in Corollary \ref{cor:stat-effect-mon-cats} that $\Stat$, and hence $\BLens{\cat{C}}$, are additionally fibrewise monoidal.
In this section, we establish analogous results for copy-composite Bayesian lenses, and statistical games and loss models in turn, as well as demonstrating that each of our loss models is accordingly monoidal.
This monoidal structure on loss models can then be used to measure the error obtained by inverting channels independently with respect to the marginals of a joint prior.

Because statistical games are defined over copy-composite channels, our starting point must be to establish a monoidal structure on $\ccopara(\cat{C})$.

\begin{prop} \label{prop:mon-ccopara}
  If the copy-discard category $\cat{C}$ is symmetric monoidal, then $\ccopara(\cat{C})$ inherits a monoidal structure $(\otimes,I)$, with the same unit object $I$ as in $\cat{C}$.
  On 1-cells $f:A\xto[M]{}B$ and $f':A'\xto[M']{}B'$, the tensor $f\otimes f':A\otimes A'\xto[M\otimes M']{}B\otimes B'$ is defined by
  \[ \scalebox{1.0}{\input{img/f-copara-tensor.tikz}} \quad . \]
  On 2-cells $\varphi:f\Rightarrow g$ and $\varphi':f'\Rightarrow g'$, the tensor $\varphi\otimes\varphi':(f\otimes f')\Rightarrow(g\otimes g')$ is given by the string diagram
  \[ \scalebox{1.0}{\input{img/copara2-2cell-tensor.tikz}} \quad . \]
  \begin{proof}
        To establish that $(\ccopara(\cat{C}),\otimes,I)$ is a monoidal bicategory, we need to show that $\otimes$ is a pseudofunctor $\ccopara(\cat{C})\times\ccopara(\cat{C})\to\ccopara(\cat{C})$ and that $I$ induces a pseudofunctor $\Cat{1}\to\ccopara(\cat{C})$, such that the pair of pseudofunctors satisfies the relevant coherence data.
    We will omit the coherence data, and only sketch that the pseudofunctor $\otimes$ is well defined, leaving a full proof for later work.
    (In the sequel here, we will not make very much use of this tensor.)

    First, we confirm that $\otimes$ is locally functorial, meaning that our definition gives a functor on each pair of hom categories.
    We begin by noting that $\otimes$ is well-defined on 2-cells, that $\varphi\otimes\varphi'$ satisfies that change of coparameter axiom for $f\otimes f'$; this is immediate from instantiating the axiom's string diagram.
    Next, we note that $\otimes$ preserves identity 2-cells; again, this is immediate upon substituting identities into the defining diagram.
    We therefore turn to the preservation of composites, which requires that $(\gamma\odot\varphi)\otimes(\gamma'\odot\varphi')=(\gamma\otimes\gamma')\odot(\varphi\otimes\varphi')$, and which translates to the following graphical equation:
    \[ \scalebox{1.0}{\input{img/copara2-tensor-func-1.tikz}} \quad = \quad \scalebox{1.0}{\input{img/copara2-tensor-func-2.tikz}} \]
    It is easy to see that this equation is satisfied: use the naturality of the symmetry of $(\cat{C},\otimes,I)$.
    This establishes that $\otimes$ is locally functorial.

    Next, we confirm that $\otimes$ is horizontally (pseudo) functorial.
    First, we note that $\id_f\otimes\id_{f'} = \id_{f\otimes f'}$ by the naturality of the symmetry of $(\cat{C},\otimes,I)$.
    Second, we exhibit a multiplication natural isomorphism, witnessing pseudofunctoriality, with components $\mu_{g,g',f,f'}:(g\otimes g')\circ(f\otimes f')\Rightarrow(g\circ f)\otimes(g'\circ f')$ for all composable pairs of 1-cells $g,f$ and $g',f'$.
    Let these 1-cells be such that $(g\otimes g')\circ(f\otimes f')$ has the underlying depiction
    \[ \scalebox{1.0}{\input{img/copara2-1cell-tensor-1.tikz}} \]
    and so $(g\circ f)\otimes(g'\circ f')$ has the depiction
    \[ \scalebox{1.0}{\input{img/copara2-1cell-tensor-2.tikz}} \quad . \]
    It is then easy to see that defining $\mu_{g,g',f,f'}$ and its inverse $\mu^{-1}_{g,g',f,f'}$ as the 2-cells with the following respective underlying depictions gives us the desired isomorphism:
    \[ \scalebox{1.0}{\input{img/copara2-1cell-tensor-lax-1.tikz}} \qquad\text{and}\qquad \scalebox{1.0}{\input{img/copara2-1cell-tensor-lax-2.tikz}} \quad . \]
    The naturality of this definition is a consequence of the naturality of the symmetry of $(\cat{C},\otimes,I)$.

    That this tensor satisfies the monoidal bicategory axioms --- of associativity, unitality, and coherence --- follows from the fact that the monoidal structure $(\otimes,I)$ satisfies correspondingly decategorified versions of these axioms; we leave the details to subsequent exposition.
  \end{proof}
\end{prop}

Following the monoidal Grothendieck recipe, establishing that $\BLens{}_2$ is monoidal entails establishing that $\SStat$ is a monoidal indexed bicategory.
But first we must define the latter concept, by categorifying Definition \ref{def:mon-idx-cat}.

\begin{defn} \label{def:mon-idx-bicat}
  Suppose $(\cat{B},\otimes,I)$ is a monoidal bicategory.
  We will say that $F:\cat{B}\coop\to\Cat{Bicat}$ is a \textit{monoidal indexed bicategory} when it is equipped with the structure of a weak monoid object in the 3-category of indexed bicategories, indexed pseudofunctors, indexed pseudonatural transformations, and indexed modifications.

  More explicitly, we will take $F$ to be a monoidal indexed bicategory when it is equipped with
  \begin{enumerate}[label=(\roman*)]
  \item an indexed pseudofunctor $\mu:F(-)\times F(=)\to F(-\otimes=)$ called the \textit{multiplication}, \textit{i.e.},
    \begin{enumerate}[label=(\alph*)]
    \item a family of pseudofunctors $\mu_{X,Y}: FX\times FY \to F(X\otimes Y)$, along with
    \item for any 1-cells $f:X\to X'$ and $g:Y\to Y'$ in $\cat{B}$, a pseudonatural isomorphism $\mu_{f,g}:\mu_{X',Y'}\circ(Ff\times Fg)\Rightarrow F(f\otimes g)\circ\mu_{X,Y}$;
    \end{enumerate}
  \item a pseudofunctor $\eta:\Cat{1}\to FI$ called the \textit{unit};
  \end{enumerate}
  as well as three indexed pseudonatural isomorphisms --- an associator, a left unitor, and a right unitor --- which satisfy weak analogues of the coherence conditions for a monoidal indexed category \parencite[\S3.2]{Moeller2018Monoidal}, up to invertible indexed modifications.
\end{defn}

\begin{rmk}
  Because it is not our main purpose, and because the coherence data for higher-dimensional structures rapidly becomes cumbersome, the preceding definition only suggests the form of this coherence data.
  Unfortunately, we are not presently aware of a full explicit definition in the literature of the concept of monoidal indexed bicategory.
\end{rmk}

Using this notion, we can establish that $\SStat$ is monoidal.

\begin{thm} \label{thm:monoidal-sstat}
  $\SStat$ is a monoidal indexed bicategory, in the explicit sense of Definition \ref{def:mon-idx-bicat}.
  \begin{proof}[Proof sketch]
    We only check the explicit requirements of the preceding definition, and expect that the higher coherence data is satisfied by the fact that each of our high-dimensional structures is obtained from a well-behaved lower-dimensional one using canonical categorical machinery.

    In this way, the multiplication $\mu$ is given first by the family of pseudofunctors $\mu_{X,Y}:\SStat(X)\times\SStat(Y)\to\SStat(X\otimes Y)$ which are defined on objects simply by tensor
    \[ \mu_{X,Y}(A,B) = A\otimes B \]
    since the objects do not vary between the fibres of $\SStat$, and on hom categories by the functors
    \begin{align*}
      & \SStat(X)(A,B) \times \SStat(Y)(A',B') \\
      &= \Cat{Cat}\bigl(\disc\cat{C}(I,X),\ccopara[r](\cat{C})(A,B)\bigr) \times \Cat{Cat}\bigl(\disc\cat{C}(I,Y),\ccopara[r](\cat{C})(A',B')\bigr) \\
      &\cong \Cat{Cat}\bigl(\disc\cat{C}(I,X)\times\disc\cat{C}(I,Y),\ccopara[r](\cat{C})(A,B)\times\ccopara[r](\cat{C})(A',B')\bigr) \\
      &\xto{\Cat{Cat}(\disc\cat{C}(I,\proj_X)\times\disc\cat{C}(I,\proj_Y), \otimes)} \Cat{Cat}\bigl(\disc\cat{C}(I,X\otimes Y)^2, \ccopara[r](\cat{C})(A\otimes A',B\otimes B')\bigr) \\
      &\xto{\Cat{Cat}(\bcopier,\id)} \Cat{Cat}(\disc\cat{C}(I,X\otimes Y),\ccopara[r](C)(A\otimes A', B\otimes B') \\
      &= \SStat(X\otimes Y)(A\otimes A', B\otimes B') \; .
    \end{align*}
    where $\Cat{Cat}\left(\bcopier,\id\right)$ indicates pre-composition with the universal (Cartesian) copying functor.
    For all $f:X\to X'$ and $g:Y\to Y'$ in $\ccopara[l](\cat{C})$, the pseudonatural isomorphisms
    \[ \mu_{f,g}:\mu_{X',Y'}\circ\bigl(\SStat(f)\times\SStat(g)\bigr) \Rightarrow \SStat(f\otimes g)\circ\mu_{X,Y} \]
    are obtained from the universal property of the product $\times$ of categories.
    The unit $\eta:\Cat{1}\to\SStat(I)$ is the pseudofunctor mapping the unique object of $\Cat{1}$ to the monoidal unit $I$.
    Associativity and unitality of this monoidal structure follow from the functoriality of the construction, given the monoidal structures on $\cat{C}$ and $\Cat{Cat}$.
  \end{proof}
\end{thm}

Just as the monoidal Grothendieck construction induces a monoidal structure on categories of lenses for monoidal pseudofunctors \parencite{Moeller2018Monoidal}, we obtain a monoidal structure on the bicategory of copy-composite bayesian lenses.

\begin{cor} \label{cor:mon-blens}
  The bicategory of copy-composite Bayesian lenses $\BLens{}_2$ is a monoidal bicategory.
  The monoidal unit is the object $(I,I)$.
  The tensor $\otimes$ is given on 0-cells by $(X,A)\otimes(X',A') := (X\otimes X', A\otimes A')$, and on hom-categories by
  \begin{align*}
    & \BLens{}_2\bigl((X,A),(Y,B)\bigr) \times \BLens{}_2\bigl((X,A),(Y,B)\bigr) \\
    &= \ccopara[l](\cat{C})(X,Y)\times\SStat(X)(B,A) \times \ccopara[l](\cat{C})(X',Y')\times\SStat(X')(B',A') \\
    &\xto{\sim} \ccopara[l](\cat{C})(X,Y)\times\ccopara[l](\cat{C})(X',Y') \times \SStat(X)(B,A)\times\SStat(X')(B',A')\\
    &\xto{\otimes\, \times\, \mu_{X,X'}\op} \ccopara[l](\cat{C})(X\otimes X', Y\otimes Y') \times \SStat(X\otimes X')(B\otimes B',A\otimes A') \\
    &= \BLens{}_2\bigl((X,A)\otimes(X',A'), (Y,B)\otimes(Y',B')\bigr) \; .
  \end{align*}
\end{cor}

And similarly, we obtain a monoidal structure on statistical games.

\begin{prop}
  The bicategory of copy-composite statistical games $\SGame{}$ is a monoidal bicategory.
  The monoidal unit is the object $(I,I)$.
  The tensor $\otimes$ is given on 0-cells as for the tensor of Bayesian lenses, and on hom-categories by
  \begin{align*}
    & \SGame{}\bigl((X,A),(Y,B)\bigr) \times \SGame{}\bigl((X',A'),(Y',B')\bigr) \\
    &= \BLens{}_2\bigl((X,A),(Y,B)\bigr) \times \Stat(X)(B,I) \\
    &\qquad \times \BLens{}_2\bigl((X',A'),(Y',B')\bigr) \times \Stat(X')(B',I) \\
    &\xto{\sim} \BLens{}_2\bigl((X,A),(Y,B)\bigr) \times \BLens{}_2\bigl((X',A'),(Y',B')\bigr) \\
    &\qquad \times \Stat(X)(B,I) \times \Stat(X')(B',I) \\
    &\xto{\otimes\,\times\,\mu_{X,X'}} \BLens{}_2\bigl((X,A)\otimes(X',A'),(Y,B)\otimes(Y',B')\bigr) \times \Stat(X\otimes X')(B\otimes B', I\otimes I) \\
    &\xto{\sim} \SGame{}\bigl((X,A)\otimes(X',A'),(Y,B)\otimes(Y',B')\bigr)
  \end{align*}
  where here $\mu$ indicates the multiplication of the monoidal structure on $\Stat$ (\textit{cf.} Proposition \ref{prop:stat-lax}).
\end{prop}

Having obtained a monoidal structure on statistical games, we are in a position to ask for monoidal structures on inference systems and loss models:

\begin{defn}
  A \textit{monoidal inference system} is an inference system $(\cat{D},\ell)$ for which $\ell$ is a lax monoidal pseudofunctor.
  A \textit{monoidal loss model} is a loss model $L$ which is a lax monoidal lax functor.
\end{defn}

To make sense of this definition, we need a notion of lax monoidal structure appropriate for strong (pseudo-) and lax functors: a lax-functor generalization of the notion of lax monoidal functor\footnote{
  Note that, although lax functors themselves generalize lax monoidal functors (as bicategories generalize monoidal categories), \textit{lax monoidal lax functors} are different again, adding another dimension (as monoidal functors add a dimension to functors):
  a lax monoidal lax functor is equivalently a homomorphism of one-object tricategories.}
from Definition \ref{def:lax-mon-func}.
Just as a lax monoidal structure on a functor is given by equipping the functor with natural transformations, a lax monoidal structure on a lax functor is given by equipping it with pseudonatural transformations.
The general structure is given by \textcite[{\S2.2}]{Moeller2018Monoidal} for the case of pseudofunctors; the lax case is similar.

In the following remark, we instantiate this structure for loss models.

\begin{rmk} \label{rmk:mon-loss-data}
  A loss model $L:\cat{B}\to\SGame{}$ is lax monoidal when it is equipped with strong transformations
  \[\begin{tikzcd}
    {\cat{B}\times\cat{B}} & {\SGame{}\times\SGame{}} \\
    {\cat{B}} & {\SGame{}}
    \arrow["{\otimes_{\cat{B}}}"', from=1-1, to=2-1]
    \arrow["{\otimes_{\Cat{G}}}", from=1-2, to=2-2]
    \arrow["{L\times L}", from=1-1, to=1-2]
    \arrow["L"', from=2-1, to=2-2]
    \arrow["\lambda"', shorten <=12pt, shorten >=12pt, Rightarrow, from=1-2, to=2-1]
  \end{tikzcd}
  \quad\text{and}\qquad
  \begin{tikzcd}
    {\Cat{1}} \\
    {\cat{B}} & {\SGame{}}
    \arrow[""{name=0, anchor=center, inner sep=0}, "{(I,I)}", curve={height=-20pt}, from=1-1, to=2-2]
    \arrow["{(I,I)}"', from=1-1, to=2-1]
    \arrow["L"', from=2-1, to=2-2]
    \arrow["{\lambda_0}"', shorten <=8pt, shorten >=6pt, Rightarrow, from=0, to=2-1]
  \end{tikzcd}\]
  where $\otimes_{\cat{B}}$ and $\otimes_{\Cat{G}}$ denote the monoidal products on $\cat{B}\hookrightarrow\BLens{}_2$ and $\SGame{}$ respectively, and when $\lambda$ and $\lambda_0$ are themselves equipped with invertible modifications satisfying coherence axioms, as in \textcite[{\S2.2}]{Moeller2018Monoidal}.

  Note that, because $L$ must be a (lax) section of the 2-fibration $\pLoss|_{\cat{B}}:\SGame{}|_{\cat{B}}\to\cat{B}$, the unitor $\lambda_0$ is forced to be trivial, picking out the identity on the monoidal unit $(I,I)$.
  Likewise, the laxator $\lambda:L(-)\otimes L(=)\Rightarrow L({-}\otimes{=})$ must have 1-cell components which are identities:
  \[ L(X,A)\otimes L(X',A) = (X,A)\otimes(X',A') = (X\otimes X', A\otimes A') = L\bigl((X,A)\otimes L(X',A)\bigr) \]
  The interesting structure is therefore entirely in the 2-cells.
  We follow the convention of \parencite[{Def. 4.2.1}]{Johnson20202Dimensional} that a strong transformation is a lax transformation with invertible 2-cell components.
  Supposing that $(c,c'):(X,A)\lensto(Y,B)$ and $(d,d'):(X',A')\lensto(Y',B')$ are 1-cells in $\cat{B}$, the corresponding 2-cell component of $\lambda$ has the form $\lambda_{c,d}:L\bigl((c,c')\otimes(d,d')\bigr)\Rightarrow L(c,c')\otimes L(d,d')$, hence filling the following square in $\SGame{}$:
  \[\begin{tikzcd}[row sep=scriptsize]
	{(X,A)\otimes(X',A')} && {(Y,B)\otimes(Y',B')} \\
	\\
	{(X,A)\otimes(X',A')} && {(Y,B)\otimes(Y',B')}
	\arrow["{L(c,c')\otimes L(d,d')}", from=1-1, to=1-3]
	\arrow[Rightarrow, no head, from=1-3, to=3-3]
	\arrow[Rightarrow, no head, from=1-1, to=3-1]
	\arrow["{L((c,c')\otimes(d,d'))}"', from=3-1, to=3-3]
	\arrow["{\lambda_{c,d}}", shorten <=18pt, shorten >=18pt, Rightarrow, from=3-1, to=1-3]
  \end{tikzcd}\]
  Intuitively, these 2-cells witness the failure of the tensor $L(c,c')\otimes L(d,d')$ of the parts to account for correlations that may be evident to the ``whole system'' $L\bigl((c,c')\otimes(d,d')\bigr)$.
\end{rmk}

Just as there is a notion of monoidal natural transformation accompanying the notion of monoidal functor (recall Definition \ref{def:mon-nat-trans}), there is a notion of monoidal icon between lax monoidal lax functors\footnote{
The notion of monoidal icon can be obtained by weakening the notion of monoidal pseudonatural transformation given by \textcite[{\S2.2}]{Moeller2018Monoidal}.},
from which we obtain a symmetric monoidal category of monoidal loss models.

\begin{prop}
  Monoidal loss models and monoidal icons form a subcategory $\MonLoss(\cat{B})$ of $\Loss(\cat{B})$, and the symmetric monoidal structure $(+,0)$ on the latter restricts to the former.
\end{prop}

\subsection{Examples}

In this section, we present the monoidal structure on the loss models considered above.
Because loss models $L$ are (lax) sections, following Remark \ref{rmk:mon-loss-data}, this monoidal structure is given in each case by a lax natural family of 2-cells $\lambda_{c,d}:L\bigl((c,c')\otimes(d,d')\bigr)\Rightarrow L(c,c')\otimes L(d,d')$, for each pair of lenses $(c,c'):(X,A)\lensto(Y,B)$ and $(d,d'):(X',A')\lensto(Y',B')$.
Such a 2-cell $\lambda_{c,d}$ is itself given by a loss function of type $B\otimes B'\xklto{X\otimes X'}I$ satisfying the equation $L\bigl((c,c')\otimes(d,d')\bigr) = L(c,c')\otimes L(d,d') + \lambda_{c,d}$;
we can think of it as measuring the difference between the joint game $L\bigl((c,c')\otimes(d,d')\bigr)$ and the ``mean field'' games $L(c,c')$ and $L(d,d')$ taken together.

Following \textcite[{Eq. 4.2.3}]{Johnson20202Dimensional}, lax naturality requires that $\lambda$ satisfy the following equation of 2-cells, where $K$ denotes the laxator (with respect to horizontal composition $\diamond$) with components $K(e,c):Le\diamond Lc \Rightarrow L(e\lenscirc c)$:
\begin{gather*}
  \begin{tikzcd}[ampersand replacement=\&,row sep=scriptsize]
    \& {(Y,B)\otimes(Y',B')} \\
    \\
    {(X,A)\otimes(X',A')} \&\& {(Z,C)\otimes(Z',C')}
    \arrow["{L(c\otimes d)}", curve={height=-12pt}, from=3-1, to=1-2]
    \arrow["{L(e\otimes f)}", curve={height=-12pt}, from=1-2, to=3-3]
    \arrow[""{name=0, anchor=center, inner sep=0}, "{L\bigl((e\lenscirc c)\otimes(f\lenscirc d)\bigr)}"{description}, from=3-1, to=3-3]
    \arrow[""{name=1, anchor=center, inner sep=0}, "{L(e\lenscirc c)\otimes L(f\lenscirc d)}"', curve={height=60pt}, from=3-1, to=3-3]
    \arrow["{K(e\otimes f,c\otimes d)\,}"', shorten >=10pt, Rightarrow, from=1-2, to=0]
    \arrow["{\lambda(e\lenscirc c,f\lenscirc d)\,}"', shorten <=9pt, shorten >=6pt, Rightarrow, from=0, to=1]
  \end{tikzcd}
  \\ = \\
  \begin{tikzcd}[ampersand replacement=\&,row sep=scriptsize]
    \&\& {(Y,B)\otimes(Y',B')} \\
    \\
    {(X,A)\otimes(X',A')} \&\& {(Y,B)\otimes(Y',B')} \&\& {(Z,C)\otimes(Z',C')}
    \arrow[""{name=0, anchor=center, inner sep=0}, "{L(c\otimes d)}"{pos=0.667}, curve={height=-12pt}, from=3-1, to=1-3]
    \arrow[""{name=1, anchor=center, inner sep=0}, "{L(e\otimes f)}"{pos=0.333}, curve={height=-12pt}, from=1-3, to=3-5]
    \arrow[""{name=2, anchor=center, inner sep=0}, "{L(e\lenscirc c)\otimes L(f\lenscirc d)}"', curve={height=60pt}, from=3-1, to=3-5]
    \arrow[Rightarrow, no head, from=1-3, to=3-3]
    \arrow[""{name=3, anchor=center, inner sep=0}, "{Lc\otimes Ld}"{pos=0.6}, from=3-1, to=3-3]
    \arrow[""{name=4, anchor=center, inner sep=0}, "{Le\otimes Lf}"{pos=0.4}, from=3-3, to=3-5]
    \arrow["{\lambda(c,d)}", shift left=3, shorten <=4pt, shorten >=12pt, Rightarrow, from=0, to=3]
    \arrow["{\lambda(e,f)}"', shift right=3, shorten <=4pt, shorten >=12pt, Rightarrow, from=1, to=4]
    \arrow["{K(e,c)\otimes K(f,d)}"', shorten >=8pt, Rightarrow, from=3-3, to=2]
  \end{tikzcd}
\end{gather*}
Since vertical composition in $\SGame{}$ is given on losses by $+$, we can write this equation as
\begin{align}
  & \lambda(e\lenscirc c,f\lenscirc d) + K(e\otimes f,c\otimes d) \nonumber \\
  &\;= \lambda(e,f)\diamond \lambda(c,d) + K(e,c)\otimes K(f,d) \nonumber \\
  &\;= \lambda(e,f)_{c\otimes d} + \lambda(c,d)\circ(e'\otimes f')_{c\otimes d}  + K(e,c)\otimes K(f,d) \; . \label{eq:laxa-nat}
\end{align}
In each of the examples below, therefore, we establish the definition of the laxator $\lambda$ and check that it satisfies equation \ref{eq:laxa-nat}.

We will often use the notation $(-)_X$ to denote projection onto a factor $X$ of a monoidal product.

\subsubsection{Relative entropy}

\begin{prop} \label{prop:kl-lax-mon}
  The loss model $\KL$ of Proposition \ref{prop:kl-loss-model} is lax monoidal.
  Supposing that $(c,c'):(X,X)\lensto(Y,Y)$ and $(d,d'):(X',X')\lensto(Y',Y')$ are lenses in $\cat{B}$, the corresponding component $\lambda^{\KL}(c,d)$ of the laxator is given, for $\omega:I\klto X\otimes X'$ and $(y,y'):Y\otimes Y'$, by
  \[ \lambda^{\KL}(c,d)_\omega(y,y') := \E_{\substack{(x,x',m,m') \, \sim \\ (c'_{\omega_X}\otimes\, d'_{\omega_{X'}})(y,y')}} \left[ \log \frac{p_{\omega_X\otimes\omega_{X'}}(x,x')}{p_\omega(x,x')} \right] + \log \frac{p_{(c\otimes d)^{\smallground}\klcirc\omega}(y,y')}{p_{(c\otimes d)^{\smallground}\klcirc(\omega_X\otimes\omega_{X'})}(y,y')} \; . \]
  (Note that the first term has the form of a ``posterior mutual information'' and the second a log-likelihood ratio.)
  \begin{proof}
    We have
    \begin{align*}
      &\bigl(\KL(c)\otimes\KL(d)\bigr)_\omega(y,y') \\
      &\;= \E_{(x,m)\sim c'_{\omega_X}(y)} \left[ \log p_{c'_{\omega_X}}(x,m|y) - \log p_{c^\dag_{\omega_X}}(x,m|y) \right] \\ &\hspace*{1.25cm} + \E_{(x',m')\sim d'_{\omega_{X'}}(y')} \left[ \log p_{d'_{\omega_{X'}}}(x',m'|y') - \log p_{d^\dag_{\omega_{X'}}}(x',m'|y') \right] \\
      &\;= \E_{\substack{(x,x',m,m') \, \sim \\ (c'_{\omega_X}\otimes\, d'_{\omega_{X'}})(y,y')}} \left[ \log p_{c'_{\omega_X}\otimes\, d'_{\omega_{X'}}}(x,x',m,m'|y,y') - \log p_{c^\dag_{\omega_X}\otimes\, d^\dag_{\omega_{X'}}}(x,x',m,m'|y,y') \right]
    \end{align*}
    and
    \begin{align*}
      &\bigl(\KL(c\otimes d)_\omega(y,y') \\
      &\;= \E_{\substack{(x,x',m,m') \, \sim \\ (c'_{\omega_X}\otimes\, d'_{\omega_{X'}})(y,y')}} \left[ \log p_{c'_{\omega_X}\otimes\, d'_{\omega_{X'}}}(x,x',m,m'|y,y') - \log p_{(c\otimes d)^\dag_\omega}(x,x',m,m'|y,y') \right] \; .
    \end{align*}
    Using Bayes' rule, we can rewrite the exact inversions in these expressions, obtaining
    \begin{align*}
      &\bigl(\KL(c)\otimes\KL(d)\bigr)_\omega(y,y') \\
      &\;= \E_{\substack{(x,x',m,m') \, \sim \\ (c'_{\omega_X}\otimes\, d'_{\omega_{X'}})(y,y')}} \Big[ \log p_{c'_{\omega_X}\otimes\, d'_{\omega_{X'}}}(x,x',m,m'|y,y') - \log p_c(y,m|x) - \log p_d(y',m'|x') \\&\hspace{2cm} - \log p_{\omega_X}(x) - \log p_{\omega_{X'}}(x') + \log p_{c^{\smallground}\klcirc\omega_X}(y) + \log p_{d^{\smallground}\klcirc\omega_{X'}}(y') \Big]
    \end{align*}
    and
    \begin{align*}
      &\bigl(\KL(c\otimes d)_\omega(y,y') \\
      &\;= \E_{\substack{(x,x',m,m') \, \sim \\ (c'_{\omega_X}\otimes\, d'_{\omega_{X'}})(y,y')}} \Big[ \log p_{c'_{\omega_X}\otimes\, d'_{\omega_{X'}}}(x,x',m,m'|y,y') - \log p_c(y,m|x) - \log p_d(y',m'|x') \\&\hspace{2cm} - \log p_\omega(x,x') + \log p_{(c\otimes d)^{\smallground}\klcirc\omega}(y,y') \Big]
      \; .
    \end{align*}
    We define $\lambda^{\KL}(c,d)_\omega(y,y')$ as the difference from $\bigl(\KL(c\otimes d)_\omega(y,y')$ to $\bigl(\KL(c)\otimes\KL(d)\bigr)_\omega(y,y')$, and so, with a little rearranging, we obtain the expression above:
    \begin{align*}
      \lambda^{\KL}(c,d)_\omega(y,y')
      := \;& \bigl(\KL(c\otimes d)_\omega(y,y') - \bigl(\KL(c)\otimes\KL(d)\bigr)_\omega(y,y') \\
      = \;& \E_{\substack{(x,x',m,m') \, \sim \\ (c'_{\omega_X}\otimes\, d'_{\omega_{X'}})(y,y')}} \left[ \log \frac{p_{\omega_X\otimes\omega_{X'}}(x,x')}{p_\omega(x,x')} \right] + \log \frac{p_{(c\otimes d)^{\smallground}\klcirc\omega}(y,y')}{p_{(c\otimes d)^{\smallground}\klcirc(\omega_X\otimes\omega_{X'})}(y,y')} \; .
    \end{align*}

    Next, we need to validate lax naturality.
    Since $\KL$ is strict on losses, we need only check that
    \[ \lambda^{\KL}(e\lenscirc c,f\lenscirc d) = \lambda^{\KL}(e,f)_{c\otimes d} + \lambda^{\KL}(c,d)\circ(e'\otimes f')_{c\otimes d} \, . \]
    By definition, we have
    \begin{align*}
      & \bigl(\lambda^{\KL}(e,f)_{c\otimes d}\bigr)_\omega(z,z') \\
      &\;= \E_{\substack{(y,y',n,n') \, \sim \\ (e'_c\otimes\, f'_d)_\omega(z,z')}} \left[ \log \frac{p_{(c\otimes d)^{\smallground}\klcirc(\omega_X\otimes\omega_{X'})}(y,y')}{p_{(c\otimes d)^{\smallground}\klcirc\omega}(y,y')} \right] + \log \frac{p_{(e\otimes f)^{\smallground}\klcirc(c\otimes d)^{\smallground}\klcirc\omega}(z,z')}{p_{(e\otimes f)^{\smallground}\klcirc(c\otimes d)^{\smallground}\klcirc(\omega_X\otimes\omega_{X'})}(z,z')}
    \end{align*}
    and
    \begin{align*}
      & \bigl(\lambda^{\KL}(c,d)\circ(e'\otimes f')_{c\otimes d}\bigr)_\omega(z,z') \\
      &\;= \E_{\substack{(y,y',n,n') \, \sim \\ (e'_c\otimes\, f'_d)_\omega(z,z')}} \left[ \E_{\substack{(x,x',m,m') \, \sim \\ (c'_{\omega_X}\otimes\, d'_{\omega_{X'}})(y,y')}} \left[ \log \frac{p_{\omega_X\otimes\omega_{X'}}(x,x')}{p_\omega(x,x')} \right] + \log \frac{p_{(c\otimes d)^{\smallground}\klcirc\omega}(y,y')}{p_{(c\otimes d)^{\smallground}\klcirc(\omega_X\otimes\omega_{X'})}(y,y')} \right] \, .
    \end{align*}
    And so we also have
    \begin{align*}
      & \lambda^{\KL}(e\lenscirc c,f\lenscirc d)_\omega(z,z') \\
      &\;= \E_{\substack{(x,x',m,m') \, \sim \\ \bigl((c'\circ e'_c)\otimes(d'\circ f'_d)\bigr)_\omega(z,z')}} \left[ \log \frac{p_{\omega_X\otimes\omega_{X'}}(x,x')}{p_\omega(x,x')} \right] + \log \frac{p_{(e\otimes f)^{\smallground}\klcirc(c\otimes d)^{\smallground}\klcirc\omega}(z,z')}{p_{(e\otimes f)^{\smallground}\klcirc(c\otimes d)^{\smallground}\klcirc(\omega_X\otimes\omega_{X'})}(z,z')} \\
      &\;= \bigl(\lambda^{\KL}(c,d)\circ(e'\otimes f')_{c\otimes d}\bigr)_\omega(z,z') + \bigl(\lambda^{\KL}(e,f)_{c\otimes d}\bigr)_\omega(z,z')
    \end{align*}
    thereby establishing the lax naturality of $\lambda^{\KL}$, by the commutativity of $+$.
  \end{proof}
\end{prop}

\begin{rmk}
  Although $\KL$ is lax monoidal, its laxness arises from the state-dependence of the inversions, and we saw in the opening of this chapter, and then more formally in Remark \ref{rmk:kl-chain-rule}, that in its simplest form the relative entropy does not depend on the inversions; in some sense, the statistical game structure is extraneous.

  In Remark \ref{rmk:kl-chain-rule}, we saw that $D_{KL}$ defines a strict section of a 2-fibration $\int K \xto{\pi_K} \cat{B}$, attaching relative entropies to parallel pairs of channels and capturing their chain rule compositionally.
  Since this section does not involve any inversions, we may thus wonder whether it is more than lax monoidal: and indeed it is!
  $D_{KL}$ is in fact a \textit{strong} monoidal section which is moreover strict monoidal on the losses themselves.
  The laxator simply maps $(c,c',D_{KL}(c,c'))$ and $(d,d',D_{KL}(d,d'))$ to $(c\otimes d, c'\otimes d', D_{KL}(c,c') + D_{KL}(d,d'))$; and indeed it is easy to verify that $D_{KL}(c\otimes d, c'\otimes d') = D_{KL}(c,c') + D_{KL}(d,d')$.
\end{rmk}

\subsubsection{Maximum likelihood estimation}

\begin{prop} \label{prop:mle-lax-mon}
  The loss model $\MLE$ of Proposition \ref{prop:mle-loss-model} is lax monoidal.
  Supposing that $(c,c'):(X,X)\lensto(Y,Y)$ and $(d,d'):(X',X')\lensto(Y',Y')$ are lenses in $\cat{B}$, the corresponding component $\lambda^{\MLE}(c,d)$ of the laxator is given, for $\omega:I\klto X\otimes X'$ and $(y,y'):Y\otimes Y'$, by
  \[ \lambda^{\MLE}(c,d)_\omega(y,y') := \log \frac{p_{(c\otimes d)^{\smallground}\klcirc(\omega_X\otimes\omega_{X'})}(y,y')}{p_{(c\otimes d)^{\smallground}\klcirc\omega}(y,y')} \; . \]
  \begin{proof}
    To obtain the definition of $\lambda^{\MLE}(c,d)$, we consider the difference from $\MLE(c\otimes d)$ to $\MLE(c)\otimes\MLE(d)$:
    \begin{align*}
      \lambda^{\MLE}(c,d)_\omega(y,y') := \;& \MLE(c\otimes d)_\omega(y,y') - \bigl(\MLE(c)\otimes\MLE(d)\bigr)_\omega(y,y') \\
      =\;& -\log p_{(c\otimes d)^{\smallground}\klcirc\omega}(y,y') + \log p_{c^{\smallground}\klcirc\omega_X}(y) - \log p_{d^{\smallground}\klcirc\omega_{X'}}(y') \\
      =\;& \log \frac{p_{(c\otimes d)^{\smallground}\klcirc(\omega_X\otimes\omega_{X'})}(y,y')}{p_{(c\otimes d)^{\smallground}\klcirc\omega}(y,y')} \; .
    \end{align*}
    To demonstrate lax naturality, recall that $\MLE$ is a lax section, so we need to consider the corresponding $\diamond$-laxator.
    From Proposition \ref{prop:mle-loss-model}, the laxator $K^{\MLE}(e,c):\MLE(e)\diamond\MLE(c) \Rightarrow \MLE(e\lenscirc c)$ is given by $K^{\MLE}(e,c) := \MLE(c)\circ e'_c$.
    Next, observe that
    \begin{align*}
      \lambda^{\MLE}(e\lenscirc c, f\lenscirc d)_\omega(z,z')
      &= \log \frac{p_{\bigl((e\klcirc c)^{\smallground}\otimes(f\klcirc d)^{\smallground}\bigr)\klcirc(\omega_X\otimes\omega_{X'})}(z,z')}{p_{\bigl((e\klcirc c)^{\smallground}\otimes(f\klcirc d)^{\smallground}\bigr)\klcirc\omega}(z,z')} \\
      &= \log \frac{p_{(e\otimes f)^{\smallground}\klcirc(c\otimes d)^{\smallground}\klcirc(\omega_X\otimes\omega_{X'})}(z,z')}{p_{(e\otimes f)^{\smallground}\klcirc(c\otimes d)^{\smallground}\klcirc\omega}(z,z')} \\
      &= \lambda^{\MLE}(e,f)_{(c\otimes d)\klcirc\omega}(z,z') \, .
    \end{align*}
    Consequently, we need to verify the equation
    \[ \MLE(c\otimes d)\circ (e\otimes f')_{c\otimes d} = \lambda^{\MLE}(c,d)\circ(e'\otimes f')_{c\otimes d} + \bigl(\MLE(c)\otimes\MLE(d)\bigr)\circ(e'\otimes f')_{c\otimes d} \]
    which, by bilinearity of effects, is equivalent to verifying
    \[ \MLE(c\otimes d) = \lambda^{\MLE}(c,d) + \MLE(c)\otimes\MLE(d) \, . \]
    But, since $+$ is commutative, this is satisfied by the definition of $\lambda^{\MLE}(c,d)$ as a 2-cell of type $\MLE(c\otimes d)\Rightarrow\MLE(c)\otimes\MLE(d)$.
  \end{proof}
\end{prop}

\subsubsection{Free energy}

Since $\KL$ and $\MLE$ are both lax monoidal, it follows that so is $\FE$.

\begin{cor} \label{cor:fe-lax-mon}
  The loss model $\FE$ of Definition \ref{def:fe-loss-model} is lax monoidal.
  Supposing that $(c,c'):(X,X)\lensto(Y,Y)$ and $(d,d'):(X',X')\lensto(Y',Y')$ are lenses in $\cat{B}$, the corresponding component $\lambda^{\FE}(c,d)$ of the laxator is given, for $\omega:I\klto X\otimes X'$ and $(y,y'):Y\otimes Y'$, by
  \[ \lambda^{\FE}(c,d)_\omega(y,y') := \E_{(x,x')\sim(c'_{\omega_X}\otimes d'_{\omega_{X'}})(y,y')} \left[ \log \frac{p_{\omega_X\otimes\omega_{X'}}(x,x')}{p_\omega(x,x')} \right] \; . \]
  \begin{proof}
    $\FE$ is defined as $\KL + \MLE$, and hence $\lambda^{\FE}$ is obtained as $\lambda^{\KL} + \lambda^{\MLE}$.
    Since $+$ is functorial, it preserves lax naturality, and so $\lambda^{\FE}$ is also lax natural.
    $\lambda^{\FE}$ is thus a strong transformation $\FE(-)\otimes\FE(=)\Rightarrow\FE({-}\otimes{=})$, and hence $\FE$ is lax monoidal by Remark \ref{rmk:mon-loss-data}.
  \end{proof}
\end{cor}

\subsubsection{Laplacian free energy}

In order to demonstrate that the lax monoidal structure on $\FE$ is not destroyed by the Laplace approximation, we prove explicitly that $\LFE$ is also lax monoidal.

\begin{prop} \label{prop:lfe-lax-mon}
  The loss model $\LFE$ of Propositions \ref{lemma:laplace-approx} and \ref{prop:lfe-loss-model} is lax monoidal.
  Supposing that $(c,c'):(X,X)\lensto(Y,Y)$ and $(d,d'):(X',X')\lensto(Y',Y')$ are lenses in $\cat{B}$, the corresponding component $\lambda^{\LFE}(c,d)$ of the laxator is given, for $\omega:I\klto X\otimes X'$ and $(y,y'):Y\otimes Y'$, by
  \[ \lambda^{\LFE}(c,d)_\omega(y,y') := \log \frac{p_{\omega_X\otimes\omega_{X'}}(\mu_{(c\otimes d)'_\omega}(y,y')_{XX'})}{p_\omega(\mu_{(c\otimes d)'_\omega}(y,y')_{XX'})} \]
  where $\mu_{(c\otimes d)'_\omega}(y,y')_{XX'}$ is the $(X\otimes X')$-mean of the Gaussian distribution $(c'_{\omega_X}\otimes d'_{\omega_{X'}})(y,y')$.
  \begin{proof}
    We have
    \begin{align*}
      & \LFE(c\otimes d)_\omega(y,y') \\
      &\;= -\log p_{c\otimes d}(\mu_{(c'\otimes d')_\omega}(y,y'),y,y') - \log p_\omega(\mu_{(c'\otimes d')_\omega}(y,y')_{XX'}) \\&\hspace*{1cm} - S_{XX'MM'}\left[(c'\otimes d')_\omega(y,y')\right] \\
      &\;= -\log p_c(\mu_{c'_{\omega_X}}(y),y) - \log p_d(\mu_{d'_{\omega_{X'}}}(y'),y') - p_\omega(\mu_{(c'\otimes d')_\omega}(y,y')_{XX'}) \\&\hspace*{1cm} - S_{XM}\left[c'_{\omega_X}(y)\right] - S_{X'M'}\left[d'_{\omega_{X'}}(y')\right]
    \end{align*}
    and
    \begin{align*}
      & \bigl(\LFE(c)\otimes\LFE(d)\bigr)_\omega(y,y') \\
      &\;= \LFE(c)_{\omega_X}(y) + \LFE(d)_{\omega_{X'}}(y') \\
      &\;= -\log p_c(\mu_{c'_{\omega_X}}(y),y) - p_{\omega_X}(\mu_{c_{\omega_X}}(y)_{X}) - S_{XM}\left[c'_{\omega_X}(y)\right] \\&\hspace*{0.6cm} - \log p_d(\mu_{d'_{\omega_{X'}}}(y'),y') - p_{\omega_{X'}}(\mu_{d'_{\omega_{X'}}}(y')_{X'}) - S_{X'M'}\left[d'_{\omega_{X'}}(y')\right]
    \end{align*}
    so that
    \begin{align*}
      \lambda^{\LFE}(c,d)_\omega(y,y') &= \LFE(c\otimes d)_\omega(y,y') - \bigl(\LFE(c)\otimes\LFE(d)\bigr)_\omega(y,y') \\
      &= \log \frac{p_{\omega_X\otimes\omega_{X'}}(\mu_{(c\otimes d)'_\omega}(y,y')_{XX'})}{p_\omega(\mu_{(c\otimes d)'_\omega}(y,y')_{XX'})}
    \end{align*}
    as given above.

    We need to verify lax naturality, which means checking the equation
    \[ \lambda^{\LFE}(e\lenscirc c,f\lenscirc d) + \kappa(e\otimes f,c\otimes d) = \lambda^{\LFE}(e,f)_{c\otimes d} + \lambda^{\LFE}(c,d)\circ(e'\otimes f')_{c\otimes d} + \kappa(e,c)\otimes \kappa(f,d) \]
    where $\kappa$ is the $\diamond$-laxator with components $\kappa(e,c):\LFE(e)\diamond\LFE(c) \Rightarrow \LFE(e\lenscirc c)$ given by
    \begin{align*}
      \kappa(e,c)_\pi(z) &= \frac{1}{2} \tr\left[ \left(\partial_{y}^2 E^\mu_{(c,\pi)}\right)\bigl(\mu_{e'_{c\klcirc\pi}}(z)_Y\bigr) \, \Sigma_{e'_{c\klcirc\pi}}(z)_{YY} \right] - \log p_{c^{\smallground}\klcirc\pi}\bigl(\mu_{e'_{c\klcirc\pi}}(z)_Y\bigr) \; .
    \end{align*}
    (see Proposition \ref{prop:lfe-loss-model}).
    We have
    \begin{align*}
      \lambda^{\LFE}(e\lenscirc c,f\lenscirc d)
      &= \log \frac{p_{\omega_X\otimes\omega_{X'}}(\mu_{(c\otimes d)'_\omega}(\mu_{(e\otimes f)'_{(c\otimes d)\klcirc\omega}}(z,z')_{YY'})_{XX'})}{p_\omega(\mu_{(c\otimes d)'_\omega}(\mu_{(e\otimes f)'_{(c\otimes d)\klcirc\omega}}(z,z')_{YY'})_{XX'})} \\
      &= \lambda^{\LFE}(c,d)_\omega(\mu_{(e\otimes f)'_{(c\otimes d)\klcirc\omega}}(z,z')_{YY'})_{XX'})
    \end{align*}
    and, by the Laplace approximation,
    \begin{align*}
      & \bigl(\lambda^{\LFE}(c,d)\circ(e'\otimes f')_{c\otimes d})_\omega(z,z') \\
      &=\; \E_{\substack{(y,y',n,n') \, \sim \\ (e'_c\otimes\, f'_d)_\omega(z,z')}} \left[ \lambda^{\LFE}(c,d)_\omega(y,y') \right] \\
      &\approx\; \lambda^{\LFE}(c,d)_\omega(\mu_{(e\otimes f)'_{(c\otimes d)\klcirc\omega}}(z,z')_{YY'}) \\&\hspace*{1cm} + \frac{1}{2} \tr \left[ \left(\partial^2_{(y,y')}\lambda^{\LFE}(c,d)_\omega\right)\left(\mu_{(e\otimes f)'_{(c\otimes d)\klcirc\omega}}(z,z')_{YY'}\right) \, \Sigma_{(e\otimes f)'_{(c\otimes d)\klcirc\omega}}(z,z')_{(YY')(YY')} \right] \, .
    \end{align*}
    We also have
    \begin{align*}
      & \bigl(\kappa(e,c)\otimes \kappa(f,d)\bigr)_\omega(z,z') \\
      &=\; \kappa(e,c)_{\omega_X}(z) + \kappa(f,d)_{\omega_{X'}}(z') \\
      &=\; \frac{1}{2} \tr\left[ \left(\partial_{y}^2 E^\mu_{(c,\omega_X)}\right)\bigl(\mu_{e'_{c\klcirc\omega_X}}(z)_Y\bigr) \, \Sigma_{e'_{c\klcirc\omega_X}}(z)_{YY} \right] - \log p_{c^{\smallground}\klcirc\omega_X}\bigl(\mu_{e'_c\klcirc\omega_X}(z)_Y\bigr) \\
      &\hspace{1cm} + \frac{1}{2} \tr\left[ \left(\partial_{y'}^2 E^\mu_{(d,\omega_{X'})}\right)\bigl(\mu_{f'_{d\klcirc\omega_{X'}}}(z')_{Y'}\bigr) \, \Sigma_{f'_{d\klcirc\omega_{X'}}}(z')_{Y'Y'} \right] - \log p_{d^{\smallground}\klcirc\omega_{X'}}\bigl(\mu_{f'_{d\klcirc\omega_{X'}}}(z')_{Y'}\bigr) \\
      &=\; \frac{1}{2} \tr\Big[ \left(\partial_{(y,y')}^2 E^\mu_{(c\otimes d,\omega_X\otimes\omega_{X'})}\right)\bigl(\mu_{(e\otimes f)'_{(c\otimes d)\klcirc(\omega_X\otimes\omega_{X'})}}(z,z')_{YY'}\bigr) \\&\hspace*{2cm} \Sigma_{(e\otimes f)'_{(c\otimes d)\klcirc(\omega_X\otimes\omega_{X'})}}(z,z')_{(YY')(YY')} \Big] \\&\hspace*{1cm} - \log p_{(c\otimes d)^{\smallground}\klcirc(\omega_X\otimes\omega_{X'})}\bigl(\mu_{(e\otimes f)'_{(c\otimes d)\klcirc(\omega_X\otimes\omega_{X'})}}(z,z')_{YY'}\bigr) \, .
    \end{align*}
    The left-hand side of the lax naturality equation is therefore given by
    \begin{align*}
      & \bigl(\lambda^{\LFE}(e\lenscirc c,f\lenscirc d) + \kappa(e\otimes f,c\otimes d)\bigr)_\omega(z,z') \\
      &=\; \lambda^{\LFE}(c,d)_\omega(\mu_{(e\otimes f)'_{(c\otimes d)\klcirc\omega}}(z,z')_{YY'})
      \\&\hspace*{1cm} + \frac{1}{2} \tr\left[ \left(\partial_{(y,y')}^2 E^\mu_{(c\otimes d,\omega)}\right)\bigl(\mu_{(e\otimes f)'_{(c\otimes d)\klcirc\omega}}(z,z')_{YY'}\bigr) \, \Sigma_{(e\otimes f)'_{(c\otimes d)\klcirc\omega}}(z,z')_{(YY')(YY')} \right]
      \\&\hspace*{1cm} - \log p_{(c\otimes d)\klcirc\omega}\bigl(\mu_{(e\otimes f)'_{(c\otimes d)\klcirc\omega}}(z,z')_{YY'}\bigr)
    \end{align*}
    while the right-hand side is given by
    \begin{align*}
      & \bigl(\lambda^{\LFE}(e,f)_{c\otimes d} + \lambda^{\LFE}(c,d)\circ(e'\otimes f')_{c\otimes d}  + \kappa(e,c)\otimes \kappa(f,d)\bigr)_\omega(z,z') \\
      &=\; \log \frac{p_{(c\otimes d)^{\smallground}\klcirc(\omega_X\otimes\omega_{X'})}(\mu_{(e\otimes f)'_{(c\otimes d)\klcirc\omega}}(z,z')_{YY'})}{p_{(c\otimes d)^{\smallground}\klcirc\omega}(\mu_{(e\otimes f)'_{(c\otimes d)\klcirc\omega}}(z,z')_{YY'})} \\&\hspace*{1cm} + \lambda^{\LFE}(c,d)_\omega(\mu_{(e\otimes f)'_{(c\otimes d)\klcirc\omega}}(z,z')_{YY'}) \\&\hspace*{1cm} + \frac{1}{2} \tr \left[ \left(\partial^2_{(y,y')}\lambda^{\LFE}(c,d)_\omega\right)\left(\mu_{(e\otimes f)'_{(c\otimes d)\klcirc\omega}}(z,z')_{YY'}\right) \, \Sigma_{(e\otimes f)'_{(c\otimes d)\klcirc\omega}}(z,z')_{(YY')(YY')} \right] \\&\hspace*{1cm} + \frac{1}{2} \tr\Big[ \left(\partial_{(y,y')}^2 E^\mu_{(c\otimes d,\omega_X\otimes\omega_{X'})}\right)\bigl(\mu_{(e\otimes f)'_{(c\otimes d)\klcirc(\omega_X\otimes\omega_{X'})}}(z,z')_{YY'}\bigr) \\&\hspace*{2.5cm} \Sigma_{(e\otimes f)'_{(c\otimes d)\klcirc(\omega_X\otimes\omega_{X'})}}(z,z')_{(YY')(YY')} \Big] \\&\hspace*{1cm} - \log p_{(c\otimes d)^{\smallground}\klcirc(\omega_X\otimes\omega_{X'})}\bigl(\mu_{(e\otimes f)'_{(c\otimes d)\klcirc(\omega_X\otimes\omega_{X'})}}(z,z')_{YY'}\bigr) \\
      &\;= -\log p_{(c\otimes d)^{\smallground}\klcirc\omega}(\mu_{(e\otimes f)'_{(c\otimes d)\klcirc\omega}}(z,z')_{YY'}) + \lambda^{\LFE}(c,d)_\omega(\mu_{(e\otimes f)'_{(c\otimes d)\klcirc\omega}}(z,z')_{YY'}) \\&\hspace*{1cm} + \frac{1}{2} \tr \left[ \left(\partial^2_{(y,y')}\lambda^{\LFE}(c,d)_\omega\right)\left(\mu_{(e\otimes f)'_{(c\otimes d)\klcirc\omega}}(z,z')_{YY'}\right) \, \Sigma_{(e\otimes f)'_{(c\otimes d)\klcirc\omega}}(z,z')_{(YY')(YY')} \right] \\&\hspace*{1cm} + \frac{1}{2} \tr\Big[ \left(\partial_{(y,y')}^2 E^\mu_{(c\otimes d,\omega_X\otimes\omega_{X'})}\right)\bigl(\mu_{(e\otimes f)'_{(c\otimes d)\klcirc(\omega_X\otimes\omega_{X'})}}(z,z')_{YY'}\bigr) \\&\hspace*{2.5cm} \Sigma_{(e\otimes f)'_{(c\otimes d)\klcirc(\omega_X\otimes\omega_{X'})}}(z,z')_{(YY')(YY')} \Big] \, .
    \end{align*}
    The difference from the left- to the right-hand side is thus
    \begin{align*}
      & \frac{1}{2} \tr\left[ \left(\partial_{(y,y')}^2 E^\mu_{(c\otimes d,\omega)}\right)\bigl(\mu_{(e\otimes f)'_{(c\otimes d)\klcirc\omega}}(z,z')_{YY'}\bigr) \, \Sigma_{(e\otimes f)'_{(c\otimes d)\klcirc\omega}}(z,z')_{(YY')(YY')} \right] \\
      &\; - \frac{1}{2} \tr\Big[ \left(\partial_{(y,y')}^2 E^\mu_{(c\otimes d,\omega_X\otimes\omega_{X'})}\right)\bigl(\mu_{(e\otimes f)'_{(c\otimes d)\klcirc(\omega_X\otimes\omega_{X'})}}(z,z')_{YY'}\bigr) \\&\hspace*{2.5cm} \Sigma_{(e\otimes f)'_{(c\otimes d)\klcirc(\omega_X\otimes\omega_{X'})}}(z,z')_{(YY')(YY')} \Big] \\
      &\; - \frac{1}{2} \tr \left[ \left(\partial^2_{(y,y')}\lambda^{\LFE}(c,d)_\omega\right)\left(\mu_{(e\otimes f)'_{(c\otimes d)\klcirc\omega}}(z,z')_{YY'}\right) \, \Sigma_{(e\otimes f)'_{(c\otimes d)\klcirc\omega}}(z,z')_{(YY')(YY')} \right] \, .
    \end{align*}
    Now, by definition $\Sigma_{(e\otimes f)'_{(c\otimes d)\klcirc\omega}} = \Sigma_{(e\otimes f)'_{(c\otimes d)\klcirc(\omega_X\otimes\omega_{X'})}}$, and so by the linearity of the trace and of derivation, this difference simplifies to
    \begin{align*}
      & \frac{1}{2} \tr\Big[ \left(\partial_{(y,y')}^2 \left( E^\mu_{(c\otimes d,\omega)} - E^\mu_{(c\otimes d,\omega_X\otimes\omega_{X'})} - \lambda^{\LFE}(c,d)_\omega \right) \right) \\&\hspace*{1.25cm} \bigl(\mu_{(e\otimes f)'_{(c\otimes d)\klcirc\omega}}(z,z')_{YY'}\bigr) \, \Sigma_{(e\otimes f)'_{(c\otimes d)\klcirc\omega}}(z,z')_{(YY')(YY')} \Big] \, .
    \end{align*}
    Recall from the proof of Proposition \ref{prop:lfe-loss-model} that $E^\mu_{(c,\pi)}(y) := E_{(c,\pi)}\bigl(\mu_{c'_\pi}(y),y\bigr)$, and hence
    \begin{align*}
      & \bigl(E^\mu_{(c\otimes d,\omega)} - E^\mu_{(c\otimes d,\omega_X\otimes\omega_{X'})}\bigr)(y,y') \\
      &\;= \bigl(E_{(c\otimes d,\omega)} - E_{(c\otimes d,\omega_X\otimes\omega_{X'})}\bigr)\bigl(\mu_{(c\otimes d)'_\omega}(y,y'),y,y'\bigr) \\
      &\;= -\log p_\omega(\mu_{(c\otimes d)'_\omega}(y,y')_{XX'}) + \log p_{\omega_X\otimes\omega_{X'}}(\mu_{(c\otimes d)'_\omega}(y,y')_{XX'}) \\
      &\;= \log \frac{p_{\omega_X\otimes\omega_{X'}}(\mu_{(c\otimes d)'_\omega}(y,y')_{XX'})}{p_\omega(\mu_{(c\otimes d)'_\omega}(y,y')_{XX'})} \\
      &\;= \lambda^{\LFE}(c,d)_\omega(y,y')
    \end{align*}
    so that $E^\mu_{(c\otimes d,\omega)} - E^\mu_{(c\otimes d,\omega_X\otimes\omega_{X'})} - \lambda^{\LFE}(c,d)_\omega = 0$.
    This establishes that $\lambda^{\LFE}$ is lax natural.
  \end{proof}
\end{prop}

\section{Discussion} \label{sec:sgame-discuss}

Having established the basic structure of statistical games and a handful of examples, there is much more to be done, and so in this section we discuss a number of seemingly fruitful avenues of future research.

An important such avenue is the link between this structure and the similar structure of diegetic open (economic) games \parencite{Capucci2022Diegetic}, a recent reformulation of compositional game theory \parencite{Ghani2016Compositional}, which can also be understood as a constituting a fibration over lenses.
Accordingly, the close connection between game theory and reinforcement learning \parencite{Bellman1954Theory,Hedges2022Value} suggests that algorithms for approximate inference (such as expectation-maximization) and reinforcement learning (such as dynamic programming) are more than superficially similar.
More broadly, we expect all three of active inference, game theory, and reinforcement learning to fit into the general programme of categorical systems theory \parencite{Myers2022Categorical} (with cybernetic extensions \parencite{Capucci2021Towards,Smithe2020Cyber}), and we expect that reframing these disciplines in this way will elucidate their relationships.
In Chapter \ref{chp:brain}, we supply functorial dynamical semantics for approximate inference --- a form of approximate inference algorithm --- but we leave the expression of this in systems-theoretic terms to future work.
Likewise, we leave to the future the study of the performance and convergence of algorithms built upon these compositional foundations\footnote{
It is not clear that the fixed points of jointly optimizing the factors of a composite statistical game are the same as those of the optimization of the composite.
If one is only concerned with optimizing the inversions, then the lens-like composition rule tells us that we may proceed by backward induction, first optimizing the factor nearest the codomain, and then optimizing each remaining factor in turn back towards the domain.
But the problem is harder if we also wish to optimize the forward channels, as the inversion nearest the codomain still depends on the forward channel nearest the domain.}.

Another avenue for further investigation concerns mathematical neatness.
First, we seek an abstract characterization of copy-composition and $\ccopara$: Owen Lynch has suggested to us that the computation by compilers of ``static single-assignment form'' (SSA) \parencite{Kelsey1995Correspondence} by compilers may have a similar structure, and so we expect an abstract characterization to capture both SSA and our examples; we also hope that a more abstract approach will alleviate some of the higher-categorical complexity resulting from the weakness of copy-composition.
Second, the explicit constraint defining simple coparameterized Bayesian lenses is inelegant; as indicated in Remark \ref{rmk:bl-dep-opt}, we expect that using dependent optics \parencite{Braithwaite2021Fibre,Vertechi2022Dependent,Capucci2022Seeing} may help to encode this constraint in the type signature, at the cost of higher-powered mathematical machinery.

Finally, we seek further examples of loss models, and more abstract (and hopefully universal) characterizations of those we already have; for example, it is known that the Shannon entropy has a topological origin \parencite{Bradley2021Entropy} via a ``nonlinear derivation'' \parencite{Leinster2022Entropy}, and we expect that we can follow this connection further.
In following this path, we expect to make use of the duality between algebra and geometry \parencite{Maruyama2016Meaning,Nestruev2020Smooth} (and their intersection in (quantitative) categorical logic \parencite{Jacobs2017Recipe,Caramello2011Topostheoretic}), for as we have already noted, loss functions have a natural algebraic structure.
We consider such investigations part of the nascent field of categorical information geometry.

\chapter{Open dynamical systems, coalgebraically} \label{chp:coalg}

In Chapter \ref{chp:algebra}, we saw how to compose neural circuits together using an algebraic approach to connectomics.
These neural circuits are dynamical systems, formalized as sets of ordinary differential equations.
However, simply specifying these sets obscures the general compositional structure of dynamical systems themselves, the revelation of which supports a subtler intertwining of syntax and semantics, form and function---or, as it happens, algebra and coalgebra.
In this chapter we begin by introducing categorical language for describing general dynamical systems `behaviourally'.
These systems will be `closed' (non-interacting), and so we then explain how the language of coalgebra, and specifically polynomial coalgebras, can be used to open them up.

However, traditional coalgebraic methods are restricted to discrete-time dynamical systems, whereas we are also interested in the continuous-time systems that are commonly used in science, such as our earlier neural circuits.
This motivates the development of a class of generalized polynomial coalgebras that model open systems governed by a general time monoid, and which therefore encompass systems of dynamically interacting ordinary differential equations.
In order to account for stochastic dynamics, we generalize the situation still further, by redefining the category of polynomial functors so that it can be instantiated in a nondeterministic setting.
This will show us how to define open Markov processes coalgebraically, and we also demonstrate related categories of open random dynamical systems.

Finally, we use the polynomial setting to package these systems into monoidal bicategories of `hierarchical' cybernetic systems, of which some are usefully generated differentially.
In the next chapter, these bicategories will provide the setting in which we cast the dynamical semantics of approximate inference.

\begin{rmk}
  The story told in this chapter is of a form similar to that of categorical systems theory \parencite{Myers2020Double,Myers2022Categorical}, in which systems on interfaces collect into (doubly) indexed (double) categories.
  That story tells a general tale, but here we are interested in a specific case: coalgebraic systems with polynomial interfaces whose time evolution is governed by an arbitrary monoid and which may have non-determinism or side effects governed by a monad.
  Such systems appear to sit at a sweet spot of scientific utility; in particular, the next chapter will use them to formalize models of predictive coding.
  In future work, we intend to connect the two stories, expressing our generalized polynomial coalgebras in the double-categorical framework.
\end{rmk}

\section{Categorical background on dynamics and coalgebra}

In this section, we introduce the background material needed for our development of open dynamical systems as polynomial coalgebras.

\subsection{Dynamical systems and Markov chains} \label{sec:closed-sys}

We begin by recalling a `behavioural' approach to dynamical systems popularized by \textcite{Lawvere2009Conceptual} (who give a pedagogical account).
These systems are `closed' in the sense that they do not require environmental interaction for their evolution.
Later, when we consider open systems, their `closures' (induced by interaction with an environment) will constitute dynamical systems of this form.

The evolution of dynamics is measured by time, and we will take time to be represented by an arbitrary monoid $(\Tt,+,0)$.
This allows us to consider time-evolution that is not necessarily reversible, such as governed by $\nn$ or $\rr_{+}$, as well as reversible evolution that is properly governed by groups such as $\Zb$ or $\rr$.
With this in mind, we give a classic definition of dynamical system, as a $\Tt$-action.

\begin{rmk}
  We will work in an abstract category $\cat{E}$ whose objects are considered to be ``state spaces''; its morphisms will determine the nature of the dynamical evolution.
  Therefore, for deterministic systems, we can take $\cat{E}$ simply to be $\Set$, or alternatively some other Cartesian category or category of comonoid homomorphisms.
  For stochastic systems, we may take $\cat{E}$ to be a copy-discard category such as $\Kl(\Da)$ or $\Cat{sfKrn}$, or some other category whose morphisms are considered to be stochastic maps.
  For differential systems, we will require $\cat{E}$ to be equipped with a tangent bundle endofunctor $\Tan$; more on this in \secref{sec:diff-sys}.
\end{rmk}

\begin{defn}
  Let \((\Tt, +, 0)\) be a monoid, representing time.
  Let \(X : \cat{E}\) be some space, called the \textit{state space}.
  Then a \textit{closed dynamical system} \(\vartheta\) \textit{with state space} \(X\) \textit{and time} \(\Tt\) is an action of \(\Tt\) on \(X\).
  When \(\Tt\) is also an object of \(\cat{E}\), then this amounts to a morphism \(\vartheta : \Tt \times X \to X\) (or equivalently, a time-indexed family of \(X\)-endomorphisms, \(\vartheta(t) : X \to X\)), such that \(\vartheta(0) = \id_X\) and \(\vartheta(s + t) = \vartheta(s) \circ \vartheta(t)\).
  In this dynamical context, we will refer to the action axioms as the \textit{flow conditions}, as they ensure that the dynamics can `flow'.
\end{defn}

Note that, in discrete time, this definition implies that a dynamical system is governed by a single \textit{transition map}.

\begin{prop} \label{prop:transition-map}
  In discrete time \(\Tt = \nn\), any dynamical system \(\vartheta\) is entirely determined by its action at \(1 : \Tt\).
  That is, letting the state space be \(X\), we have \(\vartheta(t) = \vartheta(1)^{\circ t}\) where \(\vartheta(1)^{\circ t}\) means ``compose \(\vartheta(1) : X \to X\) with itself \(t\) times''.
  \begin{proof}
    The proof is by induction on \(t : \Tt\).
    We must have \(\vartheta(0) = \id_X\) and \(\vartheta(t+s) = \vartheta(t) \circ \vartheta(s)\).
    So for any \(t\), we must have \(\vartheta(t+1) = \vartheta(t) \circ \vartheta(1)\).
    The result follows immediately; note for example that \(\vartheta(2) = \vartheta(1+1) = \vartheta(1) \circ \vartheta(1)\).
  \end{proof}
\end{prop}

An ordinary differential equation $\dot{x} = f(x)$ defines a vector field $x\mapsto(x,f(x))$ on its state space $X$, and its solutions $x(t)$ for $t:\rr$ define in turn a closed dynamical system, as the following example sketches.

\begin{ex} \label{ex:closed-vector-field}
  Let $\Tan$ denote a tangent bundle functor $\cat{E}\to\cat{E}$ on the ambient category of spaces $\cat{E}$.
  Suppose \(X : U \to \Tan U\) is a vector field on \(U\), with a corresponding solution (integral curve) \(\chi_x : \rr \to U\) for all \(x : U\); that is, \(\chi'(t) = X(\chi_x(t))\) and \(\chi_x(0) = x\).
  Then letting the point \(x\) vary, we obtain a map \(\chi : \rr \times U \to U\).
  This \(\chi\) is a closed dynamical system with state space \(U\) and time \(\rr\).
\end{ex}

So far, we have abstained from using much categorical language.
But these closed dynamical systems have a simple categorical representation.

\begin{prop} \label{prop:closed-dyn-cat}
  Closed dynamical systems with state spaces in \(\cat{E}\) and time \(\Tt\) are the objects of the functor category \(\Cat{Cat}(\deloop{\Tt}, \cat{E})\), where \(\deloop{\Tt}\) is the delooping of the monoid \(\Tt\).
  (Recall delooping from Prop. \ref{prop:deloop}.)
  Morphisms of dynamical systems are therefore natural transformations.
  \begin{proof}
    The category \(\deloop{\Tt}\) has a single object \(\ast\) and morphisms \(t : \ast \to \ast\) for each point \(t : \Tt\); the identity is the monoidal unit \(0 : \Tt\) and composition is given by \(+\).
    A functor \(\vartheta : \deloop{\Tt} \to \cat{E}\) therefore picks out an object \(\vartheta(\ast) : \cat{E}\), and, for each \(t : \Tt\), a morphism \(\vartheta(t) : \vartheta(\ast) \to \vartheta(\ast)\), such that the functoriality condition is satisfied.
    Functoriality requires that identities map to identities and composition is preserved, so we require that \(\vartheta(0) = \id_{\vartheta(\ast)}\) and that \(\vartheta(s + t) = \vartheta(s) \circ \vartheta(t)\).
    Hence the data for a functor \(\vartheta : \deloop{\Tt} \to \cat{E}\) amount to the data for a closed dynamical system in \(\cat{E}\) with time \(\Tt\), and the functoriality condition amounts precisely to the flow condition.
    A morphism of closed dynamical systems \(f : \vartheta \to \psi\) is a map on the state spaces \(f : \vartheta(\ast) \to \psi(\ast)\) that commutes with the flow, meaning that \(f\) satisfies \(f \circ \vartheta(t) = \psi(t) \circ f\) for all times \(t : \Tt\); this is precisely the definition of a natural transformation \(f : \vartheta \to \psi\) between the corresponding functors.
  \end{proof}
\end{prop}

By changing the state space category $\cat{E}$, this simple framework can represent different kinds of dynamics.
For example, by choosing $\cat{E}$ to be a category of stochastic channels, such as $\Kl(\Dst)$ or $\Cat{sfKrn}$, we obtain categories of closed Markov processes.

\begin{ex}[Closed Markov chains and Markov processes] \label{ex:mrkv-proc}
  A closed \textit{Markov chain} is given by a stochastic transition map \(X\klto X\), typically interpreted as a Kleisli morphism $X\to \Pa X$ for some probability monad $\Pa:\cat{E}\to\cat{E}$ (\textit{cf.} \secref{sec:prob-monads} on probability monads).
  Following the discussion above, a closed Markov chain is therefore an object in $\Cat{Cat}\big(\deloop{\nn}, \Kl(\Pa)\big)$.
  With more general time \(\Tt\), one obtains closed \textit{Markov processes}: objects in \(\Cat{Cat}\big(\deloop{\Tt}, \Kl(\Pa)\big)\).
  More explicitly, a closed Markov process is a time-indexed family of Markov kernels; that is, a morphism \(\vartheta : \Tt \times X \to \Pa X\) such that, for all times \(s,t : \Tt\), \(\vartheta_{s+t} = \vartheta_s \klcirc \vartheta_t\) as a morphism in \(\Kl(\Pa)\).
  Note that composition \(\klcirc\) in \(\Kl(\Pa)\) is typically given by the Chapman-Kolmogorov equation, so this means that
  \[
  \vartheta_{s+t}(y|x) = \int_{x':X} \vartheta_s(y|x') \, \vartheta_t(\d x'|x) \, .
  \]
\end{ex}

\subsection{Coalgebra}

We saw above that a closed discrete-time deterministic dynamical system is a function $X\to X$, and that a closed discrete-time Markov chain is a function $X\to \Pa X$.
This suggests a general pattern for discrete-time dynamical systems, as morphisms $X\to FX$ for some endofunctor $F$: such a morphism is called a \textit{coalgebra} for the endofunctor $F$.

\begin{defn}
  Let $F:\cat{E}\to\cat{E}$ be an endofunctor.
  A \textit{coalgebra for} $F$, or $F$\textit{-coalgebra}, is a pair $(X, c)$ of an object $X:\cat{E}$ and a morphism $c:X\to FX$.

  A morphism of $F$-coalgebras or \textit{coalgebra morphism} $(X,c)\to(X',c')$ is a morphism $f:X\to X'$ that commutes with the coalgebra structures; \textit{i.e.}, that makes the following diagram commute:
  \[\begin{tikzcd}
	X & {X'} \\
	FX & {FX'}
	\arrow["f", from=1-1, to=1-2]
	\arrow["c"', from=1-1, to=2-1]
	\arrow["{c'}", from=1-2, to=2-2]
	\arrow["Ff"', from=2-1, to=2-2]
  \end{tikzcd}\]
  $F$-coalgebras and their morphisms constitute a category, denoted $\Coalg(F)$.
  The identity morphism on $(X,c)$ is simply the identity morphism $\id_X:X\to X$.
\end{defn}

\begin{rmk}
  In \secref{sec:comon}, we briefly discussed the notion of \textit{coalgebra for a comonad}, which is a coalgebra in the sense of the preceding definition that additionally satisfies axioms dual to those defining algebras for a monad (Definition \ref{def:monad-alg}).
  In our dynamical applications, the endofunctors not in general be comonads, and so it does not make sense to demand such axioms.
\end{rmk}

\begin{rmk}
  At the same time, the duality of algebra and coalgebra underlies the subtle powers of the field of coalgebraic logic, in which the algebraic structure of logical syntax is used to define constraints on or propositions about the behaviours of dynamical systems\parencite{Kurz2006Logic,Pavlovic2006Testing,Jacobs2017Introduction,Cirstea2000algebra,Corfield2009Coalgebraic}.
  These tools are particularly useful in setting of formal verification, where it is desirable to prove that systems behave according to a specification (for instance, for safety reasons).
\end{rmk}

With the notion of $F$-coalgebra to hand, we immediately obtain categories of closed discrete-time deterministic systems and Markov chains:

\begin{ex}
  The category of closed discrete-time deterministic dynamical systems in $\cat{E}$ is the category $\Coalg(\id)$ of coalgebras for the identity endofunctor $\id_{\cat{E}}:\cat{E}\to\cat{E}$.
\end{ex}

\begin{ex}
  Let $\Pa:\cat{E}\to\cat{E}$ be a probability monad on $\cat{E}$.
  The category of Markov chains is the category $\Coalg(\Pa)$ of $\Pa$-coalgebras.
\end{ex}

Of course, polynomial functors are endofunctors $\Set\to\Set$, so they come with a notion of coalgebra, and we may ask how such objects behave.

\begin{ex} \label{ex:poly-coalg}
  Suppose $p:\Set\to\Set$ is a polynomial functor.
  A coalgebra for $p$ is a function $c:X\to pX$ for some set $X$.
  By Definition \ref{def:poly-set}, we can write $p$ as $\sum_{i:p(1)} y^{p[i]}$, and hence the $p$-coalgebra $c$ has the form $c:X\to\sum_{i:p(1)} X^{p[i]}$.
  Such a function corresponds to a choice, for each $x:X$, of an element of $p(1)$ which we denote $c^o(x)$ and an associated function $c^u_x:p[c_1(x)]\to X$.
  We can therefore write $c$ equivalently as a pair $(c^o,c^u)$ where $c^u$ is the coproduct $\sum_xc^u_x:\sum_xp[c^o(x)]\to X$.
  We think of $p$ as defining the interface of the dynamical system represented by $c$, with $p(1)$ encoding the set of possible `outputs' or `configurations' of the system, each $p[i]$ the set of possible `inputs' for the system when it is in configuration $i:p(1)$, and $X$ as the dynamical state space.
  The coalgebra $c$ can then be understood as an \textit{open} discrete-time dynamical system: the map $c^u$ takes a state $x:X$ and a corresponding input in $p[c^o(x)]$ and returns the next state; and the map $c^o$ takes a state $x:X$ and returns the system's corresponding output or configuration $c^o(x)$.
\end{ex}

A pair of functions $c^o:X\to p(1)$ and $c^u:\sum_xp[c^o(x)]\to X$ is precisely a morphism $c:Xy^X\to p$ of polynomials, and so we have established a mapping from $p$-coalgebras $(X,c)$ to morphisms $Xy^X\to p$.
In fact, we have a stronger result.

\begin{prop} \label{prop:poly-coalg-adj}
  There is an isomorphism of hom-sets $\Poly(Ay^B,p)\cong\Set(A,pB)$ natural in $A,B,p$, and hence adjunctions $(-)y^B\dashv(-)\circ B:\Poly\to\Set$ and $Ay^{(-)}\dashv p\circ(-):\Poly\op\to\Set$.
  \begin{proof}[Proof sketch]
    In Example \ref{ex:poly-coalg}, we established a mapping $\Set(A,pB)\to\Poly(Ay^B,p)$ for the case where $A=B$; the general case is analogous.
    The inverse mapping follows directly from Proposition \ref{prop:poly-bundles}.
    Naturality in $A$ and $B$ follows from naturality of pre-composition; naturality in $p$ follows from naturality of post-composition.
  \end{proof}
\end{prop}

Polynomial coalgebras therefore constitute a type of open discrete-time dynamical systems.
But what if we want open \textit{continuous-time} dynamical systems: do these fit into the coalgebra formalism?
In a different direction, what if we want \textit{open Markov chains}?
In discrete time, we should be able to consider coalgebras for composite endofunctors $p\Pa$, but what if we want to do this in general time?

Let us turn now to answering these questions.

\section{Open dynamical systems on polynomial interfaces} \label{sec:poly-dyn}

In this section, we begin by incorporating dynamical systems in general time into the coalgebraic framework, before generalizing the notion of polynomial functor to incorporate `side-effects' such as randomness.
The resulting framework will allow us to define types of system of interest, such as open Markov processes, quite generally using coalgebraic methods, and in the subsequent sections we will make much use of the newly available compositionality.

\subsection{Deterministic systems in general time}

In this section, let us suppose for simplicity that the ambient category $\cat{E}$ is $\Set$.
We will begin by stating our general definition, before explaining the structures and intuitions that justify it.

\begin{defn} \label{def:poly-dyn}
  A \textit{deterministic open dynamical system} with interface $p:\Poly$, state space $S:\Set$ and time $\Tt:\Set$ is a morphism $\beta : Sy^S \to [\Tt y, p]$ of polynomials, such that, for any section $\sigma : p \to y$ of $p$, the induced morphism
  \begin{gather*}
    Sy^S \xto{\beta} [\Tt y, p] \xto{[\Tt y, \sigma]} [\Tt y, y] \xto{\sim} y^\Tt
  \end{gather*}
  is a $\lhd$-comonoid homomorphism.
\end{defn}

To see how such a morphism $\beta$ is like an `open' version of the closed dynamical systems of \secref{sec:closed-sys}, note that by the tensor-hom adjunction, $\beta$ can equivalently be written with the type $\Tt y \otimes Sy^S \to p$.
In turn, such a morphism corresponds to a pair $(\beta^o, \beta^u)$, where $\beta^o$ is the component `on configurations' with the type $\Tt \times S \to p(1)$, and $\beta^u$ is the component `on inputs' with the type $\sum_{t:\Tt} \sum_{s:S} p[\beta^o(t,s)] \to S$.
We will call the map $\beta^o$ the \textit{output map}, as it chooses an output configuration for each state and moment in time; and we will call the map $\beta^u$ the \textit{update map}, as it takes a state $s:S$, a quantity of time $t:\Tt$, and an input in $p[\beta^o(t,s)]$, and returns a new state.
We might imagine the new state as being given by evolving the system from $s$ for time $t$, and the input as supplied while the system is in the configuration corresponding to $(s,t)$.

It is, however, not sufficient to consider merely such pairs $\beta = (\beta^o,\beta^u)$ to be our open dynamical systems, for we need them to be like `open' monoid actions: evolving for time $t$ then for time $s$ must be equivalent to evolving for time $t+s$, given the same inputs.
It is fairly easy to prove the following proposition, whose proof we defer until after establishing the categories $\Coalg^\Tt(p)$, when we prove it in an alternate form as Proposition \ref{prop:closed-sys-in-dyn-y}.

\begin{prop} \label{prop:closed-sys-y-comon}
  Comonoid homomorphisms $Sy^S \to y^\Tt$ correspond bijectively with closed dynamical systems with state space $S$, in the sense given by functors $\deloop{\Tt}\to\Set$.
\end{prop}

This establishes that seeking such a comonoid homomorphism will give us the monoid action property that we seek, and so it remains to show that a composite comonoid homomorphism of the form $[\Tt y, \sigma]\circ \beta$ is a closed dynamical system with the ``right inputs''.
Unwinding this composite, we find that the condition that it be a comonoid homomorphism corresponds to the requirement that, for any $t:\Tt$, the \textit{closure} $\beta^\sigma : \Tt\times S\to S$ of $\beta$ by $\sigma$ given by
\[
\beta^\sigma(t) := S \xto{\beta^o(t)^\ast \sigma} \Sum_{s:S} p[\beta^o(t, s)] \xto{\beta^u} S
\]
constitutes a closed dynamical system on $S$.
The idea here is that $\sigma$ gives the `context' in which we can make an open system closed, thereby formalizing the ``given the same inputs'' requirement above.

With this conceptual framework in mind, we are in a position to render open dynamical systems on $p$ with time $\Tt$ into a category, which we will denote by $\Coalg^\Tt(p)$.
Its objects will be pairs $(S, \beta)$ with $S$ and $\beta$ an open dynamical on $p$ with state space $S$; we will often write these pairs equivalently as triples $(S, \beta^o, \beta^u)$, making explicit the output and update maps.
Morphisms will be maps of state spaces that commute with the dynamics:

\begin{prop} \label{prop:poly-dyn}
  Open dynamical systems over \(p\) with time \(\Tt\) form a category, denoted \(\Coalg^\Tt(p)\).
  Its morphisms are defined as follows.
  Let \(\vartheta := (X, \vartheta^o, \vartheta^u)\) and \(\psi := (Y, \psi^o, \psi^u)\) be two dynamical systems over \(p\).
  A morphism \(f : \vartheta \to \psi\) consists in a morphism \(f : X \to Y\) such that, for any time \(t : \Tt\) and section \(\sigma : p(1) \to \Sum_{i:p(1)} p[i]\) of \(p\), the following naturality squares commute:
  \[\begin{tikzcd}
	X & {\Sum_{x:X} p[\vartheta^o(t, x)]} & X \\
	Y & {\Sum_{y:Y} p[\psi^o(t, y)]} & Y
	\arrow["{\vartheta^o(t)^\ast \sigma}", from=1-1, to=1-2]
	\arrow["{\vartheta^u(t)}", from=1-2, to=1-3]
	\arrow["f"', from=1-1, to=2-1]
	\arrow["f", from=1-3, to=2-3]
	\arrow["{\psi^o(t)^\ast \sigma}"', from=2-1, to=2-2]
	\arrow["{\psi^u(t)}"', from=2-2, to=2-3]
  \end{tikzcd}\]
  The identity morphism \(\id_\vartheta\) on the dynamical system \(\vartheta\) is given by the identity morphism \(\id_X\) on its state space \(X\).
  Composition of morphisms of dynamical systems is given by composition of the morphisms of the state spaces.
  \begin{proof}
    We need to check unitality and associativity of composition.
    This amounts to checking that the composite naturality squares commute.
    But this follows immediately, since the composite of two commutative diagrams along a common edge is again a commutative diagram.
  \end{proof}
\end{prop}

We can alternatively state Proposition \ref{prop:closed-sys-y-comon} as follows, noting that the polynomial $y$ represents the trivial interface, exposing no configuration to any environment nor receiving any signals from it:

\begin{prop} \label{prop:closed-sys-in-dyn-y}
  \(\DynT{T}(y)\) is equivalent to the classical category \(\Cat{Cat}(\deloop{\Tt}, \Set)\) of closed dynamical systems in \(\Set\) with time \(\Tt\).
  \begin{proof}
    The trivial interface \(y\) corresponds to the trivial bundle \(\id_1 : 1\to 1\).
    Therefore, a dynamical system over \(y\) consists of a choice of state space \(S\) along with a trivial output map \(\vartheta^o = \ground : \Tt \times S \to 1\) and a time-indexed update map \(\vartheta^u : \Tt \times S \to S\).
    This therefore has the form of a classical closed dynamical system, so it remains to check the monoid action.
    There is only one section of \(\id_1\), which is again \(\id_1\).
    Pulling this back along the unique map \(\vartheta^o(t) : S \to 1\) gives \(\vartheta^o(t)^\ast \id_1 = \id_S\).
    Therefore the requirement that, given any section \(\sigma\) of \(y\), the maps \(\vartheta^u \circ \vartheta^o(t)^\ast \sigma\) form an action means in turn that so does \(\vartheta^u : \Tt \times S \to S\).
    Since the pullback of the unique section \(\id_1\) along the trivial output map \(\vartheta^o(t) = \ground : S \to 1\) of any dynamical system in \(\DynT{T}(y)\) is the identity of the corresponding state space \(\id_S\), a morphism \(f : (\vartheta(\ast), \vartheta^u, \ground) \to (\psi(\ast), \psi^u, \ground)\) in \(\DynT{T}(y)\) amounts precisely to a map \(f : \vartheta(\ast) \to \psi(\ast)\) on the state spaces in \(\Set\) such that the naturality condition \(f \circ \vartheta^u(t) = \psi^u(t) \circ f\) of Proposition \ref{prop:closed-dyn-cat} is satisfied, and every morphism in \(\Cat{Cat}(\deloop{\Tt}, \Set)\) corresponds to a morphism in \(\DynT{T}(y)\) in this way.
  \end{proof}
\end{prop}

Now that we know that our concept of open dynamical system subsumes closed systems, let us consider some more examples.

\begin{ex}
  Consider a dynamical system \((S, \vartheta^o, \vartheta^u)\) with outputs but no inputs.
  Such a system has a `linear' interface $p := Oy$ for some $O$; alternatively, we can write its interface $p$ as the `bundle' $\id_O : O\to O$.
  A section of this bundle must again be $\id_O$, and so \(\vartheta^o(t)^\ast \id_O = \id_S\).
  Once again, the update maps collect into to a closed dynamical system in \(\Cat{Cat}(\deloop{\Tt}, \Set)\); just now we have outputs \(\vartheta^o : \Tt \times S \to p(1) = O\) exposed to the environment.
\end{ex}

\begin{prop} \label{prop:open-transition-map}
  When time is discrete, as with \(\Tt = \nn\), any open dynamical system \((X, \vartheta^o, \vartheta^u)\) over \(p\) is entirely determined by its components at \(1 : \Tt\).
  That is, we have \(\vartheta^o(t) = \vartheta^o(1) : X \to p(1)\) and \(\vartheta^u(t) = \vartheta^u(1) : \sum_{x:X} p[\vartheta^o(x)] \to X\).
  A discrete-time open dynamical system is therefore a triple \((X, \vartheta^o, \vartheta^u)\), where the two maps have types \(\vartheta^o : X \to p(1)\) and \(\vartheta^u : \sum_{x:X} p[\vartheta^o(x)] \to X\).
  \begin{proof}
    Suppose \(\sigma\) is a section of \(p\).
    We require each closure \(\vartheta^\sigma\) to satisfy the flow conditions, that \(\vartheta^\sigma(0) = \id_X\) and \(\vartheta^\sigma(t+s) = \vartheta^\sigma(t) \circ \vartheta^\sigma(s)\).
    In particular, we must have \(\vartheta^\sigma(t+1) = \vartheta^\sigma(t) \circ \vartheta^\sigma(1)\).
    By induction, this means that we must have \(\vartheta^\sigma(t) = \vartheta^\sigma(1)^{\circ t}\) (compare Proposition \ref{prop:transition-map}).
    Therefore we must in general have \(\vartheta^o(t) = \vartheta^o(1)\) and \(\vartheta^u(t) = \vartheta^u(1)\).
  \end{proof}
\end{prop}

\begin{rmk}
  Note that the preceding proposition means that the objects of $\Coalg^\nn(p)$ are the objects of the traditional category $\Coalg(p)$ of $p$-coalgebras.
  In fact, we have more than this: $\Coalg^\nn(p) \cong \Coalg(p)$; \textit{cf.} Example \ref{ex:poly-coalg} and Proposition \ref{prop:poly-coalg-adj}.
\end{rmk}

\begin{ex} \label{ex:open-vect-field}
  We can express `open' vector fields in this framework.
  Suppose therefore that $X$ is a differentiable manifold (and write $X$ equally for its underlying set of points), and let \(\dot{x} = f(x, a)\) and \(b = g(x)\), with \(f : X \times A \to \Tan X\) and \(g : X \to B\).
  Then, as for the `closed' vector fields of Example \ref{ex:closed-vector-field}, this induces an open dynamical system \((X, \int f, g) : \Coalg^\rr(By^A)\), where \(\int f : \rr \times X \times A \to X\) returns the \((X,A)\)-indexed solutions of \(f\).
\end{ex}

\begin{ex} \label{ex:open-vect-field-2}
  The preceding example is easily extended to the case of a general polynomial interface.
  Suppose similarly that \(\dot{x} = f(x, a_x)\) and \(b = g(x)\), now with \(f : \sum_{x:X} p[g(x)] \to \Tan X\) and \(g : X \to p(1)\).
  Then we obtain an open dynamical system \((X, \int f, g) : \Coalg^\rr(p)\), where now \(\int f : \rr \times \sum_{x:X} p[g(x)] \to X\) is the `update' and \(g : X \to p(1)\) the `output' map.
\end{ex}

By letting the polynomial $p$ vary, it is quite straightforward to extend $\Coalg^\Tt(p)$ to an opindexed category $\Coalg^\Tt$.

\begin{prop}
  $\Coalg^\Tt$ extends to an opindexed category $\Coalg^\Tt : \Poly \to \Cat{Cat}$.
  On objects (polynomials), it returns the categories above.
  On morphisms of polynomials, we simply post-compose: given $\varphi : p\to q$ and $\beta : Sy^S \to [\Tt y, p]$, obtain $Sy^S \to [\Tt y, p] \to [\Tt y, q]$ in the obvious way.
\end{prop}

When we introduced $\Poly$ in \secref{sec:poly}, it was as a ``syntax for interacting adaptive systems'', and we know that we can understand $\Poly$ multicategorically, as it has a monoidal structure $(\otimes,y)$ allowing us to place systems' interfaces side-by-side (and which therefore gives us a multicategory, $\cat{O}\Poly$ by Proposition \ref{prop:multicat-smc}).
We motivated our development of coalgebraic dynamical systems as a compositional extension of the sets of ordinary differential equations that we used to formalize rate-coded neural circuits (Definition \ref{def:rate-circ}), and we have seen that linear circuit diagrams embed into $\Poly$ (Remark \ref{rmk:lincirc-poly}).

One may wonder, therefore, whether the opindexed categories $\Coalg^\Tt$ might supply the general ``semantics for interacting adaptive systems'' that we seek: more precisely, is $\Coalg^\Tt$ a $\Poly$-algebra?
This question can be answered affirmatively, as $\Coalg^\Tt$ is lax monoidal: more precisely, it is a strong monoidal opindexed category.

\begin{prop} \label{prop:coalg-monoidal}
  $\Coalg^\Tt$ is a monoidal opindexed category $(\Poly,\otimes,y)\to(\Cat{Cat},\times,1)$.
  \begin{proof}
    We need to define a natural family of functors $\mu_{p,q}:\Coalg^\Tt(p)\times\Coalg^\Tt(q)\to\Coalg^\Tt(p\otimes q)$ constituting the laxator, and a unit $\eta:\Cat{1}\to\Coalg^\Tt(y)$, along with associators $\alpha$ and left and right unitors $\lambda$ and $\rho$ satisfying the pseudomonoid axioms of Definition \ref{def:mon-idx-cat}.

    The unit $\eta:\Cat{1}\to\Coalg^\Tt(y)$ is given by the trivial system $(1,!,!)$ with the trivial state space and the trivial interface: the output map is the unique map $1\to 1$ (the identity); likewise, the update map is the unique map $1\times 1\to 1$.
    Note that $1\times 1\cong 1$.

    The laxator $\mu_{p,q}$ is given on objects $(X,\vartheta):\Coalg^\Tt(p)$ and $(Y,\varphi):\Coalg^\Tt(q)$ by the $\mu_{p,q}(\vartheta,\varphi) := \bigl(XY,(\vartheta\varphi)\bigr)$ where the state space $XY = X\times Y$ and $(\vartheta\varphi)$ is the system given by the right adjunct of
    \begin{align*}
      XYy^{XY}\otimes\Tt y \xto{{\sim}\otimes\bcopier y}
      Xy^X\otimes Yy^Y\otimes\Tt y\otimes\Tt y \xto{Xy^X\otimes\mathsf{swap}\otimes\Tt y}
      Xy^X\otimes\Tt y\otimes Yy^Y\otimes\Tt y \xto{\vartheta^\flat\otimes\varphi^\flat}
      p\otimes q
    \end{align*}
    under the tensor-hom adjunction in $\Poly$, where $\vartheta^\flat$ and $\varphi^\flat$ are the corresponding left adjuncts of $\vartheta$ and $\varphi$, and where $\sim$ is the isomorphism $XYy^{XY}\xto{\sim}Xy^X\otimes Yy^Y$.
    On morphisms $f:(X,\vartheta)\to(X',\vartheta')$ and $g:(Y,\varphi)\to(Y',\varphi)$, $\mu_{p,q}$ acts as $\mu_{p,q}(f,g) := f\times g$; functoriality hence follows from that of $\times$.

    Next, we need to define $\mu$ on morphisms $\zeta:p\to p'$ and $\xi:q\to q'$ of polynomials, giving natural isomorphisms $\mu_{\zeta,\xi}:\mu_{p',q'}\circ\bigl(\Coalg^\Tt(\zeta)\times\Coalg^\Tt(\xi)\bigr) \Rightarrow \Coalg^\Tt(\zeta\otimes\xi)\circ\mu_{p,q}$.
    But it is easy to see in fact that $\mu_{p',q'}\circ\bigl(\Coalg^\Tt(\zeta)\times\Coalg^\Tt(\xi)\bigr) = \Coalg^\Tt(\zeta\otimes\xi)\circ\mu_{p,q}$, as both sides act by post-composing $\zeta\otimes\xi$.

    The associator is defined componentwise on objects as
    \[ \alpha_{p,q,r} : \Bigl( (XY)Zy^{(XY)Z} \xto{(\vartheta\otimes\varphi)\otimes\psi} [(p\otimes q)\otimes r,\Tt y] \Bigr) \mapsto \Bigl( X(YZ)y^{X(YZ)} \xto{\vartheta\otimes(\varphi\otimes\psi)} [p\otimes(q\otimes r),\Tt y] \Bigr) \]
    and on morphisms as $\alpha_{p,q,r} : (f\times g)\times h \mapsto f\times(g\times h)$, implicitly using the associators of $\otimes$ on $\Poly$ and $\times$ on $\Set$.

    Likewise, the left unitor is defined by
    \[ \lambda_p : \Bigl( 1Xy^{1X} \xto{\mu_{y,p}(\eta,\vartheta)} [y\otimes p,\Tt y] \Bigr) \mapsto \Bigl( Xy^X\xto{\vartheta}[p,\Tt y] \Bigr) \]
    implicitly using the left unitors of $\otimes$ on $\Poly$ and $\times$ on $\Set$; and the right unitor is defined dually, using the corresponding right unitors on $\Poly$ and $\Set$.

    That the associators and unitors satisfy the indexed monoidal category axioms follows from the satisfaction of the monoidal category axioms by $(\Poly,\otimes,y)$ and $(\Set,\times,1)$. (But it is easy, though laborious, to verify this manually.)
  \end{proof}
\end{prop}

\begin{rmk}
  We emphasize that the functor $\Coalg^\Tt$ is \textit{lax} monoidal---the laxators are not equivalences---since not all systems over the parallel interface $p\otimes q$ factor into a system over $p$ alongside a system over $q$.
\end{rmk}

With this indexed monoidal structure, we can show that, as we might hope from a general semantics for interacting dynamical systems, $\Coalg$ subsumes our earlier linear circuit algebra of rate-coded neural circuits.

\begin{prop}
  There is an inclusion $\iota$ of monoidal indexed categories as in the diagram
  \[\begin{tikzcd}
    {\bigl(\LinCirc,+,(0,0)\bigr)} && {(\Set,\times,1)} \\
    \\
    {(\Poly,\otimes,y)} && {(\Cat{Cat},\times,\Cat{1})}
    \arrow[hook, from=1-1, to=3-1]
    \arrow[hook, from=1-3, to=3-3]
    \arrow[""{name=0, anchor=center, inner sep=0}, "R", from=1-1, to=1-3]
    \arrow[""{name=1, anchor=center, inner sep=0}, "{\Coalg^\rr}", from=3-1, to=3-3]
    \arrow["\iota", shorten <=12pt, shorten >=12pt, Rightarrow, from=0, to=1]
  \end{tikzcd}\]
  where $R$ is the algebra from Proposition \ref{prop:rate-circ-alg}.
  \begin{proof}[Proof sketch]
    $\iota$ is defined by a family of functors $\iota_{(n_o,n_i)}:R(n_o,n_i)\to\Coalg^\rr(\rr^{n_o}y^{\rr^{n_i}})$, where each set $R(n_o,n_i)$ is treated as the corresponding discrete category; this means that $\iota_{(n_o,n_i)}$ is trivially functorial, and needs only be defined on objects (rate-coded neural circuits).
    Each such circuit $(\lambda,\alpha,\beta,\gamma,W)$ defines an `open' ordinary differential equation by Definition \ref{def:rate-circ} with inputs $i:\rr^{n_i}$.
    $\iota_{(n_o,n_i)}$ is then defined by taking this ordinary differential equation to a corresponding open dynamical system following Example \ref{ex:open-vect-field}, where the output space is the same as the state space $\rr^{n_o}$ and the output map is $\id_{\rr^{n_o}}$.

    We then need to check that this definition of $\iota$ is natural, meaning that the following diagram commutes for each linear circuit diagram $(A,B):(n_o,n_i)\to(m_o,m_i)$, where $\Coalg^\rr(A,B)$ is defined by treating $(A,B)$ as a lens and hence a morphism of monomials of the type indicated.
    \[\begin{tikzcd}
      {R(n_o,n_i)} && {R(m_o,m_i)} \\
      \\
      {\Coalg^\rr(\rr^{n_o}y^{\rr^{n_i}})} && {\Coalg^\rr(\rr^{m_o}y^{\rr^{m_i}})}
      \arrow["{R(A,B)}", from=1-1, to=1-3]
      \arrow["{\Coalg^\rr(A,B)}", from=3-1, to=3-3]
      \arrow["{\iota_{(n_o,n_i)}}"', from=1-1, to=3-1]
      \arrow["{\iota_{(m_o,m_i)}}", from=1-3, to=3-3]
    \end{tikzcd}\]
    To see that this diagram commutes, observe that we can write a rate-coded neural circuit $\kappa$ as a morphism $\rr^{n_o}y^{\Tan\rr^{n_o}}\to\rr^{n_o}y^{\rr^{n_i}}$ of polynomials, where $\Tan$ is the tangent bundle functor; and observe that the action of $R(A,B)$ is to post-compose the lens $(A,B)$ after $\kappa$, as in $\rr^{n_o}y^{\Tan\rr^{n_o}}\xto{\kappa}\rr^{n_o}y^{\rr^{n_i}}\xto{(A,B)}\rr^{m_o}y^{\rr^{m_i}}$.
    Now, $\iota_{(n_o,n_i)}$ acts by taking $\kappa$ to a system $\rr^{n_o}y^{\rr^{n_o}}\to[\Tt y, \rr^{n_o}y^{\rr^{n_i}}]$, and $\Coalg^\rr(A,B)$ post-composes $[\Tt y,(A,B)]$, so we obtain the system
    \[ \rr^{n_o}y^{\rr^{n_o}}\xto{\iota_{(n_o,n_i)}(\kappa)}[\Tt y, \rr^{n_o}y^{\rr^{n_i}}]\xto{[\Tt y,(A,B)]}[\Tt y, \rr^{m_o}y^{\rr^{m_i}}] \; . \]
    This is precisely the system obtained by applying $\iota_{(m_o,m_i)}$ to $(A,B)\circ\kappa$, and hence $\iota$ is natural.

    Finally, it is easy to check that $\iota$ is a monoidal natural transformation (Definition \ref{def:mon-nat-trans}), which by \textcite[{Proposition 3.6}]{Moeller2018Monoidal} entails that $\iota$ is a morphism of monoidal indexed categories.
    That it is additionally an inclusion follows from the evident fact that each functor $\iota_{(n_o,n_i)}$ is an embedding.
  \end{proof}
\end{prop}

At some point during the preceding exposition, the reader may have wondered in what sense these open dynamical systems are coalgebras.
To answer this, recall from Proposition \ref{prop:poly-coalg-adj} that a polynomial morphism $Sy^S\to q$ is equivalently a function $S\to qS$ and hence by Example \ref{ex:poly-coalg} a $q$-coalgebra.
Then, by setting $q = [\Tt y, p]$, we see the connection immediately: the objects of $\Coalg^\Tt(p)$ are $[\Tt y, p]$-coalgebras that satisfy the $\lhd$-comonoid condition, and the morphisms of $\Coalg^\Tt(p)$ are coalgebra morphisms.

In the following subsection, we generalize the constructions above to allow for non-deterministic (`effectful') feedback, using a generalization of the category $\Poly$.

\subsection{Polynomials with `effectful' feedback, and open Markov processes}
\label{sec:poly-stoch}

The category $\Poly$ of polynomial functors $\Set\to\Set$ can be considered as a category of `deterministic' polynomial interaction; notably, morphisms of such polynomials, which we take to encode the coupling of systems' interfaces, do not explicitly incorporate any kind of randomness or uncertainty.
Even if the universe is deterministic, however, the finiteness of systems and their general inability to perceive the totality of their environments make it a convenient modelling choice to suppose that systems' interactions may be uncertain; this will be useful not only in allowing for stochastic interactions between systems, but also to define stochastic dynamical systems `internally' to a category of polynomials.

To reach the desired generalization, we begin by recalling that $\Poly$ is equivalent to the category of Grothendieck lenses for the self-indexing of $\Set$ (Example \ref{ex:lens-poly}).
We define our categories of generalized polynomials from this perspective, by considering Kleisli categories indexed by their ``deterministic subcategories''.
This allows us to define categories of Grothendieck lenses which behave like $\Poly$ when restricted to the deterministic case, but also admit uncertain inputs.
In order to apply the Grothendieck construction, we begin by defining an indexed category.

\begin{defn} \label{def:poly-c-idx}
  Suppose $\cat{E}$ is a category with all limits, and suppose $M:\cat{E}\to\cat{E}$ is a monad on $\cat{E}$.
  Define the indexed category $\cat{E}_M/- : \cat{E}\op \to \Cat{Cat}$ as follows.
  On objects $B : \cat{E}$, we define $\cat{E}_M/B$ to be the full subcategory of $\Kl(M)/B$ on those objects $\iota p : E\klto B$ which correspond to maps $E \xto{p} B \xto{\eta_B} MB$ in the image of $\iota$.
  Now suppose $f : C\to B$ is a map in $\cat{E}$.
  We define $\cat{E}_M/f : \cat{E}_M/B \to \cat{E}_M/C$ as follows.
  The functor $\cat{E}_M/f$ takes objects $\iota p : E\klto B$ to $\iota(f^*p) : f^*E\klto C$ where $f^*p$ is the pullback of $p$ along $f$ in $\cat{E}$, included into $\Kl(M)$ by $\iota$.

  To define the action of $\cat{E}_M/f$ on morphisms $\alpha : (E, \iota p:E\klto B) \to (F, \iota q:F\klto B)$, note that since we must have $\iota q\klcirc \alpha = \iota p$, we can alternatively write $\alpha$ as the $B$-dependent sum $\sum_{b:B}\alpha_b : \sum_{b:B}p[b] \to \sum_{b:B} Mq[b]$.
  Then we can define $(\cat{E}_M/f)(\alpha)$ accordingly as $(\cat{E}_M/f)(\alpha) := \sum_{c:C}\alpha_{f(c)} : \sum_{c:C}p[f(c)] \to \sum_{c:C}Mq[f(c)]$.
\end{defn}

\begin{defn} \label{def:poly-c}
  We define $\Poly_{M}$ to be the category of Grothendieck lenses for $\cat{E}_M/-$.
  That is, $\Poly_{M} := \int \cat{E}_M/-\op$, where the opposite is again taken pointwise.
\end{defn}

\begin{ex}
  When $\cat{E} = \Set$ and $M = \id_{\Set}$, Definition \ref{def:poly-c-idx} recovers our earlier definition of $\Poly$.
\end{ex}

\begin{ex}
  When $M$ is a monad on $\Set$, we find that the objects of $\Poly_M$ are the same polynomial functors as constitute the objects of $\Poly$.
  The morphisms $f : p \to q$ are pairs $(f_1, f^\sharp)$, where $f_1 : B\to C$ is a function in $\Set$ and $f^\sharp$ is a family of morphisms $q[f_1(x)]\klto p[x]$ in $\Kl(M)$, making the following diagram commute:
  \[\begin{tikzcd}
    \sum_{x:B} Mp[x] & \sum_{b:B} q[f_1(x)] & \sum_{y:C} q[y] \\
    B & B & C
    \arrow["{f^\sharp}"', from=1-2, to=1-1]
    \arrow[from=1-2, to=1-3]
    \arrow["q", from=1-3, to=2-3]
    \arrow["{\eta_B}^* p"', from=1-1, to=2-1]
    \arrow[from=2-1, to=2-2, Rightarrow, no head]
    \arrow["{f_1}", from=2-2, to=2-3]
    \arrow[from=1-2, to=2-2]
    \arrow["\lrcorner"{anchor=center, pos=0.125}, draw=none, from=1-2, to=2-3]
  \end{tikzcd}\]
\end{ex}

\begin{rmk}
  Consequently, we can think of $\Poly_M$ as a dependent version of the category of $M$-monadic lenses, in the sense of \textcite[{\S 3.1.3}]{Clarke2020Profunctor}.
\end{rmk}

\begin{rmk}
  Any monad $(M,\mu,\eta)$ on $\Set$ induces a comonad $(\bar{M},\delta,\epsilon)$ on the category $\Poly$ of polynomial functors $\Set\to\Set$, and $\Poly_M$ can be recovered as the coKleisli category of this comonad, $\Poly_M \cong \coKl(\bar{M})$.
  We heard of this idea from David Spivak.

  On objects (polynomial functors), $\bar{M}:\Poly\to\Poly$ acts to map $p:\sum_{i:p(1)}p[i]\to p(1)$ to $\bar{M}p:\sum_{i:p(1)}Mp[i]\to p(1)$.
  Given a morphism of polynomials $\varphi:p\to q$, $\bar{M}$ returns the morphism $\bar{M}(\varphi)$ whose forward component is again $\varphi_1$ and whose backward component is defined by $\bar{M}(\varphi)^\sharp_i := M(\varphi^\sharp_i)$ for each $i$ in $p(1)$.

  The counit $\epsilon:\bar{M}\Rightarrow\id_{\Poly}$ is defined for each $p$ as the morphism $\epsilon_p:\bar{M}p\to p$ whose forward component is $\id_{p(1)}$ and whose backward component is given for each $i:p(1)$ by the unit $\eta_{p[i]}:p[i]\to Mp[i]$ of the monad $M$.
  Similarly, the comultiplication $\delta:\bar{M}\bar{M}\Rightarrow\bar{M}$ is defined for each $p$ as the morphism $\delta_p:\bar{M}\bar{M}p\to\bar{M}p$ whose forward component is again the identity and whose backward components are given by the multiplication of the monad $\mu$, \textit{i.e.} $(\delta_p^\sharp)_i := \mu_{p[i]}$.

  Finally, the coKleisli category $\coKl(\bar{M})$ has the same objects as $\Poly$.
  A morphism $p\to q$ in $\coKl(\bar{M})$ is a morphism $\bar{M}p\to q$ in $\Poly$.
  Composition in $\coKl(\bar{M})$ is the usual composition in the forward direction and Kleisli composition in the backward direction.
\end{rmk}

\begin{rmk}
  Since $\cat{E}$ is assumed to have all limits, it must have a product structure $(\times,1)$.
  When $M$ is additionally a monoidal monad (Definition \ref{def:monoidal-monad}), then $\Poly_M$ acquires a tensor akin to that defined for $\Poly$ in Proposition \ref{prop:poly-tensor}, and which we also denote by $(\otimes, I)$:
  the definition only differs by substituting the structure $(\otimes, I)$ on $\Kl(M)$ for the product $(\times, 1)$ on $\Set$.
  This monoidal structure follows as before from the monoidal Grothendieck construction: $\cat{E}_M/-$ is lax monoidal, with laxator taking $p : \cat{E}_M/B$ and $q : \cat{E}_M/C$ to $p\otimes q : \cat{E}_M/(B\otimes C)$.

  On the other hand, for $\Poly_{M}$ also to have an internal hom $[q,r]$ requires each fibre of $\cat{E}_M/-$ to be closed with respect to the monoidal structure.
  In cases of particular interest, $\cat{E}$ will be locally Cartesian closed, and restricting $\cat{E}_M/-$ to the self-indexing $\cat{E}/-$ gives fibres which are thus Cartesian closed.
  In these cases, we can think of the broader fibres of $\cat{E}_M/-$, and thus $\Poly_{M}$ itself, as being `deterministically' closed.
  This means, for the stochastic example $\Poly_\Pa$ for $\Pa$ a probability monad, we get an internal hom satisfying the adjunction $\Poly_\Pa(p\otimes q, r) \cong \Poly_\Pa(p, [q,r])$ only when the backwards components of morphisms $p\otimes q \to r$ are `uncorrelated' between $p$ and $q$.
\end{rmk}

\begin{rmk} \label{rmk:poly-m-pi-types}
  For $\Poly_{M}$ to behave faithfully like the category $\Poly$ of polynomial functors of sets and their morphisms, we should want the substitution functors $\cat{E}_M/f : \cat{E}_M/C \to \cat{E}_M/B$ to have left and right adjoints (corresponding respectively to dependent sum and product).
  Although we do not spell it out here, it is quite straightforward to exhibit concretely the left adjoints.
  On the other hand, writing $f^*$ as shorthand for $\cat{E}_M/f$, we can see that a right adjoint only obtains in restricted circumstances.
  Denote the putative right adjoint by $\Pi_f : \cat{E}_M/B \to \cat{E}_M/C$, and for $\iota p : E\klto B$ suppose that $(\Pi_f E)[y]$ is given by the set of `partial sections' $\sigma : f^{-1}\{y\} \to ME$ of $p$ over $f^{-1}\{y\}$ as in the commutative diagram:
  \[\begin{tikzcd}
	& {f^{-1}\{y\}} & {\{y\}} \\
	ME & B & C
	\arrow[from=1-2, to=1-3]
	\arrow[from=1-3, to=2-3]
	\arrow[from=1-2, to=2-2]
	\arrow["f", from=2-2, to=2-3]
	\arrow["\lrcorner"{anchor=center, pos=0.125}, draw=none, from=1-2, to=2-3]
	\arrow["{{\eta_B}^*p}", from=2-1, to=2-2]
	\arrow["\sigma"', curve={height=12pt}, from=1-2, to=2-1]
  \end{tikzcd}\]
  Then we would need to exhibit a natural isomorphism $\cat{E}_M/B(f^*D, E) \cong \cat{E}_M/C(D, \Pi_f E)$.
  But this will only obtain when the `backwards' components $h^\sharp_y : D[y]\to M(\Pi_f E)[y]$ are in the image of $\iota$---otherwise, it is not generally possible to pull $f^{-1}\{y\}$ out of $M$.
\end{rmk}

Despite these restrictions, we do have enough structure at hand to instantiate $\Coalg^\Tt$ in $\Poly_{M}$.
The only piece remaining is the composition product $\lhd$, but for our purposes it suffices to define its action on objects, which is identical to its action on objects in $\Poly$\footnote{%
We leave the full exposition of $\lhd$ in $\Poly_{M}$ to future work.},
and then to consider $\lhd$-comonoids in $\Poly_{M}$.
The comonoid laws force the structure maps to be deterministic (\textit{i.e.}, in the image of $\iota$), and so $\lhd$-comonoids in $\Poly_{M}$ are just $\lhd$-comonoids in $\Poly_{\id_{\Set}}$.

Finally, we note that, even if the internal hom $[-,-]$ is not available in general, we can define morphisms $\beta : Sy^S \to [\Tt y, p]$:
these again just correspond to morphisms $\Tt y \otimes Sy^S \to p$, and the condition that the backwards maps be uncorrelated between $\Tt y$ and $p$ is incontrovertibly satisfied because $\Tt y$ has a trivial exponent.
Unwinding such a $\beta$ according to the definition of $\Poly_{M}$ indeed gives precisely a pair $(\beta^o, \beta^u)$ of the requisite types; and a comonoid homomorphism $Sy^S\to y^\Tt$ in $\Poly_{M}$ is precisely a functor $\deloop{\Tt} \to \Kl(M)$, thereby establishing equivalence between the objects of $\Coalg^{\Tt}(p)$ established in $\Poly_{M}$ and the objects of $\PMCoalgT{\Tt}$.

Henceforth, therefore, we will write $\Coalg^\Tt_M$ to denote the instantiation of $\Coalg^\Tt$ in $\Poly_M$.
We will call the objects of $\Coalg^\Tt_M(p)$ \(pM\)\textit{-coalgebras with time} \(\Tt\), and to get a sense of how, in the case where $M$ is a probability monad, they provide a notion of open Markov process, we can read off the definition a little more explictly.

\begin{prop} \label{prop:pM-coalg}
  A \(pM\)\textit{-coalgebra with time} \(\Tt\) consists of a triple \(\vartheta := (S,\vartheta^o,\vartheta^u)\) of a \textnormal{state space} \(S : \cat{E}\) and two morphisms \(\vartheta^o : \Tt \times S \to p(1)\) and \(\vartheta^u : \sum_{t:\Tt}\sum_{s:\mathbb{S}} p[\vartheta^o(t,s)] \to MS\), such that, for any section \(\sigma : p(1) \to \sum_{i:p(1)} p[i]\) of \(p\) in $\cat{E}$, the maps \(\vartheta^\sigma : \Tt \times S \to MS\) given by
  \[ \Sum_{t:\Tt} S \xto{\vartheta^o(t)^\ast \sigma} \Sum_{t:\Tt} \Sum_{s:S} p[\vartheta^o(t, s)] \xto{\vartheta^u} MS \]
  constitute an object in the functor category \(\Cat{Cat}\big(\deloop{\Tt},\Kl(M)\big)\), where \(\deloop{\Tt}\) is the delooping of \(\Tt\) and \(\Kl(M)\) is the Kleisli category of \(M\). (Once more, we call the closed system \(\vartheta^\sigma\), induced by a section \(\sigma\) of \(p\), the \textit{closure} of \(\vartheta\) by \(\sigma\).)
\end{prop}

Following Example \ref{ex:mrkv-proc} and the intuition of Example \ref{ex:open-vect-field-2}, we can see how this produces an open version of a Markov process.

Since stochastic dynamical systems are often alternatively presented as random dynamical systems, we now briefly consider how these can be incorporated into the coalgebraic framework.

\subsection{Open random dynamical systems} \label{sec:open-rds}

In the analysis of stochastic systems, it is often fruitful to consider two perspectives: on one side, one considers explicitly the evolution of the distribution of the states of the system, by following (for instance) a Markov process, or Fokker-Planck equation.
On the other side, one considers the system as if it were a deterministic system, perturbed by noisy inputs, giving rise to the frameworks of stochastic differential equations and associated \textit{random dynamical systems}.

Whereas a (closed) Markov process is typically given by the action of a time monoid on an object in a Kleisli category of a probability monad, a (closed) random dynamical system is given by a bundle of closed dynamical systems, where the base system is equipped with a probability measure which it preserves:
the idea being that a random dynamical system can be thought of as a `random' choice of dynamical system on the total space at each moment in time, with the base measure-preserving system being the source of the randomness \parencite{Arnold1998Random}.

This idea corresponds in non-dynamical settings to the notion of \textit{randomness pushback} \parencite[Def. 11.19]{Fritz2019synthetic}, by which a stochastic map \(f : A\to\Pa B\) can be presented as a deterministic map \(f^\flat : \Omega\times A\to B\) where \((\Omega,\omega)\) is a measure space such that, for any \(a:A\), pushing \(\omega\) forward through \(f^\flat(\mdash,a)\) gives the state \(f(a)\); that is, \(\omega\) induces a random choice of map \(f^\flat(\omega,\mdash) : A\to B\).
Similarly, under nice conditions, random dynamical systems and Markov processes do coincide, although they have different suitability in applications.

In this section, we sketch how the generalized-coalgebraic structures developed above extend also to random dynamical systems. %
We begin by defining the concept of measure-preserving dynamical system, which itself requires the notion of measure space (in order that measure can be preserved); we define the corresponding category abstractly, using a notion of slice category dual to that of Definition \ref{def:slice-cat}.

\begin{defn} \label{def:slice-under}
  Suppose $X$ is an object of a category $\cat{E}$.
  The slice of $\cat{E}$ \textit{under} $X$, denoted $X/\cat{E}$, is the category whose objects $(A,i)$ are morphisms $X\xto{i} A$ out of $X$ in $\cat{E}$, and whose morphisms $f:(A,i)\to(B,j)$ are the evident triangles:
  \[\begin{tikzcd}
    & X \\
    A && B
	  \arrow["f", from=2-1, to=2-3]
	  \arrow["i"', from=1-2, to=2-1]
	  \arrow["j", from=1-2, to=2-3]
  \end{tikzcd}\]
  There is a projection functor $F:X/\cat{E}\to\cat{E}$ mapping each object $(A,i)$ to $A$ and each morphism $f:(A,i)\to(B,j)$ to $f:A\to B$.
\end{defn}

We can use this notion to define a notion of `pointed' category.

\begin{defn} \label{def:point-spc}
  Let $(\cat{C},\otimes,I)$ be a monoidal category, $\cat{D}$ be a subcategory $\cat{D}\hookrightarrow\cat{C}$, and let $F$ denote the projection $I/\cat{C}\to\cat{C}$.
  We define the category $\cat{D}_*$ to be the pullback category over the diagram $\cat{D}\hookrightarrow\cat{C}\xfrom{F} I/\cat{C}$.
\end{defn}

The category $\cat{D}_*$ has objects `pointed' by corresponding states in $\cat{C}$, and its morphisms are those that preserve these states.
The category of measure spaces is obtained as an example accordingly.

\begin{ex} \label{ex:meas-spc}
  Consider the deterministic subcategory $\Cat{Meas}\hookrightarrow\Cat{sfKrn}$.
  The pointed category $\Cat{Meas}_*$ obtained from Definition \ref{def:point-spc} is the category whose objects are \textit{measure spaces} $(M,\mu)$ with $\mu$ a measure $1\klto M$, and whose morphisms $f:(M,\mu)\to(N,\nu)$ are measure-preserving maps; \textit{i.e.}, measurable functions $f:M\to N$ such that $\nu = f\klcirc\mu$ in $\Cat{sfKrn}$.
  Likewise, if $\Pa$ is a probability monad on $\cat{E}$, then we have $\cat{E}\hookrightarrow\Kl(\Pa)$ and hence can understand $\cat{E}_*$ as a category of abstract measure spaces.
\end{ex}

\begin{prop}
  There is a projection functor \(U:\cat{E}_* \to \cat{E}\) taking measure spaces \((B, \beta)\) to the underlying spaces \(B\) and their morphisms \(f : (A, \alpha) \to (B, \beta)\) to the underlying maps \(f : A \to \Pa B\).
  We will write \(B\) to refer to the space in \(\cat{E}\) underlying a measure space \((B, \beta)\), in the image of $U$.
  \begin{proof}
    The functor is obtained as the projection induced by the universal property of the pullback.
  \end{proof}
\end{prop}

\begin{defn} \label{def:metric-sys}
  Let \((B, \beta)\) be a measure space in \(\cat{E}\hookrightarrow\Kl(\Pa)\).
  A closed \textit{metric} or \textit{measure-preserving} dynamical system \((\vartheta,\beta)\) on \((B, \beta)\) with time \(\Tt\) is a closed dynamical system \(\vartheta\) with state space \(B : \cat{E}\) such that, for all \(t : \Tt\), \(\Pa \vartheta(t) \circ \beta = \beta\); that is, each \(\vartheta(t)\) is a \((B, \beta)\)-endomorphism in \(1/\Kl(\Pa)\).
\end{defn}

\begin{prop} \label{prop:metric-sys}
  Closed measure-preserving dynamical systems in \(\cat{E}\) with time \(\Tt\) form the objects of a category \(\Cat{Cat}(\deloop{\Tt}, \cat{E}_*)\) whose morphisms \(f : (\vartheta, \alpha) \to (\psi, \beta)\) are maps \(f : \vartheta(\ast) \to \psi(\ast)\) in \(\cat{E}\) between the state spaces that preserve both flow and measure, as in the following commutative diagram, which also indicates their composition:
  \[\begin{tikzcd}%
    && {\Pa\vartheta(\ast)} && {\Pa\vartheta(\ast)} \\
    \\
    1 && {\Pa\psi(\ast)} && {\Pa\psi(\ast)} && 1 \\
    \\
    && {\Pa\lambda(\ast)} && {\Pa\lambda(\ast)}
    \arrow["\alpha", from=3-1, to=1-3]
    \arrow["\beta"', from=3-1, to=3-3]
    \arrow["\gamma"', from=3-1, to=5-3]
    \arrow["\alpha"', from=3-7, to=1-5]
    \arrow["\beta", from=3-7, to=3-5]
    \arrow["\gamma", from=3-7, to=5-5]
    \arrow["{\Pa\vartheta(t)}", from=1-3, to=1-5]
    \arrow["{\Pa\psi(t)}"', from=3-3, to=3-5]
    \arrow["{\Pa\lambda(t)}"', from=5-3, to=5-5]
    \arrow["{\Pa f}", from=1-3, to=3-3]
    \arrow["{\Pa f}"', from=1-5, to=3-5]
    \arrow["{\Pa g}", from=3-3, to=5-3]
    \arrow["{\Pa g}"', from=3-5, to=5-5]
  \end{tikzcd}\]
  \begin{proof}
    The identity morphism on a closed measure-preserving dynamical system is the identity map on its state space.
    It is easy to check that composition as in the diagram above is thus both associative and unital with respect to these identities.
  \end{proof}
\end{prop}

As we indicated in the introduction to this section, closed random dynamical systems are bundles of deterministic systems over metric systems.

\begin{defn}
  Let \((\vartheta, \beta)\) be a closed measure-preserving dynamical system.
  A closed \textit{random dynamical system} over \((\vartheta, \beta)\) is an object of the slice category \(\Cat{Cat}(\deloop{\Tt}, \cat{E})/\vartheta\); it is therefore a bundle of the corresponding functors.
\end{defn}

\begin{ex} \label{ex:brown-sde}
  The solutions \(X(t, \omega; x_0) : \rr_+ \times \Omega \times M \to M\) to a stochastic differential equation \(\d X_t = f(t, X_t) \d t + \sigma(t, X_t) \d W_t\), where \(W : \rr_+ \times \Omega \to M\) is a Wiener process in \(M\), define a random dynamical system \(\rr_+ \times \Omega \times M \to M : (t, \omega, x) \mapsto X(t, \omega; x_0)\) over the Wiener base flow \(\theta : \rr_+ \times \Omega \to \Omega : (t, \omega) \mapsto W(s+t, \omega) - W(t, \omega)\) for any \(s : \rr_+\).
\end{ex}

We can use the same trick, of opening up closed systems along a polynomial interface, to define a notion of open random dynamical system --- although at this point we do not have an elegant concise definition.

\begin{defn} \label{def:poly-rdyn}
  Let \((\theta, \beta)\) be a closed measure-preserving dynamical system in \(\cat{E}\) with time \(\Tt\), and let \(p : \Poly_{\id_{\cat{E}}}\) be a polynomial in \(\cat{E}\).
  Write \(\Omega := \theta(\ast)\) for the state space of \(\theta\), and let \(\pi : S \to \Omega\) be an object (bundle) in \(\cat{E}/\Omega\).
  An \textit{open random dynamical system over} \((\theta,\beta)\) \textit{on the interface} \(p\) \textit{with state space} \(\pi:S \to \Omega\) \textit{and time} \(\Tt\) consists in a pair of morphisms \(\vartheta^o : \Tt \times S \to p(1)\) and \(\vartheta^u : \Sum_{t:\Tt} \Sum_{s:S} p[\vartheta^o(t, s)] \to S\), such that, for any section \(\sigma : p(1) \to \Sum_{i:p(1)} p[i]\) of \(p\), the maps \(\vartheta^\sigma : \Tt \times S \to S\) defined as
  \[ \Sum_{t:\Tt} S \xto{\vartheta^o(-)^\ast \sigma} \Sum_{t:\Tt} \Sum_{s:S} p[\vartheta^o(-, s)] \xto{\vartheta^u} S \]
  form a closed random dynamical system in \(\Cat{Cat}(\deloop{\Tt}, \cat{E})/\theta\), in the sense that, for all \(t : \Tt\) and sections \(\sigma\), the following diagram commutes:
  \[\begin{tikzcd}
    S && {\Sum_{s:S} p[\vartheta^o(t, s)]} && S \\
    \Omega &&&& \Omega
    \arrow["\pi"', from=1-1, to=2-1]
    \arrow["\pi", from=1-5, to=2-5]
    \arrow["{\theta(t)}"', from=2-1, to=2-5]
    \arrow["{\vartheta^o(t)^\ast \sigma}", from=1-1, to=1-3]
    \arrow["{\vartheta^u(t)}", from=1-3, to=1-5]
  \end{tikzcd}\]
\end{defn}

\begin{prop} \label{prop:poly-rdyn}
  Let \((\theta, \beta)\) be a closed measure-preserving dynamical system in \(\cat{E}\) with time \(\Tt\), and let \(p : \Poly_{\id_{\cat{E}}}\) be a polynomial in \(\cat{E}\).
  Open random dynamical systems over \((\theta, \beta)\) on the interface \(p\) form the objects of a category \(\RDynT{T}(p, \theta)\).
  Writing \(\vartheta := (\pi_X, \vartheta^o, \vartheta^u)\) and \(\psi := (\pi_Y, \psi^o, \psi^u)\), a morphism \(f : \vartheta \to \psi\) is a map \(f: X \to Y\) in \(\cat{E}\) making the following diagram commute for all times \(t : \Tt\) and sections \(\sigma\) of \(p\):
  \[\begin{tikzcd}
	X &&&& {\Sum_{x:X} p[\vartheta^o(t, x)]} &&&& X \\
	\\
	&& \Omega &&&& \Omega \\
	\\
	Y &&&& {\Sum_{y:Y} p[\psi^o(t, y)]} &&&& Y
	\arrow["{\pi_X}"', from=1-1, to=3-3]
	\arrow["{\pi_X}", from=1-9, to=3-7]
	\arrow["{\theta(t)}", from=3-3, to=3-7]
	\arrow["{\vartheta^o(t)^\ast \sigma}", from=1-1, to=1-5]
	\arrow["{\vartheta^u(t)}", from=1-5, to=1-9]
	\arrow["{\psi^o(t)^\ast \sigma}"', from=5-1, to=5-5]
	\arrow["{\psi^u(t)}"', from=5-5, to=5-9]
	\arrow["{\pi_Y}", from=5-1, to=3-3]
	\arrow["{\pi_Y}"', from=5-9, to=3-7]
	\arrow["f"', from=1-1, to=5-1]
	\arrow["f", from=1-9, to=5-9]
  \end{tikzcd}\]
  Identities are given by the identity maps on state-spaces.
  Composition is given by pasting of diagrams.
\end{prop}

\begin{prop} \label{prop:poly-rdyn-idx}
  The categories \(\RDynT{T}(p, \theta)\) collect into a doubly-indexed category of the form \(\RDynT{T} : \Poly_{\id_{\cat{E}}} \times \Cat{Cat}(\deloop{\Tt}, \cat{E}_*) \to \Cat{Cat}\).
  By the universal property of the product \(\times\) in \(\Cat{Cat}\), it
  suffices to define the actions of \(\RDynT{T}\) separately on morphisms of
  polynomials and on morphisms of closed measure-preserving systems.

  Suppose therefore that \(\varphi : p \to q\) is a morphism of polynomials.
  Then, for each measure-preserving system \((\theta, \beta) : \Cat{Cat}(\deloop{\Tt}, \cat{E}_*)\), we define the functor \(\RDynT{T}(\varphi, \theta) : \RDynT{T}(p, \theta) \to \RDynT{T}(q, \theta)\) as follows.
  Let \(\vartheta := (\pi_X : X \to \Omega, \vartheta^o,\vartheta^u) : \RDynT{T}(p, \theta)\) be an object (open random dynamical system) in \(\RDynT{T}(p, \theta)\).
  Then \(\RDynT{T}(\varphi,\theta)(\vartheta)\) is defined as the triple \((\pi_X, \varphi_1 \circ \vartheta^o, \vartheta^u \circ {\varphi^o}^\ast \varphi^\sharp) : \RDynT{T}(q, \theta)\), where the two maps are explicitly the following composites:
  \begin{gather*}
    \Tt \times X \xto{\vartheta^o} p(1) \xto{\varphi_1} q(1) \, ,
    \qquad
    \Sum_{t:\Tt} \Sum_{x:X} q[\varphi_1 \circ \vartheta^o(t, x)] \xto{{\vartheta^o}^\ast \varphi^\sharp} \Sum_{t:\Tt} \Sum_{x:X} p[\vartheta^o(t, x)] \xto{\vartheta^u} X \, .
  \end{gather*}
  On morphisms \(f : (\pi_X : X \to \Omega, \vartheta^o, \vartheta^u) \to (\pi_Y : Y \to \Omega, \psi^o, \psi^u)\), the image \(\RDynT{T}(\varphi, \theta)(f) : \RDynT{T}(\varphi, \theta)(\pi_X, \vartheta^o, \vartheta^u) \to \RDynT{T}(\varphi, \theta)(\pi_Y, \psi^o, \psi^u)\) is given by the same underlying map \(f : X \to Y\) of state spaces.

  Next, suppose that \(\phi : (\theta, \beta) \to (\theta', \beta')\) is a morphism of closed measure-preserving dynamical systems, and let \(\Omega' := \theta'(\ast)\) be the state space of the system \(\theta'\).
  By Proposition \ref{prop:metric-sys}, the morphism \(\phi\) corresponds to a map \(\phi : \Omega \to \Omega'\) on the state spaces that preserves both flow and measure.
  Therefore, for each polynomial \(p : \Poly_{\id_{\cat{E}}}\), we define the functor \(\RDynT{T}(p, \phi) : \RDynT{T}(p, \theta) \to \RDynT{T}(p,\theta')\) by post-composition.
  That is, suppose given open random dynamical systems and morphisms over \((p, \theta)\) as in the diagram of Proposition \ref{prop:poly-rdyn}.
  Then \(\RDynT{T}(p, \phi)\) returns the following diagram:
  \[\begin{tikzcd}
	X &&&& {\Sum_{x:X} p[\vartheta^o(t, x)]} &&&& X \\
	\\
	&& {\Omega'} &&&& {\Omega'} \\
	\\
	Y &&&& {\Sum_{y:Y} p[\psi^o(t, y)]} &&&& Y
	\arrow["{\theta'(t)}"', from=3-3, to=3-7]
	\arrow["{\vartheta^o(t)^\ast \sigma}", from=1-1, to=1-5]
	\arrow["{\vartheta^u(t)}", from=1-5, to=1-9]
	\arrow["{\psi^o(t)^\ast \sigma}"', from=5-1, to=5-5]
	\arrow["{\psi^u(t)}"', from=5-5, to=5-9]
	\arrow["f"', from=1-1, to=5-1]
	\arrow["f", from=1-9, to=5-9]
	\arrow["{\phi\circ\pi_Y}"', from=5-9, to=3-7]
	\arrow["{\phi\circ\pi_X}", from=1-9, to=3-7]
	\arrow["{\phi\circ\pi_Y}", from=5-1, to=3-3]
	\arrow["{\phi\circ\pi_X}"', from=1-1, to=3-3]
  \end{tikzcd}\]
  That is, \(\RDynT{T}(p, \phi)(\vartheta) := (\phi\circ\pi_X, \vartheta^o, \vartheta^u)\) and \(\RDynT{T}(p, \phi)(f)\) is given by the same underlying map \(f : X \to Y\) on state spaces.
\end{prop}

%
%
%

%
%

\section{Cilia: monoidal bicategories of cybernetic systems} \label{sec:mon-bicat-coalg}

Whereas it is the morphisms (1-cells) of process-theoretic categories---such as categories of lenses, or the categories of statistical games to be defined in Chapter \ref{chp:brain}---that represent open systems, it is the objects (0-cells) of the opindexed categories \(\Coalg^\Tt_{M}\)\footnote{or, more precisely, their corresponding opfibrations $\int\Coalg^\Tt_{M}$} that play this rôle; in fact, the objects of $\Coalg^\Tt_{M}$ each represent both an open system and its (polynomial) interface.
In order to supply dynamical semantics for statistical games---functors from categories of statistical games to categories of dynamical systems---we need to cleave the dynamical systems from their interfaces, making the interfaces into 0-cells and systems into 1-cells between them, thereby letting the systems' types and composition match those of the games.
Doing this is the job of this section, which we first perform in the case of the general categories $\Coalg^\Tt_M$, followed by the specific case of systems generated differentially, as in the vector-field Examples \ref{ex:open-vect-field} and \ref{ex:open-vect-field-2}.

\subsection{Hierarchical bidirectional dynamical systems}

To construct ``hierarchical bidirectional systems'', we will associate to each pair of objects $(A,S)$ and $(B,T)$ of a category of (for our purposes, Bayesian) lenses a polynomial $\blhom{Ay^S,By^T}$ whose configurations correspond to lenses and whose inputs correspond to the lenses' inputs.
The categories $\Coalg^\Tt_\Pa\bigl(\blhom{Ay^S,By^T}\bigr)$ will then form the hom-categories of bicategories of hierarchical inference systems called \textit{cilia}\footnote{%
`Cilia', because they ``control optics'', like the ciliary muscles of the eye.},
and it is in these bicategories that we will find our dynamical semantics.

Throughout this subsection, we will fix a category $\cat{C}$ of stochastic channels, defined by $\cat{C}:=\Kl(\Pa)$ as the Kleisli category of a probability monad $\Pa:\cat{E}\to\cat{E}$, which we will also take to define a category $\Poly_{\Pa}$ of polynomials with stochastic feedback.
We will assume $\Pa$ to be a monoidal monad, and we will write the monoidal structure on $\cat{C}$ as $(\otimes,I)$.
Finally, we will assume that $\cat{C}$ is enriched in its underlying category of spaces $\cat{E}$.

\begin{defn} \label{def:blhom}
  Let $\BLens{}$ be the category of Bayesian lenses in $\cat{C}$.
  Then for any pair of objects $(A,S)$ and $(B,T)$ in $\BLens{}$, we define a polynomial $\blhom{Ay^S,By^T}$ in $\Poly_{\Pa}$ by
  \[
  \blhom{Ay^S,By^T} := \sum_{l:\BLens{}\bigl((A,S),(B,T)\bigr)} y^{\cat{C}(I,A)\times T} \, .
  \]
\end{defn}

\begin{rmk} \label{rmk:ext-hom}
  We can think of $\blhom{Ay^S,By^T}$ as an `external hom' polynomial for $\BLens{}$, playing a rôle analogous to the internal hom $[p,q]$ in $\Poly_{\Pa}$.
  Its `bifunctorial' structure---with domain and codomain parts---is what enables cleaving systems from their interfaces, which are given by these parts.
  The definition, and the following construction of the monoidal bicategory, are inspired by the operad $\Cat{Org}$ introduced by \textcite{Spivak2021Learners}.
\end{rmk}

\begin{rmk} %
  Note that $\blhom{Ay^S,By^T}$ is strictly speaking a monomial, since it can be written in the form $Iy^J$ for $I = \BLens{}\bigl((A,S),(B,T)\bigr)$ and $J = \cat{C}(I,A)\times T$.
  However, we have written it in polynomial form with the view to extending it in future work to dependent lenses and dependent optics \parencite{Vertechi2022Dependent,Braithwaite2021Fibre} and these generalized external homs will in fact be true polynomials.
\end{rmk}

\begin{prop} \label{prop:blhom-funct}
  Definition \ref{def:blhom} defines a functor $\BLens{}\op\times\BLens{}\to\Poly_{\Pa}$. Suppose $c:=(c_1,c^\sharp):(Z,R)\lensto(A,S)$ and $d:=(d_1,d^\sharp):(B,T)\lensto(C,U)$ are Bayesian lenses.
  We obtain a morphism of polynomials $\blhom{c,d}:\blhom{Ay^S,By^T}\to\blhom{Zy^R,Cy^U}$ as follows.
  Since the configurations of $\blhom{Ay^S,By^T}$ are lenses $(A,S)\lensto(B,T)$, the forwards map acts by pre- and post-composition:
  \begin{align*}
    \blhom{c,d}_1 := d\lenscirc(-)\lenscirc c : \BLens{}\bigl((A,S),(B,T)\bigr) & \to \BLens{}\bigl((Z,R),(C,U)\bigr) \\
    l & \mapsto d\lenscirc l\lenscirc c
  \end{align*}

  For each such $l$, the backwards map $\blhom{c,d}^\sharp_l$ has type $\cat{C}(I,Z)\otimes U \to \cat{C}(I,A)\otimes T$ in $\cat{C}$, and is obtained by analogy with the backwards composition rule for Bayesian lenses.
  We define
  \begin{align*}
    \blhom{c,d}^\sharp_l &:= \cat{C}(I,Z)\otimes U \xto{{c_1}_*\otimes U} \cat{C}(I,A)\otimes U \xto{\bcopier\otimes U} \cat{C}(I,A)\otimes\cat{C}(I,A)\otimes U \cdots \\
    & \qquad \cdots \xto{\cat{C}(I,A)\otimes {l_1}_*\otimes U} \cat{C}(I,A)\otimes\cat{C}(I,B)\otimes U\xto{\cat{C}(I,A)\otimes d^\sharp\otimes U} \cat{C}(I,A)\otimes\cat{C}(U,T)\otimes U \cdots \\
    & \qquad \cdots \xto{\cat{C}(I,A)\otimes\Fun{ev}_{U,T}} \cat{C}(I,A)\otimes T
  \end{align*}
  where $l_1$ is the forwards part of the lens $l:(A,S)\lensto(B,T)$, and ${c_1}_* := \cat{C}(I,c_1)$ and ${l_1}_* := \cat{C}(I,l_1)$ are the push-forwards along $c_1$ and $l_1$, and $\Fun{ev}_{U,T}$ is the evaluation map induced by the enrichment of $\cat{C}$ in $\cat{E}$.

  Less abstractly, with $\cat{C} = \Kl(\Pa)$, we can write $\blhom{c,d}^\sharp_l$ as the following map in $\cat{E}$, depicted as a string diagram:
  \[
  \blhom{c,d}^\sharp_l \quad = \quad \input{img/blhom-funct-bkwd.tikz}
  \]
  Here, we have assumed that $\Kl(\Pa)(I,A) = \Pa A$, and define $d^\flat : \Pa B\times U\to\Pa T$ to be the image of $d^\sharp : \Pa B\to\Kl(\Pa)(U,T)$ under the Cartesian closure of $\cat{E}$, and $\mathsf{str}:\Pa A\times\Pa T\to\Pa\bigl(\Pa A\times T)$ the (left) strength of the monoidal monad $\Pa$.

  The morphism $\blhom{c,d}_l$ acts to `wrap' the lens $l$ by pre-composing with $c$ and post-composing with $d$.
  The backwards component  $\blhom{c,d}_l^\sharp$ therefore acts to take the inputs of the resulting composite $d\lenscirc l\lenscirc c$ to appropriate inputs for $l$; that is, it maps a pair $(\pi,u)$ to $(c_1\klcirc\pi, d^\sharp_{l_1\klcirc c_1\klcirc\pi}(u))$.
  \begin{proof}
    We need to check that the mappings defined above respect identities and composition.
    It is easy to see that the definition preserves identities: in the forwards direction, this follows from the unitality of composition in $\BLens{}$; in the backwards direction, because pushing forwards along the identity is again the identity, and because the backwards component of the identity Bayesian lens is the constant state-dependent morphism on the identity in $\cat{C}$.

    To check that the mapping preserves composition, we consider the contravariant and covariant parts separately. Suppose $b:=(b_1,b^\sharp):(Y,Q)\lensto(Z,R)$ and $e:=(e_1,e^\sharp):(C,U)\lensto(D,V)$ are Bayesian lenses.
    We consider the contravariant case first: we check that $\blhom{c\lenscirc b,By^T} = \blhom{b,By^T}\circ\blhom{c,By^T}$.
    The forwards direction holds by pre-composition of lenses.
    In the backwards direction, we note from the definition that only the forwards channel $c_1$ plays a rôle in $\blhom{c,By^T}^\sharp_l$, and that rôle is again pre-composition.
    We therefore only need to check that $(c_1\klcirc b_1)_* = {c_1}_*\circ{b_1}_*$, and this follows immediately from the functoriality of $\cat{C}(I,-)$.

    We now consider the covariant case, that $\blhom{Ay^S,e\lenscirc d} = \blhom{Ay^S,e}\circ\blhom{Ay^S,d}$.
    Once again, the forwards direction holds by composition of lenses.
    For simplicity of exposition, we consider the backwards direction (with $\cat{C} = \Kl(\Pa)$) and reason graphically.
    In this case, the backwards map on the right-hand side is given, for a lens $l:(A,S)\lensto(B,T)$ by the following string diagram:
    \[
    \input{img/blhom-funct-bkwd-proof.tikz}
    \]
    It is easy to verify that the composition of backwards channels here is precisely the backwards channel given by $e\lenscirc d$---see Theorem \ref{thm:buco}---which establishes the result.
  \end{proof}
\end{prop}

\begin{rmk} \label{rmk:strng-mnd}
  Above, we claimed that a monoidal monad $\Pa:\cat{E}\to\cat{E}$ on a symmetric monoidal category $(\cat{E},\times,1)$ is equipped with a (left) strength $\mathsf{str}_{X,Y}:X\times\Pa Y\to\Pa\bigl(X\times Y)$, in the sense of Definition \ref{def:strength}.
  This can be obtained from the unit $\eta$ and the laxator $\alpha$ of the monad as follows:
  \[ \mathsf{str}_{X,Y} : X\times\Pa Y \xto{\eta_X\times\id_{\Pa Y}} \Pa X\times\Pa Y \xto{\alpha_{X,Y}} \Pa(X\times Y) \]
  Using the monad laws, a strength obtained in this way can be shown to satisfy the following axioms (that the strength commutes with the monad structure), and so one may say that $\Pa$ is a \textit{strong monad}:
  \[\begin{tikzcd}[sep=scriptsize]
    & {A\times B} \\
    \\
    {A\times\Pa B} && {\Pa(A\times B)}
    \arrow["{\id_A\times\eta_B}"', from=1-2, to=3-1]
    \arrow["{\mathsf{str}_{A,B}}"', from=3-1, to=3-3]
    \arrow["{\eta_{A\times B}}", from=1-2, to=3-3]
  \end{tikzcd}\]
  \[\begin{tikzcd}[sep=scriptsize]
    {A\times\Pa\Pa B} && {\Pa(A\times\Pa B)} && {\Pa\Pa(A\times B)} \\
    \\
    {A\times\Pa B} &&&& {\Pa(A\times B)}
    \arrow["{A\times\mu_B}"', from=1-1, to=3-1]
    \arrow["{\mathsf{str}_{A,\Pa B}}", from=1-1, to=1-3]
    \arrow["{\Pa(\mathsf{str}_{A,B})}", from=1-3, to=1-5]
    \arrow["{\mu_{A\times B}}", from=1-5, to=3-5]
    \arrow["{\mathsf{str}_{A,B}}", from=3-1, to=3-5]
  \end{tikzcd}\]
\end{rmk}

Now that we have an `external hom', we might expect also to have a corresponding `external composition', represented by a family of morphisms of polynomials; we establish such a family now, and it will be important in our bicategorical construction.

\begin{defn} \label{def:blhom-comp}
  We define an `external composition' natural transformation $\Fun{c}$, with components
  \[ \blhom{Ay^S,By^T}\otimes\blhom{By^T,Cy^U}\to\blhom{Ay^S,Cy^U} \]
  given in the forwards direction by composition of Bayesian lenses.
  In the backwards direction, for each pair of lenses $c:(A,S)\lensto(B,T)$ and $d:(B,T)\lensto(C,U)$, we need a map
  \[ \Fun{c}^\sharp_{c,d} : \cat{C}(I,A)\otimes U \to \cat{C}(I,A)\otimes T\otimes \cat{C}(I,B)\otimes U\bigr)\]
  which we define as follows:
  \begin{align*}
    \Fun{c}^\sharp_{c,d} &:= \cat{C}(I,A) \otimes U \xto{\bcopier\otimes\bcopier} \cat{C}(I,A) \otimes \cat{C}(I,A) \otimes U \otimes U \cdots \\
    &\qquad\cdots \xto{\cat{C}(I,A)\otimes{c_1}_*\otimes U\otimes U} \cat{C}(I,A) \otimes \cat{C}(I,B) \otimes U \otimes U \cdots \\
    &\qquad\cdots \xto{\cat{C}(I,A)\otimes\bcopier\otimes\cat{C}(I,B)\otimes U\otimes U} \cat{C}(I,A) \otimes \cat{C}(I,B) \otimes \cat{C}(I,B) \otimes U \otimes U \\
    &\qquad\cdots \xto{\cat{C}(I,A)\otimes\cat{C}(I,B)\otimes d^\sharp\otimes U\otimes U} \cat{C}(I,A) \otimes \cat{C}(I,B) \otimes \cat{C}(U,T) \otimes Y \otimes U \\
    &\qquad\cdots \xto{\cat{C}(I,A)\otimes\cat{C}(I,B)\Fun{ev}_{U,T}\otimes U} \cat{C}(I,A) \otimes \cat{C}(I,B) \otimes T \otimes U \\
    &\qquad\cdots \xto{\cat{C}(I,A)\otimes\mathsf{swap}\otimes U} \cat{C}(I,A) \otimes T \otimes \cat{C}(I,B) \otimes U
  \end{align*}
  where ${c_1}_*$ and $\Fun{ev}_{U,T}$ are as in \ref{prop:blhom-funct}.

  With $\cat{C} = \Kl(\Pa)$, we can equivalently (and more legibly) define $\Fun{c}^\sharp_{c,d}$ by the following string diagram:
  \[
  \Fun{c}^\sharp_{c,d} \quad := \quad \input{img/blhom-comp-bkwd.tikz}
  \]
  where $d^\flat$ and $\Fun{str}$ are also as in Proposition \ref{prop:blhom-funct}.

  We can therefore understand $\Fun{c}^\sharp_{c,d}$ as mapping forward and backward inputs for the composite lens $d\lenscirc c$ to appropriate inputs for the constituent lenses $c$ and $d$; that is, $\Fun{c}^\sharp_{c,d}$ maps $(\pi,u)$ to $(\pi,d^\sharp_{c_1\klcirc\pi}(u),c_1\klcirc\pi,u)$.
  The resulting inputs to the lens $c$ are therefore $(\pi,d^\sharp_{c_1\klcirc\pi}(u))$, and those to $d$ are $(c_1\klcirc\pi,u)$.
  (This is precisely as the law of lens composition stipulates: the forwards input to $d$ is obtained by pushing forwards through $d$; and the backwards input to $c$ is obtained from the backwards component of $d$.)
\end{defn}

We leave to the reader the detailed proof that this definition produces a well-defined natural transformation, noting only that the argument is analogous to that of Proposition \ref{prop:blhom-funct}:
one observes that, in the forwards direction, the definition is simply composition of Bayesian lenses (which is immediately natural);
in the backwards direction, one observes that the definition again mirrors that of the backwards composition of Bayesian lenses.

Next, we establish the structure needed to make our bicategory monoidal.

\begin{defn} \label{def:blhom-tensor-dist}
  We define a distributive law $\Fun{d}$ of $\blhom{-,=}$ over $\otimes$, a natural transformation with components
  \[ \blhom{Ay^S,By^T}\otimes\blhom{A'y^{S'},B'y^{T'}}\to\blhom{Ay^S\otimes A'y^{S'},By^T\otimes B'y^{T'}} \, , \]
  noting that $Ay^S\otimes A'y^{S'} = (A\otimes A')y^{(S\otimes S')}$ and $By^T\otimes B'y^{T'} = (B\otimes B')y^{(T\otimes T')}$.
  The forwards component is given simply by taking the tensor of the corresponding Bayesian lenses, using the monoidal product (also denoted $\otimes$) in $\BLens{}$.
  Backwards, for each pair of lenses $c:(A,S)\lensto(B,T)$ and $c':(A',S')\lensto(B',T')$, we need a map
  \[ \Fun{d}^\sharp_{c,c'} : \cat{C}(I, A\otimes A')\otimes T\otimes T' \to \cat{C}(I,A)\times T\times \cat{C}(I,A')\times T' \]
  for which we choose
  \begin{align*}
    & \cat{C}(I, A\otimes A')\otimes T\otimes T' \xto{\bcopier\otimes T\otimes T'} \cat{C}(I, A\otimes A')\otimes \cat{C}(I, A\otimes A')\otimes T\otimes T' \cdots \\
    & \cdots\, \xto{\cat{C}(I,\mathsf{proj}_A)\otimes\cat{C}(I,\mathsf{proj}_{A'})\otimes T\otimes T'} \cat{C}(I,A)\otimes \cat{C}(I,A')\otimes T\otimes T' \cdots \\
    & \cdots\, \xto{\cat{C}(I,A)\otimes\Fun{swap}\otimes T'} \cat{C}(I,A)\otimes T\otimes \cat{C}(I,A')\otimes T'
  \end{align*}
  where $\Fun{swap}$ is the symmetry of the tensor $\otimes$ in $\cat{C}$.
  Note that $\Fun{d}^\sharp_{c,c'}$ so defined does not in fact depend on either $c$ or $c'$.
\end{defn}

We now have everything we need to construct a monoidal bicategory $\Hier^\Tt_{\Pa}$ of dynamical hierarchical inference systems in $\cat{C}$, following the intuition outlined at the beginning of this section.
We call systems over such external hom polynomials \textit{cilia}, as they ``control optics'', akin to the ciliary muscles of the eye.
In future work, we will study the general structure of these categories and their relationship to categorical systems theory \parencite{Myers2020Double,Myers2022Categorical} and related work in categorical cybernetics \parencite{Capucci2022Diegetic}.

\begin{defn} \label{def:hier-bicat} \label{def:cilia}
  Let $\Hier^\Tt_{\Pa}$ denote the monoidal bicategory whose 0-cells are objects $(A,S)$ in $\BLens{}$, and whose hom-categories $\Hier^\Tt_{\Pa}\bigl((A,S),(B,T)\bigr)$ are given by $\Coalg^\Tt_{\Pa}\bigl(\blhom{Ay^S,By^T}\bigr)$.
  The identity 1-cell $\id_{(A,S)} : (A,S)\to(A,S)$ on $(A,S)$ is given by the system with trivial state space $1$, trivial update map, and output map that constantly emits the identity Bayesian lens $(A,S)\lensto(A,S)$.
  The composition of a system $(A,S)\to(B,T)$ then a system $(B,T)\to(C,U)$ is defined by the functor
  \begin{align*}
    & \Hier^\Tt_{\Pa}\bigl((A,S),(B,T)\bigr)\times\Hier^\Tt_{\Pa}\bigl((B,T),(C,U)\bigr) \\
    & = \Coalg^\Tt_{\Pa}\bigl(\blhom{Ay^S,By^T}\bigr)\times\Coalg^\Tt_{\Pa}\bigl(\blhom{By^T,Cy^U}\bigr) \\
    & \xto{\lambda} \Coalg^\Tt_{\Pa}\bigl(\blhom{Ay^S,By^T}\otimes\blhom{By^T,Cy^U}\bigr) \\
    & \xto{\Coalg^\Tt_{\Pa}(\mathsf{c})} \Coalg^\Tt_{\Pa}\bigl(\blhom{Ay^S,Cy^U}\bigr)
      = \Hier^\Tt_{\Pa}\bigl((A,S),(C,U)\bigr)
  \end{align*}
  where $\lambda$ is the laxator and $\mathsf{c}$ is the external composition morphism of Definition \ref{def:blhom-comp}.

  The monoidal structure $(\otimes, y)$ on $\Hier^\Tt_{\Pa}$ derives from the structures on $\Poly_{\Pa}$ and $\BLens{}$, justifying our overloaded notation.
  On 0-cells, $(A,S)\otimes(A',S') := (A\otimes A',S\otimes S')$.
  On 1-cells $(A,S)\to(B,T)$ and $(A',S')\to(B',T')$, the tensor is given by
  \begin{align*}
    & \Hier^\Tt_{\Pa}\bigl((A,S),(B,T)\bigr)\times\Hier^\Tt_{\Pa}\bigl((A',S'),(B',T')\bigr) \\
    & = \Coalg^\Tt_{\Pa}\bigl(\blhom{Ay^S,By^T}\bigr)\times\Coalg^\Tt_{\Pa}\bigl(\blhom{A'y^{S'},B'y^{T'}}\bigr) \\
    & \xto{\lambda} \Coalg^\Tt_{\Pa}\bigl(\blhom{Ay^S,By^T}\otimes\blhom{A'y^{S'},B'y^{T'}}\bigr) \\
    & \xto{\Coalg^{\Tt}_{\Pa}(\mathsf{d})} \Coalg^\Tt_{\Pa}\bigl(\blhom{Ay^S\otimes A'y^{S'},By^T\otimes B'y^{T'}}\bigr) \\
    & = \Hier^\Tt_{\Pa}\bigl((A,S)\otimes(A',S'),(B,T)\otimes(B',T')\bigr)
  \end{align*}
  where $\mathsf{d}$ is the distributive law of Definition \ref{def:blhom-tensor-dist}.
  The same functors
  \[ \Hier^\Tt_{\Pa}\bigl((A,S),(B,T)\bigr)\times\Hier^\Tt_{\Pa}\bigl((A',S'),(B',T')\bigr) \to \Hier^\Tt_{\Pa}\bigl((A,S)\otimes(A',S'),(B,T)\otimes(B',T')\bigr) \]
  induce the tensor of 2-cells; concretely, this is given on morphisms of dynamical systems by taking the product of the corresponding morphisms between state spaces.
\end{defn}

We do not give here a proof that this makes $\Hier^\Tt_{\Pa}$ into a well-defined monoidal bicategory;
briefly, the result follows from the facts that the external composition $\mathsf{c}$ and the tensor $\otimes$ are appropriately associative and unital, that $\Coalg^\Tt_\Pa$ is lax monoidal, that $\blhom{{-},{=}}$ is functorial in both positions, and that $\blhom{{-},{=}}$ distributes naturally over $\otimes$.

Before we move on, it will be useful to spell out concretely the elements of a `cilium' (a 1-cell) $(A,S)\to(B,T)$ in $\Hier^\Tt_{\Pa}$.

\begin{prop} \label{prop:unpack-hier}
  Suppose $\Pa$ is a monad on a Cartesian closed category $\cat{E}$.
  Then a 1-cell $\vartheta:(A,S)\to(B,T)$ in $\Hier^\Tt_{\Pa}$ is given by a tuple $\vartheta := (X,\vartheta^o_1,\vartheta^o_2,\vartheta^u)$ of
  \begin{itemize}
  \item a choice of state space $X$,
  \item a forwards output map $\vartheta^o_1:\Tt\times X\times A\to\Pa B$ in $\cat{E}$,
  \item a backwards output map $\vartheta^o_2:\Tt\times X\times\Pa A\times T\to\Pa S$ in $\cat{E}$, and
  \item an update map $\vartheta^u:\Tt\times X\times\Pa A\times T\to\Pa X$ in $\cat{E}$,
  \end{itemize}
  satisfying the `flow' condition of Proposition \ref{prop:pM-coalg}.
  \begin{proof}
    The result follows immediately upon unpacking the definitions, using the Cartesian closure of $\cat{E}$.
  \end{proof}
\end{prop}

\subsection{Differential systems} %
\label{sec:diff-sys}

Approximate inference doctrines describe how systems play statistical games, and are particularly of interest when one asks how systems' performance may improve during such game-playing.
One prominent method of performance improvement involves descending the gradient of a statistical game's loss function, and we will see below that this method is adopted by both the Laplace and the Hebb-Laplace doctrines.
The appearance of gradient descent prompts questions about the connections between such statistical systems and other `cybernetic' systems such as deep learners or players of economic games, both of which may also involve gradient descent \parencite{Cruttwell2022Categorical,Capucci2022Diegetic}; indeed, it has been proposed \parencite{Capucci2021Towards} that parameterized gradient descent should form the basis of a compositional account of cybernetic systems in general\footnote{
Our own view on cybernetics is somewhat more general, since not all systems that may be seen as cybernetic are explicitly structured as gradient-descenders, and nor even is explicit differential structure always apparent.
In earlier work, we suggested that statistical inference was perhaps more inherent to cybernetics \parencite{Smithe2020Cyber}, although today we believe that a better, though more informal, definition of cybernetic system is perhaps ``an intentionally-controlled open dynamical system''.
(Slightly more formally, we can understand this as ``an open dynamical system clad in a controller'', with the possible `cladding' collected into a fibration over systems of each given type.)
Nonetheless, we acknowledge that this notion of ``intentional control'' may generally be reducible to a stationary action principle, again indicating the importance of differential structure.
We leave the statement and proof of this general principle to future work.}.

In order to incorporate gradient descent explicitly into our own compositional framework, we follow the recipes above to define here first a category of differential systems opindexed by polynomial interfaces and then a monoidal bicategory of differential hierarchical inference systems.
We then show how we can obtain dynamical from differential systems by integration, and sketch how this induces a ``change of base'' from dynamical to differential hierarchical inference systems.

Differential systems require differential structure, but we are here still concerned with statistical systems whose time evolution is stochastic.
This means that a differential system will be given by a \textit{stochastic vector field}: a stochastic section of the tangent bundle over the system's state space.
However, as we have seen, the state spaces of stochastic systems are naturally found in a category of measurable spaces, but such a categorical setting does not generally supply differential structure too, and without this we cannot define tangent bundles.
This poses our first hurdle.

We will not here entirely vault this hurdle, for the interplay of randomness and smoothness is subtle and untangling it is not our purpose in this thesis.
However, we can overcome it in a manner which is satisfactory for our present needs, by noting that all our state spaces of later interest will be Euclidean, meaning that we can equip them with their standard Borel measurable structure.
In future work, we hope to generalize this situation, possibly using the notion of \textit{relative monad} \parencite{Altenkirch2015Monads}.

\begin{defn}
  Let $\Cat{Euc}$ denote the category whose objects are finite-dimensional Euclidean spaces $\rr^n$ and whose morphisms are smooth maps between them.
\end{defn}

Euclidean spaces are trivially manifolds: the tangent space over each point $x\in\rr^n$ is again $\rr^n$.
Hence, if $X$ is a Euclidean space, then the tangent bundle $\Tan{X}\to X$ is simply the projection $X\times X\to X$ mapping $(x,v)$ to $x$.
As in general differential geometry, $\Tan$ yields a functor $\Cat{Euc}\to\Cat{Euc}$.

\begin{prop}
  The tangent bundle functor $\Tan:\Cat{Euc}\to\Cat{Euc}$ maps each Euclidean space $\rr^n$ to $\rr^n\times\rr^n$ and each smooth map $f:\rr^m\to \rr^n$ to its differential $df:\rr^m\times\rr^m \to \rr^n\times\rr^n$, which in turn maps $(x,v)$ to $\bigl(f(x), \partial_x f(v)\bigr)$, where $\partial_x f$ denotes the (total) derivative of $f$ at $x$, which can be represented by its $n\times m$ Jacobian matrix.
\end{prop}

\begin{rmk}
  Differentials compose by pushforward, which yields the chain rule of differential calculus.
  Earlier we have seen that chain rules indicate the presence of a fibration, and indeed this is also the case here: $\Tan$ is properly a functor into the fibration of vector bundles over the category of spaces; composing this functor with the projection out of the fibration yields the endofunctor we have sketched in the preceding proposition.
\end{rmk}

Ordinary differential equations define vector fields, which are (deterministic) sections of the tangent bundle over a space; these are deterministic closed differential systems.
We are interested in \textit{open} differential systems that may have effectful (\textit{e.g.} stochastic) evolution: for openness, we will use the trick of \secref{sec:poly-dyn}; for stochasticity, we will need \textit{stochastic sections}, which means transporting the tangent bundles into a category of stochastic maps and considering their sections there.

\begin{prop}
  There is a functor $J:\Cat{Euc}\to\Cat{Meas}$ that takes each Euclidean space and exhibits it as a measurable space equipped with its standard Borel $\sigma$-algebra, and which takes each smooth map and exhibits it as a measurable function.
  This functor preserves products.
\end{prop}

\begin{prop}[{\textcite[{\S}III.B]{Heunen2017Convenient}}]
  There is a functor $R:\Cat{Meas}\to\Cat{QBS}$ which is full and faithful when restricted to the subcategory $\Cat{Borel}\hookrightarrow\Cat{Meas}$ of standard Borel spaces.
\end{prop}

Using these functors, we can transport a tangent bundle $\pi_X:\Tan X\to X$ in $\Cat{Euc}$ to $\Cat{QBS}$, as $RJ\pi_X$.
Then, if we let $\Pa:\Cat{QBS}\to\Cat{QBS}$ denote the probability monad on quasi-Borel spaces introduced in Example \ref{ex:qbs}, we can take the sections of $RJ\pi_X$ in $\Kl(\Pa)$ to be the stochastic vector fields over the space $X$.
Moreover, since $\Cat{QBS}$ is finitely complete and Cartesian closed, it is sufficiently structured that we may instantiate the category $\Poly_{\Pa}$ of polynomials with $\Pa$-effectful feedback.

Using these two ideas, we may define our desired categories of stochastic differential systems.
Recall that morphisms $Ay^B\to p$ in $\Poly_{\Pa}$ correspond to morphisms $A\klto pB$ in $\Kl(\Pa)$.

\begin{notation}
  In this section, let us write $(\widetilde{-})$ to denote the functor $RJ:\Cat{Euc}\to\Cat{QBS}$.
\end{notation}

\begin{defn}
  For each $p : \Poly_{\Pa}$, define a category $\DiffSys(p)$ as follows.
  Its objects are pairs $(M,m)$ of a Euclidean space $M:\Cat{Euc}$ and a morphism $m : \widetilde{M}y^{\widetilde{\Tan M}} \to p$ of polynomials in $\Poly_{\Pa}$, such that for any section $\sigma : p\to y$ of $p$, the composite morphism $\sigma\circ m : \widetilde{M}y^{\widetilde{\Tan M}}\to y$ corresponds to a stochastic section $m^\sigma : \widetilde{M}\klto\widetilde{\Tan M}$ of the tangent bundle $\Tan M\to M$ under $RJ$.
  A morphism $\alpha:(M,m)\to(M',m')$ in $\DiffSys(p)$ is a smooth map $\alpha : M \to M'$ in $\Cat{Euc}$ such that the following diagram commutes:
  \[\begin{tikzcd}
	{\widetilde{M}} & {p\widetilde{\Tan M}} \\
	{\widetilde{M'}} & {p\widetilde{\Tan M'}}
	\arrow["m", from=1-1, to=1-2]
	\arrow["\widetilde{\alpha}"', from=1-1, to=2-1]
	\arrow["{m'}"', from=2-1, to=2-2]
	\arrow["{p\widetilde{\Tan\alpha}}", from=1-2, to=2-2]
  \end{tikzcd}\]
\end{defn}

We obtain a monoidal opindexed category from this data in much the same way as we did for $\Coalg^\Tt$.

\begin{prop}
  $\DiffSys$ defines an opindexed category $\Poly_{\Pa}\to\Cat{Cat}$.
  Given a morphism $\varphi:p\to q$ of polynomials, $\DiffSys(\varphi) : \DiffSys(p) \to \DiffSys(q)$ acts on objects by postcomposition and trivially on morphisms.
\end{prop}

\begin{prop} \label{prop:diffsys-lax}
  The functor $\DiffSys$ is lax monoidal $(\Poly_{\Pa},\otimes,y) \to (\Cat{Cat},\times,\Cat{1})$.
  \begin{proof}[Proof sketch]
    Note that $\Tan$ is strong monoidal, with $\Tan(\rr^0) \cong \rr^0$ and $\Tan(M)\times\Tan(N)\cong\Tan(M\times N)$, that $RJ$ preserves products, and that $RJ(\rr^0) = 1$.
    The unitor $\Cat{1}\to\DiffSys(y)$ is given by the isomorphism $\widetilde{\rr^0}y^{\widetilde{\Tan\rr^0}} \cong 1y^1 \cong y$ induced by the strong monoidal structure of $\Fun{T}$.
    The laxator $\lambda_{p,q} : \DiffSys(p) \times \DiffSys(q) \to \DiffSys(p\otimes q)$ is similarly determined: given objects $m:\widetilde{M}y^{\widetilde{\Tan M}}\to p$ and $n:\widetilde{N}y^{\widetilde{\Tan N}}\to q$, take their tensor $m\otimes n:(\widetilde{M}\otimes \widetilde{N})y^{\widetilde{\Tan M}\otimes\widetilde{\Tan N}}\to p\otimes q$ and precompose with the induced morphism $(\widetilde{M\times N})y^{\widetilde{\Tan(M\times N)}} \to (\widetilde{M}\otimes \widetilde{N})y^{\widetilde{\Tan M}\otimes\widetilde{\Tan M}}$; proceed similarly on morphisms of differential systems.
    The satisfaction of the unitality and associativity laws follows from the monoidality of $\Tan$.
  \end{proof}
\end{prop}

We now define a monoidal bicategory $\DiffHier$ of differential hierarchical inference systems, following the definition of $\Hier$ above.

\begin{defn} \label{def:diff-cilia}
  Let $\DiffHier$ denote the monoidal bicategory whose 0-cells are the objects $(A,S)$ of $\BLens{\Kl(\Pa)}$ and whose hom-categories $\DiffHier\bigl((A,S),(B,T)\bigr)$ are given by $\DiffSys\bigl(\blhom{Ay^S,By^T}\bigr)$.
  The identity 1-cell $\id_{(A,S)}:(A,S)\to(A,S)$ on $(A,S)$ is given by the differential system $y \to \blhom{Ay^S,By^T}$ with state space $\rr^0$, trivial backwards component, and forwards component that picks the identity Bayesian lens on $(A,S)$.
  The composition of differential systems $(A,S)\to(B,T)$ then $(B,T)\to(C,U)$ is defined by the functor
  \begin{align*}
    & \DiffHier\bigl((A,S),(B,T)\bigr)\times\DiffHier\bigl((B,T),(C,U)\bigr) \\
    & = \DiffSys\bigl(\blhom{Ay^S,By^T}\bigr)\times\DiffSys\bigl(\blhom{By^T,Cy^U}\bigr) \\
    & \xto{\lambda} \DiffSys\bigl(\blhom{Ay^S,By^T}\otimes\blhom{By^T,Cy^U}\bigr) \\
    & \xto{\DiffSys(\mathsf{c})} \DiffSys\bigl(\blhom{Ay^S,Cy^U}\bigr)
    = \DiffHier\bigl((A,S),(C,U)\bigr)
  \end{align*}
  where $\lambda$ is the laxator of Proposition \ref{prop:diffsys-lax} and $\mathsf{c}$ is the external composition morphism of Definition \ref{def:blhom-comp}.

  The monoidal structure $(\otimes,y)$ on $\DiffHier$ is similarly defined following that of $\Hier^\Tt_{\Pa}$.
  On 0-cells, $(A,S)\otimes(A',S') := (A\otimes A',S\otimes S')$.
  On 1-cells $(A,S)\to(B,T)$ and $(A',S')\to(B',T')$ (and their 2-cells), the tensor is given by the functors
  \begin{align*}
    & \DiffHier\bigl((A,S),(B,T)\bigr)\times\DiffHier\bigl((A',S'),(B',T')\bigr) \\
    & = \DiffSys\bigl(\blhom{Ay^S,By^T}\bigr)\times\DiffSys\bigl(\blhom{A'y^{S'},B'y^{T'}}\bigr) \\
    & \xto{\lambda} \DiffSys\bigl(\blhom{Ay^S,By^T}\otimes\blhom{A'y^{S'},B'y^{T'}}\bigr) \\
    & \xto{\DiffSys(\mathsf{d})} \DiffSys_{\Pa}\bigl(\blhom{Ay^S\otimes A'y^{S'},By^T\otimes B'y^{T'}}\bigr) \\
    & = \DiffHier\bigl((A,S)\otimes(A',S'),(B,T)\otimes(B',T')\bigr)
  \end{align*}
  where $\mathsf{d}$ is the distributive law of Definition \ref{def:blhom-tensor-dist}.
\end{defn}

Following Proposition \ref{prop:unpack-hier}, we have the following characterization of a differential hierarchical inference system $(A,S) \to (B,T)$ in $\Kl(\Pa)$.

\begin{prop} \label{prop:unpack-diff}
  A 1-cell $\delta : (A,S) \to (B,T)$ in $\DiffHier$ is given by a tuple $\delta := (X,\delta^o_1,\delta^o_2,\delta^\sharp)$ of
  \begin{itemize}
  \item a choice of state space $X:\Cat{Euc}$;
  \item a forwards output map $\delta^o_1 : \widetilde{X}\times A \to \Pa B$,
  \item a backwards output map $\delta^o_2 : \widetilde{X}\times \Pa A \times T \to \Pa S$,
  \item a stochastic vector field $\delta^\sharp : \widetilde{X}\times \Pa A\times T \to \Pa\widetilde{\Tan X}$.
  \end{itemize}
\end{prop}

At least for deterministic differential systems, we can obtain continuous-time dynamical systems from differential systems by integration.
We may then discretize these flows to give discrete-time dynamical systems.

\begin{prop} \label{prop:diff-flow}
  For the purposes of this proposition, let $\Pa$ be the identity monad on a finitely complete category $\cat{E}$ of manifolds, let $(\widetilde{-})$ be the corresponding inclusion $\Cat{Euc}\hookrightarrow\cat{E}$, and let $\DiffSys$ be instantiated accordingly.
  Then integration induces an indexed functor $\Fun{Flow} : \DiffSys \to \Coalg^\rr_{\Pa}$.
  \begin{proof}
    Suppose $(M,m)$ is an object in $\DiffSys(p)$.
    The morphism $m : \widetilde{M}y^{\widetilde{\Tan M}} \to p$ consists of functions $m_1 : \widetilde{M}\to p(1)$ and $m^\sharp : \sum_{x:\widetilde{M}} p[m_1(x)] \to \widetilde{\Tan M}$.
    Since, for any section $\sigma : p\to y$, the induced map $m^\sigma : \widetilde{M}\to \widetilde{\Tan M}$ is a vector field on a compact manifold, it generates a unique global flow $\Fun{Flow}(p)(m)^\sigma : \rr\times \widetilde{M}\to \widetilde{M}$ \parencite[{Thm.s 12.9,12.12}]{Lee2012Smooth}, which factors as
    \[ \sum_{t:\rr} \widetilde{M} \xto{m_1^*\sigma} \sum_{t:\rr} \sum_{x:\widetilde{M}} p[m_1(x)] \xto{\Fun{Flow}(p)(m)^u} \widetilde{M} \, . \]
    We therefore define the system $\Fun{Flow}(p)(m)$ to have state space $\widetilde{M}$, output map $m_1$ (for all $t:\rr$), and update map $\Fun{Flow}(p)(m)^u$.
    Since $\Fun{Flow}(p)(m)^\sigma$ is a flow for any section $\sigma$, it immediately satisfies the monoid action condition.
    On morphisms $\alpha : m\to m'$, we define $\Fun{Flow}(p)(\alpha)$ by the same underlying map on state spaces; this is again well-defined by the condition that $\alpha$ is compatible with the tangent structure.
    Given a morphism $\varphi:p\to q$ of polynomials, both the reindexing $\DiffSys(\varphi)$ and $\Coalg^\rr_{\Pa}(\varphi)$ act by postcomposition, and so it is easy to see that $\Coalg^\rr_{\Pa}(\varphi)\circ\Fun{Flow}(p) \cong \Fun{Flow}(q)\circ\DiffSys_{\Pa}(\varphi)$ naturally.
  \end{proof}
\end{prop}

\begin{rmk}
  The question of integration of \textit{stochastic} systems is more vexed and we will not treat it in this thesis.
\end{rmk}

Not only may we integrate a differential system to obtain a continuous-time dynamical system, we can also variously discretize the continuous-time system to obtain a discrete-time one.

\begin{prop} \label{prop:time-change}
  Any map $f:\Tt'\to\Tt$ of monoids induces an indexed functor (a natural transformation) $\Coalg^\Tt_{\Pa} \to \Coalg^{\Tt'}_{\Pa}$.
  \begin{proof}
    We first consider the induced functor $\Coalg^\Tt_{\Pa}(p) \to \Coalg^{\Tt'}_{\Pa}(p)$, which we denote by $\Delta_f^p$.
    Note that we have a morphism $[f y,p] : [\Tt y, p] \to [\Tt' y, p]$ of polynomials by substitution (precomposition).
    A system $\beta$ in $\Coalg^\Tt_{\Pa}$ is a morphism $Sy^S \to [\Tt y, p]$ for some $S$, and so we define $\Delta_f^p(\beta)$ to be $[f,p]\circ\beta : Sy^S \to [\Tt y, p] \to [\Tt' y, p]$.
    To see that this satisfies the monoid action axiom, consider that the closure $\Delta_f^p(\beta)^\sigma$ for any section $\sigma : p\to y$ is given by
    \[ \Sum_{t:\Tt'} S \xto{\beta^o(f(t))^\ast \sigma} \Sum_{t:\Tt'} \Sum_{s:S} p[\beta^o(f(t), s)] \xto{\beta^u} S \]
    which is an object in the functor category $\Cat{Cat}\bigl(\deloop{\Tt'},\Kl(\Pa)\bigr)$ since $f$ is a monoid homomorphism.
    On morphisms of systems, the functor $\Delta_f^p$ acts trivially.

    To see that $\Delta_f$ collects into an indexed functor, consider that it is defined on each polynomial $p$ by the contravariant action $[f,p]$ of the internal hom $[-,=]$, and that the reindexing $\Coalg^\Tt(\varphi)$ for any morphism $\varphi$ of polynomials is similarly defined by the covariant action $[\Tt y,\varphi]$.
    By the bifunctoriality of $[-,=]$, we have $[\Tt' y,\varphi]\circ[f y,p] = [f y,\varphi] = [f y,q]\circ[\Tt y,\varphi]$, and so $\Coalg^{\Tt'}_{\Pa}(\varphi)\circ\Delta_f^p = \Delta_f^q\circ\Coalg^\Tt_{\Pa}$.
  \end{proof}
\end{prop}

\begin{cor} \label{cor:coalg-disc}
  For each $k : \rr$, the canonical inclusion $\iota_k : \nn \hookrightarrow \rr : i \mapsto ki$ induces a corresponding `discretization' indexed functor $\Fun{Disc}_k := \Delta_\iota : \Coalg^\rr_{\Pa} \to \Coalg^{\nn}_{\Pa}$.
\end{cor}

\begin{rmk} \label{rmk:diff-disc}
  From Proposition \ref{prop:diff-flow} and Corollary \ref{cor:coalg-disc} we obtain a family of composite indexed functors $\DiffSys \xto{\Fun{Flow}} \Coalg^\rr_{\Pa} \xto{\Fun{Disc}_k} \Coalg^\nn_{\Pa}$ taking each differential system to a discrete-time dynamical system in $\cat{C}$.
  Below, we will define approximate inference doctrines in discrete time that arise from processes of (stochastic) gradient descent, and which therefore factor through differential systems, but the form in which these are given---and in which they are found in the informal literature (\textit{e.g.}, \textcite{Bogacz2017tutorial})---is not obtained via the composite $\Fun{Disc}_k \circ \Fun{Flow}$ for any $k$, even though there is a free parameter $k$ that plays the same rôle (intuitively, a `learning rate').
  Instead, one typically adopts the following scheme, sometimes known as \textit{Euler integration} or the \textit{Euler method}.

  Euler integration induces a family of indexed functors $\Fun{Euler}_k : \DiffSys \to \Coalg^\nn_{\Pa}$, for $k:\rr$, which we illustrate for a single system $(\rr^n,m)$ over a fixed polynomial $p$, with $m : \rr^n y^{\rr^n\times\rr^n} \to p$.
  This system is determined by a pair of morphisms $m_1 : \rr^n \to p(1)$ and $m^\sharp : \sum_{x:\rr^n} p[m_1(x)] \klto \rr^n\times \rr^n$, and we can write the action of $m^\sharp$ as $(x,y) \mapsto (x, v_x(y))$.

  Using these, we define a discrete-time dynamical system $\beta$ over $p$ with state space $\rr^n$.
  This $\beta$ is given by an output map $\beta^o$, which we define to be equal to $m_1$, $\beta^o := m_1$, and an update map $\beta^u : \sum_{x:\rr^n} p[\beta^o(x)] \klto \rr^n$, which we define by $(x,y) \mapsto x + k\, v_x(y)$.
  Together, these define a system in $\Coalg^\nn_{\Pa}(p)$, and the collection of these systems $\beta$ produces an indexed functor by the definition $\Fun{Euler}_k(p)(m) := \beta$.

  By contrast, the discrete-time system obtained via $\Fun{Disc}_k \circ \Fun{Flow}$ involves integrating a continuous-time system for $k$ units of real time for each unit of discrete time: although this in general produces a more accurate simulation of the trajectories implied by the vector field, it is computationally more arduous; to trade off simulation accuracy against computational feasibility, one may choose a more sophisticated discretization scheme than that sketched above, or at least choose a ``sufficiently small'' timescale $k$.
\end{rmk}

Finally, we can use the foregoing ideas to translate differential hierarchical inference systems to dynamical hierarchical inference systems.

\begin{cor}
  The indexed functors $\Fun{Disc}_k : \Coalg^\rr_{\Pa} \to \Coalg^\nn_{\Pa}$, $\Fun{Flow} : \DiffSys \to \Coalg^\rr_{\Pa}$, and $\Fun{Euler}_k : \DiffSys \to \Coalg^\nn_{\Pa}$ induce functors (respectively) $\H\Fun{Disc}_k : \Hier^\rr_{\Pa} \to \Hier^\nn_{\Pa}$, $\H\Fun{Flow} : \DiffHier \to \Hier^\rr_{\Pa}$ and $\H\Fun{Euler}_k : \DiffHier \to \Hier^\nn_{\Pa}$ by change of base of enrichment.
\end{cor}

\chapter{Approximate inference doctrines for predictive coding} \label{chp:brain}

The construction of the predictive coding models that underlie the theory of the Bayesian brain involves mapping a (`generative') statistical model, representing how the modeller believes the brain to understand the world, to a dynamical system which plays the rôle of the neural circuits which are hypothesized to instantiate that model.
This dynamical system is then simulated and the resulting trajectories studied: for instance, to compare with experimental neural or psychological data, or to judge against a synthetic benchmark.

Typically, both the generative model and the resulting dynamical systems are `modular', and the mapping from the former to the latter preserves this structure: that is to say, predictive coding forms an example of functorial semantics, of which we saw a rudimentary example in \secref{sec:sys-circ}, when we considered an algebra of rate-coded neural circuits.
This chapter makes this functoriality explicit, which we hope to have a useful scientific consequence: it often seems to be the case that researchers manually derive complicated dynamical systems from their statistical models \parencite{Buckley2017free,DaCosta2020Active,Tschantz2019Learning,Kaplan2018Planning,Ueltzhoeffer2018Deep,Friston2010Action,Bastos2012Canonical} \parencite[{Chapter 5}]{Parr2022Active}, but once functoriality is established, this manual labour is unnecessary; the functor represents a machine with which the process may be automated.

We call such functors \textit{approximate inference doctrines}.
In defining them, we bring together the statistical games of Chapter \ref{chp:sgame} (which supply the `syntax' of generative models) and the open dynamical systems of Chapter \ref{chp:coalg} (which supply the `semantics'), and we explain precisely how these doctrines may factorize through the various components we have seen: the stochastic channels, the inference systems, the loss models, the differential systems, and the cilia.
This is the work of \secref{sec:doctrines}, which also establishes the functoriality of predictive coding under the (Laplacian) free energy principle.
Before we get there, we construct some final pieces of technical machinery, aspects of which we have seen before: stochastic channels with Gaussian noise, to model functions of the form $f(x)+\omega$ with $\omega$ Gaussian-distributed (\secref{sec:doctrines-gauss}); and externally parameterized Bayesian lenses, so that our constructions have the freedom to learn (\secref{sec:doctrines-para}).

\section{Channels with Gaussian noise}
\label{sec:doctrines-gauss}

Our motivating examples from the predictive coding literature in computational neuroscience are defined over a subcategory of channels between Cartesian spaces with additive Gaussian noise \parencite{Friston2007Variational,Bogacz2017tutorial,Buckley2017free}; typically one writes $x \mapsto f(x) + \omega$, with $f : X\to Y$ a deterministic map and $\omega$ sampled from a Gaussian distribution over $Y$.
This choice is made, as we saw in \secref{sec:laplace}, because it permits some simplifying assumptions, and the resulting dynamical systems resemble known neural circuits.

In this section, we develop some categorical language in which we can express such Gaussian channels, expanding on the informal definition given in Remark \ref{rmk:gauss-chan}.
We do so by thinking of $x \mapsto f(x) + \omega$ as a map parameterized by a noise source, and so to construct a category of such channels, we can use the $\para$ construction, following Proposition \ref{prop:para-bicat}.
Because the noise comes from the parameter, we need a category whose objects are spaces equipped with measures.
For this, we can use the `pointing' construction introduced in \secref{sec:open-rds}; as we saw in Example \ref{ex:meas-spc}, this gives us a category of measure spaces.
The next step is to spell out an actegory structure that induces the parameterization we seek.

\begin{prop} \label{prop:probstoch-act}
  Suppose $(\cat{C},\otimes,I)$ is a monoidal category, and suppose $\cat{D}\hookrightarrow\cat{C}$ is a subcategory to which the monoidal structure restricts.
  Then there is a $\cat{D}_*$-actegory structure $\cat{D}_*\to\Cat{Cat}(\cat{D},\cat{D})$ on $\cat{D}$ as follows.
  For each $(M,\mu):\cat{D}_*$, define $(M,\mu)\ast(-):\cat{D}\to\cat{D}$ by $(M,\mu)\ast X := M\otimes(-)$.
  For each morphism $f:(M,\mu)\to(N,\nu)$ in $\cat{D}_*$, define $f\ast(-) := f\otimes(-)$.
  \begin{proof}[Proof sketch]
    The action on morphisms is well-defined because each morphism $f:(M,\mu)\to (N,\nu)$ in $\cat{D}_*$ projects to a map $f:M\to N$ in $\cat{D}$; it is clearly functorial.
    The unitor and associator of the actegory structure are inherited from the monoidal structure, and they satisfy the actegory axioms for the same reason.
\end{proof}
\end{prop}

\begin{rmk}
  Note that the construction of $\ast$ is easily extended to an action on the whole of $\cat{C}$.
  We will however be concerned only with the action of $\cat{D}_*$ on $\cat{D}$.
\end{rmk}

When we instantiate $\ast$ in the context of $\Cat{Meas}\hookrightarrow\Cat{sfKrn}$, the resulting $\para$ bicategory  $\para(\ast)$ can be thought of as a bicategory of maps each of which is equipped with an independent noise source; the composition of maps takes the product of the noise sources, and 2-cells are noise-source reparameterizations.

In this case, the actegory structure $\ast$ is moreover symmetric monoidal, and the 1-categorical truncation $\para(\ast)_1$ (\textit{cf.} Proposition \ref{prop:para-1cat}) is a copy-delete category, as we now sketch.

\begin{prop}
  Suppose $(\cat{C},\otimes,I)$ is a symmetric monoidal copy-discard category, and let the symmetry and copy-discard structure restrict to $\cat{D}\hookrightarrow\cat{C}$.
  Then \(\para(\ast)_1\) is also a symmetric monoidal copy-delete category.
  \begin{proof}[Proof sketch]
    The monoidal structure is defined following Proposition \ref{prop:para-tensor}.
    We need to define a right costrength $\rho$ with components $(N,\nu)\ast(X\otimes Y) \xto{\sim} X\otimes((N,\nu)\ast Y)$.
    Since $\ast$ is defined by forgetting the pointing and taking the monoidal product, the costrength is given by the associator and symmetry in $\cat{D}$:
    \[ (N,\nu)\ast(X\otimes Y) = N\otimes(X\otimes Y) \xto{\sim} N\otimes(Y\otimes X) \xto{\sim} (N\otimes Y)\otimes X \xto{\sim} X\otimes(N\otimes Y) = X\otimes((N,\nu)\ast Y) \]
    As the composite of natural isomorphisms, this definition gives again a natural isomorphism; the rest of the monoidal structure follows from that of the monoidal product on $\cat{C}$.

    We now need to define a symmetry natural isomorphism $\beta_{X,Y}:X\otimes Y \xto{\sim} Y\otimes X$ in $\para(\ast)$.
    This is given by the symmetry of the monoidal product in $\cat{D}$, under the embedding of $\cat{D}$ in $\para(\ast)$ that takes every map to its parameterization by the monoidal unit.

    The rest of the copy-delete structure is inherited similarly from $\cat{C}$ via $\cat{D}$.
  \end{proof}
\end{prop}

When $\cat{C}$ is a category of Markov kernels, we will typically think of the morphisms of $\para(\ast)_1$ as kernels whose uncertainty arises from a noisy parameter.
To formalize this we can push forward the noise to obtain again a morphism in $\cat{C}$.
This yields a functor $\Fun{Push}:\para(\ast)_1\to\cat{C}$.

\begin{prop} \label{prop:embed-para-ast-kl}
  There is a strict monoidal functor \(\Fun{Push}:\para(\ast)_1\to\cat{C}\).
  Given a morphism in \(\para(\ast)_1\) represented by \(f : X \xto{(\Omega, \mu)} Y\), let $\Fun{Push}(f)$ be the composite \(f \klcirc (\mu\otimes\id_X) : X \klto Y\) in \(\cat{C}\).
  \begin{proof}[Proof sketch]
    First, the given mapping preserves identities: the identity in $\para(\ast)$ is trivially parameterized, and is therefore taken to the identity in $\cat{C}$.
    The mapping also preserves composites, by the naturality of the unitors of the symmetric monoidal structure on $\cat{C}$.
    That is, given $f : X\xto{(\Omega,\mu)}Y$ and $g : Y\xto{(\Theta,\nu)}Z$, their composite $g\circ f:X\xto{(\Theta\otimes\Omega,\nu\otimes\mu)}Z$ is taken to
    \[ X \xklto{\sim} 1\otimes 1\otimes X \xklto{\nu\otimes\mu\otimes\id_X} \Theta\otimes\Omega\otimes X \xklto{g\circ f} Z \]
    where here $g\circ f$ is treated as a morphism in $\cat{C}$.
    Composing the images of $g$ and $f$ under the given mapping gives
    \[ X \xklto{\sim} 1\otimes X \xklto{\mu\otimes\id_X} \Omega\otimes X \xklto{f} Y \xklto{\sim} 1\otimes Y \xklto{\nu\otimes Y} \Theta\otimes Y \xklto{g} Z \]
    which is equal to
    \[ X \xklto{\sim} 1\otimes 1\otimes X \xklto{\nu\otimes\mu\otimes\id_X} \Theta\otimes\Omega\otimes X \xklto{\id_\Theta\otimes f} \Theta\otimes Y \xklto{g} Z \]
    which in turn is equal to the image of the composite above.

    Since the monoidal structure on $\para(\ast)$ is inherited from that on $\cat{C}$ (with identical objects), the embedding is strict monoidal.
  \end{proof}
\end{prop}

\begin{rmk}
  Note that $\Fun{Push}$ is not an embedding, since the mapping on hom sets need not be injective: pushing forward the noise of two parallel morphisms in $\para(\ast)_1$ may yield equal morphisms in $\cat{C}$ without the noise sources being isomorphic, and hence without the original morphisms being equivalent in $\para(\ast)$; that is to say, the parameterization of noise sources is not generally unique.
\end{rmk}

We now restrict our attention to Gaussian morphisms in $\cat{C} = \Cat{sfKrn}$.

\begin{defn} \label{def:gauss}
  We say that \(f : X \klto Y\) in \(\Cat{sfKrn}\) is \textit{Gaussian} if, for any \(x : X\), the measure \(f(x)\) is Gaussian\footnote{We admit Dirac delta distributions, and therefore deterministic morphisms, as Gaussian, since delta distributions can be seen as Gaussians with infinite precision.}.
  Similarly, we say that \(f : X \xto{(\Omega, \mu)} Y\) in \(\para(\ast)\) is Gaussian if its image under $\Fun{Push}$ is Gaussian.
  We will write $\gauss$ to denote the subcategory of $\Cat{sfKrn}$ generated by Gaussian kernels and their composites; likewise, we will write $\gauss_*$ to denote the Gaussian subcategory of $\para(\ast)$.
  Given a separable Banach space $X$, we will write $\gauss(X)$ for the space of Gaussian states on $X$.
\end{defn}

\begin{ex}
  Random functions of the form $x\mapsto f(x) + \omega$, where $\omega:\Omega$ is distributed according to a Gaussian, are therefore morphisms in $\gauss_*$.
  Under the embedding into $\gauss$, the corresponding kernel emits, for each $x:X$, a Gaussian distribution with mean $f(x) + \mu_\omega$, where $\mu_\omega$ is the mean of the Gaussian random variable $\omega$, and variance the same as that of $\omega$.
\end{ex}

\begin{rmk} \label{rmk:gauss-not-closed}
  In general, Gaussian morphisms are not closed under composition:
  pushing a Gaussian distribution forward along a nonlinear transformation will not generally result in another Gaussian.
  For instance, consider the Gaussian functions $x\mapsto f(x) + \omega$ and $y\mapsto g(y) + \omega'$.
  Their composite in $\gauss_*$ is the morphism $x\mapsto g\bigl(f(x)+\omega)\bigr)+\omega'$; even if $g\bigl(f(x)+\omega)\bigr)$ is Gaussian-distributed, the sum of two Gaussians is in general not Gaussian, and so $g\bigl(f(x)+\omega)\bigr)+\omega'$ will not be Gaussian.
  This non-closure underlies the power of statistical models such as the variational autoencoder, which are often constructed by pushing a Gaussian forward along a learnt nonlinear transformation \parencite{Kingma2017Variational}, in order to approximate an unknown distribution;
  since sampling from Gaussians is relatively straightforward, this method of approximation can be computationally tractable.
  The $\gauss$ construction here is an abstraction of the Gaussian-preserving transformations of \textcite{Shiebler2020Categorical}, and is to be distinguished from the category with the same name introduced by \textcite{Fritz2019synthetic}, whose morphisms are affine transformations (which do preserve Gaussianness) and which \textit{are} therefore closed under composition;
  there is nonetheless an embedding of Fritz's $\gauss$ into our $\gauss$.
\end{rmk}

For Laplacian statistical games (in the image of $\LFE$), and for the associated approximate inference doctrines, we are interested only in Gaussian channels between finite-dimensional Cartesian spaces $\rr^n$ for $n:\nn$.

\begin{defn} \label{def:fd-gauss}
  Denote by $\FdGauss$ the full subcategory of $\gauss$ spanned by the objects $\rr^n$ for $n:\nn$.
\end{defn}

\begin{prop} \label{prop:gauss-stat-param}
  Every channel \(c : X \klto Y\) in $\FdGauss$ admits a density function \(p_c : Y \times X \to [0, 1]\) with respect to the Lebesgue measure on \(Y\).
  Moreover, since \(Y = \rr^n\) for some \(n : \nn\), this density function is determined by two maps: the \textit{mean} \(\mu_c : X \to \rr^n\), and the \textit{covariance} \(\Sigma_c : X \to \rr^{n\times n}\) in \(\cat{E}\).
  We call the pair \((\mu_u, \Sigma_c) : X \to \rr^n \times \rr^{n\times n}\) the \textit{statistical parameters} of \(c\) (to be distinguished from any parameterization in the category-theoretic sense of \secref{sec:para-sys}).
  \begin{proof}
    The density function \(p_c : Y \times X \to [0, 1]\) is defined by
    \[
    \log p_c (y | x) = \frac{1}{2} \innerprod{\epsilon_c(y,x)}{{\Sigma_c(x)}^{-1} \, \epsilon_c(y,x)} - \log \sqrt{(2 \pi)^n \det \Sigma_c(x) }
    \]
    where the `error' function $\epsilon_c : Y \times X \to Y$ is defined by $\epsilon_c(y, x) := y - \mu_c(x)$.
  \end{proof}
\end{prop}

\section{Externally parameterized Bayesian lenses and statistical games} \label{sec:doctrines-para}

The statistical games of Chapter \ref{chp:sgame} are simply Bayesian lenses equipped with loss functions.
Given a statistical game, its lens is therefore fixed, and the only way to a high score on its loss is through its openness to the environment---the dependence on a prior and an observation.
But this seems like a strange model of adaptive or cybernetic systems, which should also be free to change themselves in order to improve their performance.

Indeed, this changing-oneself is at the heart of the construction of approximate inference doctrines, and in order to incorporate it into the structure, there must be some more freedom in the model: the freedom to choose the lens.
This freedom is afforded by the use of parameterized statistical games, and in particular, \textit{externally} parameterized statistical games, in the sense of \secref{sec:ext-para}.

\begin{rmk}
  It is of course possible to define an actegorical (internal) parameterization of statistical games, but this seems to prove more complicated than necessary for our purposes.
\end{rmk}

In advance of our use of external parameterization in the construction of approximate inference doctrines, recall that we denote the external parameterization of an enriched category $\cat{C}$ in its base of enrichment $\cat{E}$ by $\eP\cat{C}$.
This section is dedicated to exhibiting the external parameterizations $\eP\BLens{}_2$ and $\eP\SGame{}$ of Bayesian lenses and statistical games, and the notion of parameterized loss model.

\begin{rmk}
  Because $\BLens{}_2$ and $\SGame{}$ are both bicategories, they are weakly enriched in $\Cat{Cat}$.
  Consequently, following Remark \ref{rmk:ext-para-enrch}, $\eP$ has the type $\Cat{Cat}\mdash\Cat{Cat} \to (\Cat{Cat}\mdash\Cat{Cat})\mdash\Cat{Cat}$, or equivalently, $\Cat{Bicat}\to\Cat{Tricat}$.
  This means that, in full generality, $\eP\BLens{}_2$ and $\eP\SGame{}$ are tricategories: if $\cat{B}$ is a bicategory, then the hom-bicategory $\eP\cat{B}(a,b)$ is the bicategory $\Cat{Cat}/\cat{B}(a,b)$.
  Because we are now working with weakened structures (weak enrichment, bicategories, lax loss models), we take this to be a \textit{lax} slice of $\Cat{Cat}$.
\end{rmk}

We pause to define this new notion, generalizing our earlier Definition \ref{def:slice-cat} (slice category).

\begin{defn} \label{def:lax-slice}
  Suppose $X$ is a 0-cell in a bicategory $\cat{B}$.
  The \textit{lax slice} of $\cat{B}$ over $X$, denoted $\cat{B}/X$, is the following bicategory.
  Its 0-cells are pairs $(A,p)$ where $A$ is a 0-cell and $p$ is a 1-cell $A\to X$ in $\cat{B}$.
  A 1-cell $(A,p)\to(B,q)$ is a pair $(f,\phi)$ where $f$ is a 1-cell $A\to B$ and $\phi$ is a 2-cell $p\Rightarrow q\circ f$ in $\cat{B}$, as in the diagram
  \[\begin{tikzcd}[row sep=scriptsize]
    A && B \\
    \\
    & X
    \arrow["f", from=1-1, to=1-3]
    \arrow[""{name=0, anchor=center, inner sep=0}, "p"', from=1-1, to=3-2]
    \arrow[""{name=1, anchor=center, inner sep=0}, "q", from=1-3, to=3-2]
    \arrow["\phi", shorten <=6pt, shorten >=6pt, Rightarrow, from=0, to=1]
  \end{tikzcd} \; . \]
  A 2-cell $(f,\phi)\Rightarrow(g,\gamma)$ is a 2-cell $\alpha:f\Rightarrow g$ in $\cat{B}$ such that
  \[
  \begin{tikzcd}[sep=small]
    p && {q\circ f} && {g\circ g}
    \arrow["\phi", Rightarrow, from=1-1, to=1-3]
    \arrow["q\circ\alpha", Rightarrow, from=1-3, to=1-5]
  \end{tikzcd}
  \quad = \quad
  \begin{tikzcd}[sep=small]
    p && {q\circ g}
    \arrow["\gamma", Rightarrow, from=1-1, to=1-3]
  \end{tikzcd} \; . \]
  (In this definition, $\circ$ denotes horizontal composition in $\cat{B}$.)
  The horizontal composition in $\cat{B}/X$ is given in the obvious way by the horizontal composition of the relevant 1- and 2-cells.
  Likewise, vertical composition in $\cat{B}/X$ is vertical composition in $\cat{B}$.
  (It is easy to check that these definitions all satisfy the relevant axioms, hence constituting a valid bicategory.)
\end{defn}

We will see how this structure works in practice in our examples of parameterization below.

\begin{rmk}
  To avoid venturing into 3- and 4-dimensional category theory, we will restrict the hom-bicategories of $\eP\BLens{}_2$ and $\eP\SGame{}$ to be \textit{locally discrete}, with the parameterizing objects being mere sets (treated as discrete categories).
  Strictly speaking, our parameterizing sets will be the underlying sets of differential manifolds --- specifically, the trivial manifolds $\rr^n$ --- and we could treat them properly as parameterizing categories by using their groupoidal (path) structure, but we do not pursue this here.
  (Alternatively, we could follow the idea of Proposition \ref{prop:para-1cat} and truncate the hom-categories by quotienting by connected components; but this turns the 1-cells into equivalence classes of functors, which are again more complicated than we have the need or appetite for here.)
\end{rmk}

Restricting $\eP$ to discrete parameterization means that we instantiate $\eP\BLens{}_2$ and $\eP\SGame{}$ as follows.
Both being constructed over $\ccopara(\cat{C})$, we build up from $\eP\ccopara(\cat{C})$, after first sketching the horizontal composition of externally parameterized bicategories.

\begin{rmk} \label{rmk:ext-para-comp}
  Given a bicategory $\cat{B}$, horizontal composition in $\eP\cat{B}$ is obtained from the strong monoidal structure of the covariant self-indexing (which follows from the universal property of the product of categories) and the horizontal composition in $\cat{B}$.
  For each triple of 0-cells $a,b,c:\cat{B}$, the composition pseudofunctor is given by
  \begin{align*}
    & \eP\cat{B}(b,c)\times\eP\cat{B}(a,b) = \Cat{Cat}/\cat{B}(b,c) \times \Cat{Cat}/\cat{B}(a,b) \; \cdots \\ & \cdots \xto{\sim} \Cat{Cat}/\bigl(\cat{B}(b,c)\times\cat{B}(a,b)\bigr) \xto{\Cat{Cat}/\circ_{a,b,c}} \Cat{Cat}/\cat{B}(a,c) = \eP\cat{B}(a,c) \; .
  \end{align*}
  Observe that this generalizes the lower-dimensional case of Definition \ref{def:ext-para}: first, we take the product of the parameterizing functors, and then we compose in their codomain.
\end{rmk}

\begin{ex} \label{ex:ext-para-ccopara}
  The 0-cells of $\eP\ccopara(\cat{C})$ are the 0-cells of $\ccopara(\cat{C})$, which are in turn the objects of $\cat{C}$.
  A 1-cell from $X$ to $Y$ is a choice of (discrete) parameterizing category (hence a set) $\Theta$, along with a functor $\Theta\to\ccopara(\cat{C})(X,Y)$.
  More intuitively, we can think of such a 1-cell as a morphism in $\cat{C}$ that is both (externally) parameterized and (internally) coparameterized, and write it as $f:X\xto[M]{\Theta}Y$, denoting a 1-cell with parameter $\Theta$ (in the base of enrichment of $\cat{C}$), domain $X$, codomain $Y$, and coparameter $M$.

  A 2-cell from $f:X\xto[M]{\Theta}Y$ to $f':X\xto[M']{\Theta'}Y$ is a pair $(\phi,\varphi)$ of a functor $\phi:\Theta\to\Theta'$ and a natural transformation $\varphi:f\Rightarrow f'\circ\phi$.
  The functor $\phi$ changes the parameterization; and the natural transformation $\varphi$ permits additionally a compatible change of \textit{co}parameterization, being given by a natural family of 2-cells in $\ccopara(\cat{C})$
  \[ \varphi^\theta : \bigl( f^\theta:X\xto[M]{\theta}Y \bigr) \; \Rightarrow \; \bigl( f'^{\phi(\theta)}:X\xto[M']{\phi(\theta)}Y \bigr) \]
  indexed by the parameters $\theta:\Theta$.
  (With discrete parameterization, such a family is trivially natural.)
  Recalling the definition of $\ccopara$ in Theorem \ref{thm:copara2}, this means that each component $\varphi^\theta$ corresponds to a morphism $X\otimes M\otimes Y\to N$ in $\cat{C}$ satisfying the change of coparameter axiom with respect to $f^\theta$ and $f'^{\phi(\theta)}$.

  Horizontal composition in $\eP\ccopara(\cat{C})$ is as sketched in Remark \ref{rmk:ext-para-comp}: given 1-cells $f:X\xto[M]{\Theta}Y$ and $g:Y\xto[N]{\Omega}Z$, their composite is the evident $g\circ f: X\xto[M]{\Theta}Y\xto[N]{\Omega}Z$ whose parameter is the product $\Omega\times\Theta$ and whose coparameter is the tensor of $M$ and $N$.
  The horizontal composition of 2-cells is likewise by first forming the product of their parameters.
  Vertical composition in $\eP\ccopara(\cat{C})$ is given by the horizontal composition in each lax slice hom (bi)category.
\end{ex}

The structure of $\eP\BLens{}_2$ and $\eP\SGame{}$ follows the same pattern.

\begin{ex} \label{ex:ext-para-blens}
  The 0-cells of $\eP\BLens{}_2$ are the same pairs $(X,A)$ as in $\BLens{}_2$.
  A 1-cell from $(X,A)$ to $(Y,B)$ is a biparameterized Bayesian lens: a pair $(c,c')$ of a biparameterized forwards channel $c:X\xklto[M]{\Theta}Y$ and a biparameterized inversion (state-dependent) channel $c':B\xklto[M']{\Theta;X}A$; here we have denoted the state-dependence and the parameterization together as $\Theta;X$.
  (Note that in all our examples, the forwards and backwards coparameters will be equal, \textit{i.e.}, $M=M'$; \textit{cf.} Remark \ref{rmk:bl-dep-opt} on dependent optics.)

  A 2-cell from $(c,c'):(X,A)\xlensto[M,M']{\Theta}(Y,B)$ to $(d,d'):(X,A)\xlensto[N,N']{\Omega}(Y,B)$ is a triple $(\alpha,\alpha_1,\alpha')$ such that $\alpha$ is a functor $\Theta\to\Omega$, $(\alpha,\alpha_1)$ is a 2-cell $c\Rightarrow d$ in $\eP\ccopara(\cat{C})$ (\textit{cf.} Example \ref{ex:ext-para-ccopara}), and $(\alpha,\alpha')$ is a 2-cell $c'\Rightarrow d'$ in $\eP\SStat(X)(B,A)$.
  The latter means that $\alpha'$ is a family of 2-cells in $\ccopara(\cat{C})(B,A)$
  \[ \alpha'^\theta_\pi : \bigl( c'^\theta_\pi : B\xto[M']{\theta;\pi}A \bigr) \; \Rightarrow \; \bigl( d'^{\alpha(\theta)}_\pi : B\xto[N']{\alpha(\theta);\pi}A \bigr) \]
  natural in $\theta:\Theta$ and indexed by $\pi:\cat{C}(I,X)$.
  (The preceding example shows how this corresponds to an indexed natural family of change-of-coparameter morphisms in $\cat{C}$.)

  Horizontal composition in $\eP\BLens{}_2$ is of course by taking the product of the parameters and then applying horizontal composition in $\BLens{}_2$; and vertical composition is horizontal composition in the lax slices making up each hom (bi)category.
\end{ex}

\begin{ex} \label{ex:ext-para-sgame}
  Statistical games are obtained by attaching loss functions to Bayesian lenses, and hence to understand parameterized statistical games having elaborated parameterized Bayesian lenses in the preceding example, it suffices to exhibit parameterized loss functions.

  A parameterized statistical game $(X,A)\xto[M,M']{\Theta}(Y,B)$ consists of a parameterized Bayesian lens $(X,A)\xlensto[M,M']{\Theta}(Y,B)$ along with a parameterized loss function $B\xklto{\Theta;X}I$ in $\eP\Stat(X)(B,I)$.
  Since $\Stat(X)(B,I)$ is a discrete category, such a loss function is given by a function $\Theta_0 \to \Stat(X)(B,I)$, or equivalently (by the Cartesian closure of $\Set$) a function $\Theta_0\times\cat{C}(I,X)\to\cat{C}(B,I)$.
  In the case where $\cat{C} = \Cat{sfKrn}$, this means a function $\Theta_0\times\Cat{sfKrn}(1,X)\times B\to\rr_+$ which is measurable in $B$.

  A 2-cell from the parameterized statistical game $(c,c',K):(X,A)\xto[M,M']{\Theta}(Y,B)$ to $(d,d',L):(X,A)\xto[N,N']{\Omega}(Y,B)$ is a quadruple $(\alpha,\alpha_1,\alpha',\tilde{\alpha})$ where $(\alpha,\alpha_1,\alpha')$ is a 2-cell of Bayesian lenses and $\tilde{\alpha}$ is a family of parameterized loss functions $B\xklto{\Theta;X}I$ such that $K^\theta = L^{\alpha(\theta)} + \tilde{\alpha}^\theta$, naturally in $\theta:\Theta$.

  Horizontal and vertical composition of parameterized statistical games and their 2-cells follow the pattern of the preceding examples.
\end{ex}

Because $\eP$ is functorial, we can consider parameterized versions of the inference systems and loss models that we introduced in \secref{sec:loss-models}.
We can think of parameterization as introducing a `hole' in a structure (such as an extra input to a system), and parameterized inference systems and loss models are inference systems and loss models that account for (and possibly modulate) such holes.

\begin{ex}
  Suppose $(\cat{D},\ell)$ is an inference system in $\cat{C}$.
  $\eP$ acts on the canonical inclusion $(-)^{\smallbcopier}:\cat{D}\hookrightarrow\ccopara[l](\cat{D})$ to return the inclusion $\eP(-)^{\smallbcopier}:\eP\cat{D}\hookrightarrow\eP\ccopara[l](\cat{D})$, which maps a parameterized channel $d:X\xklto{\Theta}Y$ to its trivially coparameterized form $d:X\xklto[I]{\Theta}Y$.

  $\ell$ then maps a channel $d$ to a lens $(d,\ell d)$.
  If $d$ is parameterized by $\Theta$, then its inversion $\ell d$ under $\ell$ will be parameterized accordingly, so that the whole lens $(d,\ell d)$ has parameter $\Theta$.
  This mapping is the action of the pseudofunctor $\eP\ell:\eP\cat{D}^{\smallbcopier}\to\eP\BLens{}_2|_{\cat{D}}$, induced by the parameterization of $\ell$.
\end{ex}

However, in the next section, we will want approximate inference systems that do not just preserve an existing parameterization, but which also add to it, equipping (possibly parameterized) morphisms with inversions that may have their own distinct capacity for improvement or learning.
For this reason, we make the following definition.

\begin{defn}
  Suppose $(\cat{C},\otimes,I)$ is a copy-delete category.
  A \textit{parameterized inference system} in $\cat{C}$ is a pair $(\cat{B},\ell)$ of a sub-bicategory $\cat{B}\hookrightarrow\eP\cat{C}$ along with a (strong functorial) section $\ell:\cat{B}^{\smallbcopier}\to\eP\BLens{}_2|_{\cat{B}}$ of the restriction $\eP(\pLens)|_{\cat{B}}$ to $\cat{B}$ of the parameterized 2-fibration $\eP(\pLens):\eP\BLens{}_2\to\eP\ccopara[l](\cat{C})$, where $\cat{B}^{\smallbcopier}$ is the essential image of the restriction to $\cat{B}$ of the (parameterized) canonical lax inclusion $\eP(-)^{\smallbcopier}:\eP\cat{C}\hookrightarrow\eP\ccopara[l](\cat{C})$.
  We say \textit{lax parameterized inference system} when $\ell$ is a lax functor.
\end{defn}

A trivial example of a lax parameterized inference system is obtained by taking the parameters to be hom categories, and the choice functor to be the identity, as the following example shows.

\begin{ex}
  The following data define a lax parameterized inference system $\ell$ acting on the entirety of $\eP\cat{C}$.
  First, let $\Rho(X,Y,M)$ denote the full subcategory of $\SStat(X)(Y,X)$ on those objects (state-dependent morphisms) with coparameter $M$.
  Then $\ell$ is defined as follows.
  \begin{enumerate}[label=(\roman*)]
  \item Each 0-cell $X$ is mapped to the 0-cell $(X,X)$.
  \item Each 1-cell $c:X\xklto[M]{\Theta}Y$ is mapped to the parameterized lens $(c,c'):(X,X)\xlensto[M]{\Theta\times\Rho(X,Y,M)}(Y,Y)$ whose forward channel is chosen by
    \[
    \Theta\times\Rho(X,Y,M) \xto{\proj_1} \Theta \xto{c} \ccopara[l](\cat{C})(X,Y)
    \]
    and whose inverse channel $c':Y\xklto[M]{\Rho(X,Y,M);X}X$ is chosen by
    \[
    \Theta\times\Rho(X,Y,M) \xto{\proj_2} \Rho(X,Y,M) \hookrightarrow \SStat(X)(Y,X)
    \]
  \item Each 2-cell $(a,\alpha):\bigl(c:X\xklto[M]{\Theta}Y\bigr)\Rightarrow\bigl(d:X\xklto[M']{\Theta'}Y\bigr)$ is mapped to the 2-cell $(a\times\alpha_\ast,\alpha,\alpha)$, where $\alpha_\ast$ is the functor defined by post-composing with $\alpha$ taken as a family of 2-cells in $\ccopara[r](\cat{C})$ and hence in $\Rho(X,Y,M)$.
  \end{enumerate}
  \begin{proof}
    First, we confirm that the mapping is well-defined on 1-cells (taking it to be evidently so on 0-cells): in general, the coparameters in $\eP\ccopara(\cat{C})$ may depend on the parameters, but here the parameters arise from the embedding $(-)^{\smallbcopier}:\eP\cat{C}\to\eP\ccopara(\cat{C})$.
    The only coparameters are therefore those that arise by copy-composition, and their type is thus not parameter-dependent.
    It is therefore legitimate to map a 1-cell $c:X\xto[M]{\Theta}Y$ to a lens with type $(X,X)\xlensto[M]{\Theta\times\Rho(X,Y,M)}(Y,Y)$.
    
    Next, we check well-definedness on 2-cells.
    Note that the 2-cell $(a,\alpha):\bigl(c:X\xklto[M]{\Theta}Y\bigr)\Rightarrow\bigl(d:X\xklto[M']{\Theta'}Y\bigr)$ in $\eP\ccopara[l](\cat{C})$ is constituted by a family of morphisms $\alpha^\theta:X\otimes M\otimes Y \klto M'$, and that a 2-cell $\bigl(Y\xklto[M]{}X\bigr)\Rightarrow\bigl(Y\xklto[M']{}X\bigr)$ in $\ccopara[r](\cat{C})$ has an underlying morphism of the same type; hence each $\alpha^\theta$ witnesses such a 2-cell in $\ccopara[r](\cat{C})$.
    In particular, for each $\pi:I\klto X$ in $\cat{C}$, and for each state-dependent $\rho:Y\xklto[M]{X}X$, $\alpha^\theta$ yields a 2-cell from $\rho_\pi$ to
    \[ \alpha_\ast(\theta,\rho)_\pi \; := \; \input{img/copara2-r-2cell.tikz} \quad . \]
    The functor $\alpha_\ast$ is thus defined by mapping $(\theta,\rho):\Theta\times\Rho(X,Y,M)$ to $\alpha_\ast(\theta,\rho):\Rho(X,Y,M')$; its own action on 2-cells is likewise by parameterized post-composition.
    Finally, note that $d'$ is also given by evaluation, and so $\alpha$ also defines an indexed natural family of 2-cells
    \[ \alpha^{\theta,\rho}_\pi : \bigl( c'^\rho_\pi = \rho_\pi : Y\xklto[M]{}X\bigr) \Rightarrow \bigl( d'^{\alpha_\ast(\theta,\rho)}_\pi = \alpha_\ast(\theta,\rho)_\pi : Y\xklto[M']{}X\bigr) \]
    as required (\textit{cf.} Example \ref{ex:ext-para-blens}).
    Therefore, $(a\times\alpha_\ast,\alpha,\alpha)$ defines a 2-cell in $\eP\BLens{}_2$.
    This is compatible with $\ell$ being a section of $\eP\pLens$, as $(a\times\alpha_\ast,\alpha,\alpha)\mapsto(a,\alpha)$.

    To establish lax unity, we need to exhibit a family of 2-cells $(i_X,{i_X}_1,i_X'):\id_{(X,X)}\Rightarrow(\id_X,\id_X')$ natural in $X$, where $\id_{(X,X)}$ is the identity lens on $(X,X)$ in $\eP\BLens{}_2$ with trivial parameter $\Cat{1}$, $\id_X$ is the likewise trivially parameterized identity on $X$ in $\eP\cat{C}$, and $\id_X'$ is the parameterized state-dependent channel $\id_x':X\xklto[1]{\Cat{1}\times\Rho(X,X,1)}X$ defined by the inclusion
    \[ \Cat{1}\times\Rho(X,X,1) \xto{\sim} \Rho(X,X,1) \hookrightarrow \SStat(X)(X,X) \; . \]
    Clearly $\id_X'$ is not constantly the identity morphism, and this is why $\ell$ is only laxly unital.
    By defining the functor $i_X:\Cat{1}\to\Cat{1}\times\Rho(X,X,1)$ to pick the element $\id_X$, the 2-cell ${i_X}_1$ to be the identity on $\id_X$, and likewise $i_x'$, we obtain the required family of witnesses.

    Lax functoriality is witnessed by a family of 2-cells
    \[ (f_{dc},{f_{dc}}_1,f_{dc}'):(d,d')\lenscirc(c,c')\Rightarrow(d\klcirc c,(d\klcirc c)') \]
    natural in $c:X\xklto[M]{\Theta}Y$ and $d:Y\xklto[N]{\Phi}Z$.
    We define the functor
    \[ f_{dc} : \Theta\times\Rho(X,Y,M)\times\Phi\times\Rho(Y,Z,N) \to \Theta\times\Phi\times\Rho(X,Z,M\otimes N) \]
    by composition, $f_{dc}(\theta,\rho,\phi,\chi) := (\theta,\phi,\rho\circ \chi_{c^\theta})$; it is the fact that not all morphisms $X\klto Z$ factor through $Y$ that makes $\ell$ lax functorial.
    With $f_{dc}$ so defined, we can set both ${f_{dc}}_1$ and $f_{dc}'$ to be identity 2-cells, and thus obtain the requisite witnesses.
  \end{proof}
\end{ex}

On the other hand, the only parameterized loss models we encounter will be those of \secref{sec:sgame-examples} under the action of $\eP$.
This is because the ability to change is part of the system itself, rather than part of how we measure the system\footnote{
Physically speaking, we adopt the `Schrödinger' perspective on change rather than the `Heisenberg' perspective.}:
we do not seek to ``move the goalpoasts''.
(In future work, we may consider systems whose performance is dependent on some broader context; but not here.)
Therefore, our parameterized loss models will all be of the following form.

\begin{ex}
  If $L$ is a loss model for $\cat{B}$, then its parameterization $\eP L$ assigns to a parameterized Bayesian lens $(c,c'):(X,A)\xlensto[M,M']{\Theta}(Y,B)$ the correspondingly parameterized statistical game $\bigl(c,c',L(c)\bigr)$.
  The parameterized loss function $L(c)$ thus also has parameter $\Theta$ and depends accordingly on it, with type $L(c):B\xklto{\Theta;X}I$.
  For each $\theta:\Theta$, its component is the loss function $L(c)^\theta:B\xklto{X}I$ which is assigned to the lens $(c^\theta,c'^\theta)$ by $L$ (as a loss model applied to an unparameterized lens).
\end{ex}

\begin{rmk}
  Before we move on to examples of approximate inference doctrines, let us note the similarity of the notions of externally parameterized lens (Example \ref{ex:ext-para-blens}), cilia (Definition \ref{def:cilia}), and differential cilia (Definition \ref{def:diff-cilia}):
  both of the latter can be considered as externally parameterized lenses with extra structure, where the extra structure is a morphism or family of morphisms back into (an algebra of) the parameterizing object:
  in the case of differential cilia, this `algebra' is the tangent bundle; for (dynamical) cilia, it is trivial; and forgetting this extra structure returns a mere external parameterization.
  Notably, the `input' on the external hom polynomial defining both types of cilia (Definition \ref{def:blhom}) corresponds precisely to the domain of the loss function of a statistical game; and so the domains of the update maps of either type of cilia correspond to the domains of parameterized loss functions.
  We will make use of this correspondence in defining approximate inference doctrines in the next section.
\end{rmk}

\section{Approximate inference doctrines} \label{sec:doctrines}

We are at last in a position to build the bridge between abstract statistical models and the dynamical systems that play them: functors from a copy-discard category of parameterized channels to a category of cilia that factorize through an inference system (modelling how the system inverts the channels) and possibly a loss model (encoding how well the system is doing).
In the round, we can think of the resulting approximate inference doctrines as ``dynamical algebras'' for categories of parameterized stochastic channels (considered as statistical models), which take the parameters as part of the dynamical state space so that they might improve themselves.
This line of thinking leads us to the following definitions.

\begin{defn}
  Let $(\cat{C},\otimes,I)$ be a copy-discard category, and let $(\cat{B},\ell)$ be a parameterized inference system in $\cat{C}$.
  \begin{enumerate}[label=(\alph*)]
  \item An \textit{approximate inference doctrine} is a pseudofunctor $\cat{B}\to\Cilia^\Tt$ that factors through $\ell$, as
    \[ \cat{B} \xto{\eP(-)^{\smallbcopier}|_{\cat{B}}} \cat{B}^{\smallbcopier} \xto{\ell} \im(\ell) \xto{\Fun{D}} \Cilia^\Tt \; . \]
    We say \textit{lax approximate inference doctrine} if $\Fun{D}$ is instead a lax functor.
  \item An \textit{approximate inference doctrine with loss} $L$ is an approximate inference doctrine along with a loss model $L$ for $\im(\ell)$, a pseudofunctor $\Fun{D}^L:\im(L)\to\Cilia^\Tt$, and an icon $\lambda:\Fun{D}\Rightarrow \Fun{D}^L\circ L$, as in the diagram
    \[\begin{tikzcd}[sep=scriptsize]
      {\cat{B}} && {\cat{B}^{\smallbcopier}} && {\im(\ell)} &&&& {\Cilia^\Tt} \\
      \\
      &&&&&& {\im(L)}
      \arrow[""{name=0, anchor=center, inner sep=0}, "{\Fun{D}}", from=1-5, to=1-9]
      \arrow["\ell", from=1-3, to=1-5]
      \arrow["{\eP(-)^{\smallbcopier}|_{\cat{B}}}", from=1-1, to=1-3]
      \arrow["L", from=1-5, to=3-7]
      \arrow["{\Fun{D}^L}"', from=3-7, to=1-9]
      \arrow["\lambda", shorten <=7pt, shorten >=7pt, Rightarrow, from=0, to=3-7]
    \end{tikzcd} \quad . \]
    We say \textit{lax approximate inference doctrine with loss} if $L$ and $\Fun{D}^L$ are lax functors.
  \item A \textit{differential approximate inference doctrine with loss} $L$ is an approximate inference doctrine with loss $L$ such that $\Fun{D}^L$ factors through a differential system, as in the diagram
    \[\begin{tikzcd}[sep=scriptsize]
      {\cat{B}} && {\cat{B}^{\smallbcopier}} && {\im(\ell)} &&& {} & {\Cilia^\Tt} \\
      \\
      &&&&&& {\im(L)} & {} & {\Cilia^\Tt} \\
      \\
      &&&&&&& \DiffCilia
      \arrow["{\Fun{D}}", from=1-5, to=1-9]
      \arrow["\ell", from=1-3, to=1-5]
      \arrow["{\eP(-)^{\smallbcopier}|_{\cat{B}}}", from=1-1, to=1-3]
      \arrow[""{name=0, anchor=center, inner sep=0}, "{\Fun{D}^L}", from=3-7, to=3-9]
      \arrow[Rightarrow, no head, from=1-9, to=3-9]
      \arrow["\nabla"', from=3-7, to=5-8]
      \arrow["\int"', from=5-8, to=3-9]
      \arrow["L"', from=1-5, to=3-7]
      \arrow["\lambda"{pos=0.4}, shorten <=4pt, shorten >=12pt, Rightarrow, from=1-8, to=3-8]
      \arrow["\delta", shorten <=7pt, shorten >=7pt, Rightarrow, from=0, to=5-8]
    \end{tikzcd} \quad . \]
    We say \textit{lax differential approximate inference doctrine} when $L$,$\nabla$ and $\int$ are lax functors.
  \end{enumerate}
\end{defn}

The different factors of a differential approximate inference doctrine with loss encode the different stages by which a dynamical system is constructed from a statistical model: the parameterized inference system $\ell$ equips a parameterized channel with a parameterized inversion; the loss model $L$ equips the resulting lens with a loss function; the functor $\nabla$ translates this statistical game to a differential system, possibly representing gradient descent on the loss; and finally the functor $\int$ turns this differential system into a dynamical system that `flows', possibly by integration.

With these definitions to hand, we come to our motivating neuroscientific examples.
First (\secref{sec:doctrines-laplace}), we formalize predictive coding using the Laplace approximation to the free energy \parencite{Friston2009Predictive,Bastos2012Canonical,Bogacz2017tutorial}, which we saw in \secref{sec:laplace} forms a lax loss model for Gaussian lenses.
This approximation allows the resulting dynamical systems to exhibit some biological plausibility, with prediction errors computed linearly and the dynamical updates obtained as affine transformations of prediction errors.
We call this the \textit{Laplace doctrine}.

Apart from requiring Gaussian channels, the Laplace doctrine is agnostic about how predictions are actually generated, and it does not produce systems which are able to improve their predictions; they have no `synaptic' plasticity, and thus do not learn.
To remedy this, our second example of an approximate inference doctrine (\secref{sec:doctrines-hebb-laplace}) is more opinionated about the predictive forward channels, restricting them to be of the form $x\mapsto \theta\,h(x) + \omega$ where $\theta$ is a square matrix on $Y$, $h$ is a differentiable function $X\to Y$, and $\omega$ is distributed according to a Gaussian on $Y$; compare this with the form of the firing rate dynamics of rate-coded neural circuits in Definition \ref{def:rate-circ}.
The `synaptic' parameter (or weight matrix) $\theta$ can then be learnt, and this is incorporated into the state space of the systems produced by the corresponding \textit{Hebb-Laplace doctrine}, which formalizes another standard scheme in the neuroscience literature \parencite{Bogacz2017tutorial}.
The name of this doctrine indicates another aspect of the biological plausibility of this scheme: the $\theta$-updates can be seen as a form of Hebbian learning \parencite{Hebb1949organization}.

\begin{rmk}
  In what follows, in order to focus on exemplification, we omit a full treatment of all the higher-categorical structure.
  This means that we do not consider the action of the doctrines on 2-cells, and leave leave the full elaboration of the 2-dimensional structure to future work.
  Our main concern in this final part is the scientific question of the compositional structure of predictive coding, and one further mathematical consequence of this is that the inference systems on which the doctrines are based will not be unital: the schemes that are presented in the literature involve mappings which do not preserve identity channels.
\end{rmk}

\subsection{Predictive coding circuits and the Laplace doctrine}
\label{sec:doctrines-laplace}

\begin{notation}
  Any category $\cat{C}$ embeds into its external parameterization $\eP\cat{C}$ by mapping every morphism to its trivially parameterized form; in a mild abuse of notation, we will denote the image of this embedding simply by $\cat{C}$.
  In this section, we will work with the trivial parameterization of the subcategory $\FdGauss$ of $\Cat{sfKrn}$ of Gaussian kernels between finite-dimensional Cartesian spaces (Definition \ref{def:fd-gauss}).
  Hence, when we write $\FdGauss^{\smallbcopier}$, it denotes the image of $\FdGauss$ under $\eP(-)^{\smallbcopier}$.
\end{notation}

We begin by presenting the action of the Laplace doctrine on a (non-coparameterized\footnote{
Note that all coparameterized channels of interest are obtained as the copy-composites of non-coparameterized channels.
}) Gaussian channel $c$.
Below, we will see how the resulting system is obtained from a differential approximate inference doctrine with the Laplacian free energy loss.

\begin{prop} \label{prop:laplace-sys}
  Suppose $c:X\klto Y$ is a morphism in $\FdGauss$, and fix a ``learning rate'' $\lambda:\rr$.
  Then the following data define a system $\Laplace_\lambda(c):(X,X)\to(Y,Y)$ in $\Cilia^\nn$, following the representation of Proposition \ref{prop:unpack-hier}.
  \begin{enumerate}[label=(\roman*)]
  \item the state space is $X$;
  \item the forwards output map $\Laplace_\lambda(c)^o_1 : X\times X \to \gauss(Y)$ is defined by
    \[ \Laplace_\lambda(c)^o_1 : X\times X \xto{\proj_2} X \xto{c} \gauss(Y) \; ; \]
  \item the inverse output map $\Laplace_\lambda(c)^o_2 : X\times \gauss(X) \times Y \to \gauss(X)$ is defined by
    \begin{align*}
      \Laplace_\lambda(c)^o_2 : X\times \gauss(X) \times Y &\to \rr^{|X|}\times \rr^{|X|\times|X|} \hookrightarrow \gauss(X) \\
      (x,\pi,y)&\mapsto\bigl(x,\Sigma_{c'}(x,\pi,y)\bigr)
    \end{align*}
    where the inclusion picks the Gaussian state with the indicated statistical parameters, whose covariance $\Sigma_{c'}(x,\pi,y) := \left(\partial_{x}^2 E_{(c,\pi)}\right)(x, y)^{-1}$ is defined following equation \eqref{eq:laplace-sigma-rho-pi} of Lemma \ref{lemma:laplace-approx} (with trivial coparameterization $M=1$);
  \item the update map $\Laplace_\lambda(c)^u:X\times \gauss(X)\times Y\to \Ga(X)$ is defined by
    \begin{align*}
      \Laplace_\lambda(c)^u : X\times \gauss(X) \times Y &\to X \hookrightarrow \Ga(X) \\
      (x,\pi,y) &\mapsto x + \lambda \, \partial_{x} \mu_c(x,y)^T \eta_c(x,y) - \lambda \, \eta_\pi(x)
    \end{align*}
  \end{enumerate}
  where the inclusion $X\hookrightarrow\Ga(X)$ is given by the unit of the Giry monad $\Ga$ which takes each $x:X$ to the corresponding delta distribution, and where $\eta_c(x,y) := \Sigma_c(x)^{-1}\,\epsilon_c(y,x)$ and $\eta_\pi(x) := \Sigma_\pi^{-1}\,\epsilon_\pi(x)$.
\end{prop}

\begin{rmk}
  Note that the update map $\Laplace_\lambda(c)^u$ is actually deterministic, in the sense that it is defined as a deterministic map followed by the unit of the probability monad.
  However, the general stochastic setting is necessary, because the composition of system depends on the composition of Bayesian lenses; recall Definition \ref{def:blhom-comp}, which defines the bidirectional composition of cilia.
  Intuitively, we can consider a composite system $\Laplace_\lambda(d)\circ\Laplace_\lambda(c)$ and note that the forward inputs to the $d$ component and the backward inputs to the $c$ component will be sampled from the stochastic outputs of $c$ and $d$ respectively.
  Because these inputs are passed to the corresponding update maps, the updates inherit this stochasticity.
\end{rmk}

\begin{rmk}
  The terms $\eta_c(x,y) = \Sigma_c(x)^{-1}\,\epsilon_c(y,x)$ in the update map of the Laplace doctrine can be understood as \textit{precision-weighted error} terms: the inverse covariance $\Sigma_c(x)^{-1}$ encodes the `precision' of the distribution (consider the univariate case); and the term $\epsilon_c(y,x) = y - \mu_c(x)$ encodes the `error' between the observation $y$ and the predicted mean $\mu_c(x)$.
  The representation of prediction errors is a hallmark of predictive coding schemes.
\end{rmk}

To define an approximate inference doctrine, we need a (parameterized) inference system.
For predictive coding, this will be obtained by assigning to each channel $c$ an inversion whose parameter represents the mean of the emitted posterior; this parameter will later be learned by the resulting doctrine.
In order for this assignment to be functorial, we restrict the posteriors emitted by this inference system to have diagonal covariance, meaning that there will be no correlations between dimensions.
This formalizes what is known in the literature as a \textit{mean field} assumption \parencite{Buckley2017free,Friston2007Variational}, without which those earlier works would not have been able to make implicit use of functoriality.

\begin{prop}[Mean field Laplace] \label{prop:laplace-inf-sys}
  As long as $\klcirc$ denotes copy-composition, the following data define a (non-unital) strong parameterized inference system $\ell$ on $\FdGauss$.
  Each 0-cell $X$ is mapped to $(X,X)$.
  Each 1-cell $c:X\xklto[M]{}Y$ is mapped to the parameterized lens $(c,c'):(X,X)\xlensto[M]{X\times M}(Y,Y)$ whose forward channel is $c$ and whose parameterized backward channel $c':Y\xklto[M]{X\times M;X}X$ emits the Gaussian with mean $(x,m):X\times M$ determined by the parameter and which minimizes the (mean-field) Laplacian free energy.
  Thus, writing $\bigl(\mu_{c'_\pi}^{x,m}(y), \Sigma_{c'_\pi}^{x,m}(y)\bigr)$ for the statistical parameters of $c'^{x,m}_\pi(y)$, $\ell$ assigns
  \[
  \mu_{c'_\pi}^{x,m}(y) := (x,m)
  \qquad\text{and}\qquad
  \Sigma_{c'_\pi}^{x,m}(y) :=
  \begin{pmatrix*}
    \bigl(\dhess{x} E_{(c,\pi)}\bigr)(x,m,y)^{-1} & 0 \\
    0 & \bigl(\dhess{m} E_{(c,\pi)}\bigr)(x,m,y)^{-1}
  \end{pmatrix*} \; .
  \]
  where $\dhess{}$ denotes the diagonal Hessian\footnote{
  That is, $\dhess{x} f(x)$ can be represented as the matrix with diagonal equal to the diagonal of the Hessian matrix $\partial^2_x f(x)$ and with all other coefficients 0.}
  It is the diagonal structure of $\Sigma_{c'_\pi}^{x,m}$ that justifies the `mean-field' moniker.
  \begin{proof}
    First, we note that $\ell$ fails to be unital because, for each identity channel $\id_X:X\xklto X$, the mean of the assigned inversion $\id_X'$ is determined by the parameter $X$, rather than the input.
    If this parameter happens to equal to the input, then $\id_X'$ will actually act as the identity channel.
    This is because we can understand the identity channel as the limit as $\sigma\to 0$ of a Gaussian with mean equal to the input $x$ and variance $\sigma \Cat{1}_X$ (where $\Cat{1}_X$ is the identity matrix on $X$).
    Informally, we have $\Sigma^x_{(\id_X')_\pi}(x') = \bigl(\dhess{x} E_{(\id_X,\pi)}(x,x')^{-1} = 0$, and so $(\id_X')^x$ acts as the Dirac delta distribution on the parameter $x$; but of course in general the parameter $x$ need not equal the forward input.

    Next, we show that $\ell$ is strongly functorial (as long as $\klcirc$ is always interpreted as copy-composition).
    If $c:X\xklto[M]{}Y$ and $d:Y\xklto[N]{}Z$ are composable Gaussian channels, then the statistical parameters of the composite approximate inversion $c'\circ d'_c:Z\xklto[MYN]{YNXM;X}X$ are $\mu_{(c'\circ d'_c)_\pi}^{y,n,x,m}(z) = (x,m,y,n)$ and
    \[ \Sigma_{(c'\circ d'_c)_\pi}^{y,n,x,m}(z) = \mathsf{diag} \begin{bmatrix*}
      \bigl(\dhess{x} E_{(c,\pi)}\bigr)(x,m,y)^{-1} \\
      \bigl(\dhess{m} E_{(c,\pi)}\bigr)(x,m,y)^{-1} \\
      \bigl(\dhess{y} E_{(d,c\klcirc\pi)}\bigr)(x,m,y,n,z)^{-1} \\
      \bigl(\dhess{n} E_{(d,c\klcirc\pi)}\bigr)(x,m,y,n,z)^{-1}
    \end{bmatrix*} \; . \]
    Note that, by interpreting $\klcirc$ as copy-composition, we have
    \[ E_{(d,c\klcirc\pi)}(x,m,y,n,z) = - \log p_d(n,z|y) - \log p_c(m,y|x) - \log p_\pi(x) \; . \]
    On the other hand, $\ell$ assigns to $d\klcirc c:X\xklto[MYN]{}Z$ the lens $\bigl(d\klcirc c, (d\klcirc c)'\bigr)$ whose inversion $(d\klcirc c)':Z\xklto[MYN]{XMYN;X}X$ is defined by the statistical parameters $\mu_{(d\klcirc c)'_\pi}^{x,m,y,n}(z) = (x,m,y,n)$ and
    \[ \Sigma_{(d\klcirc c)'_\pi}^{x,m,y,n}(z) = \mathsf{diag} \begin{bmatrix*}
      \bigl(\dhess{x} E_{(d\klcirc c,\pi)}\bigr)(x,m,y,n,z)^{-1} \\
      \bigl(\dhess{m} E_{(d\klcirc c,\pi)}\bigr)(x,m,y,n,z)^{-1} \\
      \bigl(\dhess{y} E_{(d\klcirc c,\pi)}\bigr)(x,m,y,n,z)^{-1} \\
      \bigl(\dhess{n} E_{(d\klcirc c,\pi)}\bigr)(x,m,y,n,z)^{-1}
    \end{bmatrix*} \]
    where
    \begin{align*}
      E_{(d\klcirc c,\pi)}(x,m,y,n,z) &= - \log p_d(n,z|y) - \log p_c(m,y|x) - \log p_\pi(x) \\
      &= E_{(d,c\klcirc\pi)}(x,m,y,n,z) \; .
    \end{align*}
    Consequently, $\Sigma_{(d\klcirc c)'_\pi}^{x,m,y,n}(z) = \Sigma_{(c'\circ d'_c)_\pi}^{y,n,x,m}(z)$.
    It therefore suffices to take the laxator $\ell(d)\lenscirc \ell(c) \Rightarrow \ell(d\klcirc c)$ to be defined by the isomorphism $(Y\times N)\times (X\times M) \xto{\sim} (X\times M) \times (Y \times N)$.
  \end{proof}
\end{prop}

\begin{rmk}
  Note that the preceding inference system requires $\klcirc$ to be interpreted as copy-composition everywhere, which is not strictly in accordance with our earlier usage (which mixed copy-composition with ordinary composition in the state-dependence).
  Resolving this irregularity is the subject of ongoing work.
\end{rmk}

\begin{prop} \label{prop:sgd-laplace}
  Stochastic gradient descent with respect to the mean parameter of Laplacian free energy games in the image of $\ell$ yields a strong functor $\nabla:\cat{L}\to\DiffCilia$, where $\cat{L}$ is the essential image of $\LFE$ restricted to the essential image of $\ell$.
  If $c := (c,c',L^c):(X,X)\xto[M]{X\times M}(Y,Y)$ is such a game (a 1-cell) in $\cat{L}$, then $\nabla c$ is the differential cilium $(X,X)\xto{X\times M}(Y,Y)$ with state space equal to the parameter $X\times M$ defined as follows.

  For each $(x,m):X\times M$, $\nabla c$ outputs the non-coparameterized Bayesian lens $\ell(c)^{x,m}_{\smallground}:(X,X)\lensto(Y,Y)$ obtained by taking the dynamical state $(x,m)$ as the parameter of the lens and discarding any coparameters.

  The `update' vector field $(\nabla c)^u : (X\times M) \to \gauss(X) \to Y \klto \Tan (X\times M)$ is obtained by taking the negative gradient of the loss function $L^c:Y\xklto{\gauss(X\times M);X}I$ with respect to the posterior mean parameter, evaluated at the posterior mean:
  \begin{align*}
    (X\times M) \to \gauss(X) \to Y &\to \Tan (X\times M) \\
    (x,m,\pi,y) &\mapsto - \left( \partial_{(x,m)} E_{(c,\pi)} \right) (x,m,y) \; .
  \end{align*}
  (This yields a morphism in $\Cat{sfKrn}$ via the embedding $\Cat{Meas}\hookrightarrow\Cat{sfKrn}$; it is clearly measurable as it is a continuous function between Cartesian spaces.)
  \begin{proof}
    Since the state space $X\times M$ is the space of means of the Laplacian posteriors, the `update' action of $\nabla c$, the open vector field $(\nabla c)^u$, is defined as the (negative) gradient of $L^c$ with respect to these means (so that the associated flow performs gradient descent).
    The parameterized loss function $L^c:Y\xklto{X\times M;X}I$ encodes the Laplacian free energy associated to the parameterized lens $(c,c')$, and corresponds (by Example \ref{ex:sfkrn-bilin-eff}) to the function
    \begin{align*}
      X\times M\to\gauss(X)\to Y &\to[0,\infty] \\
      (x,m,\pi,y)&\mapsto \LFE(c,c'^{x,m})_\pi(y)
    \end{align*}
    where
    \[ \LFE(c,c'^{x,m})_\pi(y) = E_{(c,\pi)}(x, m, y) - S_{X\otimes M}[c'^{x,m}_\pi(y)] \; . \]
    The entropy $S_{X\otimes M}[c'^{x,m}_\pi(y)]$ does not depend on the mean of $c'^{x,m}_\pi(y)$, and so the gradient of $\LFE(c,c'^{x,m})_\pi(y)$ with respect to $(x,m)$ is simply $\left(\partial_{(x,m)} E_{(c,\pi)}\right)(x,m,y)$.
    Hence defining $(\nabla c)^u$ as stated yields
    \[ (\nabla c)^u : (x,m,\pi,y) \mapsto -\left(\partial_{(x,m)} E_{(c,\pi)}\right)(x,m,y) \; . \]

    We now show that $\nabla$ is strongly functorial with respect to composition of 1-cells in $\cat{L}$.
    First, we check that $\nabla$ satisfies the strong unity axiom, which means we need a 2-isomorphism $\id_{(X,X)} \Rightarrow \nabla(\id_{(X,X)})$ in $\DiffCilia$.
    Note that the cilium $\id_{(X,X)}$ has trivial state space $1$, trivial update map, and outputs the identity lens $(X,X)\lensto(X,X)$.
    Likewise, the identity game $\id_{(X,X)}$ has trivial parameter $1$, loss function equal to $0$, and lens being the (trivially coparameterized copy-composite) identity lens $(X,X)\lensto(X,X)$.
    Since the loss function is constantly $0$ with trivial parameter, $\nabla$ acts to return a cilium $(X,X)\xto{1}(X,X)$ again with trivial state space and which constantly outputs the identity lens; its update map is likewise trivial.
    Therefore we take the 2-cell $\id_{(X,X)} \Rightarrow \nabla(\id_{(X,X)})$ to be witnessed by the identity $\id_1 : 1\to 1$, which satisfies strong unity \textit{a fortiori}.

    Finally, we check that $\nabla$ satisfies the strong functoriality axiom, meaning that we seek a 2-isomorphism $\nabla(d,d',L^d)\circ\nabla(c,c',L^c) \Rightarrow \nabla\bigl((d,d',L^d)\circ(c,c',L^c)\bigr)$ for each pair of composable Laplacian free energy games $(c,c',L^c):(X,X)\xlensto[M]{X\times M}(Y,Y)$ and $(d,d',L^d):(Y,Y)\xlensto[N]{Y\times N}(Z,Z)$.
    Note that the composite game has the type $(X,X)\xlensto[MYN]{(Y\times N)\times(X\times M)}(Z,Z)$, that by the universal property of $\times$ we have an isomorphism $(Y\times N)\times(X\times M) \cong (X\times M)\times(Y\times N)$, and that the product of Gaussians is again Gaussian.
    Note also that the parameterized loss function $L^d\circ L^c$ equals
    \[\begin{matrix}
      (Y\times N)\times(X\times M) & \to & \gauss(X) &\to& Z &\to& [0,\infty] \\
      (y,n,x,m, && \pi, && z)& \mapsto & (L^c)^{x,m}_\pi\circ d'^{y,n}_{c\klcirc\pi}(z) + (L^d)^{y,n}_{c\klcirc\pi}(z) \; .
    \end{matrix}\]
    On the other hand, the update map of the composite cilium $\bigl(\nabla(d,d',L^d)\circ\nabla(c,c',L^c)\bigr)^u$ equals
    \[\begin{matrix}
    (X\times M)\times(Y\times N) &\to& \gauss(X) &\klto& Z &\to& \Tan\bigl((X\times M)\times(Y\times N)\bigr) \\
    (x,m,y,n, && \pi, && z) &\mapsto& \bigl((\nabla c^u)^{x,m}_\pi\klcirc d'^{y,n}_{c\klcirc\pi}(z),\, (\nabla d^u)^{y,n}_{c\klcirc\pi}(z)\bigr)
    \end{matrix} \; .\]

    The desired 2-isomorphism $\nabla(d,d',L^d)\circ\nabla(c,c',L^c) \Rightarrow \nabla\bigl((d,d',L^d)\circ(c,c',L^c)\bigr)$ is thus witnessed by a map $(Y\times N)\to(X\times M) \to (X\times M)\times(Y\times N)$, which we take to be the symmetry $\mathsf{swap}$ of the categorical product.
    Computing the gradient of the $L$ terms in $L^d\circ L^c$ with respect to the mean of the joint Gaussian $(\chi,\rho)$ yields the update map
    \[\begin{matrix}
    (Y\times N)\times(X\times M) &\to& \gauss(X) &\klto& Z &\to& \Tan\bigl((Y\times N)\times(X\times M)\bigr) \\
    (y,n,x,m, && \pi, && z) &\mapsto& \bigl((\nabla d^u)^{y,n}_{c\klcirc\pi}(z),\, (\nabla c^u)^{x,m}_\pi\klcirc \bkwd{d}^{y,n}_{c\klcirc\pi}(z)\bigr)
    \end{matrix}\]
    which is clearly equal to $\bigl(\nabla(d,d',L^d)\circ\nabla(c,c',L^c)\bigr)^u$ upon composition with $\mathsf{swap}$.
    It therefore only remains to check that the two cilia output the same Bayesian lenses $(X,X)\lensto(Z,Z)$, up to $\mathsf{swap}$.
    This follows from the strong functoriality of $\ell$.
  \end{proof}
\end{prop}

\begin{rmk}
  Although we have defined $\nabla$ manually, we expect that it can alternatively be obtained more abstractly, from a proper treatment of stochastic gradient descent applied to statistical games.
  We leave this to future work.
\end{rmk}

Finally, to obtain the dynamical systems with which we started this subsection (in Proposition \ref{prop:laplace-sys}), we use Euler integration, using the functor $\Fun{Euler}_\lambda$ of Remark \ref{rmk:diff-disc}.

\begin{cor} \label{cor:laplace-doctrine}
  Fix a real number $\lambda:\rr$. By defining $\Laplace_\lambda := \Fun{Euler}_\lambda\circ\nabla\circ\LFE$ one obtains Laplacian predictive coding as a differential approximate inference doctrine, the \textit{Laplace doctrine} for the mean field Laplace inference system $\ell$.
  The systems of Proposition \ref{prop:laplace-sys} are obtained in its image.
  \begin{proof}
    Suppose $c:X\klto Y$ is a morphism in $\FdGauss$.
    It is not coparameterized, so $\ell$ assigns to it the parameter space $X$, which becomes the state space of the cilium $\Laplace_\lambda(c)$.
    By definition, this cilium emits the same lens --- and therefore has the same output maps --- as those given in Proposition \ref{prop:laplace-sys}.
    We therefore only need to check that
    \begin{align*}
      \bigl(\nabla(c)^u\bigr)^x_\pi(y)
      &= - \bigl(\partial_x E_{(c,\pi)}\bigr)(x,y) \\
      &= \partial_x \mu_c(x,y)^T \eta_c(x,y) - \eta_\pi(x) \; .
    \end{align*}
    Recall from Proposition \ref{prop:laplace-approx} that
    \begin{align*}
      E_{(c,\pi)}(x, y)
      &= - \log p_c(y|x) - \log p_\pi(x) \\
      &= - \frac{1}{2} \innerprod{\epsilon_c(y,x)}{{\Sigma_c}^{-1} {\epsilon_c(y,x)}}
      - \frac{1}{2} \innerprod{\epsilon_\pi(x)}{{\Sigma_\pi}^{-1} {\epsilon_\pi(x)}} \\
      &\quad
      + \log \sqrt{(2 \pi)^{|Y|} \det \Sigma_c }
      + \log \sqrt{(2 \pi)^{|X|} \det \Sigma_\pi }
      \; .
    \end{align*}
    It is then a simple exercise in vector calculus to show that
    \[ - \bigl(\partial_x E_{(c,\pi)}\bigr)(x,y) = \partial_x \mu_c(x,y)^T \eta_c(x,y) - \eta_\pi(x) \]
    as required.
  \end{proof}
\end{cor}

\subsection{Synaptic plasticity with the Hebb-Laplace doctrine}
\label{sec:doctrines-hebb-laplace}

The Laplace doctrine constructs dynamical systems that produce progressively better posterior approximations given a fixed forwards channel, but natural adaptive systems---brains in particular---do more than this:
they also refine the forwards channels themselves, in order to produce better predictions.
In doing so, these systems better realize the abstract nature of free energy games, for which improving performance means improving both prediction as well as inversion.
To be able to improve the forwards channel requires allowing some freedom in its choice, which means giving it a nontrivial parameterization.

The Hebb-Laplace doctrine that we introduce in this section therefore modifies the Laplace doctrine by fixing a class of parameterized forwards channels and performing stochastic gradient descent with respect to both these parameters as well as the posterior means;
we call it the \textit{Hebb}-Laplace doctrine as the particular choice of forwards channels results in their parameter-updates resembling the `local' Hebbian plasticity known from neuroscience, in which the strength of the connection between two neurons is adjusted according to their correlation \parencite{Hebb1949organization,Gerstner2002Mathematical,Shouval2007Models,Rolls1998Neural,Dayan2001Theoretical}.
(Here, we could think of the `neurons' as encoding the level of activity along a basis vector.)

We begin by defining the category of these parameterized forwards channels, after which we proceed by modifying the mean-field Laplace inference system and the Laplace doctrine accordingly.

\begin{defn}[`Neural' channels]
  Let $\cat{H}$ denote the subbicategory of $\eP\FdGauss_\ast$ generated by 1-cells $X\klto Y$ of the form
  \begin{align*}
    \Theta_X & \to \gauss_\ast(X,Y) \\
    \theta\;\; & \mapsto \;\; \Bigl( x \mapsto \theta\, h(x) + \omega \Bigr)
  \end{align*}
  where $X$ and $Y$ are finite-dimensional Cartesian spaces, $h$ is a differentiable map $X\to Y$, $\Theta_X$ is the vector space of square matrices on $X$, and $\omega$ is sampled from a Gaussian distribution on $Y$.
\end{defn}

\begin{prop}[Mean field Hebb-Laplace]
  Taking $\klcirc$ as copy-composition, the following data define a (non-unital) strong parameterized inference system $\ell$ on $\cat{H}$.
  Each 0-cell $X$ is mapped to $(X,X)$.
  Each (parameterized) 1-cell $c:X\xklto[M]{\Theta}Y$ is mapped to the parameterized lens $(c,c'):(X,X)\xlensto[M]{\Theta\times(X\times M)}(Y,Y)$ whose forward channel is given by projecting $\Theta$ from $\Theta\times(X\times M)$ and applying $c$, and whose backward channel is defined as in Proposition \ref{prop:laplace-inf-sys} (mean-field Laplacian inference).
  \begin{proof}
    The only difference from Proposition \ref{prop:laplace-inf-sys} is in the forward channel; but these are just taken from $\cat{H}$, and so they compose strongly by assumption.
  \end{proof}
\end{prop}

Like the Laplace doctrine, the Hebb-Laplace doctrine is obtained by stochastic gradient descent with respect to the parameters.

\begin{prop}
  Let $\cat{L}$ denote the essential image of $\LFE$ restricted to the essential image of $\ell$.
  Let $c := (c,c',L^c):(X,X)\xto[M]{\Theta\times(X\times M)}(Y,Y)$ be a 1-cell in $\cat{L}$.
  Then stochastic gradient descent yields an identity-on-objects strong functor $\nabla:\cat{L}\to\DiffCilia$ mapping $c$ to the differential cilium $\nabla(c):(X,X)\xto{\Theta\times(X\times M)}(Y,Y)$ defined as follows.

  For each triple of parameters $(\theta,x,m):\Theta\times(X\times M)$, $\nabla c$ outputs the non-coparameterized Bayesian lens $\ell(c)^{\theta,x,m}_{\smallground}:(X,X)\lensto(Y,Y)$ obtained by taking the dynamical state $(\theta,x,m)$ as the parameter of the lens and discarding any coparameters.

  The vector field $(\nabla c)^u$ is obtained by taking the gradient of the loss function $L^c$ with respect to the `synaptic' parameter $\theta:\Theta$ and the posterior mean $(x,m):X\times M$, evaluated at $(\theta,x,m)$:
  \begin{align*}
    \Theta\times (X\times M) \to \gauss(X) \to Y &\to \Tan (\Theta\times (X\times M)) \\
    (\theta,x,m,\pi,y) &\mapsto - \left( \partial_{(\theta,x,m)} E_{(c,\pi)} \right) (\theta,x,m,y) \; .
  \end{align*}
  \begin{proof}
    The proof is almost identical to that of Proposition \ref{prop:sgd-laplace}: the sole difference is that now we also take gradients with respect to the synaptic parameter $\theta:\Theta$, but the reasoning is otherwise the same.
  \end{proof}
\end{prop}

Finally, we obtain dynamical systems by Euler integration.

\begin{defn}
  Fix a real number $\lambda:\rr$.
  The \textit{Hebb-Laplace doctrine} is obtained as the composite $\Hebb_\lambda := \Fun{Euler}_\lambda\circ\nabla\circ\LFE$, yielding a differential approximate inference doctrine for the mean field Hebb-Laplace inference system $\ell$.
\end{defn}

\begin{cor}
  Suppose $c:X\xklto{\Theta}Y$ is a channel in $\cat{H}$ defined by $c^\theta(x) = \theta\, h(x) + \omega$, for some differentiable $h$ and Gaussian noise $\omega$.
  Then the update map $\Hebb_\lambda(c)^u$ is given by
  \[\begin{matrix}
    \Theta\times X &\to& \gauss(X) &\to& Y &\to& \Theta\times X \\
    (\theta,x, && \pi, && y) &\mapsto& \begin{pmatrix} \theta - \lambda\, \eta_{c^\theta}(x,y) \, h(x)^T \\ x + \lambda \, \partial_x h(x)^T \theta^T \eta_{c^\theta}(x,y) - \lambda\, \eta_\pi(x) \end{pmatrix}
  \end{matrix}\]
  where $\eta_{c^\theta}(x,y) = \Sigma_{c^\theta}^{-1}\,\epsilon_{c^\theta}(y,x)$ and $\eta_\pi(x) = \Sigma_\pi^{-1}\,\epsilon_\pi(x)$ are the precision-weighted error terms.
  \begin{proof}
    Following Corollary \ref{cor:laplace-doctrine} (the Laplace doctrine), We just need to check that
    \[ \bigl( \partial_{(\theta,x)} E_{(c^\theta,\pi)} \bigr)(x,y)
    = \begin{pmatrix*}
      \partial_{\theta} E_{(c^\theta,\pi)} \\
      \partial_{x} E_{(c^\theta,\pi)}
    \end{pmatrix*}(x,y)
    = \begin{pmatrix*}
      \eta_{c^\theta}(x,y) \, h(x)^T \\
      - \partial_x h(x)^T \theta^T \eta_{c^\theta}(x,y) + \lambda\, \eta_\pi(x)
    \end{pmatrix*} \; . \]
    This amounts to verifying that $\partial_x \mu_{c^\theta}(x) = \theta\,\partial_x h(x)$ and that $\partial_\theta E_{(c^\theta,\pi)}(x,y) = \eta_{c^\theta}(x,y) \, h(x)^T$.
    The former holds by the linearity of derivation since $\mu_{c^\theta}(x) = \theta\, h(x)$ by definition; and the latter holds because
    \begin{align*}
      \partial_\theta E_{(c^\theta,\pi)}(x,y)
      &= \frac{-\partial_\theta}{2} \innerprod{\epsilon_{c^\theta}(y,x)}{\Sigma_{c^\theta}^{-1}\, \epsilon_{c^\theta}(y,x)} \\
      &= \frac{-\partial_\theta}{2} \innerprod{y - \theta\, h(x)}{\Sigma_{c^\theta}^{-1}\, \bigl(y - \theta\, h(x)\bigr)} \\
      &= \Sigma_{c^\theta}^{-1} \, \bigl(y - \theta\, h(x)\bigr) \, h(x)^T \\
      &= \Sigma_{c^\theta}^{-1} \, \epsilon_{c^\theta}(y,x) \, h(x)^T \\
      &= \eta_{c^\theta}(x,y) \, h(x)^T
    \end{align*}
    as required.
  \end{proof}
\end{cor}

\begin{rmk} \label{rmk:biophys-bundles}
  From a biophysical point of view, the Hebb-Laplace doctrine so defined has a notably suboptimal feature: the `synaptic' forwards parameter $\theta:\Theta$ is updated on the same timescale $\lambda$ as the parameter $x:X$ that encodes the posterior mean, even though the latter parameter is typically interpreted as encoding the activity of a population of neurons, which therefore changes on a faster timescale than those neurons' synapses.
  Not only is this important for reasons of biological plausibility, but also for mathematical reasons: we should understand the backwards activity as bundled over the forwards synapses, and a change in the parameter $\theta$ should induce a corresponding `transport' of $x$.
  An appropriately geometric treatment of compositional approximate inference and predictive coding, resulting in bundles of open dynamical systems, is again something that we leave to future work.
\end{rmk}

\chapter{Future directions} \label{chp:future}

A powerful motivation propelling the development of this thesis was the belief that science, and particularly the cognitive sciences, will benefit from being supplied with well-typed compositional foundations.
In this final chapter, we survey a number of new vistas that we have glimpsed from the vantage point of our results, and indicate routes that we might climb in order to see them better.

One important benefit of the categorical framework is that it helps us express ideas at a useful level of abstraction, and thereby compare patterns across systems and phenomena of interest.
As a result, although our primary system of interest is the brain, we are aware that much of our work is more diversely applicable, and so our survey here is similarly not restricted to neural systems.
At the same time, as neural systems are our finest examples of natural intelligence, we attempt to stay grounded in current neuroscience.

Beyond the evident shortcomings of the work that we have presented---which we review momentarily---we first consider how to use the categorical language of structure to incorporate structure better into our models themselves (\secref{sec:worlds}), with a particular focus on the brain's ``cognitive maps'' (\secref{sec:cog-cart}).
We will see that the compositional consideration of the structure of open systems naturally leads us to consider \textit{societies} of systems (\secref{sec:society}), and hence the relationships between compositional active inference and single- and multi-agent reinforcement learning and economic game theory (\secref{sec:rl}), although our first priority in this section is the incorporation of action (\secref{sec:action}) and planning (\secref{sec:planning}) into the framework of statistical games.
From our abstract vantage point, there is little difference between societies of agents and collective natural systems such as ecosystems\footnote{%
After all, a single multicellular organism is itself a kind of society of agents.},
and so we then consider the prospects for a compositional mathematics of life (\secref{sec:life}).
Finally, we close with some thoughts on matters of fundamental theory (\secref{sec:theory}).

Before we wade into the thick of it, let us note three prominent examples of the aforementioned evident shortcomings.

Firstly, the current presentation of copy-composite stochastic channels, and the bicategories of lenses and statistical games that result from them, is quite inelegant: the necessity of coparameters introduces substantial complexity that is never repaid, because all coparameters arise from the copy-composition of ordinary channels.
This complexity infects the notion of approximate inference doctrine, which could benefit both from simplification and from further exemplification, ideally by examples drawn from beyond neuroscience.

Secondly, the generative models that we have considered are somehow `static', despite our interest in dynamical systems, and this warrants a satisfactory exploration of \textit{dynamical} generative models.

Thirdly, although we considered ``lower level'' neural circuit models in \secref{sec:sys-circ}, we did not explicitly connect our approximate inference doctrines to these more `biological' models.
A satisfactory account of the Bayesian brain would of course span from abstract principles to detailed biology, a relationship the elaboration of which we sadly we leave to future work.

Fortunately, although these three shortcomings may be pressing, we expect that the pursual of a research programme akin to that sketched below would result in overcoming them.

\section{Structured worlds} \label{sec:worlds}

\subsection{Bayesian sensor fusion}

A situation that is common in natural embodied systems but which is not yet well treated by current statistical and machine learning methods\footnote{%
This is beginning to change: recently, the use of sheaf-theoretic and other applied-topological devices has started to penetrate machine learning \parencite{Bodnar2022Neural,Unyi2022Utility}.},
particularly those that are most popular in computational neuroscience, is that of sensor fusion.
In this situation, one has a number of sensors (such as cameras or retinal ganglion cells) which report spatially situated data, and where the sensor fields \emph{overlap} in the space;
the problem is then how to combine these ``local views'' of the space into a coherent global picture.
Mathematically, fusing `local' data into a `global' representation is the job of \emph{sheaves}:
a sheaf is a ``spatially coherent data type''---something like a bundle for which `local' sections can always be uniquely glued together into a global section---and sheaf theory and the related fields of applied topology and cohomology allow us to judge when it is possible to form a consensus, and quantify the obstructions to the formation of consensus; recent work has also begun to suggest algorithms and dynamics by which we can construct consensus-forming distributed sensor systems \parencite{Hansen2020Laplacians}.

Sheaves therefore allow us to construct and to measure spatially structured data types, but missing from the current sheaf-theoretic understanding of sensor fusion is a thorough treatment of belief and uncertainty, especially from a Bayesian perspective.
Since biological systems contain many distributed sensor types, and each of these systems is constituted by many cells, the mathematics of neural representations may be expected to be sheaf-theoretic.
A first possible extension of the work presented here, therefore, is to extend statistical games and approximate inference doctrines (and hence the classes of model that they encompass) to structured data types such as sheaves.
Because statistical games and approximate inference doctrines are defined using lenses over an abstract category of stochastic channels, we expect that the first step will be to consider categories of channels between sheaves; recently, there has been work on categorifying lenses \parencite{Clarke2020Internal,Clarke2022Introduction}, and we expect that this may prove relevant here.
We also expect that at this point the fibrational structure of statistical games will again prove utile in order that losses may be correctly counted on any overlaps.
Fortunately, being similarly structured, sheaves and fibrations are natural partners, and so we expect that a second spatial extension of the present work will be to exploit the latent geometric structure of fibrations of statistical games.

In this context, we may also encounter connections to sheaf-theoretic approaches to `contextuality', in which answers to questions depend on (the topology of) how they are asked, and which seems to lie at the heart of quantum nonlocality.
It is notable that lenses originated in database theory \parencite{Bohannon2006Relational,Foster2007Combinators} and that contextuality can also be observed in database systems \parencite{Abramsky2019Non,Caru2019Logical}, and so at this point, it may be possible to uncover the mathematical origins of `quantum-like' psychological effects \parencite{Busemeyer2012Quantum,Khrennikov2014Quantum,Aerts2017Quantum}, and relate them formally to other kinds of perceptual bistability that have been interpreted in a Bayesian context \parencite{Jardri2013Computational,Leptourgos2020Circular}.
Sheaves come with a cohomology theory that permits the quantification of the `disagreements' that underlie such paradoxes \parencite{Abramsky2015Contextuality,Curry2013Sheaves,Brown1973Abstract}, and dynamical systems can be designed accordingly to minimize disagreements and thus seek consensus \parencite{Hansen2020Laplacians,Hansen2019Learning,Hansen2020Opinion}.
We hope that these tools will supply new and mathematically enlightening models of these psychological phenomena, while at the same time also suggesting new connections to work on quantum-theoretic formulations of the free-energy framework itself \parencite{Fields2022free,Fields2022Physical}.

The adoption of a sheaf-theoretic framework in this way may furthermore illuminate connections between computational neuroscience and machine learning.
Graph neural networks \parencite{Kipf2016Variational,Kipf2017SemiSupervised,Zhou2018Graph}, and their generalization in `geometric' deep learning \parencite{Bronstein2021Geometric}, are increasingly used to apply the techniques of deep learning to arbitrarily structured domains, and, as indicated above, recent work has found sheaves to supply a useful language for their study \parencite{Bodnar2022Neural}.
In a similar fashion, we expect connections here to the structure of message passing algorithms
\parencite{Pearl1982Reverend,Yedidia2005Constructing,Zhou2018Graph,Morton2014Belief,Vries2017Factor} (also hinted at by \textcite{Sergeant-Perthuis2022Regionalized}) and less conventional structured machine learning architectures such as capsule networks \parencite{Sabour2017Dynamic}.
Finally, each category of sheaves is naturally a topos \parencite{MacLane1992Sheaves}, and hence comes with its own rich internal language, modelling dependent type theory (\textit{cf.} \secref{sec:closed-cats}).

\subsection{Learning structure and structured learning}

Having considered the incorporation of structured data into the process of inference, we can consider the incorporation of structure into the process of learning, and here we make an important distinction between \textit{structured learning} and \textit{learning structure}.
By the former, we mean extending the process of learning to a structured setting (such as the sheaf-theoretic one of the preceding section), whereas by the latter, we mean learning the underlying structures themselves.
This latter process is also known in the literature as \textit{structure learning} \parencite{Tervo2016neural,Summerfield2019Structure,Jafarian2019Structure}, but in order to avoid ambiguity, we swap the order of the two words.

The observation at the end of the preceding section, that each category of sheaves forms a topos, is pertinent here, as dependent type theory formalizes a notion of logical `context', containing the ``axioms that are valid in the present situation'', and determining which (non-tautological) statements can be derived.
In the categorical semantics of dependent type theory, the context is witnessed by the object over which a slice category is defined, and so in some sense it defines the ``shape of the universe''.
By the Grothendieck construction, there is a correspondence between sheaves and certain bundles (objects of slice categories), and so (very roughly speaking) we can think of structured inference and learning as taking place in appropriate slice categories.

In the same way that we motivated random dynamical systems (\textit{qua} bundles, \secref{sec:open-rds}) through ``parameterization by a noise source'', we can think of bundle morphisms as generalized parameterized maps.
The problem of learning structure then becomes a problem of generalized parameter-learning, and much like this can be formalized by a `reparameterization' in the $\para$ construction (\secref{sec:int-para}), in this more general setting it is formalized by the ``generalized reparameterization'' of base-change between topoi (\textit{cf.} Remark \ref{rmk:base-change}).
Base-change synthesizes notions of parallel transport, allowing us to translate spatially-defined data coherently between spaces---and, in particular, along changes of structure; recall our earlier remark about the importance of parallel transport to a biophysically-plausible Hebb-Laplace doctrine (Remark \ref{rmk:biophys-bundles}).
In this setting therefore, we expect that work on functorial lenses \parencite{Clarke2020Internal}, as well as work on functorial data migration \parencite{Spivak2012Functorial,Spivak2022Functorial}, may prove relevant.

At the same time, we expect this line of enquiry to clarify the relationships between our formalism of approximate inference and other related work on the categorical foundations of cybernetics \parencite{Capucci2022Diegetic,Capucci2021Towards}, which have typically been studied in a differential rather than probabilistic setting \parencite{Cruttwell2022Categorical}.
We expect the connection to be made via information geometry \parencite{Amari1997Information,Amari2016Information,Nielsen2018elementary}, where Bayesian inference can be understood both using gradient descent \parencite{Ollivier2019Extended} and as a kind of parallel transport \parencite{Sakthivadivel2022Constraint}.

\subsection{Compositional cognitive cartography} \label{sec:cog-cart}

Natural systems such as animals learn the structure of their environments as they explore them.
We will come below (\secref{sec:action}) to the question of how to incorporate action---and hence exploration---into the compositional framework that we have developed here, but meanwhile we note that the topos-theoretic developments sketched above may provide a suitable setting in which to understand the neural basis for navigation, and help explain how ostensibly `spatial' navigation processes and circuits are invariably involved in more abstract problem solving \parencite{Garvert2017map,Behrens2018What,Mark2020Transferring,Bernardi2018geometry,Bellmund2018Navigating}.

There are two key observations underlying this proposal.
Firstly, a topos is not only a richly structured category of spaces (or spatial types), but it can also be understood as a categorified space itself \parencite{Shulman2017Homotopy}: in this context, we can call each categorified space a `little' topos, and the category of spaces itself is the corresponding `big' topos; changes in spatial structure---witnessed by base-change between little topoi---thus correspond to trajectories within the space represented by the big topos.

Secondly, under the free energy principle, there is a close relationship between beliefs about the geometry of an environment and beliefs about expected future trajectories in that environment \parencite{Kaplan2018Planning}: fundamentally, this is also the idea underlying the ``successsor representation'' \parencite{Dayan1993Improving} of the cognitive map, which says roughly that the brain's representation of where it is is equivalently a representation of where it soon expects to be \parencite{Stachenfeld2014Design,Stachenfeld2016hippocampus,Brunec2019Predictive}.
Although there have been studies in the informal scientific literature attempting to connect free-energy models of navigation, exploration, and the cognitive map with the successor representation \parencite{Millidge2022Successor}, and to place both of these in less overtly spatial contexts \parencite{Brunec2019Predictive,Parr2022Active}, there has not yet been a comprehensive mathematical treatment explaining the structures that underlie this nexus.

By placing such a mathematical treatment in a topos-theoretic context, it may be possible to make sense of the ``logic of space'' of topoi to explain why animals' abstract problem-solving makes use of their abilities for spatial navigation: in particular, proving a proposition is mathematically analogous to finding a path from premise to conclusion.
Moreover, in a spatial topos, the ``truth values'' are no longer simply binary, but encode \textit{where} a proposition is (believed to be) true; the (sub)object classifier of a spatial topos encodes something like the ``internal universe'' of that topos, or ``the universe according to the system''.

To be successful, this mathematical treatment should be attentive to the results and proposals of computational and theoretical neuroscience, and so we now turn to our second key observation: the relationship between (believed) geometry and (expected) dynamics.
This will require an extension of statistical games and approximate inference to \textit{dynamical} generative models; until this point, our treatment has merely supplied inference (or `recognition' \parencite{Buckley2017free}) dynamics to static models.
Through this extension, we should expect a connection to other work on dynamical inference, such as filtering \parencite{Friston2010Generalised,Kalman1960New} and particularly its emerging compositional treatment \parencite{vanderMeulen2022Automatic,vanderMeulen2022Introduction}.

Under the free-energy principle, and similarly under the successor representation, the expected dynamics is a geodesic flow, which is by geodesy determined by beliefs about the spatial geometry.
But these beliefs in turn are not static: they depend on what the agent believes will happen \parencite{DaCosta2020Active,Millidge2020Whence}, and this has been suggested as an explanation for the `predictive' nature of the cognitive map \parencite{Kaplan2018Planning}.
The cognitive map has its central locus in the hippocampus \parencite{Epstein2017cognitive,OKeefe1976Place,Nadel1980Hippocampus}, which we may therefore understand as representing the base space over which the big topos is sliced; and since changes-of-plan seem therefore to induce changes-of-base, we might see the `functional' connectivity of the brain \parencite{Sporns2007Brain} as witnessing this mathematical structure.

Because the internal universe of the topos represented by the cognitive map is inherently context-dependent, it seems to fit naturally with the subjectivist metaphysics implied by the free energy framework---that the universe as experienced by an agent is a construction of that agent's internal model, as updated by approximate inference---and thus to provide a natural setting for the mathematical study of phenomenology.
Moreover, as categories of sheaves, agents' internal topoi encode the consensus formed by the distributed circuits and sensors that constitute their beliefs, and this points a way towards understanding how societies of agents are able to inhabit shared spaces about which they form a consensus between themselves: the mathematics of this shared universe should be little different from the mathematics of a single agent's internal universe.

Such multi-agent adaptive systems have been studied in the context of reinforcement learning (of which more below), but this potential for the formation of `nested' systems with shared universes implied by consensus is not the only connection between cognitive cartography and reinforcement learning, as it is in reinforcement learning that the successor representation originates.
We therefore hope that this line of enquiry may illuminate the relationship between reinforcement learning and compositional active inference, to the basis of which we now turn.

\section{Societies of systems} \label{sec:society}

Adaptive agents being necessarily in interaction with an external environment, we saw in the previous section how consideration of the compositional structure of agents' internal maps of their worlds easily leads to the consideration of societies of agents.
However, in order for us to study these, we first need to make the more basic step of incorporating action into the compositional framework: a collection of purely passive agents is no society.

\subsection{\textit{Active} inference} \label{sec:action}

The doctrines of approximate inference introduced in this thesis are inherently perceptual.
As we described in Remark \ref{rmk:cinema}, the forwards channel of a statistical game points ``towards the environment'', predicting the expected incoming sense-data, whereas the backwards channel points from the environment into the agent, terminating in the agent's most causally abstract beliefs.
In other contemporary work on categorical cybernetics, the orientation appears different: the forwards channel of an open (economic) game, for instance, points along the direction of interaction in the environment, in the direction of causality, from observations to actions \parencite{Ghani2016Compositional,Bolt2019Bayesian}; there is no room for prediction and its inversion, and the two kinds of game seem somehow perpendicular.

In resolution of this apparent disagreement, we can observe that an open economic game does have a perpendicular direction: a second\footnote{Or third, if one remembers the monoidal structure.} dimension inhabited by the strategies.
That is to say, an open economic game is a lens externally parameterized by strategies, a function from the latter to the former, and therefore formally much like our cilia (\secref{sec:mon-bicat-coalg}).
This resemblence becomes even closer when one considers the recent `diegetic' formulation of open games, in which strategies themselves can be updated using a backwards map from the arena of the game back into strategies (or rather, strategy updates).

This suggests one way in which we can incorporate action and thereby shape the framework of this thesis into a framework for \textit{active} inference:
the forwards channel should predict not only sense-data incoming from the environment, but also the actions to be taken by the agent.
Indeed this matches the usual informal presentation of active inference, which adopts a channel of the form $X\klto S\otimes A$, where $S$ is the space of sense-data, $A$ the space of possible actions, and $X$ the `latent' space.

Yet at this point the formal similarity between compositional active inference and compositional game theory again begins to recede, as a channel $X\klto S\otimes A$ is more like a ``stochastic span'' than an open economic game's player model $\Sigma\to[S,A]$.
Moreover, we expect our active inference systems to have a richer variety of patterns of interaction, being embodied in a world---in part, this motivated our adoption of polynomial functors for structuring interaction.
We therefore expect the compositional theory of active inference to have forwards channels rather of the form $X\klto\sum_{a:A}S[a]$, so that an agent's sensorium depends on the configuration (or `action') that it has chosen.

This was the approach we sketched in our earlier work-in-progress on \textit{Polynomial Life} \parencite{Smithe2021Polynomial}, where we suggested that polynomial functors supply a formalization of the notion of ``Markov blanket'' used in the informal active inference literature to characterize the interaction boundaries of adaptive systems \parencite{Palacios2017Biological,Kirchhoff2018Markov,Friston2019free} (a formalization that is situated at a useful level of technical flexibility, being neither as abstract as the very general notion of interface adopted by categorical systems theory \parencite{Myers2022Categorical}, nor as concrete as simple products of spaces).
In this way, we believe that a fruitful direction in which to pursue a compositional theory of active inference is, like our theory of open dynamical systems, as a $\Poly$-algebra of statistical games.
Fortunately, although the types prove somewhat different, the structural resemblence between active inference and economic games is maintained: in both cases, one has categories of lenses into the arena of interaction, indexed by a category of interfaces, and thus in philosophical (and thus we expect also mathematical) concordance with Myers' double-categorical view of systems theory \parencite{Myers2022Categorical}.

Once again, this line of enquiry naturally leads on to the consideration of multi-agent systems.
But before we come to that, there remain important questions about single-agent systems, and the connection between single-agent active inference and the cousin of economic games, reinforcement learning.

\subsection{What is the type of a plan?} \label{sec:planning}

Each active inference system has an internal `latent' state space equipped (by its generative model) with a prior distribution, which represents the systems's initial beliefs about the likelihood of those states.
As we have seen, the system can perceive, changing that distribution better to match incoming sense data.
And as we hope to see, it should also be able to act, affecting its environment so that future states better match its initial beliefs.
Perception and action are thus in general the two dual ways in which a system can minimize its free energy, akin to the two degrees of freedom available in a free energy game.

But a system that acts must necessarily be motivated towards some goal, even if that goal is simply ``stay alive'' or ``perform action $a$'', and even though this goal may be adjusted by the system's perceptions.
In order to realize its goal, whatever it may be, the system must enact a plan, however trivial---and the informal literature on active inference encodes the plan into the system's latent prior.
When it comes to static models, the prior may be simply a (static) distribution over the state space itself; but in the dynamical case, it is typically a distribution over \emph{trajectories} of states.

Such a distribution is often \parencite{DaCosta2020Active,Kaplan2018Planning} taken to encode likelihoods of hypothetical courses of action, which one might call a \emph{policy}\footnote{%
In the language of polynomial functors, this seems to be something like a distribution over the cofree comonad on the system's polynomial interface.}:
the system then perceives and acts in order to implement its policy.
But the construction of this policy may involve a lot of data, such as the specification of goal states and the accumulation of the ``expected free energy'' of trajectories in the context of those goals, and so it seems unnecessarily crude to hide all of this data inside a single undifferentiated choice of prior distribution.

This prompts us to ask, what is the form of this data, and how can we incorporate it into the compositional framework?
In other words, what is the type of a plan?
These seem to us to be key questions for future work.

\subsection{Reinforcement learning, open games, and ecosystems} \label{sec:rl}

There is known to be a close relationship between active inference in Markov decision problems (MDPs) and reinforcement learning \parencite{Costa2022Reward}, and it is through this relationship that one sees particularly clearly the strangeness of encoding all the data of an agent's policy in a single `prior' state.
This relationship is seemingly not superficial, as there are hints of a deep structural connection.

First, recall that the standard algorithm for obtaining a Bellman-optimal policy for an MDP is \emph{backward induction} (otherwise known as \emph{dynamic programming}) \parencite{Puterman1994Markov,Zermelo1913Uber}\footnote{%
Also see \parencite{Sutton2018Reinforcement,Blumensath2017Compositional,Hedges2022Value,Friston2009Reinforcement,Escardo2010Selection} for other presentations.}.
It is now known that backward induction is structured according a similar bidirectional pattern (the \emph{optic} pattern) to that of both Bayesian inference and reverse differentiation \parencite{Hedges2022Value}, and that MDPs themselves fit into the associated general framework of open games \parencite{Bolt2019Bayesian} (which are governed by the same pattern).
Second, in the informal active inference approach to MDPs, the system in question counterfactually evaluates policies using a backward-induction-like process, accumulating free energies in order to score them \parencite{Costa2022Reward}.
It is this process that results in the prior discussed above, which is then updated by the agent's inference process.
Future work will need to untangle this knot of interrelated bidirectional processes; and as usual in categorical modelling, this means first writing them all down precisely.
We hope that, having done so, we will see how the whole picture emerges, and how it relates to the developing geometric (or `diegetic') framework in categorical cybernetics \parencite{Capucci2022Diegetic} (possibly involving the further development of our notion of `cilia' from \secref{sec:mon-bicat-coalg}).
In particular, since the free energy principle underlying active inference asserts a certain informal universality (on which more in \secref{sec:bayes-mech}), we might also hope that the satisfactory development of compositional active inference might exhibit a universal property: that any other doctrine of cybernetic systems factors uniquely through it.

The story of these connections will initially be told from the perspective of a single agent, as backward induction only considers how to find a single policy for a single MDP; although this policy may involve multiple agents, the implied global search entails a common controller: the procedure doesn't consider the factorisation of the agents.
But casting this account into the emerging framework of compositional active inference will point towards a bridge to multi-agent reinforcement learning.
For example, multi-agent RL often studies the emergence of collaboration, and we might expect to see this represented in the formal structure, thereby understanding how to incorporate the factorisation of agents into the compositional framework for backward induction (which in turn may be helpful for designing collaborative `edge' AI systems).

The resulting general account of multi-agent intelligence will encompass both reinforcement learning and active inference, allowing us to understand their relative strengths and differences.
One seeming difference (at this early stage, and following our thinking above) is that compositional active inference envisages the latent state spaces of agents as their ``internal universes'', which come along with sufficient structure that we might consider them as Umwelten (\emph{i.e.}, their subjective worlds, in the sense of biosemiotics; see \secref{sec:biosemiot} below).
Consequently, we should be able to study how agents might come to consensus, thereby resolving their disagreements.
And because agents are embodied in a shared world within which they act, this process might involve planning cooperation, at which point the teleological structure of compositional game theory may become important, as cooperating agents will have to bet on spatiotemporally distributed actions.
We hope therefore that one distal outcome of this work will be a new and beneficial understanding of corporate activity.

Below, in \secref{sec:life}, we will discuss how active inference and the free energy principle aim not only to be theories of brains or other prominent intelligent systems, but rather universal theories of all adaptive things.
Consequently, their compositional treatment should extend in the `multi-agent' case not just to corporate activity, but to ecosystems more broadly.
And, following the multicategorical algebra latent throughout this thesis, it will undoubtedly prove natural, once we have considered a single level of nesting of systems into ecosystems, to let the hierarchy continue to infinity, producing a fractal-like structure.
At this point, we should expect once more to make contact with topics such as higher categories and type theory, particularly in operadic or opetopic (\textit{i.e.}, `directed') forms; synthetic approaches to mathematical physics; and iterated parameterization in categorical systems theory.

It almost goes without saying that we should expect any framework resulting from this work to capture existing models of collective active inference, such as recent work on spin glasses \parencite{Heins2022Spin}.

\section{The mathematics of life} \label{sec:life}

We move on to consider the relationships between compositional active inference and the contemporary mathematics of life.
We hope that compositional active inference may supply part of the story of a modern theory of autopoiesis, the ability for life to recreate itself \parencite{Varela1991Autopoiesis}.

\subsection{Bayesian mechanics and the free energy principle} \label{sec:bayes-mech}

Recently, it has been suggested in various venues \parencite{Friston2019free,Parr2019Markov} that the free energy framework provides a `universal' way to understand the behaviour of adaptive systems, in the sense that, given a random dynamical system, it may be possible to write down a generative model such that the dynamics of the system can be modeled as performing inference on this model.
In the language of the conjectured compositional framework for active inference, we may be able to describe a canonical statistical game that each given random dynamical system can be seen as playing.

If this is true, we should be able to express this canonicity precisely: in particular, it should correspond to a \emph{universal property}.
Since approximate inference doctrines already gives us functorial ways to turn statistical games into dynamical systems, this suggests we should seek functors that associate to each random dynamical system a statistical game; and we should expect these functors to be adjoint (as morphisms of categories indexed by the systems' interfaces).
The desired universal property would then be expressed by the adjunction.
(Notably, adjunctions are at the heart of recent synthetic approaches to mathematical physics \parencite{Schreiber2013Differential}.)
This would constitute an important mathematical step to establishing the universality of the free energy principle, or to establishing the conditions that must be satisfied by any satisfactory successor.

\textit{Bayesian mechanics} promises to build upon the nascent understanding of random dynamics via inference \parencite{Sakthivadivel2022Towards} to supply a new theory of mechanics for statistical systems \parencite{Ramstead2022Bayesian}.
The present formulation of Bayesian mechanics is constructed using mathematical tools from physics, but not (yet) the kinds of compositional tool promoted in this thesis and further described above.
We expect that developments along the lines sketched here will unify the on-going development of Bayesian mechanics (and the resulting understanding of non-equilibrium systems) with the new synthetic understanding of mathematical physics.
By casting all dynamics as abstract inference, we should also expect this line of enquiry to begin to quantify the persistence of things and imbue much of physics with an \textit{élan vital}.

\subsection{Biosemiotics} \label{sec:biosemiot}

It is increasingly acknowledged that biological systems are characterized not only by information-processing, but by \emph{communication} \parencite{Barbieri2007Introduction}:
an often overlooked fact about `information' in the strict mathematical sense is that it is only meaningful in context.
In the original Nyquist-Hartley-Shannon conception of information, this context is the communication of a predefined message over a noisy channel \parencite{Nyquist1928Certain,Hartley1928Transmission,Shannon1948Mathematical}; but more generally, we might think of this context as simply ``a question'', in which case it is easy to see that information answering one question may not be useful in answering another; or, in a more computational setting, we can see that the bits of an encrypted signal are only useful in revealing the message if one has the decryption key.

Still, one often encounters descriptions of signals as containing \(n\) bits of information, without a clear specification of \emph{about what}.
Mathematically, the confusion arises because information theory is framed by classical probability, and the assumed context is always the problem of trying to communicate a probability distribution over a pre-defined space \(X\); and once the space is fixed, the only question that can be asked is ``what is the distribution?'' (Mathematically, this is to say that in the Markov category of classical stochastic channels, there are no non-trivial effects or costates.)

Yet, in the shared universe that we inhabit, there are more questions than this: in quantum theory, for instance, one can ask many questions of the state of a system, by projecting the state onto the subspace representing one's question. (These projections are the non-trivial effects or costates of quantum probability.)
This act of projection is an act of \emph{interpretation} of the message encoded by the state at hand.

The emerging `biosemiotic' reconceptualization of life explicitly acknowledges the importance and universality of communication in context \parencite{Barbieri2007Introduction}, proposing that in any such situation the interpreting system necessarily has an internal representation of the external world (its \emph{Umwelt}) which is updated by interpreting incoming signals.
We can in principle reconstruct the external world by understanding it as ``that which a collection of systems agrees about'': perhaps, then, the shared universe (as determined topos-theoretically) of a fusion of active inference agents is a good model of this `semiosphere'.
It seems therefore that the mathematics resulting from our work on internal universes and their interactions --- and, more broadly, many of the formal ingredients of compositional active inference --- is well aligned with the informal structures of biosemiotics, and so it may be desirable to re-express biosemiotics accordingly.
In doing so, perhaps the mathematics for a modern Bayesian subjectivist metaphysics will be found\footnote{%
Perhaps getting to the structural heart of the theory known as \textit{QBism} \parencite{Fuchs2013Introduction,Fuchs2017Notwithstanding}.}:
for instance, by expressing communication and its phenomenology as a geometric morphism (a generalized base-change) between agents' internal universes.
More pragmatically, perhaps we will be able to say precisely when some object may act as a symbol, and how systems may (learn to) manipulate such symbols.

\section{Fundamental theory} \label{sec:theory}

Future work connected to this thesis need not only be in applications; a number of purely theoretical questions raise themselves, too.

\subsection{Geometric methods for (structured) belief updating}

The mathematics of `belief' is in large part about replacing definite points with `fuzzier' distributions over them.
In dependent type theory, we replace points with `terms' (non-dependent terms are exactly points): so a type theory with belief should somehow encompass ``fuzzy terms''.
Just as we can replace points with distributions, we can replace dependent points with dependent distributions.
However, the standard replacement (moving from a category of functions to a category of stochastic channels) obscures some of the `universal' categorical structure that underpins the rules of type theory.
This standard replacement also misses something else: while it does allow for fuzzy terms, it omits a model of fuzzy \emph{types}; and we might well want to express beliefs about things whose identity we are not quite sure.
(This omission also seems to be related to the loss of universal structure.)

There seem to be a couple of related resolutions to this puzzle.
The first is to notice that replacing points by distributions yields another space: the space of distributions over the original space; this is akin to the move in dynamics from working with the random motion of states to working with the deterministic motion of the distribution over states.
This space of distributions has a particular geometry (its \emph{information geometry}), and hence we should expect corresponding flavours of topos and type theory.
As we have indicated above, there is a move in fundamental mathematical physics (\emph{cf.} \textcite{Schreiber2013Differential}) to work `synthetically', expressing concepts using the universal structures of higher topoi.
This has proven particularly fruitful in the context of differential systems, but it is interesting that stochastic and differential structures bear a number of similarities\footnote{%
Both conditional probability and differential calculus exhibit ``chain rules'' of similar types, which give rise to backwards actions that compose via the lens rule: in the former case, Bayesian inversion; in the latter, reverse differentiation.
Categories that admit a differentiation operation have begun to be axiomatized (as differential categories \parencite{Blute2006Differential} and reverse-derivative categories \parencite{Cockett2020Reverse}), and categories whose morphisms behave like stochastic channels are also presently being axiomatized (in the framework of Markov categories \parencite{Fritz2019synthetic}), but the connections between these various formalisms are not yet clear.
The similar structures indicate that the two families of axiomatisation may have a common generalization.}:
what are we to make of this?
Does Bayesian inversion induce a canonical geometric morphism, by which structured models may be coherently updated?
We have already indicated above signs of a relationship between inference and parallel transport; it seems that it may at least be fruitful to consider `metric' topoi, appropriately enriched.

The second resolution is to work with topoi as we work with random dynamical systems, by noticing that randomness is often like ``an uncertain parameterization''.
By parameterizing a topos with a category of noise sources, we may obtain a notion of ``stochastic topos'' in which the standard operations of dependent type theory are available, but where each type and term may depend on the realization of the noise source, thereby giving us notions of fuzzy term and fuzzy type.
The mathematics of such uncertainly parameterized topoi is as yet undeveloped, although we expect that they should bear a relationship to the ``topoi of beliefs'' of the foregoing first resolution similar to the relationship of Fokker-Planck to random dynamical systems.

Finally, we note that higher topoi behave abstractly somewhat like vector spaces (with sheaves like categorified functionals).
Since distributions are themselves like vectors, perhaps this observation is a first step towards relating the resolutions.

\subsection{Dynamics}

Chapter \ref{chp:coalg} has supplied the beginnings of a compositional coalgebraic theory for open stochastic and random dynamical systems in general time, and we hope that this theory could provide a home for a modern account of non-equilibrium systems, with the category of polynomial functors supplying a satisfactory account of these systems' interfaces (\emph{i.e.}, the boundaries across which information flows, along which they compose, and through which they interact).

In this context, and in parallel to the abstract questions above, there are similar questions to be asked specifically of dynamical systems.
For instance, what is the precise relationship between the category of Markov processes on an interface, and the category of random dynamical systems on that interface?
We know that categories of deterministic discrete-time polynomial coalgebras are topoi \parencite{Spivak2021Learners}, so does the same hold in general time?
To what extent is the logic of our systems related to coalgebraic logics \parencite{Corfield2009Coalgebraic,Kurz2006Logic,Venema2006Automata,Pavlovic2006Testing,Jacobs2017Introduction}?

Besides these `parallel' questions, there are a number of more technical ones.
For instance, our current definition of ``Markov process on a polynomial interface'' is somewhat inelegant, and we seek to simplify it.
Similarly, we believe that there is a better definition of ``random dynamical system on a polynomial interface'' that may be obtained by a (different) generalization of the category of polynomial functors, using random variables.
And we know that a topology for the cofree comonoid on an interface can be generated by the corresponding free monoid, which may be relevant for understanding the topological structure of open systems.
An important set of open questions about open random dynamical systems in this framework comes from attempting to import notions about random systems from the classical `closed' setting: fundamentally, we ask, does this framework indeed supply a satisfactory setting in which to understand stochastic systems away from equilibrium?

\subsection{Computation}

The early 21\textsuperscript{st} century understanding of biological systems as information-processing involves treating them as \emph{computational}, but remarkably lacks a precise concept of what it means for a system to compute, other than in the context of artificial machines.
To us, it seems that a crisper understanding of computation in general might begin with the slogan that ``computation is dynamics plus semantics'', which is philosophically aligned with the semiotic understanding of biological information-processing sketched above: for example, we know that attractor networks in the brain can informally be understood as computational \parencite{Amit1992Modeling}, but these are `continuous' systems for which we do not yet have a good corresponding concept of \emph{algorithm} (and it is upon algorithms that our current understanding is built).
But what more is an algorithm than a description of a discrete-time open dynamical system?
The quality that makes an algorithm computational is that its states or its outputs correspond to some quantity of interest, and that it reaches a fixed point (it halts) at the target quantity when the computation is complete.
If this intuition is correct, then a new understanding of computation may follow the semiotic understanding of information-processing that we propose above: perhaps we could say more precisely that computation is the dynamics of semiosis?
The time is right for such a reconceptualization, as human-made systems increasingly move away from von Neumann architectures towards more biosimilar ones (such as memristors, optical processors, neuromorphic technology, graph processors, or even many-core and mesh-based evolutions of classical processors).

\appendix
\chapter{Auxiliary material}

\section{From monads to multicategories} \label{sec:a-monad-operad}

The assignment of domain and codomain to the morphisms of a small category $\cat{C}$ constitutes a pair of functions $\cat{C}_1\to\cat{C}_0$, which we can write as a \textit{span}, $\cat{C}_0\xfrom{\cod}\cat{C}_1\xto{\dom}\cat{C}_0$.
Similarly, the assignment of domain and codomain to the morphisms of a multicategory $\cat{M}$ constitutes a span $\cat{M}_0\xfrom{\cod}\cat{M}_1\xto{\dom}\List(\cat{M}_0)$.
This observation was used by \textcite{Leinster2004Higher} to construct a general framework for constructing multicategories, replacing $\List$ with an arbitrary `Cartesian' monad $T$, which opens the way to a connection between monad algebras and multicategory algebras.
In this section, we explore this connection, starting by defining categories of spans.

\begin{defn}
  Suppose $A$ and $B$ are two objects in a category $\cat{C}$.
  We will write a \textit{span} from $A$ to $B$ as $(X,x):A\xleftarrow{x_A}X\xrightarrow{x_B}B$, and call $X$ the \textit{apex} of the span and $x_A,x_B$ its \textit{legs} or \textit{projections}.
  The \textit{category of spans from} $A$ \textit{to} $B$, denoted $\Span(A,B)$ has spans $(X,x)$ as objects, and the morphisms $f:(X,x)\to(X',x')$ are morphisms $f:X\to X'$ in $\cat{C}$ that commute with the spans, as in the following diagram:
  
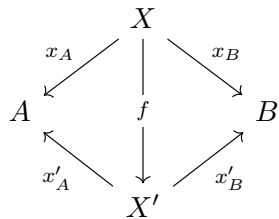
\[\begin{tikzcd}
	  & X \\
    A && B \\
	  & X'
	  \arrow["{x_A}"', from=1-2, to=2-1]
	  \arrow["{x'_A}", from=3-2, to=2-1]
	  \arrow["{x_B}", from=1-2, to=2-3]
	  \arrow["{x'_B}"', from=3-2, to=2-3]
	  \arrow["f"{description}, from=1-2, to=3-2]
  \end{tikzcd}\]
\end{defn}

We can treat the categories $\Span(A,B)$ as the hom categories of a bicategory.

\begin{defn}
  Suppose $\cat{C}$ is a category with all pullbacks.
  The \textit{bicategory of spans} in $\cat{C}$, denoted $\Span$, has for objects the objects of $\cat{C}$, and for hom-categories the categories $\Span(A,B)$ of spans from $A$ to $B$.
  Given spans $(X,x):A\to B$ and $(Y,y):B\to C$, their horizontal composite $(Y,y)\circ(X,x):A\to C$ is the pullback span defined by
  
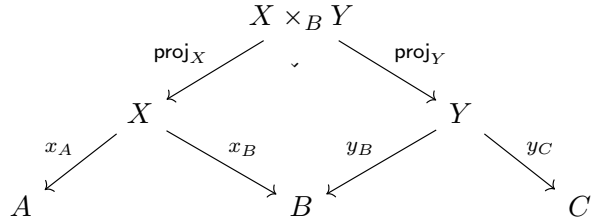
\[\begin{tikzcd}
    && {X\times_BY} \\
    & X && Y \\
    A && B && C
	  \arrow["{\proj_X}"', from=1-3, to=2-2]
	  \arrow["{x_A}"', from=2-2, to=3-1]
	  \arrow["{\proj_Y}", from=1-3, to=2-4]
	  \arrow["{y_C}", from=2-4, to=3-5]
	  \arrow["{x_B}", from=2-2, to=3-3]
	  \arrow["{y_B}"', from=2-4, to=3-3]
	  \arrow["\lrcorner"{anchor=center, pos=0.125, rotate=-45}, draw=none, from=1-3, to=3-3]
  \end{tikzcd} \, .\]
  If $(X',x'):A\to B$ and $(Y',y'):B\to C$ are also spans with $f:(X,x)\Rightarrow(X',x')$ and $g:(Y,y)\Rightarrow(Y',y')$ vertical morphisms, the horizontal composite of $f$ and $g$ is also defined by pullback as $f\times_B g:(Y,y)\circ(X,x)\Rightarrow(Y',y')\circ(X',x')$.
  The identity span on an object $A$ is $(A,\id):A\xlongequal{}A\xlongequal{}A$.

  If the ambient category $\cat{C}$ is not clear from the context, we will write $\Span_{\cat{C}}$ to denote the bicategory of spans in $\cat{C}$.
\end{defn}

\begin{rmk}
  Note that $\Span$ really is a bicategory rather than a 2-category: since the horizontal composition of spans is defined by pullback, it is only defined up to isomorphism.
  Consequently, the composition of spans can in general only be associative and unital up to isomorphism, rather than the strict equality required by a 2-category.
\end{rmk}

Now, recall that `monad' is another name for ``monoid in a bicategory'', where the bicategory has so far been taken to be $\Cat{Cat}$: but it need not be.

\begin{rmk} \label{rmk:3-monad-bicat}
  Since $\cat{C}^{\cat{C}}$ is the endomorphism monoid on $\cat{C}$ in the bicategory $\Cat{Cat}$, we can generalize the preceding definition of monad to any bicategory $\cat{B}$: a \textit{monad in a bicategory} $\cat{B}$ is simply a monoid in $\cat{B}$, as defined in Remark \ref{rmk:monoid-bicat}.
  That is, a monad in $\cat{B}$ is a monoid object in the monoidal category $\bigl(\cat{B}(b,b),\circ,\id_b\bigr)$ for some choice of 0-cell $b:\cat{B}$, where $\circ$ denotes horizontal composition.
  Explicitly, a monad $(t,\mu,\eta)$ in $\cat{B}$ is a 1-cell $t:b\to b$, a multiplication 2-cell $\mu:t\circ t\Rightarrow t$, and a unit 2-cell $\eta:\id_b\Rightarrow t$, such that the associativity and unitality diagrams commute in $\cat{B}(b,b)$:
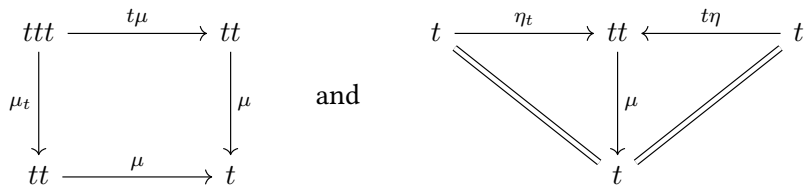
  \[
  \begin{tikzcd}
	  ttt && tt \\
	  \\
	  tt && t
	  \arrow["{\mu_t}"', from=1-1, to=3-1]
	  \arrow["{t\mu}", from=1-1, to=1-3]
	  \arrow["{\mu}", from=1-3, to=3-3]
	  \arrow["{\mu}", from=3-1, to=3-3]
  \end{tikzcd}
  \qquad\text{and}\qquad
  \begin{tikzcd}
	  t && tt && t \\
	  \\
	  && t
	  \arrow["{\mu}", from=1-3, to=3-3]
	  \arrow["{\eta_t}", from=1-1, to=1-3]
	  \arrow["{t\eta}"', from=1-5, to=1-3]
	  \arrow[Rightarrow, no head, from=1-1, to=3-3]
	  \arrow[Rightarrow, no head, from=1-5, to=3-3]
  \end{tikzcd}
  \]
\end{rmk}

With this more general notion of monad, we obtain another monadic definition of ``small category'', to add to the explicit Definition \ref{def:small-cat} and the monad-algebraic Example \ref{ex:path-algebras}.

\begin{prop}
  Small categories are monads in $\Span_{\Set}$.
  \begin{proof}[Proof]
    A monad in $\Span_{\Set}$ is a choice of object $\cat{C}_0$ and monoid in $\Span_{\Set}(\cat{C}_0,\cat{C}_0)$.
    Such a monoid is a span of sets $\cat{C}:\cat{C}_0\xleftarrow{\cod}\cat{C}_1\xrightarrow{\dom}\cat{C}_0$ along with functions $\bullet:\cat{C}_1\times_{\cat{C}_0}\cat{C}_1\to\cat{C}_1$ and $\id:\cat{C}_0\to\cat{C}_1$.
    The set $\cat{C}_1\times_{\cat{C}_0}\cat{C}_1$ is the apex of the pullback span $\cat{C}\circ\cat{C}$ as in
    
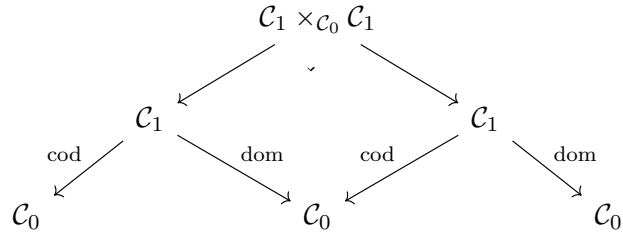
\[\begin{tikzcd}
	    && {\cat{C}_1\times_{\cat{C}_0}\cat{C}_1} \\
	    & {\cat{C}_1} && {\cat{C}_1} \\
	    {\cat{C}_0} && {\cat{C}_0} && {\cat{C}_0}
	    \arrow[from=1-3, to=2-2]
	    \arrow["\cod"', from=2-2, to=3-1]
	    \arrow["\dom", from=2-2, to=3-3]
	    \arrow[from=1-3, to=2-4]
	    \arrow["\dom", from=2-4, to=3-5]
	    \arrow["\cod"', from=2-4, to=3-3]
	    \arrow["\lrcorner"{anchor=center, pos=0.125, rotate=-45}, draw=none, from=1-3, to=3-3]
    \end{tikzcd}\]
    so that $\bullet$ and $\id$ make the following diagrams commute:
    \[\begin{tikzcd}
      & {\cat{C}_1\times_{\cat{C}_0}\cat{C}_1} \\
      {\cat{C}_1} && {\cat{C}_1} \\
	    {\cat{C}_0} & {\cat{C}_1} & {\cat{C}_0}
	    \arrow[from=1-2, to=2-1]
	    \arrow["\cod"', from=2-1, to=3-1]
	    \arrow[from=1-2, to=2-3]
	    \arrow["\dom", from=2-3, to=3-3]
	    \arrow["\cod", from=3-2, to=3-1]
	    \arrow["\dom"', from=3-2, to=3-3]
	    \arrow["\bullet"{description}, from=1-2, to=3-2]
    \end{tikzcd}
    \quad\text{and}\quad
    \begin{tikzcd}
	    && {\cat{C}_0} \\
	    \\
      {\cat{C}_0} && {\cat{C}_1} && {\cat{C}_0}
	    \arrow["\id"{description}, from=1-3, to=3-3]
	    \arrow[Rightarrow, no head, from=1-3, to=3-1]
	    \arrow[Rightarrow, no head, from=1-3, to=3-5]
	    \arrow["\cod", from=3-3, to=3-1]
	    \arrow["\dom"', from=3-3, to=3-5]
    \end{tikzcd}\]
    This means that $\cod(g\bullet f) = \cod(g)$ and $\dom(g\bullet f) = \dom(f)$, and $\cod(\id_x) = \dom(\id_x) = x$.
    It is easy to check that $(\cat{C},\bullet,\id)$ therefore constitutes the data of a small category; moreover, the functions $\bullet$ and $\id$ must satisfy the monoid axioms of associativity and (right and left) unitality, which correspond directly to the categorical axioms of associativity and unitality.
  \end{proof}
\end{prop}

As we indicated at the opening of this section, by generalizing to a category of `spans' of the form $A\from X\to TB$, we can use the preceding result to produce generalized multicategories whose morphisms have domains ``in the shape of $T$''.
Since the horizontal composition of spans is by pullback, we need an extra condition on the monad $T$ to ensure that pullbacks of $T$-spans are well-defined.
This condition is known as `Cartesianness'.

\begin{defn}
  A \textit{Cartesian natural transformation} between functors $F$ and $G$ is a natural transformation $\alpha:F\Rightarrow G$ for which every naturality square is a pullback:
  
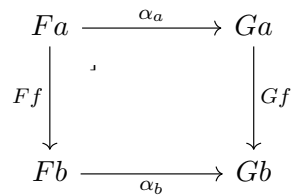
\[\begin{tikzcd}
    Fa && Ga \\
    \\
    Fb && Gb
	  \arrow["{\alpha_a}", from=1-1, to=1-3]
	  \arrow["Gf", from=1-3, to=3-3]
	  \arrow["Ff"', from=1-1, to=3-1]
	  \arrow["{\alpha_b}"', from=3-1, to=3-3]
	  \arrow["\lrcorner"{anchor=center, pos=0.125}, draw=none, from=1-1, to=3-3]
  \end{tikzcd}\]
  A \textit{Cartesian monad} is a monad $(T:\cat{C}\to\cat{C},\mu,\eta)$ such that $\cat{C}$ has all pullbacks, $T$ preserves these pullbacks (sending pullback squares to pullback squares), and $\mu$ and $\eta$ are Cartesian natural transformations.
\end{defn}

\begin{defn}
  Suppose $T$ is a monad on $\cat{C}$. A $T$\textit{-span} from $A$ to $B$ is a span from $A$ to $TB$ in $\cat{C}$.
  The \textit{category of} $T$\textit{-spans} from $A$ to $B$, denoted $\Span_T(A,B)$ has $T$-spans as objects, and the morphisms $f:(X,x)\to(X',x')$ are morphisms $f:X\to X'$ in $\cat{C}$ that commute with the spans, as in the diagram
  
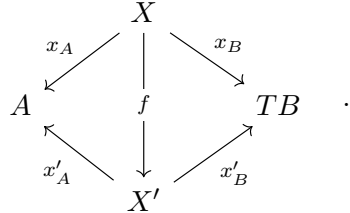
\[\begin{tikzcd}
	  & X \\
    A && TB \\
	  & X'
	  \arrow["{x_A}"', from=1-2, to=2-1]
	  \arrow["{x'_A}", from=3-2, to=2-1]
	  \arrow["{x_B}", from=1-2, to=2-3]
	  \arrow["{x'_B}"', from=3-2, to=2-3]
	  \arrow["f"{description}, from=1-2, to=3-2]
  \end{tikzcd} \quad .\]
\end{defn}

\begin{defn}
  Suppose $(T,\mu,\eta)$ is a Cartesian monad on $\cat{C}$.
  The \textit{bicategory of} $T$\textit{-spans} in $\cat{C}$, denoted $\Span_T$, has for objects the objects of $\cat{C}$, and for hom-categories the categories $\Span_T(A,B)$ of $T$-spans from $A$ to $B$.
  Given $T$-spans $(X,x):A\to B$ and $(Y,y):B\to C$, their horizontal composite $(Y,y)\circ(X,x):A\to C$ is the outer $T$-span in the diagram
  
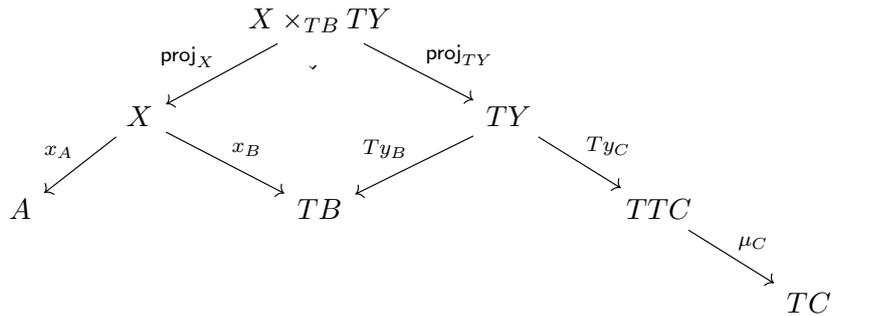
\[\begin{tikzcd}
	  && {X\times_{TB}TY} \\
	  & X && TY \\
	  A && TB && TTC \\
	  &&&&& TC
	  \arrow["{\proj_X}"', from=1-3, to=2-2]
	  \arrow["{x_A}"', from=2-2, to=3-1]
	  \arrow["{\proj_{TY}}", from=1-3, to=2-4]
	  \arrow["{Ty_C}", from=2-4, to=3-5]
	  \arrow["{x_B}", from=2-2, to=3-3]
	  \arrow["{Ty_B}"', from=2-4, to=3-3]
	  \arrow["\lrcorner"{anchor=center, pos=0.125, rotate=-45}, draw=none, from=1-3, to=3-3]
	  \arrow["{\mu_C}", from=3-5, to=4-6]
  \end{tikzcd} \quad . \]
  If $(X',x'):A\to B$ and $(Y',y'):B\to C$ are also $T$-spans with $f:(X,x)\Rightarrow(X',x')$ and $g:(Y,y)\Rightarrow(Y',y')$ vertical morphisms, the horizontal composite of $f$ and $g$ is defined as $f\times_{TB}Tg$ accordingly.
  The identity span on an object $A$ is $A\xleftarrow{\id_A}A\xrightarrow{\eta_A}TA$.
\end{defn}

With these notions to hand, the general concept of $T$-multicategory is easy to define.

\begin{defn}[{\textcite[{Def. 4.2.2}]{Leinster2004Higher}}]
  Suppose $T$ is a Cartesian monad on $\cat{C}$.
  A $T$\textit{-multicategory} is a monad in the bicategory $\Span_T$ of $T$-spans.
\end{defn}

And of course we can recover our earlier examples of category shapes accordingly.

\begin{ex}
  The identity monad on a category with all pullbacks is trivially a Cartesian monad.
  Therefore, taking $T = \id_{\Set}$ to be the identity monad on $\Set$, we immediately see that an $\id_{\Set}$-multicategory is a small category.
\end{ex}

\begin{ex}[{\textcite[{Examples 4.1.4 and 4.2.7}]{Leinster2004Higher}}]
  The free monoid monad $\List:\Set\to\Set$ is Cartesian.
  Unpacking the definitions, we find that a $\List$-multicategory is precisely a multicategory as in Definition \ref{def:multicat}.
\end{ex}

At this point, we can sketch how multicategory algebras correspond to monad algebras, referring the reader to \textcite[{\S4.3}]{Leinster2004Higher} for the details.
The basic picture is that, if $T:\cat{C}\to\cat{C}$ is a Cartesian monad and $\cat{M}$ is a $T$-multicategory, then one can obtain functorially a monad $T_{\cat{M}}$ on the slice $\cat{C}/\cat{M}_0$ of $\cat{C}$ over the object $\cat{M}_0$ of $\cat{M}$-objects.
The algebras $\alpha:T_{\cat{M}}(X,p)\to(X,p)$ of this monad are morphisms $\alpha:T_{\cat{M}}X\to X$ as in the commuting diagram
\[\begin{tikzcd}
	& {T_{\cat{M}}X} && X \\
	TX && {\cat{M}_1} \\
	& {T\cat{M}_0} && {\cat{M}_0}
	\arrow["p", from=1-4, to=3-4]
	\arrow["\cod"', from=2-3, to=3-4]
	\arrow["\dom", from=2-3, to=3-2]
	\arrow["Tp"', from=2-1, to=3-2]
	\arrow[from=1-2, to=2-1]
	\arrow[from=1-2, to=2-3]
	\arrow["\lrcorner"{anchor=center, pos=0.125, rotate=-45}, draw=none, from=1-2, to=3-2]
	\arrow["\alpha", from=1-2, to=1-4]
\end{tikzcd}\]
where $T_{\cat{M}}(X,p)$ is defined as the bundle $T_{\cat{M}}X\to\cat{M}_0$ on the right leg of the pullback square.
To get a sense for how this works, consider the case where $T=\id_{\Set}$: a $T$-multicategory is then simply a small category $\cat{C}$, and as \textcite[{Example 4.3.2}]{Leinster2004Higher} shows, its algebras are functors $\cat{C}\to\Set$.

%
%
%
%

%

%
%
%
%
%
%

\chapter{Bibliography}
\printbibliography[heading=none]

\end{document}